\documentclass[11pt]{article}

\newif\ifheader 
\headerfalse

\makeatletter
\g@addto@macro\bfseries{\boldmath}
\makeatother

\usepackage[table,usenames,dvipsnames]{xcolor}
\usepackage{amssymb,amsmath,amsfonts}
\usepackage{times}
\usepackage{fancyhdr}
\usepackage{lastpage}
\usepackage[colorlinks=true, breaklinks=true, urlcolor  = blue, linkcolor = blue, citecolor=blue]{hyperref}
\usepackage{graphicx}
\usepackage{floatflt}
\usepackage{lscape}
\usepackage{pgfgantt}
\usepackage[normalem]{ulem} 
\usepackage{etaremune}
\usepackage{tikz}
\usepackage[gen]{eurosym}
\usepackage{multicol}										
\usepackage{paralist}
\usepackage{wrapfig}
\usepackage{color,soul}
\usepackage{tabularx}
\usepackage{array}
\usepackage{bm}
\usepackage{multirow}

\usepackage{tikz}

\usepackage{scalerel}
\usepackage{stackengine,wasysym}
\newcommand\reallywidetilde[1]{\ThisStyle{%
  \setbox0=\hbox{$\SavedStyle#1$}%
  \stackengine{-.1\LMpt}{$\SavedStyle#1$}{%
    \stretchto{\scaleto{\SavedStyle\mkern.2mu\AC}{.5150\wd0}}{.6\ht0}%
  }{O}{c}{F}{T}{S}%
}}

\makeatletter
\newcommand{\setword}[2]{%
  \phantomsection
  #1\def\@currentlabel{\unexpanded{#1}}\label{#2}%
}
\makeatother

\newcommand{\MYhref}[3][violet]{\href{#2}{\color{#1}{#3}}}%

\newcommand{\ie}{\textsl{i.e.~}}

\newcommand{\eg}{\textsl{e.g.~}}

\newcommand{\etc}{\textsl{etc.~}}

\newcommand{\SR}{{{}_\mathrm{SR}}}
\newcommand{\USR}{{{}_\mathrm{USR}}}

\newcommand{\E}{E_{\mathrm{B}}}

\newcommand{\phiend}{\phi_{\text{end}}}
\newcommand{\phiwell}{\phi_{\text{well}}}		
\newcommand{\phiuv}{\phi_{\text{uv}}}			

\newcommand{\N}{\mathcal{N}}
\newcommand{\zetacg}{\zeta_{\mathrm{cg}}}
\newcommand{\zetac}{\zeta_{\mathrm{c}}}
\newcommand{\dphiwell}{\Delta\phi_\mathrm{well}}

\let\oldsqrt\sqrt
\def\sqrt{\mathpalette\DHLhksqrt}
\def\DHLhksqrt#1#2{%
\setbox0=\hbox{$#1\oldsqrt{#2\,}$}\dimen0=\ht0
\advance\dimen0-0.2\ht0
\setbox2=\hbox{\vrule height\ht0 depth -\dimen0}%
{\box0\lower0.4pt\box2}}

\newcommand{\order}[1]{\mathcal{O}\!\left(#1\right)}

\DeclareMathOperator{\Ei}{Ei}

\DeclareMathOperator{\erf}{erf}
\DeclareMathOperator{\erfc}{erfc}
\newcommand{\tr}{\mathrm{Tr}}

\newcommand{\dd}{\mathrm{d}}
\newcommand{\ee}{e}

\newcommand{\sss}[1]{{\scriptscriptstyle{#1}}}
\newcommand{\boldmathsymbol}[1]{{\ensuremath{\boldsymbol{#1}}}}

\newcommand{\uPl}{\mathrm{Pl}}
\newcommand{\uin}{\mathrm{in}}

\newcommand{\umax}{\mathrm{max}}
\newcommand{\uend}{\mathrm{end}}

\newcommand{\ueff}{\mathrm{eff}}

\newcommand{\ucl}{\mathrm{cl}}

\newcommand{\uc}{\mathrm{c}}

\newcommand{\uS}{\mathrm{S}}

\newcommand{\usssS}{\sss{\uS}}

\newcommand{\lo}{\sss{\mathrm{LO}}}
\newcommand{\nlo}{\sss{\mathrm{NLO}}}
\newcommand{\nnlo}{\sss{\mathrm{NNLO}}}

\newcommand{\usssPl}{\sss{\uPl}}

\newcommand{\nS}{n_\usssS}

\newcommand{\alphaS}{\alpha_\usssS}

\newcommand{\uNL}{\mathrm{NL}}

\newcommand{\calH}{\mathcal{H}}
\newcommand{\calP}{\mathcal{P}}

\newcommand{\Mp}{M_\usssPl}

\newcommand{\fnl}{f_\uNL}
\newcommand{\gnl}{g_\uNL}

\newcommand{\efolds}{$e$-folds}
\newcommand{\efold}{$e$-fold}

\newcommand{\beq}{\begin{equation}}
\newcommand{\eeq}{\end{equation}}
\newcommand{\bea}{\begin{eqnarray}}
\newcommand{\eea}{\end{eqnarray}}
\newcommand{\lp}{\left(}
\newcommand{\rp}{\right)}
\newcommand{\lb}{\left[}
\newcommand{\rb}{\right]}

\newlength{\wsingfig}
\newlength{\wdblefig}
\newlength{\wquadfig}
\newlength{\wtriplefig}
\newcommand{\Eq}[1]{Eq.~(\ref{#1})}
\newcommand{\Eqs}[1]{Eqs.~(\ref{#1})}
\newcommand{\Fig}[1]{Fig.~{\ref{#1}}}
\newcommand{\Figs}[1]{Figs.~{\ref{#1}}}
\newcommand{\Refa}[1]{Ref.~{\cite{#1}}}
\newcommand{\Refs}[1]{Refs.~{\cite{#1}}}
\newcommand{\Sec}[1]{Sec.~\ref{#1}}
\newcommand{\Secs}[1]{Secs.~\ref{#1}}

\newcommand{\vev}{\textit{vev}}
\newcommand{\sr}{\mathrm{sr}}
\newcommand{\nsr}{\mathrm{usr}}

\newcommand{\longspace}{$\quad\quad\quad\quad\quad\quad\quad\quad\quad\quad\quad\quad\quad\quad\quad\quad\quad\quad\quad\ $}

\newlength{\temp}
  {\begin{list}{}{
    \settowidth{\temp}{#2}%
    \settowidth{\labelwidth}{#1}%
    \setlength{\leftmargin}{\labelwidth}}}%
  {\end{list}}

\newcommand{\Sep}{\vspace{1.5em}}
\newcommand{\SmallSep}{\vspace{0.5em}}

\newcommand{\CVpaper}[3]
	{\indent \MYhref[violet]{#2}{#3}}

\begin{document}


\baselineskip 14.5pt


$\ $
\vskip 3.cm

{\huge
\begin{center}
{\bf Manuscrit d'habilitation \`a diriger des recherches
}\\
{\bf {\Large Universit\'e Paris Saclay}}
\end{center}

\vskip 3.cm
 
\begin{center}
Stochastic Inflation and Primordial Black Holes
\end{center}

\vskip 3cm

\begin{center}
{\huge
{\bf Vincent Vennin}}
\vskip 1cm
{\Large
{\bf
{\it
Charg\'e de recherches au Centre National de la Recherche Scientifique, \\Laboratoire Astroparticule et Cosmologie}}}
\end{center}

\vskip 5cm

\begin{center}
{\Large
Th\`ese d'habilitation soutenue le 30 Juin 2020
}
\end{center}

\clearpage

\addcontentsline{toc}{section}{\textsc{Abstract}}
\baselineskip 11pt
\normalsize
{\bf \noindent Abstract}\\

Inflation is a phase of accelerated expansion that occurs at extremely high energy in the very early universe. During this epoch, vacuum quantum fluctuations are amplified and stretched to astrophysical distances. They give rise to fluctuations in the cosmic microwave background temperature and polarisation, and to large-scale structures in our universe.\\

They can also trigger the formation of primordial black holes. Such objects could provide the progenitors of the black-hole mergers recently detected through their gravitational-wave emission, and constitute part or all of the dark matter. Their observation would give invaluable access to parts of the inflationary sector that are unconstrained by the cosmic microwave background, at energy scales far beyond those accessible in particle physics experiments.\\

Since primordial black holes require large inhomogeneities to form, they are produced in scenarios where vacuum quantum fluctuations substantially modify the large-scale dynamics of the universe. In the present habilitation thesis, this ``backreaction'' effect is investigated by means of the stochastic inflation formalism, an effective theory for the long-wavelengths of quantum fields during inflation, which can be described in a classical but stochastic way once the small wavelengths have been integrated out. It describes an inflating background that gets randomly and constantly corrected by the vacuum quantum fluctuations as they get stretched to large distances. \\

After a brief review of the stochastic inflation formalism, we explain how it can be combined with standard techniques of cosmological perturbation theory (the $\delta N$ formalism) to provide a framework in which the full probability density function of curvature perturbations can be computed in the presence of non-perturbative quantum diffusion (the so-called ``stochastic-$\delta N$ formalism''). These results are then applied to the calculation of primordial black holes, where we show that quantum diffusion effects can change the expected abundance by several orders of magnitude. Finally, since inflationary models giving rise to cosmologically relevant primordial black holes often feature violations of slow roll, the stochastic and stochastic-$\delta N$ formalisms are generalised to non slow-roll dynamics. We conclude by highlighting several research directions that remain to be explored.

\clearpage

{
  \hypersetup{linkcolor=black}
  \tableofcontents
}

\vskip 2cm

\begin{center}
\noindent \textit{Texts highlighted in \MYhref[violet]{http://inspirehep.net/author/profile/V.Vennin.1}{violet} contain url links to webpages.}
\end{center}

\newpage


\renewcommand{\baselinestretch}{0.} 

\section{Foreword}

My first steps into cosmology and scientific research took place nine years ago, in 2011, as a PhD student at the {\textit{Institut d'Astrophysique de Paris}}. At that time, accurate measurements of the Cosmic Microwave Background (CMB) temperature anisotropies, by the Planck satellite mission, were about to be released. The main project of my PhD, supervised by J\'er\^ome Martin, was thus to build a numerical pipeline that would systematically compute the predictions of all singe-field models that had been proposed in the literature (more than 200 models), and compare them with the CMB data using Bayesian model comparison techniques. This was done in collaboration with Christophe Ringeval. When the Planck data came out, they confirmed the main predictions of inflation (small spatial curvature; almost Gaussian, close-to-scale-invariant, and phase-coherent primordial perturbations) and showed that the simplest models of inflation, in which a single scalar field (the inflaton) slowly rolls down a quasi-flat potential, are enough to account for all observations. This made the tool we had developed, \MYhref[violet]{http://cp3.irmp.ucl.ac.be/~ringeval/aspic.html}{ASPIC}, of particular interest (it led to the results now presented in the ``inflation'' section of the \textit{particle data group} official review), but also contrived access to other degrees of freedom during inflation, since no evidence for multiple-field effects was found. When embedded in high-energy frameworks, single-field models however often come with additional degrees of freedom, and an important question is therefore to explain the emergence of an effective single-field phenomenology from a likely multiple-field setup.\\

After my PhD, in 2014, I was hired as a postdoctoral fellow by David Wands at the {\textit{Institute of Cosmology and Gravitation}} of the University of Portsmouth, in England. While in Portsmouth, I studied the curvaton models, in which an additional light spectator field during inflation (the curvaton) comes to dominate the energy budget of the universe afterwards, providing the main source of cosmological perturbations. These models can be made in excellent agreement with the data, the main difference with single-field models being the presence of local non-Gaussianities, yet at a level that is still undetectable (though it may not remain so forever). Incorporating these multiple-field scenarios in \MYhref[violet]{http://cp3.irmp.ucl.ac.be/~ringeval/aspic.html}{ASPIC}, we realised that a crucial parameter to determine the predictions of the models is the vacuum expectation value (\vev) of the curvaton field at the end of inflation. This field being light, its \vev~is mostly set by the accumulation of vacuum quantum fluctuations of small wavelengths, as they get amplified and stretched to large distances during inflation. This can be calculated in the stochastic inflation formalism, where we have shown that the details of the entire expansion history of inflation (\ie not only the last $\sim 50$ \efolds~when observable scales are produced) can play an important role, even if inflation occurs in the slow-roll regime. This for instance led to observational constraints on the overall duration of inflation, in the favoured curvaton models.\\

Stochastic inflation can also be applied to non-test fields, in particular to the inflaton itself. It then describes an inflating background that gets randomly and constantly corrected by the vacuum quantum fluctuations as they cross out the Hubble radius. While stochastic effects on the background dynamics had been widely studied in the literature, we realised that stochastic inflation also provides a tool to compute properties of the cosmological perturbations, and to study how they are modified by quantum diffusion, in a non-perturbative way. This led us to develop the so-called stochastic-$\delta N$ formalism, which has then be applied to study various problems, by myself and various other research groups. Since large quantum diffusion is associated to large perturbations, when it takes place, it may lead to the production of primordial black holes. This is why we have then applied the stochastic-$\delta N$ formalism to the calculation of the abundance of primordial black holes, and shown that quantum diffusion effects can change the expected abundance by several orders of magnitude.\\

In the mean time, in 2017, I was awarded a Marie Curie fellowship at the {\textit{Laboratoire Astroparticules et Cosmologie}} in Paris, and immediately after, a permanent CNRS position. In Paris, I have kept working on primordial black holes, in particular when produced by the inevitable preheating instability arising from the inflaton oscillations around the minimum of its potential at the end of inflation, and on the stochastic gravitational wave background these primordial black holes produce. The increase in the interest in primordial black holes in the community over the last few years has two main motivations: they provide a possible explanation to the dark matter in the universe, while direct detection experiments have not observed dark-matter candidate particles yet, and supersymmetry has not been discovered at the LHC; and the recent LIGO/VIRGO detection of gravitational waves emitted by black-hole mergers with progenitors having a few solar masses unveiled a population of black holes in this mass range, which is precisely one of the remaining windows where primordial black holes could constitute an appreciable fraction of the dark matter.\\

A topic of continuous interest in my research has also been the quantum nature of cosmological perturbations, and whether they retain a genuine quantum signature that could be looked for experimentally and that would prove their quantum origin. This is why I have worked on the quantum discord contained in the quantum state of inflationary perturbations, on how this state can violate Bell and Leggett-Garg inequalities, and how it is subject to quantum decoherence (and what observational imprints this leaves). I have also applied continuous collapse models of the wavefunction, proposed to solve the quantum measurement problem inherent to the Copenhaguen interpretation of quantum mechanics, to primordial cosmological perturbations dynamics, and shown that the CMB places very competitive constraints on these alternative formulations of quantum mechanics.\\

I have also worked on reheating, the epoch during which the energy contained in the field(s) that drive inflation decay into the degrees of freedom of the standard model of particle physics, and the universe thermalises. Since the amount of expansion during reheating determines the location of the CMB observational window along the inflationary potential, it can be indirectly constrained by CMB observations. While in Portsmouth I have become an active member of the COrE collaboration, a CMB $B$-mode satellite proposed to ESA, and have produced the official forecasts for inflationary model comparison for that mission. Other contributions include investigations of the issue of initial conditions for inflation, formal aspects of canonical transformations in scalar-field cosmology, primordial magnetogenesis, the geometrical destabilisation of inflation,  Feebly Interacting Massive Particle (FIMP) models of dark matter, high-precision calculations of the power spectra in models of inflation with non-canonical kinetic terms, and constraints on the Lorentz factor in gamma-ray bursts.\\

In this manuscript, I choose to focus on my work on the stochastic-$\delta N$ formalism, which I have been developing with collaborators (Jose Mar\'ia Ezquiaga, Hassan Firouzjahi, Juan Garc\'ia-Bellido, Mahdiyar Noorbala, Chris Pattison, Alexei Starobinsky, Yuichiro Tada, David Wands) since 2015. I provide a self-contained summary of the research articles listed in \Sec{sec:hdr:papers}, organised in a way that makes connections between different results more apparent, and presented in a form that, I hope, benefits from the deeper understanding I have acquired by working on this topic for a few years. Not all technical details or scientific discussions are covered, and the interested reader is invited to check the articles listed in \Sec{sec:hdr:papers} for further content.\\

Over the last 9 years, I have shared my excitement, doubts and head scratching moments with many colleagues, some of whom became dear friends. Even if research in theoretical physics is about unveiling fundamental laws of nature, the existence and manifestations of which do not rely on the existence of human beings (though this depends on how quantum mechanics is interpreted), research is conducted by human beings, with other human beings, and these human interactions are what makes it a truly delightful and fulfilling activity. This is why I want to warmly thank all my collaborators:
Kenta Ando,
Hooshyar Assadullahi, 
Robert Brandenberger,
Cliff Burgess,
Chris Byrnes,
Sebastien Clesse,
Dries Coone,
Kari Enqvist,
Jose Mar\'ia Ezquiaga,
Hassan Firouzjahi,
Juan Garc\'ia-Bellido,
Julien Grain,
Robert Hardwick,
Richard Holman,
uncompromising (but always fair) EvaluatorIAP,
Kazuya Koyama,
Laurence Perreault Levasseur,
Tommi Markkanen,
Jerome Martin,
Mahdiyar Noorbala,
Sami Nurmi,
Theodoros Papanikolaou,
Chris Pattison,
Patrick Peter,
Syksy Rasanen, 
S\'ebastien Renaux-Petel,
Christophe Ringeval,
Diederik Roest,
Alexei A. Starobinsky,
Yuichiro Tada,
Tommi Tenkanen,
Jes\'us Torrado,
Thomas Tram,
Roberto Trotta,
Krzysztof Turzy\'nski and
David Wands.

Finally, I would like to thank the members of the jury, who kindly accepted to use their great expertise in the field to assess my research work:\\
\begin{center}
Clifford P. Burgess (rapporteur)\\
Julien Grain (rapporteur)\\
Christophe Ringeval (rapporteur)\\
David Langlois (examiner)\\
Patrick Peter (examiner)\\
David Polarski (examiner)\\
\end{center}

\clearpage
\section{Scientific context}
\label{sec:Context}
With the advent of high-precision cosmological and astrophysical surveys, we have entered the ``precision cosmology'' era. The anatomy of the universe on the largest observable scales has been unveiled with an unprecedented accuracy, thanks to a wealth of data probing the distribution of matter and energy at different epochs. In particular, the recent \MYhref[violet]{https://www.cosmos.esa.int/web/planck}{Planck} satellite mission, in combination with small-scale ground-based experiments, have provided us with extremely high-quality measurements of the Cosmic Microwave Background (CMB) anisotropies, shedding new light on the physical processes that took place in the early universe. The precision of this picture will further increase by orders of magnitude in the near future, with several major experiments aiming at measuring the CMB polarisation or the distribution of galaxies. 

These observations constitute a fantastic opportunity to constrain the physical conditions that prevailed at early times, where inflation is believed to have taken place. Inflation~\cite{Starobinsky:1980te, Guth:1980zm} is a phase of accelerated expansion that occurred at very high energy and that was first introduced 40 years ago as a possible solution to the hot Big Bang model problems. During inflation, vacuum quantum fluctuations are stretched to astrophysical scales and parametrically amplified (see \Fig{fig:background:inflation}). This gives rise to primordial cosmological perturbations, leading to CMB anisotropies (see \Fig{fig:Cl}) and large-scale structures in our universe. Inflation predicts that these perturbations should be almost Gaussian, close to scale invariance and phase coherent, predictions that have been remarkably well confirmed. Further, their detailed statistics allow one to constrain the microphysics of inflation and the details of its dynamics. Inflation has thus become a very active field of research, since the energy scales involved during this early epoch are many orders of magnitude larger than those accessible in particle physics experiments. This is why the early universe is certainly the most promising probe, and possibly the only one, to test far-beyond-standard-model physics.

\subsection{Open issues in inflationary cosmology}
\label{sec:inflation}

Despite its great phenomenological success, inflation leaves open a number of fundamental issues that we summarise below. As we will see, one of the main limitations for learning more about these issues is that the CMB only gives access to a restricted range of scales, hence it constrains a limited time interval of the inflationary phase. Ultimately, to truly uncover the physics at play in the early universe, one needs to probe scales that are beyond the reach of current cosmological surveys. This will naturally lead us to considering Primordial Black Holes (PBHs). We will show that they would give access to the missing scales of inflation, and have the potential to unveil some of its mysteries. 
\begin{figure}[!t]
\begin{center}
\includegraphics[width=0.48\textwidth]{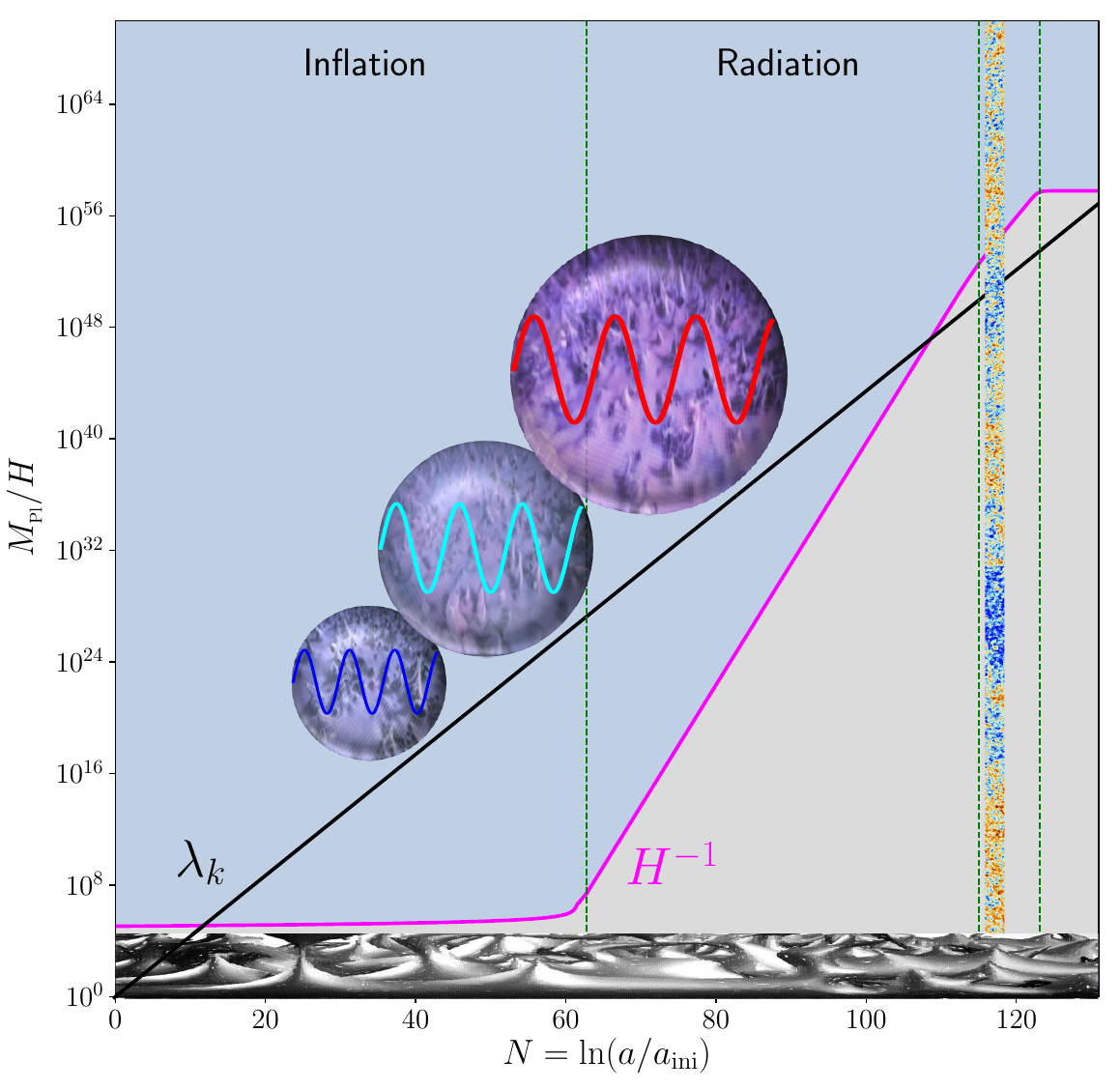}
\caption{Hubble radius $H^{-1}=a/\dot{a}$ (inverse expansion rate, magenta line) and wavelength $\lambda_k\propto a$ (black line) as a function of time measured by the number of $e$-folds, $\mathrm{ln}\, a$, where $a$ is the scale factor of the universe. During inflation, $H$ is almost constant, hence $\lambda_k$ crosses out the Hubble radius for wavelengths of astrophysical interest today. At early time during inflation, $\lambda_k$ lies inside the Hubble radius where space-time is effectively flat, and initial conditions are taken in the quantum adiabatic, Minkowski vacuum. At the end of inflation, $\lambda_k$ lies outside the Hubble radius, where it is parametrically amplified by the curvature of space-time. After inflation, during the radiation era, $\lambda_k$ crosses the Hubble radius back in. At the beginning of the matter-dominated epoch, the universe becomes transparent: this is the last-scattering surface of photons, which gives rise to the cosmic microwave background. Temperature and polarisation inhomogeneities in the cosmic microwave background are imprinted by the quantum vacuum fluctuations amplified during inflation, hence they allow us to constrain the microphysics of this early epoch.}
\label{fig:background:inflation}
\end{center}
\end{figure}

\subsubsection{Model building in high-energy theories}

Inflation can proceed at energy scales as high as $10^{16}$ GeV, where particle physics remains elusive. The nature of the fields driving inflation, and their connection to the rest of the standard model of particle physics, is therefore still unknown. Various implementations of inflation have been put forward, embedded in different extensions of the standard model of particle physics, but no uncontroversial UV-complete model of inflation has been proposed so far. The main issue is that inflation is sensitive to the physics at the Planck scale, in the sense that order-one changes in the interactions of the field(s) responsible for inflation with Planck-scale degrees of freedom generically have significant effects on the inflationary dynamics. 

For instance, adding a Planck-suppressed correction to the inflationary potential, $V\to V(1+g \phi^2/\Mp^2)$, where $\phi$ denotes the inflaton and $V$ its potential energy, $\Mp$ is the reduced Planck mass and $g$ a dimensionless constant of order one, leads to a negligible modification to the potential energy value if  $\phi\ll \Mp$, but it results in a significant correction to the mass $m$ that becomes of order the Hubble parameter $H$, since $\Delta m^2 = \Delta V'' \ni 2gV/\Mp^2\simeq 6 g H^2$, preventing inflation from occurring. This effect can also be illustrated in an effective-field-theory approach, where the radiative corrections to the mass are given by $m^2\rightarrow m^2+gM^2\ln({\Lambda}/{\mu})$, where $\mu$ is the normalisation scale, $\Lambda>H$ the cut-off of the effective theory in which the model in embedded, $M>\Lambda$ the energy scale of heavy fields, and $g$ the coupling constant. This again leads to $m/H > 1$, which spoils inflation, unless $g$ is extremely small. 

This problem is known as the $\eta$-problem of inflation~\cite{Baumann:2014nda}. It can be fixed by requiring symmetries to be preserved (the prototypical example being shift symmetry, see \eg \Refa{Kawasaki:2000yn}), but in general, when building a model of inflation, one should ensure that high-energy interactions are under control and remain harmless to inflation, which is a highly non-trivial task. This however also means that our ability to see through the inflationary window can turn the early universe into a laboratory for ultra-high energy physics at energies entirely inaccessible to conventional experimentation.

Another issue that is often discussed concerns the values of the parameters that one needs to assume in order for a given model to fit the data. The prototypical example is the single-field model $V(\phi)=\lambda \phi^4$, where $\lambda$ is a dimensionless constant. In order to predict the correct amplitude of the temperature fluctuation power spectrum in the CMB, one needs to assume $\lambda \sim 10^{-12}$, which may be viewed as problematic from the standard lore that a model is ``natural'' if all dimensionless quantities are of order one. For $V(\phi)= m^2 \phi^2/2$, one finds $m \sim 10^{-6}\Mp$, which may be viewed as better. Both models are now disfavoured by the data anyway, so it is more relevant to consider a plateau potential that is observationally favoured, such as Higgs inflation~\cite{Bezrukov:2007ep}, where one assumes that the inflaton field is the Higgs boson, non minimally coupled to gravity. In this case, the non-minimal coupling constant, $\xi$, must be given by $\xi \sim 46000 \sqrt{\lambda}$, where $\lambda$ is the self-interacting Higgs coupling constant. The fact that $\xi/\sqrt{\lambda} \gg 1$ can again be viewed as problematic for model building, although defining the naturalness of the value of a parameter is not an obvious task. 
\begin{figure}[!t]
\begin{center}
\includegraphics[width=0.68\textwidth]{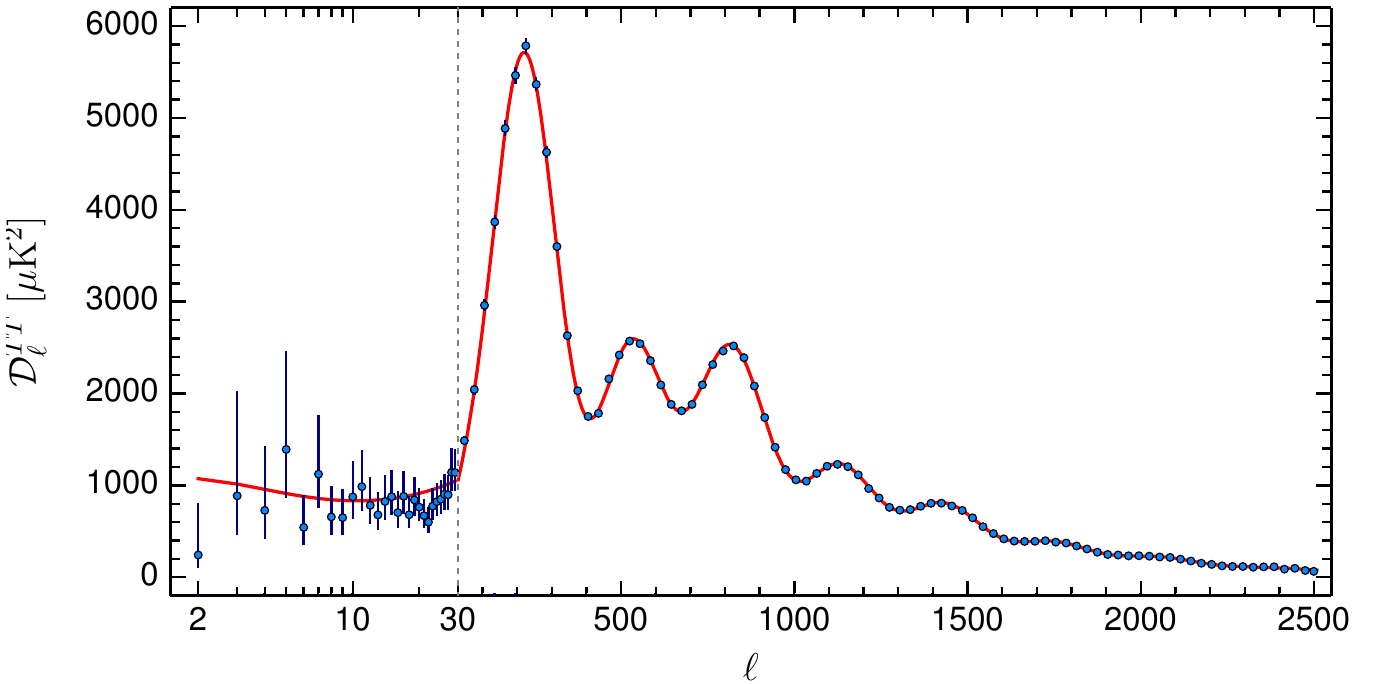}
\caption{CMB temperature fluctuation power spectrum as a function of the multiple moment $\ell$, as measured by \MYhref[violet]{https://www.cosmos.esa.int/web/planck}{Planck}. The base-$\Lambda$CDM theoretical spectrum that best fits the likelihood is plotted in red. From \Refa{Akrami:2018vks}.}
\label{fig:Cl}
\end{center}
\end{figure}

This explains why many attempts to build models of inflation in various extensions of the standard model of particle physics have been pursued. Even if one restricts to single-field models, which are the simplest scenarios compatible with observations, hundreds of possibilities have been investigated. Some of them have been excluded (using the \MYhref[violet]{http://cp3.irmp.ucl.ac.be/~ringeval/aspic.html}{ASPIC} library~\cite{Martin:2013tda, Martin:2013nzq, Martin:2014nya}), one third of the single-field models of inflation are now ruled out with very strong evidence.} since they cannot account for the CMB anisotropies as recently mapped by \MYhref[violet]{https://www.cosmos.esa.int/web/planck}{Planck}~\cite{Akrami:2018odb}, but many possibilities still remain. Although future experiments should improve this unsatisfying state of affairs (see section~\ref{sec:experiments}), the CMB only gives access to a limited range of scales, and the time frame during which these scales are generated during inflation is therefore limited as well, and cannot encompass more than $\sim$ 7 $e$-folds (sketched with the orange interval in \Fig{fig:pot:inflation}), over the $\sim 60$ $e$-folds elapsed between the generation of these scales and the end of inflation. This means that the constraints on the inflationary potential that the CMB can place are restricted to a small region. Ultimately, to learn more about the early universe, one thus needs to combine the CMB with other data sets that probe different scales.
\subsubsection{Reheating the universe}
After inflation, the energy contained in the fields driving inflation needs to decay into the other degrees of freedom of the standard model of particle physics, and the universe needs to thermalise. This epoch is known as ``reheating'' and is driven by the interactions between the inflaton and the other fundamental fields. By constraining reheating, one can thus learn about these couplings, and probe the inflationary potential in a field regime that is different from where inflation takes place. 

\begin{figure}[!ht]
\begin{center}
\includegraphics[width=0.6\textwidth]{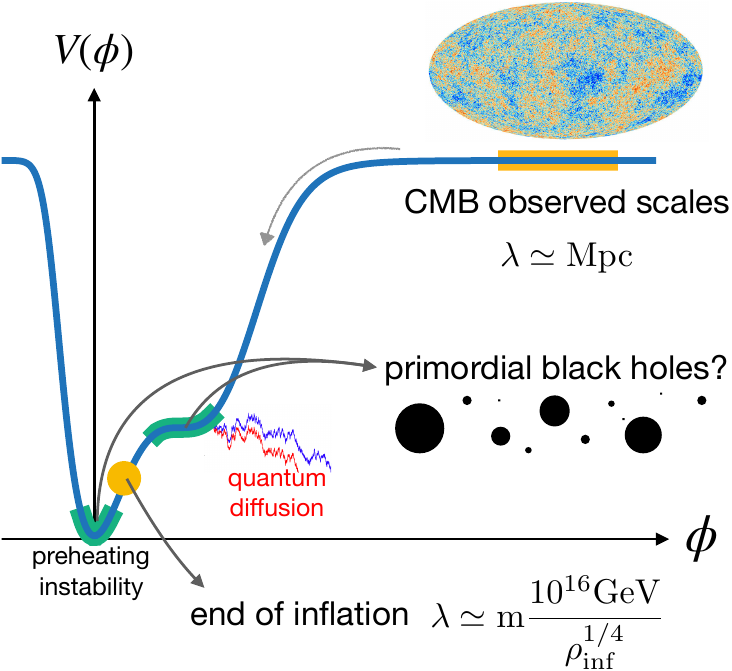}
\caption{Sketch of an inflationary potential leading to primordial black holes. Inflation starts from large values of $\phi$, \ie from the right end of the figure. Around $50$ \efolds~before the end of inflation (denoted with an orange dot), the potential has a plateau shape, and the scales probed in the CMB, of the order of Mpc today, are produced. The range of scales accessed in the CMB corresponds to a time frame of about $7$ \efolds, depicted with the orange line. Later on, if the potential contains a very flat region (highlighted in green), large cosmological perturbations are produced, which can give rise to the subsequent formation of primordial black hole. If this is the case, quantum diffusion plays an important role during this epoch, and needs to be taken into account. This is the topic of the present manuscript. Let us notice that after inflation, the oscillations of the inflaton around a local minimum of its potential (also highlighted in green) trigger a parametric instability for scalar perturbations at sub-Hubble scales, which can also give rise to primordial black holes. I have studied this mechanism in \Refs{Martin:2019nuw, Martin:2020fgl} but I do not report on it here.}
\label{fig:pot:inflation}
\end{center}
\end{figure}

The expansion history of the universe during reheating determines the location of the observational window along the inflationary potential, \ie it allows one to relate physical scales as measured today with the time frame (sketched with the orange interval in \Fig{fig:pot:inflation}) during inflation when they are produced. It can therefore be indirectly constrained by CMB observations. For instance, using the \MYhref[violet]{http://cp3.irmp.ucl.ac.be/~ringeval/aspic.html}{ASPIC} library, one can show that with the \MYhref[violet]{https://www.cosmos.esa.int/web/planck}{Planck} satellite results, $\sim 40$\% of the possible reheating scenarios can be rejected~\cite{Martin:2016oyk}. 

Beyond these constraints, the physics of reheating is however poorly known, since a direct observational access to the reheating epoch would require to probe scales that exit the Hubble radius around the end of inflation, with a wavelength today that can be as small as one meter (if inflation proceeds at $\rho_{\mathrm{inf}}=10^{16}\mathrm{GeV}$; it otherwise scales with $\rho_{\mathrm{inf}}^{-1/4}$). Contrary to the scales accessed in the CMB (of the order of $10^6$ parsecs), such distances fall far into the non-linear regime and cannot be directly studied with cosmological surveys.
\subsubsection{Quantum origins of cosmological structures}
Since the energy scale at which inflation occurs can be as high as only three orders of magnitude below the Planck scale, it is, in some sense, the most violent phenomenon observed in Nature, and the one which operates the closest to the scale of Quantum Gravity. Moreover, the inflationary mechanism for the production of cosmological perturbations explicitly makes use of General Relativity (GR) and Quantum Mechanics (QM), two theories that are notoriously difficult to combine. Since this mechanism leads to theoretical predictions for the CMB anisotropies, inflation is probably the only case in Physics where, given our present day technological capabilities, an effect based on GR and QM can be tested experimentally. 

This makes it an ideal playground to discuss fundamental questions related to the interplay between these two theories. For instance, due to the accelerated expansion that takes place during inflation, the quantum ground state in which cosmological perturbations are initially placed evolves into a two-mode squeezed state. Since this is not an eigenstate of the temperature fluctuation operator $\widehat{\delta T}/T$~\cite{Sudarsky:2009za, Goldstein:2015mha}, a non-unitary process needs to be invoked, during which the state evolves from the two-mode squeezed state into an eigenstate of $\widehat{\delta T}/T$. This can be rephrased by noticing that since the two-mode squeezed state is invariant under spatial translations, and given that the Hamiltonian generating its dynamics commutes with spatial translation operators, it cannot give rise to a configuration that contains spatial inhomogeneities, unless a non-unitary mechanism projects the two-mode squeezed state onto such a configuration. This problem is no more than the celebrated measurement problem of Quantum Mechanics, which, in the Copenhagen approach, is ``solved'' by the collapse of the wavefunction. In the context of Cosmology, however, the use of the Copenhagen interpretation appears to be problematic~\cite{Hartle:2019hae}. Indeed, it requires the existence of a classical domain, exterior to the system, which performs a measurement on it.  In quantum cosmology, one calculates the wavefunction of the entire universe and there is, by definition, no classical exterior domain at all. In the context of inflation, one could argue that the perturbations do not represent all degrees of freedom and that some other classical degrees of freedom could constitute the exterior domain, but they do not qualify as ``observers'' in the Copenhagen sense. The transition to an eigenstate of the temperature fluctuation operator, which necessarily occurs in the early universe before structure formation starts, thus proceeds in the absence of any observer, something at odds with the Copenhagen interpretation.

There are several ways to address this problem. One possibility is to resort to the many-world interpretation together with decoherence~\cite{Kiefer:2006je}. It can also be understood if one uses alternatives to the Copenhagen interpretation such as dynamical collapse models, see \Refs{Perez:2005gh, DeUnanue:2008fw, Leon:2010fi, Canate:2012ua, Das:2013qwa}. In this case, one obtains different predictions, that can be confronted with CMB measurements~\cite{Martin:2012pea, Martin:2019jye}. Other solutions involve the Bohmian interpretation of Quantum Mechanics~\cite{Peter:2006hx, PintoNeto:2011ui, Peter:2016kan}.

Let us also note that highly squeezed states are sometimes described as ``classical'', since most of the corresponding quantum correlation functions can be obtained using a classical distribution in phase space~\cite{Polarski:1995jg, Kiefer:1998qe, Martin:2015qta}. However, they also possess properties usually considered as highly non classical. They are indeed entangled states, very similar to the Einstein-Podolsky-Rosen (EPR) state, with a large quantum discord~\cite{Martin:2015qta}, which allows one to construct observables for which the Bell inequality~\cite{Martin:2016tbd, Martin:2017zxs} and the Leggett-Garg inequalities~\cite{Martin:2016nrr} are violated. However, the presence of quantum decoherence makes it still unclear whether or not a genuinely quantum signal can be seen in the CMB, that would confirm the quantum origin of cosmological perturbations~\cite{Martin:2018zbe}. Such a signal would also be difficult to detect in practice since it is hidden in the decaying modes of cosmological perturbations. The main reason is that the growing mode and the decaying mode are two non-commuting observables at the quantum-mechanical level, and that only by measuring both can one access this non-vanishing commutator. 

The amplitude of the decaying mode decreases on super-Hubble scales (hence its name), while the quantum squeezing (responsible for the presence of genuine quantum correlations), increases. On CMB scales, quantum squeezing is extremely large (and much larger than what can be achieved in laboratory experiments, \eg in quantum optics setups), but the amplitude of the decaying mode is also too small to be measured. The best place to look for genuine quantum signatures may therefore be at wavelengths that exit the Hubble radius only a few $e$-folds before the end of inflation, such that squeezing operates but does not entirely suppress the decaying mode. Since these are precisely the scales at which PBHs are expected to form, they may provide a natural candidate to look for quantum imprints~\cite{Espinosa-Portales:2019peb}.
 \subsubsection{Initial conditions}
Inflation was originally proposed as a solution to the hot big-bang problems, in particular the horizon (why is the observable universe so homogeneous on large scales, given that, in the absence of inflation, it spans several causally disconnected regions?) and the flatness problems (why is the universe so flat given that, in the absence of inflation, the contribution of the spatial curvature to the overall energy budget can only increase with time?). Both are initial condition problems, and for inflation to solve them, it should not be flawed with initial conditions issues itself. In other words, one should check that the initial conditions required for a successful phase of inflation to proceed are not fine tuned. There are in fact several aspects to this question.

\paragraph{Initial conditions in physical space}$ $\\
First, one may ask whether inflation naturally starts from generic inhomogeneous initial conditions. If the size of the inhomogeneities is initially much smaller than the Hubble radius, the so-called ``effective-density approximation''~\cite{Goldwirth:1989pr,Goldwirth:1991rj} can be employed, and it can be shown that the energy density contained in the field fluctuation $\delta\phi$ decays as $\rho_{\delta\phi}\propto a^{-4}$ (see also \Refa{Chowdhury:2019otk}). This is why, after a phase where the universe effectively behaves as radiation dominated, inflation may start. If the size of the inhomogeneities is initially much larger than the Hubble radius, it can be locally absorbed in a re-normalisation of the background energy density and does not prevent inflation either. In between, \ie for inhomogeneities with wavelength of the order of the Hubble radius, and/or populating various length scales, a full numerical treatment is compulsory. The first numerical solutions~\cite{Goldwirth:1989pr, Goldwirth:1989vz, Goldwirth:1991rj, Goldwirth:1990pm} were obtained under the assumption that space-time is spherically symmetric. This simplifies the calculations since then the problem only depends on time and on one radial coordinate. This analysis was improved in \Refs{KurkiSuonio:1987pq, Laguna:1991zs, KurkiSuonio:1993fg} in which the spherical symmetry assumption was relaxed. More recently, \Refs{East:2015ggf, Clough:2016ymm, Bloomfield:2019rbs} have run new simulations (and seem to confirm the validity of the behaviour $\rho_{\delta\phi}\propto a^{-4}$ found in the effective-density approximation, even when $k\sim a H$).  All these works have technical restrictions and, at this stage, it is difficult to draw a completely general conclusion. However, it seems that large-field and plateau models work better than small-field inflation, and that although the size of the initial homogeneous patch is an important parameter of the problem, strong gradients may also help in starting inflation (see also \Refs{Calzetta:1992gv, Perez:2012pn}).

\paragraph{Initial conditions in phase space}$ $\\
Let us now assume that space has been homogenised. For a given inflationary potential, a successful, long enough phase of inflation does not proceed from any initial value of the inflaton field and its velocity, and the question is whether these initial values need to be fine tuned. Due to the presence of a dynamical attractor, the so-called ``slow-roll'' attractor~\cite{Liddle:1994dx}, the dependence on the initial velocity is efficiently erased. The remaining dependence on the initial field value depends on the potential. If inflation proceeds close to a local maximum, and if the width of the ``hill'' is sub-Planckian, then the inflaton field value needs to be fine tuned very close to the top of the hill. However, such models are now strongly disfavoured by observations, since they predict a too low value for the spectral index of scalar perturbations. In fact, the models favoured by the data, namely plateau potentials, are precisely those that do not suffer from phase-space initial fine tuning. It therefore seems that this issue has been greatly alleviated by the recent CMB measurements. An extensive discussion can be found in \Refa{Chowdhury:2019otk}, where various potentials have been studied in details, and  several phase-space measures have been incorporated in the analysis. 

\paragraph{Initial conditions of the perturbations}$ $\\
The above discussions deal with initial conditions of the classical background. Similar issues exist for the initial quantum state of perturbations. If inflation lasts long enough, modes of astrophysical interest today lie deep inside the Hubble radius at the onset of inflation, where the effect of space-time expansion on their dynamics can be neglected. One therefore usually assumes their quantum state to be in the adiabatic vacuum state of Minkowski, \ie flat, space-times, the so-called ``Bunch-Davies vacuum''~\cite{Bunch:1978yq}. This prescription is however not invariant under changes of the canonical variables used to describe the system~\cite{Grain:2019vnq}, and other initial configurations are possible.

Ultimately, if inflation is preceded by another cosmological epoch, such as a contracting phase followed by a bounce for instance, the state of the universe at the onset of inflation is inherited from this earlier epoch. If inflation lasts arbitrarily long, the Bunch-Davies vacuum does feature some specific properties, such as \eg being de Sitter invariant, but it is also the case for any state obtained from the Bunch-Davies vacuum through a $k$-independent Bogolyubov transformation~\cite{PhysRevD.32.3136}. It is also sometimes argued that the Bunch-Davies vacuum is a local attractor~\cite{Brandenberger:1985fc, Kaloper:2018zgi}, which is mainly due to the explosive particle production mechanism taking place on super-Hubble scales: for any state that differs from the Bunch-Davies vacuum by a finite number of particles, this difference is quickly overtaken by the vast number of particles created on super-Hubble scales, and becomes negligible. However, this remains true for other reference states~\cite{Grain:2019vnq}.

Furthermore, if inflation lasts even longer, the length scales of cosmological interest today are smaller than the Planck length at the onset of inflation. This is the case in most models of inflation, and it is in that regime that the Bunch-Davies vacuum state is set. But one can wonder whether this is legitimate and whether quantum field theory in curved space-time is still valid in this case. Notice that the energy density of the background remains much less than the Planck energy density, so that the use of a classical background is well justified, and it is only the wavelengths of the perturbations that can be smaller than the Planck length. This issue is known as the trans-Planckian problem of inflation~\cite{Martin:2000xs, Brandenberger:2000wr, Brandenberger:2004kx}.

In the absence of a final theory of quantum gravity, it is difficult to predict what would be the modifications to the behaviour of the perturbations if physics beyond the Planck scale were taken into account. Most approaches to modelling the modifications originating from the space-time foam lead to corrections to observables that scale as $(H/M_{\mathrm{c}})^p$~\cite{Martin:2000xs, Brandenberger:2000wr}, where $H$ is the Hubble scale during inflation and $M_{\mathrm{c}}$ the energy scale at which new physical effects appear (typically the Planck scale or, possibly, the string scale); $p$ is a model-dependent index. Those corrections are therefore typically small, although there are other ways of modelling the new physics that could lead to more drastic modifications~\cite{Martin:2002kt}.
\subsection{Primordial black holes as a probe of the missing small scales of inflation}
\label{sec:PBHs}
From the above considerations, it becomes clear that accessing scales that are smaller than the ones probed in the CMB is of paramount importance to learn more about the early universe: it would 1) extend the part of the inflationary potential one can probe (and better constrain the nature of the fields that drive inflation); 2) give a direct access to the physics of reheating; and 3) provide a place to look for genuine quantum effects that would prove or disprove the quantum origins of cosmological perturbations. Accessing the small scales emerging close to the end of inflation is however observationally challenging, since they fall far into the non-linear regime and are difficult to reconstruct from surveys of the large-scale structure of our universe.

Fortunately however, an additional probe of these scales may exist in the form of primordial black holes~\cite{Carr:1974nx}. PBHs are expected to form from rare large density perturbations produced during inflation, when they re-enter the cosmological horizon and collapse into black holes. For the scales probed in the CMB, the amplitude of the fluctuations is too small to yield a substantial abundance of PBHs. At smaller scales however, where the amplitude of the fluctuations is less constrained, inhomogeneities produced during inflation could be large enough, and PBHs thus open up a new observational window (see \Fig{fig:pot:inflation}).
\begin{figure}[!t]
\begin{center}
\includegraphics[width=0.68\textwidth]{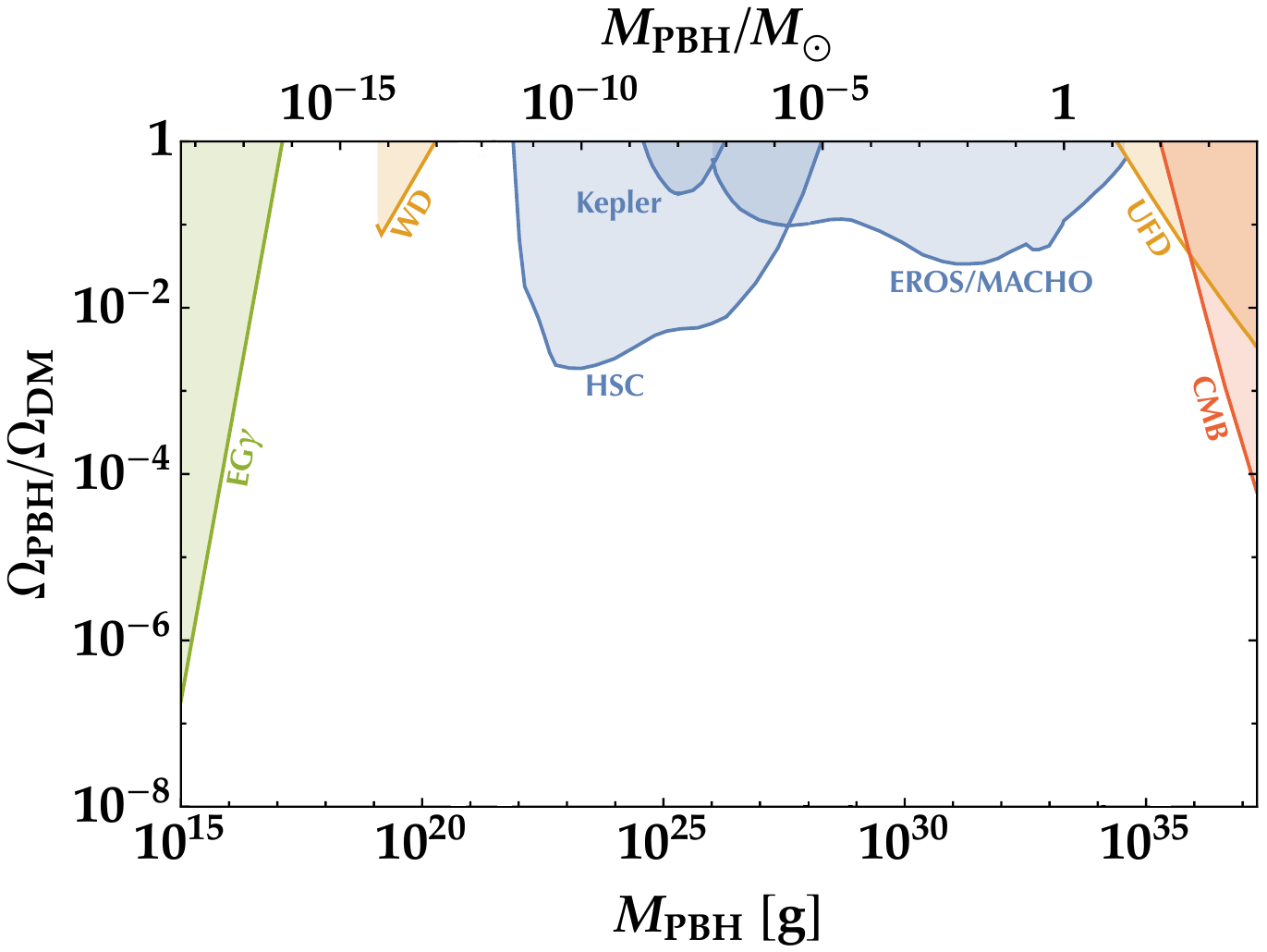}
\caption{Observational constraints on the fraction of dark matter comprised in primordial black holes, if they all form with the same mass labeled in the horizontal axis. The figure is adapted from \Refa{Tada:2019amh}. EG$\gamma$ stands for the the extragalactic gamma-ray background that small-mass PBHs would induce by Hawking evaporation and that is not observed~\cite{Carr:2009jm}. WD corresponds to the existence of white dwarfs in our local galaxy, that would explode as supernovae if PBHs transit through them~\cite{Graham:2015apa}. It leaves a first window around $M\simeq 10^{-15}M_\odot$ and $M\simeq 10^{-12}M_\odot$ for PBHs to constitute all of dark matter. Subaru HSC~\cite{Niikura:2017zjd}, Kepler~\cite{Griest:2013esa} and EROS/MACHO~\cite{Tisserand:2006zx} are constraints from the absence of detection of microlensing from PBHs. UFD stands for  the observation of a star cluster near the centre of the Ultra-Faint Dwarf (UFD) galaxy Eridanus II, that would be tidally disrupted by passing PBHs if they were too abundant~\cite{Brandt:2016aco}. This constraint is however debated and, if relaxed, it leaves a second window around $M\simeq 10 M_\odot$ for PBHs to constitute all of dark matter. Finally, CMB refers to the Planck satellite constraints: large PBHs accrete matter, and the release of accretion luminosity heats up the medium and ionises hydrogen. The spectral distortions this induces in the CMB are too small to be constraining, but it changes the recombination history that is tightly constrained by CMB anisotropies~\cite{Ali-Haimoud:2016mbv, Blum:2016cjs, Horowitz:2016lib}.
}
\label{fig:PBHs:constraints}
\end{center}
\end{figure}

Moreover, there has been renewed and ever increasing interest in PBHs since the \MYhref[violet]{https://www.ligo.caltech.edu/}{LIGO}/\MYhref[violet]{http://www.virgo-gw.eu/}{VIRGO} collaboration reported the first detection of gravitational waves associated to black-hole mergers in 2015~\cite{Abbott:2016blz}. They may indeed explain the existence of progenitors for these events, that would differ from astrophysical black holes originating from stellar collapse. PBHs may also solve a number of problems currently encountered in astrophysics and cosmology, such as explaining the seeding of the supermassive black holes in galactic nuclei~\cite{Bean:2002kx}, the generation of large-scale structures~\cite{Meszaros:1975ef, Carr:2018rid} (either individually through the ``seed'' effect or collectively through the ``Poisson'' effect), the minimum radius and the large mass-to-light ratios of ultra-faint dwarf galaxies~\cite{Clesse:2017bsw}, and the generation of correlations between the soft X-ray and infrared backgrounds~\cite{Kashlinsky:2016sdv} (see \Refa{Clesse:2017bsw} for other hints in favour of the existence of PBHs). They could also produce gravitational relics (Planck relics or gravitinos)~\cite{Markov:1984xd, Coleman:1991ku, Barrau:2003xp, Khlopov:2004tn}.

Tight constraints on the abundance of PBHs have been placed in various mass ranges (see \Fig{fig:PBHs:constraints} and \eg \Refs{Carr:2009jm, Carr:2017jsz} for reviews), from the gravitational lensing, production of gravitational waves by merging, Hawking evaporation or disruption of various astrophysical objects they should induce. This leaves three mass windows open for PBHs to constitute an appreciable fraction, and possibly all, of dark matter~\cite{Blais:2002nd}, around $M\sim 10^{-15}M_\odot$, $M\sim 10^{-12}M_\odot$ and $M\sim 10-100 M_\odot$. Interestingly, the third window precisely falls within the \MYhref[violet]{https://www.ligo.caltech.edu/}{LIGO}/\MYhref[violet]{http://www.virgo-gw.eu/}{VIRGO} detection band.
\subsection{Observational prospects}
\label{sec:experiments}
\begin{figure}[!t]
\begin{center}
\includegraphics[width=0.6\textwidth]{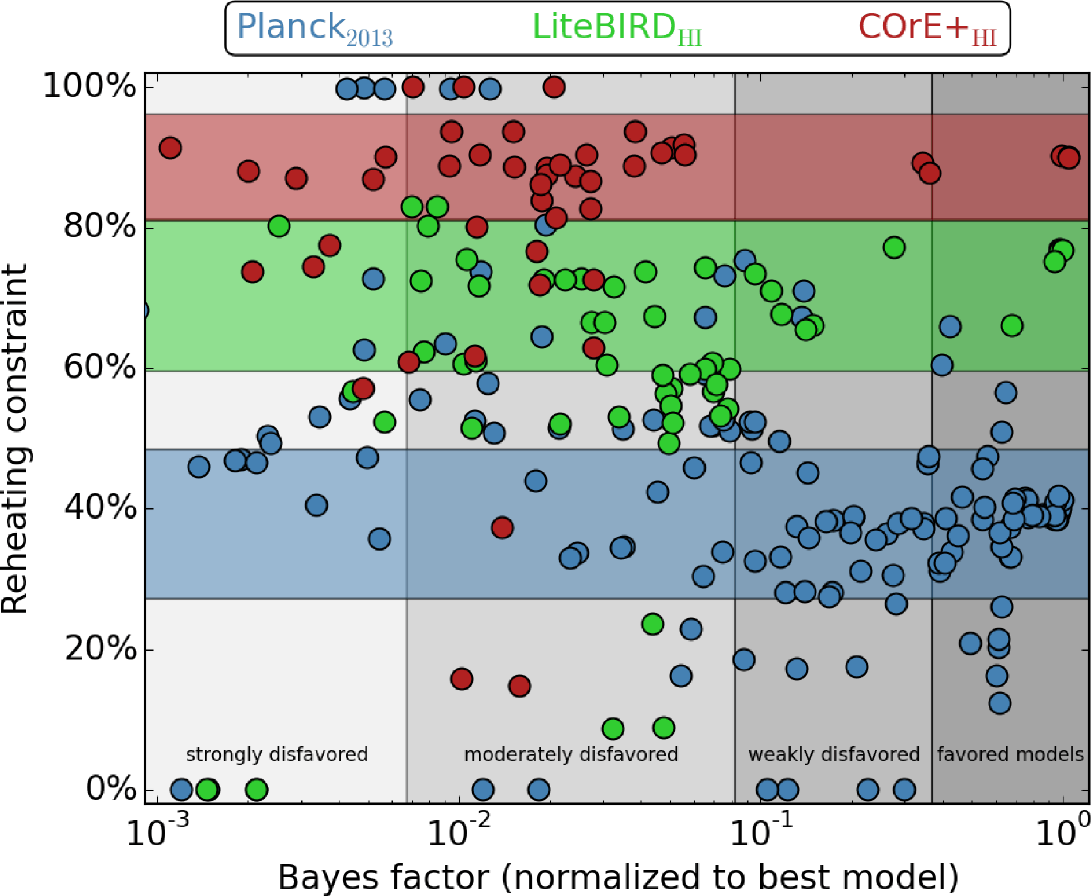}
\caption{Posterior-prior reduction in the reheating parameter plotted against Bayesian evidence for single-field inflation models using \MYhref[violet]{https://www.cosmos.esa.int/web/planck}{Planck} data, \MYhref[violet]{http://litebird.jp/eng/}{LiteBIRD} and \MYhref[violet]{http://www.core-mission.org/}{COrE} forecasts (prepared by S. Clesse, C. Ringeval and V. Vennin for the \MYhref[violet]{http://www.core-mission.org/}{COrE} proposal to ESA).}
\label{fig:forecast}
\end{center}
\end{figure}

An important prediction of inflation, untested so far, is that primordial gravitational waves should be produced in the early universe, leaving an imprint in the B-mode polarisation of the CMB anisotropies. Their detection would be a key experimental advance, as they would lead to a determination of the energy scale of inflation and of the inflaton field excursion (at least in the simplest models). It would also allow us to test the consistency relation that relates the amplitude of the tensor power spectrum with its tilt. This is why several experiments aiming at detecting B-mode CMB polarisation are currently operating and/or are on their way. Ground-based experiments that are currently operating include \MYhref[violet]{https://www.cfa.harvard.edu/CMB/bicep2/}{BICEP3, Keck} and \MYhref[violet]{https://pole.uchicago.edu/}{SPT} in Antarctica,  \MYhref[violet]{https://sites.krieger.jhu.edu/class/}{CLASS}, \MYhref[violet]{https://act.princeton.edu/}{ACTPol} and \MYhref[violet]{https://bolo.berkeley.edu/polarbear/}{POLARBEAR}/\MYhref[violet]{https://simonsobservatory.org/}{SIMONS} in the Atacama desert, and \MYhref[violet]{http://research.iac.es/proyecto/cmb/pages/en/quijote-cmb-experiment.php}{QUIJOTE} in the Canary islands. Within the next five years, these experiments aim at reaching the target $r=10^{-2}$ or even $r=10^{-3}$, where $r$ is the tensor-to-scalar ratio, which corresponds to the signal amplitude predicted by the most favoured plateau models. In space, the \MYhref[violet]{http://litebird.jp/eng/}{LiteBIRD} satellite has just been recently selected  as the strategic large mission by the Institute of Space and Astronautical Science of the Japan Aerospace Exploration Agency. This ensures that the mission will fly  and that one should expect, and get prepared for, new observational insight in the coming years (see \Fig{fig:forecast}). 

These experiments will also greatly improve the constraints on the curvature power spectrum, with sensitivities on the spectral index and its running shrinking by a factor of 5 compared to \MYhref[violet]{https://www.cosmos.esa.int/web/planck}{Planck}. Eventually, measurements of the spectral distortions of the CMB by  \MYhref[violet]{https://asd.gsfc.nasa.gov/pixie/}{PIXIE}-like experiments will probe the primordial power spectrum on smaller scales, extending the part of the inflationary potential being probed to $\sim  17$ $e$-folds~\cite{Chluba:2012gq}, and will reach higher levels of sensitivity on local non-Gaussianity~\cite{Pajer:2012vz}.

In addition, the drastic increase in the accuracy of astrophysical surveys (such as \MYhref[violet]{https://www.euclid-ec.org/}{Euclid} or \MYhref[violet]{https://www.skatelescope.org/}{SKA}) mapping the large-scale structure makes them competitive to constrain inflation. For instance, by increasing the lever arm in scales, one can improve the constraints on the tilts of the scalar and tensor primordial power spectra and their running. By measuring the scale dependence of the galaxy biases, one can also constrain primordial non Gaussianities~\cite{Dalal:2007cu}. Combination of CMB data with galaxy clustering, weak lensing or 21 cm tomography at high redshift surveys will thus open up new avenues to constrain the early universe in the following decade.\\

In addition to these cosmological, large-scale surveys, various observational perspectives should confirm (or exclude) the presence of PBHs in our universe. For instance, a straightforward way to distinguish between stellar and primordial origins for the black holes is to detect a merger involving a black hole with a mass smaller than the Chandrasekhar limit of 1.4 $M_\odot$, which is within the reach of the upcoming runs of \MYhref[violet]{https://www.ligo.caltech.edu/}{LIGO}/\MYhref[violet]{http://www.virgo-gw.eu/}{VIRGO}. Another way is to measure the spin and mass distribution of the BHs (since PBHs should form with negligible spins~\cite{Chiba:2017rvs} contrary to astrophysical ones~\cite{Kinugawa:2016ect}, and due to the existence of universal conditions on the mass distributions of PBHs~\cite{Kocsis:2017yty}), which should soon be better constrained with improved statistics. The PBH scenario can also be tested with the stochastic gravitational wave background associated with PBH binaries. In the future, the \MYhref[violet]{https://www.elisascience.org/}{LISA} project~\cite{Bartolo:2016ami}, which has been selected as an L3 mission of ESA's ``Cosmic Vision'' program, will vastly increase the sensitivity and frequency coverage of currently running experiments and will give access to such backgrounds. This will be complemented by the pulsar timing arrays constraints from the \MYhref[violet]{https://www.skatelescope.org/}{SKA} project~\cite{Zhao:2013bba}. Finally, laser interferometers could also detect the bursts of gravitational waves coming from hyperbolic encounters of PBHs in dense clusters~\cite{Garcia-Bellido:2017qal}. 

This ambitious observational program, in the era of multi-messenger astronomy, provides promising prospects for learning more about the physics of inflation and reheating, and the conditions under which primordial black holes may have formed in the early universe.
\subsection{Quantum diffusion}
As they get stretched beyond the Hubble radius during inflation, vacuum quantum fluctuations at small scales modify the large-scale dynamics of the universe. This ``backreaction'' effect is usually negligible in the perturbative regime, where density fluctuations are small. This is however not the case in regimes leading to PBHs, where one requires large inhomogeneities to be produced. Therefore, in order to properly assess the amount of PBHs arising in a given inflationary model, the effect of quantum diffusion on the background dynamics must be taken into account. 

The goal of the present manuscript is to show how it can be done, making use of the stochastic inflation formalism~\cite{Starobinsky:1982ee, Starobinsky:1986fxa}. It is an effective theory for the long-wavelengths part of quantum fields during inflation, which can be described with a classical, stochastic theory once the small wavelengths have been integrated out. In the ``separate universe approach'', the resulting theory describes a set of inflating backgrounds that get randomly and constantly shaken by the vacuum quantum fluctuations as they cross out the Hubble radius. While the stochastic effects on the background dynamics during inflation have been widely studied in the literature, how they affect the properties of cosmological perturbations has not been investigated. This is however necessary to properly assess the amount of PBHs arising from a given model of inflation. This is why part of my research has been devoted to developing a new scheme that enables the proper inclusion of quantum diffusion, and on which this manuscript reports. In the next sections, I will show how one can combine standard methods of cosmological perturbations theory (the so-called $\delta N$-formalism), and tools developed in other areas of stochastic analysis (``first-passage time'' techniques, first developed to tackle financial analysis problems~\cite{Bachelier1900}), to derive the ``stochastic-$\delta N$ formalism''~\cite{Vennin:2015hra}, in which observable quantities (such as the CMB temperature and polarisation anisotropies, or the PBH abundance) can be directly computed in the presence of non-perturbative quantum diffusion.
\section{Stochastic inflation}
\label{sec:StochasticInflation}
During inflation, scalar field perturbations are placed in two-mode squeezed states~\cite{Grishchuk:1990bj,Grishchuk:1992tw}, which undergo a ``quantum-to-classical transition''~\cite{Polarski:1995jg,Lesgourgues:1996jc,Polarski:2001yn,Kiefer:2008ku,Burgess:2014eoa,Martin:2015qta} in the sense that on super-Hubble scales, the non-commutative parts of the fields become small compared to their anti-commutative parts (see \Sec{sec:quantum_to_classical}). It gives rise to the stochastic inflation formalism~\cite{Starobinsky:1982ee, Starobinsky:1986fx, Nambu:1987ef, Nambu:1988je, Kandrup:1988sc, Nakao:1988yi, Nambu:1989uf, Mollerach:1990zf, Linde:1993xx, Starobinsky:1994bd, Finelli:2008zg, Finelli:2010sh}, consisting of an effective theory~\cite{Burgess:2015ajz} for the long-wavelength parts of the quantum fields, which are ``coarse grained'' at a fixed physical scale (\ie non-expanding), larger than the Hubble radius during the whole inflationary period. In this framework, the small wavelength fluctuations behave as a classical noise acting on the dynamics of the super-Hubble scales as they cross the coarse-graining scale, and the coarse-grained fields can thus be described by a stochastic classical theory, following Langevin equations.

Stochastic inflation can be obtained in an effective-field-theory approach, using the Schwinger-Keldysh formalism~\cite{Morikawa:1989xz,Hu:1994dka,Matarrese:2003ye}. Here we present a heuristic derivation in phase space, adapted from \Refa{Grain:2017dqa}, which is useful for generalising it beyond slow roll, see \Sec{sec:BeyondSlowRoll}. The role played by the so-called ``quantum-to-classical'' transition, and the precise sense according to which this transition occurs, are carefully discussed. The case of a test, light scalar field on an inflating background is first analysed, before extending the formalism to non-test fields. In this later case, gauge corrections have to be accounted for, which we explain following \Refa{Pattison:2019hef}. The stochastic inflation program is finally specified to the case where inflation proceeds along the slow-roll attractor, which provides the formalism mainly employed in the three following sections, \Sec{sec:stochastic:delta:N} (at least from \Sec{sec:FirstMoments:N} onwards), \Secs{sec:PBHs} and~\ref{sec:tail:expansion}.
\subsection{Cosmology in the Hamiltonian framework}
\label{ssec:cosmo}
Let us first review the Hamiltonian framework for studying the dynamics of a scalar field $\phi$ minimally coupled to gravity in a 4-dimensional curved space-time with metric $g_{\mu\nu}$, described by the action
\bea
	S=\displaystyle\int \dd^4x\sqrt{-g}\left[\frac{\Mp^2}{2}R-\frac{1}{2}g^{\mu\nu}\partial_\mu\phi\partial_\nu\phi-V(\phi)\right] .
	\label{eq:action}
\eea
In this expression, $R$ is the Ricci scalar curvature and $V(\phi)$ is the potential of the scalar field. The Hamiltonian formulation is obtained through the ADM formalism~\cite{thieman_book,Langlois:1994ec}, which provides a foliation of 4-dimensional space-times into a set of 3-dimensional space-like hypersurfaces. The foliation is determined by the lapse function $N(\tau,x^i)$ and the shift vector $N^i(\tau,x^i)$, which enter the covariant line element as
\bea
\label{eq:line_element}
	\dd s^2=-N^2\dd \tau^2+\gamma_{ij}\left(N^i\dd \tau+\dd x^i\right)\left(N^j\dd \tau+\dd x^j\right),
\eea
where $\gamma_{ij}(t,x^i)$ is the induced metric on the 3-dimensional space-like hypersurfaces. 

The canonical variables for the gravitational sector are $\gamma_{ij}$ and $\pi^{ij}=\delta S/\delta\dot{\gamma}_{ij}$, where a dot means a derivation with respect to the time variable $\tau$. Their associated Poisson bracket is $\left\{\gamma_{ij}(\vec{x}),\pi^{kl}(\vec{y})\right\}= (\delta^{k}_i\delta^{l}_j+\delta^{l}_i\delta^{k}_j ) \delta^3(\vec{x}-\vec{y})/2$. Similarly, for the scalar field, the canonical variables are $\phi$ and $\pi_\phi=\delta S/\delta\dot{\phi}$, and their Poisson bracket reads $\left\{\phi(\vec{x}),\pi_\phi(\vec{y})\right\}=\delta^3(\vec{x}-\vec{y})$. The dynamics is thus described by the total Hamiltonian
\bea
\label{eq:total_Hamiltonian}
	C=\displaystyle\int \dd^3x\left[N\left(\mathcal{C}_G+\mathcal{C}_\phi\right)+N^i\left(\mathcal{C}^G_i+\mathcal{C}^\phi_i\right)\right].
\eea
In this expression, $G$ and $\phi$ stand for the gravitational and the scalar field sectors respectively, $\mathcal{C}=\mathcal{C}_G+\mathcal{C}_\phi$ is the scalar constraint and $\mathcal{C}_i=\mathcal{C}^G_i+\mathcal{C}^\phi_i$ is the spatial-diffeomorphism constraint. For the scalar field, they read
\bea
\label{eq:Cphi}
	\mathcal{C}_\phi&=&\frac{1}{2\sqrt{\gamma}}\pi^2_\phi+\frac{\sqrt{\gamma}}{2}\gamma^{ij}\partial_i\phi\partial_j\phi+\sqrt{\gamma}V(\phi) , \\
	\mathcal{C}^\phi_i&=&\pi_\phi\partial_i\phi ,
\label{eq:Ciphi}
\eea
where $\gamma$ denotes the determinant of $\gamma_{ij}$, and similar expressions can be found for the gravitational sector~\cite{Langlois:1994ec}. Any function $F$ of the phase-space variables then evolves under the Hamilton equations
\bea
\label{eq:Hamilton:constraint}
	\dot{F}(\gamma_{ij},\pi^{kl},\phi,\pi_\phi)=\left\{F,C\right\}.
\eea
Finally, variations of the action with respect to the lapse function and the shift vector show that both $\mathcal{C}$ and $\mathcal{C}_i$ are constrained to be zero.

We first study the case of a test scalar field, for which the gravitational part of the Hamiltonian does not depend on $\phi$ and $\pi_\phi$~\cite{Langlois:1994ec}. The Hamilton equations  $\dot\phi=\left\{\phi,C\right\}$ and $\dot\pi_\phi=\left\{\pi_\phi,C\right\}$ give rise to
\bea
	\dot\phi(\vec{x})&=&\displaystyle\int \dd^3y \left[N(\vec{y})\left\{\phi(\vec{x}),\mathcal{C}_\phi(\vec{y})\right\}+N^i(\vec{y})\left\{\phi(\vec{x}),\mathcal{C}^\phi_i(\vec{y})\right\}\right] , \\
	\dot\pi_\phi(\vec{x})&=&\displaystyle\int \dd^3y \left[N(\vec{y})\left\{\pi_\phi(\vec{x}),\mathcal{C}_\phi(\vec{y})\right\}+N^i(\vec{y})\left\{\pi_\phi(\vec{x}),\mathcal{C}^\phi_i(\vec{y})\right\}\right] ,
\eea	
where the time-dependence is made implicit for display convenience. Making use of \Eqs{eq:Cphi} and~(\ref{eq:Ciphi}), one obtains the local evolution equations
\bea
\label{eq:dotphigen}
	\dot\phi&=&\frac{N}{\sqrt{\gamma}}\pi_\phi+N^i\partial_i\phi ,  \\
	\dot\pi_\phi&=&-N\sqrt{\gamma}V_{,\phi}+\partial_i\left(N\sqrt{\gamma}\gamma^{ij}\partial_j\phi\right)
+N^i\partial_i\pi_\phi	
	+\pi_\phi\partial_iN^i ,
\label{eq:dotpgen}
\eea
where the space dependence is made implicit for clarity and where the last three terms in \Eq{eq:dotpgen} are obtained by integration by parts.

For simplicity, let us assume that $\phi$ is a test field sufficiently decoupled from the metric and other fields perturbations that the latter can be ignored~\cite{Langlois:1994ec, Lyth:2001nq} (this assumption will be relaxed in \Sec{sec:Stochastic:Non:test:fields}). One can drop out perturbations from the line element~(\ref{eq:line_element}) which becomes of the Friedman-Lema\^itre-Robertson-Walker (FLRW) type, and in spatially flat universes, it is given by
\bea
\label{eq:line_element:flat}
	\dd s^2=-N^2(\tau)\dd\tau^2+a^2(\tau)\delta_{ij}\dd x^i\dd x^j ,
\eea
where the lapse function $N$ and the scale factor $a$ depend on time only. In this expression, the shift vector is zero and choosing a lapse function simply means choosing a time variable. For example, $N=1$ corresponds to working with cosmic time,  $N=a$ with conformal time, and $N=1/H$  with the number of $e$-folds. In absence of a shift vector, since \Eq{eq:line_element:flat} also gives rise to $\gamma_{ij}=a^2\delta_{ij}$, the Hamiltonian for the scalar field becomes
\bea
\label{eq:calH:phi}
\mathcal{H}_\phi = \int \dd^3 x \,  \mathcal{C}_\phi = \int \dd^3 x\,  N \left[\frac{{\pi_\phi}^2}{2a^{3}}+\frac{a}{2}\delta^{ij}\partial_i\phi \partial_j\phi + a^{3} V(\phi)\right]
\eea 
and \Eqs{eq:dotphigen} and~(\ref{eq:dotpgen}) simplify to
\bea
\label{eq:dotphiflrw}
	\dot\phi&=&\frac{N}{a^{3}}\pi_\phi ,  \\
	\dot\pi_\phi&=&-Na^{3}V_{,\phi}+Na\Delta\phi , 
\label{eq:dotpflrw}
\eea
where $\Delta\equiv \delta^{ij}\partial_i\partial_j$ is the 3-dimensional flat Laplace operator. 
\subsection{Langevin equation in phase space}
\label{sec:Langevin}
The strategy of the stochastic inflation formalism consists in deriving an effective theory for the long wavelength part of the scalar field $\phi$ by integrating out the small wavelengths. This requires to introduce a time-dependent cut-off in Fourier space
\bea
\label{eq:ksigma}
k_\sigma=\sigma aH ,
\eea
where $H=\dd a/\dd t$ is the Hubble scale and $\sigma$ is the ratio between the Hubble radius and the cut-off wavelength. It disappears from all physical quantities in the limit $\sigma\ll 1$ under conditions that will be carefully discussed in \Secs{sec:quantum_to_classical}, \ref{sec:explicit} and~\ref{sec:slowroll:stochastic:attractor}, see the discussions around \Eqs{eq:cond:classicalTransition}, \eqref{eq:condition:m_over_H} and~\eqref{eq:conditions:sigma:test}. In \Sec{sec:quantum_to_classical}, we will also explain in more details why, on super-Hubble scales, the quantum state of the field is such that it can be treated as a stochastic classical process. In practice, the dynamics of the long-wavelength part of the field can be described by a Langevin equation that we now derive, in which the small-wavelength part of the field provides the noise term as modes continuously cross out $k_\sigma$.

In the Hamiltonian formalism, the coarse-graining is performed in phase space through the decomposition $\phi=\bar\phi+\phi_Q$ and $\pi_\phi=\bar\pi+\pi_Q$, where
\bea
\label{eq:phiq}
	\phi_Q&=&\displaystyle\int_{\mathbb{R}^3}\frac{\dd^3k}{(2\pi)^{3/2}}W\left(\frac{k}{k_\sigma}\right)\left[a_{\vec{k}}~\phi_{\vec{k}}(\tau)e^{-i\vec{k}\cdot\vec{x}}+a^\dag_{\vec{k}}~\phi^\star_{\vec{k}}(\tau)e^{i\vec{k}\cdot\vec{x}}\right] \\
	\pi_Q&=&\displaystyle\int_{\mathbb{R}^3}\frac{\dd^3k}{(2\pi)^{3/2}}W\left(\frac{k}{k_\sigma}\right)\left[a_{\vec{k}}~\pi_{\vec{k}}(\tau)e^{-i\vec{k}\cdot\vec{x}}+a^\dag_{\vec{k}}~\pi^\star_{\vec{k}}(\tau)e^{i\vec{k}\cdot\vec{x}}\right]
 \label{eq:piq}
\eea
are the small-wavelength parts of $\phi$ and $\pi_\phi$ defined through the window function $W$ such that $W\simeq0$ for $k \ll k_\sigma$ and $W\simeq1$ for $k \gg k_\sigma$, and $\bar{\phi}$ and $\bar{\pi}$ are the long-wavelength, or coarse-grained, parts of $\phi$ and $\pi_\phi$. In \Eqs{eq:phiq} and~(\ref{eq:piq}), $a_{\vec{k}}$ and $a^\dag_{\vec{k}}$ are annihilation and creation operators satisfying the usual commutation relations $[a_{\vec{k}},a^\dag_{\vec{k}'}]=\delta^3(\vec{k}-\vec{k}')$ and $[a_{\vec{k}},a_{\vec{k}'}]=[a^\dag_{\vec{k}},a^\dag_{\vec{k}'}]=0$. They are time independent, contrary to the mode functions $\phi_{\vec{k}}$ and $\pi_{\vec{k}}$ that are solutions of the linearised \Eqs{eq:dotphiflrw} and~(\ref{eq:dotpflrw}), which in spatial Fourier space read
\bea
\label{eq:eomphiq}
	\dot\phi_k&=&\frac{N}{a^{3}}\pi_k ,  \\
	\dot\pi_k&=&-Na^{3}V_{,\phi\phi}(\phi)\phi_k-Na k^2\phi_k . 
\label{eq:eompiq}
\eea
We note that if the initial state is statistically isotropic, because \Eqs{eq:eomphiq} and~(\ref{eq:eompiq}) only involve the norm of the wavenumber $k$, the mode functions depend only on $k$ as well. This is why hereafter, $\phi_{\vec{k}}$ and $\pi_{\vec{k}}$ are simply written $\phi_k$ and $\pi_k$. The canonical quantisation of the short-wavelength modes is made using the Klein-Gordon product as an inner product~\cite{birrell1982}. They are thus normalised according to\footnote{This normalisation is equivalent to the one performed in the Lagrangian approach. In the Hamiltonian formalism indeed, one has ${\Pi}_{\vec{k}}=\sqrt{\gamma}(\partial_\tau{\Phi}_{\vec{k}}-N^i\partial_i{\Phi}_{\vec{k}})/N$. By plugging this expression into $i\int_{\Sigma_\tau}\dd^3x({\Phi}_{\vec{k}}{\Pi}^\star_{\vec{k}'}-{\Pi}_{\vec{k}}{\Phi}^\star_{\vec{k}'})$, the standard Klein-Gordon product is obtained, $i\int_{\Sigma_\tau}\dd^3x\sqrt{\gamma}n^\mu ({\Phi}_{\vec{k}}\partial_\mu{\Phi}^\star_{\vec{k}'}-{\Phi}^\star_{\vec{k}'}\partial_\mu{\Phi}_{\vec{k}})$, where $n^\mu=(1/N,N^i/N)$ is a unit 4-vector orthogonal to $\Sigma_\tau$.}
$i\int_{\Sigma_\tau}\dd^3x({\Phi}_{\vec{k}}{\Pi}^\star_{\vec{k}'}-{\Pi}_{\vec{k}}{\Phi}^\star_{\vec{k}'})=\delta^3(\vec{k}-\vec{k'})$, where ${\Phi}_{\vec{k}}=\phi_k(\tau)e^{-i\vec{k}\cdot\vec{x}}$ and ${\Pi}_{\vec{k}}=\pi_k(\tau)e^{-i\vec{k}\cdot\vec{x}}$, and where $\Sigma_\tau$ is a space-like hypersurface of fixed time $\tau$. 

The Langevin equation for the long-wavelength part of the field is then obtained by plugging the decomposition 
\bea
\label{eq:field;decomposition:phi}
\phi=\bar\phi+\phi_Q\\
\pi_\phi=\bar\pi+\pi_Q
\label{eq:field;decomposition:pi}
\eea
into \Eqs{eq:dotphiflrw} and~(\ref{eq:dotpflrw}). Linearising these equations in $\phi_Q$ and $\pi_Q$, one obtains
\bea
\label{eq:Hamilton:split:phi}
	\dot{\bar\phi}&=&\frac{N}{a^{3}}\bar\pi-\dot\phi_Q+\frac{N}{a^{3}}\pi_Q , \\
	\dot{\bar\pi}&=&-Na^{3}V_{,\phi}(\bar\phi)-\dot\pi_Q-Na^{3}V_{,\phi\phi}(\bar\phi)\phi_Q+Na \Delta\phi_Q .
\label{eq:Hamilton:split:pi}
\eea
In these expressions, the Laplacian of $\bar\phi$ has been dropped since it is suppressed by $\sigma$. Replacing $\phi_Q$ and $\pi_Q$ by \Eqs{eq:phiq} and~(\ref{eq:piq}), and making use of the fact that the mode functions $\phi_k$ and $\pi_k$ satisfy \Eqs{eq:eomphiq} and~(\ref{eq:eompiq}), the Hamilton equations for $\bar\phi$ and $\bar\pi$ can be written as~\cite{Nakao:1988yi, PhysRevD.46.2408, Rigopoulos:2005xx, Tolley:2008na, Weenink:2011dd}
\bea
	\dot{\bar\phi}&=&\frac{N}{a^{3}}\bar\pi+\xi_\phi(\tau), \label{eq:eombarphi} \\
	\dot{\bar\pi}&=&-Na^{3}V_{,\phi}(\bar\phi)+\xi_\pi(\tau), \label{eq:eombarpi}
\eea
where the quantum noises $\xi_\phi$ and $\xi_\pi$ are given by
\bea
	\xi_\phi&=&-\displaystyle\int_{\mathbb{R}^3}\frac{\dd^3k}{(2\pi)^{3/2}}\dot{W}\left(\frac{k}{k_\sigma}\right)\left[a_{\vec{k}}\phi_k(\tau)e^{-i\vec{k}\cdot\vec{x}}+a^\dag_{\vec{k}}\phi^\star_k(\tau)e^{i\vec{k}\cdot\vec{x}}\right] , \label{eq:noisephi}\\
	\xi_\pi&=&-\displaystyle\int_{\mathbb{R}^3}\frac{\dd^3k}{(2\pi)^{3/2}}\dot{W}\left(\frac{k}{k_\sigma}\right)\left[a_{\vec{k}}\pi_k(\tau)e^{-i\vec{k}\cdot\vec{x}}+a^\dag_{\vec{k}}\pi^\star_k(\tau)e^{i\vec{k}\cdot\vec{x}}\right] . \label{eq:noisepi}
\eea
\subsection{Statistical properties of the noise}
\label{ssec:noise}
Let us now assume that the field fluctuations $\phi_k$ and $\pi_k$ are placed in their vacuum state. Since we work at linear order in perturbation theory, they thus feature Gaussian statistics with vanishing mean. The statistical properties of the quantum noises $\xi_\phi$ and $\xi_\pi$ are therefore fully characterised by their two-point correlation matrix 
\bea
 \label{eq:xidef}
	\boldsymbol{\Xi} \left (\vec{x}_1,\tau_1;\vec{x}_2,\tau_2\right )=\left(\begin{array}{cc}
		\left<0\right|\xi_\phi(\vec{x}_1,\tau_1)\xi_\phi(\vec{x}_2,\tau_2)\left|0\right> & \left<0\right|\xi_\phi(\vec{x}_1,\tau_1)\xi_\pi(\vec{x}_2,\tau_2)\left|0\right> \\
		\left<0\right|\xi_\pi(\vec{x}_1,\tau_1)\xi_\phi(\vec{x}_2,\tau_2)\left|0\right> & \left<0\right|\xi_\pi(\vec{x}_1,\tau_1)\xi_\pi(\vec{x}_2,\tau_2)\left|0\right>
	\end{array}\right) .
\eea
Hereafter, bold notations denote vector or matrix quantities. Letting the annihilation and creation operators act on the vacuum state $|0\rangle$, the entries of this matrix read
\bea
	\Xi_{f_1,g_2}=\displaystyle\int_{\mathbb{R}^3}\frac{\dd^3k}{(2\pi)^3}\dot{W}\left[\frac{k}{k_\sigma(\tau_1)}\right]\dot{W}\left[\frac{k}{k_\sigma(\tau_2)}\right]f_k(\tau_1)g^\star_k(\tau_2)e^{i\vec{k}\cdot(\vec{x}_2-\vec{x}_1)} ,
\eea
where the notation $\Xi_{f_1,g_2}=\left<0\right|\xi_f(\vec{x}_1,\tau_1)\xi_g(\vec{x}_2,\tau_2)\left|0\right>$ has been introduced for display convenience, $f$ and $g$ being either $\phi$ or $\pi$. Note that, at this stage, the order of the subscripts $f$ and $g$ does matter as a result of the non-commutativity of $\xi_\phi$ and $\xi_\pi$. The angular integral over $\vec{k}/k$ can be performed easily since, as explained below  \Eqs{eq:eomphiq} and~(\ref{eq:eompiq}), the mode functions $\phi_k$ and $\pi_k$ only depend on the norm of $\vec{k}$. One obtains
\bea
	\Xi_{f_1,g_2}=\displaystyle\int_{\mathbb{R}^+}\frac{k^2\dd k}{2\pi^2}
	\dot{W}\left[\frac{k}{k_\sigma(\tau_1)}\right]\dot{W}\left[\frac{k}{k_\sigma(\tau_2)}\right]f_k(\tau_1)g^\star_k(\tau_2)
	\frac{\sin\left(k \vert \vec{x}_2-\vec{x}_1 \vert\right)}{k \vert \vec{x}_2-\vec{x}_1 \vert } .
\label{eq:Xif1g2:kintegrated}
\eea
We now need to specify the filter function $W$, which for simplicity we choose to be a Heaviside function $W(k/k_\sigma)=\Theta\left(k/k_{\sigma}-1\right)$. Its time derivative gives a Dirac distribution, and the integrand of \Eq{eq:Xif1g2:kintegrated} contains $\delta[k-k_\sigma(\tau_1)]\delta[k-k_\sigma(\tau_2)]$ which yields $\delta(\tau_1-\tau_2)$, meaning that the noises are white. One obtains
\bea
 \label{eq:noisecorrel}
	\Xi_{f_1,g_2}=\frac{1}{6\pi^2}\left.\frac{\dd k^3_\sigma(\tau)}{\dd\tau}\right|_{\tau_1}f_{k=k_\sigma(\tau_1)}g^\star_{k=k_\sigma(\tau_1)}\frac{\sin\left[k_\sigma(\tau_1) \vert \vec{x}_2-\vec{x}_1 \vert \right]}{k_\sigma(\tau_1)\vert \vec{x}_2-\vec{x}_1 \vert}\delta\left(\tau_1-\tau_2\right) .
\eea
In the following, we will be essentially interested in the autocorrelation of the noises, $\vec{x}_1=\vec{x}_2$, for which $\sin[k_\sigma(\tau_1) \vert \vec{x}_2-\vec{x}_1 \vert ]/[k_\sigma(\tau_1) \vert \vec{x}_2-\vec{x}_1 \vert ]=1$. The noises being white, the correlations are non-zero only at equal time. We will thus write the correlation matrix of the noise as $\Xi_{f_1,g_2}\equiv \Xi_{f,g}(\tau_1)\delta(\tau_1-\tau_2)$. The correlator $ \Xi_{f,g}(\tau)$ can be expressed in terms of the power spectrum of the quantum fluctuations
\begin{equation}
\label{eq:calP:def}
	 \mathcal{P}_{f,g}(k;\tau)=\frac{k^3}{2\pi^2}f_{k}(\tau)g^\star_{k}(\tau) ,
\end{equation}
which gives rise to
\begin{equation}
\label{eq:noisecorrel_Pk}
	\Xi_{f,g}(\tau)=\frac{\dd \ln\left[k_\sigma(\tau)\right]}{\dd\tau} \mathcal{P}_{f,g}\left[k_\sigma(\tau);\tau\right] . 
\end{equation}

Let us finally notice that the noises correlator is described by an hermitian matrix, i.e $\Xi^\star_{f,g}=\Xi_{g,f}$, the antisymmetric part of which is proportional to the Klein-Gordon product of the mode functions
\bea
	\Xi_{\phi,\pi}(\tau)-\Xi_{\pi,\phi}(\tau)&=&\frac{1}{6\pi^2}\frac{\dd k^3_\sigma(\tau)}{\dd\tau}\left[\phi_{k=k_\sigma(\tau)}\pi^\star_{k=k_\sigma(\tau)}-\pi_{k=k_\sigma(\tau)}\phi^\star_{k=k_\sigma(\tau)}\right]\\
	&=&\frac{-i}{6\pi^2}\frac{\dd k^3_\sigma(\tau)}{\dd\tau} .
\label{eq:commutator:normalised}
\eea
In this expression, the second equality is obtained by using canonical quantisation of the fluctuations which sets the Klein-Gordon product to $-i$, as explained below \Eq{eq:eompiq}.
\subsection{Generic solution for a free scalar field}
\label{sec:testField}
Let us now consider the case of a test scalar field with quadratic potential $V(\phi)=\Lambda^4+m^2\phi^2/2$ (terms linear in $\phi$ can always be reabsorbed by field shift symmetry). If $m^2>0$, the potential is convex and of the large-field type, if $m^2<0$ it is concave and of the hilltop type. For such a potential, \Eqs{eq:eombarphi} and~(\ref{eq:eombarpi}) form a linear differential system where the noises $\xi_\phi$ and $\xi_\pi$ do not depend on the phase-space variables of the coarse-grained field. This yields simplifications that allow one to analytically solve the full stochastic dynamics, and clearly highlight some salient features of the stochastic approach.
\subsubsection{Probability distribution in phase space}
\label{ssec:fokker}
Since the dynamics described by \Eqs{eq:eombarphi} and~(\ref{eq:eombarpi}) is linear, it is convenient to work with the vector notation
\bea
\boldsymbol{\Phi}=\left(\begin{array}{c}
	\bar\phi \\
	\bar\pi
\end{array}\right)~~~\mathrm{and}~~~\boldsymbol{\xi}=\left(\begin{array}{c}
	\xi_\phi \\
	\xi_\pi
\end{array}\right) .
\eea
In terms of these variables, \Eqs{eq:eombarphi} and~(\ref{eq:eombarpi}) can be written as the Langevin equation
\bea
 \label{eq:inhomopb}
	\dot{\boldsymbol{\Phi}}=\boldsymbol{A}(\tau)\boldsymbol{\Phi}+\boldsymbol{\xi}(\tau)~~~\mathrm{with}~~~\boldsymbol{A}(\tau)=\left(\begin{array}{cc}
	0 & N/a^{3} \\
	-m^2Na^{3} & 0						\end{array}\right) .
\eea
\paragraph{Fokker-Planck equation}
This Langevin equation can be translated into a Fokker-Planck equation~\cite{risken1996fokker} for the probability density function (PDF hereafter) in phase space associated to the stochastic process~(\ref{eq:inhomopb}), given by
\bea
\label{eq:fokkerxi}
	\frac{\partial P(\boldsymbol{\Phi},\tau)}{\partial\tau}=-\displaystyle\sum_{i,j=1}^2\frac{\partial}{\partial\Phi}_i\left[A_{ij} \Phi_j P(\boldsymbol{\Phi},\tau)\right]+\frac{1}{2}\displaystyle\sum_{i,j=1}^2\frac{\partial^2}{\partial\Phi_i\partial\Phi_j}\left[\Xi_{ij}(\tau)P(\boldsymbol{\Phi},\tau)\right] . 
\eea
A generic derivation of the Fokker-Planck equation from the Langevin equation is given below in \Sec{sec:FP}.
The first term in the right-hand side of \Eq{eq:fokkerxi} is called the drift term and traces the deterministic part of the dynamics, and the second term is the diffusive term that traces the stochastic component of the evolution. In the latter, $\Xi_{ij}(\tau)$ can be factored out of the phase-space differential operator since it does not depend on $\boldsymbol{\Phi}$. This term can therefore be written as $\mathrm{Tr}\left[\boldsymbol{H}\,\boldsymbol{\Xi}\right]/2$, where $\mathrm{Tr}$ is the trace operator and $H_{ij}\equiv \partial^2P(\boldsymbol{\Phi},\tau)/(\partial \Phi_i\partial \Phi_j)$ is the Hessian of the PDF.
\paragraph{Losing the commutator}
\label{sec:lossing_commutator}
As noticed below \Eq{eq:noisecorrel_Pk}, the noise correlator matrix $\boldsymbol{\Xi}$ is hermitian and it can thus be decomposed on the basis $\{\boldsymbol{I},\boldsymbol{J}_x,\boldsymbol{J}_y,\boldsymbol{J}_z\}$, where $\boldsymbol{I}$ is the $2\times2$ identity matrix and the three following matrices are the Pauli matrices. The decomposition reads
\bea
\label{eq:Pauli}
	\boldsymbol{\Xi}=\frac{1}{2}\left(\Xi_{\phi,\phi}+\Xi_{\pi,\pi}\right)\boldsymbol{I}+\frac{1}{2}\left(\Xi_{\phi,\pi}+\Xi_{\pi,\phi}\right)\boldsymbol{J}_x+\frac{i}{2}\left(\Xi_{\phi,\pi}-\Xi_{\pi,\phi}\right)\boldsymbol{J}_y+\frac{1}{2}\left(\Xi_{\phi,\phi}-\Xi_{\pi,\pi}\right)\boldsymbol{J}_z .\quad
\eea
In this expression, the coefficient multiplying $\boldsymbol{J}_y$ is built from the commutator of the quantum fluctuations~(\ref{eq:commutator:normalised}) and thus traces the very quantum nature of the noise. However, its contribution to the Fokker-Planck equation vanishes. Indeed, the Hessian of the PDF, $H_{ij}$, is symmetric with respect to the indices $i$ and $j$ while the matrix $\boldsymbol{J}_y$ is antisymmetric. As a consequence, $\mathrm{Tr}\left[\boldsymbol{H}\boldsymbol{J}_y\right]=0$ and the quantum commutator disappears from \Eq{eq:fokkerxi}. This makes sense since it implies that, if one wants to describe the full quantum dynamics by a stochastic theory, one looses the information about the commutators. The reason why it provides a good approximation is because these commutators become negligible on large scales, as will be made more explicit in \Sec{sec:quantum_to_classical}. As a consequence, the symmetric terms of $\boldsymbol{\Xi}$ remaining in the Fokker-Planck equation, although drawn out from a quantum state, can be equivalently described by a classical (though correlated) distribution. Defining the diffusion matrix $\boldsymbol{D}$ as the symmetric part of $\boldsymbol{\Xi}$, \ie $ \boldsymbol{D}=(\Xi_{\phi,\phi}+\Xi_{\pi,\pi})\boldsymbol{I}/2+(\Xi_{\phi,\pi}+\Xi_{\pi,\phi})\boldsymbol{J}_x/2+(\Xi_{\phi,\phi}-\Xi_{\pi,\pi})\boldsymbol{J}_z/2$, the Fokker-Planck equation is then given by
\bea
\label{eq:fokker}
	\frac{\partial P(\boldsymbol{\Phi},\tau)}{\partial\tau}=-\displaystyle\sum_{i,j=1}^2\frac{\partial}{\partial \Phi_i}\left[A_{ij} {\Phi}_jP(\boldsymbol{\Phi},\tau)\right]+\frac{1}{2}\displaystyle\sum_{i,j=1}^2 {D}_{ij}(\tau)\frac{\partial^2P(\boldsymbol{\Phi},\tau)}{\partial {\Phi}_i\partial {\Phi}_j} . 
\eea
\paragraph{Green formalism}
Because the Langevin equation~(\ref{eq:inhomopb}), or equivalently the Fokker-Planck equation~(\ref{eq:fokker}), is linear, it can be analytically solved using the Green's matrix formalism. Let us consider the linear homogeneous system associated to the stochastic dynamics of \Eq{eq:inhomopb}, $\dot{\boldsymbol{\Phi}}=\boldsymbol{A}(\tau)\boldsymbol{\Phi}$, and let us assume that two independent solutions $(\bar\phi^{(1)},\bar\pi^{(1)})$ and $(\bar\phi^{(2)},\bar\pi^{(2)})$ are known. The so-called ``fundamental'' matrix of the system is defined as
\bea
\label{eq:U:def}
\boldsymbol{U}(\tau)=\left(\begin{array}{cc}
	\bar\phi^{(1)} & \bar\phi^{(2)} \\
	\bar\pi^{(1)} & \bar\pi^{(2)}
\end{array}\right).
\eea
By construction, one can check that $\dd \boldsymbol{U}(\tau)/\dd \tau =\boldsymbol{A}(\tau)\boldsymbol{U}(\tau)$. Let us also notice that since the two solutions are independent, $\det(\boldsymbol{U})\neq0$. The matrix $\boldsymbol{U}$ is then invertible and gives rise to the Green's matrix
\bea
\label{eq:Green:fundamental}
	\boldsymbol{G}(\tau,\tau_0)&=&\boldsymbol{U}(\tau)\left[\boldsymbol{U}(\tau_0)\right]^{-1}
	\Theta(\tau-\tau_0) ,
\eea
which satisfies $\partial \boldsymbol{G}(\tau,\tau_0)/\partial\tau=\boldsymbol{A}(\tau)\boldsymbol{G}(\tau,\tau_0)+\boldsymbol{I}\delta(\tau-\tau_0)$, where $\boldsymbol{I}$ is the identity matrix. One can also note that ${\dd}\det[\boldsymbol{U}(\tau)]/{\dd\tau}=\mathrm{Tr}\left[\boldsymbol{A}(\tau)\right]\det[\boldsymbol{U}(\tau)]$ with ``$\mathrm{Tr}$'' being the trace operation. The coefficients matrix $\boldsymbol{A}$, defined in \Eq{eq:inhomopb}, is traceless and $\det[\boldsymbol{U}(\tau)]$ is thus a conserved quantity. It is therefore sufficient to find two solutions such as $\det[\boldsymbol{U}(\tau_0)]\neq0$, and this ensures the Green's matrix to be properly defined throughout the evolution. In this case, one can always normalise the two independent solutions so that $\det[\boldsymbol{U}(\tau)]=1$. But even if this is not the case, the Green's matrix is such that $\det[\boldsymbol{G}(\tau,\tau_0)]=1$, which is easily derived from the fact that the determinant of $\boldsymbol{U}$ is conserved through evolution. 

The generic solution to the homogenous problem then reads
\bea
\label{eq:Phidet}
\boldsymbol{\Phi}_{\mathrm{det}}(\tau)=\boldsymbol{G}(\tau,\tau_0)\boldsymbol{\Phi}_0 .
\eea 
Here, $\boldsymbol{\Phi}_{\mathrm{det}}$ is the deterministic trajectory that field variables would follow in the absence of the noises and starting from the initial state $\boldsymbol{\Phi}(\tau_0)=\boldsymbol{\Phi}_0$. Solutions of the Fokker-Planck equation~(\ref{eq:fokker}) can be formally obtained introducing the Green function $\mathcal{W}(\boldsymbol{\Phi},\tau|\boldsymbol{\Phi}_0,\tau_0)$,  giving the PDF in phase space at time $\tau$ if the field and its momentum are initially at $\boldsymbol{\Phi}(\tau_0)=\boldsymbol{\Phi}_0$, through 
\bea
P\left(\boldsymbol{\Phi},\tau\right)=\displaystyle\int \dd\boldsymbol{\Phi}_0\mathcal{W}(\boldsymbol{\Phi},\tau|\boldsymbol{\Phi}_0,\tau_0)P\left(\boldsymbol{\Phi}_0,\tau_0\right) .
\eea
For the Fokker-Planck equation~(\ref{eq:fokker}), the Green function is the gaussian distribution
\bea
\label{eq:Gaussian:Green}
	\mathcal{W}\left(\boldsymbol{\Phi},\tau|\boldsymbol{\Phi}_0,\tau_0\right)=\displaystyle\frac{1}{\sqrt{2\pi^2\det\left[\boldsymbol{\Sigma}_{\boldsymbol{\Phi}}(\tau)\right]}}\exp\left\lbrace-\frac{1}{2}\left[\boldsymbol{\Phi}-\boldsymbol{\Phi}_{\mathrm{det}}(\tau)\right]^\dag\boldsymbol{\Sigma}^{-1}_{\boldsymbol{\Phi}}(\tau)\left[\boldsymbol{\Phi}-\boldsymbol{\Phi}_{\mathrm{det}}(\tau)\right]\right\rbrace, 
\eea
where $\dag$ means the conjugate-transpose. From this expression, one can check that  $\left<\boldsymbol{\Phi}(\tau)\right>=\int\dd\boldsymbol{\Phi}~\boldsymbol{\Phi}\mathcal{W}(\boldsymbol{\Phi},\tau|\boldsymbol{\Phi}_0,\tau_0)$ is equal to $\boldsymbol{\Phi}_{\mathrm{det}}$, which means that the deterministic trajectory is also the average trajectory of the stochastic field since the noises have a vanishing mean, in agreement with Ehrenfest theorem. In \Eq{eq:Gaussian:Green}, $\boldsymbol{\Sigma}_{\boldsymbol{\Phi}}$ is the covariance matrix of the field variables that captures all the diffusive processes. It is obtained as the forward propagation of the diffusion matrix,
\bea
\label{eq:Sigma}
	\boldsymbol{\Sigma}_{\boldsymbol{\Phi}}(\tau)=\displaystyle\int^\tau_{\tau_0}\dd s~\boldsymbol{G}(\tau,s)\boldsymbol{D}(s)\boldsymbol{G}^\dag(\tau,s) ,
\eea	
and is related to the two-point correlation of the coarse-grained field through $\langle\left[\boldsymbol{\Phi}(\tau)-\left<\boldsymbol{\Phi}(\tau)\right>\right]\left[\boldsymbol{\Phi}(\tau)-\left<\boldsymbol{\Phi}(\tau)\right>\right]^\dag\rangle=\boldsymbol{\Sigma}_{\boldsymbol{\Phi}}(\tau)$. 
\subsubsection{Quantum-to-classical transition in the stochastic picture}
\label{sec:quantum_to_classical}
In \Sec{sec:lossing_commutator}, it was shown that a description of the full quantum dynamics in terms of a Fokker-Planck equation necessarily drops out the commutators of the theory, and can therefore only provide an approximation to the actual results. In this section, we show that this approximation becomes accurate in the limit $\sigma\ll 1$, illustrating the ``quantum-to-classical'' transition of inflationary perturbations on super-Hubble scales.

To this end we evaluate the moments of the coarse-grained scalar field with and without assuming the noises to be in a quasiclassical state, \ie, with and without resorting to the Fokker-Planck equation. Let us consider the case where the coarse-grained field is in a classical state $\boldsymbol{\Phi}_0$ at initial time. Under \Eq{eq:inhomopb}, it becomes a mixture of a classical state and quantum operators at later time, through the contribution of the quantum noise $\boldsymbol{\xi}$. Under \Eq{eq:fokker} however, it simply becomes a random variable. The solution to \Eq{eq:inhomopb} is formally given by
\bea
\label{eq:quant_Sol}
	\boldsymbol{\Phi}_{\mathrm{quant}}(\tau)=\boldsymbol{G}(\tau,\tau_0)\boldsymbol{\Phi}_0+\displaystyle\int_{\tau_0}^\tau\dd s\boldsymbol{G}(\tau,s) \boldsymbol{\xi}(s) ,
\eea
where the subscript ``quant'' stresses that we are dealing with the solution of the full quantum-field theory, while the solution to \Eq{eq:fokker} has already be given in \Eqs{eq:Gaussian:Green}, (\ref{eq:Phidet}) and~(\ref{eq:Sigma}).
\paragraph{Linear observables} 
Evaluating \Eq{eq:quant_Sol} on the vacuum, since $\langle 0 \vert \xi \vert 0 \rangle=0$, one obtains
\bea
	\left\langle 0 \left\vert \boldsymbol{\Phi}_{\mathrm{quant}} \right\vert 0 \right\rangle=\boldsymbol{G}\left(\tau,\tau_0\right)\boldsymbol{\Phi}_0=\boldsymbol{\Phi}_\mathrm{det} ,
\eea
as follows from \Eq{eq:Phidet}. This also corresponds to the solution for the mean stochastic path as noticed below \Eq{eq:Gaussian:Green}, so the full quantum and the stochastic theories match for linear observables.
\paragraph{Quadratic observables} 
\label{sec:QuantumToClassical:Quadratic}
The initial coarse-grained state $\boldsymbol{\Phi}_0$ being classical, it commutes with the noise and evaluating the square of \Eq{eq:quant_Sol} on the vacuum gives rise to
\bea
\label{eq:quadmoment:quant}
	\left\langle 0 \left\vert  \boldsymbol{\Phi}_\mathrm{quant}\boldsymbol{\Phi}_\mathrm{quant}^\dag\right\vert 0 \right\rangle -
	\left\langle 0 \left\vert  \boldsymbol{\Phi}_\mathrm{quant}\right\vert 0 \right\rangle 
	\left\langle 0 \left\vert  \boldsymbol{\Phi}_\mathrm{quant}^\dag\right\vert 0 \right\rangle 
	= \displaystyle\int^\tau_{\tau_0}\dd s~\boldsymbol{G}(\tau,s)\boldsymbol{\Xi}(s)\boldsymbol{G}^\dag(\tau,s) .
\eea
On the other hand, the corresponding expression for the stochastic solution is given below \Eq{eq:Sigma}. It is identical to \Eq{eq:quadmoment:quant} except that $\boldsymbol{\Xi}$ is replaced by $\boldsymbol{D}$. The difference it yields can be thus quantified through
\bea
	\boldsymbol{\Delta}&\equiv&
	\left\langle 0 \left\vert  \boldsymbol{\Phi}_\mathrm{quant}\boldsymbol{\Phi}_\mathrm{quant}^\dag \right\vert 0 \right\rangle -
	\langle\boldsymbol{\Phi}\boldsymbol{\Phi}^\dag\rangle\\
	&=&\frac{i}{2}\displaystyle\int^\tau_{\tau_0}\dd s~\left[\Xi_{\phi,\pi}(s)-\Xi_{\pi,\phi}(s)\right]\boldsymbol{G}(\tau,s)\boldsymbol{J}_y\boldsymbol{G}^\dag(\tau,s),
\label{eq:quadratic:Delta}
\eea
where the decomposition of $\boldsymbol{\Xi}$ in terms of the Pauli matrices introduced in \Eq{eq:Pauli} has been used. Since $\boldsymbol{G}$ is a symplectic matrix, one has\footnote
{\label{footnote:symplectic}A symplectic matrix $M$ is a real matrix satisfying $\boldsymbol{M}^\mathrm{T}\boldsymbol{\Omega} \boldsymbol{M}=\boldsymbol{\Omega}$, where $T$ means transpose and 
\bea
\label{eq:def:Omega}
\boldsymbol{\Omega}\equiv \left(\begin{array}{cc} 0 & 1 \\ -1 & 0 \end{array}\right) .
\eea
In 2 dimensions, it is easy to show that the symplectic matrices are the real matrices with unit determinant, which is the case of the Green matrix. More generally, evolution generated by quadratic Hamiltonians can always be viewed as the action of the symplectic group on the phase-space variables which also explains why $\boldsymbol{G}$ is symplectic. Since $\boldsymbol{J}_y=-i\boldsymbol{\Omega}$, this means that $\boldsymbol{G}\boldsymbol{J}_y\boldsymbol{G}^\dag=\boldsymbol{J}_y$.
}
$\boldsymbol{G}\boldsymbol{J}_y\boldsymbol{G}^\dag=\boldsymbol{J}_y$, which can be factored out of the integral as it does not depend on time. Making use of \Eq{eq:commutator:normalised}, the time integration can be performed and one obtains
\bea
\label{eq:Delta:quad}
	\boldsymbol{\Delta}=\frac{1}{12\pi^2}\left[k_\sigma^3(\tau)-k_\sigma^3(\tau_0)\right]\boldsymbol{J}_y
	=\sigma^3\frac{a^3(\tau)H^3(\tau)}{12\pi^2}\left[1-\frac{a^3(\tau_0)H^3(\tau_0)}{a^3(\tau)H^3(\tau)}\right]\boldsymbol{J}_y ,
\eea
where in the second equality one has made use of \Eq{eq:ksigma}. Since $\boldsymbol{J}_y$ is an off-diagonal matrix, this needs to be compared to the $\phi,\pi$ component of the covariance matrix. In \Sec{sec:massless}, by integrating the mode equations~(\ref{eq:eomphiq}) and~(\ref{eq:eompiq}), it is shown that for a massless field, at leading order in $\sigma$,  one has
\bea
\left.  {\Sigma}_{\phi,\pi}\right\vert_{m=0} = \sigma^2\frac{a^3(\tau)H^3(\tau)}{12\pi^2}\left[1-\frac{a^3(\tau_0)H^3(\tau_0)}{a^3(\tau)H^3(\tau)}\right] ,
\eea
while for a light test field with a non-vanishing mass $m\ll H$, it will be shown in \Sec{sec:light} that
\bea
 {\Sigma}_{\phi,\pi} = 
 \left\lbrace
 \begin{array}{ll}
\dfrac{m^2}{H^2} \dfrac{a^3(\tau)H^3(\tau)}{12\pi^2} \ln\left[\dfrac{a(\tau_0)H(\tau_0)}{a(\tau)H(\tau)}\right] &
\mathrm{if}\   \ln\left[\dfrac{a(\tau)H(\tau)}{a(\tau_0)H(\tau_0)}\right] \ll \dfrac{3H^2}{2m^2} \\
 \dfrac{ a^3(\tau) H^3(\tau)}{8\pi^2}& 
\mathrm{if}\   \ln\left[\dfrac{a(\tau)H(\tau)}{a(\tau_0)H(\tau_0)}\right] \gg \dfrac{3H^2}{2m^2} 
 \end{array}
 \right.
  .
\eea
In all cases, one can see that $ {\Delta}_{\phi,\pi}$ is suppressed by higher powers of $\sigma$ and can therefore be neglected if $\sigma\ll 1$, with the slightly stronger condition $\sigma\ll (m/H)^{2/3}$ for a light test field at early time.

In fact, the limit $\sigma\ll 1$ does not need to be invoked if one restricts correlators to observable operators. Indeed, observables are necessarily described in terms of hermitian operators, since the outcome of a quantum measurement can only be a real number. This is why instead of $\boldsymbol{\Phi}\boldsymbol{\Phi}^\dag$ in \Eq{eq:quadmoment:quant}, one should consider $(\boldsymbol{\Phi}\boldsymbol{\Phi}^\dag+\boldsymbol{\Phi}^\star\boldsymbol{\Phi}^\mathrm{T})/2$. Since $\boldsymbol{G}$ is real, this means that $\boldsymbol{\Xi}$ in \Eq{eq:quadmoment:quant} must be replaced with $(\boldsymbol{\Xi}+\boldsymbol{\Xi}^\star)/2$, and since $\boldsymbol{\Xi}$ is hermitian, this is identical to $(\boldsymbol{\Xi}+\boldsymbol{\Xi}^\mathrm{T})/2$. This precisely corresponds to $\boldsymbol{D}$, the symmetric part of $\boldsymbol{\Xi}$. In this case, one thus recovers the predictions of the stochastic theory, even without resorting to the large-scale limit.
\paragraph{Quartic observables} 
In terms of observable correlators, the full quantum and the stochastic theories give the exact same results for linear and quadratic operators. Since cubic powers of Gaussian noises vanish, they also match for cubic correlators and one has to consider quartic observables to start probing observable deviations between the stochastic framework and the full quantum theory. When calculating such correlators, one has to evaluate
\bea
\left\langle 0 \left\vert \boldsymbol{\xi}(\tau_1)\boldsymbol{\xi}(\tau_2)\boldsymbol{\xi}(\tau_3)\boldsymbol{\xi}(\tau_4) \right\vert 0 \right\rangle = 3 \delta\left(\tau_1-\tau_2\right)\delta\left(\tau_1-\tau_3\right)\delta\left(\tau_1-\tau_4\right) \boldsymbol{\Xi}^2 ,
\eea
which can easily be derived from \Eqs{eq:noisephi} and~(\ref{eq:noisepi}) using a Heaviside window function as in \Sec{ssec:noise}, and which simply translates the Gaussian character of $\boldsymbol{\xi}$. Real correlators are therefore encoded in the real part of $\boldsymbol{\Xi}^2$, while in the stochastic framework they are given by $\boldsymbol{D}^2$. The difference between the two theories is therefore characterised by
\bea
\boldsymbol{\Xi}^2+\boldsymbol{\Xi}^{2\star} - \boldsymbol{D}^2 - \boldsymbol{D}^{2\star} &=& \frac{1}{2} \left(\boldsymbol{\Xi}-\boldsymbol{\Xi}^\mathrm{T}\right)^2 = -\frac{1}{2} \left(\Xi_{\phi,\pi}-\Xi_{\pi,\phi}\right)^2 \boldsymbol{I}\\ 
&=& \frac{1}{72\pi^4}\left(\frac{\dd k_\sigma^3}{\dd\tau}\right)^2 \boldsymbol{I} = \frac{N^2a^6H^8}{8\pi^4}  \sigma^6\boldsymbol{I}   ,
\label{eq:diff:quartic}
\eea
where we have used that $\boldsymbol{D}=(\boldsymbol{\Xi}+\boldsymbol{\Xi}^\mathrm{T})/2$. Contrary to \Eq{eq:Delta:quad} where $\boldsymbol{J}_y$ is purely imaginary, this difference is real, hence observable. Unsurprisingly, it is proportional to the antisymmetric part of $\boldsymbol{\Xi}$, which can be evaluated using \Eq{eq:Pauli}, and where we have further used that $\boldsymbol{J}^2=\boldsymbol{I}$.  In \Eq{eq:diff:quartic}, the commutator $\Xi_{\phi,\pi}-\Xi_{\pi,\phi}$ is expressed using \Eqs{eq:commutator:normalised} and~(\ref{eq:ksigma}), which is evaluated in the de-Sitter case where $H$ is a constant for simplicity. Since this matrix is proportional to $\boldsymbol{I}$, it needs to be compared to $ {D}^2_{\ \phi,\phi}+ {D}^2_{\ \pi,\pi}$, which is the component of $ \boldsymbol{D}^2 + \boldsymbol{D}^{2*}$ along $\boldsymbol{I}$ according to the decomposition~(\ref{eq:Pauli}). Using the results of \Sec{sec:massless}, at leading order in $\sigma$, it is given by
\bea
\left. {D}^2_{\ \phi,\phi}+ {D}^2_{\ \pi,\pi}\right\vert_{m=0} = \frac{N^2a^6H^8}{4\pi^4}\sigma^4
\eea
for a purely massless field, and \Eq{eq:diff:quartic} is suppressed by higher powers of $\sigma$ hence can be neglected in the limit
\bea
\left.\sigma\right\vert_{m=0}\ll 1 .
\eea
Using the results of \Sec{sec:light}, for a light test field with mass $m\ll H$, one obtains
\bea
 {D}^2_{\ \phi,\phi}+ {D}^2_{\ \pi,\pi} = \frac{9N^2a^6H^8}{4\pi^4}\frac{m^4}{H^4} ,
\eea
which dominates over \Eq{eq:diff:quartic} if
\bea
\label{eq:cond:classicalTransition}
\sigma\ll \left(\frac{m}{H}\right)^{\frac{2}{3}} ,
\eea 
and the same conditions as in \Sec{sec:QuantumToClassical:Quadratic} are recovered. One concludes that even at quartic order where the quantum and the stochastic theories start giving different results for observable correlators, these differences are suppressed on large scales by $\sigma$ and can therefore be neglected if $\sigma$ is taken to be sufficiently small.
\subsubsection{Quantum-to-classical transition in the Wigner picture}
\label{sec:quantum_to_classical:wigner}
The fact that a classical stochastic description is able to reproduce some observable predictions on large scales is often referred to as the ``quantum-to-classical'' transition. Such a transition can also be described in terms of the quantum state in which the field $\phi$ is placed, in particular by investigating the Wigner function of this state. In this section, we provide such a description, and explain how it is related to the stochastic framework.
\paragraph{Quantisation in the Schr\"odinger picture} 
The first step consists in performing a canonical transformation from the variables $\phi$ and $\pi_\phi$ to
\bea
\label{eq:u:phi}
\left(
\begin{array}{c}
u\\
\pi_u
\end{array}
\right)
= \underbrace{
\left(
\begin{array}{c c}
a & 0\\
0 & 1/a
\end{array}
\right)}_{\boldsymbol{M}}
\left(
\begin{array}{c}
\phi\\
\pi_\phi
\end{array}
\right)\, .
\eea
Since $\boldsymbol{M}^\mathrm{T}\boldsymbol{\Omega}\boldsymbol{M}=\boldsymbol{\Omega}$, where 
\bea
\boldsymbol{\Omega}=\left(\begin{array}{cc}
		0 & 1 \\
		-1 & 0
	\end{array}\right),
\eea 
$\boldsymbol{M}$ indeed defines a canonical transformation~\cite{Grain:2019vnq}. The next step is to Fourier expand the field variables $\phi$ and $\pi_\phi$, and to rewrite the Hamiltonian~(\ref{eq:calH:phi}) as
\bea
\mathcal{H}_\phi = \int_{\mathbb{R}^{3+}}\dd \bm{k}
\left(
\begin{array}{c}
\phi({\bm{k}})\\
\pi_{\phi}(\bm{k})
\end{array}
\right)^\dagger
\boldsymbol{H}_\phi(\bm{k})
\left(
\begin{array}{c}
\phi({\bm{k}})\\
\pi_{\phi}(\bm{k})
\end{array}
\right),
\eea
where the Hamiltonian kernel is given by
\bea
\boldsymbol{H}_\phi(\bm{k})= N
\left(\begin{array}{cc}
		a^3 m^2+a k^2 & 0 \\
		0 & 1/a^3
	\end{array}\right)\, ,
\eea
where the cosmological constant $\Lambda$ is not included. Notice that the integral is performed over half of the Fourier modes. Indeed, since $\phi(\bm{x})$ is a real field, one has $\phi(-\bm{k})=\phi_k^*(\bm{k})$ (and a similar relation for $\pi_\phi$), so the integral can be restricted to $\mathbb{R}^{3+}$. This also means that $\phi(\bm{k})$ and $\phi(-\bm{k})$ are not independent variables, so restricting to half of the Fourier space will also prove convenient when quantising the theory. In terms of the new canonical variables $u$ and $p_u$, the Hamiltonian kernel is given by (see equation (3.13) of \Refa{Grain:2019vnq})
\bea
\boldsymbol{H}_u(\bm{k}) = 
\left[\left(\boldsymbol{M}^{-1}\right)^\mathrm{T}\boldsymbol{H}_\phi(\bm{k})\boldsymbol{M}^{-1}+\left(\boldsymbol{M}^{-1}\right)^\mathrm{T}\boldsymbol{\Omega}\frac{\dd\boldsymbol{M}^{-1}}{\dd \tau}\right]\, .
\eea
This leads to
\bea
\mathcal{H}_\phi  = \int_{\mathbb{R}^{3+}}\dd \bm{k} N
\left\lbrace  a^2 \left(m^2+\frac{k^2}{a^2}\right) u(\bm{k}) u^*(\bm{k}) +  a \pi_u(\bm{k}) \pi_u^*(\bm{k})+H \left[u(\bm{k}) \pi_u^*(\bm{k}) + \pi_u(\bm{k}) u^*(\bm{k})  \right] \right\rbrace\, .
\eea
The field variables can then be decomposed into their real and imaginary parts,
\bea
\label{eq:uk:R}
u(\bm{k}) = \frac{u^\mathrm{R}(\bm{k})+iu^\mathrm{I}(\bm{k})}{\sqrt{2}}\\
\pi_u(\bm{k}) = \frac{\pi_u^\mathrm{R}(\bm{k})+i\pi_u^\mathrm{I}(\bm{k})}{\sqrt{2}}\, ,
\label{eq:uk:I}
\eea
in terms of which the Hamiltonian is separable, 
\bea
\label{eq:Hks}
\mathcal{H}_\phi  =  \sum_{s=\mathrm{R},\mathrm{I}}\int_{\mathbb{R}^{3+}}\dd \bm{k}
 \underbrace{ \left\lbrace\frac{N a}{2} \left(m^2+\frac{k^2}{a^2}\right)\left[ u^s(\bm{k})\right]^2 + \frac{N }{2a} \left[\pi_u^s(\bm{k})\right]^2+\frac{N  H}{2} \left[u^s(\bm{k}) \pi_u^s(\bm{k}) + \pi_u^s(\bm{k}) u^s(\bm{k})  \right]\right\rbrace}_{\mathcal{H}_{\bm{k},s}  } \, .
\eea
Deriving Hamilton's equation from $\mathcal{H}_{\bm{k},s}$, one obtains the mode equation for the Fourier moments $u^s_{\bm{k}}$,
\bea
\label{eq:mode:equation:uk}
\frac{\dd u^s_{\bm{k}}}{\dd\tau^2}+\left(NH-\frac{\dd\ln N}{\dd\tau}\right)\frac{\dd u^s_{\bm{k}}}{\dd\tau} + N^2\left[m^2+\frac{k^2}{a^2}-2H^2-\frac{1}{N}\frac{\dd H}{\dd\tau}\right] u^s_{\bm{k}}=0\, .
\eea
Let us note that, if the time variable $\tau$ is taken to be the conformal time $\eta$, then the lapse function reads $N=a$, and the term proportional to $\dd u^s_{\bm{k}}/\dd\tau$ in the above equation vanishes.

Let us now quantise the system in the Schr\"odinger picture, in the configuration representation. Because the Hamiltonian is separable in Fourier space and in the $\mathrm{R},\mathrm{I}$ decomposition, wavefunctions that are also separable,
\bea
\label{eq:Psi:Fourier}
\left\vert \Psi\left\lbrace u(\bm{x} )_{\bm{x}\in\mathbb{R}^3}\right\rbrace \right\rangle = \underset{s=\mathrm{R},\mathrm{I}}{\bigotimes}\ \underset{\bm{k}\in  \mathbb{R}^{3+}}{\bigotimes}\left\vert \Psi _{\bm{k}, s} \left(u^s_{\bm{k}}\right)\right\rangle\, ,
\eea
remain so throughout the evolution, and the Schr\"odinger equation reduces into 
\bea
\label{eq:Schrodinger:ks}
i \frac{\partial}{\partial \tau}  \Psi _{\bm{k}, s}  = \mathcal{\hat{H}}_{\bm{k},s}  \Psi _{\bm{k}, s} 
\eea
for each component of the wavefunction. Since the Hamiltonian is quadratic, the Schr\"odinger equation possesses Gaussian solutions,
\bea
\label{eq:wavefunction:Omega}
  \Psi _{\bm{k}, s} \left(u_{\bm{k},s}\right) = N_{\bm{k},s} \ee^{-\Omega_{\bm {k},s}  u_{\bm{k},s}^2  }\, .
\eea
Plugging this ansatz into the Schr\"odinger equation~(\ref{eq:Schrodinger:ks}), and given that, in the configuration representation, $\hat{u}^s(\bm{k}) \Psi _{\bm{k}, s} = u_{\bm{k},s} \Psi _{\bm{k}, s} $ and $\hat{\pi}_u^s(\bm{k}) \Psi _{\bm{k}, s} = - i \partial\Psi _{\bm{k}, s}  / \partial   u_{\bm{k},s}$, one obtains
\bea
\label{eq:dNk:dtau}
i \frac{\dd\ln N_{\bm{k},s}}{\dd\tau} & = &\frac{N}{a} \Omega_{\bm {k},s} - \frac{i}{2} NH\\
\frac{\dd \Omega_{\bm {k},s} }{\dd\tau} & = &-2i \frac{N}{a}\Omega_{\bm {k},s} ^2 -2 N H \Omega_{\bm {k},s}  + \frac{i}{2} N a \left(m^2+\frac{k^2}{a^2}\right).
\label{eq:dOmegak:dtau}
\eea
The first equation can be solved as follows. By decomposing $N_{\bm{k},s}$ into its modulus and its phase, $N_{\bm{k},s} = \rho _{\bm{k},s}  \ee^{i\theta_{\bm{k},s}} $, and by taking the real and imaginary parts of \Eq{eq:dNk:dtau}, one obtains
\bea
\label{eq:d:rho:d:tau}
\frac{\dd \ln \rho_{\bm{k},s}}{\dd\tau} &=& \frac{N}{a}\Omega_{\bm {k},s}^{\mathrm{I}}-\frac{N}{2a}\\
\frac{\dd\theta_{\bm{k},s}}{\dd\tau} &=& -\frac{N}{a}\Omega_{\bm {k},s}^{\mathrm{R}}\, .
\eea
The second of these equations sets the phase of the wavefunction in which we will not be interested in what follows. For the first equation, one notices that the real part of \Eq{eq:dOmegak:dtau} reads $\dd \Omega_{\bm {k},s}^{\mathrm{R}}/\dd\tau = 4 N \Omega_{\bm {k},s}^{\mathrm{I}}/a-2HN \Omega_{\bm {k},s}^{\mathrm{R}}$. Combined with \Eq{eq:d:rho:d:tau}, this gives rise to $4 \dd \ln \rho_{\bm{k},s}/\dd\tau = \dd\ln\Omega_{\bm {k},s}^{\mathrm{R}}/\dd\tau$, so $\ln \rho_{\bm{k},s} \propto(\Omega_{\bm {k},s}^{\mathrm{R}})^{1/4} $, which is nothing but the requirement that the norm of the wavefunction is preserved in time. Setting this norm to one, one therefore obtains
\bea
\left\vert N_{\bm{k},s} \right\vert = \left(\frac{2}{\pi}  \Omega_{\bm {k},s}^{\mathrm{R}} \right)^{1/4}.
\eea
Since \Eq{eq:dOmegak:dtau} is quadratic in $\Omega_{\bm{k},s}$ and of the first order, it is a Ricatti equation, so it can be cast into a  second order linear ordinary differential equation by a suitable redefinition of the function. More precisely, if one introduces $f_{\bm{k},s}$ related to $\Omega_{\bm{k},s}$ via
\bea
\Omega_{\bm{k},s} = \frac{i}{2}\frac{a}{N}\left(NH-\frac{\dd\ln f_{\bm{k},s}}{\dd\tau}\right),
\label{eq:Omega:f}
\eea
then one can show that \Eq{eq:dOmegak:dtau} is satisfied provided $f_{\bm{k},s}$ obeys the classical mode equation~(\ref{eq:mode:equation:uk}). This is a first indication that, for quadratic Hamiltonians, the quantum dynamics of the system can be expressed in terms of solutions to the classical dynamics.

Initial conditions are set in the Bunch-Davies vacuum. This is done by introducing the creation and annihilation operators, 
\bea
\hat{a}_{\bm{k},s} &=& \sqrt{\frac{k}{2}}\left[\hat{u}^s({\bm{k}})+\frac{i}{k}\hat{\pi }_u^s({\bm{k}})\right],\\
\hat{a}_{\bm{k},s}^\dagger &=&  \sqrt{\frac{k}{2}}\left[\hat{u}^s({\bm{k}})-\frac{i}{k}\hat{\pi }_u^s({\bm{k}})\right],
\eea
and by computing the expectation value of the number of particles operator,
\bea
\label{eq:mean:particle:number}
\left\langle \hat{n}_{\bm{k},s}\right\rangle = \left\langle \hat{a}_{\bm{k},s}^\dagger \hat{a}_{\bm{k},s}\right\rangle = \frac{k}{2}\left[\left\langle \hat{u}^s({\bm{k}})^2\right\rangle + \frac{\left\langle \hat{\pi }_u^s({\bm{k}})^2\right\rangle}{k^2}-\frac{1}{k}\right]
\eea
where we have used that the commutator between $\hat{u}^s({\bm{k}})$ and $\hat{\pi}_u^s({\bm{k}})$ reads $[\hat{u}^s({\bm{k}}),\hat{\pi}_u^s({\bm{k}})]=i$. This expression features quadratic moments of the phase-space variables, which can be expressed in terms of $\Omega_{\bm{k},s}$ using the expression~(\ref{eq:wavefunction:Omega}) for the wavefunction,
\bea
\label{eq:mean:u2}
\left\langle \hat{u}^s({\bm{k}})^2\right\rangle & \kern -0.5em = & \kern -0.5em \int_{-\infty}^\infty  \dd u_{\bm{k},s} \Psi^*_{\bm{k}, s} \left(u_{\bm{k},s}\right) \Psi _{\bm{k}, s} \left(u_{\bm{k},s}\right) u_{\bm{k},s}^2=\dfrac{1}{4\Omega_{\bm{k},s}^{\mathrm{R}}}\\
\left\langle  \hat{\pi }_u^s({\bm{k}})^2 \right\rangle & \kern -0.5em = & \kern -0.5em  -\int_{-\infty}^\infty \mathrm{d} u_{\bm{k},s}  \Psi^*_{\bm{k}, s} \left(u_{\bm{k},s}\right) \frac{\partial^2}{\partial  u_{\bm{k},s} ^2} \Psi _{\bm{k}, s}\left(u_{\bm{k},s}\right) \nonumber \\
& \kern -0.5em = & \kern -0.5em 
\int_{-\infty}^\infty \mathrm{d} u_{\bm{k},s} \frac{\partial}{\partial u_{\bm{k},s}}\Psi^*_{\bm{k}, s} \left(u_{\bm{k},s}\right) \frac{\partial}{\partial  u_{\bm{k},s}}\Psi_{\bm{k}, s} \left(u_{\bm{k},s}\right) \nonumber \\
& \kern -0.5em = & \kern -0.5em 4 \left\vert \Omega_{\bm{k},s} \right\vert ^2 \int_{-\infty}^\infty \mathrm{d}  u_{\bm{k},s}   \Psi^*_{\bm{k}, s} \left(u_{\bm{k},s}\right)\Psi_{\bm{k}, s} \left(u_{\bm{k},s}\right) u_{\bm{k},s}^2
=4 \left\vert \Omega_{\bm{k},s} \right\vert ^2 \left\langle u_{\bm{k},s}^2\right\rangle\\
\left\langle \hat{u}^s({\bm{k}}) \hat{\pi }_u^s({\bm{k}}) +  \hat{\pi }_u^s({\bm{k}}) \hat{u}^s({\bm{k}}) \right\rangle & \kern -0.5em = & \kern -0.5em
-i - 2 i \int_{-\infty}^\infty \dd u_{\bm{k},s}  \Psi^*_{\bm{k}, s} \left(u_{\bm{k},s}\right)  u_{\bm{k},s} \frac{\partial}{\partial  {u_{\bm{k},s} }} \Psi _{\bm{k}, s}\left(u_{\bm{k},s}\right)
\nonumber\\
& \kern -0.5em = & \kern -0.5em
-i + 4 i \Omega_{\bm{k},s} \left\langle \hat{u}^s({\bm{k}})^2\right\rangle = - \frac{\Omega_{\bm{k},s}^{\mathrm{I}}}{\Omega_{\bm{k},s}^{\mathrm{R}}}.
\label{eq:mean:up}
\eea
Plugging these expressions into \Eq{eq:mean:particle:number}, one obtains
\bea
\left\langle \hat{n}_{\bm{k},s}\right\rangle = \frac{1}{2}\left[ \frac{k}{4\Omega_{\bm{k},s}^{\mathrm{R}} } + \frac{\Omega_{\bm{k},s}^{\mathrm{R}}}{k} + \frac{\left(\Omega_{\bm{k},s}^{\mathrm{I}}\right)^2}{k \Omega_{\bm{k},s}^{\mathrm{R}}}-1\right]\, .
\eea
In order to minimise the mean number of particles, since $\Omega_{\bm{k},s}^{\mathrm{R}}$ has to be positive in order for the wavefunction to be normalisable, the above expression indicates that one must take $\Omega_{\bm{k},s}^{\mathrm{I}}=0$. Then, by studying  $k/(4 \Omega_{\bm{k},s}^{\mathrm{R}} ) + \Omega_{\bm{k},s}^{\mathrm{R}} /k$ as a function of $\Omega_{\bm{k},s}^{\mathrm{R}} \geq 0$, one obtains that this function has a global minimum at $\Omega_{\bm{k},s}^{\mathrm{R}}=k/2$, where its value is $1$. One concludes that the vacuum state, \ie the state with a vanishing number of particles, is such that
\bea
\label{eq:Omegak:vacuum}
\Omega_{\bm{k},s}^{\mathrm{vacuum}} = \frac{k}{2}\, .
\eea
In practice, this is the initial condition we use when solving \Eq{eq:dOmegak:dtau}, which then entirely determines the quantum state of the system. Let us note that this initial condition is independent of $s$, and only involves the modulus of the wavevector $k$. Since the equation of motion also has these two properties, $\Omega_{\bm{k}}^s$ remains independent of $s$ and of the direction of $\bm {k}$, so $\Omega_{\bm{k}}^s$ can be simply denoted $\Omega_k$. 
\paragraph{The Wigner function}
Let us consider the phase space $\lbrace \hat{u}^s({\bm{k}}) , \hat{\pi }_u^s({\bm{k}}) \rbrace $ (\ie $\bm{k}$ and $s$ are fixed), and a quantum operator $\hat{A}$. At the classical level, $A$ is given by a certain function of the phase space variables, $A=A[u^s({\bm{k}}),{\pi }_u^s({\bm{k}})]$, which is canonically quantised into $\hat{A} =A[\hat{u}^s({\bm{k}}),\hat{\pi }_u^s({\bm{k}})] $. The Weyl transform $\tilde{A}$ of the quantum operator $\hat{A}$ is then defined according to
\begin{align}
\label{eq:Weyltransform:def}
\tilde{A}[u^s({\bm{k}}),{\pi }_u^s({\bm{k}})]\equiv \int\dd x 
\ee^{-i x {\pi }_u^s({\bm{k}})  } 
 \left\langle u^s({\bm{k}})+\frac{x}{2}   
\left \vert \hat{A} \right \vert u^s({\bm{k}})
-\frac{x}{2} \right\rangle,
\end{align}
which builds a real function in phase space, $\tilde{A}$, out of the
quantum operator $\hat{A}$. It is important to stress that $\tilde{A}$
is a function (\ie not an operator) which, for instance, means that
$\tilde{A}\tilde{B}=\tilde{B}\tilde{A}$. However, $\widetilde{AB}\neq
\widetilde{BA}$ in general. It is also important to notice that, a
priori, $\tilde{A}\neq A$. In fact, introducing the vector $\bm{U}=\left[ k^{1/2} u^s({\bm{k}}), k^{-1/2} {\pi }_u^s({\bm{k}})\right]^\mathrm{T}$, one can easily show that (see Appendix F of \Refa{Martin:2015qta}, where more general expressions are also established)
\bea
\widetilde{{U}}_i={U}_i\, ,\quad \widetilde{{U}_j {U}_k}
={U}_j {U}_k+\frac{i}{2}{J}_{jk}\, ,
\label{eq:Weyl:ViVj}
\eea
where $\bm{J}$ is the anti-diagonal matrix with coefficients $\lbrace
-1,1 \rbrace$ on the antidiagonal and verifying
$i {J}_{j,k}=[\hat{{U}}_j,\hat{{U}}_k]$. 

A fundamental property of the Weyl transform is that
\begin{equation}
\tr \left(\hat{A}\hat{B}\right) 
=\int \tilde{A}(R)\tilde{B}(R) \frac{\dd^2 \bm{U}}{2\pi },
\end{equation}
where $\dd^2 \bm{U} =  \dd {U}_1\dd {U}_2 = \dd u^s({\bm{k}}) \dd {\pi }_u^s({\bm{k}})$.
Given that $\langle \tilde{A}\rangle =\tr
\left(\hat{\rho}_{\bm{k},s}\hat{A}\right)$, where $\hat{\rho}_{\bm{k},s} = \left\vert  \Psi _{\bm{k}, s}  \right\rangle \left\langle  \Psi _{\bm{k}, s}  \right\vert $ is the density matrix, this implies that
\bea
\label{eq:meanA:Wigner}
\left\langle \tilde{A} \right\rangle = \int W_{\bm{k},s}(\bm{U}) \tilde{A}(\bm{U}) \dd^2 \bm{U},
\eea
where $W_{\bm{k},s} = \tilde{\rho}_{\bm{k},s}/(2\pi)$, \ie
\begin{align}
\label{eq:Wigner:def}
W_{\bm{k},s}(\bm{U}) \equiv &  \frac{1}{2\pi}
\int\dd x \ee^{-i x {\pi }_u^s({\bm{k}}) } 
 \left\langle u^s({\bm{k}})+\frac{x}{2}  
\biggl\vert \hat{\rho}_{\bm{k},s} \biggr \vert u^s({\bm{k}})
-\frac{x}{2} \right\rangle\, .
\end{align}
For the Gaussian state~(\ref{eq:wavefunction:Omega}), \Eq{eq:Wigner:def} leads to
\begin{align}
\label{eq:Wigner:Gaussian}
W_{\bm{k},s}(\bm{U}) = \frac{1}{\pi} \exp\left[-\frac{\left\vert {\pi }_u^s({\bm{k}}) -2 i \Omega_{\bm {k},s} u^s({\bm{k}}) \right\vert^2}{2 \Omega_{\bm {k},s}^{\mathrm{R}}}\right]\, .
\end{align}
Let us highlight several salient properties of the Wigner function. 

First, the state being Gaussian, one can check explicitly on \Eq{eq:Wigner:Gaussian} that the Wigner function is positive. One can also check that $\int W_{\bm{k},s}(\bm{U}) \dd^2 \bm{U}=1$, which follows from the normalisation of the wavefunction. This means that the Wigner function can be viewed as a probability distribution in phase space. 

Second, an evolution equation for $W_{\bm{k},s}$ can be obtained by differentiating \Eq{eq:Wigner:Gaussian} with respect to time and by making use of \Eq{eq:dOmegak:dtau}. It turns out that the result is nothing but the classical Liouville equation for the distribution $W$. Indeed, if one introduces the Poisson bracket $\lbrace \rbrace_{\mathrm{PB}}$ defined by
\begin{equation}
\left\lbrace f,g\right\rbrace _{\mathrm{PB}}= \frac{\partial
  f}{\partial u^s({\bm{k}}) } \frac{\partial g}{\partial {\pi }_u^s({\bm{k}}) } -
\frac{\partial f}{\partial {\pi }_u^s({\bm{k}}) } \frac{\partial g}{\partial
  u^s({\bm{k}}) } \, ,  
\end{equation}
then it is straightforward to write the time derivative of \Eq{eq:Wigner:Gaussian} as 
\begin{equation} \frac{\dd W_{\bm{k},s}}{\dd \tau} = \left\lbrace \mathcal{H}_{\bm{k},s},W_{\bm{k},s} \right\rbrace_{\mathrm{PB}}\, ,
\end{equation} 
where $\mathcal{H}_{\bm{k},s}$ was given in \Eq{eq:Hks}. This means that $W_{\bm{k},s}$ describes a stochastic distribution of points in phase space, each of them following the classical Hamilton's equations of motion. This result holds in fact for any quadratic Hamiltonian, and we thus recover the fact that the quantum dynamics of linear systems can always be expressed in terms of solutions to their classical dynamics, as already mentioned below \Eq{eq:Omega:f}. 

Third, because of \Eq{eq:meanA:Wigner}, the Wigner phase-space distribution can be used to compute quantum expectation values, \ie observable predictions. For linear observables, thanks to the first \Eq{eq:Weyl:ViVj}, the correspondence is immediate, in agreement with the discussion of \Sec{sec:quantum_to_classical}. For quadratic observables, since $\widetilde{{U}_j {U}_k} \neq {U}_j {U}_k$ if $j\neq k$, see \Eq{eq:Weyl:ViVj}, the stochastic average procedure does not always match the quantum expectation value. It is however the case for Hermitian operators, since from \Eq{eq:Weyl:ViVj} it is straightforward to see that $\reallywidetilde{{U}_j {U}_k + {U}_k {U}_j} = {U}_j {U}_k + {U}_k {U}_j$. Moreover, the difference between the stochastic average of ${U}_1 {U}_2$ and the quantum expectation value is given by $i/2$, see \Eq{eq:Weyl:ViVj}, while the stochastic average is $-\Omega_{\bm {k},s}^{\mathrm{I}}/(2\Omega_{\bm {k},s}^{\mathrm{R}})$. In an inflationary background, if the mode $\bm{k}$ is far beyond the Hubble radius, \ie $k\ll a H$, combining \Eq{eq:Omega:f} with the considerations of \Sec{sec:explicit}, one obtains that $\Omega_{\bm {k},s}^{\mathrm{R}} \simeq k [a H/ (a_* H_*)]^{-2}/2$ and $\Omega_{\bm {k},s}^{\mathrm{I}} \simeq - k a H/ (a_* H_*)$, where $a_*$ and $H_*$ denote the value of the scale factor and of the Hubble scale when $k$ crosses out the Hubble radius, \ie at the time when $k=aH$. This gives rise to $-\Omega_{\bm {k},s}^{\mathrm{I}}/(2\Omega_{\bm {k},s}^{\mathrm{R}}) \simeq [a H/(a_* H_*)]^3$. Since this number grows to exponentially large values on super-Hubble scales, the stochastic average procedure gives a very good approximation to the quantum expectation value, again in agreement with the considerations of \Sec{sec:quantum_to_classical}. In fact, in the super-Hubble limit, since $\Omega_{\bm {k},s}^{\mathrm{R}} \ll \Omega_{\bm {k},s}^{\mathrm{I}}$, \Eq{eq:Wigner:Gaussian} reduces to
\bea
W_{\bm{k},s}(\bm{U}) \simeq \delta\left[ {\pi }_u^s({\bm{k}}) + 2   \Omega_{\bm {k},s}^{\mathrm{R}} u^s({\bm{k}}) \right]\, .
\eea
This means that the Wigner function gets very squeezed along one particular direction of phase space, which effectively becomes one-dimensional. This explains why the genuine quantum signatures, that involve non vanishing commutators between conjugated variables, are suppressed in this large squeezing limit. 

The same conclusion applies to quartic operators. For instance, one can show that (see Appendix F of \Refa{Martin:2015qta}) $\widetilde{{U}_1 ^2 {U}_2^2}={U}_1^2
{U}_1^2+2i{U}_1{U}_2-1/2$ and $\widetilde{{U}_2^2 {U}_1^2}= {U}_2^2 {U}_1^2-2i {U}_2 {U}_1-1/2$, but since $\langle {U}_1{U}_2 \rangle \propto [a H/(a_* H_*) ]^3$ on super-Hubble scales, the difference between the stochastic average procedure and the quantum expectation value become negligible in the large-squeezing limit. Let us stress again that these considerations are the exact translation, in the Wigner picture, of the discussion carried out in \Sec{sec:quantum_to_classical}. 
 
\subsubsection{Correspondence between the stochastic and the Wigner picture}

The similarity between the analyses developed in \Secs{sec:quantum_to_classical} and~\ref{sec:quantum_to_classical:wigner} suggests that the two objects considered in these sections, namely the phase-space distribution function that solves the Fokker-Planck equation of stochastic inflation in \Sec{sec:quantum_to_classical}, and the Wigner function in \Sec{sec:quantum_to_classical:wigner}, are related. In this section, the connection between these objects is made explicit. Let us first note that the distribution function obtained in \Sec{ssec:fokker} lives in the phase space of the coarse-grained fields, while the Wigner function introduced above was derived for each Fourier mode independently. A first step is therefore to derive the Wigner function of the coarse-grained phase-space variables. We thus consider the quantum state of the system comprised of modes above the coarse-graining scale,
\bea
\label{eq:Psi:IR}
\left\vert {\Psi}_{\mathrm{IR}} \right\rangle = \underset{s=\mathrm{R},\mathrm{I}}{\bigotimes}\ \ \ \underset{\bm{k}\in  \mathbb{R}^{3+}; \, k_{\mathrm{IR}}<k<k_\sigma }{\bigotimes}\left\vert \Psi _{\bm{k}, s} \left(u^s_{\bm{k}}\right)\right\rangle\, ,
\eea
which, compared to the quantum state~(\ref{eq:Psi:Fourier}) of the full system, is restricted to Fourier modes in the coarse-grained sector, $k<k_\sigma$, and where an infrared cutoff $k_{\mathrm{IR}}$ has also been introduced for later convenience. The Wigner function associated to this state lives on a space of infinite dimension, and generalising the definition given in \Eq{eq:Wigner:def} for the case of a quantum state of one dimension, it reads
\bea
\label{eq:Wigner:coarse:grained}
& & \kern -5em W_{\mathrm{IR}}\left[\left\lbrace u^s(\bm{k}),\pi_u^s(\bm{k}) \right\rbrace_{s=\mathrm{R},\mathrm{I};\  \bm{k}\in  \mathbb{R}^{3+}; \ k_{\mathrm{IR}}<k<k_\sigma }\right] =   
\int  \left (\underset{s=\mathrm{R},\mathrm{I}}{\prod}  \ \   \underset{\bm{k}\in  \mathbb{R}^{3+}; \, k_{\mathrm{IR}}<k<k_\sigma }{\prod} \frac{\dd x_{\bm{k},s}}{2\pi} \right)
\nonumber \\ & &
\exp\left[-i \sum_{s=\mathrm{R},\mathrm{I}}\int_{\bm{k}\in  \mathbb{R}^{3+}; \, k_{\mathrm{IR}}<k<k_\sigma } \dd^3 \bm{k}x_{\bm{k},s} \pi_u^s(\bm{k})\right]
\nonumber \\ & &
\Biggl[
\underset{s=\mathrm{R},\mathrm{I};\, \bm{k}\in  \mathbb{R}^{3+}; \, k_{\mathrm{IR}}<k<k_\sigma }{\bigotimes}
 \left\langle  u^s({\bm{k}})+\frac{x_{\bm{k},s}}{2} 
\biggl\vert  \Biggr] \hat{\rho}_{\mathrm{IR}}
\Biggl[
\underset{s=\mathrm{R},\mathrm{I};\, \bm{k}\in  \mathbb{R}^{3+}; \, k_{\mathrm{IR}}<k<k_\sigma }{\bigotimes}
  \biggr \vert u^s({\bm{k}})+\frac{x_{\bm{k},s}}{2} \right\rangle\Biggr]  \, .
\eea
Since the state~(\ref{eq:Psi:IR}) is factorisable, it is easy to check that
\bea
\label{eq:wigner:factored}
W_{\mathrm{IR}}\left[\left\lbrace u^s(\bm{k}),\pi_u^s(\bm{k}) \right\rbrace_{s=\mathrm{R},\mathrm{I};\  \bm{k}\in  \mathbb{R}^{3+}; \ k_{\mathrm{IR}}<k<k_\sigma }\right] = \underset{s=\mathrm{R},\mathrm{I}}{\prod}  \ \   \underset{\bm{k}\in  \mathbb{R}^{3+}; \, k_{\mathrm{IR}}<k<k_\sigma }{\prod}  W_{\bm{k},s} \left[u^s(\bm{k}),\pi_u^s(\bm{k})\right]\, .
\eea
In this expression, the individual Wigner functions $ W_{\bm{k},s} \left[u^s(\bm{k}),\pi_u^s(\bm{k})\right]$ have been calculated in \Eq{eq:Wigner:Gaussian}, which can be recast in the form
\begin{align}
\label{eq:Wigner:Gaussian:Sigma}
W_{\bm{k},s}(\bm{U}) = \frac{1}{2 \pi \sqrt{\det\left(\boldsymbol{\mathcal{S}}_u^{\bm{k},s} \right)}} \exp\left[-
\left(
\begin{array}{c}
u^s({\bm{k}})\\
{\pi }_u^s({\bm{k}})
\end{array}
\right)^\mathrm{T}
\left(\boldsymbol{\mathcal{S}}_u^{\bm{k},s}\right)^{-1}
\left(
\begin{array}{c}
u^s({\bm{k}})\\
{\pi }_u^s({\bm{k}})
\end{array}
\right)
\right]\, ,
\end{align}
where 
\bea
\label{eq:Sigma:k}
\boldsymbol{\mathcal{S}}_u^{\bm{k},s} = 
\left(
\begin{array}{cc}
\dfrac{1}{4 \Omega_{\bm{k},s}^\mathrm{R}} & - \frac{\Omega_{\bm{k},s}^\mathrm{I}}{2 \Omega_{\bm{k},s}^\mathrm{R}}\\
-\dfrac{\Omega_{\bm{k},s}^\mathrm{I}}{2 \Omega_{\bm{k},s}^\mathrm{R}} & \Omega_{\bm{k},s}^\mathrm{R}+\dfrac{\left(\Omega_{\bm{k},s}^\mathrm{I}\right)^2}{\Omega_{\bm{k},s}^\mathrm{R}}
\end{array}
\right)\, .
\eea
Since the entries of $\boldsymbol{\mathcal{S}}_u^{\bm{k},s}$ coincide with the correlators calculated in \Eqs{eq:mean:u2}-(\ref{eq:mean:up}), one can write ${\mathcal{S}}_u^{\bm{k},s}(f,g)=\mathrm{Sym} [P_{f,g}(k) ]$, where $f$ and $g$ are either $u^s({\bm{k}})$ or $\pi_u^s({\bm{k}})$, and we have defined $P_{f,g}=\langle \hat{f}\hat{g} \rangle$, the symmetric part of which reads $\mathrm{Sym} (P_{f,g} )=\langle \hat{f}\hat{g}+\hat{g}\hat{f} \rangle/2$. The expression~(\ref{eq:wigner:factored}) thus indicates that each vector $\lbrace u^s(\bm{k}), \pi_u^s(\bm{k})\rbrace$ can be seen as an independent Gaussian random variable, with covariance matrix given by \Eq{eq:Sigma:k}. 

The coarse-grained fields $\bar{u}(\bm{x})$ and $\bar{\pi}_u(\bm{x})$ correspond to particular operators living in the infrared sectors, that are linear combinations of the phase-space variables, namely
\bea
\left(
\begin{array}{c}
\bar{u}(\bm{x})\\
\bar{\pi}_u(\bm{x})
\end{array}
\right)&=&
\frac{1}{\left(2\pi\right)^{3/2}}\int_{\bm{k} \in \mathbb{R}^{3};k_{\mathrm{IR}}<k<k_\sigma } \dd^3 \bm{k} \ee^{-i\bm{k}\cdot\bm{x}}\left(
\begin{array}{c}
u(\bm{k})\\
{\pi}_u(\bm{k})
\end{array}
\right)
\\ &=&
\frac{\sqrt{2}}{\left(2\pi\right)^{3/2}}\int_{\bm{k} \in \mathbb{R}^{3+};k_{\mathrm{IR}}<k<k_\sigma } \dd^3 \bm{k} 
\left[\left(
\begin{array}{c}
u^{\mathrm{R}}(\bm{k})\\
{\pi}_u^{\mathrm{R}}(\bm{k})
\end{array}
\right) \cos\left(\bm{k}\cdot\bm{x}\right) + \left(
\begin{array}{c}
u^{\mathrm{I}}(\bm{k})\\
{\pi}_u^{\mathrm{I}}(\bm{k})
\end{array}
\right) \sin\left(\bm{k}\cdot\bm{x}\right) \right]
\eea
where in the second equality, we have used the decomposition~(\ref{eq:uk:R}) and (\ref{eq:uk:I}) together with the fact that, since $u(\bm{x})$ and $\pi_u(\bm{x})$ are real fields, $u(-\bm{k}) = u^*(\bm{k}) $ and  $\pi_u(-\bm{k}) = \pi_u^*(\bm{k}) $. Linear combinations of independent Gaussian random variables are still Gaussian random variables, where the variances simply add up. Therefore, the coarse-grained variables are Gaussian random variables, with variance
\bea
\label{eq:Sigma:u:def}
\overline{\boldsymbol{\mathcal{S}}}_u=\int_{\bm{k} \in \mathbb{R}^{3+};k_{\mathrm{IR}}<k<k_\sigma } \dd^3 \bm{k} \left[ \boldsymbol{\mathcal{S}}_u^{\bm{k},\mathrm{R}} \frac{\cos^2\left(\bm{k}\cdot\bm{x}\right)}{4\pi^3} + \boldsymbol{\mathcal{S}}_u^{\bm{k},\mathrm{I}} \frac{\sin^2\left(\bm{k}\cdot\bm{x}\right)}{4 \pi^3} \right] .
\eea
As explained below \Eq{eq:Omegak:vacuum}, $\Omega_{\bm{k}}^s$ remains independent of $s$ and of the direction of $\bm {k}$ throughout the evolution, so $\Omega_{\bm{k}}^s$ can be simply denoted $\Omega_k$. One then concludes from \Eq{eq:Sigma:k} that $\boldsymbol{\mathcal{S}}_u^{\bm{k},s} $ is also independent of $s$ and of the direction of $\bm {k}$, and can simply be denoted $\boldsymbol{\mathcal{S}}_u^k$. In \Eq{eq:Sigma:u:def}, it can thus be factored out, leaving the sum of the two trigonometric functions that trivially equals one. Since $\boldsymbol{\mathcal{S}}_k$ does not depend on the direction of $\bm{k}$, the integral over the angular degrees of freedom contained in $\bm{k}$ can then be performed (mind that, since $\bm{k} \in \mathbb{R}^{3+}$, the polar angle $\theta$ should be integrated between $-\pi/2$ and $\pi/2$ and the azimuthal angle between $0$ and $\pi$). This gives rise to
\bea
\overline{\boldsymbol{\mathcal{S}}}_u= \frac{1}{2\pi^2} \int_{k_{\mathrm{IR}}}^{k_\sigma } k^2  \boldsymbol{\mathcal{S}}_u^k \dd k\, .
\eea
Making use of the power spectrum definition~(\ref{eq:calP:def}), this can be written as $\overline{{\mathcal{S}}}_u (\bar{f},\bar{g}) = \int_{k_{\mathrm{IR}}}^{k_\sigma}  \mathrm{Sym}[\calP_{f,g}](k) \dd \ln k$, where $f$ and $g$ are either $u$ or ${\pi}_u$. The Wigner function for the coarse-grained variables is then given by
\bea
\label{eq:Wigner:coarse:grained;u}
\overline{W}_{u}= \frac{1}{2 \pi \sqrt{\det\left(\overline{\boldsymbol{\mathcal{S}}}_u \right)}} \exp\left[-
\left(
\begin{array}{c}
\bar{u}\\
\bar{\pi}_u
\end{array}
\right)^\mathrm{T}
\left(\overline{\boldsymbol{\mathcal{S}}}_u\right)^{-1}
\left(
\begin{array}{c}
\bar{u}\\
\bar{\pi}_u
\end{array}
\right)
\right]\, .
\eea
Going back to the original field variables $\phi$ and $\pi_\phi$, related to $u$ and $\pi_u$ through \Eq{eq:u:phi}, the coarse-grained Wigner function is finally given by
\bea
\label{eq:Wigner:coarse:grained;u}
\overline{W}_{\phi}= \frac{1}{2 \pi \sqrt{\det\left(\overline{\boldsymbol{\mathcal{S}}}_\phi \right)}} \exp\left[-
\left(
\begin{array}{c}
\bar{\phi}\\
\bar{\pi}_\phi
\end{array}
\right)^\mathrm{T}
\left(\overline{\boldsymbol{\mathcal{S}}}_\phi\right)^{-1}
\left(
\begin{array}{c}
\bar{\phi}\\
\bar{\pi}_\phi
\end{array}
\right)
\right]\, ,
\eea
where $\overline{\boldsymbol{\mathcal{S}}}_\phi  =  \boldsymbol{M}^{-1} \overline{\boldsymbol{\mathcal{S}}}_u (\boldsymbol{M}^{-1})^{\dagger}$ (see section 3.3 of \Refa{Grain:2017dqa}), \ie
\bea
\label{eq:S:coarse:grained}
\overline{\boldsymbol{\mathcal{S}}}_\phi(\bar{f},\bar{g}) = \int_{k_{\mathrm{IR}}}^{k_\sigma}  \dd\ln k ~ \mathrm{Sym}\left[ \calP_{f,g}(k) \right]  \, ,
\eea
where $f$ and $g$ are either $\phi$ or $\pi_\phi$. 

Let us now compare this coarse-grained Wigner function with the phase-space distribution function obtained from the Fokker-Planck equation in \Sec{ssec:fokker}. That distribution function was also found to be Gaussian, and combining \Eqs{eq:Sigma} and~(\ref{eq:noisecorrel_Pk}), its variance is given by
\bea
\label{eq:Sigma:calP}
	\boldsymbol{\Sigma}_{\boldsymbol{\Phi}}(\tau)=\displaystyle\int^\tau_{\tau_0}\dd s~
	\frac{\dd \ln\left[k_\sigma(s)\right]}{\dd s}
	\boldsymbol{G}(\tau,s) \mathrm{Sym} \left\lbrace \boldsymbol{\mathcal{P}}\left[k_\sigma(s); s\right]\right\rbrace \boldsymbol{G}^\dag(\tau,s) ,
\eea	
where we have used that the diffusion matrix $\boldsymbol{D}$ is the symmetric part of $\boldsymbol{\Xi}$, as noted above \Eq{eq:fokker}, and where $\boldsymbol{\mathcal{P}}$ is the matrix of coefficients $\mathcal{P}_{f,g}$ with $f$ and $g$ being either $u[k_\sigma(s)]$ or $\pi_u[k_\sigma(s)]$. Let us now perform a change of integration variable: instead of integrating over time $s$, let us integrate over the wavenumber that crosses the coarse-graining scale at that time, \ie $s\to k_\sigma(s)$. Denoting the inverse of the function $k_\sigma(s)$ by  $s_\sigma(k)$, one obtains
\bea
\label{eq:Sigma:calP}
	\boldsymbol{\Sigma}_{\boldsymbol{\Phi}}(\tau)=\displaystyle\int^{k_\sigma(\tau)}_{k_{\mathrm{IR}}}\dd \ln k ~
        \boldsymbol{G}\left[\tau,s_\sigma(k)\right] \mathrm{Sym} \left\lbrace \boldsymbol{\mathcal{P}}\left[k; s_\sigma(k)\right]\right\rbrace \boldsymbol{G}^\dag\left[\tau,s_\sigma(k)\right] ,
\eea	
where we have set $k_{\mathrm{IR}}=k_\sigma(\tau_0)$. The expression inside the integral evaluates the power spectra at the time when a given mode crosses out the coarse-graining radius, and evolves it with the background equations of motion, \ie with the background Green function $\boldsymbol{G}$, until the time $\tau$. By doing so, one does not exactly obtain the power spectra at time $\tau$, since in principle, the mode functions need to be evolved with the perturbation equations of motion, which, compared to the background equations of motion, contain additional gradient terms. One therefore does not obtain $ \boldsymbol{\mathcal{P}}(k,\tau)$ exactly, but a quantity that we denote $ \widetilde{\boldsymbol{\mathcal{P}}}(k,\tau)$, in terms of which the coarse-grained field variance becomes
\bea
\label{eq:Sigma:calP:tilde}
	\boldsymbol{\Sigma}_{\boldsymbol{\Phi}}(\tau)=\displaystyle\int^{k_\sigma(\tau)}_{k_{\mathrm{IR}}}\dd \ln k ~
        \mathrm{Sym} \left\lbrace \widetilde{\boldsymbol{\mathcal{P}}}\left[k; s_\sigma(k)\right]\right\rbrace .
\eea	
Up to the difference between $\boldsymbol{\mathcal{P}}$ and $\widetilde{\boldsymbol{\mathcal{P}}}$, this expression is the same as the one obtained in \Eq{eq:S:coarse:grained}. The difference between $\boldsymbol{\mathcal{P}}$ and $\widetilde{\boldsymbol{\mathcal{P}}}$ is related to the role of gradient terms on scales larger than the coarse-graining scale. Since such scales are far beyond the Hubble radius, $\sigma\ll 1$, this difference is small, and using the results of \Sec{sec:explicit}, one can check that it is of order $\sigma^2$. As a consequence, one has 
\bea
\boldsymbol{\Sigma}_{\boldsymbol{\Phi}} = \overline{\boldsymbol{\mathcal{S}}}_\phi + \order{\sigma^2},
\eea
so in the $\sigma\ll 1$ limit, the phase-space distribution that solves the Fokker-Planck equation from the stochastic inflation formalism, coincides with the coarse-grained Wigner function. 

In conclusion, and summarising the above discussions, when one uses the distribution function $P(\bar{\phi},\bar{\pi}_\phi)$ of stochastic inflation to evaluate expectation values of observables $O(\bar{\phi},\bar{\pi}_\phi)$, and makes the identification
\bea
\label{eq:stochastic:approximation}
\left\langle O\left(\hat{\bar{\phi}},\hat{\bar{\pi}}_\phi\right)\right\rangle  = 
\int \dd \bar{\phi} \dd \bar{\pi}_\phi ~ \tilde{O}\left(\bar{\phi},\bar{\pi}_\phi\right)  \overline{W}_\phi\left(\bar{\phi},\bar{\pi}_\phi\right)
\underset{\sigma\ll 1}{\longrightarrow}
\int \dd \bar{\phi} \dd \bar{\pi}_\phi ~ O\left(\bar{\phi},\bar{\pi}_\phi\right) P\left(\bar{\phi},\bar{\pi}_\phi\right),
\eea
one performs in fact two approximations:
\begin{itemize}
\item The phase-space functions $O$ and $\tilde{O}$ (the Weyl transform of $O$) do not necessarily coincide, if $O$ is of order more than two in the phase-space variables, or a hermitic function of order more than four. However, in the $\sigma\ll 1$ limit, the Wigner function $\overline{W}_\phi$, and the phase-space distribution $P$, asymptote Dirac distributions along a specific phase-space direction, this difference is suppressed. In any case, an  important point to stress is that this approximation is not inherent to the stochastic formalism, which only provides techniques to calculate $P$, but is only a limitation of the procedure~(\ref{eq:stochastic:approximation}). One could indeed perfectly integrate $\tilde{O}$ against the distribution function $P$, and be immune to this first approximation
\item The Wigner function $\overline{W}_\phi$ and the distribution function $P$ do not coincide. This approximation is inherent to the stochastic inflation formalism, which neglects the effects of gradients above the coarse-graining scale. Since it requires to work in the $\sigma\ll 1$ regime, for consistency, the first approximation can also be performed, although we stress again that it does not have to. 
\end{itemize}
\subsubsection{Quantum-to-even-more-quantum transition?}
Although the properties derived above, that can be summarised by the statement that the two approximations listed below \Eq{eq:stochastic:approximation} become valid on large scales, are often described as signalling a ``quantum-to-classical'' transition~\cite{Polarski:1995jg, Lesgourgues:1996jc, Kiefer:2008ku} on super-Hubble scales, other genuinely quantum properties also arise in the large squeezing limit, that could conversely be described as a ``quantum-to-even-more-quantum'' transition. 

For instance, the quantum state of cosmological perturbations acquires a large quantum discord on large scales~\cite{Martin:2015qta}, and this gives rise to possible Bell inequality violations, see \Refa{Martin:2016tbd, Martin:2017zxs}. This requires to work with improper operators~\cite{2005PhRvA..71b2103R}, which are operators $\widehat{O}$ for which the Weyl transform $\tilde{O}$ takes values outside the spectrum (\ie outside the set of eigenvalues of $\widehat{\mathcal{O}}$). 

In fact, even if one considers proper operators only, the emergence of a ``quantum-to-classical'' transition can be questioned. Indeed, according to the above considerations, a ``quantum-to-classical'' transition operates when the term proportional to $\boldsymbol{J}_y$ in the noise correlation matrix~(\ref{eq:Pauli}) becomes negligible, since that term comes from the non-vanishing commutator between the two phase-space variables. More precisely, when the expectation value of the anti-commutator between the two phase-space variables, which provides the contribution proportional to $\boldsymbol{J}_x$, is much larger than the expectation value of the commutator, $1/2$, which provides the contribution proportional to $\boldsymbol{J}_y$, the quantum imprint onto the correlation matrix can be neglected and the system is described as ``classical''. According to this definition,  a ``classical'' system is thus one for which maximal (anti-)correlation between the phase-space variables is reached. 

However, the amount of correlation between the two phase-space variables depends on the choice of phase-space coordinates used to describe the system. The above criterion is therefore not invariant under reparametrisation of phase space, \ie under canonical transformations. As explained in details in \Refa{Grain:2019vnq}, one can always operate a change of canonical variables such that the expectation value of the anti-commutator between the two phase-space variables vanishes, which is simply realised by working with the major and minor axes of the Wigner ellipse as the canonical variables. In that case, the two phase-space variables are entirely uncorrelated, and according to the above criterion the system would have to be said to be highly quantum. In that case indeed, the expectation values of hermitian quartic combinations of phase-space variables does strongly depend on the ordering of these phase-space variables, and the additional terms contained in the Weyl transforms of these combinations become crucial.

This implies that the first of the approximations listed below \Eq{eq:stochastic:approximation} is not valid for \emph{all} observables. However, let us stress again that this approximation is not necessary for stochastic inflation, which only assumes the second approximation in order to derive the distribution function $P$, from which expectation values could be computed by integrating this distribution function against the full Weyl transform of the operator under consideration. 

Let us also note that the major and minor axes of the Wigner ellipse correspond to what is called the ``growing'' and the ``decaying'' mode respectively. This is why the observables violating the first approximation (as well as those leading to Bell inequality violations) necessarily rely on the decaying mode, which becomes highly suppressed on super-Hubble scales, hence difficult (if not impossible) to measure in practice. In this sense the system can be said to have undergone a classical transition as far as some concretely observable quantities are concerned, although it may still contain large quantum correlations, hidden into quantities that are difficult to access observationally.

\subsubsection{Explicit solution}
\label{sec:explicit}
Let us now solve the mode equations~(\ref{eq:eomphiq}) and~(\ref{eq:eompiq}) explicitly and carry out the computational program sketched in \Sec{ssec:fokker}.
\paragraph{Massless field on a de-Sitter background}
\label{sec:massless}
We first consider the case of a massless test field evolving on a de-Sitter background where $a=-1/(H\eta)$. In terms of the conformal time $\eta$, the mode equations~(\ref{eq:eomphiq}) and~(\ref{eq:eompiq}) give rise to
\bea
\left( a \phi_k \right)^{\prime\prime}+\left(k^2-\frac{2}{\eta^2}\right)\left( a \phi_k \right) = 0 .
\eea
The solution to this equation satisfying the Klein-Gordon normalisation condition given below \Eq{eq:eompiq} reads\footnote{This solution is such that  in the remote past, \ie when $k\eta\rightarrow -\infty$, $a\phi_k=e^{ik\eta}/\sqrt{2k}$, which corresponds to the so-called Bunch-Davies vacuum discussed around \Eq{eq:Omegak:vacuum}.}
\bea
\phi_k = \frac{1}{a\sqrt{2k}} \left(1+\frac{i}{k\eta}\right)\ee^{i k \eta}  .
\eea 
From \Eq{eq:eomphiq},  the conjugated momentum is given by $\pi_\phi = a^2 \phi^\prime$, which leads to
\bea
\pi_k = a\sqrt{\frac{k}{2}} i \ee^{i k \eta} .
\eea
Making use of \Eqs{eq:ksigma} and~(\ref{eq:noisecorrel}), the symmetric part $\boldsymbol{D}$ of the noise correlator matrix $\boldsymbol{\Xi}$ is then given by
\bea
\label{eq:noisecorrelator:massless}
	\boldsymbol{D}=a\displaystyle\left(\begin{array}{cc}
		\dfrac{H^3\left(1+\sigma^2\right)}{4\pi^2} & \dfrac{H\sigma^2}{4\pi^2\eta^3} \\
		\dfrac{H\sigma^2}{4\pi^2\eta^3} & \dfrac{\sigma^4}{4H\pi^2\eta^6}
	\end{array}\right)
		 .
\eea
On the other hand, the homogeneous equation for the field is given by
\bea
\left( a \bar{\phi} \right)^{\prime\prime}-\frac{2}{\eta^2}\left( a \bar{\phi} \right) = 0
\eea
and has two independent solutions, $a \bar{\phi}^{(1)} \propto 1/\eta$ and $a \bar{\phi}^{(2)} \propto \eta^2 $. These solutions allow one to introduce the fundamental matrix $\boldsymbol{U}$ defined in \Eq{eq:U:def},
\bea
\label{eq:U:massless}
\boldsymbol{U}(\eta)=
\left(\begin{array}{ccc}
	H &\ & \dfrac{H}{3}\eta^3  \\
	0 &\ & \dfrac{1}{H}
	\end{array}\right) ,
\eea
which then gives rise to the Green matrix
\bea
	\boldsymbol{G}(\eta,\eta_0)&=&
	\left(\begin{array}{ccc}
	1 & \ &\dfrac{H^2}{3}\left(\eta^3-\eta_0^3\right) \\
	0 &\  & 1
\end{array}\right) ,
\eea
see \Eq{eq:Green:fundamental}. From here, the covariance matrix~(\ref{eq:Sigma}) can be calculated, 
\bea
\boldsymbol{\Sigma}_{\boldsymbol{\Phi}}(\eta) = 
\left(\begin{array}{ccc}
\dfrac{H^2}{4\pi^2} \ln\left(\dfrac{\eta_0}{\eta}\right) &\ & \dfrac{\sigma^2}{12\pi^2}\left(\dfrac{1}{\eta^3}-\dfrac{1}{\eta_0^3}\right) \\
\dfrac{\sigma^2}{12\pi^2}\left(\dfrac{1}{\eta^3}-\dfrac{1}{\eta_0^3}\right) &\ &  \dfrac{\sigma^4}{24\pi^2 H^2}\left(\dfrac{1}{\eta^6}-\dfrac{1}{\eta_0^6}\right)
\end{array}\right) .
\label{eq:Sigma:massless}
\eea
In this expression, according to the considerations of \Sec{sec:quantum_to_classical}, only the leading order terms in $\sigma$ have been kept. In the limit $\sigma\rightarrow 0$, one has $ {\Sigma}_{\phi,\pi}= {\Sigma}_{\pi,\phi}\simeq  {\Sigma}_{\pi,\pi} \simeq 0$, and quantum diffusion takes place in the $\phi$ direction only, growing as the logarithm of the scale factor.
\paragraph{Free field on a slow-roll background}
\label{sec:light}
Let us now see how these results generalise to a test field with mass $m$ on a slow-roll inflationary background, 
\bea
\label{eq:a:SR}
a=-\dfrac{1}{\eta H_*}\left[1+\epsilon_{1*}-\epsilon_{1*}\ln\left(\frac{\eta}{\eta_*}\right)\right] ,
\eea
where $\epsilon_1=-(\dd H/\dd t)/H^2$ denotes the first slow-roll parameter, and a star denotes the time around which the slow-roll expansion is performed (for instance, one can take $\eta_*=\eta_0$). Note that \Eq{eq:a:SR} corresponds to a first order expansion in slow roll of the background dynamics, but no assumption is made regarding the (slow-roll or non-slow-roll) dynamics of the test field. The mode equations~(\ref{eq:eomphiq}) and~(\ref{eq:eompiq}) give rise to
\bea
\label{eq:mode:light}
	\left(a\phi_k\right)^{\prime\prime}+\left(k^2-\dfrac{2+3\epsilon_{1*}-\frac{m^2}{H_*^2}}{\eta^2}\right)\left(a\phi_k\right)=0 .
\eea
Requiring the Klein-Gordon product normalisation condition again, this equation is solved by
\bea
\label{eq:modesolution:light}
\phi_k = \dfrac{\sqrt{\pi}}{2 a \sqrt{k}}\sqrt{k\eta}e^{i\frac{\pi}{4}+i\nu\frac{\pi}{2}}
H^{(1)}_\nu\left(k\eta\right) ,
\eea
where $H^{(1)}_\nu$ is the Hankel function of the first kind with index 
\bea
\label{eq:nu:def}
\nu \equiv \frac{3}{2}\sqrt{1-\frac{4m^2}{9H_*^2}+\frac{4}{3}\epsilon_{1*}} .
\eea
One can check that at leading order in background slow roll, the only effect of $\epsilon_{1*}$ is to change the effective mass of the field perturbations through \Eq{eq:nu:def}. For the conjugated momentum, one obtains
\bea
\label{eq:conjugated_momentum:light}
\pi_k =  a \frac{\ee^{i\frac{\pi}{4}+i\nu\frac{\pi}{2}}}{4}\sqrt{\frac{\pi}{\eta}}\left[2 k \eta H^{(1)}_{\nu-1}\left(k\eta\right) + \left(3-2\nu\right) H^{(1)}_\nu\left(k\eta\right) \right] .
\eea
Making use of \Eqs{eq:ksigma} and~(\ref{eq:noisecorrel}), the components of the symmetric part $\boldsymbol{D}$ of the noise correlator matrix $\boldsymbol{\Xi}$ can then be expanded in $\sigma\ll 1$. After a lengthy but straightforward calculation, one obtains
\bea
 {D}_{\phi,\phi}&=&
\frac{a H^3 }{\pi\sin^2(\nu\pi)\Gamma^2\left(1-\nu\right)}\left(\frac{\sigma}{2}\right)^{3-2\nu}
\left\lbrace
1
+\frac{2}{\nu-1}\left(\frac{\sigma}{2}\right)^{2}
\right. \nonumber \\ & & \left.
-16\frac{\Gamma\left(1-\nu\right)}{\Gamma(1+\nu)}\cos(\nu\pi)\left(\frac{\sigma}{2}\right)^{2\nu}
+\frac{2\nu-3}{\left(\nu-1\right)^2\left(\nu-2\right)}\left(\frac{\sigma}{2}\right)^{4}
\right\rbrace
+\order{\sigma^5} ,
\label{eq:massive:Dphiphi}
\\
 {D}_{\pi,\pi}&=&
\frac{a^7H^5}{4\pi\sin^2(\nu\pi)\Gamma^2\left(1-\nu\right)} \left(\frac{\sigma}{2}\right)^{3-2\nu}
\left[
\left(2\nu-3\right)^2
+\frac{2(2\nu-3)(2\nu-7)}{\nu-1}\left(\frac{\sigma}{2}\right)^{2}
\right. \nonumber \\ & & \left.
+2(2\nu-3)(2\nu+3)\frac{\Gamma\left(1-\nu\right)}{\Gamma(1+\nu)}\cos(\pi\nu)\left(\frac{\sigma}{2}\right)^{2\nu}
\right. \nonumber \\ & & \left.
+\frac{8\nu^3-68\nu^2+166\nu-131}{\left(\nu-2\right)\left(\nu-1\right)^2}\left(\frac{\sigma}{2}\right)^{4}
\right]
+\order{\sigma^5} ,
\label{eq:massive:Dpipi}
\\
 {D}_{\phi,\pi}&=& {D}_{\pi,\phi} =
\frac{(aH)^4 }{2\pi\sin^2(\pi\nu)\Gamma^2(1-\nu)}\left(\frac{\sigma}{2}\right)^{3-2\nu}
 \left[
2\nu-3
+\frac{2(2\nu-5)}{(\nu-1)}\left(\frac{\sigma}{2}\right)^{2}
\right. \nonumber \\ & & \left.
+6\frac{\Gamma(1-\nu)}{\Gamma(1+\nu)}\cos(\pi\nu)\left(\frac{\sigma}{2}\right)^{2\nu}
-\frac{\left(2\nu-3\right)\left(2\nu-7\right)}{\left(\nu-2\right)\left(\nu-1\right)^2}\left(\frac{\sigma}{2}\right)^{4}
\right]+\order{\sigma^{5}} .
\label{eq:massive:Dphipi}
\eea
This expansion in $\sigma$ has been ordered under the assumption that $1<\nu<2$, which amounts to $ -7/4<m^2/H_*^2-3\epsilon_{1*}<5/4$. In practice, only the leading terms in $\sigma$ must be kept in order to be consistent with the stochastic classical approximation as explained in \Sec{sec:quantum_to_classical}. In \Eqs{eq:massive:Dphiphi}-(\ref{eq:massive:Dphipi}) however, the first four terms of the expansion are displayed to make clear that the massless de-Sitter case of \Sec{sec:massless} is recovered in the limit $\nu=3/2$ (the non-dominant terms will be dropped in what follows). In the noise correlators involving the conjugated momentum $\pi$ indeed, one can see that the leading order contributions vanish when $\nu=3/2$. For example, in ${D}_{\pi,\pi}$ given by \Eq{eq:massive:Dpipi}, the first three terms vanish when $\nu=3/2$ and one has to go to fourth order to recover the $\pi,\pi$ component of \Eq{eq:noisecorrelator:massless}. As a consequence, the inclusion of a small mass or of a small departure from de Sitter does not only slightly modify the coefficients of the noise density matrix. It introduces new, lower order contributions in $\sigma$ that make all entries of $\boldsymbol{D}$ non-vanish in the limit $\sigma\rightarrow 0$, contrary to the massless de-Sitter case.

The homogeneous equation for the field is given by 
\bea
\label{eq:eom:classical:light}
	\left(a\bar{\phi}\right)^{\prime\prime}-\dfrac{2+3\epsilon_{1*}-\frac{m^2}{H_*^2}}{\eta^2}\left(a\bar{\phi}\right)=0
\eea
and has two independent solutions, $a\bar{\phi}^{(1)}\propto \left(-\eta\right)^{\frac{1}{2}-\nu}$ and $a\bar{\phi}^{(2)}\propto \left(-\eta\right)^{\frac{1}{2}+\nu}$, from which the fundamental matrix
\bea
\label{eq:U:massive}
\boldsymbol{U}(\eta)=
\left(\begin{array}{ccc}
	\dfrac{H_*}{\sqrt{2\nu}}\left(-\eta\right)^{\frac{3}{2}-\nu} &\ & -\dfrac{H_*}{\sqrt{2\nu}} \left(-\eta\right)^{\frac{3}{2}+\nu}\\
	\dfrac{\nu-\frac{3}{2}}{\sqrt{2\nu}H_*}\left(-\eta\right)^{-\frac{3}{2}-\nu} &\ & \dfrac{\nu+\frac{3}{2}}{\sqrt{2\nu}H_*}\left(-\eta\right)^{-\frac{3}{2}+\nu}
	\end{array}\right)
\eea
can be constructed, see \Eq{eq:U:def}. This gives rise to the Green matrix
\bea
	\boldsymbol{G}(\eta,\eta_0)&=&
	\left(\begin{array}{ccc}
	\frac{3+2\nu}{4\nu}\left(\frac{\eta}{\eta_0}\right)^{\frac{3}{2}-\nu} - \frac{3-2\nu}{4\nu}\left(\frac{\eta}{\eta_0}\right)^{\frac{3}{2}+\nu}
 &\ & \frac{H^2}{2\nu}\left(\eta\eta_0\right)^{3/2}\left[\left(\frac{\eta_0}{\eta}\right)^{\nu}-\left(\frac{\eta}{\eta_0}\right)^{\nu}\right] \\
\frac{\nu^2-\frac{9}{4}}{2\nu H^2}\left(\eta\eta_0\right)^{-3/2}\left[\left(\frac{\eta_0}{\eta}\right)^{\nu}-\left(\frac{\eta}{\eta_0}\right)^{\nu}\right]
 &\ &\frac{3+2\nu}{4\nu}\left(\frac{\eta_0}{\eta}\right)^{\frac{3}{2}-\nu} - \frac{3-2\nu}{4\nu}\left(\frac{\eta_0}{\eta}\right)^{\frac{3}{2}+\nu}
\end{array}\right) ,\quad
\eea
from which the diffusion matrix~(\ref{eq:Sigma}) can be obtained,
\bea
\label{eq:Sigma:light}
\boldsymbol{\Sigma}_{\boldsymbol{\Phi}}(\eta) = \dfrac{\left(\frac{\sigma}{2}\right)^{3-2\nu}}{\pi\sin^2(\pi\nu)\Gamma^2\left(1-\nu\right)}\left[1-\left(\frac{\eta}{\eta_0}\right)^{3-2\nu}\right]
	\left(\begin{array}{ccc}
		\dfrac{H_*^2}{3-2\nu} &\ & \dfrac{1}{2\eta^{3}} \\
		\dfrac{1}{2\eta^{3}} &\ & \dfrac{3-2\nu}{4H_*^2 \eta^6}
	\end{array}\right) .
\eea
In this expression, only the leading order contributions in $\sigma$ have been kept since only these terms are expected to be correctly described in the stochastic classical approximation. The dependence on $\sigma$ only appears through the overall $(\sigma/2)^{3-2\nu}$ factor, which can be approximated as $\sigma$-independent (and equal to one) if $3-2\nu$ is close enough to $0$, \ie if
\bea
\label{eq:condition:m_over_H}
\ee^{-\frac{1}{\left\vert 3-2\nu \right\vert }}\ll \dfrac{\sigma}{2}.
\eea
This condition matches Eq.~(81) of \Refa{Starobinsky:1994bd}. Let us also note that, from \Eq{eq:modesolution:light}, $3-2\nu$ is related to the spectral index $\nS -1 = \dd \ln \calP_{ a\delta\phi}/\dd \ln k$ of the power spectrum $\calP_{a \delta\phi} = k^3 \vert  a \delta\phi_k \vert^2/(2\pi^2)$ of the field fluctuations $a\delta\phi$ through $3-2\nu = \nS-1$.  The condition~(\ref{eq:condition:m_over_H}) therefore means that for a fixed comoving wavenumber, the amplitude of the field fluctuations should not vary much between the Hubble radius crossing time and the coarse-graining radius crossing time. It is compatible with the classical transition condition~(\ref{eq:cond:classicalTransition}), $\sigma\ll \vert 3-2\nu \vert^{1/3}$, if $\nu$ is sufficiently close to $3/2$ (for instance, if $\vert 3-2\nu\vert <0.1$, there are already 4 orders of magnitude between the two bounds). This will be further discussed in \Sec{sec:slowroll:stochastic:attractor}, but for now \Eq{eq:condition:m_over_H} allows one to expand \Eq{eq:Sigma:light} at leading order in $\nu-3/2\simeq \epsilon_{1*}-m^2/(3H_*^2)$, where one obtains
\bea
\label{eq:Sigma:light:lightlimit}
\boldsymbol{\Sigma}_{\boldsymbol{\Phi}}(\eta) = 
\frac{1-\left(\frac{\eta}{\eta_0}\right)^{3-2\nu}}{3-2\nu}
\dfrac{H_*^2}{4\pi^2}
	\left(\begin{array}{ccc}
		1 &\ & \dfrac{3-2\nu}{2H_*^2\eta^{3}} \\
		\dfrac{3-2\nu}{2H_*^2\eta^{3}}  &\ & \dfrac{(3-2\nu)^2}{4H_*^4 \eta^6}
	\end{array}\right) .
\eea
Compared to the massless case in de Sitter, one can see that the coarse-grained field now diffuses in the momentum $\pi$ direction as the cubic power of the scale factor. In the $\phi$ direction, at early time, when $\log(\eta_0/\eta)\ll 1/\vert 3-2\nu\vert$, one obtains $\Sigma_{\phi,\phi}\simeq H_*^2/(4\pi^2) \log(\eta_0/\eta)$ which coincides with the massless case~(\ref{eq:Sigma:massless}). At late time however, when $\log(\eta_0/\eta)\gg 1/\vert 3-2\nu \vert $, if $\nu<3/2$, it asymptotes to the equilibrium value $\Sigma_{\phi,\phi}\simeq H_*^2/[4\pi^2(3-2\nu)] \simeq 3H^4/(8\pi^2m^2) $ (where the second expression is valid for a light field with positive squared mass in de-Sitter~\cite{Starobinsky:1994bd}); while if $\nu>3/2$, it continues to increase as $\Sigma_{\phi,\phi}\simeq H_*^2/[4\pi^2(2\nu-3)] (\eta/\eta_0)^{3-2\nu}\simeq 3H^4/(8\pi^2\vert m^2\vert) (\eta_0/\eta)^{2\vert m^2\vert/(3H^2)} $ (where the second expression is valid for a light field with negative squared mass in de-Sitter).
\subsection{Non-test fields}
\label{sec:Stochastic:Non:test:fields}
If the field is not a test field, \eg if it is the inflaton itself, it couples to metric fluctuations and the stochastic formalism must account for those. The validity of the stochastic formalism then relies on the one of the separate universe approach~\cite{Salopek:1990jq, Sasaki:1995aw, Wands:2000dp, Lyth:2003im, Rigopoulos:2003ak, Lyth:2005fi}, or quasi-isotropic \cite{Lifshitz:1960, Starobinsky:1982mr, Comer:1994np, Khalatnikov:2002kn} picture, where on super-Hubble scales, each Hubble patch evolves forward in time independently of the other patches, and under a locally FLRW metric. In this section, which closely follows \Refa{Pattison:2019hef}, we first show why this approach is valid in general, before deriving the stochastic formalism for non-test fields. Compared to the case of test fields, we highlight the presence of gauge corrections that need to be taken into account, and explain how they can be calculated. The action we start from is still given by \Eq{eq:action}, where the perturbations of the line element~(\ref{eq:line_element}) away from the FLRW background~(\ref{eq:line_element:flat}) are parametrised as
\bea 
\label{eq:perturbedlineelement:FLRW}
\dd s^2 &= -(1+2A)\dd t^2 + 2a\partial_{i}B\dd x^{i}\dd t + a^2\left[ (1-2\psi)\delta_{ij} + 2\partial_i \partial_j E \right]\dd x^i \dd x^j \, ,
\eea 
and we restrict our analysis to scalar fluctuations only. If time is labeled by the number of \efolds~$N\equiv \ln a$,\footnote{Hereafter, the number of \efolds~$N$ should not be confused with the lapse function $N$ used in the previous sections, that equals $1/H$ in this case}, and denoting $\gamma \equiv \dd\phi/(\dd N)= \pi/(H a^3)$, the Langevin equations~(\ref{eq:eombarphi})-(\ref{eq:eombarpi}) read
\begin{align}
\frac{\dd {\bar{\phi}}}{\dd N} &= {\bar{\gamma}} + {\xi}_{\phi}(N) \label{eq:conjmomentum:langevin} \, ,\\
\frac{\dd {\bar{\gamma}}}{\dd N} &= -\left[3-\epsilon_{1}(\bar{\gamma})\right]{\bar{\gamma}} - \frac{V_{,\phi}({\bar{\phi}})}{H^2(\bar{\phi},\bar{\gamma})} +{\xi}_{\gamma}(N) \label{eq:KG:efolds:langevin} \, ,
\end{align}
The crucial difference with \Eqs{eq:eombarphi}-(\ref{eq:eombarpi}) is that, now, $V_{,\phi}$ depends on $\bar{\phi}$, the Hubble parameter $H$ depends on $\bar{\phi}$ and $\bar{\gamma}$ through the Friedmann equation,
\bea 
\label{eq:friedmann}
H^2(\bar{\phi},\bar{\gamma})= \dfrac{V(\bar{\phi})}{3\Mp^2-\bar{\gamma}^2} \, ,
\eea 
and the first slow-roll parameter depends on $\bar{\gamma}$ through
\bea
\epsilon_{1}(\bar{\gamma}) = \frac{\bar{\gamma}^2}{2\Mp^2}.
\eea
\subsubsection{Separate universes}
\label{sec:separateuniverse}
Since the spatial gradients in the Langevin equations~(\ref{eq:conjmomentum:langevin}) and~(\ref{eq:KG:efolds:langevin}) are neglected, one assumes that, on super-Hubble scales, each Hubble patch evolves forward in time independently of the other patches, and under a locally FLRW metric. This is the so-called separate universe picture~\cite{Salopek:1990jq, Sasaki:1995aw, Wands:2000dp, Lyth:2003im, Rigopoulos:2003ak, Lyth:2005fi}, or quasi-isotropic \cite{Lifshitz:1960, Starobinsky:1982mr, Comer:1994np, Khalatnikov:2002kn} picture. The validity of this approximation beyond slow roll has been questioned in \Refa{Cruces:2018cvq}, but below, we show why it is in fact still valid.

The separate universe approach is valid when each causally-disconnected patch of the universe evolves independently, obeying the same field equations locally as in a homogeneous and isotropic (FLRW) cosmology. 
Combining \Eqs{eq:dotphiflrw} and~(\ref{eq:dotpflrw}), the  Klein--Gordon equation for a homogeneous field in an FLRW cosmology, $\phi(t)$, is given by
\bea 
\label{eq:kleingordon}
\ddot{\phi} + 3 H\dot{\phi} + V_{,\phi} &= 0 \, ,
\eea 
where a dot denotes derivation with respect to cosmic time, while the Friedman equation reads
\bea
\label{eq:Friedmann:cosmic:time}
H^2 = \frac{V(\phi)+\frac{\dot{\phi}^2}{2}}{3\Mp^2} .
\eea

In this section, we derive the equation of motion for linear fluctuations about a homogeneous scalar field from (i) cosmological perturbation theory, and (ii) perturbations of the background FLRW equations of motion, \ie, the separate universe approach. We show that the two equations of motion match at leading order in a spatial gradient expansion, with or without slow roll.
\paragraph{Cosmological perturbation theory}

At linear order in perturbation theory, the perturbed Klein-Gordon equation in Fourier space, with \Eq{eq:perturbedlineelement:FLRW}, gives~\cite{Gordon:2000hv, Malik:2008im}
\bea \label{eq:pertubations:general}
\ddot{\delta\phi_{\bm{k}}} + 3H\dot{\delta\phi_{\bm{k}}} + \left(\frac{k^2}{a^2}+V_{,\phi\phi}\right)\delta\phi_{\bm{k}} = -2V_{,\phi}A_{\bm{k}} + \dot{\phi}\left[ \dot{A_{\bm{k}}} + 3\dot{\psi_{\bm{k}}} + \frac{k^2}{a^2}\left(a^2\dot{E_{\bm{k}}}-aB_{\bm{k}}\right)\right] .
\eea 
The metric perturbations that feature in the right-hand side of \Eq{eq:pertubations:general} satisfy the Einstein field equations, and in particular the energy and momentum constraints 
\begin{align}
\label{eq:energyconstraint:arbgauge} 3H\left(\dot{\psi_{\bm{k}}}+HA_{\bm{k}}\right) + \frac{k^2}{a^2}\left[\psi_{\bm{k}} + H\left(a^2\dot{E_{\bm{k}}}-aB_{\bm{k}}\right)\right] &= -\frac{1}{2\Mp^2}\left[ \dot{\phi}\left(\dot{\delta\phi_{\bm{k}}}-\dot{\phi}A_{\bm{k}}\right)+V_{,\phi}\delta\phi_{\bm{k}}\right], \\
\label{eq:momentumconstraint:arbgauge} \dot\psi_{\bm{k}} + H A_{\bm{k}} &= \frac{\dot\phi}{2\Mp^2}  \delta\phi_{\bm{k}} \, .
\end{align}
Introducing the Sasaki--Mukhanov variable~\cite{Sasaki:1986hm, Mukhanov:1988jd}
\bea \label{eq:def:Q}
Q_{\bm{k}} = \delta\phi_{\bm{k}} + \frac{\dot{\phi}}{H}\psi_{\bm{k}} \, ,
\eea
and using \Eqs{eq:energyconstraint:arbgauge} and \eqref{eq:momentumconstraint:arbgauge} to eliminate the metric perturbations, \Eq{eq:pertubations:general} can be rewritten as 
\bea \label{eq:pertubations}
\ddot{Q}_{\bm{k}} + 3H\dot{Q}_{\bm{k}} + \left[ \frac{k^2}{a^2} + V_{,\phi\phi} - \frac{1}{a^3\Mp^2}\frac{\dd}{\dd t}\left( \frac{a^3}{H}\dot{\phi}^2 \right) \right]Q_{\bm{k}} = 0 \, .
\eea 
We now compare this equation with the one coming from perturbing the background equations.

\paragraph{Perturbed background equations}

In order to easily relate the field fluctuation $\delta\phi$ to the Sasaki--Mukhanov variable, one usually chooses to work in the spatially-flat gauge where $\psi=0$, and hence $Q=\delta\phi$ according to \Eq{eq:def:Q}. In this paragraph, we will show how to perturb the background equations in that gauge, but also in the uniform-$N$ gauge that is used in stochastic inflation . 

Let us perturb the quantities appearing in \Eq{eq:kleingordon}, according to 
\bea \label{eq:perturbkleingordon}
&\phi \to \phi + \delta\phi \, , &\dd t \to (1+A) \dd t \, ,
\eea 
where $1+A$ is the lapse function introduced in \Eq{eq:perturbedlineelement:FLRW}. Let us stress that the lapse function needs to be perturbed, otherwise one is implicitly working in a synchronous gauge (where $A=0$), which in general differs from the spatially-flat and uniform-$N$ gauges, and this leads to inconsistencies~\cite{Cruces:2018cvq}. Inserting \Eq{eq:perturbkleingordon} into \Eq{eq:kleingordon} gives rise to
\bea \label{eq:perturbedKG}
\ddot{\delta\phi} + \left(3H + \frac{\dot{\phi}^2}{2\Mp^2H}\right)\dot{\delta\phi} + \left(\frac{\dot{\phi}}{2\Mp^2H}V_{,\phi} + V_{,\phi\phi}\right)\delta\phi
- \dot{\phi}\dot{A} - \left(2\ddot{\phi}+3H\dot{\phi}+ \frac{\dot{\phi}^3}{2\Mp^2H}\right)A = 0 \, ,
\eea 
where we have also used 
\bea \label{eq:deltaH} 
\delta H = \frac{V_{,\phi}\delta\phi + \dot{\phi}\dot{\delta\phi} - \dot{\phi}^2A}{6\Mp^2 H}  
\eea 
that comes from perturbing the Friedmann equation \eqref{eq:Friedmann:cosmic:time} under \Eq{eq:perturbkleingordon}.

\subparagraph{Spatially-flat gauge}
\label{sec:PertBackEOM:SFG}

In the spatially-flat gauge, the lapse function can readily be rewritten in terms of the field perturbation by imposing the momentum constraint~\eqref{eq:momentumconstraint:arbgauge}, which simplifies to 
\bea
\label{eq:flatA}
A = \frac{\dot{\phi}}{2\Mp^2 H} \delta \phi \,.
\eea
Substituting this relation into \Eq{eq:perturbedKG} gives rise to 
\bea \label{eq:perturbedKG:simplified}
\ddot{\delta\phi} + 3H\dot{\delta\phi} + \left[ V_{,\phi\phi} - \frac{1}{\Mp^2a^3}\frac{\dd}{\dd t}\left(\frac{a^3}{H}\dot{\phi}^2\right) \right] \delta\phi = 0 \, .
\eea 
Comparing \Eq{eq:perturbedKG:simplified}, obtained from the perturbed background equations, with \Eq{eq:pertubations}, obtained in linear perturbation theory in the spatially-flat gauge where $Q=\delta\phi$, we see that the two are consistent in the super-Hubble limit where $k\ll a H$. 

It is important to note that the local proper time in each patch is perturbed with respect to the cosmic time, $t$, in the background in the presence of a non-zero lapse perturbation, $A$. As can be seen from \Eq{eq:flatA} the perturbation $A$ vanishes in the spatially-flat gauge in the slow-roll limit, $\dot\phi\to0$, and the local proper time in this limit coincides with the background cosmic time. Beyond slow roll one must consistently account for local variations in the proper time interval in different patches if one wants to relate the separate universe equations to the perturbation equations written in terms of a global (background) cosmic time. This will be the aim of \Sec{sec:uniformexpansion}.

\subparagraph{Uniform-$N$ gauge}
\label{sec:PertBackEOM:UNG}
Let us introduce the expansion rate of $t=$constant hypersurfaces
\bea
\theta={n^\mu}_{;\mu}\, ,
\eea
where $n^\mu$ is the unit time-like vector, orthogonal to the constant-time hypersurfaces.
It is related to the metric perturbations in \Eq{eq:perturbedlineelement:FLRW} according to~\cite{Malik:2008im} 
\bea \label{eq:def:expansion}
\theta = \frac{3}{a}\left( \mathcal{H} - \mathcal{H}A - \psi' + \frac{1}{3}\nabla^2\sigma \right) \, ,
\eea 
where $\mathcal{H} = a'/a$ is the conformal Hubble parameter, a prime is a derivative with respect to conformal time $\eta$ defined through $\dd t = a \dd \eta$, and $\sigma = E'-B$ is the shear potential.
From the perturbed expansion rate $\theta$, one can define a perturbed integrated expansion up to first order in the metric perturbations
\bea 
\tilde{N} &= \displaystyle \frac{1}{3}\int \theta (1+A) \dd t 
= N - \psi + \frac{1}{3}\nabla^2\int\sigma\dd\eta \, .
\eea 
The last term in the right-hand side can be re-written in terms of $\E\equiv \int \sigma\dd\eta$, which corresponds to $E$ in the hypersurface-orthogonal threading where $B=0$. From now on, we work in such a spatial threading. This gives rise to
\bea  \label{eq:def:deltaN}
\delta N = -\psi + \frac{1}{3}\nabla^2\E \, ,
\eea 
\ie, the perturbation of the trace of the spatial metric on constant-time hypersurfaces.
Note, in particular, that in the spatially-flat gauge where $\psi=B=0$, we have $\delta N|_{\psi=0} = \frac{1}{3}\nabla^2\E |_{\psi=0}$. 

The uniform-$N$ gauge used in the Langevin equations \eqref{eq:conjmomentum:langevin} and \eqref{eq:KG:efolds:langevin} is defined by keeping the integrated expansion unperturbed across all patches of the universe, \ie $\delta N = 0$. From \Eq{eq:def:deltaN}, this imposes a direct relationship between $\psi$ and $E$, namely $\psi = \frac{1}{3}\nabla^2E_B$.
In the uniform-$N$ gauge, we note that the perturbation equation \eqref{eq:pertubations:general} can be written as
\bea 
\label{eq:pert:uniformN}
\ddot{\delta\phi_{\bm{k}}} + 3H\dot{\delta\phi_{\bm{k}}} + \left( \frac{k^2}{a^2} + V_{,\phi\phi} \right)\delta\phi_{\bm{k}} &=& \dot{\phi}\dot{A_{\bm{k}}} - 2V_{,\phi}A_{\bm{k}}  \\
&=& \dot{\phi}\dot{A_{\bm{k}}} + \left(2\ddot{\phi} + 6H\dot{\phi}\right)A_{\bm{k}} \, .
\eea
This can be recast in a form similar to the perturbed background equation~\eqref{eq:perturbedKG}, namely
\bea 
\label{eq:KGuniformN}
\ddot{\delta\phi_{\bm{k}}} + \left(3H + \frac{\dot{\phi}^2}{2\Mp^2H}\right)\dot{\delta\phi_{\bm{k}}} + \left( \frac{\dot{\phi}}{2\Mp^2H}V_{,\phi} +V_{,\phi\phi}\right)\delta\phi_{\bm{k}}
  - \dot{\phi}\dot{A_{\bm{k}}} - \left( 2\ddot{\phi}+3H\dot{\phi} + \frac{\dot{\phi}^3}{2\Mp^2H} \right)A_{\bm{k}} = \Delta_{\bm{k}}
\eea
where the difference between \Eqs{eq:perturbedKG} and \eqref{eq:pert:uniformN} is quantified as
\bea \label{eq:difference}
\Delta_{\bm{k}} = \frac{\dot{\phi}}{H}\left\lbrace \frac{1}{2\Mp^2}\left[\dot{\phi}\left(\dot{\delta\phi}_{\bm{k}} - \dot{\phi}\dot{A}_{\bm{k}}\right) + V_{,\phi}\delta\phi_{\bm{k}}\right] + 3H^2A_{\bm{k}}\right\rbrace - \frac{k^2}{a^2}\delta\phi_{\bm{k}} \, .
\eea 
If we now impose the energy constraint \eqref{eq:energyconstraint:arbgauge} in the uniform-$N$ gauge, and recalling that since we choose $B=0$, $\psi = \frac{1}{3}\nabla^2E$, one can show that
\bea \label{eq:UN:diff}
\Delta_{\bm{k}} = - \frac{k^2}{a^2}\left( \delta\phi_{\bm{k}} + \frac{\dot\phi}{H}\psi_{\bm{k}} \right) = - \frac{k^2}{a^2}Q_{\bm{k}} \, ,
\eea 
see \Eq{eq:def:Q}.
Hence, since we neglect $k^2/a^2$ terms in the large-scale limit, the perturbation equations and the perturbed background equations become identical on large scales. We conclude that the separate universe approach, describing the evolution of long-wavelength perturbations about an FLRW background in terms of locally FLRW patches, is valid in both the spatially-flat and uniform-$N$ gauges. This result does not rely on slow roll; we only require that we can neglect gradient terms on super-Hubble scales. 
\paragraph{Arbitrary gauge}
Let us finally see how the above arguments can be formulated without fixing a gauge. 
It is instructive to collect together metric perturbation terms in the Klein-Gordon equation from the full linear perturbation theory, \Eq{eq:pertubations:general}, which describe the perturbation of the local expansion rate \eqref{eq:def:expansion}
\bea
\delta\theta_{\bm{k}} = - 3\dot\psi_{\bm{k}} - \frac{k^2}{a^2}\left(a^2\dot{E_{\bm{k}}}-aB_{\bm{k}}\right) - 3HA_{\bm{k}} \,.
\eea
Re-writing the perturbed Klein-Gordon equation \eqref{eq:pertubations:general} in terms of $\delta\theta_{\bm{k}}$ we obtain
\bea
\label{eq:pertubations:general2}
\ddot{\delta\phi_{\bm{k}}} + 3H\dot{\delta\phi_{\bm{k}}} + \left(\frac{k^2}{a^2}+V_{,\phi\phi}\right)\delta\phi_{\bm{k}} 
= \left(2\ddot\phi+3H\dot\phi\right)A_{\bm{k}} + \dot{\phi} \dot{A_{\bm{k}}} 
- \dot{\phi}
\delta\theta_{\bm{k}}
\,.
\eea 
Finally, combining \Eq{eq:pertubations:general2} with the background equation \eqref{eq:kleingordon} and rewriting the time derivatives in terms of the local proper time rather than the coordinate time, $\partial/\partial\tau\equiv(1-A)\partial/\partial t$, one obtains
\bea
\frac{\partial^2}{\partial\tau^2}(\phi+\delta\phi) + \theta\frac{\partial}{\partial\tau}(\phi+\delta\phi) + V_{,\phi}(\phi+\delta\phi) = \frac{\nabla^2}{a^2} (\delta\phi) \,.
\eea
Thus we see that the perturbed Klein-Gordon equation~\eqref{eq:pertubations:general} from cosmological perturbation theory in an arbitrary gauge has exactly the same form, up to first order in the inhomogeneous field and metric perturbations and up to spatial gradient terms of order $\nabla^2\delta\phi$, as the Klein-Gordon equation for a homogeneous scalar field in an FLRW cosmology, \Eq{eq:kleingordon}, where we identify the local proper time, $\tau$, with the coordinate time, $t$, in an FLRW cosmology and the local expansion rate, $\theta/3$, with the Hubble rate, $H$, in an FLRW cosmology.
However to relate these local quantities to a global background coordinate system we need to fix a gauge. This cannot be determined by the local FLRW equations but requires to use additional constraint equations from the cosmological perturbation theory, as demonstrated in the preceding sub-paragraphs for the spatially-flat and uniform-$N$ gauges.
\subsubsection{Gauge corrections to the noise} 
\label{sec:uniformexpansion}
In order to derive the Langevin equations~(\ref{eq:conjmomentum:langevin}) and~(\ref{eq:KG:efolds:langevin}), only the field variables have been perturbed according to \Eqs{eq:field;decomposition:phi} and~(\ref{eq:field;decomposition:pi}), and not the entries of the metric. In particular, the lapse function, \ie $A$ in the notation of \Eq{eq:perturbedlineelement:FLRW}, has been neglected. This implies that the Langevin equations are written in a specific gauge, namely the one where the time coordinate is fixed. Since we work with the number of \efolds~as the time variable, this corresponds to the uniform-$N$ gauge. In \Eq{eq:noisecorrel}, the field perturbations $\phi_{\bm{k}}$ and $\gamma_{\bm{k}}$ must therefore be calculated in that same gauge. However, it is common to compute the field perturbations in the spatially-flat gauge, since in that gauge, they are directly related to the gauge-invariant curvature perturbation, which is quantised in the Bunch-Davies vacuum. One must therefore compute the correction to the noise amplitude that comes from translating the field fluctuations in the spatially-flat gauge to the uniform-$N$ gauge, and this is what is done in detail in this section. 

Let us note that one could work with a different time coordinate, hence in a different gauge (for instance, working with cosmic time $t$ would imply working in the synchronous gauge). This is not a problem as long as one computes gauge-invariant quantities in the end, such as the curvature perturbation $\zeta$. However, since $\zeta$ is related to the fluctuation in the number of \efolds~in the so-called ``stochastic $\delta N$ formalism''~\cite{Fujita:2013cna, Vennin:2015hra}, which will be the topic of \Sec{sec:stochastic:delta:N}, we find it convenient to work with the number of \efolds~as a time variable. Another reason is that, as will be shown in \Sec{sec:slow:roll:stochastic}, in the slow-roll regime, the spatially-flat gauge coincides with the uniform-$N$ gauge (but not, say, with the synchronous gauge), which makes the gauge correction identically vanish, and which explains why it is usually recommended~\cite{Vennin:2015hra} (but not compulsory) to work with $N$ as a time variable. 
\paragraph{Gauge transformations}
Let us denote quantities in the uniform-$N$ gauge with a tilde, \ie $\widetilde{\delta N}=0$. The transformations from the spatially-flat to the uniform-$N$ gauge can be written by means of a gauge transformation parameter $\alpha$ (that will be determined below), according to~\cite{Malik:2008im}
\begin{align} 
\label{eq:transform:phi} \delta\phi &\to \widetilde{\delta\phi} = \delta\phi + \phi'\alpha\, , \\
\label{eq:transform:psi}\psi &\to \widetilde{\psi} = \psi - \mathcal{H}\alpha\, , \\
\E &\to \widetilde{\E} = \E + \int\alpha\dd\eta \, .
\end{align}
Combining these transformation rules with \Eq{eq:def:deltaN}, the perturbed integrated expansion transforms as 
\bea \label{eq:transform:deltaN}
\delta N \to \widetilde{\delta N} = \delta N + \mathcal{H}\alpha + \frac{1}{3}\nabla^2\int\alpha \dd \eta \, .
\eea 
By definition, $\widetilde{\delta N}=0$, so one is lead to
\bea \label{eq:integraleq:alpha}
\delta N\Big|_{\psi=0} + \mathcal{H}\alpha  + \frac{1}{3}\nabla^2\int\alpha\dd\eta = 0 \, .
\eea
Taking the derivative of this expression with respect to conformal time, one obtains a differential equation for the gauge transformation parameter $\alpha$, namely
\bea \label{eq:alpha:diff}
3\mathcal{H}\alpha' + \left(3\mathcal{H}'+ \nabla^2\right)\alpha = S
 \, ,
\eea
where the source term reads
\bea
S = - 3 \delta N'\Big|_{\psi=0} = - \nabla^2 \sigma \Big|_{\psi=0} \,.
\eea

Two remarks are then in order. First, the source standing on the right-hand side of \Eq{eq:alpha:diff} remains to be calculated. In \Sec{sec:Nad:press}, we will show that for a scalar field it is related to the non-adiabatic pressure perturbation, and we will explain how it can be calculated. Below, we will provide the general solution to \Eq{eq:alpha:diff}, see \Eq{eq:alphaintegral1:general}. Second, once $\alpha$ is determined, the field fluctuations in the uniform-$N$ gauge can be obtained from those in the spatially-flat gauge via \Eq{eq:transform:phi}. The noise correlators~(\ref{eq:noisecorrel_Pk}) also involve the fluctuation in the conjugate momentum, so this needs to be transformed into the uniform-$N$ gauge as well. However, precisely since $N$ is unperturbed in the uniform-$N$ gauge, one simply has
\bea 
\label{eq:GaugeTransf:pi}
\widetilde{\delta \gamma} & = \displaystyle \frac{\dd \widetilde{\delta\phi}}{\dd N} \, ,
\eea 
and $\widetilde{\delta \gamma}$ can be inferred from $\widetilde{\delta\phi}$ by a straightforward time derivative.
\paragraph{Non-adiabatic pressure perturbation}
\label{sec:Nad:press}
Let us now show that the source function, $S(\eta)$ of \Eq{eq:alpha:diff}, coincides with the non-adiabatic pressure perturbation for a scalar field. This will prove that if inflation proceeds along a phase-space attractor (such as slow roll), where non-adiabatic pressure perturbations vanish, the source function vanishes as well; in this case \Eq{eq:alpha:diff} is solved by $\alpha=0$, and there are no gauge corrections.

Let us start by recalling the expressions for the energy constraint in an arbitrary gauge~\cite{Malik:2008im}
\bea
\label{eq:energyconstraint}
3\mathcal{H}\left(\psi' + \mathcal{H}A\right) - \nabla^2\left(\psi + \mathcal{H}\sigma\right)
= 
-\dfrac{a^2}{2\Mp^2}\delta\rho \, . 
\eea
Combining this with the momentum constraint \eqref{eq:momentumconstraint:arbgauge} gives 
\bea \label{eq:deltarhodeltaphi}
\nabla^2(\psi+\mathcal{H}\sigma) &= \dfrac{a^2}{2\Mp^2}
\delta\rho_{\mathrm{com}}
\, ,
\eea
where the comoving density perturbation for a scalar field is given by 
\bea \label{eq:def:comdensity}
\delta\rho_{\mathrm{com}} = \delta\rho - 
\frac{\rho'}{\phi'}\delta\phi \, .
\eea
This in turn can be related to the non-adiabatic pressure perturbation~\cite{Malik:2008im}
\bea \label{eq:def:deltaPnad}
\delta P_{\mathrm{nad}} = -\frac{2a^2}{3\mathcal{H}\phi'}V_{,\phi}\delta\rho_{\mathrm{com}} \, .
\eea
In particular, in the spatially-flat gauge where $\psi=0$, \Eq{eq:deltarhodeltaphi} becomes
\bea \label{eq:energyconstraint:Pnad}
\mathcal{H}\nabla^2\sigma|_{\psi=0}
= -\frac{3\mathcal{H}\phi'}{4\Mp^2V_{,\phi}} \delta P_{\mathrm{nad}} \, .
\eea
Thus the source term $S$ on the right-hand side of \Eq{eq:alpha:diff} reads
\bea
S = \frac{3\phi'}{4\Mp^2V_{,\phi}} \delta P_{\mathrm{nad}}\,,
\eea  
and it vanishes if the non-adiabatic pressure perturbation is zero, which is the case whenever inflation proceeds along a phase-space attractor, $\phi'=\phi'(\phi)$, such as during slow roll.

In order to find a general expression for $S(\eta)$, one can use the (arbitrary gauge) expression for $\delta\rho$ for a scalar field~\cite{Malik:2008im},
\bea \label{eq:constraint:rho}
\delta\rho = \frac{\phi'\delta\phi' - \phi'^2A}{a^2} + V_{,\phi}\delta\phi \, ,
\eea
and combine it with \Eq{eq:deltarhodeltaphi} to obtain
\bea 
\nabla^2(\psi+\mathcal{H}\sigma) &= \dfrac{a^2}{2\Mp^2}\left[\left(3H\dot{\phi} + V_{,\phi}\right)\delta\phi + \dot{\phi}\delta\dot{\phi} - \dot{\phi}^2A\right] \, .
\eea 
Hence, in terms of the field fluctuations in the spatially-flat gauge and using \Eq{eq:flatA} for the perturbed lapse function, one finds 
\bea
S = -\frac{1}{2\Mp^2\mathcal{H}}\left[\left(3\mathcal{H}\phi' + a^{2}V_{,\phi} - \frac{\phi'^3}{2\Mp^2\mathcal{H}} \right)Q + \phi'Q'\right]\, .
\eea
Introducing the second slow-roll parameter $\epsilon_2\equiv \dd\ln\epsilon_1/\dd N$, the source function can be rewritten in the simpler form
\bea \label{eq:sourcefunction:general}
S = \frac{Q\sqrt{2\epsilon_1}}{2\Mp} \mathrm{sign}(\dot{\phi}) \left(  \calH\frac{\epsilon_{2}}{2} - \frac{Q'}{Q}\right) \, .
\eea
\paragraph{General solution}
\label{sec:source}
When written in Fourier space, the differential equation~\eqref{eq:alpha:diff} for $\alpha_k$ has the general solution
\bea \label{eq:alphaintegral1:general}
\alpha_k = 
\displaystyle
\frac{1}{3\mathcal{H}}\int^{\eta}_{\eta_0} S_k(\eta')\exp\left[ \frac{k^2}{3}\int^{\eta}_{\eta'}\frac{\dd\eta''}{\mathcal{H}(\eta'')} \right]\dd\eta' \, .
\eea 
In this expression, $\eta_0$ is an integration constant that defines the slicing relative to which the expansion is measured. In general, one considers situations in which an attractor is reached at late times. Since, in such a regime, the gauge correction vanishes (given that the non-adiabatic pressure perturbation does), one takes $\eta_0$ in the asymptotic future, \ie $\eta_0 = 0^-$.

Finally, in \Eq{eq:sourcefunction:general} the Sasaki--Mukhanov variable, $Q$, needs to be determined, which can be done by solving the Sasaki--Mukhanov equation
\bea \label{eq:MSequn}
v_k'' + \left(k^2 - \frac{z''}{z}\right)v_k = 0 \, ,
\eea
where $v_k = aQ_k$ and $z \equiv a\sqrt{2\epsilon_{1}}\Mp $. One can show that, in full generality, 
\bea
\label{eq:zpp:z:eps}
\frac{z''}{z}=\calH^2(2-\epsilon_1+3\epsilon_2/2-\epsilon_1\epsilon_2/2+\epsilon_2^2/4+\epsilon_2\epsilon_3/2),
\eea
where we have introduced the third slow-roll parameter $\epsilon_3\equiv \dd\ln\epsilon_2/\dd\ln N$. 
\subsection{Slow-roll stochastic inflation}
\label{sec:slow:roll:stochastic}
\begin{figure}[t]
\begin{center}
\includegraphics[width=0.49\textwidth]{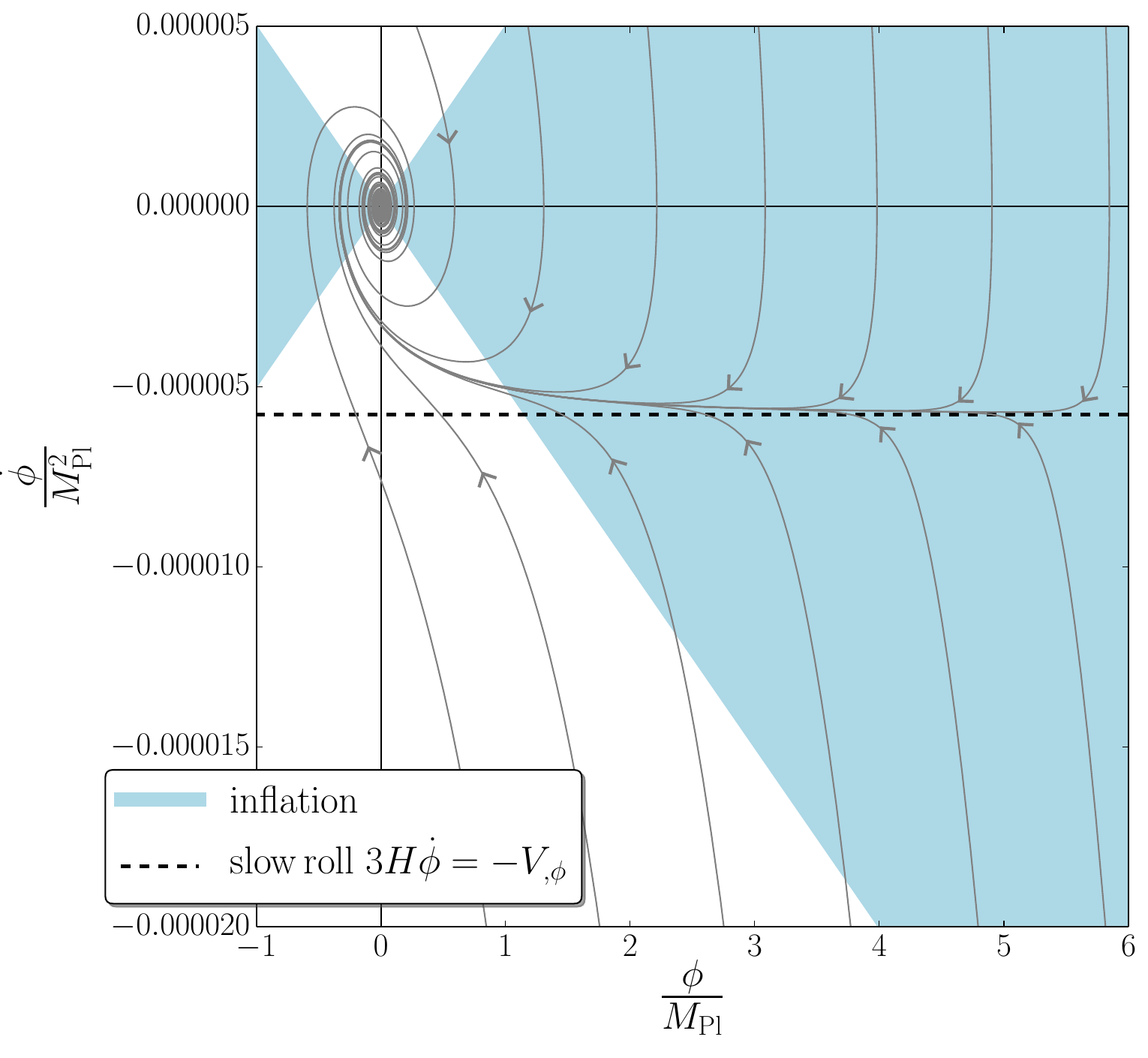}
\includegraphics[width=0.49\textwidth]{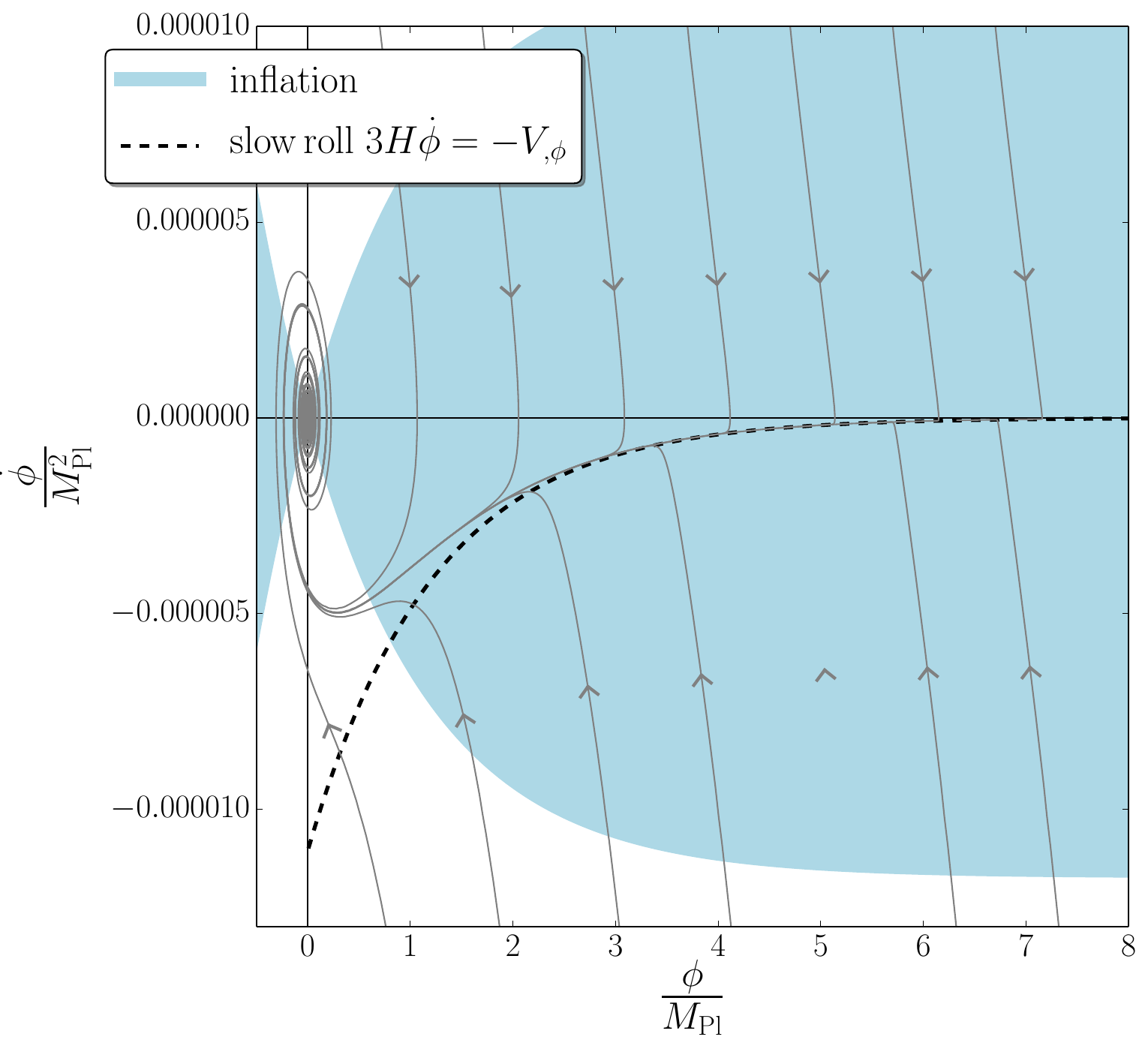}
\caption{Solutions of the the dynamical system~(\ref{eq:kleingordon})-(\ref{eq:Friedmann:cosmic:time}) displayed in the phase space $(\phi,\dot{\phi})$. The blue shaded area is where $\epsilon_1<1$, and corresponds to the region where inflation proceeds. The black dashed line stands for the slow-roll solution $3H\dot{\phi}=-V_{,\phi}$ (at leading order in slow roll), which is an attractor of the phase-space dynamics. In the left panel, the inflaton potential is $V(\phi)=m^2\phi^2/2$, where $m=7\times 10^{-6}$ in order to match the scalar power spectrum amplitude, and the right panel corresponds to the $f(R)\propto R+R^2$ Starobinsky model \cite{Starobinsky:1980te} $V(\phi)=M^4 (1-\ee^{-\sqrt{2/3}\phi/\Mp} )^2$, where $M=0.0034\Mp$ to match the power spectrum amplitude as well.}
\label{fig:classical_phase_space}
\end{center}
\end{figure}
The classical dynamics of scalar fields during inflation features an attractor known as the slow-roll regime. In \Fig{fig:classical_phase_space}, we have displayed several classical phase-space trajectories in two potentials, namely the large-field model $V(\phi)=m^2\phi^2/2$, and the $f(R)\propto R+R^2$ Starobinsky model \cite{Starobinsky:1980te} $V(\phi)=M^4 (1-\ee^{-\sqrt{2/3}\phi/\Mp} )^2$. The attractor can be clearly seen in both cases. In this section, we review the main properties of the slow-roll regime, show that it is indeed a classical attractor, derive the gauge corrections of \Sec{sec:uniformexpansion} around slow roll and show that they can be neglected at the order at which the stochastic formalism is designed, and explain why the stochastic noise is aligned with the classical field space trajectory if it proceeds along slow roll, hence does not result in departures from that solution. This leads us to the slow-roll stochastic formalism, which will be the main tool used in the following sections (with the exception of \Sec{sec:BeyondSlowRoll}).
\subsubsection{Hubble-flow parameters}
Before making any approximation, the set of Hubble-flow parameters already introduced can be completed by denoting $\epsilon_0 \equiv H_\uin/H$, and introducing
\bea
\label{def:epsilonn}
\epsilon_{n+1} \equiv \frac{\dd\ln(\epsilon_n)}{\dd N} \, .
\eea
Inserting the Klein-Gordon equation~(\ref{eq:kleingordon}) into the time derivative of the Friedmann equation~(\ref{eq:Friedmann:cosmic:time}), one obtains the first dimensionless Hubble-flow parameter
\beq
\label{eq:eps1exact}
\epsilon_1\equiv -\frac{\dot{H}}{H^2}=3\frac{\dot{\phi}^2/2}{V+\dot{\phi}^2/2} \, .
\eeq
The condition for inflation, $\ddot{a}>0$, corresponds to $\epsilon_1<1$ since $\epsilon_1=1-a \ddot{a}/\dot{a}^2$. Inserting the Klein-Gordon equation~(\ref{eq:kleingordon}) in the time derivative of \Eq{eq:eps1exact}, one obtains the second dimensionless Hubble-flow parameter
\bea
\label{eq:eps2exact}
\epsilon_2\equiv \frac{\dot{\epsilon_1}}{H\epsilon_1} =
6\left(\frac{\epsilon_1}{3}-\frac{V^\prime}{3H\dot{\phi}}-1\right)\, .
\eea
In general, there is no requirement that $\epsilon_2$ is small during inflation, in contrast to $\epsilon_1$.

We now introduce the field acceleration parameter $f$, that quantifies the relative importance of the acceleration term compared with the friction term in the Klein-Gordon equation~(\ref{eq:kleingordon}),
\bea
\label{eq:def:f}
f\equiv -\frac{\ddot{\phi}}{3H\dot{\phi}}
=1+\frac{V'}{3H\dot{\phi}}
\ .
\eea
In the second equality of the above expression, we have used the Klein-Gordon equation~(\ref{eq:kleingordon}). The field acceleration parameter can be expressed in terms of the first two Hubble-flow parameters, as can be seen from combining \Eqs{eq:eps1exact} and \eqref{eq:eps2exact} with \Eq{eq:def:f},
\bea
\label{eq:f:epsilon}
f = \frac{2\epsilon_1-\epsilon_2}{6}\, .
\eea

Since $f$ is a function of $\phi$ and $\dot{\phi}$, phase space (which is usually parametrised by $\phi $ and $\dot{\phi}$) can also be parametrised by $\phi$ and $f$. This will prove useful in the following. To this end, let us express $\dot{\phi}$ in terms of $\phi$ and $f$, which can be done by combining \Eqs{eq:Friedmann:cosmic:time} and~(\ref{eq:f:epsilon}),
\bea
\label{eq:phidot:f:phi}
\dot{\phi}^2 = V\left[\sqrt{1+\frac{2\Mp^2}{3\left(f-1\right)^2}\left(\frac{V'}{V}\right)^2}-1\right]\, .
\eea
Thus by combining \Eqs{eq:eps1exact} and~(\ref{eq:phidot:f:phi}), one can write the first Hubble flow parameter as
\bea
\label{eq:eps1:f:phi}
\epsilon_1 = \displaystyle 3\frac{\sqrt{1+\frac{2\Mp^2}{3\left(1-f\right)^2}\left(\frac{V'}{V}\right)^2}-1}{\sqrt{1+\frac{2\Mp^2}{3\left(1-f\right)^2}\left(\frac{V'}{V}\right)^2}+1}\, .
\eea
The condition for inflation to take place, $\epsilon_1<1$, then reads
\bea
\label{eq:inflation:condition}
\frac{\Mp}{\left\vert 1-f \right\vert }\left\vert \frac{V'}{V}\right\vert < \frac{3}{\sqrt{2}}\, .
\eea
One often works in the quasi-de Sitter (quasi-constant-Hubble) approximation which corresponds to $\epsilon_1\ll1$, and hence $[\Mp/\vert 1-f\vert]\vert {V'}/{V} \vert\ll1$.

Let us also note that, instead of working with $\epsilon_1$, $\epsilon_2$ and $\epsilon_3$, one can work with $\epsilon_1$, $f$ and $\eta$, where the dimensionless mass parameter $\eta$ is defined as
\bea
\label{eta:def}
\eta = \frac{V_{,\phi\phi}}{3H^2} \, .
\eea
For instance, the source term $z''/z$ in the Sasaki-Mukhanov equation~(\ref{eq:MSequn}), given in \Eq{eq:zpp:z:eps}, can be expressed as
\bea \label{eq:z''overz:general}
\frac{z''}{z} = \calH^2 \left( 2 + 5\epsilon_{1} - 3\eta - 12f\epsilon_{1} + 2\epsilon_{1}^2 \right) .
\eea 
\subsubsection{Slow-roll inflation}
\label{sec:slowroll:slowroll}
Slow-roll inflation corresponds to the regime where all Hubble-flow parameters are much smaller than one, \ie $\vert \epsilon_n\vert \ll 1$ for $n\geq 1$. From \Eq{eq:eps1exact}, this means that the kinetic energy of the inflaton field is much smaller than its potential energy. From \Eq{eq:f:epsilon}, it also means that $|f|\ll 1$, so \Eq{eq:def:f} implies that the acceleration of the inflaton field can be neglected\footnote{In this sense slow roll corresponds to quasi-equilibrium, \ie zero net force with frictional force equal and opposite to the force from the potential gradient.} compared with its friction in the Klein-Gordon equation~(\ref{eq:kleingordon}), and the dynamical system boils down to
\bea
\label{eq:slowroll}
H_\SR^2 \simeq \frac{V}{3\Mp^2}\,  \quad {\rm and} \quad 3H\dot\phi_\SR \simeq - V' \, .
\eea
In this limit, $\dot{\phi}$ is determined completely by the gradient of the potential and a single trajectory is selected out in phase space since $\dot{\phi}_\SR$ has no dependence on initial conditions. One notices that while slow roll is formally defined as $\vert\epsilon_n\vert \ll 1$ for all $n\geq 1$, the above system only relies on $\epsilon_1\ll 1$ and $\vert \epsilon_2 \vert \ll 1$.

Since $\dot{\phi}$ is an explicit function of $\phi$ through \Eq{eq:slowroll}, any phase space function can be written as a function of $\phi$ only. For the first Hubble-flow parameter and the field acceleration parameter, substituting \Eq{eq:slowroll} into \Eqs{eq:eps1exact} and \eqref{eq:def:f} respectively, one obtains
\bea
\label{eq:eps1:SR}
\epsilon_{1\SR} \simeq \frac{\Mp^2}{2}\left(\frac{V'}{V}\right)^2 \,,
\eea
\bea
f_\SR \simeq \frac{\Mp^2}{3} \left[\frac{V''}{V} - \frac{1}{2}\left(\frac{V'}{V}\right)^2\right]\, .
\label{eq:f:SR}
\eea
For a given inflationary potential $V(\phi)$, the existence of a regime of slow-roll inflation can thus be checked by verifying that  the potential slow-roll parameters $\epsilon_V$ and $\eta_V$, defined as
\bea
\label{eq:epsV}
\epsilon_{1V} \equiv \frac{\Mp^2}{2}\left(\frac{V'}{V}\right)^2 
\quad {\rm and} \quad
\eta_{V} \equiv \Mp^2 \frac{V''}{V} \, ,
\eea
remain small,
\bea
\label{eq:sr:consistency}
 \epsilon_{1V} \ll 1 \quad {\rm and} \quad |\eta_V |\ll1 \,.
\eea
\subsubsection{Slow roll as a classical attractor}
\label{sec:slowroll:classical:attractor}
Since the field acceleration parameter, $f$, quantifies the importance of the acceleration term in the Klein-Gordon equation \eqref{eq:kleingordon}, it essentially parameterises whether we are in slow roll ($\vert f \vert \ll 1$) or not.
As such, knowing the evolution of $f$ will allow us to study the stability of the slow-roll regime.

We begin by recasting \Eq{eq:kleingordon} with $\phi$ as the ``time" variable, which reduces the equation to a first-order differential equation, namely
\bea 
\label{eq:kgphi}
\frac{\dd\dot{\phi}}{\dd\phi} +3H + \frac{V'}{\dot{\phi}} = 0 \, ,
\eea 
which can also be written as
\bea
\label{eq:phiddot:f:phi}
\frac{\dd}{\dd\phi}  \left(\dot{\phi}^2\right) = -2 V'\frac{f}{f-1}\, .
\eea
Combined with \Eq{eq:phidot:f:phi}, this leads to an equation for the evolution of $f$,
 \bea
 \frac{\dd f}{\dd \phi} = \frac{3}{2\Mp^2}\frac{V}{V'}\left(f-1\right)^2\left(f+1\right)\left[\sqrt{1+\frac{2\Mp^2}{3\left(f-1\right)^2}\left(\frac{V'}{V}\right)^2}-\frac{1-f}{1+f}\right]-\left(1-f\right)\frac{V''}{V'}\, .
 \label{eq:f:dynamical}
 \eea
This can be written in terms of the potential slow-roll parameters \eqref{eq:epsV} as
\bea
\frac{\dd f}{\dd \phi} = \frac{3}{2\Mp}\frac{\left(f-1\right)^2\left(f+1\right)}{\sqrt{2\epsilon_{V}}}\left[\sqrt{1+\frac{4\epsilon_{V}}{3\left(f-1\right)^2}}-\frac{1-f}{1+f}\right]-\frac{\left(1-f\right)\eta_{V}}{2\Mp\sqrt{2\epsilon_{V}}}\, .
\label{eq:f:dynamical:srparams}
\eea
Note that this equation is exact and does not make any assumption about the smallness or otherwise of the slow-roll parameters.

If we now expand the right-hand side of \Eq{eq:f:dynamical:srparams} to first order in the potential slow-roll parameters \eqref{eq:epsV}, and take $f$ to be of first order in the slow-roll parameters as suggested by \Eq{eq:f:SR}, one obtains
\bea
\frac{V'}{V}\frac{\dd f}{\dd \phi} \simeq \frac{1}{2}\left(\frac{V'}{V}\right)^2 - \frac{V''}{V}+\frac{3}{\Mp^2}f\, .
\label{eq:f:dynamical:SR}
\eea
We see that the right-hand side of \Eq{eq:f:dynamical:SR} vanishes for the slow-roll solution \eqref{eq:f:SR}, which is consistent with the fact that $f$ is first order in the slow-roll parameters and $  V'/V \dd f/\dd \phi  \simeq \dd f/\dd N$ is therefore second order in slow roll. 

The stability of the slow-roll solution can then be studied by considering a deviation from \Eq{eq:f:SR} parametrised by
\bea
f \simeq f_{\mathrm{SR}}+\Delta\, .
\label{eq:f:fSR:epsilon}
\eea
In this expression, $f_{\mathrm{SR}}$ is given by \Eq{eq:f:SR} plus corrections that are second order in slow roll and $\Delta$ describes deviations from slow roll that are nonetheless first order in slow-roll parameters or higher. For instance, we imagine that initially, one displaces $f$ from the standard slow-roll expression given in \Eq{eq:f:SR} (\eg by adding another linear combination of some slow-roll parameters) and study how this displacement evolves in time. 
By substituting \Eq{eq:f:fSR:epsilon} into \Eq{eq:f:dynamical:SR}, one obtains
\beq
\frac{V'}{V}\frac{\dd \Delta}{\dd \phi} \simeq \frac{3}{\Mp^2}\Delta\, ,
\eeq
which at leading order in slow roll, using \Eq{eq:slowroll}, can easily be solved to give
\bea
\Delta \simeq \Delta_\uin \exp\left[-3\left(N-N_\uin\right)\right]\, ,
\eea
which is always decreasing as inflation continues. This shows that slow roll is a stable attractor solution whenever the consistency conditions \eqref{eq:sr:consistency} are satisfied. This is of course a well-known result~\cite{Salopek:1990jq, Liddle:1994dx} but is here formally proven with a formalism that will prove useful to study other regimes than slow roll, such as ultra slow roll in \Sec{sec:BeyondSlowRoll}
\subsubsection{Gauge corrections in slow roll}
\label{sec:slowroll:gauge:corrections}
Let us apply the programme sketched in \Sec{sec:uniformexpansion} to the case of slow-roll inflation. As argued before, the presence of a dynamical attractor in that case makes the non-adiabatic pressure perturbation vanish, hence we should not find any gauge correction to the field fluctuations in the uniform-$N$ gauge and thus to the correlators for the noise. This is therefore a consistency check of our formalism. 

In the slow-roll regime, combining the results of \Sec{sec:slowroll:slowroll}, one has $\eta \simeq 2\epsilon_1-\epsilon_2/2$, so at leading order in the slow-roll parameters, \Eq{eq:z''overz:general} reduces to 
\bea 
\label{eq:zpp:z:slow:roll:gen}
\frac{z''}{z} &\simeq a^2H^2 \left[ 2 - \epsilon_{1} + \frac{3}{2}\epsilon_{2}  +\order{\epsilon^2}\right] \, , 
\eea 
In order to write \Eq{eq:zpp:z:slow:roll:gen} as an explicit function of conformal time, note that 
\bea 
\label{eq:eta:integral}
\eta = \int \frac{\dd t}{a} = \int \frac{\dd a}{a^2H(a)} = -\frac{1}{aH} + \int \frac{\epsilon_{1} \dd a}{a^2H} \, ,
\eea 
where we have integrated by parts to get the last equality.

Combining \Eqs{eq:eps1exact} and~\eqref{eq:Friedmann:cosmic:time}, one has $\epsilon_{1}  = \dot{\phi}^2/(2\Mp^2H^2)$, and derivating this expression with respect to time, and making use of the Klein-Gordon equation~(\ref{eq:kleingordon}), one finds $\dot{\epsilon_{1}}/\epsilon_{1} = 2H\left( \epsilon_{1} - 3f \right)$. This allows us to integrate by parts one more time and to obtain
\bea 
\int \frac{\epsilon_{1} \dd a}{a^2H} = -\frac{\epsilon_{1}}{aH} + \int \frac{\dd a}{a^2H}\frac{\dot{\epsilon_{1}}}{\epsilon_{1}}\frac{\epsilon_{1}}{H} + \mathcal{O}\left( \epsilon_{1}^2 \right)
 = -\frac{\epsilon_{1}}{aH} + \mathcal{O}\left(\epsilon_{1}^2, f\epsilon_{1} \right) \, .
\eea 
Therefore, from \Eq{eq:eta:integral},
\bea \label{eq:tau:slowroll}
\eta \simeq -\frac{1}{aH}\left( 1 + \epsilon_{1} \right)
\eea 
at first order in slow roll, and \Eq{eq:zpp:z:slow:roll:gen} becomes
\bea \label{eq:z'':slowrolllimit}
\frac{z''}{z} &\simeq \frac{2}{\eta^2} \left( 1 + \frac{3}{2}\epsilon_{1} + \frac{3}{4}\epsilon_{2} \right)\, .
\eea 

At leading order in slow roll, the slow-roll parameters can simply be evaluated at the Hubble-crossing time $\eta_{*}\simeq -1/k$, since their time dependence is slow-roll suppressed, \ie $\epsilon_{1} = \epsilon_{1*} + \mathcal{O}(\epsilon^2)$, \etc. At that order, \Eq{eq:z'':slowrolllimit} becomes $z''/z\simeq 2(1+3\epsilon_{1*}/2 + 3 \epsilon_{2*}/4)/\eta^2$, and \Eq{eq:MSequn} is solved according to 
\bea \label{eq:MS:SRsolution}
v_k = \frac{\sqrt{-\pi\eta}}{2}\mathrm{H}_{\nu}^{(2)}\left(-k\eta\right) = aQ_k \, ,
\eea
where $\mathrm{H}_{\nu}^{(2)}$ is the Hankel function of the second kind and $\nu \equiv 3/2+\epsilon_{1*}+\epsilon_{2*}/2$, and where one has imposed Bunch-Davies initial conditions. Since the coarse-graining parameter is such that $\sigma\ll 1$, the power spectra in \Eq{eq:noisecorrel_Pk} need to be evaluated in the super-Hubble regime, \ie when $-k\eta \ll 1$. One can therefore make use of the asymptotic behaviour
\bea \label{eq:Hankel2:largescale}
H_{\nu}^{(2)}(-k\eta) \simeq \frac{i\Gamma(\nu)}{\pi}\left(\frac{2}{-k\eta}\right)^{\nu}\left[1 + \frac{1}{4(\nu-1)}\left(-k\eta\right)^{2} +\order{k^4\eta^4}\right] \, .
\eea
On the other hand, at first order in slow roll, the scale factor can also be expanded, and one finds
\bea \label{eq:scalefactor:slowroll}
a = -\frac{1}{H_{*}\eta}\left[1 + \epsilon_{1*} - \epsilon_{1*}\ln{\left(\frac{\eta}{\eta*}\right)}+\mathcal{O}(\epsilon^2)\right] \, ,
\eea
where we have used \Eq{eq:tau:slowroll}. Combining the two previous equations then leads to
\bea 
\label{eq:Q'overQ:SR}
\displaystyle
\frac{Q'_k}{Q_k} \simeq \frac{\frac{3}{2}+\epsilon_{1*}-\nu}{\eta} + \frac{\frac{7}{2}-\nu}{4(\nu-1)}k^2\eta \, ,
\eea
which is valid at next-to-leading order both in the slow-roll parameters and in $k\eta$. With the expression given above for $\nu$, one can see that $Q'_k/Q_k\simeq -\epsilon_2/(2\eta)$ at leading order in $k\eta$. Since, at leading order, $\calH\simeq -1/\eta$, the two terms in the right-hand side of \Eq{eq:sourcefunction:general} exactly cancel, and the source function $S_k$ vanishes. This confirms that the gauge corrections are indeed suppressed in that case.

In fact, the first contribution to the gauge correction comes from the decaying mode, and for completeness we now derive its value. Plugging the previous expressions into \Eq{eq:sourcefunction:general} leads to
\bea \label{eq:sourcefunction:SR}
S_k = \frac{i}{2}\frac{H_*}{\Mp}\sqrt{k\epsilon_{1*} }\eta\left(-k\eta\right)^{-\epsilon_{1*}-\frac{\epsilon_{2*}}{2}}  \mathrm{sign}\left(\dot{\phi}\right) \, .
\eea
One can then insert \Eq{eq:sourcefunction:SR}, along with $\mathcal{H} = -(1+\epsilon_{1*})/\eta$, see \Eq{eq:tau:slowroll}, into \Eq{eq:alphaintegral1:general}, and derive the gauge transformation parameter from the spatially-flat gauge to the uniform-$N$ gauge in the large-scale and slow-roll limit, 
\bea \label{eq:alphaLO:SR}
\alpha_k = \frac{iH_{*}\sqrt{\epsilon_{1*}}}{12\Mp}k^{-\frac{5}{2}}\left(-k\eta\right)^{3-\epsilon_{1*}-\frac{\epsilon_{2*}}{2}}  \mathrm{sign}\left(\dot{\phi}\right) \, .
\eea
In the uniform-$N$ gauge, according to \Eq{eq:transform:phi}, the field fluctuation thus reads
\bea 
\label{eq:GaugeCorr:SR}
\widetilde{\delta\phi}_k &= Q_k\left[ 1 - \frac{\epsilon_{1*}}{6}\left(-k\eta\right)^2 \right] ,
\eea
and its deviation from $Q$ is therefore both slow-roll suppressed and controlled by the amplitude of the decaying mode. Since it needs to be evaluated at the coarse-graining scale $k_\sigma = \sigma a H$ in \Eq{eq:noisecorrel_Pk}, the relative gauge correction to the correlations of the noises scales as $\epsilon_1\sigma^2$, which can be neglected since the stochastic formalism assumes $\sigma\rightarrow 0$.
\subsubsection{Slow roll as a stochastic attractor}
\label{sec:slowroll:stochastic:attractor}
In \Sec{sec:slowroll:classical:attractor}, it was shown that slow roll is an attractor of the classical equations of motion. In stochastic inflation, quantum noises act in phase space, and can a priori induce diffusion away from the slow roll trajectory. In this section, we examine this possibility, and conclude that in fact, the stochastic noise is always aligned with the classical attractor (if inflation proceeds along such an attractor), and that the slow-roll attractor is therefore immune to stochastic effects. We first study the case of a free scalar field for which the phase-space PDF was obtained in \Sec{sec:testField}, before extending the discussion to other types of fields (including the inflaton field).
\paragraph{Free scalar field}
For a free scalar field, in \Sec{sec:light}, it was shown that the equation of motion~(\ref{eq:eom:classical:light}) for the classical evolution of the homogeneous field has two independent solutions,
\bea
\label{eq:def:PhiSR}
	\boldsymbol{\Phi}_\sr=\left(\begin{array}{c}
		\phi_\sr \\
		\pi_\sr
	\end{array}\right)=\left(\begin{array}{c}
		\displaystyle \frac{H_*}{\sqrt{2\nu}}(-\eta)^{\frac{3}{2}-\nu} \\
		\displaystyle\frac{\nu-\frac{3}{2}}{\sqrt{2\nu}H_*}(-\eta)^{-\frac{3}{2}-\nu}
	\end{array}\right)
\eea
and
\bea
\label{eq:def:PhiNSR}
	\boldsymbol{\Phi}_\nsr=\left(\begin{array}{c}
		\phi_\nsr \\
		\pi_\nsr
	\end{array}\right)=\left(\begin{array}{c}
		\displaystyle \frac{-H_*}{\sqrt{2\nu}}(-\eta)^{\frac{3}{2}+\nu} \\
		\displaystyle\frac{\nu+\frac{3}{2}}{\sqrt{2\nu}H_*}(-\eta)^{-\frac{3}{2}+\nu}
	\end{array}\right) .
\eea
The subscripts ``SR'' and ``USR'' stand for ``slow roll'' and ``ultra slow roll'' respectively. Indeed, the field acceleration parameter~(\ref{eq:def:f}) on these two branches is given by
\bea
\label{eq:phidotdotSR}
f_\sr& = & -\frac{\ddot{\phi}_\sr}{3H\dot{\phi}_\sr}= \frac{1}{2}\left(1-\sqrt{1-\frac{4}{9}\frac{m^2}{H^2}}\right)\simeq  \frac{m^2}{9H^2} , \\
f_\nsr & = & - \frac{\ddot{\phi}_\nsr}{3H\dot{\phi}_\nsr} = \frac{1}{2}\left(1+\sqrt{1-\frac{4}{9}\frac{m^2}{H^2}}\right)\simeq   1-\frac{m^2}{9H^2} ,
\eea
where the second equalities hold for $m\ll H$. In \Sec{sec:slowroll:slowroll}, we have seen already that slow roll corresponds to $\vert f \vert\ll 1$,\footnote{\label{footnote:SR:test:field}This allows us to extend the notion of slow-roll dynamics, defined for the inflaton in \Sec{sec:slowroll:slowroll}, to all scalar fields during inflation. For a scalar field $\phi$ with potential $V(\phi)$, that may or may not substantially contribute to the total energy budget of the Universe, the fractional energy density contained in the kinetic term can indeed be still quantified by a first ``slow-roll'' parameter
\bea
\label{eq:eps1:testfield}
\epsilon_{1}^{\phi}= 3\frac{\dot{\phi}^2/2}{V(\phi)+\dot{\phi}^2/2} ,
\eea
even if there is no relationship between $\epsilon_{1}^{\phi}$ and $H$ as in \Eq{eq:eps1exact}. By using the Klein-Gordon equation~(\ref{eq:kleingordon}) only, that still holds for test fields (contrary to the Friedmann equation), the second slow-roll parameter $\epsilon_{2}^{\phi}=\dd\ln\epsilon_{1}^{\phi}/\dd N$ is given by
\bea
\label{eq:eps2:testfield}
\epsilon_{2}^{\phi}=6\left(\frac{\epsilon_{1}^{\phi}}{3}-\frac{V_{,\phi}}{3H\dot{\phi}}-1\right) ,
\eea  
even if, here again, there is no relationship between $\epsilon_{2}^{\phi}$ and $H$ as in \Eq{eq:eps2exact}. Similarly, the full hierarchy of slow-roll parameters $\epsilon_{n+1}^{\phi}=\dd\ln\vert \epsilon_{n}^{\phi} \vert/\dd N$ can be constructed, and if one defines ``slow roll'' for a generic scalar field as being the regime where $\vert \epsilon_n^\phi \vert \ll 1$ for all $n>0$, \Eq{eq:eps1:testfield} implies that the kinetic energy of a generic slowly rolling field is negligible compared to its potential energy, and \Eq{eq:eps2:testfield} means that its phase-space trajectory is $\dot{\phi}\simeq -V_{,\phi}/(3H)$ at leading order, \ie that its field acceleration parameter is small, which explains the notation in the main text. Making use of this phase-space trajectory allows one to derive approximate expressions for the slow-roll parameters in terms of the potential $V$,
\bea
\label{eq:eps1:appr:test}
\epsilon_1^\phi&\simeq&\frac{V_{,\phi}^2}{6H^2V} ,\\
\epsilon_2^\phi&\simeq& 2\epsilon_{1}+\frac{V_{,\phi}^2}{3H^2V}-\frac{2V_{,\phi\phi}}{3H^2} .
\label{eq:eps2:appr:test}
\eea  
In the case where $\phi$ is the inflaton, by plugging the Friedmann equation~(\ref{eq:Friedmann:cosmic:time}) into \Eqs{eq:eps1:appr:test} and~(\ref{eq:eps2:appr:test}), one recovers \Eqs{eq:eps1:SR} and~(\ref{eq:f:SR}), but the expressions derived here are more generic. Again, they translate the slow-roll conditions into conditions on the potential.}
and we denote by ultra slow roll the regime where $\vert f-1 \vert\ll 1$ (hence the potential gradient term in the one negligible in the Klein-Gordon equation), that will be further discussed in \Sec{sec:BeyondSlowRoll}. At present, it is enough the describe it as being the ``non-slow roll'' solution.  At late time, the SR branch of the solution dominates over the USR branch, hence the SR solution is a dynamical attractor of the system (this was proven if $\phi$ is the inflaton in \Sec{sec:slowroll:classical:attractor}, we have now shown that this property applies to test fields too).

Let us note that the solutions $\boldsymbol{\Phi}_\sr$ and $\boldsymbol{\Phi}_\nsr$ match the ones given below \Eq{eq:eom:classical:light} (where $\bar{\phi}^{(1)}$ needs to be identified with $\phi_\sr$ and $\bar{\phi}^{(2)}$ with $\phi_\nsr$), so that the fundamental matrix defined in \Eq{eq:U:massive} simply reads $\boldsymbol{U}(\eta)=\left(\boldsymbol{\Phi}_\sr,~\boldsymbol{\Phi}_\nsr\right)$. 
\paragraph{Diffusion in the interaction picture}
In the covariance matrix~(\ref{eq:Sigma:light:lightlimit}) of the free coarse-grained field, one can note that $ {\Sigma}_{\phi,\pi}=-\sqrt{ {\Sigma}_{\phi,\phi} {\Sigma}_{\pi,\pi}}$, implying that the noises in the $\phi$ and $\pi$ directions are totally anticorrelated. This also means that $\det(\boldsymbol{\Sigma}_{\boldsymbol{\Phi}})=0$, hence the covariance matrix has one non-zero eigenvalue and one vanishing eigenvalue, respectively defining a first direction where diffusion occurs, and a second non-diffusive direction. The question is how these two directions relate to the attractor (SR) and repeller (USR) directions. In order to establish this relation, let us formulate the stochastic dynamics in terms of canonical variables that are aligned with the SR and USR solutions of the homogeneous problem, which we call the interaction picture.

Any solution of the homogeneous problem can be expressed as $\boldsymbol{\Phi}_\mathrm{det}=z_\sr\boldsymbol{\Phi}_\sr+z_\nsr \boldsymbol{\Phi}_\nsr$, where $z_\sr$ and $z_\nsr$ are two constants. If the solution satisfies the initial value problem $\boldsymbol{\Phi}_\mathrm{det}(\eta_0)=\boldsymbol{\Phi}_0$, these constants are given by $\boldsymbol{z}=(z_\sr,z_\nsr)^\dagger=\boldsymbol{U}^{-1}(\eta_0)\boldsymbol{\Phi}_0$. They are also formally obtained from the Wronskian
\bea
\label{eq:def:zsr}
	z_\sr=\boldsymbol{\Phi}^\dag\boldsymbol{\Omega}\boldsymbol{\Phi}_\nsr, \\
	z_\nsr=-\boldsymbol{\Phi}^\dag\boldsymbol{\Omega}\boldsymbol{\Phi}_\sr,
\label{eq:def:znsr}
\eea
where $\boldsymbol{\Omega}$ has been defined in \Eq{eq:def:Omega} and where the subscript ``det'' has been dropped to let $\boldsymbol{z}$ describe a generic parametrisation of phase space. Note that the SR constant is obtained by projecting the general solution on the USR branch, and vice versa. This means that the set of classical solutions can be parametrised by the constants $\boldsymbol{z}$. Combining \Eqs{eq:Phidet} and~(\ref{eq:Green:fundamental}), the link between $\boldsymbol{z}$ and $\boldsymbol{\Phi}$ can also be written $\boldsymbol{\Phi}  = \boldsymbol{U}(\eta)\boldsymbol{z} $. Since $\boldsymbol{U}(\eta)$ is a $(2\times2)$ real matrix with unit determinant, as explained in footnote~\ref{footnote:symplectic}, it is symplectic. It thus defines a linear and homogeneous canonical transformation~\cite{Grain:2019vnq}. 

In the set of canonical variables $\boldsymbol{z}$, the classical dynamics is simply frozen,
\bea
\boldsymbol{z}_\mathrm{det}(\eta) = \boldsymbol{z}_\mathrm{det}(\eta_0) ,
\eea
and the averaged trajectory reduces to a single point in phase space. The deterministic part of the dynamics thus factors out and only diffusion remains, hence the name ``interaction picture''. Since the first and second entries of $\boldsymbol{z}$ correspond respectively to the attractor and repeller branches,  the averaged trajectory asymptotes to the attractor solution unless $z_{\mathrm{det}}^{\sr}=0$, and the attractor branch dominates the dynamics at times $\eta\gg -\vert z_{\mathrm{det}}^{\sr}/z_{\mathrm{det}}^{\nsr} \vert^{1/(2\nu)}$. 

We introduce the covariance matrix in the $\boldsymbol{z}$ variables similarly to what was done around \Eq{eq:Sigma}, \ie
\bea
\label{eq:Sigmaz:SigmaPhi}
\boldsymbol{\Sigma}_{\boldsymbol{z}}(\eta) &=& \langle\left[\boldsymbol{z}(\tau)-\left<\boldsymbol{z}(\tau)\right>\right]\left[\boldsymbol{z}(\tau)-\left<\boldsymbol{z}(\tau)\right>\right]^\dag\rangle
= \boldsymbol{U}^{-1}(\eta) \boldsymbol{\Sigma}_{\boldsymbol{\Phi}}(\eta)  \left[\boldsymbol{U}^{-1}(\eta)\right]^{\dagger}\\
& = & \int_{\eta_0}^\eta \dd s \underbrace{
\boldsymbol{U}^{-1}(s) \boldsymbol{D}_{\boldsymbol{\Phi}}(s) \left[\boldsymbol{U}^{-1}(s)\right]^\dagger}_{\boldsymbol{D}_{\boldsymbol{z}}(s)}
\eea
where, in the first equality, we have used that $\boldsymbol{\Phi}  = \boldsymbol{U}(\eta)\boldsymbol{z} $, and in the second equality, which defines $\boldsymbol{D}_{\boldsymbol{z}}$, we have combined \Eqs{eq:Sigma} and~\eqref{eq:Green:fundamental}. In the interaction picture, the covariance matrix is therefore nothing but the noise power cumulated over time. Combining \Eqs{eq:U:massive} and~(\ref{eq:Sigma:light:lightlimit}) into \Eq{eq:Sigmaz:SigmaPhi}, it reads
\bea
\label{eq:Sigma:z}
\boldsymbol{\Sigma}_{\boldsymbol{z}}(\eta)=
\frac{1-\left(\frac{\eta}{\eta_0}\right)^{3-2\nu}}{3-2\nu}\frac{3}{4\pi^2}
\left(\begin{array}{cc}
1 &0 \\
0  &0
\end{array}\right) .
\eea
This clearly entails that diffusion takes place in the SR direction only, and that the USR direction remains deterministic. As will be carefully shown below, this implies that the classical attractor generalises to a stochastic attractor of the free-field theory.

This result can be reformulated by noticing that, for free fields (and only for free fields), the equation of motion of field perturbations~(\ref{eq:mode:light}) on super-Hubble scales $k\ll -1/\eta$ is the same as the one for the background~(\ref{eq:eom:classical:light}). The use of the interaction picture is again convenient to make this argument, since it can be extended down to the quantum fluctuations. Using vectorial notation $\boldsymbol{\Phi}_k=(\phi_k,\pi_k)^\mathrm{T}$, their dynamics is given by \Eq{eq:mode:light}, namely
\bea
\label{eq:eom:pert:vectorial}
\boldsymbol{\Phi}'_k=\left(\boldsymbol{A}+\boldsymbol{V}_k\right)\boldsymbol{\Phi}_k ,
\eea
where $\boldsymbol{A}$ has been defined in \Eq{eq:inhomopb} and corresponds to the deterministic evolution of the coarse-grained field, to which the additional potential
\bea
	\boldsymbol{V}_k=\left(\begin{array}{cc}
	0 & 0 \\
	-a^2(\eta)k^2 & 0
	\end{array}\right)
\eea
is added. Under the canonical transformation $\boldsymbol{\Phi}_k=\boldsymbol{U}(\eta)\boldsymbol{z}_k$, making use of the relation $\boldsymbol{U}^\prime = \boldsymbol{AU}$ given below \Eq{eq:U:def}, \Eq{eq:eom:pert:vectorial} reads 
\bea
\label{eq:eom:pert:z}
\boldsymbol{z}'_k=\boldsymbol{U}^{-1}\boldsymbol{V}_k\boldsymbol{U}\boldsymbol{z}_k .
\eea
This shows that the canonical transformation sending to the interaction picture of the classical, deterministic dynamics is also sending to the interaction picture of the quantum fluctuations and hence to the interaction picture of the stochastic, coarse-grained dynamics.

In this interaction picture, the solution~(\ref{eq:modesolution:light}) of \Eq{eq:eom:pert:z} is such that, at leading order in the coarse-graining parameter $\sigma$, $\boldsymbol{z}_k\propto (1,0)^\dagger$. This implies that in the super-Hubble limit, quantum fluctuations are highly squeezed along the attractor branch of the classical theory. The reason is that on super-Hubble scales, the interaction potential $\boldsymbol{V}_k$ becomes subdominant so that the equation of motion of perturbations matches the one for the background. In the asymptotic future, perturbations therefore confine to the classical attractor and quantum diffusion takes place in phase space along this attractor direction only.

We have thus shown that quantum diffusion takes place in phase space along the slow-roll classical attractor only. This implies that, if one starts on the attractor, the subsequent stochastic dynamics remains confined to the attractor at any later time. If initial conditions are displaced from the attractor however, the stochastic dynamics explores regions of phase space that are inaccessible to the classical dynamics, and that may lie outside the slow-roll domain. This raises two questions that we now address:
\begin{itemize}
\item For which initial conditions does slow roll generalise to a stochastic attractor (in other words, what is the basin of attraction of stochastic slow-roll inflation)?
\item When this is the case, how much time does it take to relax towards slow roll and how does it compare to the classical situation?
\end{itemize}
Before answering these two questions, let us formulate them in more quantitative terms. Since quantum diffusion takes place along the SR direction only, in the interaction picture, solutions of the stochastic dynamics can be written as
\bea
\label{eq:decom:stoch:interaction}
\boldsymbol{\Phi}(\eta)=\widehat{z}_\sr\left(\eta\right)\boldsymbol{\Phi}_\sr+z_\nsr^{(0)}\boldsymbol{\Phi}_\nsr .
\eea
In this expression, $\widehat{z}_\sr$ is a Gaussian random variable with mean equal to $z_\sr(\eta_0)$ that we denote $z_\sr^{(0)}$ for simplicity, and variance equal to the $(z_\sr,z_\sr)$ component of the covariance matrix~(\ref{eq:Sigma:z}) that we denote $\Sigma_\sr(\eta)$ for simplicity. The deterministic quantity $z_\nsr^{(0)}$ is set by initial conditions, and hereafter, stochastic quantities are denoted with a hat. The averaged trajectory of the coarse-grained field is $\langle \boldsymbol{\Phi} (\eta)\rangle =z_\sr^{(0)}\boldsymbol{\Phi}_\sr+z_\nsr^{(0)}\boldsymbol{\Phi}_\nsr= \boldsymbol{\Phi} (\eta_0) $, meaning that, as already stressed, the averaged coarse-grained field evolves according to the classical dynamics (which is also the most probable trajectory, the PDF being Gaussian) and reaches the slow-roll late-time attractor unless $z_\sr^{(0)}=0$. Making use of \Eq{eq:Gaussian:Green}, the field-space PDF is given by
\bea
\label{eq:PDF:interactionPic}
\mathcal{W}\left( \boldsymbol{z},\eta \left\vert \boldsymbol{z}^{(0)},\eta_0 \right. \right)=\delta\left(z_\nsr-z_\nsr^{(0)}\right)
\dfrac{\exp\left[-\dfrac{1}{2\Sigma_\sr}\left(z_\sr-z^{(0)}_\sr\right)^2\right]}{\sqrt{2\pi\Sigma_\sr}} .
\eea
In this expression, it is clear that if the coarse-grained field is initially set in the attractor branch, \ie if $z^{(0)}_\nsr=0$, it never leaves the attractor although it diffuses along the SR direction. If $z^{(0)}_\nsr \neq 0$ however, at any finite time, the PDF never lies exactly on the attractor branch, but the question is whether it gets sufficiently close to it. To answer it, we define the ``slow-roll'' region of phase space $(z_\sr,z_\nsr)$ as being the domain where\footnote
{As explained in footnote~\ref{footnote:SR:test:field}, the notion of slow roll for a test scalar field is defined by requiring that the parameters $\epsilon_{n}\phi$ are small. Plugging the decomposition $\phi = z_\sr \phi_\sr +  z_\nsr \phi_\nsr$ into \Eqs{eq:eps1:testfield} and~(\ref{eq:eps2:testfield}), with  $\dot{\phi}_\sr/\phi_\sr = (\nu-3/2)/H$ and $\dot{\phi}_\nsr/\phi_\nsr = (\nu+3/2)/H $, on can express $\epsilon_1^\phi$ and $\epsilon_2^\phi$ as functions of the ratio $z_\nsr\phi_\nsr/(z_\sr \phi_\sr)$ only. This shows that $\epsilon_1^\phi$ and $\epsilon_2^\phi$ are small when this ratio is.
}
$\left\vert z_\nsr\phi_\nsr\right\vert<R \left\vert z_\sr\phi_\sr\right\vert$, with $R\ll 1$ a dimensionless parameter. In the following, the PDF will be said to have reached the slow-roll attractor if its overlap with this domain is close to total.
\paragraph{Probability to enter slow roll}
\begin{figure}
\begin{center}
\includegraphics[width=0.99\textwidth]{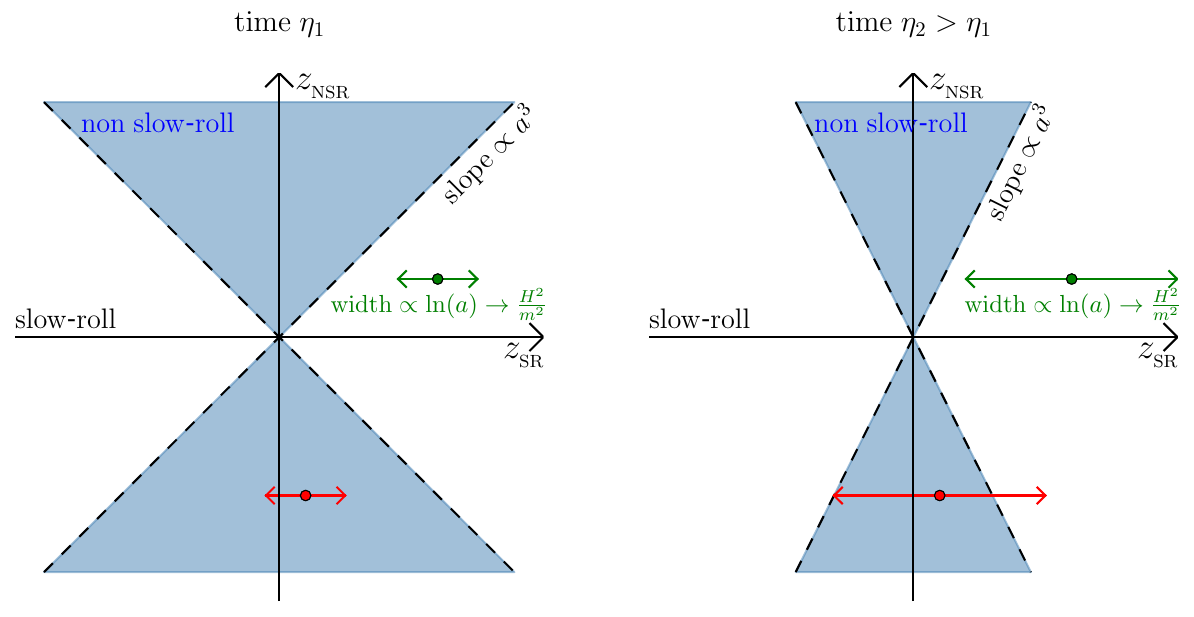} 
\caption{Schematic representation of the stochastic dynamics in phase space for a free scalar field. The right panel corresponds to a time $\eta_2$ greater than the time $\eta_1$ of the left panel. The shaded area corresponds to the non slow-roll region, which shrinks as $a^{-3}$ with time. The green and red arrows depict the width $\Sigma_\sr$ of the covariance matrix along the slow-roll direction, which increases with time as $\ln(a)$ before saturating to the value $9H^2/(8\pi^2m^2)$ at late time if the squared mass is positive [otherwise it increases as $a^{2\vert m^2\vert/(3 H^2)}$], for two different choices of the initial conditions, $z^{(0)}_\sr>z_R(\eta_0)$ in green, and $z^{(0)}_\sr<z_R(\eta_0)$ in red (note that $\Sigma_\sr$ is independent of the averaged evolution of the field). In the interaction picture parametrised by the variable $\boldsymbol{z}$, the averaged dynamics is given by a single constant point, and quantum diffusion only takes place along the slow-roll direction.}
\label{fig:phsp}
\end{center}
\end{figure}
Unless $z_\sr^{(0)}=0$, the slow roll condition $\left\vert z_\nsr\phi_\nsr\right\vert<R \left\vert z_\sr\phi_\sr\right\vert$ is always satisfied at late time for the classical dynamics since $z_\sr$ and $z_\nsr$ are frozen in this case, and the ratio $\phi_\nsr/\phi_\sr$ decreases asymptotically to $0$. In the stochastic dynamics however, $z_\nsr$ is frozen but $\widehat{z}_\sr$ undergoes quantum diffusion, so that values of $z_\sr$ such that $\left\vert z_\nsr\phi_\nsr\right\vert > R \left\vert z_\sr\phi_\sr\right\vert$, \ie values of $\vert z_\sr \vert<z_R(\eta)$ with
\bea
\label{eq:def:zR}
z_R(\eta) = \frac{\left\vert z_\nsr^{(0)} \right\vert}{R}(-\eta)^{2\nu} ,
\eea 
are not forbidden even at late time. Let us study how the probability for such values to be realised evolves in time.
\begin{figure}
\begin{center}
\includegraphics[width=0.99\textwidth]{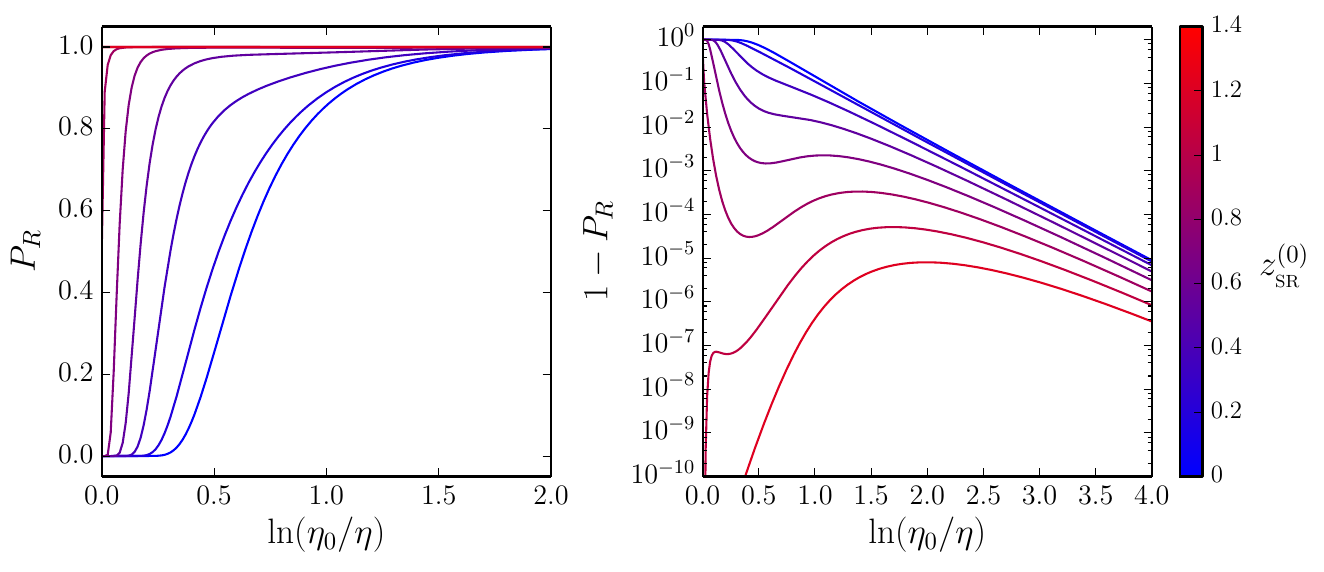} 
\caption{Probability to lie in the slow-roll region [$z_\sr>z_R(\eta)$, left panel] and probability to lie in the non slow-roll region [$z_\sr<z_R(\eta)$, right panel] as a function of conformal time for a free scalar field with mass $m/H=10^{-2}$ in de Sitter. Initial conditions are such that $z_\nsr^{(0)}=10^{-2}$ and different curves correspond to different values of $z_\sr^{(0)}$ given in the colour bar (from blue to red,  $z_\sr^{(0)}=0,\,0.2,\,0.4,\,0.6,\,0.8,\,1,\,1.2,\,1.4$). In the definition~(\ref{eq:def:zR}) of $z_R$, the parameter $R$ is taken to $R=10^{-2}$ and $\eta_0=-1$.}
\label{fig:probaSR}
\end{center}
\end{figure}

In \Fig{fig:phsp}, the slow-roll and non slow-roll regions of phase space are displayed, separated by the lines $z_\sr=\pm z_R$, and one can see that the slow-roll region expands as a power-law in conformal time. On the other hand, the width of the PDF along the SR direction is $\Sigma_\sr$ given in \Eq{eq:Sigma:z}. As noticed below \Eq{eq:Sigma:light:lightlimit}, at early time, when $\log(\eta_0/\eta)\ll 1/\vert 3-2\nu\vert$, it increases logarithmically in conformal time, $\Sigma_\sr \simeq 3\ln(\eta_0/\eta)/(4\pi^2) $; while at late time, when $\log(\eta_0/\eta)\gg 1/\vert 3-2\nu \vert$, if $\nu<3/2$ then it asymptotes to the constant value $\Sigma_\sr\simeq 3/[4\pi^2(3-2\nu)]$, while if $\nu>3/2$, it increases as $\Sigma_\sr\simeq 3/[4\pi^2(2\nu-3)](\eta_0/\eta)^{2\nu-3}$. In either case the slow-roll region always expands faster than the width of the PDF, which suggests that the overlap between the PDF and the slow-roll region increases with time. This can be checked by calculating the probability $P_R$ to be in slow roll, that is to say the probability for $\vert \widehat{z}_\sr \vert$ to be larger than $z_R$,
\bea
P_{R}\left(\eta\left\vert \boldsymbol{z}^{(0)},\eta_0\right.\right)&=&1-\displaystyle\int^{z_R}_{-z_R} \dd z_\sr 
\int^{\infty}_{-\infty} \dd z_\nsr 
\mathcal{W}\left(\boldsymbol{z},\eta \left\vert \boldsymbol{z}^{(0)},\eta_0\right. \right)
\nonumber \\ &=&
1+\dfrac{1}{2}\erf\left[\dfrac{z^{(0)}_\sr-z_R(\eta)}{\sqrt{2\Sigma_\sr(\eta)}}\right]-\dfrac{1}{2}\erf\left[\dfrac{z^{(0)}_\sr+z_R(\eta)}{\sqrt{2\Sigma_\sr(\eta)}}\right] ,
\label{eq:probaSR}
\eea
where \Eq{eq:PDF:interactionPic} has been used in the second line and the function $z_R(\eta)$ given in \Eq{eq:def:zR} implicitly depends on $z_\nsr^{(0)}$. This probability is displayed as a function of time in \Fig{fig:probaSR} for a free field with mass $m/H=10^{-2}$ in de Sitter and taking $R=10^{-2}$ and $\eta_0=-1$. Initial conditions are set to $z_\nsr^{(0)}=10^{-2}$ and different curves correspond to different values of $z_\sr^{(0)}$. Initially, $z_R(\eta_0)=1$, so that if $z_\sr^{(0)}<1$, the initial PDF entirely lies in the non slow-roll region and $P_R(\eta_0)=0$. The slow-roll probability $P_R$ then increases and asymptotes to $1$ after a few \efolds. If $z_\sr^{(0)}>1$, the initial PDF entirely lies in the slow roll region and $P_R(\eta_0)=1$. Subsequently, $P_R$ slightly decreases (which can be better seen in the right panel) but never departs much from $1$ and increases back towards $1$ after a few \efolds. From \Eq{eq:probaSR}, one can see that this is generic and that $P_R$ always tends to $1$ at late time. The answer to the first question raised above is therefore that for \emph{all} initial conditions, the classical slow roll attractor generalises to a stochastic attractor of free fields.
\paragraph{Relaxation time towards slow roll}
The classical, deterministic dynamics is frozen in the interaction picture and thus reaches the slow-roll region of phase space when $z_R(\eta) < \vert z_\sr^{(0)}\vert$, \ie at a time $\eta>\eta_\sr^\mathrm{det}$ where
\bea
\label{eq:etasr:det}
\eta_\sr^\mathrm{det} = -\left(R\left\vert \frac{z_\sr^{(0)}}{z_\nsr^{(0)}} \right\vert\right)^{\frac{1}{2\nu}} .
\eea
Let us see how the relaxation time of the stochastic dynamics towards slow roll, defined as being the time $\eta_\sr^\mathrm{stoch}$ such that for $\eta>\eta_\sr^\mathrm{stoch}$, $P_R(\eta)>1/2$, compares with this value. A first remark is that since the classical trajectory coincides with the mean stochastic one, when $\eta>\eta_\sr^\mathrm{det}$, the center of the phase-space PDF lies in the slow-roll region. In fact, when $\eta=\eta_\sr^\mathrm{det}$, the slow-roll probability~(\ref{eq:probaSR}) reads $P_R(\eta_\sr^\mathrm{det})=1-1/2\erf[z_\sr^{(0)}\sqrt{2/\Sigma_\sr(\eta_\sr^\mathrm{det})}]$ and is larger than $1/2$ since the error function is always smaller than one. This shows that 
\bea
\eta_\sr^\mathrm{stoch}<\eta_\sr^\mathrm{det} ,
\eea
\ie relaxation towards slow roll is faster in the stochastic theory than in the classical one. This answers the second question raised at the beginning of this section. In some cases, the stochastic relaxation time can even be much smaller. For instance, let us consider the situation in which initial conditions are set on the anti-attractor branch, $z_\sr^{(0)}=0$. In this case, $\eta_\sr^\mathrm{det}$ given in \Eq{eq:etasr:det} vanishes, meaning that the classical trajectory never enters the slow-roll regime. In the stochastic theory however, in the regime $\log(\eta_0/\eta)\ll 1/(3-2\nu)$ where $\Sigma_\sr \simeq 3\ln(\eta_0/\eta)/(4\pi^2)$, one obtains
\bea
\eta_\sr^\mathrm{stoch} \simeq \eta_0\exp\left( -\frac{1}{4\nu} W_0\left\lbrace \frac{8\nu}{3}\left[\frac{\pi\left(-\eta_0\right)^{2\nu}}{\erf^{-1}\left(1/2\right)}\frac{z_\nsr^{(0)}}{R}\right]^2\right\rbrace \right)
\eea
where $W_0$ is the $0^\mathrm{th}$ branch of the Lambert function and $\erf^{-1}$ is the inverse error function. In the limit where the argument of the Lambert function is large, one obtains $\eta_\sr^\mathrm{stoch} \sim \vert R/z_\nsr^{(0)}\vert^{1/(2\nu)}$, which corresponds to $\eta_\sr^\mathrm{det}$ given in \Eq{eq:etasr:det} for $z_\sr^{(0)}\sim 1$. In the opposite regime where $\log(\eta_0/\eta)\gg 1/(3-2\nu)$  and $\Sigma_\sr\simeq 3/[4\pi^2(3-2\nu)]$, one has
\bea
\eta_\sr^\mathrm{stoch} \simeq -\left[\sqrt{\frac{3}{2}}\frac{\erf^{-1}(1/2)}{\pi}\frac{R}{z_\nsr^{(0)}}\frac{1}{\sqrt{3-2\nu}}\right]^{\frac{1}{2\nu}} .
\eea
In this case, even though the mean trajectory never reaches the slow-roll domain, the overlap between the PDF and the slow-roll region becomes close to total when $\eta>\eta_\sr^\mathrm{stoch}$. 

Let us finally mention that so far, the initial state of the coarse-grained field has been assumed to be known exactly, \ie the PDF at initial time $\eta_0$ has been taken to a Dirac function. For a generic initial PDF $P(\boldsymbol{z}_0,\eta_0)$, one can check that the average field is given by $\left<\boldsymbol{z}\right>=\left<\boldsymbol{z}^{(0)}\right>_0$, where $\langle\cdot\rangle_0$ denotes average over the initial PDF, \ie $\left<f(\boldsymbol{z}^{(0)})\right>_0\equiv \int\dd\boldsymbol{z}^{(0)}f(\boldsymbol{z}^{(0)})P(\boldsymbol{z}^{(0)},\eta_0)$. This implies that $\left<\boldsymbol{\Phi}\right>=\left<z^{(0)}_\sr\right>_0\boldsymbol{\Phi}_\sr+\left<z^{(0)}_\nsr\right>_0\boldsymbol{\Phi}_\nsr$ and means that in this case too, the average coarse-grained field evolves towards the slow-roll attractor unless $\left\langle z^{(0)}_\sr\right\rangle_0=0$, with a relaxation time still given by \Eq{eq:etasr:det} if one replaces $z^{(0)}_\sr$ by $\left\langle z^{(0)}_\sr\right\rangle_0$ and  $z^{(0)}_\nsr$ by $\left\langle z^{(0)}_\nsr\right\rangle_0$. For the full stochastic dynamics, the probability to be in the slow-roll region is given by $P_R(\eta)=\left\langle P_{R}\left(\eta|\boldsymbol{z}^{(0)},\eta_0\right) \right\rangle_0$, and from \Eq{eq:probaSR} one can check that it still asymptotes to $1$ at late time.
\paragraph{Beyond free fields}
So far, we have shown that for test fields with quadratic potentials, the classical slow-roll attractor generalises to a stochastic attractor. Around \Eq{eq:eom:pert:vectorial}, it was explained why this property is related to the fact that for such free fields, the equation of motion of field perturbations on large scales coincides with the one of the background, hence shares the same attractor. 

When the potential is not quadratic, or when the field is not a mere spectator and sources metric perturbations, this stops being the case\footnote{Notice that this does not contradict the results of \Sec{sec:separateuniverse}, in which the generic validity of the separate universe approach was proven, which shows that, when field perturbations are computed in the gauge where time is unperturbed, their equation of motion still coincides, on large scales, with the \emph{perturbed} background equations of motion. } and slow roll is not an exact stochastic attractor anymore. In this paragraph, we quantify the deviation from slow roll induced by stochastic effects for non-free fields.

The stochastic phase-space dynamics of non-free fields is more challenging to study than the one of free fields for the two following reasons. First, the background equation of motion~(\ref{eq:kleingordon}) is not linear anymore, so the ability to use the Green formalism, Gaussian solutions of the Fokker-Planck equation and canonical transformations to the interaction picture is lost. Second, the equation of motion for the field perturbations~(\ref{eq:eomphiq}) and~(\ref{eq:eompiq}) now depends on the background value of $\phi$, through the term $V_{,\phi\phi}(\phi)$ in \Eq{eq:eompiq}. This means that \Eq{eq:mode:light} is still valid, but $m^2$ has to be replaced by an effective mass 
\bea
\label{eq:meff:def}
m_\ueff^2=V_{,\phi\phi}(\bar{\phi})
\eea
which depends on the background field $\bar{\phi}$. However $\bar{\phi}$ is stochastic and different for each realisation of the Langevin equation. This implies that, in principle, the diffusion matrix must be re-computed at every given time and for every given realisation of the Langevin equation by integrating the equation of motion of field perturbations sourced by this realisation. In practice, this prevents any conclusion to be drawn from analytical arguments only without resorting to some approximation.

For this reason, let us restrict the analysis to the phase-space region sufficiently close to the classical attractor so that the ``reference'' background solution about which the equation of motion of field perturbations is solved can be taken as the classical slow-roll solution $\phi_\sr$. This will allow us to assess how much stochastic effects alter the slow-roll dynamics \emph{if} one starts on the classical attractor.\footnote{This also corresponds to the first recursive level of \Refa{Levasseur:2013tja}.} Under this assumption, the effective mass term in \Eq{eq:mode:light} can be taken as 
\bea
\label{eq:meffSR:def}
m_\ueff^2=V_{,\phi\phi}(\phi_\sr)
\eea
and becomes an explicit, fixed function of time. 

In fact, this function of time can be expressed in terms of slow-roll parameters. Indeed, combining \Eqs{eq:eps1:appr:test} and~(\ref{eq:eps2:appr:test}), one obtains
\bea
\label{eq:meff:eps}
\frac{m_\ueff^2}{H^2}=3\epsilon_1+3\epsilon_1^\phi-\frac{3}{2}\epsilon_2^\phi .
\eea
If $\phi$ is the inflaton field, one can combine \Eqs{eq:eps1:SR} and~(\ref{eq:f:SR}) instead, and obtain $m_\ueff^2/H^2=6\epsilon_1-3\epsilon_2/2$, which is consistent with \Eq{eq:meff:eps} if one equates $\epsilon_1^\phi$ and $\epsilon_2^\phi$ to $\epsilon_1$ and $\epsilon_2$ respectively. This means that $m_\ueff^2/H^2$ does not substantially vary over the time scale of one \efold. Since the amplitude acquired by field perturbations is mostly determined by the background dynamics around the few \efolds~surrounding their Hubble exit time, the ratio $m_\ueff^2/H^2$ can thus be approximated as being constant during this time interval.  This implies that \Eqs{eq:modesolution:light} and~(\ref{eq:nu:def}) still provide an accurate solution to the equation of motion of field perturbations if one replaces $\nu$ by $\nu[\eta_*(k)]$, in which $\epsilon_1$ and the ratio $m_\ueff^2/H^2$ are evaluated at the time $\eta_*(k)$ when the mode $k$ crosses the Hubble radius. This adiabatic approximation is in fact the standard way cosmological perturbations are calculated in the slow-roll approximation~\cite{Stewart:1993bc}.
\begin{figure}
\begin{center}
\includegraphics[width=0.49\textwidth]{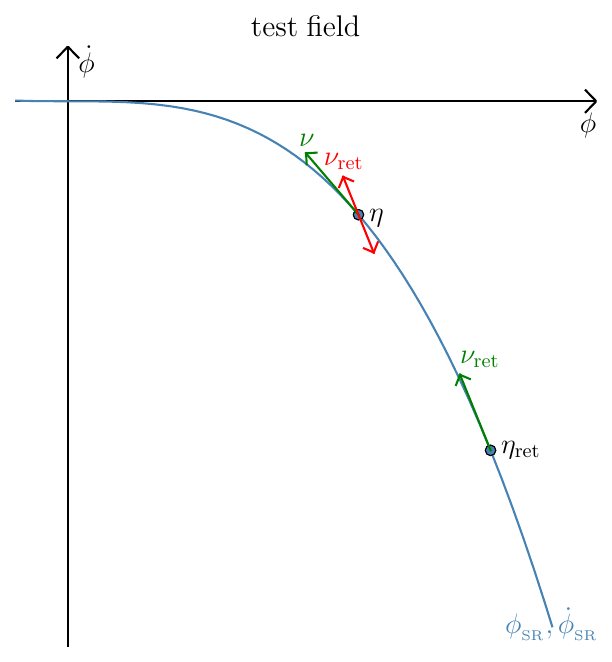} 
\includegraphics[width=0.49\textwidth]{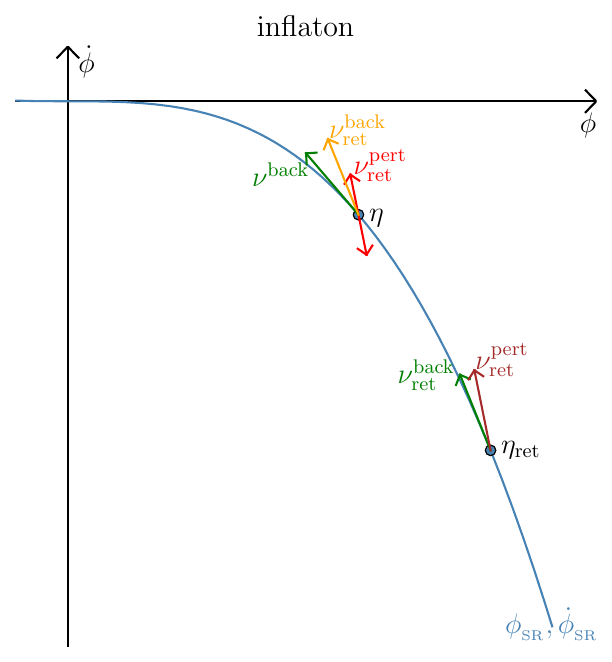} 
\caption{Quantum diffusion about the classical slow-roll trajectory $(\phi_\sr,\dot{\phi}_\sr)$ for a test field (left panel) and for the inflaton field (right panel). At time $\eta$, the direction to which the classical dynamics points (green arrow) is determined by $\nu^\mathrm{back}(\eta)$, while the direction along which quantum diffusion takes place (red double arrow) is determined by $\nu^\mathrm{pert}(\eta_\mathrm{ret})$, which corresponds to the phase-space direction of the field perturbations at the retarded time $\eta_\mathrm{ret}$ when the mode that crosses the coarse-graining radius at time $\eta$ crossed the Hubble radius. For a test field, the background and the perturbations share the same phase-space direction at a given time, $\nu^\mathrm{back}(\eta)=\nu^\mathrm{pert}(\eta)$, while this is not the case for the inflaton field that also couples to metric fluctuations. In both cases, quantum diffusion is not aligned with the classical flow and stochastic effects induce deviations from the classical attractor.}
\label{fig:retardednu}
\end{center}
\end{figure}
\paragraph{Test fields}
In the case of test fields, this approximation implies that the diffusion matrix $\boldsymbol{D}(\eta)$ is still given by the leading order terms in \Eqs{eq:massive:Dphiphi}-(\ref{eq:massive:Dphipi}), except that $\nu$ must now be evaluated at the retarded time $\eta_\mathrm{ret}(\eta)=\eta_*[k_\sigma(\eta)]$, \ie at the time when the mode $k_\sigma(\eta)$, that crosses the coarse-graining radius at time $\eta$, crosses the Hubble radius. This means that the Langevin equations~(\ref{eq:eombarphi}) and~(\ref{eq:eombarpi}), where $\xi_\phi$ and $\xi_\pi$ are totally anti-correlated, can be written as
\bea
\frac{\dd \phi}{\dd N} &=& \gamma + \frac{H}{2\pi}\xi\\
\frac{\dd \gamma}{\dd N} &=& \left(\epsilon_1-3\right)\gamma - \frac{V_{,\phi}(\phi)}{H^2} +\frac{H}{2\pi} \left[\nu_\mathrm{ret}(N_e)-\frac{3}{2}\right] \xi ,
\eea
where the number of \efolds~is used as the time variable for simplicity, phase space is parametrised by $\phi$ and $\gamma= \pi/(H a^3)$, $\xi$ is a normalised white Gaussian noise, such that $\langle \xi (N_{e}) \xi (N_{e}^\prime) \rangle = \delta(N_{e}-N_{e}^\prime)$, and the bars have been dropped to lighten the notation of the coarse-grained fields.

However, the direction of the classical trajectory in phase space is still determined by the value of $\nu$ at current time $\eta$, since at leading order in slow roll, $\dot{\phi}_\sr\simeq -V_{,\phi}(\phi_\sr)/(3H)$ gives rise to
\bea
\frac{\dd}{\dd t} 
\left(\begin{array}{c}
	\phi_\sr  \\
	\dot{\phi}_\sr
\end{array}\right) 
\propto
\left(\begin{array}{c}
	1  \\
	H_*\epsilon_1-\frac{V_{,\phi\phi}(\phi_\sr)}{3H}
\end{array}\right)
=
\left(\begin{array}{c}
	1  \\
	H_* \left[\nu\left(\eta\right)-\frac{3}{2}\right]
\end{array}\right) ,
\label{eq:fieldshift:class}
\eea
where the expression~(\ref{eq:nu:def}) relating $\nu$ to the effective mass defined in \Eq{eq:meffSR:def} and to $\epsilon_1$ has been used. The situation is summarised in the left panel of \Fig{fig:retardednu}. Because $\nu(\eta)$ and $\nu(\eta_\mathrm{ret})$ are a priori different, the classical flow and the quantum diffusion do not point to the same direction in phase space and this is why stochastic effects induce a deviation from the classical attractor, related to 
\bea
\label{eq:nuMinusnuRet}
\nu(\eta)-\nu(\eta_\mathrm{ret}) &\simeq & \alphaS(\eta) \ln\left(\sigma\right) .
\eea
In this expression, $\alphaS=\dd \nS/\dd\ln k$ denotes the running of the spectral index of the field fluctuations power spectrum introduced below \Eq{eq:condition:m_over_H}. Since $\nS=1-2(\nu-3/2)$ indeed, $\nu-\nu_\mathrm{ret}\simeq \alphaS [N(\eta_\mathrm{ret})-N(\eta)]$. The comoving wavenumber that crosses the coarse-graining radius at time $\eta$ is $k=\sigma a(\eta) H(\eta)$, see \Eq{eq:ksigma}, so it crosses the Hubble radius at time $\eta_\mathrm{ret}$ such that $k= a(\eta_\mathrm{ret}) H(\eta_\mathrm{ret})$, and one finds that $N_e(\eta_\mathrm{ret})-N_e(\eta)  = \ln[a(\eta_\mathrm{ret})/a(\eta)]\simeq \ln(\sigma)$ at leading-order in slow roll, hence \Eq{eq:nuMinusnuRet}. 

Let us now assess the effect of the misalignment between the quantum noise and the classical slow-roll direction in phase space on the relative fluctuation $\delta\rho_\phi/\rho_\phi$ in the energy density contained in $\phi$. The reason is that this ratio corresponds to the contribution of the fluctuations in $\phi$ to the total curvature perturbation $\zeta$ measured \eg in the CMB. It is therefore directly related to observable quantities, for instance in the context of curvaton scenarios~\cite{Linde:1996gt, Enqvist:2001zp, Lyth:2001nq, Moroi:2001ct, Bartolo:2002vf, Vennin:2015vfa, Vennin:2015egh}. Fourier transforming the energy density field, the amplitude of perturbations at the comoving scale $k$ is given by $\delta\rho_\phi (k)/\rho_\phi (k)$, where $\rho_\phi (k)$ is the energy density contained in $\phi$ at the time when $k$ crosses the Hubble radius, and $\delta\rho_\phi (k)$ is the fluctuation in this quantity induced by quantum diffusion over one \efold~around this time. Since $\rho_\phi=V(\phi)+H^2\gamma^2/2$, one has $\delta \rho_\phi = V_{,\phi}(\phi) \delta\phi + H^2 \gamma \delta \gamma$. If quantum diffusion was aligned with the classical slow-roll direction, one would have $\delta\gamma=(\nu-3/2)\delta\phi$ and $\delta\rho_\phi=\delta\phi[V_{,\phi}+H^2\gamma(\nu-3/2)]$. However, quantum diffusion occurs along the retarded direction $\nu_\mathrm{ret}$, which gives rise to the corrected $\delta\gamma^\mathrm{corr}=(\nu_\mathrm{ret}-3/2)\delta\phi$ and $\delta\rho_\phi^\mathrm{corr}=\delta\phi[V_{,\phi}+H^2\gamma(\nu_\mathrm{ret}-3/2)]$. At leading order in slow roll, $\gamma\simeq -V_{,\phi}/(3H^2)$, and one has
\bea
\label{eq:deltarhoCorr:test}
\frac{\delta\rho_\phi^\mathrm{corr}-\delta\rho_\phi}{\delta\rho_\phi}=\frac{\nu-\nu_\mathrm{ret}}{3} .
\eea
Making use of \Eq{eq:nuMinusnuRet}, this yields a small correction if
\bea
\label{eq:conditions:sigma:test}
\sigma\gg \ee^{-\frac{1}{\vert \alphaS \vert}} ,
\eea
which has a similar form as the condition $\sigma\gg \ee^{-\frac{1}{\vert \nS-1 \vert}}$ derived in \Eq{eq:condition:m_over_H}. It implies that between the Hubble radius crossing time and the coarse-graining radius crossing time, not only the amplitude of the field fluctuations should not vary substantially but also their spectral index. Let us note that in the slow-roll approximation, $\vert \alphaS \vert \ll \vert \nS-1 \vert$, so \Eq{eq:conditions:sigma:test} is a weaker constraint than \Eq{eq:condition:m_over_H}.
\paragraph{Inflaton field}
\label{sec:Inflaton}
In the case of the inflaton field, an additional subtlety is that even at the same fixed time, the slow-roll background flow and the perturbations dynamics have different directions $\nu^\mathrm{back}$ and $\nu^\mathrm{pert}$ in field space. In \Sec{ssec:cosmo} indeed, it was assumed that $\phi$ is a test field sufficiently decoupled from the metric perturbations that the latter can be ignored. If $\phi$ is the inflaton field, this is not the case anymore. In \Sec{sec:separateuniverse}, it was shown that in the equation of motion~(\ref{eq:mode:light}) for the field perturbations, the effective mass~(\ref{eq:meff:def}) receives a correction from gravitational coupling with metric fluctuations that reads~\cite{Langlois:1994ec, Lyth:2001nq}
\bea
\label{eq:meff:inflaton}
\widetilde{m}_{\mathrm{eff}}^2=V_{,\phi\phi}\left(\bar{\phi}\right)-\frac{1}{\Mp^2a^3}\frac{\dd}{\dd t}\left(\frac{a^3}{H}\dot{\bar{\phi}}^2\right) ,
\eea
see \Eq{eq:pertubations}.
Making use of the Friedman equation~(\ref{eq:Friedmann:cosmic:time}) and of the Klein-Gordon equation~(\ref{eq:kleingordon}) for the background field $\bar{\phi}$, this additional term can be written $-2H^2\epsilon_1(3+2\epsilon_2-\epsilon_1)$. At leading order in slow roll, the index $\nu$ given in \Eq{eq:nu:def} is then modified according to
\bea
\label{eq:nuPertMinusNuBack}
\nu^{\mathrm{pert}} = \nu^{\mathrm{back}}+2\epsilon_1 ,
\eea
where $\nu^\mathrm{back}=3/2-\epsilon_1+\epsilon_2/2$ is the value obtained in absence of gravitational coupling to metric fluctuations and corresponds to the direction of the slow-roll background dynamics. 

The situation is summarised in the right panel of \Fig{fig:retardednu}. At a given time $\eta$, not only the phase-space direction along which quantum diffusion takes place has to be evaluated at the retarded time $\eta_\mathrm{ret}$, but it is related to the dynamics of perturbations which occurs in a different direction than the one of the background. As before, let us assess the effect of this misalignment on the curvature perturbation $\zeta$, directly proportional to $\delta\rho/\rho$~\cite{Wands:2000dp}. Similarly to \Eq{eq:deltarhoCorr:test}, one has
\bea
\label{eq:deltarhoCorr:inflaton}
\frac{\delta\rho^\mathrm{corr}-\delta\rho}{\delta\rho}=\frac{\nu^\mathrm{back}-\nu^\mathrm{pert}_\mathrm{ret}}{3} .
\eea
Decomposing $\nu^\mathrm{back}-\nu_\mathrm{ret}^\mathrm{pert} =\nu^\mathrm{back}-\nu_\mathrm{ret}^\mathrm{back}   +     \nu_\mathrm{ret}^\mathrm{back}-\nu_\mathrm{ret}^\mathrm{pert} $, where $\nu^\mathrm{back}-\nu_\mathrm{ret}^\mathrm{back}$ is given by \Eq{eq:nuMinusnuRet} and $\nu_\mathrm{ret}^\mathrm{back}-\nu_\mathrm{ret}^\mathrm{pert} $ by \Eq{eq:nuPertMinusNuBack} (at retarded time $\eta_\mathrm{ret}$), one obtains the same condition~(\ref{eq:conditions:sigma:test}) as for a test field since $\epsilon_1\ll 1$. Therefore, in both cases, one finds that the observational effect of the phase-space misalignment between the classical homogeneous slow-roll dynamics and quantum diffusion remains small if the amplitude \emph{and} tilt of the field fluctuations at the Hubble radius crossing time and at the quantum-to-classical transition time are sufficiently close. 

In conclusion, we have found that provided the above-mentioned conditions on the coarse-graining scale $\sigma$ are satisfied, slow roll can be promoted to a stochastic attractor. Along this attractor, the gauge corrections of \Sec{sec:uniformexpansion} vanish (at least, at the order in $\sigma$ at which the calculation is being performed) if time is labeled by the number of \efolds, and phase space collapses to a single direction, say $\phi$, along which the Langevin equations reduce to
\bea
\label{eq:Langevin:SR:phi}
\frac{\dd\phi}{\dd N} = -\frac{V'}{3H^2}+\frac{H}{2\pi}\xi\, ,
\eea
where $\xi$ is a white Gaussian noise with vanishing mean and unit variance. The conjugate momentum to $\phi$ is simply given by the slow-roll, classical function of $\phi$. 
\section{The stochastic-$\delta N$ formalism}
\label{sec:stochastic:delta:N}
In \Sec{sec:StochasticInflation}, we have explained how the backreaction of quantum fluctuations, as they get stretched beyond the Hubble radius and source the locally FLRW separate patches forming our universe during inflation, can be incorporated in the formalism of stochastic inflation, which describes the dynamics of this ensemble of patches with stochastic, Langevin equations. In this section, we explain how cosmological perturbations can be extracted from this stochastic picture, and what observational effects quantum backreaction has. 

The link between an ensemble of inflating patches, realising different amounts of expansion, and the curvature perturbation these patches generate on large scales, can be established in the $\delta N$ formalism, which we first review in \Sec{sec:deltaN:formalism}. In this approach, the curvature perturbation is nothing but the fluctuation in the number of inflationary \efolds~realised at different spatial points. We thus explain in \Sec{sec:stochaDeltaN:stochaDeltaN} that deriving observable predictions implies to compute the statistical distribution of this number of \efolds~from the Langevin equations of stochastic inflation, which can be achieved by using the first passage time techniques presented in \Sec{sec:FirstPassageTime}. We then apply these techniques to the calculation of the first moments of the number of \efolds, which allows us to derive the mean duration of inflation, the power spectrum of the curvature perturbation, and its local non Gaussianity. This is the topic of \Sec{sec:FirstMoments:N}. Then, in \Sec{sec:Infinite:Inflation}, we highlight that, at face value, these moments can be infinite, a phenomenon that we dub ``infinite inflation'' and that strongly depends on the number of inflating fields. We show how finite predictions can be still extracted in that case. This section compiles results from \Refs{Vennin:2015hra, Assadullahi:2016gkk, Vennin:2016wnk}.
\subsection{The $\delta N$ formalism}
\label{sec:deltaN:formalism}
The starting point of the stochastic-$\delta N$ formalism is the standard, classical $\delta N$ formalism~\cite{Starobinsky:1982ee, Starobinsky:1986fxa, Sasaki:1995aw, Sasaki:1998ug, Lyth:2004gb, Lyth:2005fi}, which provides a succinct way of relating the fluctuations in the number of \efolds~of expansion during inflation for a family of homogeneous universes with the statistical properties of curvature perturbations. Starting from the unperturbed flat FLRW metric~\eqref{eq:line_element:flat}, which, in cosmic time, reads
\bea
\label{eq:metric:FLRW}
\dd s^2 = -\dd t^2 + a^2(t)\delta_{ij}\dd x^{i}\dd x^{j} \, ,
\eea
deviations from isotropy and homogeneity can be added at the perturbative level and contain scalar, vector and tensor degrees of freedom. Gauge redundancies associated with diffeomorphism invariance allow one to choose a specific gauge in which fixed time slices have uniform energy density and fixed spatial worldlines are comoving (in the super-Hubble regime this gauge coincides with the synchronous gauge supplemented by some additional conditions that fix it uniquely). Including scalar perturbations only, one obtains~\cite{Starobinsky:1982ee, Creminelli:2004yq,Salopek:1990jq}
\bea
\dd s^2 = -\dd t^2 + a^2(t)\ee^{2\zeta(t, \bm{x})}\delta_{ij}\dd x^{i}\dd x^{j} \, ,
\eea
where $\zeta$ is the adiabatic curvature perturbation. One can then introduce a local scale factor
\bea
\label{eq:alocal:def}
\tilde{a}(t, \bm{x}) = a(t)\ee^{\zeta(t, \bm{x})} \, ,
\eea
which allows us to express the amount of expansion from an initial flat space-time slice at time $t_\uin$ to a final space-time slice of uniform energy density as
\bea
N(t, \bm{x}) = \ln{\left[ \frac{\tilde{a}(t, \bm{x})}{a(t_{\mathrm{in}})} \right]} \, .
\eea
This is related to the curvature perturbation $\zeta$ via \Eq{eq:alocal:def}, which gives rise to
\bea
\label{eq:zeta:deltaN}
\zeta(t, \bm{x}) = N(t, \bm{x}) - \bar{N}(t) \equiv \delta N \, ,
\eea
where $\bar{N}(t) \equiv \ln{\left[ {a(t)}/{a(t_{\mathrm{in}})} \right]}$ is the unperturbed expansion. 
This expression forms the basis of the $\delta N$ formalism, which follows by further invoking the ``quasi-isotropic''~\cite{Lifshitz:1960, Starobinsky:1982mr, Comer:1994np, Khalatnikov:2002kn} or ``separate universe'' approach~\cite{Wands:2000dp, Lyth:2003im, Lyth:2004gb} derived in \Sec{sec:separateuniverse}, which allows us to neglect spatial gradients on super-Hubble scales. As a consequence, $N(t, \bm{x})$ is the amount of expansion in unperturbed, homogeneous universes, and $\zeta$ can be calculated from the knowledge of the evolution of a family of such universes.
\subsection{The stochastic-$\delta N$ formalism}
\label{sec:stochaDeltaN:stochaDeltaN}
\begin{figure}[t]
\begin{center}
\includegraphics[width=0.7\textwidth]{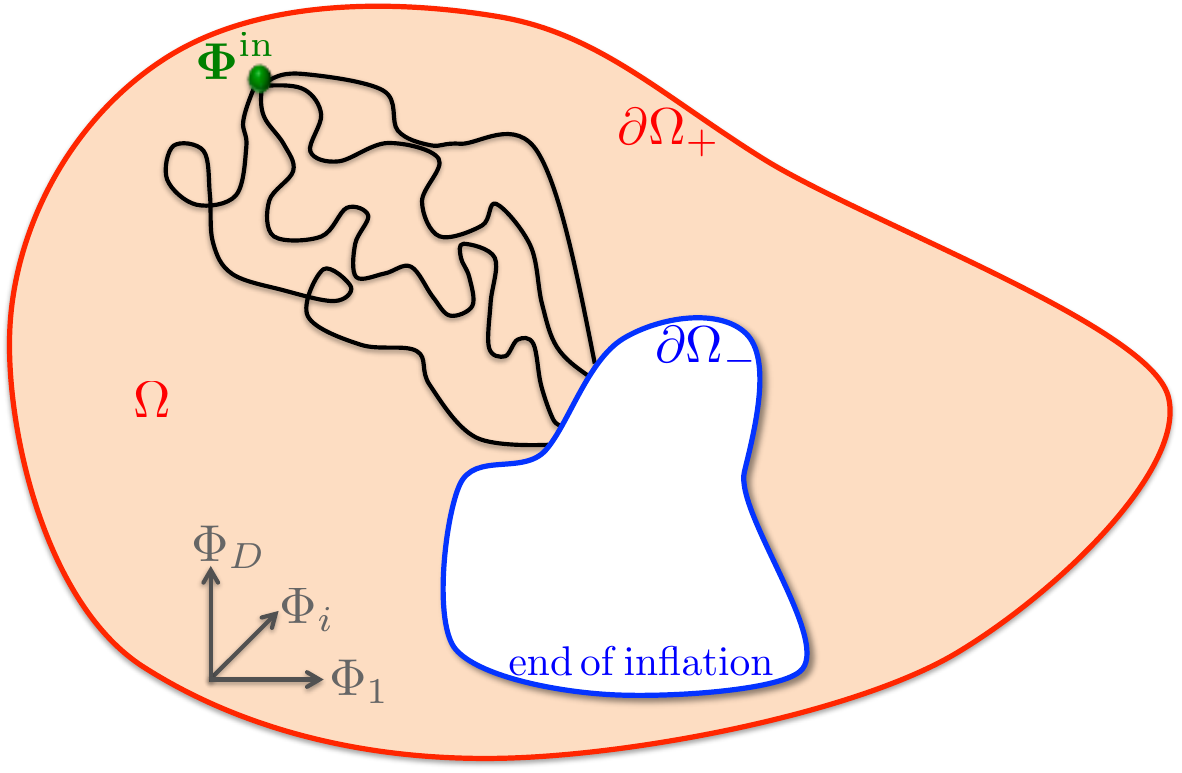}
\caption{First passage time problem. Starting from an initial field configuration $\boldmathsymbol{\Phi}^\uin$, different trajectories (black lines) realise different numbers of \efolds~$\mathcal{N}$ in the inflationary domain $\Omega$ (pale red region) until the final hypersurface $\partial\Omega_-$ (blue line) is reached. Note that $\Omega$ may also be bounded ``from above'' by $\partial\Omega_+$ (red line). In the $\delta N$ formalism, this random variable $\mathcal{N}$ corresponds to the coarse-grained curvature perturbation.}
\label{fig:sketchFirstPassageTime}
\end{center}
\end{figure}
The $\delta N$ formalism relies on the calculation of the amount of expansion realised amongst a family of homogeneous universes. When stochastic inflation is employed to describe such a family of universes and to calculate the amount of expansion realised in them, this gives rise to the stochastic-$\delta N$ formalism~\cite{Enqvist:2008kt, Fujita:2013cna, Fujita:2014tja, Vennin:2015hra, Kawasaki:2015ppx, Assadullahi:2016gkk, Vennin:2016wnk} that we now explain.

In full generality, phase space is parametrised with the field vector $\boldmathsymbol{\Phi}=(\phi_1,\pi_1,\cdots,\phi_n,\pi_n)$, which contains all fields and associated momenta. Here, we do not assume single field nor slow roll, and our considerations will be made more specific only in \Sec{sec:FirstPassageTime} and after.  As explained in \Sec{sec:StochasticInflation}, stochastic inflation is an effective theory for the long wavelength part of the fields, which are coarse-grained above a fixed physical scale $k_\sigma=\sigma a H$, 
\bea
\hat{\boldmathsymbol{\Phi}}_{\mathrm{cg}}(\bm{x}) = \frac{1}{\left(2\pi\right)^{3/2}}\int_{k<k_\sigma} \dd^3 k \hat{\boldmathsymbol{\Phi}}_k \ee^{-i k \bm{x}} ,
\eea
where $\sigma\ll 1$ is a fixed parameter setting the scale at which quantum fluctuations backreact onto the local FLRW geometry. The dynamics of the quantum operators $\hat{\boldmathsymbol{\Phi}}_{\mathrm{cg}}$ can be tracked by means of stochastic Langevin equations,
\bea
\label{eq:Langevin}
\frac{\dd {\boldmathsymbol{\Phi}}_{\mathrm{cg}}}{\dd N} = \boldmathsymbol{F}\left(\boldmathsymbol{\Phi}_{\mathrm{cg}}\right)+\boldmathsymbol{G}\left(\boldmathsymbol{\Phi}_{\mathrm{cg}}\right)\cdot \boldmathsymbol{\xi},
\eea
where $\boldmathsymbol{F}\left(\boldmathsymbol{\Phi}\right)$ encodes the classical equations of motion, $\boldmathsymbol{\xi}$ is a set of uncorrelated white Gaussian noises with vanishing mean and unit variance, \ie
\bea
\label{eq:White:Gaussian:Noises:correlators}
\left\langle \xi_i\left(\bm{x}_i,N_i\right) \xi_j\left(\bm{x}_j,N_j\right) \right\rangle = \delta_{i j} \delta\left(N_i-N_j\right),
\eea
and $\boldmathsymbol{G}\left(\boldmathsymbol{\Phi}_{\mathrm{cg}}\right)$ is a real symmetric matrix, the square of which is given by
\bea
\left(\boldmathsymbol{G}^2\right)_{i j} =  \frac{\dd\ln k_\sigma}{\dd N} \calP_{\Phi_i,\Phi_j}\left[k_\sigma(N),N\right] ,
\eea
see \Eq{eq:noisecorrel_Pk}. In this expression, $\calP_{\Phi_i,\Phi_j}[k_\sigma(N),N]$ is the cross power spectrum between the field variables $\Phi_i$ and $\Phi_j$, evaluated at the scale $k_\sigma(N)$, and at time $N$. As explained in \Sec{sec:uniformexpansion}, recall that since the time coordinate (here $N$) has not been perturbed in the Langevin equation~(\ref{eq:Langevin}), we are implicitly working in the gauge where time is unperturbed, \ie the uniform $N$-gauge in the present case. The power spectra need therefore to be computed in that gauge. 

Starting from an initial field configuration $\boldmathsymbol{\Phi}=\boldmathsymbol{\Phi}^\uin$, one can record the number of \efolds~that are realised until a certain condition is met. According to the considerations of \Sec{sec:deltaN:formalism}, this condition should define a surface $\partial\Omega_-$ in phase space on which the energy density is constant, $\partial\Omega_-=\lbrace \boldmathsymbol{\Phi} \vert \rho(\boldmathsymbol{\Phi})=\mathrm{constant} \rbrace$. In practice, $\partial\Omega_-$ is sometimes set  to be the surface on which inflation ends, $\partial\Omega_-=\lbrace \boldmathsymbol{\Phi} \vert \epsilon_{1}(\boldmathsymbol{\Phi})=1 \rbrace$, which, strictly speaking, is not a surface of constant energy density. However, since $\zeta$ freezes on large scales, and given that stochastic effects are usually negligible towards the end of inflation, this does not make a large difference. The situation is sketched in \Fig{fig:sketchFirstPassageTime}. In practice, an absorbing boundary is placed at $\partial\Omega_-$. This boundary condition defines an inflating domain $\Omega$ that is either compact in field space (think for instance of a single-field hilltop potential), or non compact (think for instance of a single-field large-field potential). If it is non compact, it is sometimes necessary to introduce another boundary condition at large-field values, $\partial\Omega_+$, the nature (absorbing, reflective, partly reflective and partly absorbing) and the precise location of which being mostly irrelevant as will be shown in \Sec{sec:Infinite:Inflation}.
 
The amount of expansion realised along a given trajectory is called $\mathcal{N}$, which is a stochastic variable. Thanks to the $\delta N$ formalism, the fluctuation in this number of \efolds, $\mathcal{N}-\langle \mathcal{N} \rangle$, is nothing but the curvature perturbation coarse grained between $k_\uin$, the scale that crosses the Hubble radius at initial time, and $k_\uend$, the scale that crosses out the Hubble radius at final time
\bea
\label{eq:deltaNcg:zeta}
\delta N_{\mathrm{cg}}\left(\bm{x}\right) = 
\mathcal{N}\left(\bm{x}\right)-\left\langle \mathcal{N} \right\rangle =
\zeta_{\mathrm{cg}}\left(\bm{x}\right) = 
\frac{1}{\left(2\pi\right)^{3/2}}
\int_{k_\uin}^{k_\uend} \dd \bm{k} \zeta_{\bm{k}} e^{i \bm{k}\cdot \bm{x}}\, .
\eea
If one is interested in primordial black holes, as will be shown in \Sec{sec:PBHs}, one is directly interested in $\zeta_{\mathrm{cg}}$ and the probability distribution function of $\zeta_{\mathrm{cg}}$ simply follows from the one of $\mathcal{N}$. If one is interested in CMB observables however, such as the power spectrum of curvature perturbations, the statistics of $\zeta$, not $\zeta_{\mathrm{cg}}$, should be derived. For the power spectrum $\calP_\zeta$ for instance, \Eq{eq:deltaNcg:zeta} implies that the coarse-grained $\delta N_{\mathrm{cg}}$ receives an integrated contribution of all modes exiting the Hubble radius during inflation, $\langle \delta N_\mathrm{cg}^2 \rangle = \int_{k_\uin}^{k_\uend} \calP_\zeta \dd k/k$, and one has~\cite{Fujita:2013cna, Vennin:2015hra}
\bea
\label{eq:Pzeta:stochaDeltaN}
\calP_\zeta = \frac{\dd \left\langle \delta N_{\mathrm{cg}}^2 \right\rangle}{\dd \left\langle \mathcal{N}\right\rangle}\, ,
\eea
where we have used the relation $\langle \mathcal{N}\rangle = \ln(a_\uend/a_*) = \ln(k_\uend/k)$, where the last equality is valid at leading order in slow roll only. In the same manner, the local bispectrum can be written as $\mathcal{B}_\zeta \propto \dd^2 \langle \delta N_\mathrm{cg}^3\rangle/\dd \langle \mathcal{N} \rangle^2$, from which the effective $\fnl^\mathrm{local}$ parameter, measuring the ratio between the bispectrum and the power spectrum squared, is given by
\bea
\label{eq:fnl:stochaDeltaN}
\fnl^\mathrm{local} = \frac{5}{72}\frac{\left\langle \delta N_\mathrm{cg}^3\right\rangle}{\dd \left\langle \mathcal{N} \right\rangle^2}\left(\frac{\dd \left\langle \delta N_{\mathrm{cg}}^2 \right\rangle}{\dd \left\langle \mathcal{N}\right\rangle}\right)^{-2}\, .
\eea

\subsection{First passage times}
\label{sec:FirstPassageTime}
From the previous considerations, it becomes clear that the main technical task for the stochastic-$\delta N$ program is to compute the statistics of the first passage time through the exit surface $\partial\Omega_-$, generated by the process~\eqref{eq:Langevin}. This is the goal of this section. In \Sec{sec:FP}, we first provide a generic derivation of the Fokker-Planck equation, which drives the probability distribution of the field values, from the Langevin equation~\eqref{eq:Langevin}. We then obtain differential equations for the moments of the first passage time, using two different techniques that both lead to the same result, and that rely on the Fokker-Planck equation (see \Sec{sec:FPT:FP}) and on the Langevin equation (see \Sec{sec:FPT:Langevin}) respectively. In passing, in \Sec{sec:FPB}, we also show how the probabilities to first hit the barrier located at $\partial\Omega_-$ before the one located at $\partial\Omega_+$ can be derived. Finally, in \Sec{sec:FPT:full:PDF}, we show how the full probability distribution function of the first passage time can be obtained.
\subsubsection{The Fokker-Planck equation}
\label{sec:FP}
Starting from the Langevin equation~\eqref{eq:Langevin}, let us derive an evolution equation for the probability $P\left(\boldmathsymbol{\Phi},N\vert \boldmathsymbol{\Phi}^\uin,N_\uin \right)$ that, starting from the field configuration $\boldmathsymbol{\Phi}^\uin$ at initial time $N_\uin$, the system is at $\boldmathsymbol{\Phi}$ at time $N$ (hereafter, for notational convenience, the subscript ``cg'' is dropped). 
\paragraph{Kramers-Moyal expansion}$ $\\
We first introduce the transition probability rate, $W_{\Delta\boldmathsymbol{\Phi}}(\boldmathsymbol{\Phi},N)$, through the relation
\bea
W_{\Delta\boldmathsymbol{\Phi}}(\boldmathsymbol{\Phi},N) \delta N = P\left(\boldmathsymbol{\Phi}+\Delta\boldmathsymbol{\Phi},N+\delta N\vert \boldmathsymbol{\Phi},N \right)\, .
\eea
This corresponds to the probability that, if the system is at $\boldmathsymbol{\Phi}$ at time $N$, it jumps to the location  $\boldmathsymbol{\Phi}+\Delta\boldmathsymbol{\Phi}$ at time $N+\delta N$, where $\delta N$ is an infinitesimal time increment. The detailed balance equation for  the probability $P\left(\boldmathsymbol{\Phi},N\vert \boldmathsymbol{\Phi}^\uin,N_\uin \right)$, which, hereafter, is simply written $P\left(\boldmathsymbol{\Phi},N \right)$ to lighten the notation, is given by
\bea
\label{eq:FP:detailed:balance}
\frac{\partial }{\partial N}P\left(\boldmathsymbol{\Phi},N \right) = \int \dd \Delta\boldmathsymbol{\Phi} \left[
W_{\Delta\boldmathsymbol{\Phi}}(\boldmathsymbol{\Phi}-\Delta\boldmathsymbol{\Phi},N)P\left(\boldmathsymbol{\Phi}-\Delta\boldmathsymbol{\Phi},N \right) 
-
W_{-\Delta\boldmathsymbol{\Phi}}(\boldmathsymbol{\Phi},N)P\left(\boldmathsymbol{\Phi},N \right) 
\right],
\eea
\ie $P\left(\boldmathsymbol{\Phi},N \right)$ increases because of realisations going from $\boldmathsymbol{\Phi}-\Delta\boldmathsymbol{\Phi}$ to $\boldmathsymbol{\Phi}$, and decreases because of realisations going from $\boldmathsymbol{\Phi}$ to $\boldmathsymbol{\Phi}-\Delta\boldmathsymbol{\Phi}$, where we integrate over $\Delta\boldmathsymbol{\Phi}$. The first term in the integrand of \Eq{eq:FP:detailed:balance} can be Taylor expanded,
\bea
W_{\Delta\boldmathsymbol{\Phi}}(\boldmathsymbol{\Phi}-\Delta\boldmathsymbol{\Phi},N)P\left(\boldmathsymbol{\Phi}-\Delta\boldmathsymbol{\Phi},N \right) =
W_{\Delta\boldmathsymbol{\Phi}}(\boldmathsymbol{\Phi},N)P\left(\boldmathsymbol{\Phi},N \right)
 +
\left(-\Delta{\Phi}_i\frac{\partial}{\partial{\Phi}_i}
+\frac{1}{2}\Delta{\Phi}_i\Delta{\Phi}_j\frac{\partial^2}{\partial{\Phi}_i\partial{\Phi}_j}
\right. \nonumber\\  \left.
+\cdots
+\frac{(-1)^\ell}{\ell!}\Delta{\Phi}_i\Delta{\Phi}_j\cdots\Delta{\Phi}_\ell\frac{\partial^\ell}{\partial{\Phi}_i\partial{\Phi}_j\cdots\partial\Phi_\ell}+\cdots
\right)
\left[W_{\Delta\boldmathsymbol{\Phi}}(\boldmathsymbol{\Phi},N)P\left(\boldmathsymbol{\Phi},N \right) \right]\, ,
\eea
where, hereafter, dummy indices are implicitly summed over. Performing the change of integration variable $\Delta\boldmathsymbol{\Phi}\to - \Delta\boldmathsymbol{\Phi}$ in the second term of the integrand of \Eq{eq:FP:detailed:balance}, one then obtain
\bea
\label{eq:detailed:balance:ai}
\frac{\partial }{\partial N}P\left(\boldmathsymbol{\Phi},N \right) =
\sum_{\ell=1}^\infty
\frac{(-1)^\ell}{\ell!}\frac{\partial^\ell}{\partial{\Phi}_i\partial{\Phi}_j\cdots\partial\Phi_\ell}
\left[a_{i j \cdots \ell}\left(\boldmathsymbol{\Phi},N \right) P\left(\boldmathsymbol{\Phi},N \right) \right]\, ,
\eea
where we have defined the moments of the field displacement
\bea
\label{eq:FP:moments:Field_Displacement:def}
a_{i j \cdots \ell}\left(\boldmathsymbol{\Phi},N \right) \equiv  \int \dd \Delta\boldmathsymbol{\Phi} \ 
\Delta{\Phi}_i\Delta{\Phi}_j\cdots\Delta{\Phi}_\ell W_{\Delta\boldmathsymbol{\Phi}}(\boldmathsymbol{\Phi},N)\, .
\eea
\paragraph{From the Langevin to the Fokker-Planck equation}$ $\\
Our goal is now to compute these field displacement moments from the Langevin equation~\eqref{eq:Langevin}. A preliminary remark is that if \Eq{eq:Langevin} is used to evolve the field value between the times $N$ and $N+\delta N$, 
\bea
\label{eq:Langevin:rule:missing}
\boldmathsymbol{\Phi} (N+\delta N) = \boldmathsymbol{\Phi} (N)+\boldmathsymbol{F}\left(\boldmathsymbol{\Phi}\right)\delta N+\boldmathsymbol{G}\left(\boldmathsymbol{\Phi}\right)\cdot \int_N^{N+\delta N}\boldmathsymbol{\xi}(\tilde{N})\dd\tilde{N},
\eea
then, there should be a prescription for where to evaluate the functions $\boldmathsymbol{F}$ and $\boldmathsymbol{G}$: either at $\boldmathsymbol{\Phi} (N)$, or at $\boldmathsymbol{\Phi} (N+\delta N)$, or somewhere in between, \etc. For non-stochastic, usual differential equations, this prescription is irrelevant since it does not affect the result in the limit $\delta N\to 0$. As we shall now see, this is not the case anymore for Langevin equations. For this reason we introduce the notation
\bea
\label{eq:FP:phi_alpha:def}
\boldmathsymbol{\Phi}_\alpha\left(N\right)=\left(1-\alpha\right)\boldmathsymbol{\Phi} (N)+\alpha\boldmathsymbol{\Phi}(N+\delta N)\, ,
\eea
which is the mean value between $\boldmathsymbol{\Phi} (N)$ and $\boldmathsymbol{\Phi}(N+\delta N)$, with weighs $1-\alpha$ and $\alpha$ respectively, where $\alpha$ is comprised between $0$ and $1$. The parameter $\alpha$ allows one to parametrise where to evaluate the field value in the argument of the $\boldmathsymbol{F}$ and $\boldmathsymbol{G}$ functions in \Eq{eq:Langevin:rule:missing}, which we now rewrite as 
\bea
\label{eq:Langevin:rule}
\boldmathsymbol{\Phi} (N+\delta N) = \boldmathsymbol{\Phi} (N)+\boldmathsymbol{F}\left[\boldmathsymbol{\Phi}_\alpha(N)\right]\delta N+\boldmathsymbol{G}\left[\boldmathsymbol{\Phi}_\alpha(N)\right]\cdot \int_N^{N+\delta N}\boldmathsymbol{\xi}(\tilde{N})\dd\tilde{N}\, .
\eea
The Langevin equation must therefore come with a prescription $\alpha$ in order to be fully defined. Common choices for the value of $\alpha$ are $\alpha=0$ (the so-called It\^o prescription) and $\alpha=1/2$ (the so-called Stratonovitch prescription).

When $\alpha$ is different from $0$, the right-hand side of \Eq{eq:Langevin:rule} depends on $\boldmathsymbol{\Phi} (N+\delta N)$, so \Eq{eq:Langevin:rule} can be seen as an equation one has to solve in order to extract $\boldmathsymbol{\Phi} (N+\delta N)$ from the value of $\boldmathsymbol{\Phi} (N)$. This can be done perturbatively in $\delta\boldmathsymbol{\Phi}\equiv \boldmathsymbol{\Phi} (N+\delta N)-\boldmathsymbol{\Phi} (N)$, by Taylor expanding the functions  $\boldmathsymbol{F}$ and $\boldmathsymbol{G}$. By rewriting \Eq{eq:FP:phi_alpha:def} as $\boldmathsymbol{\Phi}_\alpha=\boldmathsymbol{\Phi}+\alpha\delta\boldmathsymbol{\Phi}$, one obtains
\bea
F_\ell\left[\boldmathsymbol{\Phi}_\alpha(N)\right] = F_\ell(\boldmathsymbol{\Phi})+\alpha\delta\Phi_i\frac{\partial}{\partial\Phi_i}F_\ell(\boldmathsymbol{\Phi})+\frac{\alpha^2}{2}\delta\Phi_i\delta\Phi_j\frac{\partial^2}{\partial\Phi_i\partial\Phi_j}F_\ell(\boldmathsymbol{\Phi})+\cdots
\eea
and a similar expression for $\boldmathsymbol{G}\left[\boldmathsymbol{\Phi}_\alpha(N)\right] $. Plugging these expressions into \Eq{eq:Langevin:rule}, one can solve for $\delta\boldmathsymbol{\Phi}$ at iterative orders in $\delta N$, and this leads to
\bea
\delta{\Phi}_\ell& =& F_\ell(\boldmathsymbol{\Phi})\delta N+ G_{\ell i}(\boldmathsymbol{\Phi})\int_N^{N+\delta N}\xi_i(\tilde{N})\dd\tilde{N}
\nonumber \\ & & 
+\alpha G_{ij}(\boldmathsymbol{\Phi})\frac{\partial G_{\ell m}(\boldmathsymbol{\Phi})}{\partial\Phi_i}\int_N^{N+\delta N}{\xi}_j(\tilde{N})\dd\tilde{N}\int_N^{N+\delta N}{\xi}_m(\tilde{N})\dd\tilde{N}+\cdots\, ,
\eea
where ``$\cdots$'' denotes terms of higher order in $\delta N$. This expression then allows us to compute the first moments of the field displacements, \ie the $a$ numbers introduced in \Eq{eq:FP:moments:Field_Displacement:def},
\bea
a_i(\boldmathsymbol{\Phi}) &=& \lim_{\delta N\to 0}\frac{\left\langle \delta\Phi_i \right\rangle}{\delta N} = F_i(\boldmathsymbol{\Phi})+\alpha G_{mj}(\boldmathsymbol{\Phi})\frac{\partial G_{i j}(\boldmathsymbol{\Phi})}{\partial\Phi_m}\, ,\\
a_{ij}(\boldmathsymbol{\Phi}) &=& \lim_{\delta N\to 0}\frac{\left\langle \delta\Phi_i \delta\Phi_j \right\rangle}{\delta N} = G_{i m} (\boldmathsymbol{\Phi}) G_{j m}(\boldmathsymbol{\Phi})\, ,\\
a_{ij\cdots\ell}(\boldmathsymbol{\Phi})  &= &0\, ,
\eea
where we have used \Eq{eq:White:Gaussian:Noises:correlators}. This implies that only the first and second moments are non vanishing, and plugging these expressions into the detailed balance equation~\eqref{eq:detailed:balance:ai}, one obtains the Fokker-Planck equation~\cite{Risken:1984book} 
\bea
\label{eq:Fokker:Planck}
\frac{\partial }{\partial N}  P\left(\boldmathsymbol{\Phi},N\vert \boldmathsymbol{\Phi}^\uin,N_\uin \right)= \mathcal{L}_{\mathrm{FP}} \left(\boldmathsymbol{\Phi}\right) P\left(\boldmathsymbol{\Phi},N\vert \boldmathsymbol{\Phi}^\uin,N_\uin \right),
\eea
where the dependence on the initial condition $\boldmathsymbol{\Phi}^\uin,N_\uin$ has been restored in the notation, and where the Fokker-Planck operator $\mathcal{L}_{\mathrm{FP}}$ is given by
\bea
\label{eq:Fokker:Planck:operator}
\mathcal{L}_{\mathrm{FP}} \left(\boldmathsymbol{\Phi}\right) = -\frac{\partial}{\partial\Phi_i} \left[F_i\left(\boldmathsymbol{\Phi}\right) + \alpha G_{\ell j}\left(\boldmathsymbol{\Phi}\right) \frac{\partial G_{i j}\left(\boldmathsymbol{\Phi}\right)}{\partial \Phi_\ell} \right] +\frac{1}{2} \frac{\partial^2}{\partial\Phi_i\partial\Phi_j} G_{i \ell}\left(\boldmathsymbol{\Phi}\right) G_{j \ell}\left(\boldmathsymbol{\Phi}\right) .
\eea
Evaluating this expression with $\alpha=0$ and $\alpha=1/2$ gives the Fokker-Planck equation in the It\^o and Stratonovitch prescriptions respectively. One can check that, in the absence of stochastic noise ($\boldmathsymbol{G}=0$), the $\alpha$ parameter becomes irrelevant, which is expected since deterministic differential equations are not prescription dependent. 

In the situations that will be discussed below, the dependence on the prescription parameter $\alpha$ will be highly suppressed and we will thus neglect it; and in practice, work with the It\^o prescription. One should however note that it can be more important in other setups, such as in the presence of multiple inflating fields on curved field spaces~\cite{Pinol:2018euk}. The Fokker-Planck equation~\eqref{eq:Fokker:Planck} can also be written as 
\bea
\label{eq:Fokker:Planck:current}
\frac{\partial }{\partial N}  P\left(\boldmathsymbol{\Phi},N\vert \boldmathsymbol{\Phi}^\uin,N_\uin \right)= - \boldmathsymbol{\nabla}\cdot\boldmathsymbol{J},
\eea
where $\boldmathsymbol{\nabla}$ denotes the vector differential operator $\nabla_i=\partial/\partial\Phi_i$, and \boldmathsymbol{J} is the probability current,
\bea
J_i = \left[F_i\left(\boldmathsymbol{\Phi}\right) + \alpha G_{\ell j}\left(\boldmathsymbol{\Phi}\right) \frac{\partial G_{i j}\left(\boldmathsymbol{\Phi}\right)}{\partial \Phi_\ell}  -\frac{1}{2} \frac{\partial}{\partial\Phi_j} G_{i \ell}\left(\boldmathsymbol{\Phi}\right) G_{j \ell}\left(\boldmathsymbol{\Phi}\right)\right]P\left(\boldmathsymbol{\Phi},N\vert \boldmathsymbol{\Phi}^\uin,N_\uin \right)\, .
\label{eq:current:Ito}
\eea

Let us note that, since neither $\boldmathsymbol{F}$ nor $\boldmathsymbol{G}$ depends on time explicitly, $\mathcal{L}_{\mathrm{FP}}$ does not depend on time either and \Eq{eq:Fokker:Planck} describes a Markovian process, so $P(\boldmathsymbol{\Phi},N\vert \boldmathsymbol{\Phi}^\uin,N_\uin )$ depends on $N$ and $N_\uin$ only through the combination $N-N_\uin$. 
\subsubsection{First passage time from the Fokker-Planck equation}
\label{sec:FPT:FP}
Let us now derive an equation similar to \Eq{eq:Fokker:Planck}, but where the time derivative acts on the first time argument, $N_\uin$. Let us start from the Chapman-Kolmogorov relation
\begin{align} 
P\left(\boldmathsymbol{\Phi},N\vert \boldmathsymbol{\Phi}^\uin,N_\uin \right) = \int \dd\bar{\boldmathsymbol{\Phi}} P\left(\boldmathsymbol{\Phi},N\vert \bar{\boldmathsymbol{\Phi}},\bar{N} \right)P\left(\bar{\boldmathsymbol{\Phi}},\bar{N} \vert \boldmathsymbol{\Phi}^\uin,N_\uin \right)
\end{align} 
which simply states that any process starting at $\boldmathsymbol{\Phi}^\uin$ at time $N_\uin$, and ending up at $\boldmathsymbol{\Phi}$ at time $N$, goes through some configuration $\bar{\boldmathsymbol{\Phi}}$ at some intermediate time $\bar{N}$, and we integrate over all possible intermediate points $\bar{\boldmathsymbol{\Phi}}$. When one differentiates this relation with respect to $\bar{N}$, the left-hand side vanishes, since it does not depend on $\bar{N}$, and one obtains
\begin{align}
 0 &= \int \dd\bar{\boldmathsymbol{\Phi}} \left[ \frac{\partial P(\boldmathsymbol{\Phi},N\vert \bar{\boldmathsymbol{\Phi}},\bar{N})}{\partial \bar{N}}  P(\bar{\boldmathsymbol{\Phi}},\bar{N}\vert \boldmathsymbol{\Phi}^\uin,N_\uin)
 + P(\boldmathsymbol{\Phi},N\vert \bar{\boldmathsymbol{\Phi}},\bar{N}) \frac{\partial P(\bar{\boldmathsymbol{\Phi}},\bar{N}\vert \boldmathsymbol{\Phi}^\uin,N_\uin)}{\partial \bar{N}} \right]
\\ 
&= \int \dd\bar{\boldmathsymbol{\Phi}} \left[ \frac{\partial P(\boldmathsymbol{\Phi},N\vert \bar{\boldmathsymbol{\Phi}},\bar{N})}{\partial \bar{N}}  P(\bar{\boldmathsymbol{\Phi}},\bar{N}\vert \boldmathsymbol{\Phi}^\uin,N_\uin)
 + P(\boldmathsymbol{\Phi},N\vert \bar{\boldmathsymbol{\Phi}},\bar{N}) \mathcal{L}_\mathrm{FP}(\bar{\boldmathsymbol{\Phi}})\cdot P(\bar{\boldmathsymbol{\Phi}},\bar{N}\vert \boldmathsymbol{\Phi}^\uin,N_\uin)\right] 
 \, ,
 \label{eq:ChapmannKolmogorov:FP}
\end{align}
where, in the second line, we have used the Fokker-Planck equation~(\ref{eq:Fokker:Planck}). The second term in the integral of \Eq{eq:ChapmannKolmogorov:FP} can be integrated by parts making use of the adjoint Fokker-Planck operator $\mathcal{L}_\mathrm{FP}^\dagger$ defined as
\begin{align}
\label{eq:FPoperator:adjoint}
\int \dd\boldmathsymbol{\Phi} f_1 \left(\boldmathsymbol{\Phi}\right) \left[\mathcal{L}_{\mathrm{FP}}\left(\boldmathsymbol{\Phi}\right) \cdot f_2\left(\boldmathsymbol{\Phi}\right)\right]=
\int \dd\boldmathsymbol{\Phi} \left[\mathcal{L}^\dagger_{\mathrm{FP}}\left(\boldmathsymbol{\Phi}\right) \cdot  f_1 \left(\boldmathsymbol{\Phi}\right)\right] f_2\left(\boldmathsymbol{\Phi}\right),
\end{align}
and one obtains the adjoint Fokker-Planck equation
\begin{align}
\frac{\partial}{\partial N_\uin}P(\boldmathsymbol{\Phi},N\vert \boldmathsymbol{\Phi}^\uin,N_\uin) = -\mathcal{L}^\dagger_{\mathrm{FP}}\left(\boldmathsymbol{\Phi}^\uin\right)\cdot P(\boldmathsymbol{\Phi},N\vert \boldmathsymbol{\Phi}^\uin,N_\uin)\, .
\label{eq:FP:adjoint}
\end{align}
From the definition~\eqref{eq:FPoperator:adjoint}, the adjoint Fokker-Planck operator can simply be obtained from the Fokker-Planck operator~(\ref{eq:Fokker:Planck:operator}) by performing integrations by parts, and one obtains
\bea
\mathcal{L}_{\mathrm{FP}}^{\dagger, \mathrm{Stratonovitch}} \left(\boldmathsymbol{\Phi}\right) &=& 
F_i\left(\boldmathsymbol{\Phi}\right) \frac{\partial}{\partial\Phi_i}
+ \frac{1}{2} G_{i j}\left(\boldmathsymbol{\Phi}\right) \frac{\partial G_{\ell j}\left(\boldmathsymbol{\Phi}\right)}{\partial \Phi_\ell}  \frac{\partial}{\partial\Phi_i}
+\frac{1}{2}  G_{i \ell}\left(\boldmathsymbol{\Phi}\right) G_{j \ell}\left(\boldmathsymbol{\Phi}\right) \frac{\partial^2}{\partial\Phi_i\partial\Phi_j} ,\\
\mathcal{L}_{\mathrm{FP}}^{\dagger, \mathrm{\text{It\^o}}} \left(\boldmathsymbol{\Phi}\right) &=& 
F_i\left(\boldmathsymbol{\Phi}\right) \frac{\partial}{\partial\Phi_i}
+\frac{1}{2}  G_{i \ell}\left(\boldmathsymbol{\Phi}\right) G_{j \ell}\left(\boldmathsymbol{\Phi}\right) \frac{\partial^2}{\partial\Phi_i\partial\Phi_j} .
\label{eq:Fokker:Planck:adjoint:PDF:Ito}
\eea

The next step is to introduce the survival probability $S(N)$, which is the probability not to have yet crossed out the boundary $\partial\Omega_-$ at time $N$,
\begin{align}
\label{eq:S(N):def}
S(N)=\int_{\Omega} P\left(\boldmathsymbol{\Phi},N\vert \boldmathsymbol{\Phi}^\uin,N_\uin\right)\dd\boldmathsymbol{\Phi}\, .
\end{align}
This corresponds to the probability of having $\mathcal{N}>N$. If $P(\mathcal{N})$ denotes the probability distribution associated with $\mathcal{N}$, the time at which the system crosses out $\partial\Omega_-$, this means that 
\begin{align}
S(N)=\int_N^\infty P(\mathcal{N})\dd \mathcal{N}\, .
\end{align}
By differentiating this expression with respect to $N$, one obtains
\begin{align}
P(N) = -\frac{\dd}{\dd N}S(N) =- \int_{\Omega} \frac{\partial}{\partial N} P\left(\boldmathsymbol{\Phi},N\vert \boldmathsymbol{\Phi}^\uin,N_\uin\right)\dd\boldmathsymbol{\Phi}\, .
\end{align}
The $n^\mathrm{th}$ moment of $\mathcal{N}$ can therefore be expressed as
\bea
\label{eq:Nn:1}
\left\langle \mathcal{N}^n \right\rangle (\boldmathsymbol{\Phi}^\uin) &= &
\displaystyle \int_{N_\uin}^\infty N^n P(N) \dd N
= - \int_{N_\uin}^\infty N^n \dd N \int_{\Omega} \dd\boldmathsymbol{\Phi}\frac{\partial}{\partial N} P(\boldmathsymbol{\Phi},N\vert \boldmathsymbol{\Phi}^\uin,N_\uin) 
\\  
& = &
-\left[  N^n \int_\Omega \dd\boldmathsymbol{\Phi} P (\boldmathsymbol{\Phi},N\vert \boldmathsymbol{\Phi}^\uin,N_\uin)\right]_{N=N_\uin}^{N=\infty}
+
  n \displaystyle \int_{N_\uin}^\infty N^{n-1} \dd N \int_{\Omega} \dd\boldmathsymbol{\Phi} P(\boldmathsymbol{\Phi},N\vert \boldmathsymbol{\Phi}^\uin,N_\uin)
  \\  
& = &
-\left[  N^n S(N)\right]_{N=N_\uin}^{N=\infty}
+
  n \displaystyle \int_{N_\uin}^\infty N^{n-1} \dd N \int_{\Omega} \dd\boldmathsymbol{\Phi} P(\boldmathsymbol{\Phi},N\vert \boldmathsymbol{\Phi}^\uin,N_\uin) \, ,
  \label{eq:meanNn:intermediate}
\eea
where in the second equality, integration by parts has been performed, and in the third equality, we have used the definition~\eqref{eq:S(N):def}. In \Sec{sec:tail:expansion}, we will show that the distribution function of $\N$, $P(\N)$, decays exponentially at large $\N$, hence faster than any power of $\N$, and in the first term of \Eq{eq:meanNn:intermediate}, one can take $\lim_{N\to\infty}N^nS(N)=0$. Since $P(\boldmathsymbol{\Phi},N_\uin\vert \boldmathsymbol{\Phi}^\uin,N_\uin) = \delta(\boldmathsymbol{\Phi}-\boldmathsymbol{\Phi}^\uin)$ by definition, from \Eq{eq:S(N):def}, one has $S(N_\uin)=1$, hence the other contribution to the first term of \Eq{eq:meanNn:intermediate} reads $N_\uin^n S(N_\uin) = N_\uin^n$ and one obtains
\bea
\left\langle \mathcal{N}^n \right\rangle (\boldmathsymbol{\Phi}^\uin) =
N_\uin^n
+
  n \displaystyle \int_{N_\uin}^\infty N^{n-1} \dd N \int_{\Omega} \dd\boldmathsymbol{\Phi} P(\boldmathsymbol{\Phi},N\vert \boldmathsymbol{\Phi}^\uin,N_\uin) \, .
  \label{eq:meanNn:intermediate:2}
\eea
By applying the adjoint Fokker-Planck operator $\mathcal{L}_\mathrm{FP}^\dagger(\boldmathsymbol{\Phi}^\uin)$ to this relation and making use of the adjoint Fokker-Planck equation~(\ref{eq:FP:adjoint}), one obtains
\begin{align}
\mathcal{L}_\mathrm{FP}^\dagger(\boldmathsymbol{\Phi}^\uin)\cdot\left\langle \mathcal{N}^n \right\rangle (\boldmathsymbol{\Phi}^\uin)  & =-
n \int_{N_\uin}^\infty N^{n-1} \dd N \int_{\Omega} \dd\boldmathsymbol{\Phi} \frac{\partial}{\partial N_\uin} P(\boldmathsymbol{\Phi},N\vert \boldmathsymbol{\Phi}^\uin,N_\uin)\, .
\end{align}
At this stage, let us recall that one is dealing with a Markovian process, for which the transition probability depends on $N-N_\uin$ only. One then has
\begin{align}
\label{eq:FPT:Markov}
\mathcal{L}_\mathrm{FP}^\dagger(\boldmathsymbol{\Phi}^\uin)\cdot\left\langle \mathcal{N}^n \right\rangle (\boldmathsymbol{\Phi}^\uin)  & = -
n \int_{N_\uin}^\infty N^{n-1} \dd N \int_{\Omega} \dd\boldmathsymbol{\Phi} \frac{\partial}{\partial N_\uin} P(\boldmathsymbol{\Phi},N-N_\uin\vert \boldmathsymbol{\Phi}^\uin,0) \\
& = 
n \int_{N_\uin}^\infty N^{n-1} \dd N \int_{\Omega} \dd\boldmathsymbol{\Phi} \frac{\partial}{\partial N} P(\boldmathsymbol{\Phi},N-N_\uin\vert \boldmathsymbol{\Phi}^\uin,0)
\\ & 
 = n \int_{N_\uin}^\infty N^{n-1} \dd N \int_{\Omega} \dd\boldmathsymbol{\Phi} \frac{\partial}{\partial N} P(\boldmathsymbol{\Phi},N\vert \boldmathsymbol{\Phi}^\uin,N_\uin) \, .
 \label{eq:Nn:2}
\end{align}
When $n=1$, this gives rise to
\begin{align}
\mathcal{L}_\mathrm{FP}^\dagger(\boldmathsymbol{\Phi}^\uin)\cdot\left\langle \mathcal{N} \right\rangle (\boldmathsymbol{\Phi}^\uin)  & = \int_{N_\uin}^\infty \dd N \int_{\Omega} \dd\boldmathsymbol{\Phi} \frac{\partial}{\partial N} P(\boldmathsymbol{\Phi},N\vert \boldmathsymbol{\Phi}^\uin,N_\uin) 
\\ & 
=\int_{\Omega} \dd\boldmathsymbol{\Phi} \left[P(\boldmathsymbol{\Phi},\infty\vert \boldmathsymbol{\Phi}^\uin,N_\uin)-P(\boldmathsymbol{\Phi},N_\uin\vert \boldmathsymbol{\Phi}^\uin,N_\uin) \right]
\\ & 
=S(\infty)-S(N_\uin)
= -1\, .
\label{FPT:neq1}
\end{align}
When $n\geq 2$, from \Eq{eq:Nn:1}, one notices that $\langle \mathcal{N}^{n-1}\rangle$ appears in the right hand side of \Eq{eq:Nn:2}, giving rise to 
\bea
\label{eq:FPT:moment:adjoint:FP}
\mathcal{L}_\mathrm{FP}^\dagger(\boldmathsymbol{\Phi}^\uin)\cdot\left\langle \mathcal{N}^n \right\rangle (\boldmathsymbol{\Phi}^\uin) = -n \left\langle \mathcal{N}^{n-1} \right\rangle (\boldmathsymbol{\Phi}^\uin)\, .
\eea
This equation is in fact also valid for $n=1$, in which case, since $\langle \mathcal{N}^0\rangle=1$, one recovers \Eq{FPT:neq1}. It provides a hierarchy of differential equations for the moments of the first passage times, where the equation for the moment of order $n$ is sourced by the moment of order $n-1$. This hierarchy can then be solved for iteratively increasing values of $n$, which will be done explicitly in \Sec{sec:FirstMoments:N} up to $n=3$. 
\subsubsection{First passage time from the Langevin equation}
\label{sec:FPT:Langevin}
The same results, in the It\^o procedure, can also be obtained starting directly from the Langevin equation~(\ref{eq:Langevin}). In this section, we quickly sketch such a derivation for illustrative purpose. Let $f(\boldmathsymbol{\Phi})$ be a generic function of the field space coordinate $\boldmathsymbol{\Phi}$. If $\boldmathsymbol{\Phi}$ is a realisation of the stochastic process~(\ref{eq:Langevin}), using It\^o calculus, its variation is given by
\bea
\displaystyle
\dd f(\boldmathsymbol{\Phi}) &= & f_{{\Phi}_i}\dd\phi_i +\frac{1}{2}f_{\Phi_i,\Phi_j}\dd\Phi_i\dd\Phi_j+\order{\dd\boldmathsymbol{\Phi}^3}\\
&=& 
f_{{\Phi}_i} G_{i j} \xi_j \dd N
+ f_{{\Phi}_i} F_i \dd N
+ \frac{1}{2}f_{\Phi_i,\Phi_j} G_{i \ell} G_{j \ell} \dd N
+\order{\dd N^2} \\
&=& 
f_{{\Phi}_i} G_{i j} \xi_j \dd N
+\mathcal{L}^{\dagger,\mathrm{\text{It\^o}}}(\boldmathsymbol{\Phi})\cdot f(\boldmathsymbol{\Phi})
+\order{\dd N^2}
\, ,
\eea
where dummy indices are implicitly summed over, and where in the last equality, we have used \Eq{eq:Fokker:Planck:adjoint:PDF:Ito}. Integrating this relation between $N=0$ where $\boldmathsymbol{\Phi}=\boldmathsymbol{\Phi}^\uin$ and $N=\mathcal{N}$ where $\boldmathsymbol{\Phi}=\boldmathsymbol{\Phi}^\uend\in\partial\Omega_-$, one obtains the It\^o's lemma~\cite{ito1944}
\bea
f\left(\boldmathsymbol{\Phi}^\uend\in\partial\Omega_-\right) - f\left(\boldmathsymbol{\Phi}^\uin\right) & = & 
\int_0^\mathcal{N}  f_{{\Phi}_i} G_{i j} \xi_j \dd N
+ \int_0^\mathcal{N} \mathcal{L}_{\mathrm{FP}}^{\dagger,\mathrm{\text{It\^o}}}(\boldmathsymbol{\Phi})\cdot f(\boldmathsymbol{\Phi}) \dd N
\, .
\label{eq:Ito}
\eea
Let us now apply this lemma to the function $f$ that satisfies the differential equation $\mathcal{L}_{\mathrm{FP}}^{\dagger,\mathrm{\text{It\^o}}}(\boldmathsymbol{\Phi})\cdot f(\boldmathsymbol{\Phi})=-1$, with a boundary condition $ f(\boldmathsymbol{\Phi}\in\partial\Omega_-)=0$ (another boundary condition can sometimes be needed to entirely fix $f$ but we do not need it at this stage, see the discussion below \Eq{eq:FPT:h:intermediate}). By definition, the first term in the left hand side of \Eq{eq:Ito} vanishes, and the integrand of the second integral of the right hand side is $-1$. This gives rise to
\begin{align}
\mathcal{N}=f\left(\boldmathsymbol{\Phi}^\uin\right)+\int_0^\mathcal{N}  f_{{\Phi}_i} G_{i j} \xi_j \dd N \, .
\label{eq:Ito:f1}
\end{align}
By taking the stochastic average of this equation, one is led to 
\begin{align}
\left\langle \mathcal{N} \right\rangle=f\left(\boldmathsymbol{\Phi}^\uin\right)\, .
\end{align}
Note that the fact that the stochastic average of the integral term in \Eq{eq:Ito:f1} vanishes is not trivial a priori since not only the integrand but the upper bound of the integral itself is stochastic, but because the noises $\xi_j$ are uncorrelated at different times, this can be shown rigorously~\cite{Risken:1984book}. This demonstrates \Eq{eq:FPT:moment:adjoint:FP} for $n=1$. 

Larger values of $n$ can be dealt with in a similar manner. Indeed, by squaring \Eq{eq:Ito:f1} and taking the stochastic average of it, one obtains
\begin{align}
\left\langle \mathcal{N}^2 \right\rangle=f^2\left(\boldmathsymbol{\Phi}^\uin\right)
+\left\langle \int_0^\mathcal{N} f_{\Phi_i} \left(G^2\right)_{i j}  f_{\Phi_j}\dd N \right\rangle  \, ,
\label{eq:Ito:f2}
\end{align}
where we have used \Eq{eq:White:Gaussian:Noises:correlators}. 
Let us now apply It\^o's lemma~(\ref{eq:Ito}) to the function $h\equiv g-f^2$, where $f$ is still the solution of the differential equation $\mathcal{L}^{\dagger,\mathrm{\text{It\^o}}}(\boldmathsymbol{\Phi})\cdot f(\boldmathsymbol{\Phi})=-1$ with a boundary condition $ f(\boldmathsymbol{\Phi}\in\partial\Omega_-)=0$, and $g$ is the solution of the differential equation $\mathcal{L}^{\dagger,\mathrm{\text{It\^o}}}(\boldmathsymbol{\Phi})\cdot g(\boldmathsymbol{\Phi})=-2 f(\boldmathsymbol{\Phi})$ with a boundary condition $ f(\boldmathsymbol{\Phi}\in\partial\Omega_-)=0$. By definition, the first term in the left hand side of \Eq{eq:Ito} vanishes, and the integrand of the second integral of the right hand side is given by $ \mathcal{L}_{\mathrm{FP}}^{\dagger,\mathrm{\text{It\^o}}} \cdot h =  \mathcal{L}_{\mathrm{FP}}^{\dagger,\mathrm{\text{It\^o}}} \cdot g -  \mathcal{L}_{\mathrm{FP}}^{\dagger,\mathrm{\text{It\^o}}}\cdot  f^2 = -2 f - 2 f \mathcal{L}_{\mathrm{FP}}^{\dagger,\mathrm{\text{It\^o}}}\cdot  f - G_{i\ell} G_{j\ell} f_{\Phi_i} f_{\Phi_j} = - G_{i\ell} G_{j\ell} f_{\Phi_i} f_{\Phi_j} $. The stochastic average of It\^o's lemma thus gives rise to
\bea
\label{eq:FPT:h:intermediate}
h\left(\boldmathsymbol{\Phi}^\uin\right) = \left\langle \int_0^\mathcal{N} f_{\Phi_i} \left(G^2\right)_{i j}  f_{\Phi_j}\dd N \right\rangle .
\eea
By identification with \Eq{eq:Ito:f2}, one obtains that $\langle \mathcal{N}^2\rangle = f^2(\boldmathsymbol{\Phi}^\uin)+h(\boldmathsymbol{\Phi}^\uin) = g(\boldmathsymbol{\Phi}^\uin)$, which proves \Eq{eq:FPT:moment:adjoint:FP} for $n=2$. Applying the same method, one can iteratively proceed and extend the result to any value of $n$.

A few words are finally in order regarding the boundary conditions. The moments of the first passage time were shown to satisfy a second-order differential equation, so boundary conditions should be set everywhere on $\partial\Omega = \partial\Omega_-\cup \partial\Omega_+$. As explained above, if $\partial\Omega_-$ defines a compact inflating field space (as in single-field hilltop inflation for instance), there is no need to introduce $\partial\Omega_+$, and requiring that all moments of the first passage time vanish on $\partial\Omega_-$ is enough to set the boundary conditions. Otherwise, imposing additional boundary conditions on $\partial\Omega_+$ may be necessary: if absorbing boundary conditions are set, one simply requires that the moments vanish on $\partial\Omega_+$; if reflective boundary conditions are set, one requires that the gradient of the moments, projected onto the vector orthogonal to the tangent surface of $\partial\Omega_+$, vanishes; if partly absorbing / partly reflective boundary conditions are set, one imposes a combination between these two conditions. 
\subsubsection{First Passage Boundary}
\label{sec:FPB}
In this section, we consider the case where $\partial\Omega$ is made of two (or more) disconnected pieces (say $\partial\Omega_-$ and $\partial\Omega_+$) and one wants to determine with which probability $p_+$ the system exits $\Omega$ by crossing out $\partial\Omega_+$ first (or respectively, with which probability $p_-=1-p_+$ the system exits $\Omega$ by crossing out $\partial\Omega_-$ first). The result is useful to determine the probability according to which the field explores large-field, classically forbidden regions of the potential for instance, as well as in order to compute tunnelling probabilities and rates through local maxima of the potential~\cite{Noorbala:2018zlv}.  

By definition, $p_+$ corresponds to the probability that the system is somewhere along $\partial\Omega_+$ at time $N$, where $N$ is integrated over all possible values between $N_\uin$ and $\infty$,
\bea
p_+(\boldmathsymbol{\Phi}^\uin)=\int_{\boldmathsymbol{\Phi}\in\partial\Omega_+}\int_{N_\uin}^\infty\dd N P\left(\boldmathsymbol{\Phi},N\vert \boldmathsymbol{\Phi}^\uin,N_\uin\right)\, .
\eea
Let us now apply the adjoint Fokker-Planck operator to this relation. Making use of \Eq{eq:FP:adjoint}, one obtains
\begin{align}
\mathcal{L}_{\mathrm{FP}}^\dagger\left(\boldmathsymbol{\Phi}^\uin\right) \cdot p_+(\boldmathsymbol{\Phi}^\uin)&=\int_{\boldmathsymbol{\Phi}\in\partial\Omega_+}\int_{N_\uin}^\infty\dd N \mathcal{L}_{\mathrm{FP}}^\dagger\left(\boldmathsymbol{\Phi}^\uin\right) \cdot P\left(\boldmathsymbol{\Phi},N\vert \boldmathsymbol{\Phi}^\uin,N_\uin\right)\\
&=-\int_{\boldmathsymbol{\Phi}\in\partial\Omega_+}\int_{N_\uin}^\infty\dd N \frac{\partial}{\partial N_\uin} P\left(\boldmathsymbol{\Phi},N\vert \boldmathsymbol{\Phi}^\uin,N_\uin\right)\\
&=-\int_{\boldmathsymbol{\Phi}\in\partial\Omega_+}\int_{N_\uin}^\infty\dd N \frac{\partial}{\partial N_\uin} P\left(\boldmathsymbol{\Phi},N-N_\uin\vert \boldmathsymbol{\Phi}^\uin,0\right)\\
&=\int_{\boldmathsymbol{\Phi}\in\partial\Omega_+}\int_{N_\uin}^\infty\dd N \frac{\partial}{\partial N} P\left(\boldmathsymbol{\Phi},N-N_\uin\vert \boldmathsymbol{\Phi}^\uin,0\right)\label{eq:FPB:markov}\\
&=\int_{\boldmathsymbol{\Phi}\in\partial\Omega_+}\int_{N_\uin}^\infty\dd N \frac{\partial}{\partial N} P\left(\boldmathsymbol{\Phi},N\vert \boldmathsymbol{\Phi}^\uin,N_\uin\right)\\
&=\int_{\boldmathsymbol{\Phi}\in\partial\Omega_+} \left[ P\left(\boldmathsymbol{\Phi},\infty\vert \boldmathsymbol{\Phi}^\uin,N_\uin\right) - P\left(\boldmathsymbol{\Phi},{N_\uin}\vert \boldmathsymbol{\Phi}^\uin,N_\uin\right)\right]
=0\, .
\label{eq:FPB:0}
\end{align}
In \Eq{eq:FPB:markov}, similarly to what was performed in \Eq{eq:FPT:Markov}, one has used the fact that the stochastic process under consideration is Markovian, hence the transition probability depends on $N-N_\uin$ only. To obtain the final result~(\ref{eq:FPB:0}), one has also used the fact that all realisations have crossed out $\partial\Omega$ in the infinite future hence $P(\boldmathsymbol{\Phi},\infty\vert \boldmathsymbol{\Phi}^\uin,N_\uin=0)$, together with the initial condition $P\left(\boldmathsymbol{\Phi},{N_\uin}\vert \boldmathsymbol{\Phi}^\uin,N_\uin\right)=\delta(\boldmathsymbol{\Phi}-\boldmathsymbol{\Phi}^\uin)$ (and the assumption that $\boldmathsymbol{\Phi}^\uin \not\in \partial\Omega_+$, otherwise we already know that $p_+=1$ by definition). 

The probability $p_+(\boldmathsymbol{\Phi})$ that the system first reaches $\partial\Omega_+$ starting from $\boldmathsymbol{\Phi}^\uin=\boldmathsymbol{\Phi}$ is therefore given by the solution of the ordinary differential equation
\begin{align}
\mathcal{L}_{\mathrm{FP}}^\dagger\left(\boldmathsymbol{\Phi}^\uin\right)p_+ \left(\boldmathsymbol{\Phi}^\uin\right)=0\, ,
\label{eq:FPB:equadiff:app}
\end{align}
with boundary conditions $p_+=1$ on $\partial\Omega_+$ and $p_+=0$ on $\partial\Omega_-$. The probability $p_-=1-p_+$ satisfies the same differential equation, but with boundary conditions $p_-=1$ on $\partial\Omega_-$ and $p_-=0$ on $\partial\Omega_+$.
\subsubsection{Extracting the full probability distribution function}
\label{sec:FPT:full:PDF}
In the previous sections, we have seen how the moments of the first passage time could be derived, by solving differential equations involving the (adjoint) Fokker-Planck operator. Since a distribution function is entirely determined by the set of all its moments, one can use these results to design a method to obtain the full probability distribution function (PDF) of the first passage time. This is the goal of this section, which follows \Refa{Pattison:2017mbe}. 

In order to relate the PDF of $\N$ to its statistical moments, let us introduce its characteristic function
\bea
\label{eq:characteristicFunction:def}
\chi_\N(t,\boldmathsymbol{\Phi}) \equiv \left\langle \ee^{it \N(\boldmathsymbol{\Phi})} \right\rangle \, ,
\eea 
which depends on $\boldmathsymbol{\Phi}$, the initial field coordinate (hereafter, the subscript ``in'' is dropped for notational convenience), and a dummy parameter $t$. By Taylor expanding $\chi_\N(t,\boldmathsymbol{\Phi})$ around $t=0$, one has $\chi_\N(t,\boldmathsymbol{\Phi}) = \sum_{n=0}^\infty (it)^n \langle \N^n(\boldmathsymbol{\Phi})\rangle/n!$. If one applies the adjoint Fokker-Planck operator to this expansion, and uses \Eq{eq:FPT:moment:adjoint:FP} to replace each term on the right-hand side, one obtains
\bea
\label{eq:ODE:chi}
\mathcal{L}_{\mathrm{FP}}^\dagger \cdot  \chi_\N(t,\boldmathsymbol{\Phi}) 
= -i t \chi_\N(t,\boldmathsymbol{\Phi}) \, .
\eea
At fixed $t$, this is a differential equation in $\boldmathsymbol{\Phi}$, so instead of the hierarchy of coupled differential equations~(\ref{eq:FPT:moment:adjoint:FP}) one now has a set of uncoupled differential equations to solve, which can improve the tractability of the problem in some cases. These equations~(\ref{eq:ODE:chi}) have to be solved with the same boundary conditions as the ones imposed on the moments and discussed at the end of \Sec{sec:FPT:Langevin}, \ie $\chi_\N(t,\boldmathsymbol{\Phi})=1$ on absorbing surfaces, and the gradient of the characteristic function, projected onto the vector orthogonal to the tangent surface to reflective boundaries, vanishes.

Let us note that the characteristic function of the \emph{fluctuation} in the number of \efolds, $\zeta_\mathrm{cg} = \delta N_\mathrm{cg} = \N - \langle \N \rangle $, can be found by plugging this expression into \Eq{eq:characteristicFunction:def}, which gives rise to 
\bea
\label{eq:chideltaN:chiN}
\chi_{\zeta_\mathrm{cg}}\left(t,\boldmathsymbol{\Phi}\right) = e^{-i\langle \N \rangle t} \chi_\N(t,\boldmathsymbol{\Phi})\, .
\eea
One also notices, from \Eq{eq:characteristicFunction:def}, that the characteristic function $\chi_\N$ can be rewritten as
\bea
\label{eq:chi:P}
\chi_{\N}(t,\boldmathsymbol{\Phi})=\int^{\infty}_{-\infty} \ee^{it\N} P\left(\N, \boldmathsymbol{\Phi}\right) \dd \N\, ,
\eea
that is to say, the characteristic function is the Fourier transform of the PDF of curvature perturbations. Therefore, the PDF is the inverse Fourier transform of the characteristic function, \ie
\bea
\label{eq:PDF:chi}
P\left(\zeta_{\mathrm{cg}}, \boldmathsymbol{\Phi}\right) = \frac{1}{2\pi} \int^{\infty}_{-\infty} \ee^{-it\left[\zeta_{\mathrm{cg}}+\langle \N\rangle\left(\boldmathsymbol{\Phi}\right)\right]} \chi_{\N}\left(t,\boldmathsymbol{\Phi}\right)\dd t\, ,
\eea
where we have used \Eq{eq:chideltaN:chiN}. The calculational programme is thus the following: solve \Eq{eq:ODE:chi} with the appropriate boundary conditions, calculate $\langle \N \rangle$ either by solving \Eq{eq:FPT:moment:adjoint:FP} with $n=1$ or by noting that 
\bea
\label{eq:meanN:chi}
\left\langle \N \right\rangle \left(\boldmathsymbol{\Phi}\right) = -i \left. \frac{\partial\chi_\N \left(t,\boldmathsymbol{\Phi}\right)} {\partial t} \right\vert_{t=0},
\eea 
and calculate the PDF of curvature perturbations with \Eq{eq:PDF:chi}.

In passing, let us also note that the problem can be reformulated in terms of a heat equation for the PDF $P(\N,\boldmathsymbol{\Phi})$ directly. Indeed, if one plugs \Eq{eq:chi:P} into \Eq{eq:ODE:chi}, one obtains
\bea
\int \dd \N \ee^{i t \N} \mathcal{L}_{\mathrm{FP}}^\dagger\left(\boldmathsymbol{\Phi}\right) \cdot P\left(\N, \boldmathsymbol{\Phi}\right) \dd \N 
&=& - i t \int \dd \N \ee^{i t \N} P\left(\N, \boldmathsymbol{\Phi}\right) \\
&=& -  \int \dd \N \frac{\partial}{\partial\N} \left(\ee^{i t \N}\right) P\left(\N, \boldmathsymbol{\Phi}\right) \\
&=&    \int \dd \N  \ee^{i t \N} \frac{\partial}{\partial\N} P\left(\N, \boldmathsymbol{\Phi}\right),
\eea
where in the last expression, we have integrated by parts. By identifying the first and the last expression, one obtains
\bea
\label{eq:Fokker:Planck:adjoint:PDF}
\frac{\partial}{\partial\N} P\left(\N, \boldmathsymbol{\Phi}\right) = \mathcal{L}_{\mathrm{FP}}^\dagger\left(\boldmathsymbol{\Phi}\right) \cdot P\left(\N, \boldmathsymbol{\Phi}\right) .
\eea
This differential equation, which has the structure of a heat equation, needs again to be solved with the same boundary conditions as the ones imposed on the moments and discussed at the end of \Sec{sec:FPT:Langevin}. Notice the strong similarity, yet the crucial difference, with the Fokker-Planck equation~\eqref{eq:Fokker:Planck}: the Fokker-Planck equation drives the probability for the system to be in a certain configuration at a certain time, starting from a given initial configuration at a given initial time, while the adjoint Fokker-Planck equation~\eqref{eq:Fokker:Planck:adjoint:PDF} drives the probability for the duration of the process, starting from a given initial configuration. These two PDFs, while very different in nature, can thus be thought of as ``adjoint'' in the sense that they obey adjoint evolution equations. 
\subsection{First moments of the number of \efolds}
\label{sec:FirstMoments:N}
In \Sec{sec:FirstPassageTime}, we have shown how the moments of the first passage time can be derived by solving a hierarchy of differential equations. In this section, we solve these equations for the first few moments. For simplicity, we consider the situation where all scalar fields $\phi_1,\phi_2,\cdots,\phi_D$ relevant during inflation have reached the slow-roll attractor, such that the results of \Sec{sec:slow:roll:stochastic} can be applied. In particular, phase space is of dimension $D$ only (since the conjugated momenta to the fields simply rest on their slow-roll configuration), and the Langevin equations are given by \Eq{eq:Langevin:SR:phi}, which, in the language of \Eq{eq:Langevin}, reads $F_i = -V_{\phi_i}/(3 H^2)$ and $G_{ij}=H/(2\pi)\delta_{ij}$. Introducing the dimensionless potential 
\bea
v=\frac{V}{24\pi^2\Mp^4}\, ,
\eea
and working with the It\^o procedure for concreteness, the Fokker-Planck and adjoint Fokker-Planck operators are given by
\bea
\mathcal{L}_{\mathrm{FP}} = \Mp^2  \sum_{i=1}^D \left( \frac{\partial}{\partial\phi_i}\frac{v_{\phi_i}}{v}+  \frac{\partial^2}{\partial\phi_i^2} v\right)\\
\mathcal{L}_{\mathrm{FP}}^\dagger = \Mp^2 \sum_{i=1}^D  \left(- \frac{v_{\phi_i}}{v}  \frac{\partial}{\partial\phi_i}+v  \frac{\partial^2}{\partial\phi_i^2} \right)
\label{eq:Fokker:Planck:adjoint:PDF:Ito:SlowRoll}
\eea
see \Eqs{eq:Fokker:Planck:operator} and~\eqref{eq:Fokker:Planck:adjoint:PDF:Ito} respectively. We will first focus on the case if single-field inflation, and will derive the three first moments of the first passage time, which lead us to computing the power spectrum and local non-Gaussianity of the curvature perturbation. In \Sec{sec:Infinite:Inflation}, the analysis will be extended to multiple-field setups, and more attention will be given to the role played by the ``UV'' boundary condition $\delta\Omega_+$, which strongly depends on the number of fields.

Let us note that, if a single scalar field is at play, one can introduce a change of field coordinate
\bea
u(\phi) = \int_{\phiend}^\phi \ee^{-\frac{1}{v(\tilde{\phi})}} \frac{\dd \tilde{\phi}}{\Mp}\, ,
\eea
which allows us to rewrite \Eq{eq:Fokker:Planck:adjoint:PDF:Ito:SlowRoll} as 
\bea
\label{eq:Fokker:Planck:adjoint:PDF:Ito:SlowRoll:u}
\mathcal{L}_{\mathrm{FP}}^\dagger = v \ee^{-\frac{2}{v}}\frac{\partial^2}{\partial u^2} ,
\eea 
and to get rid of the drift term. Although we will not make much use of this reformulation, it immediately shows that the quantity $\ee^{-1/v}$ will play a crucial role. Since $v$ measures the potential energy in Planck units, it is a parametrically small number, hence $\ee^{-1/v}$ is exponentially small. This means that, in the pure diffusion problem of \Eq{eq:Fokker:Planck:adjoint:PDF:Ito:SlowRoll:u}, the ``temperature'' of the process is very low, and this will be at the basis of the classical expansions performed below.

Let us also notice that the stationary solution to the Fokker-Planck equation~\eqref{eq:Fokker:Planck}, $\partial P(\phi,N\vert\phi_\uin,N_\uin) = \mathcal{L}_{\mathrm{FP}}\cdot P(\phi,N\vert\phi_\uin,N_\uin) =0$, is the one for which the probability current is uniform in field space, see \Eq{eq:Fokker:Planck:current}, and \Eq{eq:current:Ito} with the It\^o prescription ($\alpha=0$) gives rise to
\bea
\label{eq:Pstat}
P_\mathrm{stat}(\phi) \propto\frac{\ee^{\frac{1}{v}}}{v},
\eea
where there is an overall multiplicative constant such that $P_\mathrm{stat}$ is properly normalised. 
\begin{figure}[t]
\begin{center}
\includegraphics[width=\wdblefig]{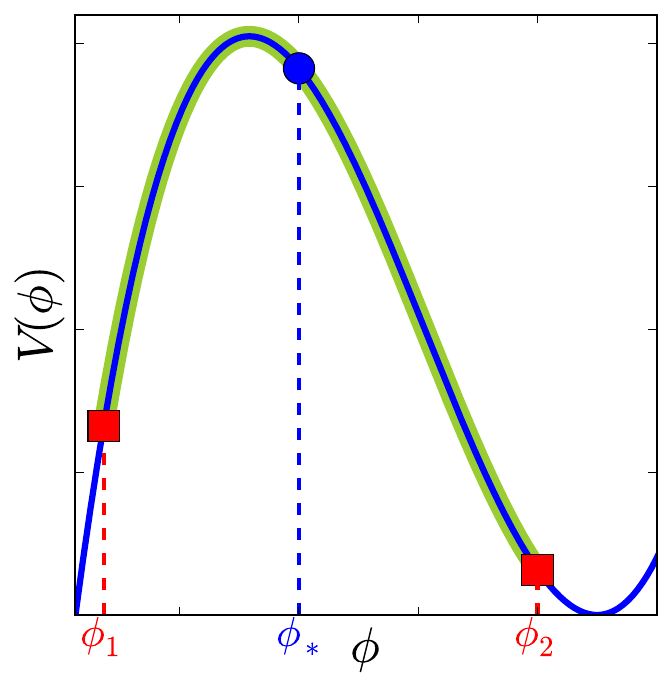}
\includegraphics[width=\wdblefig]{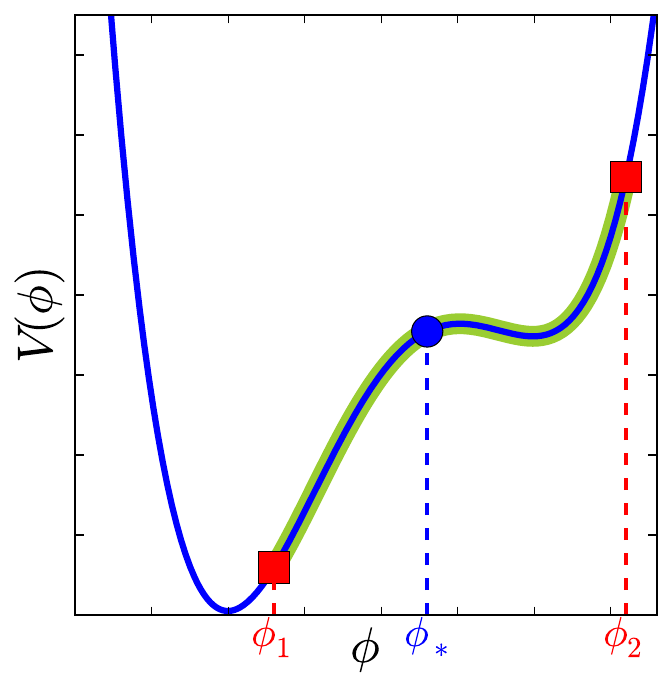}
\caption[Sketch of the Langevin bounded dynamics]{Sketch of the dynamics solved in section~\ref{sec:FirstMoments:N}. The inflaton is initially located at $\phi_*$ and evolves along the potential $V(\phi)$ under the stochastic Langevin equation~(\ref{eq:Langevin:SR:phi}). In the left panel, inflation terminates by slow-roll violation when the inflaton reaches one of the two ending values $\phi_1$ or $\phi_2$. In the right panel, another boundary is introduced at $\phi_2$ (it could correspond for instance to where $V\sim \Mp^4$), which can be of different natures: absorbing, reflective, \etc).}
\label{fig:sketch:pot:stochaDeltaN}
\end{center}
\end{figure}
\subsubsection{Ending Point Probability}
\label{sec:WallHittProba}
We consider the situation depicted in \Fig{fig:sketch:pot:stochaDeltaN}. As a first warm-up, let us calculate the probability $p_1$ that the inflaton field first reaches the ending point located at $\phi_1$ [\ie $\phi\left(\mathcal{N}\right)=\phi_1$], or, equivalently the probability $p_2=1-p_1$ that the inflaton field first reaches the ending point located at $\phi_2$ [\ie $\phi\left(\mathcal{N}\right)=\phi_2$]. This will also allow us to determine when the ending point located at $\phi_2$ plays a negligible role.

Making use of the results of \Sec{sec:FPB}, one finds that the generic solution to \Eq{eq:FPB:equadiff:app} is given by $p_1(\phi) \propto u$, \ie $p_1(\phi) = A\int_{B}^{\varphi}\exp\left[-1/v\left(x\right)\right]\dd x$ where $A$ and $B$ are two integration constants that need to be set by means of the boundary conditions $p_1(\phi_1)=1$ and $p_1(\phi_2)=0$, which gives rise to
\bea
p_1
=\dfrac{\displaystyle\int_{\phi_*}^{\phi_2}\exp\left[-\frac{1}{v\left(x\right)}\right]\dd x}{\displaystyle\int_{\phi_1}^{\phi_2}\exp\left[-\frac{1}{v\left(x\right)}\right]\dd x}
\, ,
\label{stocha:hittproba:p1}
\eea
and a symmetric expression for $p_2$.\footnote{This is in agreement with Eq. (29) of \Refa{Starobinsky:1986fx}, derived in the case where $H$ is constant, hence $v^{-1}\approx v_*^{-1} - (v-v_*)\,v_*^{-2}$, where $\phi_2$ and $\phi_1$ lie at $\pm\infty$ correspondingly, and where the initial condition for \Eq{eq:Fokker:Planck} is chosen to be $P(\phi,0)=\delta(\phi - \phi_*)$.}

A few remarks are in order about this result. First, one can check that, since $\phi_*$ lies between $\phi_1$ and $\phi_2$, the probability~(\ref{stocha:hittproba:p1}) is ensured to be comprised between $0$ and $1$. Second, one can also verify that when $\phi_*=\phi_1$, $p_1=1$, and when $\phi_*=\phi_2$, $p_1=0$, as one would expect. Third, in the case depicted in the right panel of \Fig{fig:sketch:pot:stochaDeltaN}, in the limit where $\phi_2\rightarrow\infty$, one is sure to first reach the ending point located at $\phi_1$, that is, $p_2=\int_{\phi_1}^{\phi_*}e^{-1/v}/\int_{\phi_1}^{\phi_2}e^{-1/v}=0$. Indeed, the numerator of the expression for $p_2$ is finite, since a bounded function is integrated over a bounded interval. If the potential is maximal at $\phi_2$, and if it is monotonous over an interval of the type $\left[\phi_0,\phi_2\right[$, its denominator is on the contrary larger than the integral of a function bounded from below by a strictly positive number, over an unbounded interval $\left[\phi_0,\phi_\infty\right[$. This is why it diverges, and why $p_2$ vanishes. This means that if $\phi_2$ is sufficiently large, the probability to ``explore'' $\phi_2$ can be made very small. As we will further discuss in \Sec{sec:Infinite:Inflation}, this will not be always true if more than one field are at play.
\subsubsection[Mean Number of \efolds]{Mean Number of \texorpdfstring{$\bm{e}$}{$e$}-folds}
\label{sec:meanN}
\begin{figure}[t]
\begin{center}
\includegraphics[width=\wdblefig]{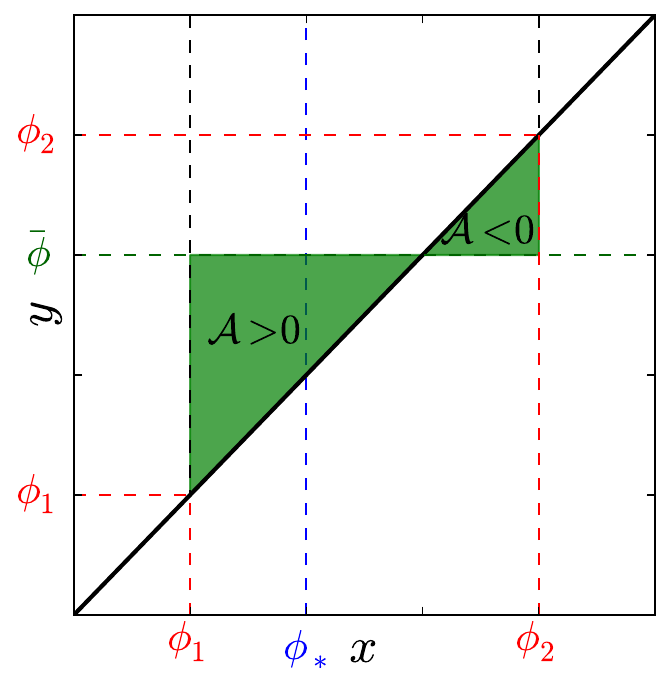}
\caption[Integration domain of the mean number of \efolds]{Integration domain of \Eq{eq:f:sol} when evaluated at $\phi=\phi_2$, in the case $\phi_1<\phi_2$ (the opposite case proceeds the same way). The discrete parameter $x$ is integrated between $\phi_1$ and $\phi_2$, while $y$ varies between $x$ and $\bar{\phi}$. The resulting integration domain is displayed in green. When $x<\bar{\phi}$, one has $\dd x\dd y >0$ and one integrates a positive contribution to the mean number of $e$-folds. Conversely, when $x>\bar{\phi}$, one has $\dd x\dd y <0$ and one integrates a negative contribution. This is necessary in order for the overall integral to vanish. This is why $\bar{\phi}$ must lie between $\phi_1$ and $\phi_2$.}
\label{fig:NmeanIntegrationDomain}
\end{center}
\end{figure}
Let us now turn to the calculation of the mean number of \efolds~$\langle\mathcal{N}\rangle$. By making use of the results of \Sec{sec:FirstPassageTime}, more precisely, by plugging \Eq{eq:Fokker:Planck:adjoint:PDF:Ito:SlowRoll} into \Eq{eq:FPT:moment:adjoint:FP} for $n=1$, one obtains $\langle \N \rangle '' - v' \langle \N \rangle '/v^2 = -1/(v\Mp^2)$, which can be solved according to\footnote{This is again in agreement with Eq. (35) of~\Refa{Starobinsky:1986fx} if $H$ is constant and $\phi_*=\bar{\phi}=0$, while $\phi_\uend =\infty$.}
\bea
\langle \N \rangle \left(\phi\right)=\int^{\phi}_{\phi_1}\frac{\dd x}{\Mp}\int^{\bar{\phi}\left(\phi_1,\phi_2\right)}_{x}\frac{\dd y}{\Mp}\frac{1}{v\left(y\right)}\exp \left[\frac{1}{v\left(y\right)}-\frac{1}{v\left(x\right)}\right]\, .
\label{eq:f:sol}
\eea
In this expression, one integration constant has been set such an absorbing boundary condition is placed at $\phi_1$, $\langle \N \rangle (\phi_1)=0$, and $\bar{\phi}$ is an integration constant set to satisfy the boundary condition at $\phi_2$. If a reflective boundary is placed at $\phi_2$, $\langle \N \rangle'(\phi_2)=0$, one simply takes $\bar{\phi}=\phi_2$. If an absorbing boundary is placed at $\phi_2$, $\langle \N \rangle(\phi_2)=0$, there is no generic expression for it $\bar{\phi}$,\footnote{Alternatively, one can write \Eq{eq:f:sol} in the explicit form~\cite{Starobinsky:1986fx}
\begin{align*}
\langle \N \rangle\left(\varphi\right)=
\int_{\phi_1}^{\phi_2} \frac{\dd y}{\Mp} \int_y^{\phi_2} \frac{\dd x}{\Mp}\frac{1}{v(y)}\exp\left[\frac{1}{v(y)}- \frac{1}{v(x)}\right]\left[\theta(x-x_*)
- p_1\right]\, ,
\end{align*}
where $p_1$ is given by \Eq{stocha:hittproba:p1} and, in the configuration of \Fig{fig:sketch:pot:stochaDeltaN}, $\theta(x-x_*)=1$ when $x>x_*$ and $0$ otherwise.} but one can be more specific. First of all, as can be seen in \Fig{fig:NmeanIntegrationDomain}, $\bar{\phi}$ must be such that, when $\langle \N \rangle$ is evaluated at $\phi_2$, the integration domain of \Eq{eq:f:sol} possesses a positive part and a negative part, which are able to compensate for each other. This implies that $\bar{\phi}$ must lie between $\phi_1$ and $\phi_2$. A second generic condition comes from splitting the $x$-integral in \Eq{eq:f:sol} into $\int_{\phi_1}^{\phi}\dd x = \int_{\phi_1}^{\phi_2} \dd x+ \int_{\phi_2}^{\phi}\dd x$. The first integral vanishes because $\langle \N \rangle(\phi_2)=0$, which means that in order for $\langle N \rangle$ to be symmetrical in $\phi_1\leftrightarrow \phi_2$, $\bar{\phi}(\phi_1,\phi_2)$ must satisfy this symmetry too, that is to say, $\bar{\phi}\left(\phi_1,\phi_2\right)=\bar{\phi}\left(\phi_2,\phi_1\right)$. Third, in the case where the potential is symmetric about a local maximum $\phi_\umax$ close to which inflation proceeds, the integrand in \Eq{eq:f:sol} is symmetric with respect to the first bisector in \Fig{fig:NmeanIntegrationDomain}. The two green triangles must therefore have the same surface, which readily leads to $\bar{\phi}=\phi_\umax$. Fourth, finally, in the case displayed in the right panel of \Fig{fig:sketch:pot:stochaDeltaN}, if $\phi_2$ is sufficiently large, we have established in section~\ref{sec:WallHittProba} that $p_2\simeq 0$ and the quantity we compute is essentially the mean number of \efolds~between $\phi_*$ and $\phi_1=\phi_\uend$. For explicitness, let us assume that $v^\prime>0$ (the same line of arguments applies in the case $v^\prime <0$). Inflation proceeds at $\phi<\phi_2$. In the domain of negative contribution in \Fig{fig:NmeanIntegrationDomain}, the argument of the exponential in \Eq{eq:f:sol} is positive. As a consequence, if $\bar{\phi}$ is finite and $\phi_2\rightarrow \infty$, the negative contribution to the integral is infinite while the positive one remains finite, which is impossible. In order to avoid this, one must then have $\bar{\phi} = \phi_2$. In practice, almost all cases boil down to one of the two previous ones and $\bar{\phi}$ is specified accordingly. 

\begin{figure}[t]
\begin{center}
\includegraphics[width=0.474\textwidth]{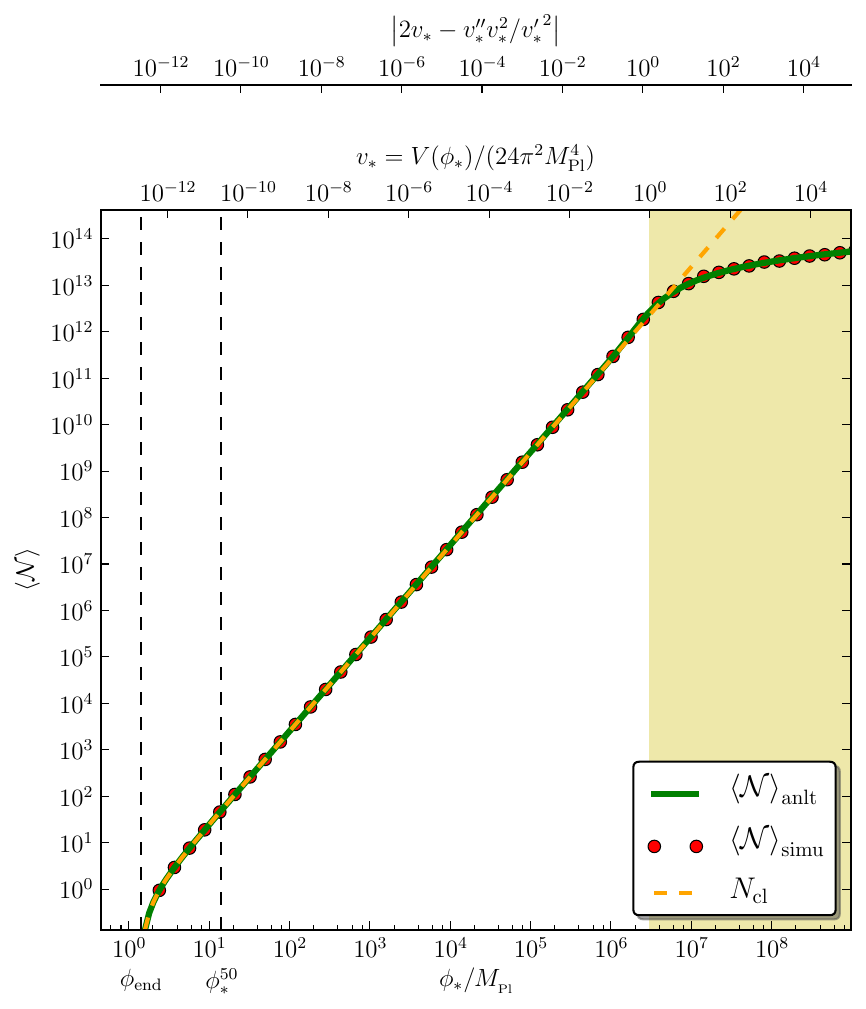}
\includegraphics[width=0.509\textwidth]{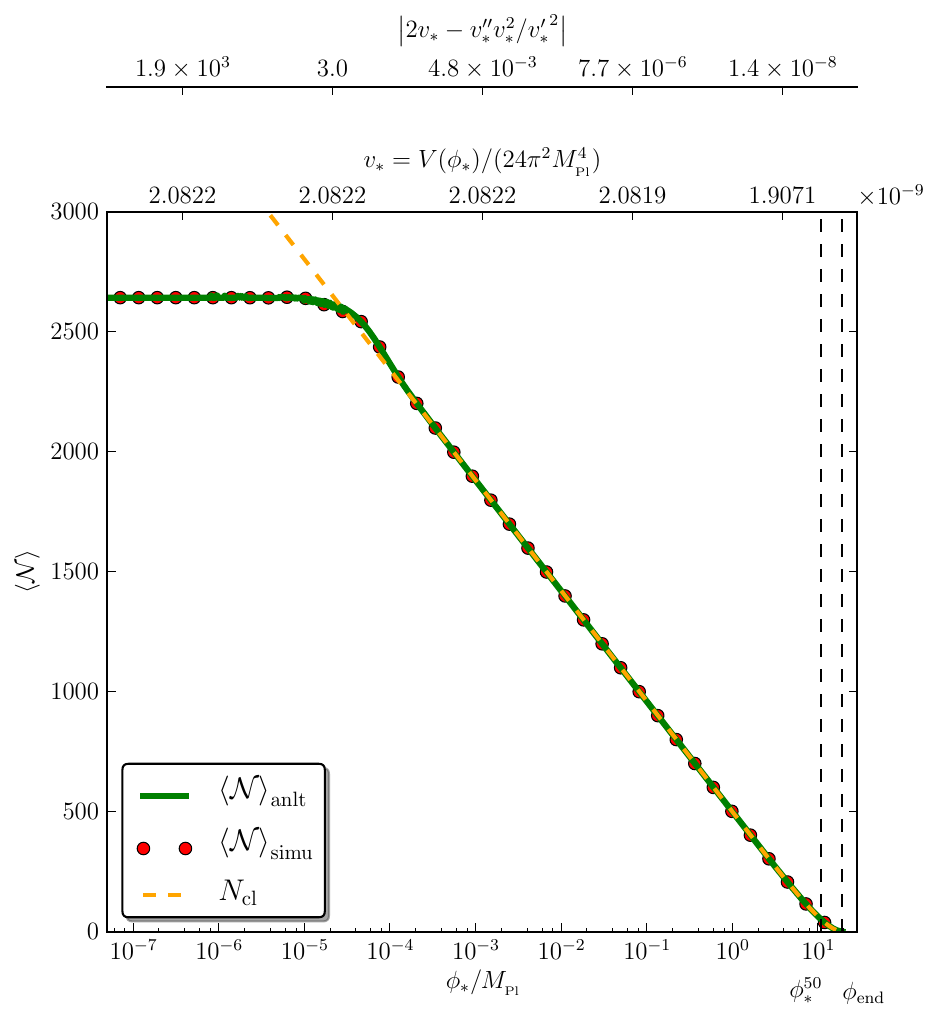}
\caption[mean number of \efolds]{Mean number of \efolds~$\langle \mathcal{N}\rangle(\phi_*) $ realised in the large field $V\propto\phi^2$ (left panel) and small field $V\propto 1-\phi^2/\mu^2$ (where $\mu=20\Mp$, right panel) potentials, as a function of the initial field value $\phi_*$. The label $\phi_*^{50}$ refers to the value of $\phi_*$ for which the classical number of \efolds~$N_\ucl=50$ and $\phi_\uend$ is where $\epsilon_1=1$. In both panels, the overall mass scale in the potential is set to the value that fits the observed amplitude of the power spectrum $\calP_\zeta\sim 2\times 10^{-9}$ when calculated $50$~\efolds before the end of inflation. 
The green line corresponds to the analytical exact result~(\ref{eq:f:sol}), and the red circles are provided by a numerical integration of the Langevin equation~(\ref{eq:Langevin:SR:phi}) for a large number of realisations over which the mean value of $\mathcal{N}$ is computed. The orange dashed line corresponds to the classical limit~(\ref{eq:stocha:meanN:classtraj}). The top axes display $v$ and the classicality criterion $\vert 2v-v^{\prime\prime}v^2/{v^\prime}^2\vert$. The yellow shaded area stands for $v > 1$, where the potential energy density becomes super-Planckian and our calculation cannot be trusted anymore.}
\label{fig:Nmean}
\end{center}
\end{figure}

The mean number of \efolds~\eqref{eq:f:sol} is plotted for large and small field potentials in \Fig{fig:Nmean}, where it is compared with the results of a numerical integration of the Langevin equation~(\ref{eq:Langevin:SR:phi}) for a large number of realisations over which the mean value of $\mathcal{N}$ is computed. One can check that the agreement is excellent.
\paragraph{Classical Limit}
$ $\\
Let us now verify that the above formula~\eqref{eq:f:sol} boils down to the classical, standard result in some ``classical limit'', of which we are also going to determine the regime of applicability. This can be done by performing a saddle-point expansion of the integrals appearing in \Eq{eq:f:sol}. Let us first work out the $y$-integral, that is to say, $\int^{\bar{\phi}}_x\dd y/v(y)\exp[1/v(y)]$. Since the integrand varies exponentially with the potential, the strategy is to evaluate it close to its maximum, \ie where the potential is minimum. The potential being maximal at $\bar{\phi}$ in most cases (see the discussion above), the integrand is clearly maximal\footnote{Strictly speaking, this is only true if the potential is a monotonous function of the field, but this is most often the case in the part of the potential that is relevant to the inflationary phase.} at $x$. Taylor expanding $1/v$ at first order around $x$, $1/v(y)\simeq 1/v(x)-v^\prime(x)/v^2(x)(y-x)$, one obtains, after integrating by parts,\footnote{Since $v(\bar{\phi})\gg v(x)$ and if $v$ is monotonous, one can also show that $\exp\left[-v^\prime(x)/v^2(x)(\bar{\phi}-x)\right]$ is exponentially vanishing and this term can be neglected.} $\int^{\bar{\phi}}_x\dd y/v(y)\exp\left[1/v(y)\right]\simeq v(x)/v^\prime(x)\exp\left[1/v(x)\right]$. Plugging back this expression into \Eq{eq:f:sol}, one finally obtains
\beq
\left.\left\langle\mathcal{N}\right\rangle\right\vert_\ucl (\phi)= \int_{\phi_\uend}^{\phi}\frac{\dd x}{\Mp^2}\frac{v(x)}{v^\prime(x)}\, ,
\label{eq:stocha:meanN:classtraj}
\eeq
which exactly matches the classical result, \ie the deterministic number of \efolds~one obtains by setting the stochastic noise to zero in \Eq{eq:Langevin:SR:phi}. The classical trajectory thus appears as a saddle-point limit of the mean stochastic trajectory, analogously to what happens \eg in the context of path integral calculations.

This calculation also allows us to identify under which conditions the classical limit is recovered. A priori, the Taylor expansion of $1/v$ can be trusted as long as the difference between $1/v(x)$ and $1/v(y)$ is not too large, say $\vert 1/v(y)-1/v(x)\vert < R$, where $R$ is some small number. If one uses the Taylor expansion of $1/v$ at first order, this means that $\vert y-x\vert<Rv^2/v'$. Requiring that the second order term of the Taylor expansion is small at the boundary of this domain yields the condition $\vert 2v-v^{\prime\prime}v^2/{v^\prime}^2\vert \ll 1$. For this reason, we define the classicality criterion
\beq
\eta_\ucl = \left\vert 2v-\frac{v^{\prime\prime}v^2}{{v^\prime}^2}\right\vert\, .
\label{eq:classicalcriterion:def}
\eeq
This quantity is displayed in the top axes in \Fig{fig:Nmean} and one can check that indeed, the classical trajectory is a good approximation to the mean stochastic one if and only if $\eta_\ucl\ll 1$. In the following, we will see that $\eta_\ucl$ is the relevant quantity to discuss the strength of the stochastic effects in general. Notice that, since $v$ measures the potential energy density in Planckian units, it has to be small for the entire formalism to be valid, hence the second term in \Eq{eq:classicalcriterion:def} is the most relevant one. 

For now, and for future use, let us give the first correction to the classical trajectory. This can be obtained going one order higher in the saddle-point approximation, that is to say, using a Taylor expansion of $1/v$ at second order. One obtains
\beq
\left.\left\langle \mathcal{N} \right\rangle\right\vert_{\eta_\ucl\ll 1} (\phi)\simeq  \int_{\phi_\uend}^{\phi}\frac{\dd x}{\Mp^2}\frac{v(x)}{v^\prime(x)}\left[1+v\left(x\right)-\frac{v^{\prime\prime}\left(x\right) v^2\left(x\right)}{{v^{\prime}}^2\left(x\right)}+\cdots\right]\, ,
\label{eq:Nmean:vll1limit}
\eeq
where the dots stand for higher order terms. In the brackets of \Eq{eq:Nmean:vll1limit}, the two last terms stand for the first stochastic correction and one should not be surprised that, when $\eta_\ucl$ is small, it is small. It is also interesting to notice that it is directly proportional to $\dd\epsilon_1/\dd N$. When $\epsilon_1$ increases as inflation proceeds, the stochastic leading correction is therefore positive and the stochastic effects tend to increase the realised number of $e$-folds, while when $\epsilon_1$ decreases as inflation proceeds, the correction is negative and the stochastic effects tend to decrease the number of $e$-folds, at least in the perturbative regime.
\subsubsection[Mean number of \efolds~squared and power spectrum]{Mean number of \texorpdfstring{$\bm{e}$}{$e$}-folds squared and power spectrum}
\label{sec:MeanNefDisp}
\begin{figure}[t]
\begin{center}
\includegraphics[width=0.474\textwidth]{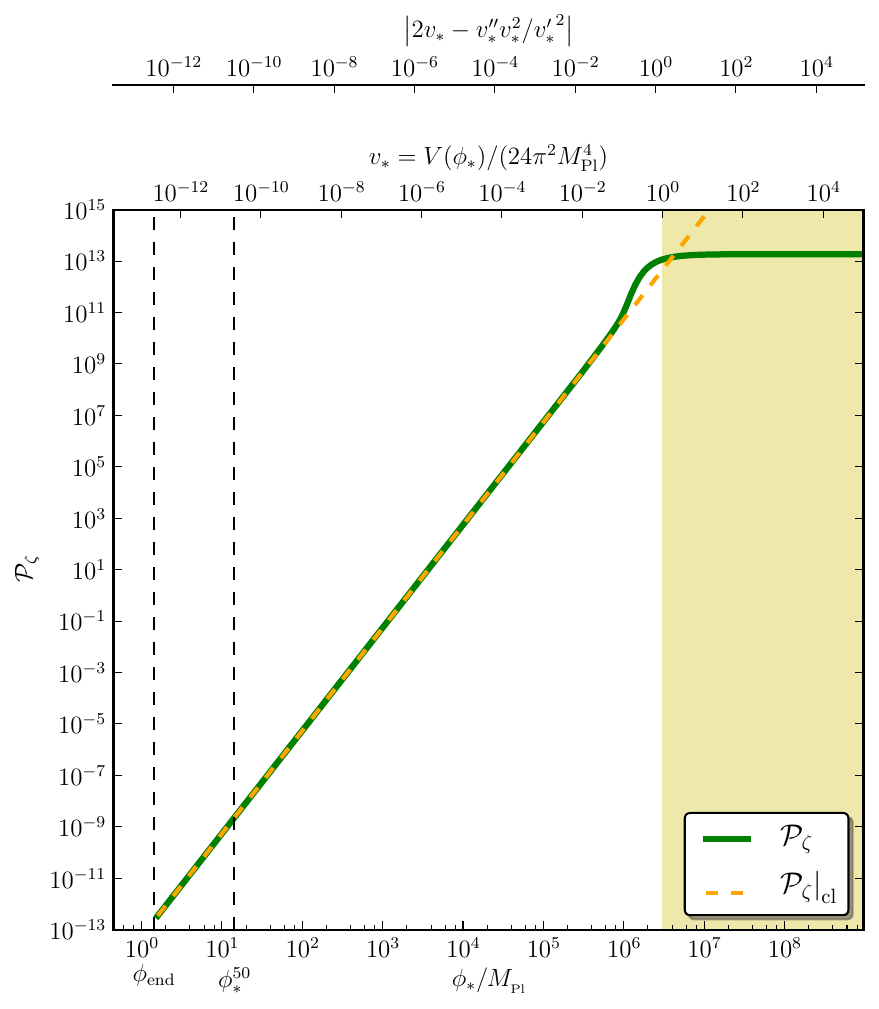}
\includegraphics[width=0.509\textwidth]{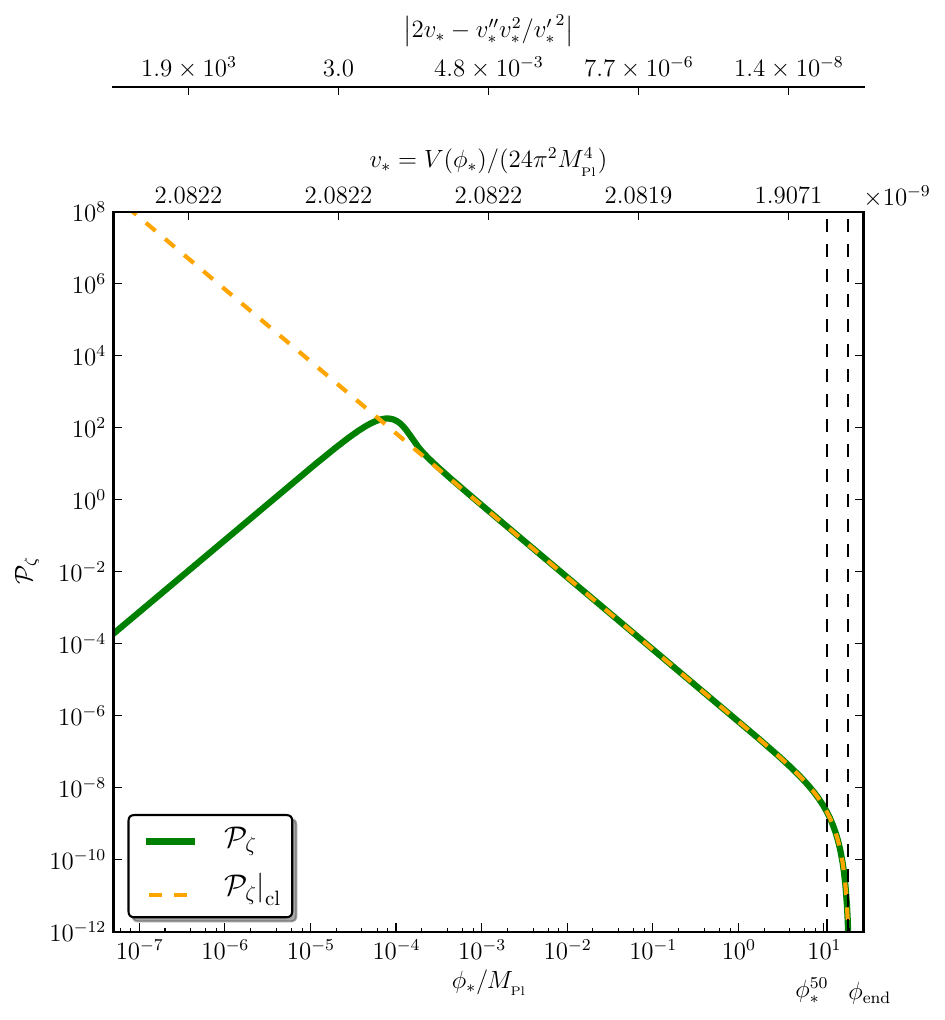}
\caption[Power spectrum]{Scalar power spectrum $\calP_\zeta (\phi_*)$ for  the large field $V\propto\phi^2$ (left panel) and small field $V\propto 1-\phi^2/\mu^2$ (where $\mu=20\Mp$, right panel) potentials, as a function of $\phi_*$. The conventions are the same as in \Fig{fig:Nmean}. The green line corresponds to the analytical exact result~(\ref{eq:PS:fullstocha}), and the orange dashed line to the classical limit $\left.\calP_\zeta\right\vert_{\ucl}$ given in \Eq{eq:PS:vll1}.}
\label{fig:Pzeta}
\end{center}
\end{figure}
Let us now turn to the calculation of the mean number of \efolds~squared, $\langle\mathcal{N}^2\rangle$. By making use of the results of \Sec{sec:FirstPassageTime}, more precisely, by plugging \Eq{eq:Fokker:Planck:adjoint:PDF:Ito:SlowRoll} into \Eq{eq:FPT:moment:adjoint:FP} for $n=2$, one obtains $\langle \N^2 \rangle '' - v' \langle \N^2 \rangle '/v^2 = -\langle \N \rangle/(v\Mp^2)$, which can be solved according to
\begin{align}
\label{eq:mean:N2}
\langle \N^2 \rangle (\phi)=2 \int_{\phi_1}^\phi\frac{\dd x}{\Mp}\int^{{\bar{\phi}}_2}_x\frac{\dd y}{\Mp}\frac{\langle \N \rangle (y)}{v(y)}\exp\left[\frac{1}{v(y)}-\frac{1}{v(x)}\right]\, ,
\end{align}
where the function $\langle \N \rangle$ has been calculated in \Eq{eq:f:sol}. Here, $\bar{\phi}_2\left(\phi_1,\phi_2\right)$ is again an integration constant that must be set in order to satisfy the boundary condition at $\phi_2$, and which shares the same properties as those of $\bar{\phi}_2\left(\phi_1,\phi_2\right)$ that we have discussed in section~\ref{sec:meanN}. 

The power spectrum can then be obtained following the considerations of \Sec{sec:stochaDeltaN:stochaDeltaN}.
The dispersion in the number of \efolds, $\left\langle \delta N_{\mathrm{cg}}^2 \right\rangle$, is simply given by $\langle \N^2\rangle - \langle \N\rangle^2$, so \Eq{eq:Pzeta:stochaDeltaN} gives rise to
\bea
\label{eq:PS:fullstocha:moments}
\calP_\zeta(\phi) &=& \frac{\langle \N^2\rangle'(\phi) - 2 \langle \N\rangle'(\phi)\langle \N\rangle(\phi)}{\langle \N\rangle'(\phi)}\\
&=&2
\left\lbrace \int_{\phi}^{\bar{\phi}}\frac{\dd x}{\Mp}\frac{1}{v\left(x\right)}\exp\left[\frac{1}{v\left(x\right)}-\frac{1}{v\left(\phi_*\right)}\right] \right\rbrace^{-1}
\times\nonumber\\ & &
\int_{\phi}^{\bar{\phi}_2}\frac{\dd x}{\Mp}\left\lbrace\int_{x}^{\bar{\phi}} \frac{\dd y}{\Mp} \frac{1}{v\left(y\right)}\exp\left[\frac{1}{v\left(y\right)}-\frac{1}{v\left(x\right)}\right] \right\rbrace^2\exp\left[\frac{1}{v\left(x\right)}-\frac{1}{v\left(\phi_*\right)}\right]\, .
\label{eq:PS:fullstocha}
\eea
where \Eq{eq:f:sol} and~\eqref{eq:mean:N2} have been used. This provides the power spectrum calculated at a scale $k$ in a patch where, when it crosses out the Hubble radius, the inflaton field value is $\phi$. This formula is plotted for large and small field potentials in \Fig{fig:Pzeta}.

From this, a generic expression for the spectral index can also be given. Since, at leading order in slow roll, $\partial/\partial\ln(k)\simeq-\partial\phi/\partial\langle\mathcal{N}\rangle\times\partial/\partial\phi$, one has
\beq
\nS\simeq1-\frac{\calP_\zeta '}{\langle \N \rangle ' \calP_\zeta}\, .
\label{eq:stocha:ns:exact}
\eeq
Here, for conciseness, we do not expand this expression in terms of integrals of the potential, but it is straightforward to do so with \Eqs{eq:f:sol} and~(\ref{eq:PS:fullstocha}).
\paragraph{Classical Limit}
$ $\\
As was done for the mean number of \efolds~in section~\ref{sec:meanN}, let us derive the classical limit of \Eq{eq:PS:fullstocha}. Obviously, in the classical setup the trajectories are not stochastic and $\delta\mathcal{N}^2=0$, and what we are interested in here is the non-vanishing leading order contribution to $\delta\mathcal{N}^2$ in the limit $\eta_\ucl\ll 1$.  As before, the $y$-integral can be worked out with a saddle-point approximation, and also making use of \Eq{eq:Nmean:vll1limit}, one is led to
\bea
\left. \delta\mathcal{N}^2 \right\vert_{\eta_\ucl\ll 1}\simeq  \frac{2}{\Mp^4}\int_{\phi_1}^{\phi}\dd x\frac{v^4\left(x\right)}{{v^\prime}^3\left(x\right)}\left[1+6v\left(x\right)-5\frac{v^2\left(x\right)v^{\prime\prime}\left(x\right)}{{v^{\prime}}^2\left(x\right)}+\cdots\right]\, ,
\label{eq:deltaN2:classlim}
\eea
which gives rise to  
\bea
\left.\mathcal{P}_\zeta\right\vert_{\eta_\ucl\ll 1}\left(\phi_*\right)\simeq
\underbrace{
\frac{2}{\Mp^2}\frac{v^3\left(\phi_*\right)}{{v^\prime}^2\left(\phi_*\right)}}_{\left.\calP_\zeta\right\vert_{\ucl}}
\left(\phi_*\right)\left[1+5v\left(\phi_*\right)-4\frac{v^2\left(\phi_*\right)v^{\prime\prime}\left(\phi_*\right)}{{v^\prime}^2\left(\phi_*\right)}\right]\, ,
\label{eq:PS:vll1}
\eea
where the leading-order result exactly matches the classical formula at leading order in slow roll.\footnote{This agreement was shown, by means of other techniques, in the specific case where the Hubble parameter varies linearly with $\phi$, and for a noise with constant amplitude, in \Refa{Fujita:2013cna}.} One can see that the stochastic correction is small precisely when the classical criterion introduced in \Eq{eq:classicalcriterion:def} also is. 
For the spectral index, one gets
\bea
\left.\nS\right\vert_{\eta_\ucl\ll 1}\left(\phi_*\right)\simeq\left.\nS\right\vert_{\ucl}\left(\phi_*\right)+\Mp^2\left[3v^{\prime\prime}\left(\phi_*\right)-2\frac{{v^\prime}^2\left(\phi_*\right)}{v\left(\phi_*\right)}-6\frac{{v^{\prime\prime}}^2\left(\phi_*\right)v\left(\phi_*\right)}{{v^\prime}^2\left(\phi_*\right)}+4\frac{v\left(\phi_*\right)v^{\prime\prime\prime}\left(\phi_*\right)}{v^\prime\left(\phi_*\right)}\right]\, .
\label{eq:ns:vll1}
\eea
\subsubsection{Higher moments and non-Gaussianities}
\label{sec:Higher:Moments:NG}
Higher moments of the first passage time can be worked out in a similar way. Indeed, the combination of \Eqs{eq:Fokker:Planck:adjoint:PDF:Ito:SlowRoll} and~\eqref{eq:FPT:moment:adjoint:FP} can be solved according to
\bea
\label{eq:fn:sol:onefield}
\langle \N^n \rangle(\phi)=n \int_{{\phi}_1}^\phi\frac{\dd x}{\Mp}\int^{{\bar{\phi}}_n}_x\frac{\dd y}{\Mp}\frac{1}{v(y)}\exp\left[\frac{1}{v(y)}-\frac{1}{v(x)}\right]\langle \N^{n-1}\rangle(y)\, .
\eea
Interestingly, the kernel of the integral over $y$, against which the moment of order $n-1$ is integrated, is nothing but the stationary distribution~\eqref{eq:Pstat}. For instance, for $n=3$, the cubic moment can be derived, and making use of \Eq{eq:fnl:stochaDeltaN}, the local non-Gaussianity parameter $\fnl$ can be obtained too. We do not display the full resulting formula since it is not particularly instructive. However, one can check that a saddle-point expansion gives rise to
\bea
\left.\delta\mathcal{N}^3\right\vert_{\eta_\ucl\ll 1}\simeq
\frac{12}{\Mp^6}\int_{\phi_1}^{\phi}\dd x\frac{v^7}{{v^\prime}^5}\left(1+14v -11\frac{v^2v^{\prime\prime}}{{v^\prime}^2}+\cdots\right)\, ,
\label{eq:skewness:class}
\eea
and then
\beq
\left.\fnl\right\vert_{\eta_\ucl\ll 1}=\underbrace{ \frac{5}{24}\Mp^2\Big[6\frac{{v^\prime}^2}{v^2}-4\frac{v^{\prime\prime}}{v}}_{\left.\fnl\right\vert_{\ucl}}+v\left(25\frac{{v^{\prime}}^2}{v^2}-34\frac{v^{\prime\prime}}{v}-10\frac{v^{\prime\prime\prime}}{v^\prime}+24\frac{{v^{\prime\prime}}^2}{{v^\prime}^2}\right)+\mathcal{O}\left(v^2\right)\Big]\ .
\label{eq:fnl:classappr}
\eeq
The first two terms in the brackets match the usual result~\cite{Maldacena:2002vr}. In contrast, it is important to stress that within the usual $\delta N$ formalism, the standard result cannot be obtained because of the intrinsic non-Gaussianity of the fields at Hubble exit~\cite{Maldacena:2002vr, Allen:2005ye}. Such effects are automatically taken into account in our formalism, which readily gives rise to the correct formula. \\

\par

An important consequence of \Eqs{eq:PS:fullstocha}, and \Eq{eq:fn:sol:onefield} in general, is the correctness of their classical limits. They show the validity of our computational programme for calculating correlation functions in general. Let us mention that within the CMB observable window, corrections to the classical results are always small, since one has 
\beq
\eta_\ucl\simeq \calP_\zeta\left(\epsilon_1+\frac{\epsilon_2}{4}\right)\, .
\eeq
More precisely, \Eqs{eq:PS:vll1} and~(\ref{eq:fnl:classappr}) can be recast as $\left.\mathcal{P}_\zeta\right\vert_{\eta_\ucl\ll 1}\simeq  \left.\calP_\zeta\right\vert_{\ucl}\left[1+\left.\calP_\zeta\right\vert_{\ucl}(\epsilon_1+\epsilon_2)\right]$ and $\left.\fnl\right\vert_{\eta_\ucl\ll 1}\simeq  \left.\fnl\right\vert_{\ucl}- \frac{5}{12}\left.\mathcal{P}_\zeta\right\vert_{\ucl}\left(38\epsilon_1^2+\frac{51}{4}\epsilon_2\epsilon_1+\frac{9}{8}\epsilon_2\epsilon_3-\frac{59}{8}\epsilon_2^2\right)$. For the scales of astrophysical interest today, in standard single-field slow-roll inflation, these corrections are therefore tiny.

However, even if the stochastic effects within the CMB observable window need to be small, let us stress that the location of the observable window along the inflationary potential can be largely affected by stochastic effects. This notably happens when the potential has a flat region between the location where the observed modes exit the Hubble radius and the end of inflation, as is the case \eg in hybrid inflation~\cite{Martin:2011ib} or in potentials with flat inflection points.

Another point to note is that, contrary to what one may have expected, the corrections we obtained are not controlled by the ratio $\Delta\phi_{\mathrm{qu}}/\Delta\phi_\ucl$ extensively used in the literature, where $\Delta\phi_{\mathrm{qu}} = H/(2\pi)$ is the mean quantum kick received over one \efold~and $\Delta\phi_\ucl = V'/(3H^2)$ is the classical drift over the same period, but rather by the classicality criterion $\eta_\ucl$ derived in \Eq{eq:classicalcriterion:def}. This has two main consequences. 

First, $\eta_\ucl$ has dimension $v$, which means that it is Planck suppressed.\footnote{This remark also sheds some new light on the old debate~\cite{Linde:2005ht, Smolin:1979ca, Bardeen:1983st} whether quantum gravitational corrections should affect inflationary predictions through powers of $\phi/\Mp$ or $V/\Mp^4$. This analysis reveals $V/\Mp^4$ corrections only, regardless of the value of $\phi/\Mp$.} This makes sense, since some of the corrections we obtained physically correspond to the self- and gravitational interactions of the inflaton field.\footnote{For this reason, one may think that performing the calculation in Fourier space as we did does not allow us to properly account for self-interaction effects and that a real space calculation should be carried out instead. However, since the stochastic inflation formalism relies on the separate universe approximation on large scales, this is not the case. Making use of the same formalism as in \Refa{Starobinsky:1994bd}, we have indeed explicitly checked that performing the calculation in real space leads to  the same results as the ones presented here.} This is why it can be useful to compare our results with loop calculations performed in the literature by means of other techniques. In particular, the self-loop correction to the power spectrum is derived in \Refa{Seery:2007we}, and graviton loop corrections are obtained in \Refa{Dimastrogiovanni:2008af} (for a nice review, see also \Refa{Seery:2010kh}). A diagrammatic approach based on the $\delta N$ formalism is also presented in \Refa{Byrnes:2007tm} where the power spectrum and the bispectrum are calculated up to two loops. In all these cases, the obtained corrections are of the form $\calP_\zeta^{1\mathrm{-loop}}=\calP_\zeta^\mathrm{tree}(1+\alpha\calP_\zeta^\mathrm{tree}\epsilon^2 N)$. Here, $\alpha$ is a numerical factor of order one that depends on the kind of loops one considers, and $\epsilon^2$ stands for second order combinations of slow-roll parameters. When the number of \efolds~$N$ is of the order $1/\epsilon$, this is exactly the kind of leading corrections we obtained. This feature is therefore somewhat generic. Obviously, it remains to understand which loops exactly our approach allows one to calculate, and how our results relate to the above mentioned ones. 
\begin{table}[t]
\begin{center}
\begin{tabular}{{|l||c|c|}}
\hline
\textbf{Potential type} & $\boldsymbol{v(\phi)}$ & $\boldsymbol{\eta_\ucl}$\\
\hline
Large field & $\propto\phi^p$ & $\left(1+\frac{1}{p}\right)v$ \\
\hline
Hilltop & $ v_0\left[1-\left(\frac{\phi}{\mu}\right)^p\right]$ & $\frac{v_0}{p}\left(\frac{\mu}{\phi}\right)^p$ \\
\hline
Polynomial plateau & $ v_\infty\left[1-\left(\frac{\phi}{\mu}\right)^{-p}\right]$ & $\frac{v_\infty}{p}\left(\frac{\phi}{\mu}\right)^p$\\
\hline
Exponential plateau & $ v_\infty\left[1-\alpha\exp\left(-\frac{\phi}{\mu}\right)\right]$ & $\frac{v_\infty}{\alpha}\exp\left(\frac{\phi}{\mu}\right)$ \\
\hline
Inflection point & $v_0\frac{n\left(n-1\right)}{\left(n-1\right)^2}\left[\left(\frac{\phi}{\phi_0}\right)^2-\frac{4}{n}\left(\frac{\phi}{\phi_0}\right)^n+\frac{1}{n-1}\left(\frac{\phi}{\phi_0}\right)^{2n-2}\right]$ & $\frac{v_0}{n\left(n-1\right)}\left\vert\frac{\phi}{\phi_0}-1\right\vert^{-3}$\\
\hline
\end{tabular}
\end{center}
\caption[Classicality Criterion for a few Potentials] {Classicality criterion $\eta_\ucl$ defined in \Eq{eq:classicalcriterion:def} for a few types of inflationary potentials. Except for ``large field'', the expression given for $\eta_\ucl$ is valid close to the flat point of the potential.} \label{tab:etaclass} 
\end{table}

Second, $\eta_\ucl$ contains $1/v'^2$ terms. This means that, even if $v$ needs to be very small,\footnote{Since $v$ can only decrease during inflation, the CMB power spectrum amplitude measurement, and the upper bound on the tensor-to-scalar ratio, imply that $v<10^{-10}$ for all observable modes.} if the potential is sufficiently flat, $\eta_\ucl$ may be large. In table~\ref{tab:etaclass}, we have summarised the shape of $\eta_\ucl$ for different prototypical inflationary potentials. For large field potentials, $\eta_\ucl$ is directly proportional to $v$. This is why, in the left panels of \Figs{fig:Nmean} and~\ref{fig:Pzeta}, departure of the stochastic results from the standard formulas occur only when $v\gg 1$, in a regime where our calculation cannot be trusted anyway. However, for potentials with flat points, different results are obtained. If the flat point is of the hilltop type, $\eta_\ucl$ diverges at the maximum of the potential. This is why, in the right panels of \Figs{fig:Nmean} and~\ref{fig:Pzeta}, even if $v$ saturates to a small maximal value, the stochastic result differs from the classical one close to the maximum of the potential. However, in most models, this happens many \efolds~before the scales probed in the CMB cross out the Hubble radius, that is to say, at extremely large, non-observable scales. The same conclusion holds for plateau potentials (either of the polynomial or exponential type) where stochastic effects lead to non-trivial modifications in far, non observable regions of the plateau. On the other hand, if the potential has a flat inflection point, $\eta_\ucl$ can be large at intermediate wavelengths, too small to lie in the CMB observable window but still of astrophysical interest. This could have important consequences in possible non-linear effects at those small scales, such as the formation of primordial black holes (PBHs). In such models, the production of PBHs is calculated making use of the standard classical formulas for the amount of scalar perturbations. However, we have shown that in such regimes, stochastic effects largely modify its value. An important question is therefore how this changes the production of PBHs in these models. In particular, it is interesting to notice that if the potential is concave ($v^{\prime\prime} < 0$), which is the case favoured by observations~\cite{Martin:2013tda, Martin:2013nzq}, the leading correction in \Eq{eq:PS:vll1} is an enhancement of the power spectrum amplitude. However, as can be seen in the right panel of \Fig{fig:Pzeta}, as soon as one leaves the perturbative regime, this can be replaced by the opposite trend: at the flat point, the classical result accounts for a diverging power spectrum while the stochastic effects make it finite. Moreover, in stochastic dominated regimes, the PDF of curvature perturbations is not Gaussian anymore, and a discussion in terms of the power spectrum only may not be sufficient. This will be the topic of \Sec{sec:Infinite:Inflation}.
\subsection{Infinite inflation}
\label{sec:Infinite:Inflation}
Let us consider again the situation depicted in the right panel of \Fig{fig:sketch:pot:stochaDeltaN}, and let us examine more carefully the role played by the UV boundary condition at $\phi_2$. In \Sec{sec:WallHittProba}, it was shown that the probability to explore the region around $\phi_2$ asymptotes zero when $\phi_2$ is sent to infinity. This suggests that, when $\phi_2$ is sufficiently large, its precise value does not matter (since the probability that a given realisation of the Langevin equation bounces against, or is absorbed by, this boundary condition becomes negligible), and that one can safely set $\phi_2=\infty$ and not worry about imposing a field UV cutoff. As we will see in this section, this is only almost true.

As a first indication of why this issue is a priori more problematic that it seems, let us consider \Eq{eq:f:sol}. If $\phi_2$ is sent to infinite values, as argued in \Sec{sec:meanN}, regardless of the nature of the boundary condition at $\phi_2$, the integration constant $\bar{\phi}$ becomes infinite too. If the potential energy asymptotes a constant at large-field values, one can see that the integral over $y$ diverges. In fact, even if the potential goes to infinity at large-field values, it has to do so at a sufficiently quick rate, namely faster than $v\propto \phi$, in order for the integral over $y$ to converge. Otherwise, the mean value of $\N$ is infinite, a phenomenon that we dub ``infinite inflation'', and the UV cutoff $\phi_2$ has to remain finite [the same then applies to all higher-order moments, since they all involve the same integration kernel, see \Eq{eq:fn:sol:onefield}]. This might seem to point out that the result does depend on the precise value of the cutoff field value in those cases, which would be an undesired feature, but we will see that, in fact, fortunately, it does not.  

Since the appearance of the phenomenon of infinite inflation strongly depends on the number of fields being present, the discussion will be carried out in the context of multiple-field inflation. Opening up the number of dimensions of field space can lead to very non-trivial effects, and usually makes the analysis more complicated. For instance, in a stationary distribution $P_\mathrm{stat}$, by definition, the divergence of the probability current $\bm{J}$ vanishes, see \Eq{eq:Fokker:Planck:current}, corresponding to incompressible flows. If only one field is present, this means that $J$ is uniform in field space, and that, in most interesting situations, the probability current itself vanishes. For example, if field space is unbounded, the normalisation condition $\int P_\mathrm{stat}\dd\phi = 1$ requires that $P_\mathrm{stat}$ decreases at infinity strictly faster than $\vert \phi \vert^{-1}$. In this case, both $P_\mathrm{stat}(\phi)$ and $\partial P_\mathrm{stat}(\phi)/\partial\phi$ vanish at infinity, hence everywhere, and this gives \Eq{eq:Pstat}. If more than one field is present, one can already see that the analysis is much less trivial. The distribution~(\ref{eq:Pstat}) is still \emph{a} solution of the stationarity problem, but non-uniform probability currents are also allowed (since only their divergence must vanish), yielding other solutions~\cite{Hardwick:2018sck}. 

\subsubsection{Harmonic Potentials}
\label{sec:harmonic}
For a fully generic multi-field potential, \Eqs{eq:FPT:moment:adjoint:FP} have no analytical solutions and one needs to resort to numerical analysis. Alternatively, in this section we identify a subclass of inflationary potentials for which \Eqs{eq:FPT:moment:adjoint:FP} can be solved exactly and the effects associated with the inclusion of multiple fields can be studied analytically.
\paragraph{Polar Coordinates}
Let us first note that \Eqs{eq:FPT:moment:adjoint:FP} are diffusion equations, akin to the Laplace equation. This suggests that some insight may be gained by reparameterising field space with polar-type coordinates. Since the slow-roll trajectory follows the gradient of the potential at the classical level [\ie without including the diffusion term in \Eq{eq:Langevin}], a natural choice\footnote{This choice of coordinates is also similar to the adiabatic-entropic decomposition of \Refa{Gordon:2000hv}, if one interprets Eqs.~(31),~(32) and~(35) of this reference by replacing the field derivatives by their classical equations of motion.} is to take the potential $v$ itself for the radial coordinate, completed by $D-1$ angular coordinates $\theta_j$ (with $1 \leq j \leq D-1$).\footnote{Strictly speaking, this procedure is well-defined only if the level lines of $v(\phi_i)$ form simply connected hyper surfaces in field space. This is implicitly assumed in what follows, even if more complicated situations can also be studied, either making use of symmetries in the potential function $v(\phi_i)$ as in hybrid inflation, or paving field space with several maps.} By expanding $\partial/\partial\phi_i = v_{\phi_i} \partial/\partial v + \sum_j (\theta_j)_{\phi_i}\partial/\partial\theta_j$ in the new coordinates system, the adjoint Fokker-Planck operator~\eqref{eq:Fokker:Planck:adjoint:PDF:Ito:SlowRoll} can be written as 
\begin{align}
&
\frac{1}{\Mp^2}\mathcal{L}_\mathrm{FP}^\dagger= 
v \left\vert  \bm{\nabla}(v)\right\vert^2 \frac{\partial^2}{\partial v^2}
+v\sum_{j,\ell = 1}^{D-1} 
\bm{\nabla}(\theta_j)\cdot \bm{\nabla}(\theta_\ell)
\frac{\partial^2}{\partial\theta_j\partial\theta_\ell}
\nonumber\\ 
&+2 v\sum_{j=1}^{D-1}
\bm{\nabla}(\theta_j)\cdot \bm{\nabla}(v)
\frac{\partial^2}{\partial v \partial\theta_j}
+\left[v \Delta v-\frac{1}{v} \left\vert  \bm{\nabla}(v)\right\vert^2\right]\frac{\partial}{\partial v}
+\sum_{j=1}^{D-1}\left[v\Delta\theta_j-\frac{1}{v}\bm{\nabla}(\theta_j)\cdot \bm{\nabla}(v)\right]\frac{\partial}{\partial\theta_j}\, .
\label{eq:Fpop:radial:1}
\end{align}
In this expression, recall that the vectorial notation (and the differential operators $\bm{\nabla}$ and $\Delta=\vert\bm{\nabla}^2\vert$) refer to field space. For example, $\bm{\nabla}(v)=\sum_{i=1}^D v_{\phi_i} \bm{\mathrm{e}}_{\phi_i}$, where $\lbrace \bm{\mathrm{e}}_{\phi_i}\rbrace $ stands for the field space basis. One can always choose the angular variables $\theta_j$ to form a system of orthogonal variables\footnote{For example~\cite{Malik:1998gy}, one can start from $\bm{\nabla}(v)$ and use Gram-Schmidt orthogonalisation procedure to iteratively derive $\bm{\nabla}(\theta_1)$, $\bm{\nabla}(\theta_2)$, etc.} and one obtains
\begin{align}
 \frac{1}{\Mp^2}\mathcal{L}_\mathrm{FP}^\dagger= 
v \left\vert  \bm{\nabla}(v)\right\vert^2 
\left\lbrace
\frac{\partial^2}{\partial v^2}
+\sum_{j = 1}^{D-1} 
\frac{\left\vert  \bm{\nabla}(\theta_j)\right\vert^2}{\left\vert  \bm{\nabla}(v)\right\vert^2}
\frac{\partial^2}{\partial\theta_j^2}
+\left[\frac{\Delta v}{\left\vert  \bm{\nabla}(v)\right\vert^2}-\frac{1}{v^2} \right]\frac{\partial}{\partial v}
+\sum_{j=1}^{D-1}\frac{\Delta\theta_j}{\left\vert  \bm{\nabla}(v)\right\vert^2} \frac{\partial}{\partial\theta_j}
\right\rbrace \, .
\label{eq:Fpop:radial}
\end{align}
\paragraph{Harmonic Potentials}
We now restrict the analysis to potentials for which separable solutions (in the basis $\lbrace v,\,\theta_j\rbrace$) of \Eqs{eq:FPT:moment:adjoint:FP} exist. An important remark is that purely radial (\ie independent of $\theta_j$) solutions of \Eqs{eq:FPT:moment:adjoint:FP} can be found if the coefficient in front of $\partial/\partial v$ is a function of $v$ only. For this reason, we define ``harmonic potentials'' as being such that 
\begin{align}
g\equiv  \frac{\Delta v}{ \left\vert  \bm{\nabla}(v)\right\vert^2}
\label{eq:harmonic:g}
\end{align}
is a function of $v$ only.

In order to understand to which extent harmonic potentials allow one to proceed analytically, let us discuss the case where $D=2$ fields are present. For two-field potentials, one has a single angular variable $\theta$, and the orthogonality condition $\bm{\nabla}(v)\perp \bm{\nabla}(\theta)$ mentioned between \Eqs{eq:Fpop:radial:1} and~(\ref{eq:Fpop:radial}) implies that $\bm{\nabla}(\theta)=h (- v_{\phi_2}\bm{\mathrm{e}}_{\phi_1}+  v_{\phi_1}\bm{\mathrm{e}}_{\phi_2})$, where $h$ is an overall factor that is left unspecified at this stage. Let us simply note that, in order for $\theta$ to be globally defined~\cite{Saffin:2012et}, the curl of $\bm{\nabla}(\theta)$ must vanish, $\theta_{\phi_1\phi_2}=\theta_{\phi_2\phi_1}$, which translates into $ h_{\phi_1} v_{\phi_1} + h_{\phi_2} v_{\phi_2} +h(v_{\phi_1\phi_1}+v_{\phi_2\phi_2})=0$. This is the only condition $h$ needs to satisfy, and for harmonic potentials where $g$ depends on $v$ only, it is interesting to notice that it can be fulfilled if $h$ is taken as depending on $v$ only as well, according to\footnote{Indeed, in this case, one has $h_{\phi_i}=v_{\phi_i}\dd h/\dd v = -g v_{\phi_i} h $ and one can easily check that $ h_{\phi_1} v_{\phi_1} + h_{\phi_2} v_{\phi_2} +h(v_{\phi_1\phi_1}+v_{\phi_2\phi_2})=0$.} $h(v)=\exp[-\int^v g(v')\dd v']$. One can also check that in this case, $\Delta\theta = h_{\phi_2} v_{\phi_1}-h_{\phi_1} v_{\phi_2}=0$. The adjoint Fokker-Planck operator then takes the simple form
\begin{align}
\frac{1}{\Mp^2}\mathcal{L}_\mathrm{FP}^\dagger = &
v \left\vert  \bm{\nabla}(v)\right\vert^2 
\left\lbrace
\frac{\partial^2}{\partial v^2}
+
\exp\left[-2\!\!\int^v\!\!\!g(v')\dd v'\right]
\frac{\partial^2}{\partial\theta^2}
+\left[g(v)-\frac{1}{v^2} \right]\frac{\partial}{\partial v}
\right\rbrace \, .
\label{eq:Fpop:radial:2D}
\end{align}
In this case, up to the overall $\vert  \bm{\nabla}(v)\vert^2$ factor, all coefficients of the adjoint Fokker-Planck operator are explicit functions of the radial coordinate $v$ only, and the problem boils down to solving ordinary differential equations after Fourier transforming the angular coordinate, as will be done explicitly below. Let us first give a few concrete examples of harmonic potentials.
\paragraph{$v(r)$ Potentials}
A subclass of harmonic potentials is provided by potentials $v(r)$ that depend on
\begin{align}
r^2\equiv\sum_i\phi_i^2
\end{align} 
only. Indeed, in this case, one can show that $g=(D-1)/[rv^\prime(r)]+v^{\prime\prime}(r)/{v^\prime}^2(r)$ depends on $r$, hence on $v$, only. In fact, $ \left\vert  \bm{\nabla}(v)\right\vert^2 = {v^\prime}^2(r)$ depends on $v$ only as well. The angular coordinates can be chosen to match the ones of the usual spherical coordinates system in $D$ dimension, and this gives rise to
\begin{align}
\frac{1}{\Mp^2}\mathcal{L}_\mathrm{FP}^\dagger= &
v(r)
\left\lbrace
\frac{\partial^2}{\partial r^2}
+\sum_{j = 1}^{D-1} 
\left[r\displaystyle\prod_{\ell=1}^{j-1}\sin(\theta_\ell)\right]^{-2}
\frac{\partial^2}{\partial\theta_j^2}
\right. \nonumber\\ & \left. 
+\left[\frac{D-1}{r} - \frac{v^\prime(r)}{v^2(r)} \right]\frac{\partial}{\partial r}
+\sum_{j=1}^{D-1}\frac{D-1-j}{\tan(\theta_j)}\left[r\displaystyle\prod_{\ell=1}^{j-1}\sin(\theta_\ell)\right]^{-2}\frac{\partial}{\partial\theta_j}
\right\rbrace \, ,
\label{eq:Fpop:radial:v(r)}
\end{align}
where, for simplicity, $r$ has been used as the radial coordinate instead of $v$. It is interesting to notice that, compared to the single-field case where $D=1$, angular terms involving $\partial/\partial\theta_j$ and $\partial^2/\partial\theta_j^2$ are obviously present, but the radial term proportional to $\partial/\partial r$ also receives a new contribution. One can also check that when $D=2$, \Eq{eq:Fpop:radial:2D} is recovered. These $v(r)$ potentials are further studied in \Sec{sec:InfiniteInflation}.
\paragraph{Linear Potentials}
Another subclass of harmonic potentials is provided by potentials $v(u)$ that depend on a linear combination of the fields
\begin{align}
\label{eq:linear:pot}
u=\sum_i\alpha_i\phi_i
\end{align}
only. Here, one can choose the $\alpha_i$, which are constant, to be normalised so that $\sum\alpha_i^2=1$. In this case, one has $\vert\bm{\nabla}(v)\vert^2={v^\prime}^2(u)$ and $\Delta v=v^{\prime\prime}(u)$, so that $g=v^{\prime\prime}(u)/{v^\prime}^2(u)$. The ``angular'' coordinates can then be defined with constant gradients so that $\lbrace \bm{\alpha},\bm{\nabla}(\theta_j)\rbrace$ form an orthonormal basis of field space (here, the $\theta_j$ variables are unbounded and should not be viewed as geometrical angles, and ``angular'' must be understood in a generic way). In this case, one obtains
\begin{align}
\frac{1}{\Mp^2}\mathcal{L}_\mathrm{FP}^\dagger=  &
v \frac{\partial^2}{\partial u^2}
+v \sum_{j = 1}^{D-1} 
\frac{\partial^2}{\partial\theta_j^2}
-\frac{v^\prime(u)}{v(u)}\frac{\partial}{\partial u}
 \, ,
\label{eq:Fpop:linear}
\end{align}
where, for simplicity, $u$ has been used as the radial coordinate instead of $v$. From this expression, it is clear that the situation is very close to a single-field setup, since inflation is only driven by the ``scalar field'' $u$. The only difference with a purely single-field setup arises if the boundary conditions discussed around \Fig{fig:sketchFirstPassageTime} depend on the other fields, and introduce some ``angular'' dependence in the solutions of \Eqs{eq:FPT:moment:adjoint:FP}. This situation is further investigated in \Sec{sec:InhomogeneousEnd}.
\paragraph{Straight Potentials}
\begin{figure}[t]
\begin{center}
\includegraphics[width=0.5\textwidth]{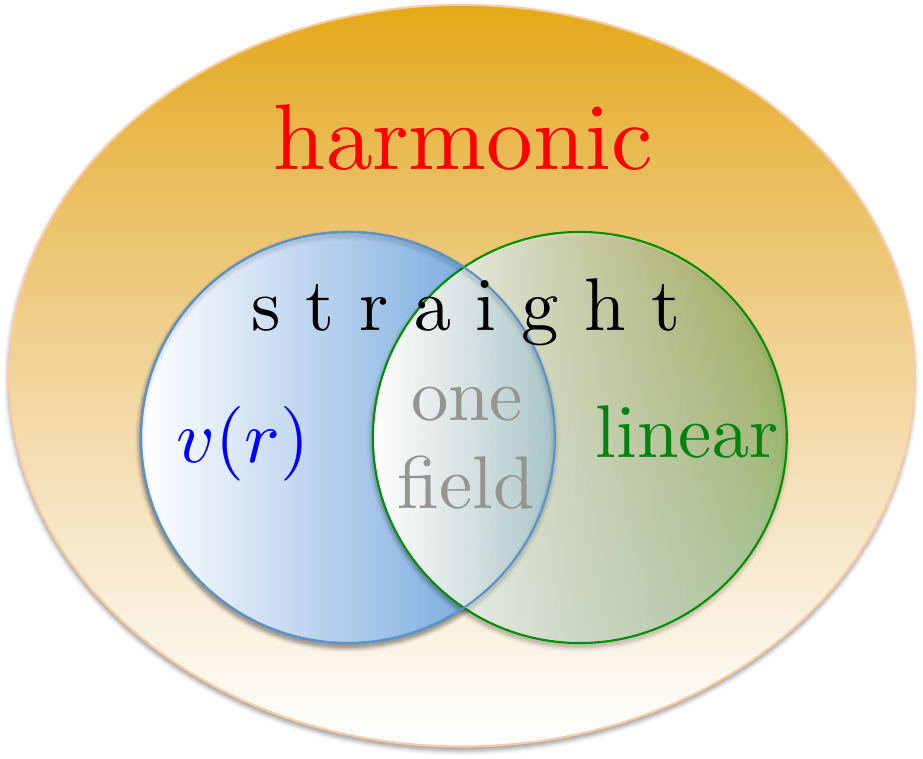}
\caption{``Harmonic Potentials'' are defined through the condition~(\ref{eq:harmonic:g}) that $\Delta v/\vert \bm{\nabla}(v)\vert^2$ is a function of $v$ only and are such that the statistical moments of the number of inflationary \efolds~can be worked out analytically. A specific class of harmonic potentials is provided by ``straight potentials'' for which the slow-roll classical trajectories are straight lines in field space. Those are made of ``linear potentials'', \ie potentials that depend on a linear combination of the fields only, see \Eq{eq:linear:pot}, and ``$v(r)$ potentials'', \ie potentials that depend on $r^2=\sum\phi_i^2$ only. Single-field potentials lie at the intersection between these two.}
\label{fig:sketchHarmonic}
\end{center}
\end{figure}
We have showed above that $v(r)$ and linear potentials are ``harmonic'' in the sense of the condition~(\ref{eq:harmonic:g}) that $\Delta v/\vert \bm{\nabla}(v)\vert^2$ is a function of $v$ only. In fact, as we are now going to see, such potentials share the property that the slow-roll classical trajectories are straight lines in field space. We call such potentials ``straight potentials''. On large scales, entropy perturbations can source adiabatic perturbations only if the background solution follows a curved trajectory in field space~\cite{Gordon:2000hv}, which is why these potentials are exactly the ones for which adiabatic perturbations are conserved on large scales, at least at the classical level.  We therefore expect them to play a special role in the present analysis. For this reason, we now try to better characterise them.

Since the slow-roll classical trajectories follow the local gradients of the potential, starting from some point $\bm{\phi}$, the next point on the classical trajectory has coordinates $\bm{\phi}+\epsilon\,\bm{\nabla}(v)$, $\epsilon$ being an infinitesimal number. The gradients evaluated at these two points must be parallel for straight potentials. In other words, the variation in the gradients between $\bm{\phi}$ and $\bm{\phi}+\epsilon\,\bm{\nabla}(v)$ must be aligned with the gradient at $\bm{\phi}$, that is to say
\begin{align}
\left[{\bm{H}}(v)\cdot \bm{\nabla}(v)\right] \wedge \bm{\nabla}(v) =\bm{0}\, ,
\end{align}
where ${\bm{H}}(v)=\bm{\nabla}^2(v)=\sum_{i,k}\partial^2 v/(\partial\phi_i\partial\phi_k)\bm{e}_{\phi_i} \otimes \bm{e}_{\phi_k}$ is the Hessian matrix of $v$. By expanding this relation into its components $\bm{e}_{\phi_i}$, it is easy to show that it leads to
\begin{align}
\left\lbrace \left\vert \bm{\nabla}(v) \right\vert^2,v\right\rbrace_{\phi_i,\phi_{i+1}}=0
\label{eq:grad(v):PB}
\end{align}
for all $1\leq i < D$, where $\left\lbrace  a,b \right\rbrace_{\phi_i,\phi_{i+1}}\equiv a_{\phi_i} b_{\phi_{i+1}} - a_{\phi_{i+1}} b_{\phi_{i}}$ stands for the $i^\mathrm{th}$ Poisson bracket between $a$ and $b$. Straight potentials are therefore such that all Poisson brackets between $ \left\vert \bm{\nabla}(v) \right\vert$ and $v$ vanish, meaning that  $ \left\vert \bm{\nabla}(v) \right\vert$ depends on $v$ only.\footnote{A first remark is that if the Poisson brackets $\lbrace , \rbrace_{\phi_{i},\phi_{i+1}}$ vanish, all Poisson brackets vanish. For example, it is easy to show that
\begin{align}
\frac{\partial b}{\partial\phi_2}\left\lbrace a,b\right\rbrace_{\phi_1,\phi_3} = \frac{\partial b}{\partial\phi_3}\left\lbrace a,b\right\rbrace_{\phi_1,\phi_2} + \frac{\partial b}{\partial\phi_1}\left\lbrace a,b\right\rbrace_{\phi_2,\phi_3}
\end{align}
so that if $\left\lbrace a,b\right\rbrace_{\phi_1,\phi_2} = \left\lbrace a,b\right\rbrace_{\phi_2,\phi_3}=0$, then $\left\lbrace a,b\right\rbrace_{\phi_1,\phi_3}=0$, so on and so forth. Then, in the basis $\lbrace v,\theta_1,\cdots \theta_{D-1}\rbrace$, the Poisson bracket $\lbrace a,b \rbrace_{v,\theta_j}$ is given by
\begin{align}
\lbrace a,b \rbrace_{v,\theta_j} &= \frac{\partial a}{\partial v}\frac{\partial b}{\partial \theta_j} -  \frac{\partial a}{\partial \theta_j}\frac{\partial b}{\partial v} 
= \sum_{i,k}  \frac{\partial a}{\partial \phi_i}\frac{\partial\phi_i}{\partial v}\frac{\partial b}{\partial\phi_k}\frac{\partial\phi_k}{\partial \theta_j}
 - \sum_{i,k}   \frac{\partial a}{\partial\phi_i}\frac{\partial\phi_i}{\partial \theta_j}\frac{\partial b}{\partial\phi_k}\frac{\partial\phi_k}{\partial v}\\
 &=\sum_{i,k} \frac{\partial\phi_i}{\partial v}\frac{\partial\phi_k}{\partial \theta_j} \left( \frac{\partial a}{\partial \phi_i}\frac{\partial b}{\partial\phi_k} - \frac{\partial a}{\partial \phi_k}\frac{\partial b}{\partial\phi_i}\right)
 =\sum_{i,k} \frac{\partial\phi_i}{\partial v}\frac{\partial\phi_k}{\partial \theta_j} \left\lbrace a,b\right\rbrace_{\phi_i,\phi_k}\, .
\end{align} 
Therefore, if all Poisson brackets between $a$ and $b$ vanish, then $\lbrace a,b \rbrace_{v,\theta_j} $ vanishes as well. If one takes $b=v$, this means that $\partial a/\partial\theta_j=0$, for all $1\leq j \leq D-1$, hence $a$ depends on $v$ only, as is the case for $ \left\vert \bm{\nabla}(v) \right\vert$ in \Eq{eq:grad(v):PB}.
\label{footnote:Poisson}
}
Since this quantity appears in various places in \Eq{eq:Fpop:radial}, we understand why straight potentials play a special role in the present context. In particular, above, it was shown that for $v(r)$ potentials, $\left\vert \bm{\nabla}(v) \right\vert^2 = {v^\prime}^2(r)$ is a function of $v$ only, which confirms that $v(r)$ potentials are straight potentials. Similarly, we saw that for linear potentials, $\left\vert \bm{\nabla}(v) \right\vert^2 = {v^\prime}^2(u)$ is a function of $v$ only, and linear potentials also are straight potentials, as announced above.

Reciprocally, one can show that straight potentials can only be of one of these two types: $v(r)$ potentials or linear potentials. Indeed, let us consider a straight potential $v$ and its (straight) gradient lines in dimension $D=2$. We first assume that its gradient lines never intersect in field space. This means that they all are parallel, and one can write $\bm{\nabla}(v) = a(\bm{\phi})\sum_i\alpha_i\bm{e}_{\phi_i}$, hence $v_{\phi_i}=a(\bm{\phi})\alpha_i$. One then has $\lbrace v, \sum_i\alpha_i\phi_i \rbrace_{\phi_k,\phi_\ell} = \alpha_\ell v_{\phi_k} - \alpha_k v_{\phi_\ell} = 0$, hence $v$ depends on $\sum_i\alpha_i\phi_i$ only (see footnote~\ref{footnote:Poisson}) and is therefore linear. Let us now assume that there is exactly one intersection point in the gradient lines of $v$, which, after performing a constant field shift, we set at the origin of field space. It is easy to see that any gradient line not passing through the origin would produce a second intersection point at least, hence all gradient lines go through the origin and one can write $\bm{\nabla}(v)=a(\bm{\phi}) \bm{e}_r$, where $\bm{e}_r$ is the unit vector pointing to the radial direction $r=\sqrt{\sum\phi_i^2}$. This means that $v_{\phi_i}=a(\bm{\phi})\phi_i/r$. Since $r_{\phi_i}=\phi_i/r$, one has $\lbrace v,r\rbrace_{\phi_k,\phi_\ell}= v_{\phi_k} r_{\phi_\ell} - v_{\phi_\ell} r_{\phi_k}=0$, hence $v$ depends on $r$ only and is of the $v(r)$ type. Finally, let us assume that the gradient lines of $v$ intersect at two or more points. Then, one can convince oneself that an infinite number of other intersection points can be obtained, that fill the entire (or a dense subset of the) field space. Since the gradient of $v$ must vanish when two non-parallel lines intersect (otherwise its direction would be ill-defined), this means that $v$ is constant, and this case is in fact trivial. This result can be generalised to $D>2$ where one finds that the potential is of the $v(r)$ type within the field subspace that is orthogonal to the one containing the fields of which $v$ is independent.

The situation is schematically summarised in \Fig{fig:sketchHarmonic}. Straight potentials are a specific class of harmonic potentials. They are either linear or $v(r)$ potentials, and single-field potentials lie at the intersection between these two. Let us finally notice that not all harmonic potentials are straight. For example, let us consider a ``loop corrected'' potential of the form $v=v_0[1+\alpha\sum_{i=1}^D\log\left(\phi_i/\Mp\right)]$. The function $g$ defined in \Eq{eq:harmonic:g} is constant, $g=-1/(v_0\alpha)$, and such potentials are therefore harmonic. However, one has $\lbrace\vert \bm{\nabla}(v)\vert^2 , v \rbrace_{\phi_k,\phi_\ell} = 2 / (v_0 \alpha \phi_k \phi_\ell) (1 / \phi_\ell^2 - 1 / \phi_k^2)$ which is not a vanishing function, hence loop corrected potentials are not straight. More generally, this is the case for all potentials $v(w)$ that are functions of $w=\prod_i\phi_i$ only, for which $g=v''(w)/{v'}^2(w)$.
\subsubsection{$v(r)$ potentials and infinite inflation}
\label{sec:InfiniteInflation}
\begin{figure}[t]
\begin{center}
\includegraphics[width=0.49\textwidth]{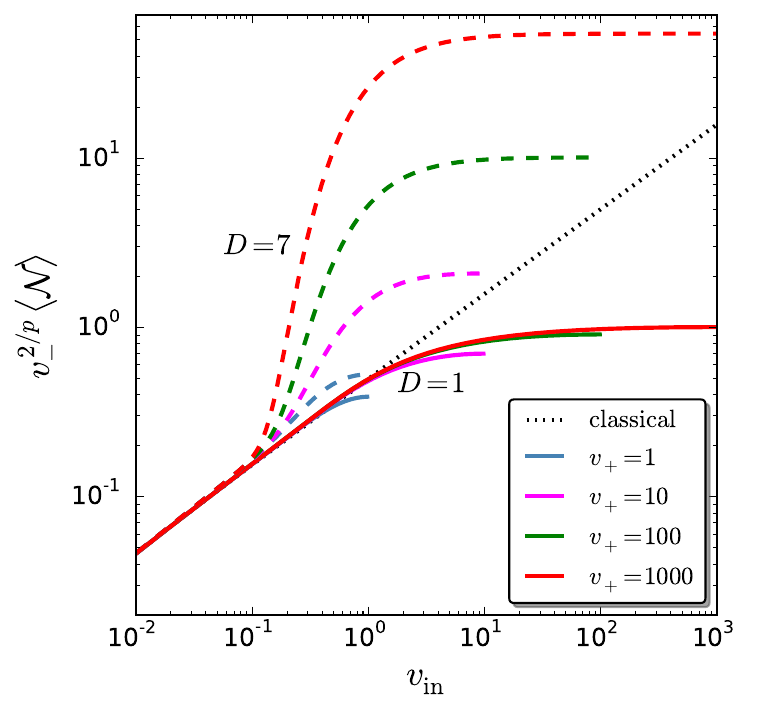}
\includegraphics[width=0.478\textwidth]{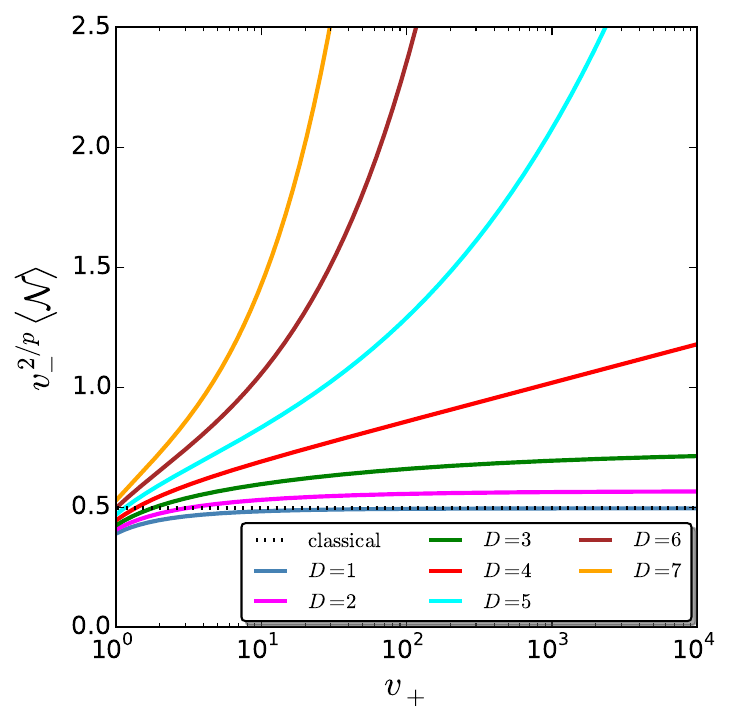}
\caption{Mean number of \efolds~(\ref{eq:v(r):sol}) for monomial potentials $v(r)\propto r^p$ with $p=4$ (rescaled by $v_-^{2/p}$, with $r_-=p/\sqrt{2}$ where inflation ends by slow-roll violation). In the left panel, $\langle\mathcal{N}\rangle$ is displayed as a function of the initial condition $v_\uin$, for $D=1$ (solid lines) and $D=7$ (dashed lines) and for different values of $v_+>v_\uin$ at which a reflecting boundary condition is placed. The black dotted line stands for the classical limit~\eqref{eq:stocha:meanN:classtraj}, $N_\ucl$ (where $\phi$ is simply replaced by $r$), towards which the stochastic results asymptote when $v\ll 1$. In the opposite regime, $\langle \mathcal{N} \rangle$ deviates from $N_\ucl$ in a way that depends on $v_+$ and $D$, and that is further discussed in the main text. One should note that, in principle, $v_\uin>1$ lies outside the validity range of the present calculation since it corresponds to initial super-Planckian energy density, but it is displayed to make clear the asymptotic behaviour of the mean number of \efolds~at large initial field value. In the right panel, $v_\uin=1$ is fixed, but $v_+$ varies, and different values of $D$ are displayed. One can check that when $D < p$, a finite asymptotic value is reached when $v_+\rightarrow\infty$, while when $D \geq p$, $\langle N\rangle$ diverges in this limit.}
\label{fig:Nmean:v(r)}
\end{center}
\end{figure}
In \Sec{sec:harmonic}, it was shown that $v(r)$ potentials provide a subclass of harmonic potentials, for which analytical solutions of the diffusion equations~(\ref{eq:FPT:moment:adjoint:FP}) can be found. In this section, we derive such solutions and use these potentials to illustrate the physical implications of including more than one scalar field in the analysis.

If one sets boundary conditions to be angular independent, that is to say if one assumes that inflation ends at $r=r_-$ and that a reflecting wall is placed at $r=r_+$, angular independent solutions of \Eq{eq:FPT:moment:adjoint:FP} can be obtained. More precisely, plugging \Eq{eq:Fpop:radial:v(r)} into \Eqs{eq:FPT:moment:adjoint:FP}, one obtains
\begin{align}
\langle \N^n\rangle (r)=
n\displaystyle\int_{r_-}^r\frac{\dd r^\prime}{\Mp}
\displaystyle\int_{r^\prime}^{r_+}\frac{\dd r^{\prime\prime}}{\Mp}
\dfrac{\ee^{\frac{1}{v(r^{\prime\prime})}-\frac{1}{v(r^{\prime})}}}{v(r^{\prime\prime})}\left(\frac{r^{\prime\prime}}{r^\prime}\right)^{D-1}\langle \N^{n-1}\rangle(r^{\prime\prime})
\label{eq:v(r):sol}
\end{align}
where we recall that, obviously, $\langle \N^0\rangle=1$. One can check that $\langle \N^n\rangle(r_-)=0$ (absorbing boundary condition) and that $\langle \N^n\rangle^\prime(r_+)=0$ (reflecting boundary condition). If one wanted to place an absorbing boundary condition at $r_+$ instead, as explained in \Sec{sec:meanN}, one would have to change the upper bound of the second integral to a smaller value, but this would not affect the following considerations. In the same manner, if, instead of the situation depicted in \Fig{fig:sketchFirstPassageTime} where the fields classically decrease during inflation, one considered a hilltop potential symmetric about $r=0$, one would have to replace $r_+$ by $0$ in the above formula, and this case is also discussed in what follows. One can also check that, when $D=1$, \Eq{eq:fn:sol:onefield} is recovered.

A preliminary remark is that both the number of fields $D$ and the location of the reflecting boundary condition $r_+$ explicitly appear in \Eq{eq:v(r):sol}, and are therefore expected to play a role. For illustration, in the left panel of \Fig{fig:Nmean:v(r)}, the mean number of \efolds~(\ref{eq:v(r):sol}) is displayed as a function of the initial condition $v_\uin$ for a quartic potential $v\propto r^4$, for different values of $r_+$, and for $D=1$ (solid lines) and $D=7$ (dashed lines). One can check that, in some regimes at least, the result strongly depends on $r_+$ and $D$ indeed (see also the right panel of \Fig{fig:Nmean:v(r)}), in a way that we now analyse in more details.
\paragraph{Classical Limit}
As discussed above in the case of single-field models, a first important consistency check consists of verifying that the correct classical limit is recovered. In the classical limit, energy densities are sub-Planckian $v\ll 1$ and the integral over $r^{\prime\prime}$ in \Eq{eq:v(r):sol} is dominated by its contribution close to the lower bound $r_-$, near which one can Taylor expand $1/v$ at first order, $1/v(r^{\prime\prime})\simeq  1/v(r^{\prime}) - v^\prime(r^\prime)/v^2(r^\prime) (r^{\prime\prime}-r^\prime)$. This gives rise to
\begin{align}
\displaystyle\int_{r^\prime}^{r_+}{\dd r^{\prime\prime}}
\dfrac{\ee^{\frac{1}{v(r^{\prime\prime})}}}{v(r^{\prime\prime})}\left(r^{\prime\prime}\right)^{D-1}
\simeq 
\ee^{\frac{1}{v(r^\prime)}}\displaystyle\int_{r^\prime}^{r_+}\frac{\dd r^{\prime\prime}}{\Mp}
\left[\frac{1}{v(r^\prime)}-\frac{v^\prime(r^\prime)}{v^2(r^\prime)}\left(r^{\prime\prime}-r^\prime\right)\right]
\ee^{-\frac{v^\prime(r^\prime)}{v^2(r^\prime)}(r^{\prime\prime}-r^\prime)}
\left(r^{\prime\prime}\right)^{D-1}\, .
\end{align}
This integral can be performed through $D-2$ integrations by parts. If one keeps contributions from the upper bound of the integral only and expands the result at leading order in $v$, one obtains
\begin{align}
\displaystyle\int_{r^\prime}^{r_+}{\dd r^{\prime\prime}}
\dfrac{\ee^{\frac{1}{v(r^{\prime\prime})}}}{v(r^{\prime\prime})}\left(r^{\prime\prime}\right)^{D-1}
\simeq \frac{v(r^\prime)}{v^\prime(r^\prime)}\ee^{\frac{1}{v(r^\prime)}}\left(r^\prime\right)^{D-1}\, .
\end{align}
By plugging this formula into \Eq{eq:v(r):sol}, one obtains
\begin{align}
\langle \mathcal{N}\rangle \simeq N_\ucl
\end{align}
in the classical limit, where $N_\ucl$ is given by \Eq{eq:stocha:meanN:classtraj} where $\phi$ is simply replaced with $r$. In the left panel of \Fig{fig:Nmean:v(r)}, the classical formula~(\ref{eq:stocha:meanN:classtraj}) is displayed and one can check that, indeed, when $v_\uin\ll 1$, it provides a good approximation to the full stochastic results indeed. Let us also notice that the classical limit~(\ref{eq:stocha:meanN:classtraj}) depends neither on the number of fields $D$ nor on the location of the upper boundary condition $r_+$, and matches the single-field result. However, as we are now going to see, stochastic corrections break this classical $D$ and $r_+$ invariance and introduce dependence on both the number of fields and the location of the upper boundary condition.
\paragraph{Infinite Inflation}
The validity of the classical limit relies on the assumption that the second integral in \Eq{eq:v(r):sol} is dominated by its contribution close to the lower bound $r^\prime$. If this is correct, this means that the upper bound, $r_+$, can be removed to infinity without affecting the leading order result, providing a well-defined regularisation procedure. In the left panel of \Fig{fig:Nmean:v(r)}, one can see that for $D=1$, the curves saturate to an asymptotic behaviour when $v_+$ increases, and such a procedure seems therefore to be well justified. However, for $D=7$, the result does not seem to converge when $v_+$ increases. This is why in the right panel of \Fig{fig:Nmean:v(r)}, the mean number of \efolds~is displayed as a function of $v_+$ for quartic $v(r)$-potentials, for a fixed $v_\uin=1$ and a few values of $D$. One can see that when $D<4$, $\langle\mathcal{N}\rangle$ goes to a constant value when $v_+\rightarrow\infty$, while when $D\geq4$, it diverges. This confirms that the number of fields plays an important role in determining whether the limit $r_+\rightarrow\infty$ is finite or not.

More precisely, the mean number of \efolds~is finite if the integrand of \Eq{eq:v(r):sol} is integrable when $r^{\prime\prime}\rightarrow\infty$, that is to say if $r^{D-1}/v(r)$ is an integrable function. This criterion depends on the number of fields $D$, as already noticed, but also on the large-field behaviour of the potential. Let us distinguish the three following cases:
\begin{itemize}
\item If, at large-field value, the potential is of the ``Plateau'' type and $v$ goes to a constant value $v_\infty>0$, then $r^{D-1}/v(r)$ is never integrable and an infinite mean number of \efolds~is always realised, regardless of the number of fields. 
\item If, at large-field value, the potential is of the monomial type $v\propto r^p$, then $r^{D-1}/v(r)$ is integrable only when $D<p$, and an infinite mean number of \efolds~is realised as soon as more than $p$ fields are present. This is consistent with the previous discussion about the right panel of \Fig{fig:Nmean:v(r)}.
\item If the potential is of the ``hilltop'' type and symmetric around $0$, as explained above, $r_+$ has to be replaced by $0$ in \Eq{eq:v(r):sol}. In this case, the integrability of $r^{D-1}/v(r)$ needs to be checked around $0$ instead of infinity. If $v$ is finite at $r=0$ this is always the case, hence the mean number of \efolds~is never infinite in such potentials.
\end{itemize}
The situation is summarised in table~\ref{table:v(r):Nmean}. In a large class of potentials (plateau potentials and some monomial potentials), the mean number of \efolds~is infinite, and we call this phenomenon ``infinite inflation''. Let us notice that this is different from ``eternal inflation''~\cite{Steinhardt:1982kg, Vilenkin:1983xq, Guth:1985ya, Linde:1986fc} where volume weighting is included and the diverging quantity is the physical volume of the inflating part of the Universe, not $\langle \mathcal{N} \rangle$. Infinite inflation implies eternal inflation but is a stronger statement. For example, eternal inflation can be realised in hilltop models~\cite{Vilenkin:1983xq, Barenboim:2016mmw} while, as we have just shown, infinite inflation never occurs in such potentials. 

\begin{figure}[t]
\begin{center}
\includegraphics[width=0.49\textwidth]{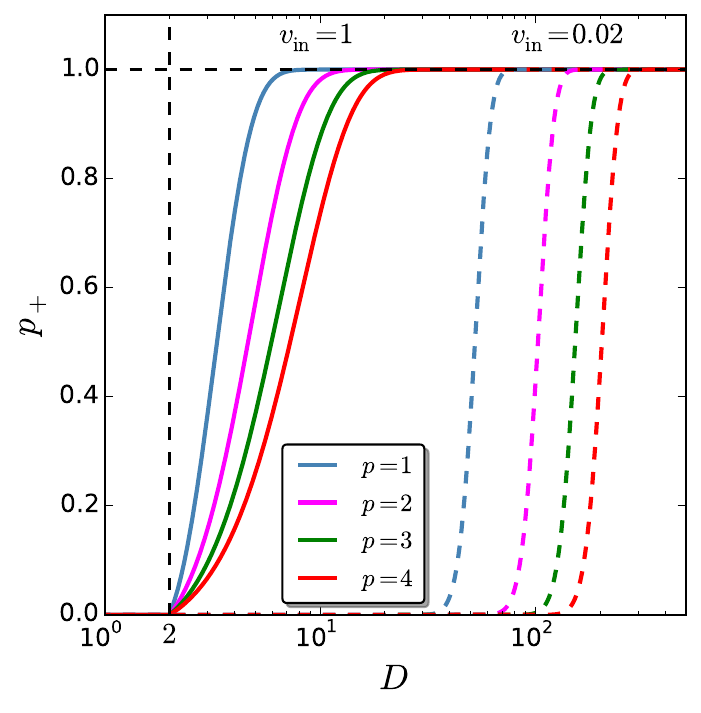}
\includegraphics[width=0.478\textwidth]{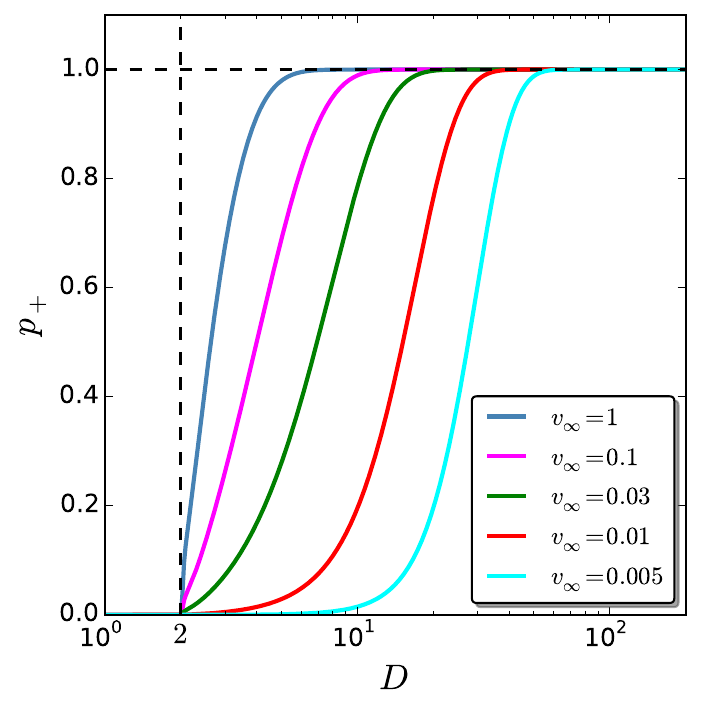}
\caption{Probability~(\ref{eq:v(r):pwall}) of exploring large-field regions of the potential $r_+$ when $r_+\rightarrow \infty$, as a function of the numbers of fields $D$. The left panel stands for monomial potentials $v(r)\propto r^p$, where a few values of $p$ are displayed. The first set of curves (solid lines) correspond to choosing the initial value of $r$ such that $v_\uin=1$, while $v_\uin=0.1$ in the second set of curves (dashed lines). The lower boundary condition is taken to be such that $r_-/\Mp=p/\sqrt{2}$ (end of inflation by slow-roll violation) but one can check that its precise value plays a negligible role. This probability is always non-zero when $D> 2$, but in practice, it is non-negligible only when $D\gtrsim 1+p/v_\uin$. The right panel stands for a plateau potential, the Starobinsky model, for which $v=v_{\infty}[1-\exp(-\sqrt{2/3}r/\Mp)]^2$.  The lower boundary condition is taken to be such that $r_-/\Mp=\sqrt{3/2}\ln(1+2/\sqrt{3})$ (end of inflation by slow-roll violation), and the initial value of $r$ is taken $50$ (classical) \efolds~before the end of inflation. Several values of $v_\infty$ are displayed, and the probability $p_+$ is non-negligible only when $D\gtrsim 2+\mathcal{O}(0.1)/v_\infty$.}
\label{fig:pwall:v(r)}
\end{center}
\end{figure}

Another important remark is that for monomial potentials, whether infinite inflation occurs or not crucially depends on the number of fields, which therefore plays the role of an ``order parameter'' (as illustrated below in \Fig{fig:pwall:v(r)}). The number of dimensions is a critical parameter for many stochastic processes (for instance in recurrence problems~\cite{Polya:1921}) and this may therefore not be so surprising. The key feature is that the more fields, the larger the volume in field space to realise inflation and the more common infinite inflation.

Beyond the physical implications related to the possibility of realising arbitrarily large number of \efolds~\cite{Polarski:1992dq, Ringeval:2010hf, Enqvist:2012xn, Kitajima:2019ibn}, infinite inflation raises the issue of practical calculability of observables. Indeed, since the correlation functions of scalar adiabatic fluctuations are related to the moments of the number of $e$-folds, see \Sec{sec:stochaDeltaN:stochaDeltaN}, it is not clear what the predictions for these observables are when those moments are infinite. How these infinities regularise is in fact a non-trivial question that we investigate below. At this stage however, let us notice that infinite inflation may suggest that the system explores regions of the potential that are far away from what its classical trajectory would allow it to reach, and that observables may be sensitive to the physics at play in these remote regions. For this reason, we now study how likely it is to explore large-field regimes in stochastic multiple field inflation.
\begin{center}
\begin{table}[t]
\definecolor{grey}{RGB}{50,50,50}
\centering
\begin{tabular}{|c|c|c|}
\hline
Potential & Mean Number of \efolds  & Probability of large field exploration  \\ \hline \hline
\multirow{2}{*}{Plateau 
}          &       \multirow{2}{*}{always infinite}      &  $0$ if $D\leq 2$, finite if $D>2$            \\ 
& & non-negligible if $D\gtrsim 2+\mathcal{O}(0.1)/ v_{\infty}$ \\
\hline
\multirow{2}{*}{Monomial $v\propto r^p$}    &      finite if $D<p$   &  $0$ if $D\leq 2$, finite if $D>2$  \\
 & $\ $infinite if $D\geq p$ & non-negligible if $D\gtrsim 1+p/ v_\uin$
\\ \hline
Hilltop          &     always finite    & \cellcolor{grey!25} \\ \hline
\end{tabular}  
\caption{Mean number of \efolds~realised in $v(r)$ potentials and probability of exploring arbitrarily large-field regions of the potential, when $D$ fields are present.}   
\label{table:v(r):Nmean}         
\end{table}
\end{center}
\paragraph{Large-field exploration}
\label{sec:wallproba}
Let us study the probability $p_+(r)$ that, starting from $r$, the system bounces at least once against the reflecting wall located at $r_+$ before exiting inflation at $r_-$ (or alternatively, if an absorbing wall is located at $r_+$, the probability that the system exits inflation at $r_+$ rather than $r_-$). In \Sec{sec:FPB}, it was explained that $p_+$ is given by the solution of \Eq{eq:FPB:equadiff:app} with boundary conditions $p_+(r_-)=0$ and $p_+(r_+)=1$. Making use of \Eq{eq:Fpop:radial:v(r)}, one obtains
\begin{align}
\label{eq:v(r):pwall}
p_+\left(r\right)=\displaystyle\dfrac{\displaystyle\int_{r_-}^r {r^\prime}^{1-D}\ee^{-\frac{1}{v({r^\prime})}}\dd {r^\prime}}{\displaystyle\int_{r_-}^{r_+} {r^\prime}^{1-D}\ee^{-\frac{1}{v({r^\prime})}}\dd {r^\prime}}\, .
\end{align}
When the upper boundary condition $r_+$ is sent to infinity, one obtains a non-vanishing probability $p_+$ if the function $r^{1-D}$ is integrable (assuming that $v$ has a positive limit at infinity). Contrary to the case of infinite inflation, this condition is independent of the shape of the potential at large-field value, and $p_+>0$ as soon as strictly more than $2$ fields are present. This information is added in table~\ref{table:v(r):Nmean}, and in \Fig{fig:pwall:v(r)}, \Eq{eq:v(r):pwall} is displayed for monomial potentials $v\propto r^p$ (left panel, for different values of $p$ and $v_\uin$) and a plateau potential, the Starobinsky model~\cite{Starobinsky:1980te}, $v=v_\infty[1-\exp(-\sqrt{2/3}r/\Mp)]^2$ (right panel, for different values of $v_\infty$), when $r_+$ is removed to infinity. One can check that when $D\leq 2$, $p_+=0$. When $D>2$, strictly speaking, $p_+>0$, but one can see that $p_+$ is non-negligible only when $D$ is larger than some value that depends on the parameters of the potential and on the initial field value. Schematically, this value is realised when the integrand of the integrals in \Eq{eq:v(r):pwall} is maximal at $v_\uin$. In monomial potentials, this leads to the conclusion that $p_+$ is non-negligible when 
\begin{align}
D\gtrsim 1+\frac{p}{v_\uin}\, ,
\end{align}
while in plateau potentials, one obtains the condition
\begin{align}
D\gtrsim 2+\frac{\mathcal{O}(0.1)}{v_\infty}\, .
\end{align}
One can numerically check that, indeed, these expressions provide good estimates of the point where $p_+$ starts to be non-negligible. This shows that including more fields increases the probability to explore large-field regions of the potential, but for sub-Planckian energy scales, one needs a very large number of fields to obtain a substantial probability. For example, if one normalises the overall mass scale of the potentials to fit the measured amplitude of the scalar power spectrum~\cite{Ade:2015xua} and starts the evolution $50$ (classical) \efolds~before the end of inflation, one finds $v_\uin\sim10^{-11}p$ for monomial models and $v_{\infty}\sim 10^{-12}$ for the Starobinsky potential, so that $10^{11}$ fields would be required to obtain appreciable values of $p_+$, a very large number indeed.

Of course, from a model building perspective, the shape of the potential may be very different at very large-field value outside the observational window than what cosmological observations constrain at smaller field values (for example~\cite{Broy:2014sia, Coone:2015fha}, the potential may be of the Plateau type where the scales probed in the CMB cross the Hubble radius, but of the monomial type at larger field), and if inflation starts high enough in the potential, large-field exploration, enhanced by the presence of multiple fields, may become likely. But the above results suggest that, in the simplest setups, cosmological observations at small (\ie sub-Planckian) energies carry limited information about the physics taking place at much higher energy (at least through stochastic effects). This question was also further investigated in \Refa{Hardwick:2017qcw}.
\paragraph{Regularisation}
\begin{figure*}
\begin{center}
\includegraphics[width=0.49\textwidth]{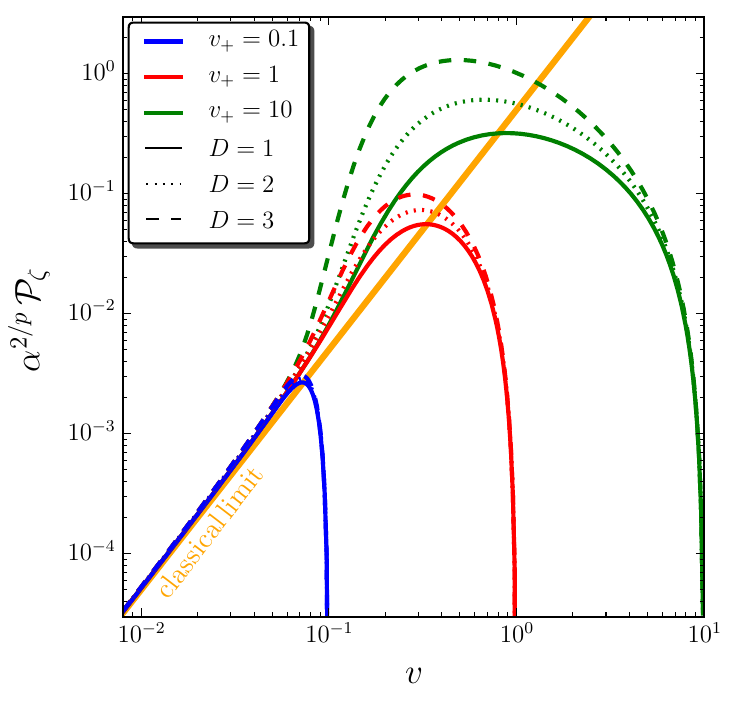}
\includegraphics[width=0.49\textwidth]{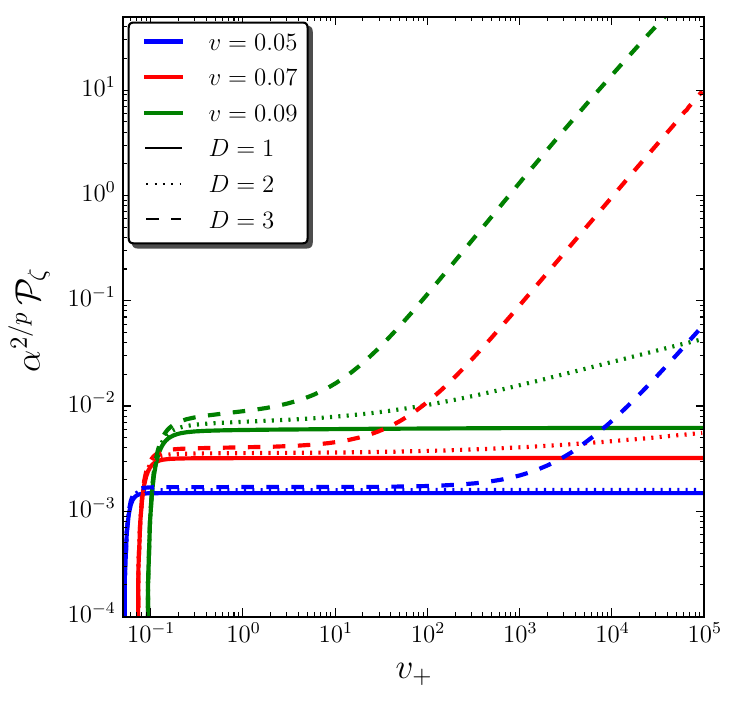}
\caption{Scalar power spectrum amplitude $\calP_\zeta$ in $v = \alpha \sum_{i=1}^D\phi_i^2$ potentials, as a function of the potential energy $v$ at which the scales for which  $\calP_\zeta$ is calculated exit the Hubble radius during inflation (left panel), and as a function of the upper reflecting wall location $v_+$ in the right panel, for a few values of the number of fields $D$.}
\label{fig:Pzeta}
\end{center}
\end{figure*}
Let us now see how the infinite inflation mechanism affects the observable quantities of the problem; namely, the correlation functions of scalar adiabatic perturbations, and how it can be regularised away. Combining \Eqs{eq:Pzeta:stochaDeltaN} and~(\ref{eq:v(r):sol}), the scalar power spectrum $\calP_\zeta$ in $v(r) = \alpha r^2$ potentials is plotted in \Fig{fig:Pzeta} for illustrative purposes. In the left panel, $\calP_\zeta$ is displayed as a function of $v$ (where the scale at which the power spectrum is calculated exits the Hubble radius), for a few values of $D$ and $v_+$ (where a reflecting wall is located). In the sub-Planckian limit where $v \ll 1$, all curves approach the classical formula $\calP_{\zeta,\ucl}=v^2/(2\alpha\Mp^2)$, which is independent of the number of fields $D$. When $v$ is of order $0.1$ or greater, the full result deviates from the classical prediction, in a way that depends on $v_+$ and $D$. Therefore, stochastic effects introduce dependences on these parameters that do not exist in the classical picture otherwise. Let us discuss the role played by both quantities.

In the right panel of \Fig{fig:Pzeta}, $\calP_\zeta$ is displayed as a function of $v_+$ for a few values of $v$ and $D$. When $D=1$ (solid lines), $\calP_\zeta$ converges to a finite value when $v_+\rightarrow\infty$, which is reached soon after $v_+\gg v$ and therefore provides a well-defined prediction when the reflecting wall is removed to infinite energy. However, when $D\geq 2$ (dashed and dotted lines), $\calP_\zeta$ diverges when $v_+\rightarrow\infty$, as a consequence of the phenomenon of infinite inflation discussed below \Eq{eq:v(r):sol}. In this case, a reflecting (or absorbing) wall at large-field value is compulsory to make the power spectrum (as well as higher correlators) finite. For plateau potentials, let us recall that this happens regardless of the number of fields. 

In such cases, how much does the result depend on the precise location of this large-field wall? In the right panel of \Fig{fig:Pzeta}, one can see that when $v_+$ increases, the power spectrum amplitude reaches a plateau the width of which decreases with $v$, before diverging. More precisely, one can show that the contribution from the upper bound $r_+$ of the second integral in \Eq{eq:v(r):sol} is subdominant when
\begin{align}
v\ll v_+\ll  
\ee^{\frac{\mathcal{O}(1)}{v}}
\end{align}
for $v(r)\propto r^p$ potentials, where the $\order{1}$ constant depends on $p$ and $D$. For example, if one takes $D=2$ and $p=3$, $v_\uin\sim 10^{-10}$ leads to $v_+\ll 10^{3,474,355,825}$. This is an extremely large, ``ultra super-Planckian'' value, way below which quantum gravity effects are expected to come into play anyway. For plateau potentials, one finds 
\begin{align}
r\ll r_+\ll \ee^{\order{1}\left(\frac{1}{v}-\frac{1}{v_\infty}\right)}\, ,
\end{align}
where the $\order{1}$ constant depends on the exact shape of the plateau and on the number of fields. In the Starobinsky model~\cite{Starobinsky:1980te} with a single field for instance, one obtains $r_+ \ll 10^{6,166,453,090} \Mp$. This value is again huge and one typically expects~\cite{Broy:2014sia, Coone:2015fha} monomial corrections to spoil the plateau potential way before then, which would bring us back to the previous monomial case. As a consequence, if inflation proceeds at sub-Planckian energy, predictions are independent of the location of the large-field wall, provided it is placed below the ultra super-Planckian values just quoted. 

Stochastic effects and infinite inflation therefore require modifying the super-Planckian limit of inflationary models to make them consistent, but this modification does not impact their predictions, up to corrections typically of order $\ee^{-\order{1}/v}$. If we neglect these, performing a saddle-point approximation of \Eq{eq:v(r):sol} in the $v\ll 1$ limit, the scalar power spectrum for $v(r)$ potentials is given by
\begin{align}
\calP_\zeta = \frac{2}{\Mp^2}\frac{v^3}{{v^\prime}^2}\left[1+v\left(5-4\frac{v v^{\prime\prime}}{{v^\prime}^2}+2\frac{D-1}{r}\frac{v}{v^\prime}\right)+\mathcal{O}(v^2)\right]\, ,
\label{Pzeta:v(r):classLim}
\end{align}
while the non-Gaussianity parameter $\fnl$ reads 
\begin{align}
&\fnl=\frac{5}{24}\Mp^2\left\lbrace 6\frac{{v^\prime}^2}{v^2}-4\frac{v^{\prime\prime}}{v}
+v\left[25\frac{{v^\prime}^2}{v^2}-34 \frac{v^{\prime\prime}}{v}
-10\frac{v^{\prime\prime\prime}}{v^\prime}+24\frac{{v^{\prime\prime}}^2}{{v^\prime}^2}
+2\frac{D-1}{r^2}\left(r\frac{v^\prime}{v}-2\right)\right]
+\mathcal{O}\left(v^2\right)\right\rbrace\, .
\label{fnl:v(r):classLim}
\end{align}
When $D=1$, one recovers \Eqs{eq:PS:vll1} and~(\ref{eq:fnl:classappr}). In these expressions, the $D$-dependent terms are typically suppressed by $v$, as the other single-field stochastic corrections. This may suggest that stochastic corrections to correlation functions are always small in the observational window. However, as already mentioned at the end of \Sec{sec:Higher:Moments:NG}, the presence of $v'$ in the denominator of the above expressions make it possible to have large stochastic effects even at sub-Planckian energy densities, even if the potential is sufficiently flat, and this possibility will be studied in \Sec{sec:PBHs}. There is also the possibility, that only arises in multiple-field models, that the inflationary potential possesses features at scales smaller than the typical quantum diffusion amplitude $H/(2\pi)$. This is the topic of the next section.
\subsubsection{Inhomogeneous end of inflation}
\label{sec:InhomogeneousEnd}
In \Sec{sec:InfiniteInflation}, we have considered the case of $v(r)$ potentials where the dynamics is governed by the ``radial'' field $v$ only, and angular independent solutions can be found. In this section, we study situations where both $v$ and $\theta_j$ play a role, and study the simple setup where the inflationary dynamics is effectively driven by a single field $\phi$ while extra fields $\chi_j$ (to be identified with the ``angles'' $\theta_j$) only appear at the surface defining the end of inflation. This notably allows one to describe ``inhomogeneous end of inflation''~\cite{Dvali:2003ar} where inhomogeneities induced from the additional light fields on the end-surface make additional contributions to curvature perturbations on super-Hubble scales. 

In the terminology introduced in \Sec{sec:harmonic}, this case is called ``linear potential'' (\ie $v$ depends on a linear combination of the fields only). In \Eq{eq:Fpop:linear}, one can see that the ``angular'' sector is only affected by a pure diffusion term, which suggests that some insight may be gained by Fourier transforming the functions
\begin{align}
\langle \N^n\rangle\left(\phi, \bm{\chi} \right)&=\int \dd^{D-1} \bm{k} \ee^{-i  \bm{k}\cdot\bm{\chi}} \langle \N^n\rangle_{\bm{k} }(\phi)\, .
\label{eq:linearpot:FourierExpansion}
\end{align}
Plugging this ansatz into \Eq{eq:Fpop:linear}, \Eq{eq:FPT:moment:adjoint:FP} gives rise to the set of recursive ordinary differential equations
\begin{align}
v \langle \N^n\rangle_{\bm{k} }^{\prime\prime}(\phi)-\frac{v^\prime(\phi)}{v(\phi)} \langle \N^n\rangle_{\bm{k} }^{\prime}(\phi) - k^2 v  \langle \N^n\rangle_{\bm{k} }(\phi)=-\frac{n}{\Mp^2}\langle \N^{n-1}\rangle_{\bm{k} }(\phi)\, ,
\label{eq:linearpot:ode}
\end{align}
where $k^2\equiv \left\vert \bm{k}\right\vert^2$. If the end-surface $\partial\Omega_-$ is parametrised by the function $\phi_-(\bm{\chi})$, and if the inflationary domain is limited from above at the reflecting surface $\partial\Omega_+$ defined by $\phi=\phi_+$, these equations need to be solved with the boundary conditions 
\begin{align}
\langle \N^n \rangle \left [\phi_-\left(\bm{\chi}\right),\bm{\chi}\right]=0\, ,\quad\quad \frac{\partial \langle \N^n\rangle}{\partial\phi}\left(\phi_+,\bm{\chi}\right)=0\, .
\label{eq:linearpot:boundarycondition}
\end{align}
The procedure one needs to carry out is therefore the following: solve \Eq{eq:linearpot:ode} for all $\bm{k}$'s in terms of two integration constants each, plug the solutions into \Eq{eq:linearpot:FourierExpansion}, and use \Eq{eq:linearpot:boundarycondition} to set all integration constants simultaneously. In practice, such a calculation may need to partly rely on numerical analysis, but it is still more straightforward (and numerically less expensive) than having to solve the full partial differential equations~(\ref{eq:FPT:moment:adjoint:FP}).
\paragraph{The example of exponential potentials}
\begin{figure}[t]
\begin{center}
\includegraphics[width=0.48\textwidth]{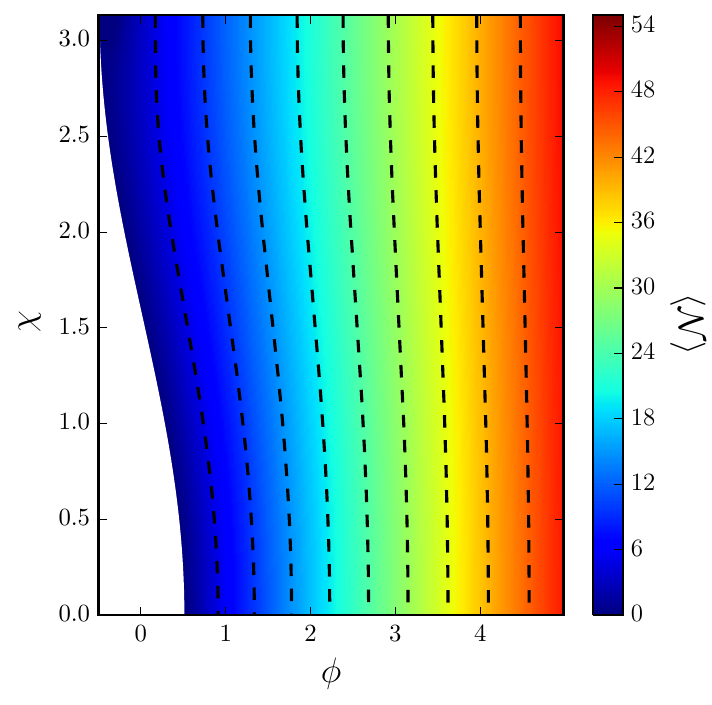}
\includegraphics[width=0.48\textwidth]{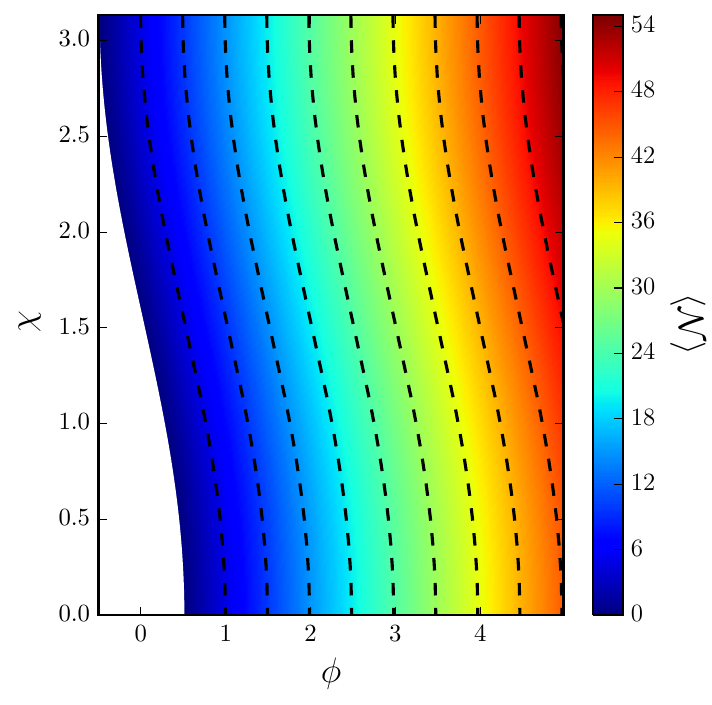}
\caption{Mean number of \efolds~$\langle\mathcal{N}\rangle$ realised in the single-field ``power-law'' potential $v=v_\uend \ee^{\alpha\phi}$, when the end of inflation is modulated by an extra field $\chi$ through the function $\phi_-(\chi)=\mu\cos(\chi/\chi_0)$. The left panel corresponds to the full stochastic result~(\ref{eq:exponential:sol}), where the integration constants $A_0$, $A_k$, $B_0$, $B_k$ are obtained imposing \Eqs{eq:linearpot:boundarycondition}. The right panel corresponds to the classical limit~(\ref{eq:modulated:classicalLimit}), which provides a good approximation to the stochastic result when $v\ll 1$. The parameter values in both panels are $\alpha=0.1$, $\chi_0/\Mp=1$, $\mu/\Mp=0.5$ and $v_\uend=0.05$. The black dashed lines correspond to various level lines of $\langle\mathcal{N}\rangle$, and help to better compare the two results. In particular, when $v$ increases, one can see that the dependence on the initial value of $\chi$ gets smeared out by the stochastic effects.}
\label{fig:modulatedReheating}
\end{center}
\end{figure}
In order to illustrate how the above procedure works in practice, let us consider the case of ``power-law inflation''~\cite{Abbott:1984fp} where the potential is of the exponential type $v\propto\ee^{\alpha\phi/\Mp}$. To be explicit, we study the situation where one extra field $\chi$ modulates the end-surface through 
\begin{align}
\phi_-\left(\chi\right) = \phi_\uend +\mu \cos\left(\frac{\chi}{\chi_0}\right)\, ,
\end{align} 
where $\mu$ is a modulation parameter (when $\mu\rightarrow 0$, one recovers the standard single-field setup), and $\chi_0$ is the scale over which the modulation takes place. In this case, solutions of  \Eq{eq:linearpot:ode} that are $2\pi\chi_0$-periodic in $\chi$ can be found, and one can replace the continuous Fourier transform of \Eq{eq:linearpot:FourierExpansion} by a discrete Fourier sum over integer numbers $k$, $\langle \N^n \rangle(\phi,\chi)=\sum_k\ee^{-ik\chi/\chi_0} \langle \N^n \rangle^k(\phi)$. Let us also note that since the exponential potential is conformally invariant, the shift symmetry in the inflaton field value allows us to take $\phi_\uend=0$ without loss of generality, and write $v=v_\uend\ee^{\alpha\phi/\Mp}$. For the mean number of \efolds $\langle \N \rangle$, \Eq{eq:linearpot:ode} then gives rise to
\begin{align}
 \langle \N \rangle_{\bm{k}}^{\prime\prime}(\phi)-\frac{\alpha}{\Mp v\left(\phi\right)}\langle \N \rangle_{\bm{k}}^{\prime}(\phi) - \frac{k^2}{\chi_0^2} \langle \N \rangle_{\bm{k}}(\phi)=-\frac{\delta_{k,0}}{\Mp^2 v\left(\phi\right)}\, .
\label{eq:linearpot:ode:exppot}
\end{align}
When $k=0$, the solution can be expressed in terms of the exponential integral function $\Ei$, while when $k\neq 0$, the solution is given in terms of confluent hypergeometric functions ${}_1F_1$. Requiring that $\langle \N \rangle$ is real, one obtains
\begin{align}
\langle \N \rangle(\phi,\chi)=&A_0+\frac{\phi}{\alpha\Mp}+B_0\Ei\left[-\frac{1}{v(\phi)}\right]
\nonumber\\ &+
\sum_{k=1}^\infty \left\lbrace A_k v^{\frac{k \Mp}{\alpha\chi_0}}(\phi)\, {}_1F_1\left[-\frac{k \Mp}{\alpha\chi_0},1-2\frac{k \Mp}{\alpha\chi_0},-\frac{1}{v\left(\phi\right)}\right]
\right.\nonumber \\ & \left.
+B_k v^{-\frac{k \Mp}{\alpha\chi_0}}(\phi)\, {}_1F_1\left[\frac{k \Mp}{\alpha\chi_0},1+2\frac{k \Mp}{\alpha\chi_0},-\frac{1}{v\left(\phi\right)}\right]
\right\rbrace\cos\left({k\frac{\chi}{\chi_0}}\right)\, ,
\label{eq:exponential:sol}
\end{align}
where $A_0$, $A_k$, $B_0$, $B_k$ are integration constants, that must be fixed making use of \Eqs{eq:linearpot:boundarycondition}. At this stage, one has to proceed numerically. In practice, if the summation over $k$ in \Eq{eq:exponential:sol} is truncated at order $k_\umax$, one has $2(k_\umax+1)$ integration constants to fix. One can choose $k_\umax+1$ values of $\chi$ uniformly distributed in $[0,\pi\chi_0]$, and evaluate both parts of \Eqs{eq:linearpot:boundarycondition} at these values. This gives rise to $2(k_\umax+1)$ linear equations for the $2(k_\umax+1)$ integration constants, that one can solve with standard matrix inversion methods. One then increases $k_\umax$ until a sufficient level of convergence is reached.

The result of such a procedure is displayed in the left panel of \Fig{fig:modulatedReheating}, where the mean number of \efolds~is plotted as a function of the initial values of $\phi$ and $\chi$, with $\alpha=0.1$, $\chi_0/\Mp=1$, $\mu/\Mp=0.5$ and $v_\uend=0.05$ (this last value does not lead to the right scalar power spectrum amplitude~\cite{Ade:2015xua}, but it is used to make clearer the effects we want to comment on). The value of $\phi_+$ has been taken to be sufficiently large so that it does not play any role, which is possible since the model is effectively single-field and non-plateau during inflation, as follows from the discussion in \Sec{sec:InfiniteInflation}. One has taken $k_\umax=100$, but the result is already very well converged when $k_\umax \gtrsim 8$. The black dashed lines are various level lines of $\langle\mathcal{N}\rangle$ and have been superimposed to guide the eye. The white region at the bottom left corresponds to $\phi<\phi_-(\chi)$, which lies outside of the inflationary domain. One may also note that only the region $0\leq\chi\leq\pi\chi_0$ is displayed, since the result for other values of $\chi$ can easily be inferred using the symmetry and periodicity properties of $\langle \N \rangle$. 
\paragraph{Smearing out the modulating field}
In the left panel of \Fig{fig:modulatedReheating}, one can notice that, going from the left to the right, the level lines of $\langle\mathcal{N}\rangle$ at first follow the modulation of the end-surface, and then become more and more straight. This result can be understood in terms of the two limits $v\ll 1$ and $v\gg 1$.

In the classical limit $v\ll 1$, the diffusion term acting on the extra field $\chi$ in \Eq{eq:Fpop:linear} is negligible, and $\chi$ freezes to its initial value. In this limit, the point where the system exits inflation in field space becomes deterministic and is simply given by $\phi_-(\chi_\uin)$. A similar expression to \Eq{eq:stocha:meanN:classtraj} can therefore be obtained, except that the lower bound now explicitly depends on $\chi$,
\begin{align}
N_\ucl = \frac{1}{\Mp^2}\int_{\phi_-(\chi)}^\phi\frac{v\left(\phi^\prime\right)}{v^\prime\left(\phi^\prime\right)}\dd\phi^\prime\, .
\label{eq:modulated:classicalLimit}
\end{align}
For the power-law model under consideration, this gives rise to $N_\ucl=[\phi-\mu\cos(\chi/\chi_0)]/(\Mp\alpha)$. This quantity is displayed in the right panel of \Fig{fig:modulatedReheating} where one can check that, at small field value where $v$ is not too large, it provides a good approximation to the full stochastic result given in the left panel.

In the stochastic dominated limit where $v\gg 1$, the diffusion term acting on $\chi$ becomes large, and quickly randomises the value of this extra field. In this regime, memory of the initial conditions on $\chi$ is quickly erased, which explains why the result becomes dependent on $\phi$ only and the level lines of $\langle \mathcal{N} \rangle$ in the left panel of \Fig{fig:modulatedReheating} tend to be merely vertical. Technically, one can check that the first line in \Eq{eq:exponential:sol}, which is the $0^\mathrm{th}$ (\ie $\chi$-independent) mode, provides the dominant contribution in the limit $v\gg 1$. This term exactly matches the one obtained in \Eq{eq:fn:sol:onefield} in a purely single-field setup. 
\begin{figure*}
\begin{center}
\includegraphics[width=0.49\textwidth]{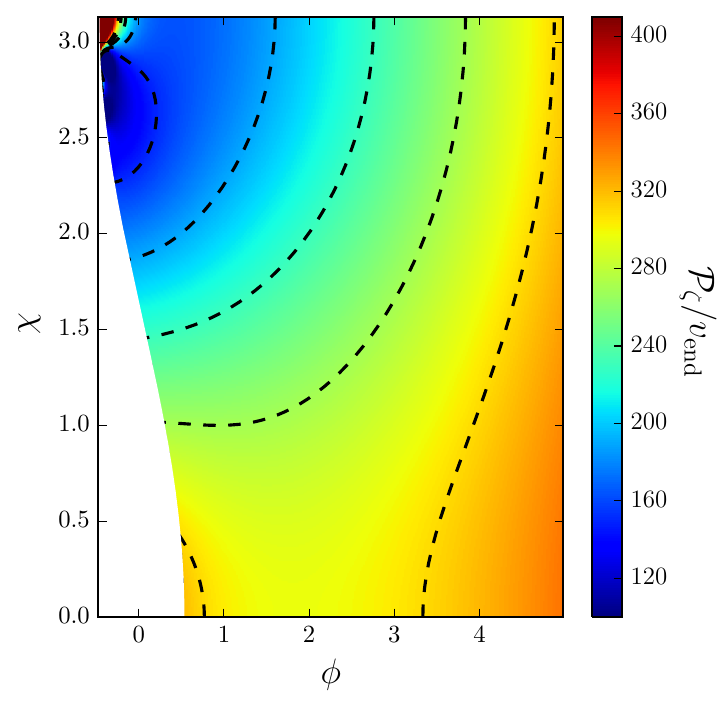}
\includegraphics[width=0.49\textwidth]{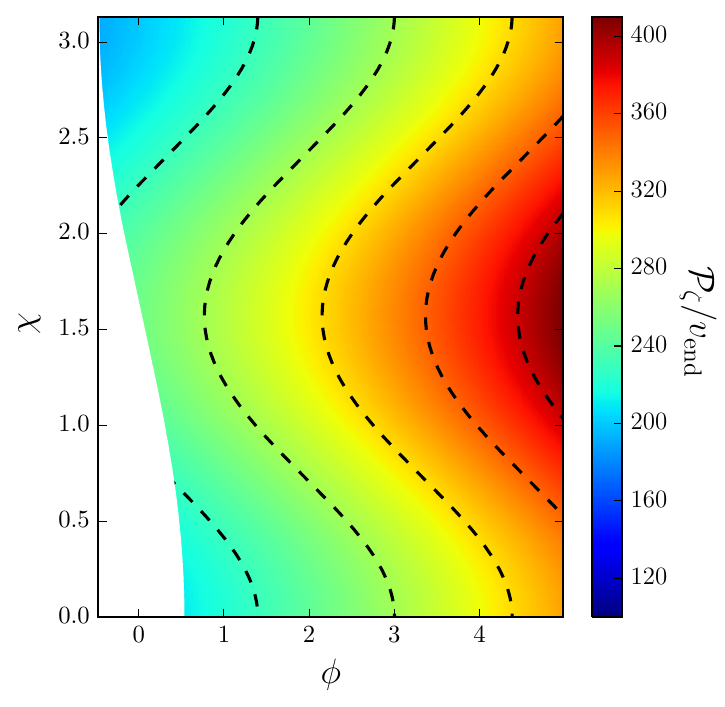}
\caption{Scalar power spectrum amplitude $\calP_\zeta$ (left panel: full stochastic results, right panel: classical formula) for the single-field potential $v=v_\uend \ee^{\alpha\phi}$ when the end of inflation is modulated by an extra field $\chi$ through $\phi_-(\chi)=\mu\cos(\chi/\chi_0)$, for $\alpha=0.1$, $\chi_0=\Mp$, $\mu=0.5\Mp$ and $v_\uend=0.05$ (this last value does not lead to the right scalar power spectrum amplitude~\cite{Ade:2015xua}, but it is used for computational convenience).
The black dashed lines are various level lines of $\calP_\zeta$, and the white regions correspond to $\phi<\phi_-(\chi)$, which is outside the inflationary domain. 
}
\label{fig:inhomoegenous}
\end{center}
\end{figure*}

In \Fig{fig:inhomoegenous}, the scalar power spectrum corresponding to the situation of \Fig{fig:modulatedReheating} is displayed. One can see that the amplitude is generically smaller than the classical prediction (since the additional contribution from the inhomogeneous end of inflation tends to be smeared out), but that the entire shape of the power spectrum is also substantially modified. 

Therefore, stochastic effects tend to erase the presence of modulating fields by blurring their initial values and averaging the exit point over the end-surface. In practice, the size of the effect depends on how the scale over which modulation takes place (denoted $\chi_0$ in the present model) compares to the dispersion acquired by the modulating field at the end of inflation, but not so much on the size of the modulation itself (here denoted $\mu$). In the simple toy model discussed in this section, if one sets $v\simeq 10^{-10}$, corresponding to the value that would fit the measured scalar power spectrum amplitude, one finds\footnote{Here, the dispersion acquired by the freely diffusing $\chi$ field is given by $\sqrt{50}H/2\pi\simeq \sqrt{100 v}\Mp$, where $50$ is taken to be the number of \efolds~between Hubble exit time of the scales probed in the CMB and the end of inflation, and $H/2\pi$ is the amplitude of the noise term in \Eq{eq:Langevin:SR:phi} which we assume to be roughly constant over these last $50$ \efolds~of inflation.} that the effect is large if $\chi_0/\Mp\lesssim 10^{-4}$. The situation is therefore different from the purely single-field case~\cite{Vennin:2015hra} where, unless the potential becomes very flat, the stochastic effects are subdominant at small values of $v$. Here, large stochastic corrections can be obtained even if $v\ll 1$, depending on the scales of the features in the end-surface.
\section{Primordial black holes and quantum diffusion}
\label{sec:PBHs}
In the previous section, \Sec{sec:stochastic:delta:N}, we have seen how the statistics of the curvature perturbation can be extracted from the stochastic-$\delta N$ formalism. We have applied this technique to compute the first moments of the probability distribution functions, which give rise to the mean duration of inflation, the power spectrum of curvature perturbations and its bispectrum. They can be constrained by measurements of the CMB anisotropies or surveys of the large-scale structure of the universe. In this section, we study another possible consequence of inflation, namely the production of large fluctuations at small scales, seeding the formation of PBHs. We will see that since PBHs require rare, large fluctuations, their abundance is determined by the tail of the distribution function of curvature perturbations, rather than by the neighbourhood of its maximum. Since the tail cannot be properly described by the few first moments of the distribution only, this requires to be beyond the results of \Sec{sec:stochastic:delta:N} and characterise the full distribution function. This is the goal of this section, which follows \Refa{Pattison:2017mbe}. We analyse two limiting regimes in detail, namely the ``classical limit'' in which stochastic effects provide a small correction, see \Sec{sec:ClassicalLimit} and the ``stochastic limit'' in which the potential gradient can be neglected and the field dynamics is only driven by the stochastic noise, see \Sec{sec:StochasticLimit}. In \Sec{sec:stocha:limit:condition}, we make explicit the regimes of applicability of these two limits, which will be a key result for the analysis of \Sec{sec:tail:expansion}. Finally, in \Sec{sec:example:1_plus_phi_to_the_p}, we study a toy-model example that interpolates between these two limits, where inflation proceeds towards a local uplifted minimum of its potential. 

In the range of scales accessible to CMB experiments~\cite{Ade:2015xua, Ade:2015lrj}, cosmological perturbations are constrained to be small, at the level $\zeta\simeq 10^{-5}$ until they re-enter the Hubble radius during the radiation era. At smaller scales however, they may be sufficiently large so that when they re-enter the Hubble radius, they overcome the pressure forces and make some Hubble patches collapse into PBHs~\cite{Hawking:1971ei, Carr:1974nx, Carr:1975qj}.
For causality reasons, it is often argued that whether or not this occurs can only depend on the value of the curvature perturbation inside the Hubble patch under consideration. This is why the coarse-grained curvature perturbation, defined as the mean value of the curvature perturbation over a Hubble patch, of comoving volume $(aH)^3$, is usually considered,
\bea
\zeta_{\mathrm{cg}} (\bm{x}) = (aH)^3 \int\dd\bm{y} \zeta(\bm{y}) W\left(aH\left\vert \bm{y} - \bm{x}\right\vert \right),
\eea
where $W$ is a window function such that $W(x)\simeq 1$ if $x\ll 1$ and $W(x)\simeq 0 $ if $x\gg 1$, and normalised in the sense that $ 4\pi \int_0^\infty x^2 W(x)\dd x=1$, such that after coarse graining, a constant field remains a constant field of the same value. A usual criterion for PBH formation is that when $\zetacg(\bm{x})$ exceeds a certain threshold $\zeta_\uc\simeq 1$~\cite{Zaballa:2006kh, Harada:2013epa} (see \Refa{Young:2014ana} for an alternative criterion based on the density contrast rather than the curvature perturbation), the Hubble patch centred on $\bm{x}$ collapses and forms a black hole.

The Fourier transform of this coarse-grained curvature perturbation is given by
\bea
\label{eq:def:tilde:W}
\zeta_{\mathrm{cg}} (\bm{k}) = 
\zeta(\bm{k})
\underbrace{
4\pi
\left(\frac{aH}{k}\right)^3 \int_0^\infty W\left(\frac{aH}{k} u \right)  \sin(u) u\dd u}_{ \widetilde{W}\left(\frac{k}{aH}\right)}
,
\eea
which defines $\widetilde{W}$, that shares similar properties with $W$. Indeed, when $aH/k \gg 1$, the values of $u$ such that $W\left(\frac{aH}{k} u \right)$ is not close to zero are very small, so one can replace $\sin(u)\simeq u$ in the integral over $u$, and using the normalisation condition stated above, one obtains $\widetilde{W}\left[k/(aH)\right] \simeq 1$ in that limit. In the opposite limit, when $aH/k \ll 1$, since $W$ is roughly $1$ until $u\sim k/(aH)$, the integral over $u$ in \Eq{eq:def:tilde:W} is of order $k/(aH)$, hence $\widetilde{W}\left[k/(aH)\right]\propto (aH/k)^2\ll 1 $. 

The details of $\widetilde{W}$ between these two limits depend on those of $W$. For instance, if $W$ is a Heaviside step function,
\bea
W(x) = \frac{3}{4\pi} \theta(1-x),
\eea
where $\theta(x)=1$ if $x>0$ and $0$ otherwise, and where the pre-factor is set in such a way that the above normalisation condition is satisfied, \Eq{eq:def:tilde:W} gives rise to
\bea
\label{eq:Fourier:transform:Heaviside}
\widetilde{W}\left(\frac{k}{aH}\right) = 3 \left(\frac{a H}{k}\right)^3 \left[ \sin \left(\frac{k}{a H}\right)-\frac{k}{a H}\cos\left(\frac{k}{a H}\right)\right],
\eea
which verifies the two limits given above. 

There is some freedom in the choice of the window function, and different window functions can lead to rather different results for the PBH abundance~\cite{Blais:2002gw, Ando:2018qdb}. In fact, comparing the locally coarse-grained curvature perturbation with a certain threshold is only an approximated procedure. More realistic approaches incorporate the full real-space profile of the density contrast across the inhomogeneity, either numerically or by studying the compaction function~\cite{Musco:2004ak, Polnarev:2006aa, Musco:2018rwt, Kalaja:2019uju}. Although such analyses reveal the existence of a variety of different situations, depending on the details of the density profile, in most cases the scales that contribute most to forming a PBH of mass $M$ are those around the Hubble radius when it contains that mass: much smaller scales average out inside the inhomogeneity, as the calculation above indicates, and much larger scales simply rescale the local amplitude of the background density. For this reason, one can consider a coarse-grained field made up of scales ``around'' the Hubble radius only,
\bea
\label{eq:zetacg:def:1}
\zeta_{\mathrm{cg}}(\bm{x}) = \left(2\pi\right)^{-3/2}\int_{k\sim a_{\mathrm{form}}H_{\mathrm{form}}}\dd {\bm{k}}\zeta_{\bm{k}} e^{i\bm{k}\cdot\bm{x}}\, .
\eea
In this expression, ``$k\sim a_{\mathrm{form}}H_{\mathrm{form}}$'' implies the existence of a window function that we now need to specify. For simplicity, we consider
\bea
\label{eq:zetacg:def}
\zeta_{\mathrm{cg}}(\bm{x}) = \left(2\pi\right)^{-3/2}\int_{a_{\mathrm{form}}H_{\mathrm{form}}<k<a_\uend H_\uend}\dd {\bm{k}}\zeta_{\bm{k}} e^{i\bm{k}\cdot\bm{x}}\, ,
\eea
\ie a top hat window function in Fourier space, that selects out modes comprised between the Hubble scale at the time of formation and the one at the end of inflation. Since we consider scales that are generated a few \efolds~before the end of inflation, we are integrating over a few \efolds~of scales as we should. Obviously, the details of the window function (its shape and the precise range of scales) are arbitrary, but what makes this choice convenient is that \Eq{eq:zetacg:def} coincides with \Eq{eq:deltaNcg:zeta}. Therefore, the statistics of $\zeta_{\mathrm{cg}}$ is precisely the one of $\delta N_{\mathrm{cg}}$ computed in the stochastic $\delta N$ formalism. 

Obviously, one could use another choice of window function and define the coarse-grained curvature perturbation differently. This would imply to coarse grain the fields differently in the stochastic inflation formalism. For instance, if a smooth $\widetilde{W}$ function is employed, the stochastic noise contains contributions from different modes, so the same mode contributes to the realisation of the noise at different times, and the noise becomes coloured. While coloured noises can be dealt with in the stochastic inflation formalism (see \eg \Refs{Casini:1998wr, Winitzki:1999ve, Matarrese:2003ye, Liguori:2004fa}), they are technically more challenging, which explains our choice.

The abundance of PBHs is usually stated in terms of the mass fraction of the universe contained within PBHs at the time of formation, $\beta_{\mathrm{f}}$. If the coarse-grained curvature perturbation $\zeta_{\mathrm{cg}}$ follows the probability distribution function (PDF) $P(\zeta_{\mathrm{cg}})$, $\beta_{\mathrm{f}}$ is given by~\cite{1975ApJ...201....1C}
\bea
\label{eq:beta:def}
\beta_{\mathrm{f}}\left(M\right) = \int_{\zeta_\uc}^\infty P\left(\zeta_{\mathrm{cg}}\right) \dd \zeta_{\mathrm{cg}}\, .
\eea
Here, $M$ is (a fraction of) the mass contained in a Hubble patch at the time of formation~\cite{Choptuik:1992jv, Niemeyer:1997mt, Kuhnel:2015vtw}, $M=3\Mp^2/H_{\mathrm{form}}$

As explained in \Sec{sec:Context}, see the discussion around \Fig{fig:PBHs:constraints}, observational constraints on $\beta_{\mathrm{f}}$ depend on the masses PBHs have when they form. For masses between $10^9\mathrm{g}$ and $10^{16} \mathrm{g}$, the constraints mostly come from the effects of PBH evaporation on big bang nucleosynthesis and the extragalactic photon background, and typically range from $\beta_{\mathrm{f}}<10^{-24}$ to $\beta_{\mathrm{f}}<10^{-17}$. Heavier PBHs, with mass between $10^{16} \mathrm{g}$ and $10^{50} \mathrm{g}$, have not evaporated yet and can only be constrained by their gravitational and astrophysical effects, at the level $\beta_{\mathrm{f}}<10^{-11}$ to $\beta_{\mathrm{f}}<10^{-5}$ (see \Refs{Carr:2009jm, Carr:2017jsz} for summaries of constraints).

Compared to the CMB anisotropies that allow one to measure $\zeta$ accurately in the largest $\sim 7$ \efolds~of scales in the observable Universe, PBHs only provide upper bounds on $\beta_{\mathrm{f}}(M)$, and hence on $\zeta$. However, these constraints span a much larger range of scales and therefore yield valuable additional information. This is why PBHs can be used to constrain the shape of the inflationary potential beyond the $\sim 7$ \efolds~that are accessible through the CMB. 

In practice, one usually assumes $P(\zeta_{\mathrm{cg}})$ to be a Gaussian PDF with standard deviation given by the integrated power spectrum $\left\langle \zeta_{\mathrm{cg}}^2 \right\rangle = \int_{k}^{k_\uend} \calP_\zeta(\tilde{k})\dd \ln \tilde{k}$, which follows from \Eq{eq:zetacg:def}, and where $k$ is related to the time of formation through $k=aH_{\mathrm{form}}$, and where $k_\uend$ corresponds to the wavenumber that exits the Hubble radius at the end of inflation. Combined with \Eq{eq:beta:def}, this gives rise to
\bea
\label{eq:beta:erfc}
\beta_{\mathrm{f}}\left(M\right) = \frac{1}{2} \erfc\left[\frac{\zeta_\uc}{\sqrt{2  \int_{k}^{k_\uend} \calP_\zeta(\tilde{k})\dd \ln \tilde{k}}}\right]\, ,
\eea
where $\erfc$ is the complementary error function. In the limit $\beta_{\mathrm{f}} \ll 1$, this leads to $\int_{k}^{k_\uend} \calP_\zeta(\tilde{k})\dd \ln \tilde{k} \simeq \zeta_\uc^2/(-2\ln\beta_{\mathrm{f}})$. Assuming the power spectrum to be scale invariant, one has $\int_{k}^{k_\uend} \calP_\zeta(\tilde{k})\dd \ln \tilde{k} \simeq \calP_\zeta \ln(k_\uend/k)  \simeq \calP_\zeta \Delta N$, where $\Delta N $ is the number of \efolds~elapsed between the Hubble radius exit times of $k$ and $k_\uend$ during inflation. This leads to
\bea
\label{eq:Pzetaconstraint:standard}
\calP_\zeta \Delta N \simeq - \frac{\zeta_\uc^2}{2\ln\beta_{\mathrm{f}}}\, .
\eea
For instance, with $\zeta_\uc=1$, the bound $\beta_{\mathrm{f}}<10^{-22}$ leads to the requirement that $\calP_\zeta \Delta N< 10^{-2}$. This can be translated into constraints on the inflationary potential $V=24\pi^2\Mp^4 v$ and its derivative $V^\prime$ with respect to the inflaton field $\phi$ using the single-field slow-roll formulae~\cite{Mukhanov:1985rz, Mukhanov:1988jd}, $\calP_\zeta = 2v^3/(\Mp^2 {v^\prime}^2)$ and $\Delta N=\int_{\phi_\uend}^\phi v/({\Mp^2v'})   \dd \tilde{\phi}$, see \Eqs{eq:stocha:meanN:classtraj} and~\eqref{eq:PS:vll1}.

These consideration however rest on very strong assumptions, namely the use of a Gaussian PDF for $P(\zeta_{\mathrm{cg}})$ together with the classical slow-roll formula for the curvature power spectrum $\calP_\zeta$ and number of \efolds~$\Delta N$, which are valid only in the regime where quantum diffusion provides a subdominant correction to the classical field dynamics during inflation. However, producing curvature fluctuations of order $\zeta\sim \zeta_\uc\sim 1$ or higher precisely corresponds to the regime where quantum diffusion dominates the field dynamics over a typical time scale of one \efold. The validity of the standard approach that is summarised above is therefore questionable and this is why, in this section, we address the problem in full stochastic inflation, making use of the techniques developed in \Sec{sec:stochastic:delta:N}, and more precisely the tools introduced in \Sec{sec:FPT:full:PDF} that allow one to derive the full PDF of curvature perturbations. 

We restrict our analysis to single-field slow-roll inflation, and the results presented below follow \Refa{Pattison:2017mbe}. We first derive the classical limit of the procedure outlined in \Sec{sec:FPT:full:PDF}, and check that the above considerations are recovered at leading order in the classical expansion. We however show that this expansion is under control only for observables probing the neighbourhood of the maximum of the PDF (such as the mean number of \efolds~or the power spectrum), but not for observables that depend on the tail of the PDF, such as the PBH abundance. We then study the opposite limit, where stochastic effects dominate, and where the tail of the PDF can be fully characterised. The regimes of validity of both these limits is discussed. A concrete example is finally studied, where inflation proceeds while approaching a local minimum of the potential. We show how the results obtained in the stochastic limit can be used to characterise the abundance of PBHs in this model, where we find that PBHs are overproduced unless slow roll is violated.
\subsection{Expansion about the classical limit}
\label{sec:ClassicalLimit}
In order to exemplify the calculational program outlined in \Sec{sec:FPT:full:PDF}, and also to check its consistency in the regime of low stochastic diffusion, we first work out the classical limit of the PDF of $\zeta_{\mathrm{cg}}$. More precisely, we want to check that, in the ``classical'' limit, our formulation allows one to recover the standard results recalled above. This is the goal of this section, where we also calculate the leading order deviation from the standard result in order to best determine its range of validity. 
\subsubsection{The characteristic function approach}
\label{sec:classicalAppr:characteristicMethod}
In single-field slow-roll inflation, the characteristic function satisfies an ordinary differential equation, \Eq{eq:ODE:chi}, where the adjoint Fokker-Planck operator is given by \Eq{eq:Fokker:Planck:adjoint:PDF:Ito:SlowRoll}, \ie 
\bea
\label{eq:ODE:chi:SF:SR}
\left(v \frac{\partial^2}{\partial\phi^2}-\frac{v'}{v}\frac{\partial}{\partial\phi}+\frac{i t}{\Mp^2}\right)\chi_\N(t,\phi) =0.
\eea
In this setup, an absorbing boundary is placed at $\phi_\uend$, so $\chi(t,\phi_\uend)=1$, and a reflective boundary is placed at $\phiuv$, so $\chi'(t,\phiuv)=0$. From \Eq{eq:ODE:chi:SF:SR}, one can see that an expansion in $v$ is equivalent to an expansion in the diffusion term, involving $\partial^2/\partial\phi^2$.
\paragraph{Leading order}
At leading order (LO) in the classical limit, the diffusion term in \Eq{eq:ODE:chi:SF:SR} can be simply neglected, and one has
\bea
\label{eq:ODE:chi:classical}
\left(- \frac{v^\prime}{v} \frac{\partial}{\partial\phi} + \frac{it}{\Mp^2}\right)\chi^{\lo}_\N(t,\phi)=0\, .
\eea 
Making use of the absorbing boundary condition at $\phi_\uend$,\footnote{In the expansion about the classical limit, there is no need to introduce a second boundary condition at $\phiuv $.} this equation can be solved as
\bea
\label{eq:chi:classical:LO}
\chi_\N^\mathrm{\lo}(t,\phi) = \exp\left[it\int_{\phiend}^{\phi} \frac{v(x)}{\Mp^2v^\prime(x)}\dd x\right]\, .
\eea
Note that the integral in the argument of the exponential is the classical number of \efolds~obtained in \Eq{eq:stocha:meanN:classtraj}, which is also the mean number of \efolds~at leading order in the classical limit, see the discussion around \Eq{eq:Nmean:vll1limit}. This is consistent with \Eq{eq:meanN:chi}. 

\begin{figure}[t]
\begin{center}
\includegraphics[width=0.6\textwidth]{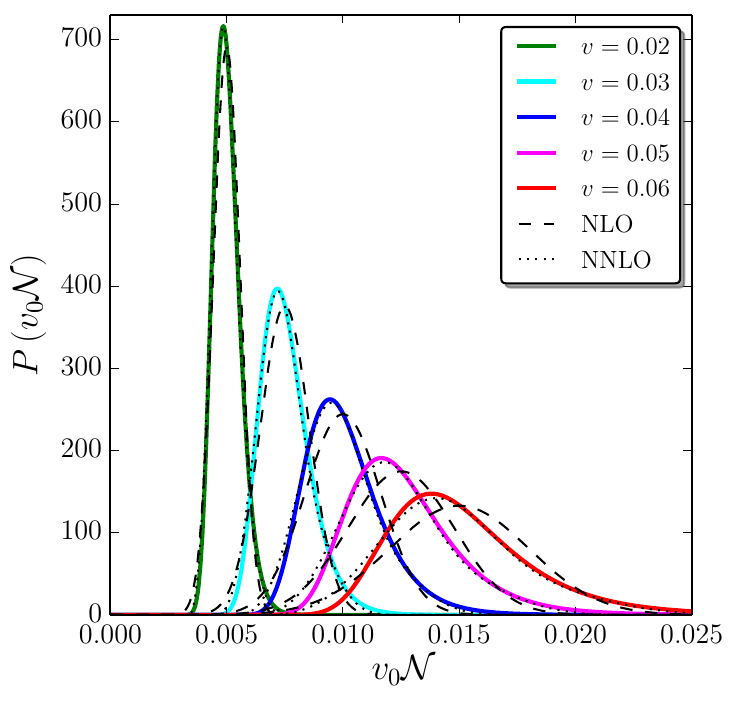}
\caption{Probability distributions of the number of \efolds~$\N$, rescaled by $v_0$, realised in the quadratic potential $v=v_0 (\phi/\Mp)^2$, between an initial field value $\phi$ parametrised by $v(\phi)$ given in the legend, and $\phi_\uend = \sqrt{2}\Mp$ where inflation ends by slow-roll violation. The coloured lines stand for a numerical integration of \Eq{eq:ODE:chi:SF:SR}, which is then Fourier transformed to obtain the PDF. The black dashed lines correspond to the NLO (Gaussian) approximation~(\ref{eq:PDF:classical:NLO}), while the dotted lines stand for the NNLO approximation~(\ref{eq:PDF:NNLO}). The smaller $v$ is, the better these approximations are, and the NNLO approximation is substantially better than the NLO one.} 
\label{fig:pdf_quadratic_solution_nnlo}
\end{center}
\end{figure}

As a consequence, \Eq{eq:chideltaN:chiN} implies that $\chi_{\delta N_{\mathrm{cg}}} = 1$, and hence its inverse Fourier transform is $P^\lo\left(\delta N_\mathrm{cg}, \phi\right) = \delta \left(\delta N_\mathrm{cg}\right)$, \ie a Dirac distribution centred around $\delta N_\mathrm{cg} = 0$. 
Thus, at leading order in the classical limit, one simply shuts down quantum diffusion, the dynamics are purely deterministic, $\delta\N \equiv 0$ and there are no curvature perturbations. 
\paragraph{Next-to-leading order}
One thus needs to go to next-to-leading order (NLO) to incorporate curvature perturbations. At NLO, the LO solution~(\ref{eq:chi:classical:LO}) can be used to evaluate the term $\chi^{-1}\partial^2\chi/\partial\phi^2$ in \Eq{eq:ODE:chi:SF:SR}, which then becomes
\bea
\frac{\partial}{\partial\phi}\chi_\N^\nlo -\frac{v^2}{v^\prime} \left(\frac{it}{v\Mp^2}+ \frac{1}{\chi_\N^\lo}\frac{\partial^2 \chi_\N^\lo}{\partial\phi^2} \right)
\chi_\mathcal{N}^\nlo  =0\, .
\eea
Making use of the boundary condition at $\phi_\uend$, the solution of this first order ordinary differential equation is
\bea
\label{eq:chi:classical:interativeSolution}
\chi_\N^\nlo (t,\phi) = \exp\left\lbrace \int_{\phi_\uend}^\phi \left[ \frac{itv(x)}{\Mp^2v' (x)} + \frac{v^2(x)}{v'(x)} \frac{1}{\chi_\N^\lo(x)}\frac{\partial^2\chi_\N^\lo}{\partial\phi^2}(x) \right] \dd x\right\rbrace\, .
\eea
Notice that if ones replaces $\lo$  by an arbitrary n${}^\mathrm{th}$ order and $\nlo$ by the n+1${}^\mathrm{th}$ order of the classical expansion, this equation is valid at any order since it is nothing but the iterative solution of \Eq{eq:ODE:chi:SF:SR}. At NLO, plugging \Eq{eq:chi:classical:LO} into \Eq{eq:chi:classical:interativeSolution}, one obtains
\bea
\label{eq:chi:classical:NLO}
\chi_{\N}^\nlo(t,\phi)  = \exp\left[ it \langle \N \rangle^\nlo\left(\phi\right) - \gamma_{1}^\nlo v t^2  \right] \, ,
\eea
where $\langle \N \rangle^\nlo$ is the mean number of \efolds~at NLO given in \Eq{eq:Nmean:vll1limit}, and we have defined
\bea
\label{eq:gamma1:nlo:def}
\gamma_{1}^\nlo = \frac{1}{v\Mp^4} \int_{\phiend}^\phi \dd x \frac{v^4}{{v^\prime}^3} \, .
\eea
From this expression, \Eq{eq:chideltaN:chiN} implies that $\chi_{\delta N_\mathrm{cg}}^\nlo\left(t,\phi\right) = \ee^{-\gamma_{1}^\nlo v t^2}$, that is to say $\chi_{\delta\N}^\nlo$ is a Gaussian and hence its inverse Fourier transform $P^\nlo\left(\zeta_\mathrm{cg}, \phi\right)$ is also a Gaussian and is given by
\bea
\label{eq:PDF:classical:NLO}
P^\nlo(\zeta_{\mathrm{cg}}, \phi) = \frac{1}{\sqrt{4\pi\gamma_{1}^\nlo v}} \exp\left(-\frac{\zeta_{\mathrm{cg}}^2}{4\gamma_{1}^\nlo v}\right) \, .
\eea
A crucial remark is that at this order, the power spectrum~(\ref{eq:Pzeta:stochaDeltaN}) is given by $\calP_\zeta = 2v^3/(\Mp^2 {v^\prime}^2)$, see \Eq{eq:PS:vll1}, so  the variance of the Gaussian distribution~(\ref{eq:PDF:classical:NLO}) reads $2\gamma_{1}^\nlo v = \int_{\phiend}^{\phi} \calP_\zeta^{\nlo} \langle \N \rangle^{'\lo} \dd x$. This precisely matches the standard result recalled above \Eq{eq:beta:erfc}, namely that $P(\zeta_{\mathrm{cg}})$ is a Gaussian PDF with standard deviation given by the integrated power spectrum $\left\langle \zeta_{\mathrm{cg}}^2 \right\rangle = \int_{k}^{k_\uend} \calP_\zeta(\tilde{k})\dd \ln \tilde{k}$, since $\dd \ln k \simeq \dd N = \langle \N \rangle'(\phi) \dd \phi$ at leading order in slow roll.
\paragraph{Next-to-next-to-leading order}
In order to study the first non-Gaussian corrections to the standard result, one needs to go to next-to-next-to-leading order (NNLO). As explained above, one can simply increment the order of the iterative relation~(\ref{eq:chi:classical:interativeSolution}), \ie replace $\lo$ by $\nlo$ and $\nlo$ by $\nnlo$. Plugging in \Eq{eq:chi:classical:NLO}, and making use of \Eq{eq:chideltaN:chiN}, this gives rise to
\bea
\label{eq:chiNNLO}
\chi_{\delta N_{\mathrm{cg}}}^\nnlo\left(t,\phi\right)  = \exp\left( - \gamma_{1}^\nnlo v t^2 - i\gamma_{2}^\nnlo v^2t^3 
 \right)  \, ,
\eea
where we have only kept the terms that are consistent at that order and where we have defined
\bea
\label{eq:gamma:nnlo:def}
\gamma_{1}^\nnlo &=& \frac{1}{v\Mp^4 }  \int_{\phiend}^\phi\dd x\left(\frac{v^4 }{v'^3}+6\frac{v^5 }{v'^3}-5\frac{v^6 v''}{v'^5}\right) \, ,\\
\gamma_{2}^\nnlo &=& \frac{2}{v^2 \Mp^6}  \int_{\phiend}^\phi\dd x \frac{v^7}{v'^5} \, .
\eea
One can already see that since the characteristic function is not a Gaussian, the PDF is not a Gaussian distribution. Using \Eq{eq:PDF:chi}, it is given by
\bea
\label{eq:PDF:NNLO:chi}
P^{\nnlo}\left( \delta N_{\mathrm{cg}}, \phi \right) = \frac{1}{2\pi}\int_{-\infty}^{\infty} \dd t \exp\left( -it\delta N_{\mathrm{cg}} - \gamma_{1}^\nnlo vt^2 + i\gamma_{2}^\nnlo v^2t^3  \right) \, .
\eea
In this integral, the second term in the argument of the exponential makes the integrand become negligible when $\gamma_1^\nnlo v t^2 \gg 1$, \ie for $\vert t \vert \gg t_\uc$ where $t_\uc = (\gamma_{1}^\nnlo v )^{-1/2}$. When $t=\pm t_\uc$, the ratio between the third and the second terms in the argument of the exponential of \Eq{eq:PDF:NNLO:chi} is of order $(\gamma_2^\nnlo / \gamma_1^\nnlo) \sqrt{v/\gamma_1^\nnlo}$, \ie of order $\sqrt{v}$ in an expansion in $v$ since the $\gamma_i$ parameters have been defined to carry no dimension of $v$ (at least at their leading orders). This is why, over the domain of integration where most of the contribution to the integral comes from, the third term is negligible and can be Taylor expanded. One obtains 
\bea
\label{eq:PDF:NNLO}
P^{\nnlo}\left( \zeta_\mathrm{cg}, \phi \right)  = \frac{1}{\sqrt{4 \pi \gamma_{1}^\nnlo v}} \exp\left(-\frac{\zeta_\mathrm{cg}^2}{4\gamma_{1}^\nnlo v}\right)\left[ 1 - \frac{\gamma_{2}^\nnlo}{8\left({\gamma_{1}^\nnlo}\right)^3 v} \zeta_\mathrm{cg} \left( 6 \gamma_{1}^\nnlo v - \zeta_\mathrm{cg}^2 \right) \right] \, .
\eea 
In \Fig{fig:pdf_quadratic_solution_nnlo}, we compare the result of a numerical integration of \Eq{eq:ODE:chi:SF:SR}, which is then Fourier transformed to obtain the PDF, with the NLO approximation~(\ref{eq:PDF:classical:NLO}) and the NNLO approximation~(\ref{eq:PDF:NNLO}). One can check that these approximations become better at smaller values of $v$ as expected, and that the NNLO approximation always provides a better fit than the NLO one. 

As a consistency check, one can verify that the distribution~(\ref{eq:PDF:NNLO}) yields the same moments at NNLO order as the ones derived in \Sec{sec:FirstMoments:N}. For the second moment, one has $\langle \delta N_{\mathrm{cg}}^2 \rangle = \int^{\infty}_{-\infty} \zeta_{\mathrm{cg}}^2 P^{\nnlo} ( \zeta_\mathrm{cg}, \phi ) \dd \zeta_{\mathrm{cg}} = 2 \gamma_1^\nnlo v$, which coincides with \Eq{eq:deltaN2:classlim}. Similarly for the third moment, $\langle \delta N_{\mathrm{cg}}^3 \rangle = \int^{\infty}_{-\infty} \zeta_{\mathrm{cg}}^3 P^{\nnlo} ( \zeta_\mathrm{cg}, \phi ) \dd \zeta_{\mathrm{cg}} = 6 \gamma_2^\nnlo v^2$, which coincides with \Eq{eq:skewness:class}. The two methods, \ie the iterative solution~(\ref{eq:chi:classical:interativeSolution}) of the characteristic function equation and the saddle-point expansion of the moment integrals performed in \Sec{sec:FirstMoments:N}, are therefore equivalent.

Let us also note that the characteristic function, $\chi_{\cal N}(t,\phi)$ defined in \Eq{eq:characteristicFunction:def}, is closely related to the cumulant generating function for the probability distribution
\bea
\label{defKtau}
K_{\cal N}(\tau,\phi) = \ln \langle e^{\tau {\cal N}(\phi)} \rangle = \sum_{n=1}^\infty \frac{\kappa_n(\phi)}{n!} \tau^n \,.
\eea
By comparing \Eqs{eq:characteristicFunction:def} and~(\ref{defKtau}) indeed, one simply has $\chi_{\cal N}(t,\phi)=\exp\left[K_{\cal N}(it,\phi)\right]$. If we now compare \Eqs{eq:chiNNLO} and~(\ref{defKtau}), we can read off the first cumulants
\bea
\kappa_2(\phi) = 2 v \gamma_{1} \,,  \quad \kappa_3(\phi) = 6v^2 \gamma_{2} \,.
\eea
One measure of the deviation from a Gaussian distribution is the skewness of the distribution which is determined by the ratio of these cumulants
\bea
\label{eq:gamma_skew}
\gamma_{\mathrm{skew}} \equiv \frac{\kappa_3}{\kappa_2^{3/2}} = \frac{3v^{1/2}\gamma_{2}}{\sqrt{2}(\gamma_{1})^{3/2}}
\,.
\eea
Since $\gamma_2$ is non-vanishing at next-to-next-to-leading (and higher) order only, the NNLO term thus represents the first non-Gaussian correction to the standard Gaussian result obtained at NLO in the classical limit. 

At this order, the distribution is positively skewed, which is indeed the case for all the distributions displayed in \Fig{fig:pdf_quadratic_solution_nnlo}. One can also note that the parameter introduced below \Eq{eq:PDF:NNLO:chi}, that must be small in order for  the classical expansion to be valid at NNLO, exactly coincides with $\gamma_{\mathrm{skew}}$. The above formulae are therefore correct in the limit $\gamma_{\mathrm{skew}}\ll 1$ only. Finally, the correcting term in the brackets of \Eq{eq:PDF:NNLO} can be expressed as $\gamma_{\mathrm{skew}} \left(\zeta_\mathrm{cg}^2/\left\langle \zeta_\mathrm{cg}^2 \right\rangle\right)^{3/2}$, where we have used the relation $\langle \delta N_{\mathrm{cg}}^2 \rangle =  2 \gamma_1^\nnlo v$ given above together with \Eq{eq:gamma_skew}. This shows that $\gamma_{\mathrm{skew}}\ll 1$ only ensures the correcting term to be small when $\zeta_\mathrm{cg}^2$ is of order $\left\langle \zeta_\mathrm{cg}^2 \right\rangle$, \ie in the neighbourhood of the maximum of the distribution. The classical approximation is therefore an expansion that is valid in the neighbourhood of the maximum of the distribution, and that has no reason to be reliable in the tail of the distribution. Since the PBH threshold $\zeta_\uc$ is usually in the far tail of the distribution (even in the standard calculation recalled above, at the level of the observational bounds, one has $\left\langle \zeta_\mathrm{cg}^2 \right\rangle \sim 10^{-2} \ll \zeta_\uc^2\sim 1$), one needs to go beyond the classical approximation to properly assess the abundance of PBHs.
\subsubsection{The heat equation approach}
Before moving on to the stochastic limit, let us briefly explain how the heat equation approach proceeds in the classical limit. Plugging the adjoint Fokker-Planck operator~\eqref{eq:Fokker:Planck:adjoint:PDF:Ito:SlowRoll} into \Eq{eq:Fokker:Planck:adjoint:PDF}, one has to solve
\bea
\label{eq:Fokker:Planck:adjoint:PDF:SF:SR}
\left(v \frac{\partial^2}{\partial\phi^2}-\frac{v'}{v}\frac{\partial}{\partial\phi}-\frac{1}{\Mp^2}\frac{\partial}{\partial\N}\right)P\left( \N, \phi \right)=0\, .
\eea
At LO, neglecting the diffusion term (\ie the one proportional to $\partial^2/\partial\phi^2$), and imposing an absorbing boundary at $\phi_\uend$, $P( \N, \phi_\uend )=\delta(\N)$, \Eq{eq:Fokker:Planck:adjoint:PDF:SF:SR} can be solved using the method of characteristics, and one obtains
\bea
\label{eq:heat:classical:LO:solution}
P^{\lo} \left( \N, \phi \right) = \delta \left[ \N - \langle \N \rangle^{\lo}\left(\phi\right) \right] \, ,
\eea
where $ \langle \N \rangle^{\lo}$ has been defined in \Eq{eq:stocha:meanN:classtraj} and corresponds to the classical number of \efolds, which is also the mean number of e-folds at leading order in the classical limit. One therefore recovers the result of \Sec{sec:classicalAppr:characteristicMethod}. At NLO, one can use \Eq{eq:heat:classical:LO:solution} to calculate the diffusive term in \Eq{eq:Fokker:Planck:adjoint:PDF:SF:SR} and iterate the procedure. However, by doing so, one has to solve a first-order partial differential equation with a source term that involves derivatives of the Dirac distribution. This makes the solving procedure technically complicated, and we therefore do not pursue this direction further since a simpler way to obtain the solution was already presented in \Sec{sec:classicalAppr:characteristicMethod}. One can already see the benefit of having two solving procedures at hand, which will become even more obvious in what follows.
\subsubsection{Primordial black holes}
\label{sec:PBH:classical}
Let us now see what these considerations imply for the production of PBHs. At NLO, the PDF is approximately Gaussian, see \Eq{eq:PDF:classical:NLO}, and the considerations presented at the beginning of \Sec{sec:PBHs} apply. Plugging \Eq{eq:PDF:classical:NLO} into \Eq{eq:beta:def}, one has $\beta_{\mathrm{f}} = \erfc[\zeta_\uc/(2\sqrt{v\gamma_1})]/2$, which is consistent with \Eq{eq:beta:erfc} as noted below \Eq{eq:PDF:classical:NLO}. In the $\beta_{\mathrm{f}}\ll 1$ limit, this leads to 
\bea 
\label{eq:constraint:classical}
v \gamma_1 \simeq -\frac{\zeta_\uc^2}{4 \ln \beta_{\mathrm{f}}}\, ,
\eea
where from now on, the order at which the $\gamma_i$ parameters are calculated is omitted for simplicity. Approximating $\gamma_1$ given in \Eq{eq:gamma1:nlo:def} by $\gamma_1\simeq (v/v')^3 \Delta\phi/\Mp^4$, where $\Delta\phi = \vert \phi-\phiend\vert$ is the field excursion, one obtains
\bea
\label{eq:Vconstraint:standard}
\left\vert  \frac{\Delta \phi v^4}{ {v^{\prime}}^3 \Mp^4} \right\vert \simeq - \frac{\zeta_\uc^2}{4\ln\beta_{\mathrm{f}}(M)}\, .
\eea
In this expression, let us recall that the left-hand side must be evaluated at a value $\phi$ which is related to the PBH mass $M$ by identifying the wavenumber that exits the Hubble radius during inflation at the time when the inflaton field equals $\phi$, with the one that re-enters the Hubble radius during the radiation-dominated era when the mass contained in a Hubble patch equals $M$. For instance, with $\zeta_\uc=1$, the bound $\beta_{\mathrm{f}}<10^{-22}$ leads to the requirement that the left-hand side of \Eq{eq:Vconstraint:standard} be smaller than $0.005$, which constrains the shape of the inflationary potential.

At NNLO, plugging \Eq{eq:PDF:NNLO} into \Eq{eq:beta:def}, one obtains 
\bea
\beta_{\mathrm{f}}(M) = \frac{1}{2}\erfc\left(\frac{\zeta_\uc}{2\sqrt{v\gamma_1}}\right) + \frac{\gamma_2}{8\sqrt{v\pi\gamma_1^5}}\ee^{-\frac{\zeta_\uc^2}{4 v \gamma_1}}\left(\zeta_\uc^2-2v\gamma_1\right)\, .
\eea
In the $\beta_{\mathrm{f}}\ll 1$ limit, \ie in the $\zeta_\uc^2 \gg v \gamma_1$ limit, this reads $\beta_{\mathrm{f}}\simeq  \ee^{-\zeta_\uc^2/(4 v \gamma_1)}\sqrt{v\gamma_1/\pi}/\zeta_\uc[1+\gamma_2 \zeta_\uc^3/(8 v \gamma_1^{3})]$. In this regime, one can see that the non-Gaussian correction is in fact larger than the Gaussian leading order, which signals that the non-Gaussian expansion breaks down on the far tail of the distribution. This confirms that non-Gaussianities cannot be simply treated at the perturbative level when it comes to PBH mass fractions~\cite{Young:2015cyn}.

\subsection{The stochastic limit}
\label{sec:StochasticLimit}
\begin{figure}[t]
\begin{center}
\includegraphics[width=0.6\textwidth]{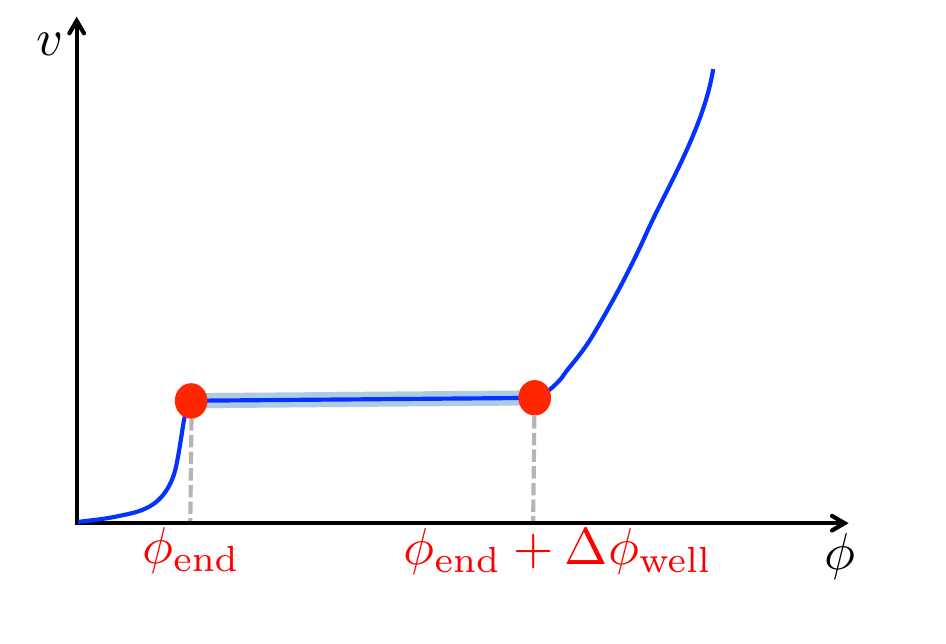}
\caption{
Schematic representation of the single-field stochastic dynamics solved in \Sec{sec:StochasticLimit}, where the potential may be taken to be 
exactly constant over the ``quantum well'' regime delimited by $\phi_\uend$ and $\phi_\uend+\Delta\phiwell$. Inflation terminates at $\phi_\uend$, where either the potential becomes very steep or a mechanism other than slow-roll violation ends inflation, and a reflective wall is placed at $\phi_\uend+\Delta\phiwell$, which can be seen as the point where the dynamics become classically dominated and the classical drift prevents the field from escaping the quantum well.}  
\label{fig:sketch2}
\end{center}
\end{figure}
We now consider the opposite limit where the inflaton field dynamics are dominated by quantum diffusion. This is the case if the potential is exactly flat, since then the slow-roll classical drift vanishes. We thus consider a potential that is constant\footnote{Obviously, if the potential is exactly constant, then slow-roll inflation cannot take place anymore and inflation proceeds along the ultra slow roll regime, which we will study in \Sec{sec:BeyondSlowRoll}. In the present case, a ``constant'' potential must simply be understood as being sufficiently flat such that the drift term can be neglected in the slow-roll Langevin equation. In \Sec{sec:example:1_plus_phi_to_the_p}, we will derive an explicit criterion for when this is the case.} between the two values $\phi_\uend$ and $\phi_\uend+\Delta\phiwell$, where $\Delta\phiwell$ denotes the width of this ``quantum well''. Inflation terminates when the field reaches $\phi_\uend$ (where either the potential is assumed to become very steep, or a mechanism other than slow-roll violation must be invoked to end inflation), and a reflective wall is located at  $\phi_\uend+\Delta\phiwell$, which can be seen as the point where the dynamics become classically dominated so that the probability for field trajectories to climb up this part of the potential and escape the quantum well can be neglected. The situation is depicted in \Fig{fig:sketch2}, and in \Secs{sec:stocha:limit:condition} and~\ref{sec:tail:expansion} we will see why this simple calculation can be used to study most cases of interest. 
\subsubsection{The characteristic function approach}
If the potential $v=v_0$ is constant, the potential gradient term vanishes in \Eq{eq:ODE:chi:SF:SR} and making use of the boundary condition at $\phi_\uend$ and at $\phiuv$, where $\phiuv$ is replaced by $\phi_\uend+\Delta\phiwell$, one obtains
\bea
\label{eq:chiN:cosh}
\chi_\N\left(t, \phi \right) = \frac{\cosh\left[\alpha \sqrt{t} \mu \left(x-1\right)\right]}{\cosh\left(\alpha \sqrt{t} \mu\right)}\, .
\eea
In this expression, $x\equiv (\phi-\phi_\uend)/\Delta\phiwell$, $\alpha\equiv (i-1)/\sqrt{2}$, and we have introduced the parameter
\bea
\label{eq:def:mu}
\mu^{2} = \frac{\Delta\phiwell^2}{v_{0}\Mp^2} 
\eea
which is the ratio between the squared width of the quantum well and its height, in Planck mass units, and which is the only combination through which these two quantities appear.

\begin{figure}[t]
\begin{center}
\includegraphics[width=0.6\textwidth]{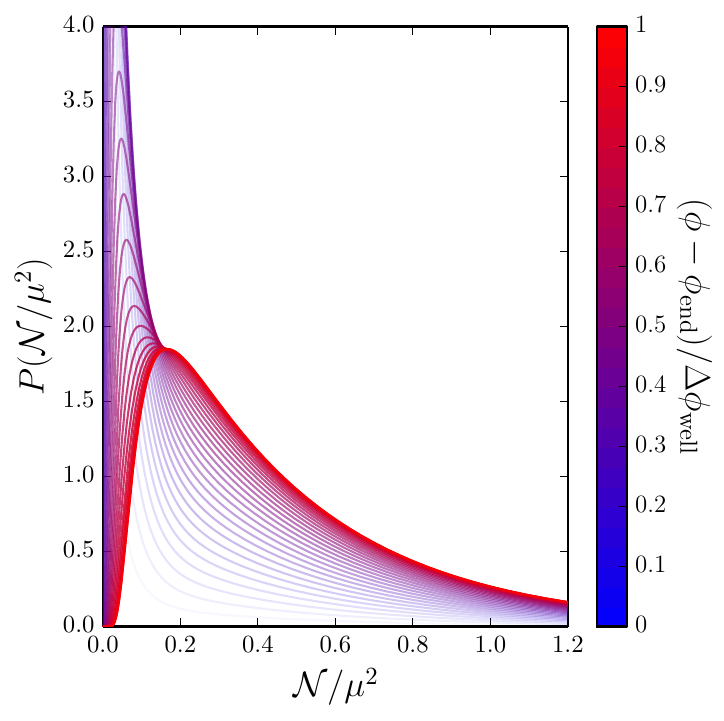}
\caption{Probability distributions of the number of \efolds~$\N$, rescaled by $\mu^2$, realised in the constant potential depicted in \Fig{fig:sketch2} between $\phi$ and $\phi_\uend$, where different colours correspond to different values of $\phi$. When $\phi$ approaches $\phi_\uend$, the distribution becomes more peaked and the transparency of the curves is increased for display purposes.} 
\label{fig:pdf_stochastic}
\end{center}
\end{figure}

\begin{figure}[t]
\begin{center}
\includegraphics[width=0.496\textwidth]{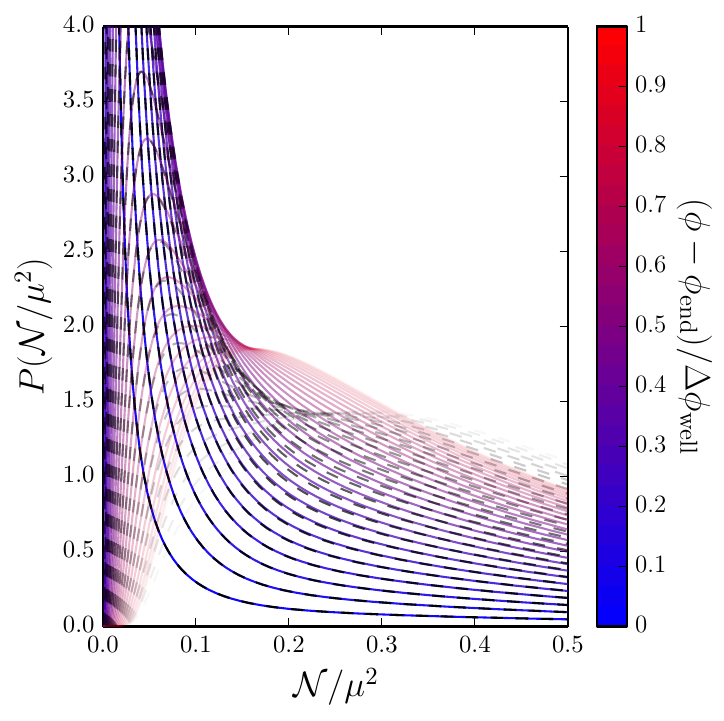}
\includegraphics[width=0.496\textwidth]{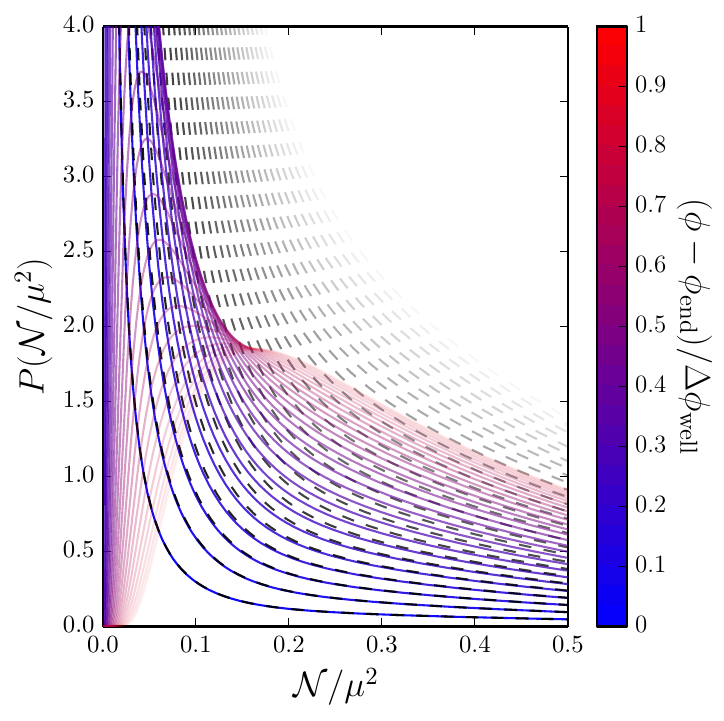}
\caption{Probability distributions of the number of \efolds~$\N$, rescaled by $\mu^2$, realised in the constant potential depicted in \Fig{fig:sketch2} between $\phi$ and $\phi_\uend$. In both panels, different colours correspond to different values of $\phi$, and the black dashed lines correspond to approximations. Left panel: the approximation~(\ref{eq:PDF:stocha:phiend:appr}) is displayed with the black dashed lines. Right panel: the approximation~(\ref{eq:PDF:stocha:phiend:appr:2}) is displayed with the black dashed lines. These approximations are valid close to the absorbing boundary of the quantum well where inflation ends. When $\phi$ increases, the approximation becomes worse, and the transparency of the curves is increased for displayed purposes, but one can see that the approximation~(\ref{eq:PDF:stocha:phiend:appr}) is excellent up to $(\phi-\phiend)/\Delta\phiwell \sim 0.3$, and slightly better than the approximation~(\ref{eq:PDF:stocha:phiend:appr:2}).} 
\label{fig:pdf_stochastic_appr}
\end{center}
\end{figure}

The PDF can be obtained by inverse Fourier transforming \Eq{eq:chiN:cosh}, see \Eq{eq:PDF:chi}, which can be done after Taylor expanding the characteristic function~(\ref{eq:chiN:cosh}) and inverse Fourier transforming each term in the sum. This leads to
\bea
\label{eq:stocha:CharacteristicFunctionMethod:PDF}
 P\left(\N, \phi \right) = \frac{1}{2\sqrt{\pi}}\frac{\mu}{\N^{3/2}}
 \left\lbrace \sum_{n=0}^\infty (-1)^n \left[2(n+1) - x \right]
 \ee^{- \frac{\mu^2}{4\N} \left[2(n+1) - x \right]^2} 
+ \sum_{n=0}^\infty (-1)^n \left[2n + x \right]
 \ee^{- \frac{\mu^2}{4\N} \left[2n + x \right]^2} 
\right\rbrace\, .
\eea
This PDF is displayed in \Fig{fig:pdf_stochastic} for different values of $x$. Interestingly, \Eq{eq:stocha:CharacteristicFunctionMethod:PDF} can be resummed to give a closed form when combined with the result from the heat equation approach presented below in \Sec{sec:StochasticLimit:HeatEquationApproach}. For now, we can derive closed form expressions at both boundaries of the quantum well, \ie in the two limits $\phi\simeq \phi_\uend+\Delta\phiwell$ and $\phi\simeq\phiend$.
\paragraph{Reflective boundary of the quantum well}
In the case where $\phi =  \phi_\uend+\Delta\phiwell$, or $x=1$, \ie at the reflective boundary of the quantum well, \Eq{eq:stocha:CharacteristicFunctionMethod:PDF} reduces to $P(\N, \phiwell) =\mu/\sqrt{\pi} \N^{-3/2}  \sum_{n=0}^\infty (-1)^n (2n+1) \ee^{- \frac{\mu^2}{4\N}  (2n+1)^2}$. Making use of the elliptic theta functions~\cite{Olver:2010:NHM:1830479:theta, Abramovitz:1970aa:theta},\footnote{There are four elliptic theta functions, defined as~\cite{Olver:2010:NHM:1830479:theta, Abramovitz:1970aa:theta}
\bea
\vartheta_1\left(z,q\right) &=& 2 \sum_{n=0}^\infty (-1)^n q^{\left(n+\frac{1}{2}\right)^2}\sin\left[\left(2n+1\right)z\right]\, ,\\
\vartheta_2\left(z,q\right) &=& 2 \sum_{n=0}^\infty  q^{\left(n+\frac{1}{2}\right)^2}\cos\left[\left(2n+1\right)z\right]\, ,\\
\vartheta_3\left(z,q\right) &=& 1+2 \sum_{n=1}^\infty q^{n^2}\cos\left(2nz\right)\, ,\\
\vartheta_4\left(z,q\right) &=& 1+2 \sum_{n=1}^\infty (-1)^nq^{n^2}\cos\left(2nz\right)\, .
\eea
By convention, $\vartheta_i^\prime$ denotes the derivative of $\vartheta_i$ with respect to its first argument $z$. For instance, one has
\bea
\label{eq:theta1prime:def}
\vartheta_1^\prime\left(z,q\right) = 2 \sum_{n=0}^\infty (-1)^n q^{\left(n+\frac{1}{2}\right)^2}\left(2n+1\right)\cos\left[\left(2n+1\right)z\right]\, ,
\eea
which appears in \Eq{eq:PDF:phiwall:thetaElliptic}. As another example, one has
\bea
\label{eq:theta4primeprime:def}
\vartheta_4^{\prime\prime}\left(z,q\right) = -8\sum_{n=1}^\infty (-1)^nq^{n^2}n^2\cos\left(2nz\right)\, ,
\eea
which is used in \Eq{eq:PDF:stocha:phiend:appr}. As a third example, one has
\bea
\label{eq:theta2prime:def}
\vartheta_2^{\prime}\left(z,q\right) = -2 \sum_{n=0}^\infty  q^{\left(n+\frac{1}{2}\right)^2}\left(2n+1\right)\sin\left[\left(2n+1\right)z\right]\, ,
\eea
which appears in \Eq{eq:PDF:thetatwo}.  As a last example, one has
\bea
\label{eq:theta2primeprime:def}
\vartheta_2^{\prime\prime}\left(z,q\right) = -2 \sum_{n=0}^\infty  q^{\left(n+\frac{1}{2}\right)^2}\left(2n+1\right)^2\cos\left[\left(2n+1\right)z\right]\, ,
\eea
which is used in \Eq{eq:PDF:stocha:phiend:appr:2}.}
this can be rewritten as 
\bea
\label{eq:PDF:phiwall:thetaElliptic}
P\left(\N, \phi =  \phi_\uend+\Delta\phiwell\right) = \frac{\mu}{2 \sqrt{\pi} \N^{3/2}}  
\vartheta^\prime_1\left(0,\ee^{-\frac{\mu^2}{\N}}\right) \, ,
\eea
where $\vartheta_1^\prime$ is the derivative (with respect to the first argument) of the first elliptic theta function
\paragraph{Absorbing boundary of the quantum well}
In the case where $\phi \simeq  \phi_\uend$, or $x\ll 1$, \ie at the absorbing boundary of the quantum well, an approximated formula can be obtained by noting that \Eq{eq:stocha:CharacteristicFunctionMethod:PDF} can be rewritten as $P(\N, \phi) = \mu /(2\sqrt{\pi}\N^{3/2})[x \ee^{-\mu^2 x^2/(4\N)}+F(-x)-F(x)]$, with $F(x)\equiv \sum_{n=0}^\infty (-1)^n [2(n +1)+ x]
 \ee^{- \frac{\mu^2}{4 \N} [2(n+1)+x]^2}$. In the limit where $x\ll 1$, $F(-x)-F(x)\simeq -2 x F'(0)$, where $F'(0)=1/2 - 1/2 \vartheta_4(0,\ee^{-\mu^2/\N})-\mu^2/(4\N)\vartheta_4''(0,\ee^{-\mu^2/4})$, and this gives rise to
\bea
\label{eq:PDF:stocha:phiend:appr}
P\left(\N, \phi \simeq \phi_\uend \right) \simeq \frac{\mu x}{2\sqrt{\pi}\N^{3/2}}
\left[ \ee^{- \frac{\mu^2 x^2}{4 \N }} 
- 1 + \vartheta_4\left(0,\ee^{- \frac{\mu^2}{\N} } \right) + \frac{\mu^2}{2 \N} \vartheta_4^{\prime\prime}\left(0,\ee^{- \frac{\mu^2}{\N} } \right) 
\right] \, .
\eea
This approximation is superimposed to the full result~(\ref{eq:stocha:CharacteristicFunctionMethod:PDF}) in the left panel of \Fig{fig:pdf_stochastic_appr}, where one can check that the agreement is excellent even up to $x\sim 0.3$.
\subsubsection{The heat equation approach}
\label{sec:StochasticLimit:HeatEquationApproach}
Let us now move on to the heat equation approach, since combined with the results of the characteristic function approach, this will allow us to derive a closed form for the PDF at arbitrary values of $x$. In the case of a constant potential, the heat equation~(\ref{eq:Fokker:Planck:adjoint:PDF:SF:SR}) becomes $(v_0\Mp^2\partial^2/\partial\phi^2 - \partial/\partial \N)P(\N, \phi ) = 0$. The boundary condition at $\phiuv$,  $\partial P/\partial\phi (\N, \phi_\uend+\Delta\phiwell) = 0$, leads to a Fourier decomposition of the form
\bea
\label{eq:stochastic:heat:FourierDecomposition}
P\left(\N, \phi \right) = \sum_{n=0}^\infty \left\lbrace A_n\left(\N\right) \sin\left[\left(\frac{\pi}{2}+ n\pi\right)x\right]
+B_n\left(\N\right) \cos\left(n\pi x \right) \right\rbrace \, ,
\eea
where, by plugging \Eq{eq:stochastic:heat:FourierDecomposition} into the heat equation~(\ref{eq:Fokker:Planck:adjoint:PDF:SF:SR}), the coefficients $A_n$ and $B_n$ must satisfy
\bea
\frac{\partial A_n}{\partial \mathcal{N}} = -\frac{\pi^2}{\mu^2} \left(n+\frac{1}{2}\right)^2 A_n\, , \quad
\frac{\partial B_n}{\partial \mathcal{N}} = -\frac{\pi^2}{\mu^2} n^2 B_n\, .
\eea
This leads to
\bea
A_n(\mathcal{N})  = a_n \exp\left[ -\frac{\pi^2}{\mu^2} \left(n+\frac{1}{2}\right)^2 \mathcal{N}\right]\, , \quad
B_n(\mathcal{N})  = b_n  \exp\left(-\frac{\pi^2}{\mu^2} n^2 \mathcal{N} \right)\, ,
\eea
where $a_n$ and $b_n$ are coefficients that depend only on $n$. They can be calculated by identifying \Eqs{eq:stocha:CharacteristicFunctionMethod:PDF} and~(\ref{eq:stochastic:heat:FourierDecomposition}) in the $\N \to 0$ limit. In this limit, in \Eq{eq:stocha:CharacteristicFunctionMethod:PDF}, the term with $n=0$ of the second sum is the dominant contribution, and using the fact that $\ee^{-x^2/(4\sigma)}/ (2\sqrt{\pi \sigma}) \to \delta(x)$ when $\sigma\to 0$, hence $- x\ee^{-x^2/(4\sigma)}/ (4\sigma \sqrt{\pi \sigma}) \to \delta'(x)$ when $\sigma\to 0$, one has 
\bea
\label{eq:stochastic:PDF:Neq0}
P(\N,\phi)\underset{\N \to 0}{\longrightarrow} -2 v_0 \Mp^2 \delta'(\phi-\phi_\uend)\, .
\eea
In passing, one notes that this expression implies that $P(\N=0,\phi)=0$ when $\phi \neq \phi_\uend$, which is consistent with the continuity of the distribution when $\N=0$ and with the fact that the probability to realise a negative number of \efolds~obviously vanishes. The case $\phi=\phiend$ is singular because of the boundary condition at $\phi_\uend$, which explains the singularity in \Eq{eq:stochastic:PDF:Neq0}. The coefficients $a_n$ and $b_n$ can then be expressed as $a_n = \int_{-1}^1 \dd x P(\N=0,\phi) \sin[(n+1/2)\pi x]$ for $n\geq 0$ and $b_n = \int_{-1}^1 \dd x P(\N=0,\phi) \cos[n\pi x]$ for $n\geq 0$, where we recall that the link between $\phi$ and $x$ is given above \Eq{eq:def:mu}. This gives rise to $a_n=2\pi(n+1/2)/\mu^2$ and $b_n=0$, hence 
\bea
\label{eq:stocha:HeatMethod:PDF:expansion}
P\left(\N, \phi \right) =  \frac{2 \pi}{\mu^2} \sum_{n=0}^\infty \left( n + \frac{1}{2} \right) \exp\left[ -\frac{\pi^2}{\mu^2} \left(n+\frac{1}{2}\right)^2 \N\right] \sin\left[x \pi\left(n+\frac{1}{2}\right)\right] \, ,
\eea
which can be written as
\bea
\label{eq:PDF:thetatwo}
P\left( \N, \phi \right) =  -\frac{\pi}{2 \mu^2} \vartheta_{2}' \left( \frac{\pi}{2}x, \ee^{-\frac{\pi^2}{\mu^2} \N} \right) \, .
\eea
This formula is compared to a numerical integration of the Langevin equation below, in \Fig{fig:USR:Sto:PDF}, where an excellent agreement is found. A few comments are in order.

First, let us stress that the results from both methods, the characteristic function method and the heat equation method, have been necessary to derive this closed form, since the expression coming from the characteristic function has allowed us to calculate the coefficients $a_n$ and $b_n$ in the heat equation solution. This further illustrates how useful it is to have two approaches at hand.

Second, the expansion~(\ref{eq:stocha:HeatMethod:PDF:expansion}) is an alternative to the one given in \Eq{eq:stocha:CharacteristicFunctionMethod:PDF} for the PDF. One can numerically check that they are identical, and in \Fig{fig:pdf_stochastic}, $P(\N,\phi)$ is displayed as a function of $\N$ for various values of $\phi$. The difference between \Eqs{eq:stocha:CharacteristicFunctionMethod:PDF} and~(\ref{eq:stocha:HeatMethod:PDF:expansion}) is that they correspond to expansions around different regions of the PDF. In \Eq{eq:stocha:CharacteristicFunctionMethod:PDF}, since one is summing over increasing powers of $\ee^{-1/\N}$, one is expanding around $\N=0$, \ie on the ``left'' tail of the distribution. In \Eq{eq:stocha:HeatMethod:PDF:expansion} however, since one is summing over increasing powers of $\ee^{-\N}$, one is expanding around $\N=\infty$, \ie on the ``right'' tail of the distribution. Therefore, if one wants to study the PDF by truncating the expansion at some fixed order $n$, one should choose to work with the expression that better describes the part of the distribution one is interested in, and both expressions can a priori be useful (let us stress again that, in the limit where all terms in the sums are included, both expressions match exactly for all values of $\N$). 

Third,  by plugging $x=1$ in \Eq{eq:PDF:thetatwo}, one obtains an expression for $P(\N,\phi=\phiend+\Delta\phiwell)$ that is an alternative to \Eq{eq:PDF:phiwall:thetaElliptic} even if both formulae involve elliptic theta functions. In fact, a third expression for $P(\N,\phi=\phiend+\Delta\phiwell)$ can even be obtained by plugging $x=1$ into \Eq{eq:stocha:HeatMethod:PDF:expansion}. Due to identities satisfied by the elliptic theta functions, these expressions can be shown to be all equivalent.\footnote{Different expressions for $P(\N,\phi=\phiend+\Delta\phiwell)$ can be obtained, either from \Eq{eq:PDF:phiwall:thetaElliptic}, or by plugging $x=1$ into \Eq{eq:stocha:HeatMethod:PDF:expansion} and making use of \Eq{eq:theta1prime:def}, or by plugging $x=1$ in \Eq{eq:PDF:thetatwo}. The three formulae are equivalent if
\bea
\label{eq:theta:identity}
\left(\frac{\mu}{\sqrt{\pi\N}}\right)^3 \vartheta^\prime_1\left(0,\ee^{-\frac{\mu^2}{\N}}\right) = 
\vartheta_{1}'\left( 0, \ee^{ -\frac{\pi^2}{\mu^2} \N} \right)  = 
-\vartheta_{2}'\left( \frac{\pi}{2}, \ee^{ -\frac{\pi^2}{\mu^2} \N } \right) \, .
\eea
The first equality in \Eq{eq:theta:identity} can be shown from the Jacobi identity for a modular transformation of the first elliptic theta function, see Eq.~(20.7.30) of \Refa{NIST:DLMF},
\bea
\label{app:identity:modulartrans}
\left( -i \tau \right)^{\frac{1}{2}} \vartheta_{1} \left( z, \ee^{i\pi \tau} \right) = -i \ee^{ -\frac{z^2}{\pi \tau}} \vartheta_{1} \left( -\frac{z}{\tau}, -\ee^{-\frac{i\pi}{\tau}} \right) \, .
\eea
By taking $\tau = i/(a \pi)$ and differentiating \Eq{app:identity:modulartrans} with respect to $z$, one obtains
\bea
\label{app:id:diff}
\left( \pi a \right)^{\frac{1}{2}} \vartheta_{1}' \left( z, \ee^{-\frac{1}{a}} \right) = - \frac{2iz}{a}\ee^{az^2} \vartheta_{1} \left( - i \pi a z, \ee^{-\pi^2 a} \right) + a \pi \ee^{az^2} \vartheta_{1}' \left( - i \pi a z, \ee^{-\pi^2 a} \right) \, .
\eea
Taking $z = 0$, one recovers the first equality in \Eq{eq:theta:identity}. The second equality in \Eq{eq:theta:identity} simply follows from \Eqs{eq:theta1prime:def} and~(\ref{eq:theta2prime:def}).}

Fourth, an approximated formula for the PDF in the limit $\phi\sim\phiend$ can be derived by Taylor expanding \Eq{eq:PDF:thetatwo}, 
\bea
\label{eq:PDF:stocha:phiend:appr:2}
P\left(\N, \phi \simeq \phiend \right) \simeq -\frac{\pi^2}{4 \mu^2} x \vartheta_{2}'' \left(0, \ee^{-\frac{\pi^2}{\mu^2}\N} \right )\, ,
\eea
see \Eq{eq:theta2primeprime:def}. This provides an alternative to the approximation~(\ref{eq:PDF:stocha:phiend:appr}), that is displayed in the right panel of \Fig{fig:pdf_stochastic_appr}. Numerically, one can check that \Eq{eq:PDF:stocha:phiend:appr} is slightly better.

Fifth, the PDF of coarse-grained curvature perturbations decays exponentially as $\ee^{-\zeta_{\mathrm{cg}}}$, \ie much slower than the Gaussian decay $\ee^{-\zeta_{\mathrm{cg}}^2}$. Since PBHs form along the tail of these distributions, we expect their mass fraction to be greatly affected by this highly non-Gaussian behaviour. More precisely, on the tail, one has 
\bea
\label{eq:PDF:stoch:tail}
P(\zeta_{\mathrm{cg}},\phi)\propto \ee^{-\frac{\pi^2}{4\mu^2}\zeta_{\mathrm{cg}}}\, ,
\eea
which is given by the dominant mode $n=0$ in the expansion~(\ref{eq:stocha:HeatMethod:PDF:expansion}). Interestingly, the decay rate of the distribution is independent of $\phi$ (this will be clarified in \Sec{sec:tail:expansion}). Let us also note that another case where the PDF decays exponentially is in presence of large local non-Gaussianities, when the PDF is a $\chi^2$ distribution~\cite{Young:2013oia, Young:2015cyn}.

\subsubsection{Primordial black holes}
\label{sec:pbh:stochastic:limit:pbh}
\begin{figure}[t]
\begin{center}
\includegraphics[width=0.496\textwidth]{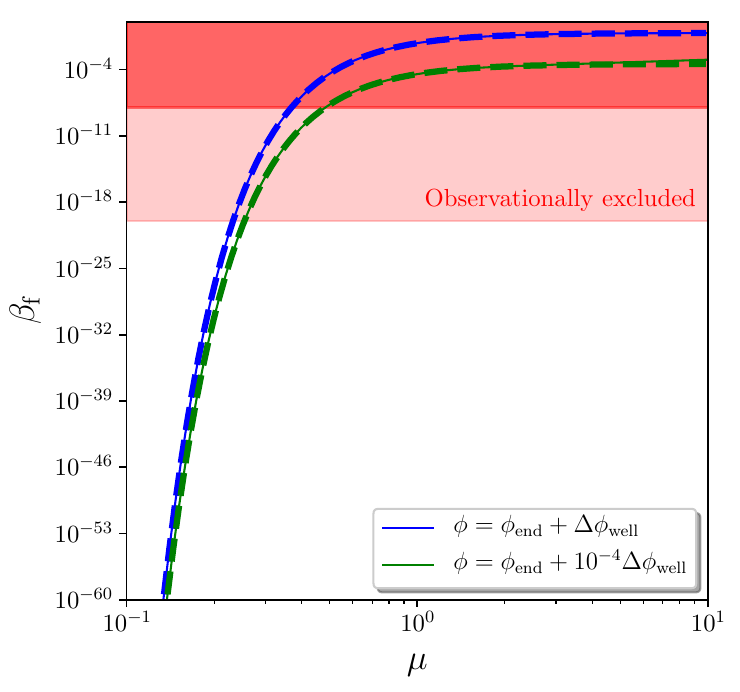}
\includegraphics[width=0.496\textwidth]{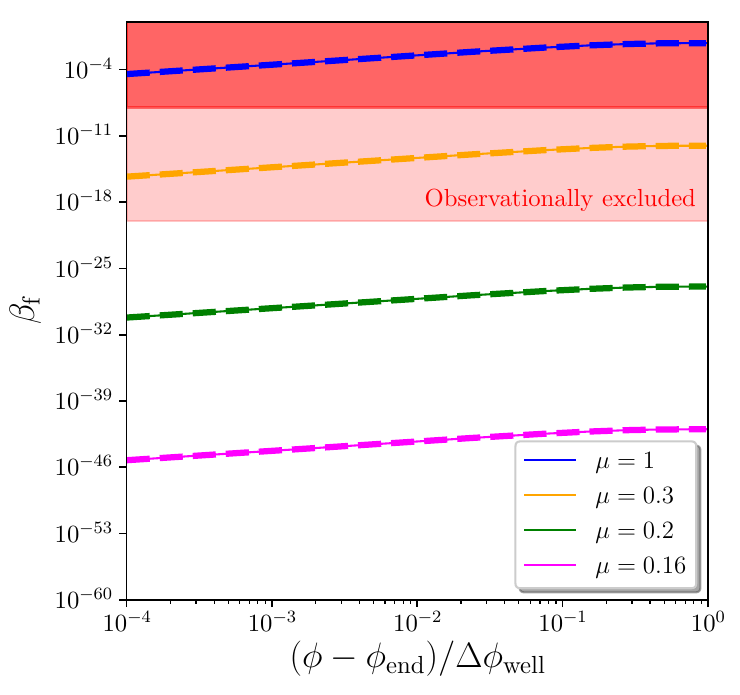}	
\caption{Mass fraction $\beta_{\mathrm{f}}$ of primordial black holes in the quantum diffusion dominated regime. The left panel displays $\beta_{\mathrm{f}}$ evaluated at $\phi=\phi_\uend+\Delta\phiwell$ (blue), \ie at the reflective boundary of the quantum well, and at $\phi=\phi_\uend+10^{-4}\Delta\phiwell$, \ie close to the absorbing boundary of the quantum well, as a function of $\mu=\Delta\phiwell/(\sqrt{v_0}\Mp)$. In the right panel, $\beta_{\mathrm{f}}$ is plotted as a function of $\phi$ for a few values of $\mu$. One can see that the mass fraction depends very weakly on $\phi$ but very strongly on $\mu$. In both panels, we have taken $\zeta_{\uc} = 1$, the solid lines correspond to the full expression~(\ref{eq:beta:full}) and the dashed line to the approximation~(\ref{eq:beta:stocha:appr}). The shaded region is excluded by observations, the light shaded area roughly corresponds to constraints for PBH masses between $10^{9}\mathrm{g}$ and $10^{16}\mathrm{g}$, the dark shaded area for PBH masses between $10^{16}\mathrm{g}$ and $10^{50}\mathrm{g}$.} 
\label{fig:beta}
\end{center}
\end{figure}
Let us now see how the constraint~(\ref{eq:Vconstraint:standard}) changes in the presence of large quantum diffusion. Plugging \Eq{eq:stocha:HeatMethod:PDF:expansion} into \Eq{eq:beta:def}, the PBH mass fraction is given by
\bea
\label{eq:beta:full}
\beta_{\mathrm{f}}(M) = \frac{2}{\pi} \sum^{\infty}_{n=0} \frac{1}{\left( n+\frac{1}{2} \right)}\sin{\left[ \pi \left( n + \frac{1}{2} \right)x \right]} \exp{\left\lbrace -\pi^2 \left( n + \frac{1}{2}\right)^2 \left[ x \left( 1 - \frac{x}{2} \right) + \frac{\zeta_{c}}{\mu^2}\right] \right\rbrace} \, .
\eea 
In this expression, we have replaced $\langle \N \rangle = \mu^2 x(1-x/2)$, which can be obtained by plugging \Eq{eq:chiN:cosh} into \Eq{eq:meanN:chi}. Let us recall that $x=(\phi-\phiend)/\Delta\phiwell$ and that $M$ and $\phi$ are related as recalled below \Eq{eq:Vconstraint:standard}. When $x=0$, \ie when $\phi=\phi_\uend$, \Eq{eq:beta:full} yields $\beta_{\mathrm{f}}=0$, which is consistent with the fact that the PDF of $\zeta_{\mathrm{cg}}$ is a Dirac distribution in this case.

The mass fraction~(\ref{eq:beta:full}) depends only on $x$, $\mu$ and $\zeta_\uc$. It is displayed in \Fig{fig:beta} for $\zeta_\uc=1$, as a function of $\mu$ for $x=1$, \ie $\phi=\phi_\uend+\Delta\phiwell$, and $x=10^{-4}$, \ie $\phi=\phi_\uend+10^{-4}\Delta\phiwell$, in the left panel, and as a function of $\phi$ for a few values of $\mu$ in the right panel. One can see that $\beta_{\mathrm{f}}$ depends only weakly on $\phi$ but very strongly on $\mu$, which is constrained to be at most of order one. More precisely, if one assumes that $\zeta_\uc\gg \mu^2$ so that $\zeta_\uc $ is well within the tail of the distribution and one can keep only the mode $n=0$ in \Eq{eq:beta:full}, as was done when deriving \Eq{eq:PDF:stoch:tail}, one has
\bea
\label{eq:beta:stocha:appr}
\beta_{\mathrm{f}}(M) \simeq \frac{4}{\pi}\sin\left(\frac{\pi x}{2}\right)\ee^{-\frac{\pi^2}{8}\left[x(2-x)+\frac{2\zeta_\uc}{\mu^2}\right]}\, .
\eea
This expression is superimposed to the full result~(\ref{eq:beta:full}) in \Fig{fig:beta} where one can see that it provides a very good approximation even when the condition $\zeta_\uc\gg \mu^2$ is not satisfied. This is because, in \Eq{eq:beta:full}, higher terms in the sum are not only suppressed by higher powers of $\ee^{-\zeta_\uc^2/\mu^2}$ but also by higher powers of $\ee^{-\pi^2x(1-x/2)}$, so that \Eq{eq:beta:stocha:appr} is an excellent proxy for all values of $\mu$ except if $x$ is tiny. With $x=1$, it gives rise to
\bea
\label{eq:stocha:constraint:mu}
\mu^2 =  -\frac{2\zeta_\uc}{1+\frac{8}{\pi^2}\ln\left(\frac{\pi}{4}\beta_{\mathrm{f}}\right)}\, ,
\eea
where $\mu$ was defined in \Eq{eq:def:mu}.

Several comments are in order regarding this result. First, with $\zeta_\uc=1$, $\beta_{\mathrm{f}}<10^{-24}$ gives rise to $\mu<0.21$ and $\beta_{\mathrm{f}}<10^{-5}$ gives rise to $\mu<0.48$. The requirement that $\mu$ be smaller than one is therefore very generic and rather independent of the level of the constraint on $\beta_{\mathrm{f}}$ or the precise value chosen for $\zeta_\uc$. Since $v_0$ needs to be smaller than $10^{-10}$ to satisfy the upper bound~\cite{Ade:2015lrj} on the tensor-to-scalar ratio in the CMB observational window, this also means that $\Delta\phiwell$ cannot exceed $\sim 10^{-5}\Mp$.

Second, \Eq{eq:stocha:constraint:mu} should be compared with its classical equivalent, \Eq{eq:Vconstraint:standard}. In the left-hand sides of these formulae, the scalings with $\Delta\phi$ and $v$ are not the same. In particular, while the PBH mass fraction increases with the energy scale $v$ in the classical picture, in the stochastic limit, it goes in the opposite direction. One should also note that when the potential is exactly flat, $v'=0$, the classical result diverges, but the stochastic one remains finite. In the right-hand sides, the scaling with $\zeta_\uc$ is also different, since the shape of the PDF $P(\zeta_{\mathrm{cg}})$ is not the same (it has a Gaussian decay in the classical case and an exponential decay in the stochastic one). The expressions~(\ref{eq:Vconstraint:standard}) and~(\ref{eq:stocha:constraint:mu}) are therefore very different, and thus translate into very different constraints on the inflationary potential. 

Third, as mentioned below \Eq{eq:beta:full}, the mean number of \efolds~realised across the quantum well is of order $\mu^2$, 
\bea
\langle \N \rangle = \mu^2 x \left(1-\frac{x}{2}\right)\, .
\eea
The conclusion one reaches is therefore remarkably simple: either the region dominated by stochastic effects is less than one \efold~long and PBHs are not overproduced ($\mu\ll 1$), or it is more than one \efold~long and PBHs are overproduced ($\mu\gg 1$). Interestingly, heuristic arguments lead to a similar conclusion in \Refa{GarciaBellido:1996qt}, in the context of hybrid inflation.

Fourth, in terms of the power spectrum, since \Eq{eq:PS:fullstocha:moments} gives $\calP_\zeta = \langle \N^2\rangle'/\langle \N \rangle'- 2 \langle \N \rangle$, with $\langle \N \rangle$ given above and $\langle \N^2 \rangle = \mu^4 x(1 - x^2/2 + x^3/8)/3$ as can be obtained by setting the potential to a constant in \Eq{eq:mean:N2}, one has 
\bea
\label{eq:Pzeta:stochasticLimit}
\calP_\zeta = \frac{\mu^2}{3}\left(2x^2-4x+2\right)\, ,
\eea
so $\mu^2$ also controls the amplitude of the power spectrum. With $\beta_{\mathrm{f}}<10^{-22}$, the constraint~(\ref{eq:stocha:constraint:mu}) on $\mu$ translates into $\calP_\zeta<1.6\times 10^{-2}$ for the value of the power spectrum close to the end of inflation. However, contrary to the classical condition $\calP_\zeta \Delta N<10^{-2}$ recalled below \Eq{eq:Pzetaconstraint:standard}, this constraint does not involve the number of \efolds~since here, a single parameter, $\mu$, determines everything: the mean number of \efolds, the power spectrum amplitude, and the mass fraction. 
\subsection{Regimes of applicability of both limits}
\label{sec:stocha:limit:condition}
So far, we have calculated the PBH mass fraction produced in the classical limit and when the inflaton field dynamics are dominated by quantum diffusion. In order to analyse a generic potential, it remains to determine where both limits apply. This can be done by comparing the NLO and NNLO results in the classical limit to estimate the conditions under which the classical expansion is under control. For instance, comparing \Eqs{eq:gamma1:nlo:def} and~(\ref{eq:gamma:nnlo:def}) for $\gamma_1$, which gives the mass fraction $\beta_{\mathrm{f}}$ at NLO as explained in \Sec{sec:PBH:classical}, one can see that $\vert \gamma_1^\nlo - \gamma_1^\nnlo \vert \ll \gamma_1^\nlo $ if $v\ll 1$ and $\vert v^2 v''/{v'}^2 \vert \ll 1$. The first condition is always satisfied, since as already pointed out, $v$ needs to be smaller than $10^{-10}$ to satisfy the upper bound~\cite{Ade:2015lrj} on the tensor-to-scalar ratio in the CMB observational window. The second condition exactly coincides with our ``classicality criterion''~\eqref{eq:classicalcriterion:def}. 

When $\eta_{\mathrm{cl}}\ll 1$, the classical expansion is under control, at least at NNLO, and one can use the results of  \Sec{sec:ClassicalLimit}. Of course, the classical expansion could a priori break down at NNNLO even with $\eta_{\mathrm{cl}}\ll 1$, but since higher-order corrections are suppressed by higher powers of $v$, such a situation is in practice very contrived, and $\eta_{\mathrm{cl}}$ provides a rather generic criterion. However, as shown and discussed in \Sec{sec:ClassicalLimit}, the classical expansion is only useful to compute quantities that probe the neighbourhood of the maximum of the PDF, such as the mean number of \efolds~or the power spectrum. For observables that depend on the shape of the tail of the PDF, such as PBHs, it breaks down. This is why, to characterise the abundance of PBHs in the ``classical'' regime (in the sense that $\eta_\ucl \ll 1$), other techniques than the classical expansion have to be used, and this will be the topic of \Sec{sec:tail:expansion}. 

When $\eta_\ucl\gg 1$, one is far from the classical regime, quantum diffusion dominates the inflaton field dynamics and the results of \Sec{sec:StochasticLimit} apply. When $\eta_\ucl$ is of order one, a full numerical treatment is required. In the next section, we see how these two regimes arise and can be interpolated in a concrete example.
\subsection{Example: inflation towards a local minimum}
\label{sec:example:1_plus_phi_to_the_p}
We consider the case where PBHs can form at scales that exit the Hubble radius towards the end of inflation, where the potential can be approximated by a Taylor expansion around $\phi=0$ where inflation is assumed to end ($\phi_\uend=0$), so
\bea
\label{eq:pot:expansEnd}
v=v_0\left[1+\left(\frac{\phi}{\phi_0}\right)^p\right]\, .
\eea
In this model, inflation does not end by slow-roll violation but another mechanism must be invoked~\cite{Linde:1991km, Linde:1993cn, Copeland:1994vg, Renaux-Petel:2015mga, Renaux-Petel:2017dia}. We also assume that the potential is in the vacuum-dominated regime for the range of field values relevant for PBH formation, so that $\phi\ll \phi_0$. In order to describe the model~(\ref{eq:pot:expansEnd}) in terms of the situation depicted in \Fig{fig:sketch2}, one has to assess $\Delta\phiwell$, which marks the boundary between the classical and the stochastic regimes. In the vacuum-dominated approximation, \Eq{eq:classicalcriterion:def} gives rise to $\eta_\ucl \simeq (p-1) v_0 (\phi/\phi_0)^{-p}/p$, which is of order one when $\phi = \Delta\phiwell$ with
\bea
\label{eq:phiwell:expansEnd}
\Delta\phiwell \simeq  \phi_0 v_0^{\frac{1}{p}}\, .
\eea
Since $(\Delta\phiwell/\phi_0)^p=v_0\ll 1$, the vacuum-dominated condition is always satisfied at this transition point. However, the slow-roll conditions~\eqref{eq:sr:consistency} are not always met, and one can show that slow roll is indeed violated at $\phi = \Delta\phiwell$ if $\phi_0/\Mp < v_0^{(p-2)/(2p)}$, unless $p=1$ for which slow-roll is violated if $\phi_0< \Mp$. In such cases, the expansion~(\ref{eq:pot:expansEnd}) fails to cover the whole quantum well and higher-order terms in the potential must be included for a consistent analysis. Otherwise, we can keep following the recipe given above.

In the classical regime, $\phi\gg \Delta\phiwell$, let us recall that the classical expansion is under control only close to the maximum of the PDF. If one were still to make use of \Eq{eq:constraint:classical} (which is a priori not licit, since the more refined techniques developed in \Sec{sec:tail:expansion} have to be used instead), where $v \gamma_1$ is given by \Eq{eq:gamma1:nlo:def}, one would obtain, in the vacuum-dominated approximation, $v\gamma_1 \simeq v_0 (\phi_0/\Mp)^4/(4p^3-3p^4)[(\phi/\phi_0)^{4-3p}-(\phi_\uend/\phi_0)^{4-3p}]$. Neglecting the contribution from $\phi_\uend$, which lies outside the validity range of the classical formula anyway, one can evaluate this expression at $\phi = \Delta\phiwell$ where the power spectrum is maximal, and combining this with \Eq{eq:constraint:classical} leads to
\bea
\label{eq:expansEnd:classConstraint}
\frac{v_0^{\frac{2}{p}-1}}{\sqrt{\vert 4p^3-3p^4 \vert }}\left(\frac{\phi_0}{\Mp}\right)^2 \simeq  \frac{\zeta_\uc}{2\sqrt{\vert \ln \beta_{\mathrm{f}} \vert}}\, .
\eea
In the stochastic regime, combining \Eqs{eq:stocha:constraint:mu} and~(\ref{eq:phiwell:expansEnd}), one has
\bea
\label{eq:expansEnd:stochConstraint}
v_0^{\frac{2}{p}-1}\left(\frac{\phi_0}{\Mp}\right)^2 \simeq  \frac{2\zeta_\uc}{\left\vert 1+\frac{8}{\pi^2}\ln\left(\frac{\pi}{4}\beta_{\mathrm{f}}\right)\right\vert}\, .
\eea
It is interesting to notice that up to an overall factor of order one, the two constraints~(\ref{eq:expansEnd:classConstraint}) and~(\ref{eq:expansEnd:stochConstraint}) are very similar, even though they are obtained in very different regimes, that yield very different PDFs for the curvature perturbations, and even if the classical expansion is not under control on the tail. This coincidence is however specific to that particular potential. 

It is also important to note that the slow-roll conditions given above imply that $v_0^{2/p-1}(\phi_0/\Mp)^2\gg 1$ except if $p=1$. Therefore, if $p$ is different from $1$, either PBHs are too abundant and the model is ruled out, or slow roll is strongly violated before one exits the classical regime and one needs to go beyond the present formalism to calculate PBH mass fractions. The case $p=1$ is subtle, since \Eq{eq:classicalcriterion:def} gives $\eta_{\mathrm{class}}=0$, and will be studied in detail in \Sec{sec:linear_potential}.

\begin{figure}[t]
\begin{center}
\includegraphics[width=0.6\textwidth]{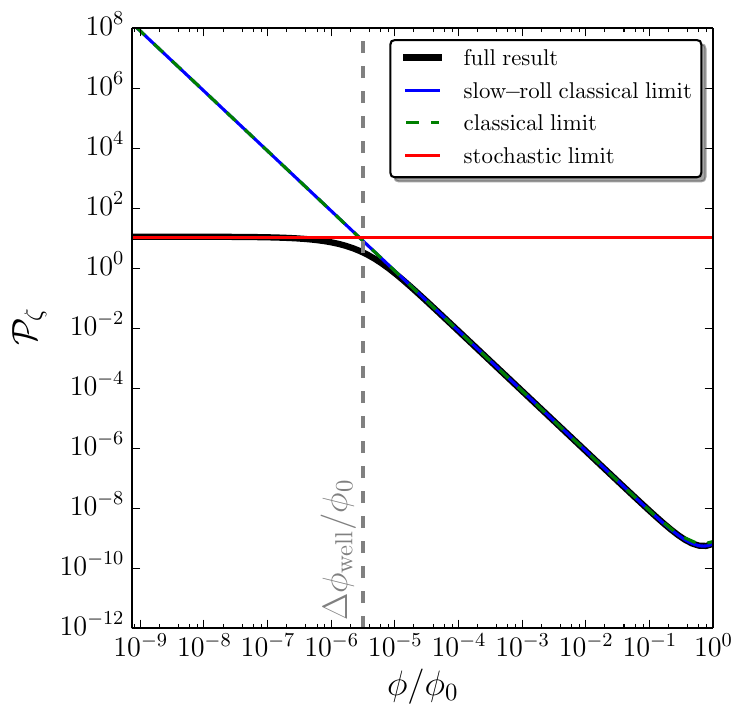}
\caption{Power spectrum of curvature perturbations $\calP_\zeta$ produced in the potential~(\ref{eq:pot:expansEnd}) with  $p=2$, $v_0 = 10^{-11}$, $\phi_0=4 \Mp$ and $\phiuv=10^{4}\phi_0$ (solid black line). The blue line corresponds to the slow-roll classical limit $\calP_\zeta = 2v^3/(\Mp^2 {v^\prime}^2)$, see \Eq{eq:PS:vll1}, while the green dashed line is obtained from solving the full Klein-Gordon and Mukhanov-Sasaki equations. The red line corresponds to the stochastic limit assuming the potential is exactly flat for $\phi<\Delta\phiwell$ and that a reflective wall is located at $\phi = \Delta\phiwell$.  The value of $\Delta\phiwell$ obtained from requiring $\eta_{\mathrm{cl}}=1$ is displayed with the grey vertical dotted line and delimitates the classical and stochastic regimes.} 
\label{fig:quad:power:spectrum}
\end{center}
\end{figure}
In passing, let us check that approximating the full potential~(\ref{eq:pot:expansEnd}) as a piecewise function consisting of a constant piece and a classical one, separated at $\phi=\Delta\phiwell$, is numerically justified. In \Fig{fig:quad:power:spectrum}, we show the full power spectrum computed from \Eq{eq:PS:fullstocha}, in the potential~(\ref{eq:pot:expansEnd}) with $p=2$, $v_0 = 10^{-11}$, $\phi_0=4 \Mp$ and $\phiuv=10^{4}\phi_0$ (solid black line). The blue line corresponds to the slow-roll classical limit  $\calP_\zeta = 2v^3/(\Mp^2 {v^\prime}^2)$, see \Eq{eq:PS:vll1}, and the green dashed line is obtained from solving the full Klein-Gordon and Mukhanov-Sasaki equations. The agreement of this solution with the slow-roll formula confirms that the slow-roll conditions are satisfied for the parameters used in this example, and that one stays on the slow-roll attractor (no regime of ultra slow roll besides the fact that the potential becomes asymptotically flat). The red line corresponds to the stochastic limit~(\ref{eq:Pzeta:stochasticLimit}) $\calP_\zeta = 2\mu^2/3$ at $\phi=0$, where $\mu$ is given by \Eqs{eq:def:mu} and~(\ref{eq:phiwell:expansEnd}), which yields $\calP_\zeta \sim 2 (\phi_0/\Mp)^2 v_0^{2/p-1}/3$. One can see that both limits are correctly reproduced, and that the value of $\Delta\phiwell$ obtained in \Eq{eq:phiwell:expansEnd} from our classicality criterion $\eta_{\mathrm{class}}<1$, and displayed with the grey vertical dotted line, indeed separates the two regimes. In fact, by expanding \Eq{eq:PS:fullstocha} in the regime $\phi\ll\Delta\phiwell$ (\ie in the stochastic limit), one finds $\calP_\zeta = 2\Gamma^2(1/p)v_0^{2/p-1}(\phi_0/\Mp)^2/p^2$, where $\Gamma$ is the gamma function. Up to an overall numerical constant of order one, one recovers the result obtained from simply assuming the potential to be exactly flat until $\phi=\Delta\phiwell$, where $\eta_{\mathrm{class}}=1$, and setting a reflective wall there. This confirms the validity of this approach, that will be further tested and corroborated in the next section.
\section{Primordial black holes from the tails}
\label{sec:tail:expansion}
As explained in \Sec{sec:PBHs}, the mass fraction of PBHs with mass $M$ corresponds to the probability that the curvature perturbation $\zetacg$, coarse-grained at the scale $k$ for which a Hubble patch weighs $M$ when $k$ re-enters the Hubble radius, exceeds a certain threshold of order one, see \Eq{eq:beta:def}. Since the typical values of $\zetacg$ are typically much smaller than one (perturbations remain in the linear regime on super-Hubble scales), \Eq{eq:beta:def} involves an integral over the high-curvature tail of the PDF.

If perturbations have Gaussian statistics, which is consistent with current observations, these tails are fully characterised by the two-point correlation function of curvature perturbations, and the mass fraction can be directly determined from the power spectrum, see \Eq{eq:beta:erfc}. However, current constraints on the amount of non-Gaussianities come from upper bounds on the amplitude of the bispectrum and the trispectrum~\cite{Akrami:2019izv}, and therefore only capture deviations from Gaussian statistics close to the maximum of the distribution functions, leaving the tails unconstrained. Moreover, these constraints apply to scales probed in the CMB, while PBHs could form at much smaller scales where little is known about the (non)-Gaussian nature of the statistics. 

Non-Gaussianities may therefore play a crucial role in determining the abundance of PBHs~\cite{Byrnes:2012yx, Young:2013oia, Young:2015cyn, Garcia-Bellido:2016dkw, Garcia-Bellido:2017aan, Ezquiaga:2018gbw, Franciolini:2018vbk, Cai:2018dig, Passaglia:2018ixg, Young:2019yug, DeLuca:2019qsy, Panagopoulos:2019ail, Yoo:2019pma,Carr:2019hud}. Nothing really limits the amount of non-Gaussianities on the tail from the observational side, and from the theoretical side, there is also little known about them. Indeed, most techniques developed in the literature to compute the PDF of $\zetac$ from inflation are designed to provide approximations modelling the neighbourhood of the maximum of the PDF, not its tail. This is for instance the case for the expansion in terms of the non-linearity parameters $f{{}_\mathrm{NL}}$ and $g{{}_\mathrm{NL}}$~\cite{Gangui:1993tt}.

This is why, in this section, which closely follows \Refa{Ezquiaga:2019ftu}, we build on the formalism developed in \Sec{sec:PBHs} to characterise the tails of the PDFs of curvature perturbations. In \Sec{sec:ClassicalLimit}, a classical expansion was performed, which was shown to correspond to an expansion around the maximum of the PDF, that matches the standard approach. Here a similar type of expansion is developed but on the tail of the distribution. In \Sec{sec:StochasticLimit} the tail was computed for the case of a quasi-flat potential, which is exactly solvable. Here we generalise these results to arbitrary potentials.

We will find that the tails are always highly non-Gaussian, and can never be described with standard non-Gaussian expansions. More precisely, we will show that tails always have an exponential, rather than Gaussian, decay. These exponential tails are inevitable, and do not require any non-minimal feature as they simply result from the quantum diffusion of the inflaton field along its potential. We will apply our formalism to a few relevant single-field, slow-roll inflationary potentials, and discuss the implications for the expected abundance of primordial black holes in these models. They can differ from standard results by several orders of magnitude. In particular, we will find that potentials with an inflection point overproduce primordial black holes, unless slow roll is violated, similarly to what was found in \Sec{sec:example:1_plus_phi_to_the_p} for models where inflation proceeds towards an uplifted minimum of its potential.
\subsection{Tail expansion}
\label{sec:tail_curvature}
Let us consider the distribution function of the coarse-grained curvature perturbation $\zetacg$, which, from the considerations developed in \Sec{sec:stochastic:delta:N}, is nothing but the distribution function $P(\N, \boldmathsymbol{\Phi})$ of the number of inflationary \efolds~$\N$, realised from a certain field configuration $\boldmathsymbol{\Phi}$. This distribution function satisfies the adjoint Fokker-Planck equation~\eqref{eq:Fokker:Planck:adjoint:PDF}, and in what follows, we will show that it admits an expansion of the form
\bea
\label{eq:tail_expansion}
P_{\boldmathsymbol{\Phi}}(\N)=\sum_{n}a_n(\boldmathsymbol{\Phi})e^{-\Lambda_n\,\N}\, .
\eea
In this expression, the functions $a_n(\boldmathsymbol{\Phi})$ determine the amplitude of the tail, and the coefficients $\Lambda_n$, which we will show do \emph{not} depend on $\boldmathsymbol{\Phi}$, set the exponential decay rates. Let us recall that in the classical, Gaussian picture, the PDF is given by
\bea
\label{eq:tail_expansion:classical}
\left. P_{\boldmathsymbol{\Phi}}(\N)\right\vert_{\ucl} 
\underset{\mathcal{N}\gg 1}{\propto}
 \exp\left[-\frac{1}{2}\frac{\mathcal{N}^2}{\int_{\bar{k}}^{k_\uend} \calP_{\zeta,\mathrm{cl}}(k)\dd\ln k}\right]\,,
\eea
where $\calP_{\zeta,\mathrm{cl}}$ is the classical value of the power spectrum [in single-field slow-roll inflation, it is given by $\calP_{\zeta,\mathrm{cl}}=2v^3/(\Mp^2v_\phi^2)$], and $\bar{k}$ and  $k_\uend$ are the scales that cross out the Hubble radius when the system is at location $\boldmathsymbol{\Phi}$, and at the end of inflation, respectively. If non-Gaussianities are perturbatively introduced by the means of the usual non-linearity parameters $f{{}_\mathrm{NL}}$, $g{{}_\mathrm{NL}}$ \etc , \Eq{eq:tail_expansion:classical} is modified with polynomial corrections in $\N$~\cite{Vennin:2015hra, Vennin:2016wnk}, which cannot capture the exponential decay of \Eq{eq:tail_expansion}. 

We now present two complementary techniques to compute $a_n(\boldmathsymbol{\Phi})$ and $\Lambda_n$, before applying them to concrete examples in \Sec{sec:applications}.
\subsubsection{Poles of the characteristic function }
\label{sec:pole:chi}
In order to analyse the solution of \Eq{eq:Fokker:Planck:adjoint:PDF} in the large-$\N$ limit, it is convenient to introduce the characteristic function as in \Eq{eq:characteristicFunction:def}. From \Eq{eq:chi:P}, let us recall that the characteristic function is nothing but the Fourier transform of the PDF, hence the PDF can be obtained by inverse Fourier transforming the characteristic function, see \Eq{eq:PDF:chi}, 
\bea
\label{eq:pdf:chi}
P_{\boldmathsymbol{\Phi}}\left(\N\right)=\frac{1}{2\pi}\int_{-\infty}^{\infty}e^{-it\N}\chi_\N\left(t,\boldmathsymbol{\Phi}\right)\dd t\, .
\eea
The characteristic function satisfies the differential equation~\eqref{eq:ODE:chi}, with the boundary conditions given below \Eq{eq:ODE:chi}. 
\begin{figure}[t]
\centering 
\includegraphics[width=.49\textwidth]{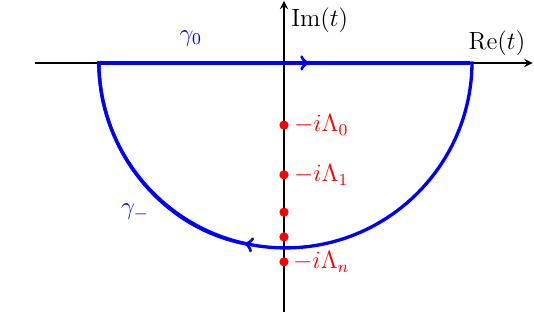}
 \caption{Schematic representation of the pole structure of the characteristic function. In order to compute the PDF, $P_{\boldmathsymbol{\Phi}}(\N)$, from the characteristic function, making use of the residue theorem, the real axis integral of \Eq{eq:pdf:chi} can be obtained from the integral over the contour $\gamma_0\cup\gamma_{-}$ in the complex plane.}
 \label{fig:contour_poles}
\end{figure}
The idea is to perform the integral of \Eq{eq:pdf:chi} by means of the residue theorem.\footnote{If $f(z)$ is a regular function in the complex plane, and $\gamma$ a close contour that circles in a certain point $z_p$ (the winding number of $\gamma$ around $z_p$ is one), one has
\bea
\oint_{\gamma}\frac{f(z)\,\dd z}{(z-z_p)^{n+1}}=\frac{2\pi i}{n!}\lb\frac{\dd ^n}{\dd z^n}f(z)\rb_{z=z_p}\, .
\eea
}
This is done by expanding the characteristic function according to
\bea
\label{eq:chi:pole:expansion}
\chi_\N\left(t,\boldmathsymbol{\Phi}\right)=\sum_{n}\frac{a_n(\boldmathsymbol{\Phi})}{\Lambda_n-it}+g(t,\boldmathsymbol{\Phi})\,,
\eea
where $g(t,\boldmathsymbol{\Phi})$ is a regular function of $t$, and the $\Lambda_n$ are positive numbers that do not depend on $\boldmathsymbol{\Phi}$. The form of this expansion can be justified as follows. First, $t$ is only involved through the combination $i t$ in \Eq{eq:ODE:chi}, which is the only place where a complex number appears (the rest of \Eq{eq:ODE:chi}, and the boundary conditions, only involve real quantities). The characteristic function is therefore a real function of $it$, which explains why its poles are necessarily located on the imaginary axis. Second, \Eq{eq:ODE:chi} is a second-order linear differential equation, which is further linear in $t$, so there exist independent solutions that are regular in $t$, and the poles in $t$ only appear when enforcing the boundary conditions. This explains why the $\Lambda_n$ do not depend on $\boldmathsymbol{\Phi}$, but only on the location of the surfaces $\partial\Omega_0$ and $\partial\Omega_+$. Third, the fact that \Eq{eq:ODE:chi} is linear in $t$ explains why there are only simple poles. Fourth, since the Fokker-Planck operator and its adjoint are positive operators, the $\Lambda_n$ are all positive. In what follows, the $\Lambda_n$ are ordered such that $0<\Lambda_0<\Lambda_1<\Lambda_2<\cdots<\Lambda_n$. 

Following \Fig{fig:contour_poles}, the integration over the real axis $\gamma_0$ can be complemented by an integral over $\gamma_-$, which, thanks to the term $\ee^{-i t \N}$ in \Eq{eq:pdf:chi}, asymptotically vanishes (assuming that $g$, which appears in \Eq{eq:chi:pole:expansion}, does not increase exponentially or faster at large $\vert t\vert$). This leads precisely to \Eq{eq:tail_expansion}. This form is always valid, but at large $\N$, only the first terms in the sum dominate, and it provides a tail expansion in terms of decaying exponentials. The dominant term is given by the lowest pole of the characteristic function $\Lambda_0$ and its residue $a_0(\boldmathsymbol{\Phi})$. In practice, the decay rates $\Lambda_n $ can be found by solving the characteristic function from \Eq{eq:ODE:chi} and finding the zeros of its inverse. The residues $a_n(\boldmathsymbol{\Phi})$ can then be obtained from evaluating the derivative of the inverse characteristic function at $t=-i\Lambda_0$, \ie
\begin{align}
\label{eq:alphan:flat:generic}
a_n(\phi) &= -i \left[\frac{\partial}{\partial t}\chi_\N^{-1}\left(t=-i\Lambda_n,\phi\right)\right]^{-1}\, .
\end{align}
Let us stress again that, while the amplitude of the tail, controlled by $a_0$, depends on the initial field value, the decay rates $\Lambda_n$ are universal for a given potential. 
\subsubsection{An equivalent eigenvalue problem}
\label{sec:eigenvalues}
Let us now present an alternative method that leads to the same tail expansion, but that can be of complementary practical convenience. This relies on viewing \Eq{eq:Fokker:Planck:adjoint:PDF} as a heat equation, and employing well-known late-time limit techniques designed for heat or diffusion equations to solve it. Formally, \Eq{eq:Fokker:Planck:adjoint:PDF} can be solved as
\bea
\label{eq:adjoint:FokkerPlanck:formal:solution}
P_\boldmathsymbol{\Phi}\left(\mathcal{N}\right) = \exp\left[\mathcal{N}\mathcal{L}^\dagger_{\mathrm{FP}}\left(\boldmathsymbol{\Phi}\right)\right] P_\boldmathsymbol{\Phi}\left(\mathcal{N}=0\right)\, .
\eea
One then introduces an orthonormal set of eigenfunctions $\Psi_n$ of the operator $\mathcal{L}^\dagger_{\mathrm{FP}}$, 
\bea
\label{eq:eigen:value:problem}
\mathcal{L}^\dagger_{\mathrm{FP}}\cdot \Psi_n\left(\boldmathsymbol{\Phi}\right) = - \Lambda_n \Psi_n\left(\boldmathsymbol{\Phi}\right)
\eea
(here a minus sign is introduced for notational convenience), with boundary conditions $\Psi_n(\boldmathsymbol{\Phi})=0$ when $\boldmathsymbol{\Phi} \in \partial\Omega_-$, and $[\bm{u}(\boldmathsymbol{\Phi})\cdot\bm{\nabla}]\Psi_n(\boldmathsymbol{\Phi})=0$ when $\boldmathsymbol{\Phi} \in \partial\Omega_+$ with $\bm{u}$ orthogonal to $\partial\Omega_+$. Decomposing $P_\boldmathsymbol{\Phi} \left(\mathcal{N}=0\right)$ on the basis formed by these functions,
\bea
\label{eq:EigenProblem}
P_\boldmathsymbol{\Phi}\left(\mathcal{N}=0\right) = \sum_n \alpha_n \Psi_n\left(\boldmathsymbol{\Phi}\right)\, ,
\eea
\Eq{eq:adjoint:FokkerPlanck:formal:solution} gives rise to
\bea
\label{eq:tail_expansion:2}
P_\boldmathsymbol{\Phi}\left(\mathcal{N}\right) = \sum_n \alpha_n \Psi_n\left(\boldmathsymbol{\Phi}\right) \ee^{-\Lambda_n \mathcal{N}}\, .
\eea
This expression is nothing but the tail expansion~(\ref{eq:tail_expansion}), if one identifies $a_n(\boldmathsymbol{\Phi})=\alpha_n\Psi_n(\boldmathsymbol{\Phi})$.

Let us note that the first boundary condition given below \Eq{eq:eigen:value:problem} comes from the requirement that $P_{\boldmathsymbol{\Phi}}(\N) = \delta(\N)$ when $\boldmathsymbol{\Phi} \in \partial\Omega_-$, so all eigen-components should be identically zero for $\boldmathsymbol{\Phi} \in \partial\Omega_+$ except when $\Lambda_n=\infty$. The second boundary condition simply comes from the reflective surface located at $\partial\Omega_+$. 

One can also notice that \Eq{eq:eigen:value:problem} for the eigenfunctions $\Psi_n$ is the same as \Eq{eq:ODE:chi} for the characteristic function, if one identifies $t$ with $-i\Lambda_n$. However, the boundary conditions are different, which makes the two problems technically different (and one can be more convenient to solve than the other), although perfectly equivalent. In particular, solving one problem automatically gives the solution for the other. Indeed, if \Eq{eq:ODE:chi} has been solved and the functions $a_n(\boldmathsymbol{\Phi})$ derived, then the coefficients $\alpha_n$ can be obtained as follows. Making use of the fact that the eigenfunctions $\Psi_n$ form an orthonormal set, \ie
\bea
\label{eq:Psin:normalisation}
\left\langle \Psi_n , \Psi_m \right\rangle =
\int_{\Omega} \Psi_n(\boldmathsymbol{\Phi}) \Psi_m(\boldmathsymbol{\Phi}) \dd \boldmathsymbol{\Phi} = \delta_{n,m},
\eea
where we recall that $\Omega$ is the field-space domain located between $\partial\Omega_-$ and $\partial\Omega_+$ (see \Fig{fig:sketchFirstPassageTime}), \Eq{eq:tail_expansion} leads to
\bea
\left\langle \Psi_n ,P_\boldmathsymbol{\Phi}\left(\mathcal{N}\right)\right\rangle = \left[
\int_{\Omega} \Psi_n\left(\boldmathsymbol{\Phi}\right)
a_n(\boldmathsymbol{\Phi})
\dd\boldmathsymbol{\Phi}
\right]\,e^{-\Lambda_n\, \N}\, ,
\eea
while \Eq{eq:tail_expansion:2} gives rise to
\bea
\label{eq:flat:<Phi,P>:1}
\left\langle \Psi_n ,P_\boldmathsymbol{\Phi}\left(\mathcal{N}\right)\right\rangle = \alpha_n  \ee^{-\Lambda_n \mathcal{N}}\, .
\eea
By identifying the two expressions, one obtains
\bea
\label{eq:alphan:from:an}
\alpha_n = 
\int_{\Omega} \Psi_n\left(\boldmathsymbol{\Phi}\right)
a_n(\boldmathsymbol{\Phi})
\dd\boldmathsymbol{\Phi}\, .
\eea
Conversely, if the decomposition~(\ref{eq:EigenProblem}) has been performed and the coefficients $\alpha_n$ are known, then the functions $a_n(\boldmathsymbol{\Phi})$ can be obtained from the relation $a_n(\boldmathsymbol{\Phi})=\alpha_n\Psi_n(\boldmathsymbol{\Phi})$ given below \Eq{eq:tail_expansion:2}.

\subsection{Applications}
\label{sec:applications}
We now apply the techniques developed in the previous section to concrete examples. We investigate four single-field, slow-roll inflationary potentials: an exactly flat potential in \Sec{sec:flat_potential}, a potential with a constant slope in \Sec{sec:linear_potential}, a potential with a cubic flat inflection point in \Sec{sec:inflection_potential}, and a potential with a linearly-tilted cubic inflection point in \Sec{sec:inflection_linear_potential}. The adjoint Fokker-Planck operator is given by \Eq{eq:Fokker:Planck:adjoint:PDF:Ito:SlowRoll}, and the boundary conditions simply consist in an absorbing wall located at $\phi_\uend$ and a reflective wall located at $\phiuv$.
\subsubsection{Flat potentials}
\label{sec:flat_potential}
\begin{figure}[t]
\centering 
\includegraphics[width=.49\textwidth]{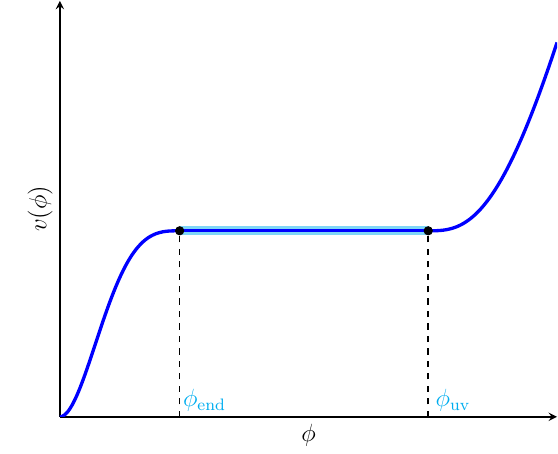}
\includegraphics[width=.49\textwidth]{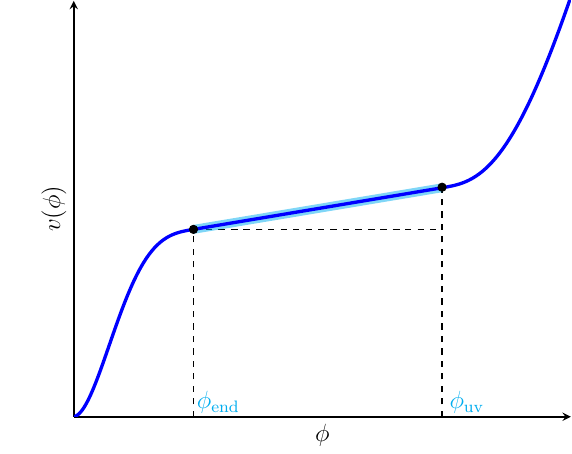}
 \caption{Schematic representation of the flat (left) and linear (right) potentials studied in sections \ref{sec:flat_potential} and \ref{sec:linear_potential} respectively. We only consider the region between $\phiend$ and $\phiuv$.}
 \label{fig:flat_linear_potential}
\end{figure}
Let us begin by considering the flat potential discussed in \Sec{sec:StochasticLimit} and displayed again in the left panel of \Fig{fig:flat_linear_potential}, namely $v=v_0$ between $\phiend = 0$ and $\phiuv=\dphiwell$. Let us recall that, in principle, if the potential is exactly flat, slow roll is violated since there is no potential gradient. However, in \Sec{sec:example:1_plus_phi_to_the_p}, it was shown that for a potential of the form $v = v_0 [1+(\phi/\phi_0)^p]$, where $\phi_0\gg \Mp$ such that slow roll is never violated, in the region of the potential located between $\phi=0$ and $\phi=\dphiwell = \phi_0 v_0^{1/p}$, the potential gradient term  in \Eq{eq:Fokker:Planck:adjoint:PDF:Ito:SlowRoll} (\ie the first term on the right-hand side) can be neglected, and the classical part of the potential above $\dphiwell$ acts as a reflective wall (see also the discussion around \Eq{eq:classicality_criterion} below). The full results were thus shown to be very accurately reproduced if one places a reflective boundary condition at $\phi=\dphiwell $ and considers a pure diffusion process between $\phi=0$ and $\phi=\dphiwell$. 

\begin{figure}[t!]
\centering 
\includegraphics[width=.59\textwidth]{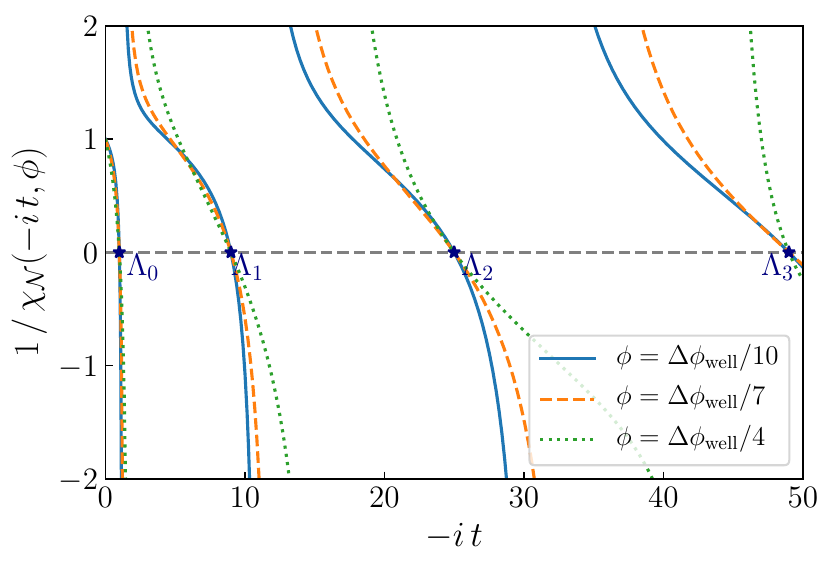}
\caption{Zeros of the inverse characteristic function for a flat potential. We have chosen $\mu^2=\pi^2/4$ so that zeros are located at $\Lambda_n=(2n+1)^2$. We have evaluated $\chi^{-1}_\N(t,\phi)$ at different field values $\phi$. Although the characteristic functions for each $\phi$ are different, the location of the poles $\Lambda_n$ (which determine the decay rates of the PDF) is universal for a given potential.}
 \label{fig:ichi_flat}
\end{figure}

This is the situation we consider here, where ``flat'' potential has to be taken in the sense of that specific limit, in which slow roll is not violated because the initial field velocity is also taken to zero, more quickly than the potential slope. The problem was entirely solved in \Sec{sec:StochasticLimit}, so here we only want to check the consistency of this approach sketched in \Sec{sec:tail_curvature} with our previous results.
\paragraph{Poles of the characteristic function.} $ $ \\
In this simple example, the equation for the characteristic function \eqref{eq:ODE:chi} reads
\bea
\label{eq:eom:chi:flat}
\chi''_\N(t,\phi) + \frac{i\,t}{v_0\Mp^2} \chi_\N(t,\phi) = 0\,,
\eea
where a prime denotes derivation with respect to the field value $\phi$, with boundary conditions $\chi_\N(t,0)=1$ and $\chi_\N^\prime(t,\phiuv)=0$. It can be solved as
\bea 
\label{eq:chi_flat}
\chi_\N(t,\phi)=\frac{\cos\lb(it)^{1/2}\mu(x-1)\rb}{\cos\lb(it)^{1/2}\mu\rb}\,,
\eea
where we use the dimensionless field variable $x=\phi/\dphiwell$ as in \Sec{sec:StochasticLimit}, and the quantity $\mu$ defined in \Eq{eq:def:mu}. The poles of \Eq{eq:chi_flat} correspond to when the argument of the $\cos$ function in the denominator equals $(n+1/2)\pi$, where $n$ is an integer number, and calling $\Lambda_n$ the value of $it$ at these poles, one has
\bea \label{eq:flat_poles}
\Lambda_n=\frac{\pi^2}{\mu^2}\lp n+\frac{1}{2}\rp^2\,.
\eea
One can check that, in agreement with the discussion of \Sec{sec:pole:chi}, the $\Lambda_n$'s are all real, positive and independent of $\phi$.
The exponential decay rate of the tail of the PDF therefore depends both on the width of the quantum well, $\dphiwell$, and its scale $v_0$, through the combination $\mu$. We have plotted the inverse characteristic function for a flat potential in \Fig{fig:ichi_flat}, for a few field values, where the zeros of $\chi^{-1}_\N(t,\phi)$ correspond to $-i\Lambda_n$. This illustrates again that, although the details of the characteristic functions depend on $\phi$, the location of their poles $\Lambda_n$ is universal for a given potential.

Finally, making use of \Eq{eq:alphan:flat:generic}, the coefficients $a_n$ are given by
\begin{align}
a_n(\phi) = (-1)^n \frac{\pi}{\mu^2}\left(2n+1\right) \cos\left[\frac{\pi}{2}\left(2n+1\right)\left(x-1\right)\right] .
\label{eq:an:flat}
\end{align}

\begin{figure}[t!]
\centering 
\includegraphics[width=.59\textwidth]{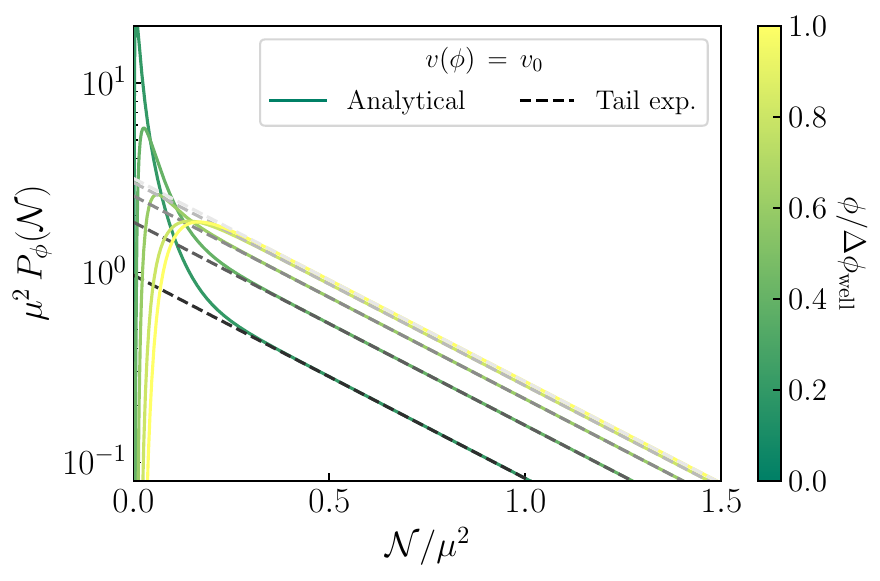}
 \caption{Probability distribution function of the number of \efolds~$\N$ in a flat potential, starting from different initial field values $\phi$. We compare the full PDF, \Eq{eq:stocha:HeatMethod:PDF:expansion} (solid lines), with the leading term in the tail expansion~(\ref{eq:tail_expansion}) (dashed lines). We rescale the axes by $\mu^2=\dphiwell ^2/(v_0 \Mp^2)$, such that, using the self-similarity of \Eq{eq:stocha:HeatMethod:PDF:expansion}, the result does not depend on~$\mu$.} 
 \label{fig:PDF_flat}
\end{figure}

\paragraph{Eigenvalue problem.}$ $ \\
In the case of a flat potential, the eigenvalue problem \eqref{eq:eigen:value:problem} reads
\bea
\label{eq:Psin:eom:flat}
\Psi_n''\left(\phi\right) + \frac{\Lambda_n}{v_0\Mp^2} \Psi_n\left(\phi\right) = 0,
\eea
with boundary conditions $\Psi_n(0) = \Psi_n'(\dphiwell)=0$.  
The generic solution of \Eq{eq:Psin:eom:flat} is
\bea
\Phi_n\left(\phi\right) = A_n \exp\left(i\sqrt{\frac{\Lambda_n}{v_0\Mp^2}}\phi\right) + B_n \exp\left(-i\sqrt{\frac{\Lambda_n}{v_0\Mp^2}}\phi\right)\, ,
\eea
where the first boundary condition imposes that $B_n=-A_n$, hence $\Psi_n(\phi) \propto \sin[\sqrt{{\Lambda_n}/({v_0\Mp^2})}\phi]$. The second boundary condition then implies that $\cos[\sqrt{{\Lambda_n}/({v_0\Mp^2})}\dphiwell]=0$, which precisely gives rise to \Eq{eq:flat_poles}. Normalising the functions $\Psi_n$ as in \Eq{eq:Psin:normalisation}, one then has
\bea
\label{eq:phin_flat}
\Psi_n\left(\phi\right)= \sqrt{\frac{2}{\dphiwell}} \sin\left[\pi\left(n+\frac{1}{2}\right)\frac{\phi}{\dphiwell}\right] .
\eea
The coefficients $\alpha_n$ can be computed from \Eq{eq:alphan:from:an}, and \Eqs{eq:an:flat} and~(\ref{eq:phin_flat}) give rise to
\bea
\label{eq:alphan:flat}
\alpha_n = \frac{2\pi}{\mu^2}\left(n+\frac{1}{2}\right) \sqrt{\frac{\dphiwell}{2}}\, .
\eea
Altogether,  for the PDF in the constant potential, one obtains \Eq{eq:stocha:HeatMethod:PDF:expansion} exactly, which confirms the validity of our approach. In \Fig{fig:PDF_flat} (see also \Fig{fig:pdf_stochastic}), we plot both the full PDF~(\ref{eq:stocha:HeatMethod:PDF:expansion}) and the leading term in the tail expansion~(\ref{eq:tail_expansion}), $a_0(\phi)e^{-\Lambda_0\N}$. As it can be observed, at large-$\N$ values, the tail expansion provides an excellent approximation to the full result. Note also that \Eq{eq:stocha:HeatMethod:PDF:expansion} is such that the PDF of the quantity $\N/\mu^2$ is independent of $\mu$, which is why this quantity is displayed in \Fig{fig:PDF_flat}. This shows that increasing $v_0$, or decreasing $\dphiwell$, decreases the typical values of $\N$.

\subsubsection{Potentials with constant slope}
\label{sec:linear_potential}

Let us now consider a potential of the type
\bea
\label{eq:pot:constant:slope}
v\left(\phi\right) = v_0\left(1+\alpha\frac{\phi}{\Mp}\right),
\eea
with a constant slope $\alpha$, which we will assume is positive without loss of generality. The model~\eqref{eq:pot:constant:slope} is bounded between $\phi=0$ where the potential is supposed to become steeper and/or inflation ends; and $\phi=\phiuv$ where the potential is supposed to become steeper and the dynamics of $\phi$ dominated by classical drift, which acts as a reflective wall as discussed at the beginning of \Sec{sec:flat_potential}. Since $\epsilon_1\simeq \alpha^2/2$, one should have $\alpha\ll 1$ in order for slow roll to be valid. Moreover, we will consider only scenarios where $\phi_{\mathrm{uv}}\ll \Mp/\alpha$, such that the potential is almost constant, $v\simeq v_0$, between $\phi=0$ and $\phi=\phiuv$ (see the right panel of \Fig{fig:flat_linear_potential}). 

\paragraph{Poles of the characteristic function.}$ $ \\
The equation for the characteristic function \eqref{eq:ODE:chi} is given by
\bea \label{eq:chi_linear_general}
\chi_\N''(t,\phi) - \frac{v_0\alpha}{\Mp v(\phi)^2}\chi_\N'(t,\phi)+ \frac{i\,t}{\Mp^2v(\phi)} \chi_\N(t,\phi) = 0\,,
\eea
which, compared to \Eq{eq:eom:chi:flat} for a flat potential, contains an additional friction term. The other difference is that now, the coefficients of the differential equation depend on $\phi$, so there is no generic analytic solution. 
There is, however, an analytic solution in the ``almost-constant'' regime, $\phi_{\mathrm{uv}}\ll \Mp/\alpha$, where $v(\phi)$ can be replaced with $v_0$ in \Eq{eq:chi_linear_general}. This solution reads
\bea
\label{eq:chiN:cosh:ext}
\chi_\N\left(t, \phi \right) = \ee^{\frac{\alpha \mu x}{2  \sqrt{v_0}}}\frac{2\gamma\sqrt{it v_0} \cos \left[ \mu \gamma \sqrt{it}(x-1)\right] - \alpha  \sin \left[ \mu \gamma \sqrt{it}(x-1)\right]}{2\gamma \sqrt{ i t v_0} \cos \left({\mu} \gamma\sqrt{it }\right)+\alpha  \sin \left({\mu} \gamma\sqrt{it }\right)}\, ,
\eea
with 
\bea
\gamma = \sqrt{1-\frac{\alpha^2}{4itv_0}}\, ,
\eea
$x=\phi/\phiuv$, and $\mu$ is given by \Eq{eq:def:mu} where $\dphiwell$ is replaced by $\phiuv$. When $\alpha=0$, this boils down to the flat potential solution~\eqref{eq:chi_flat}.

\begin{figure}[t!]
\centering 
\includegraphics[width=.59\textwidth]{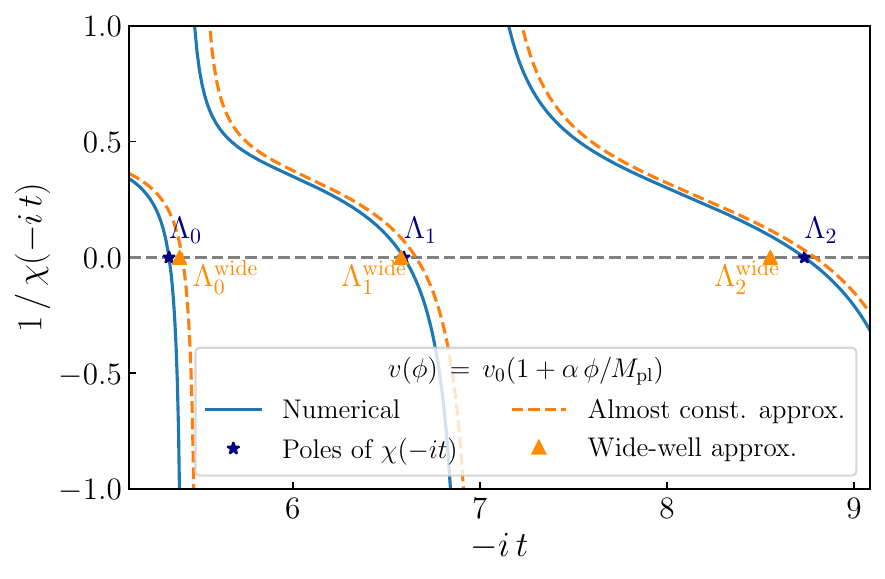}
\caption{Inverse characteristic function for a linear potential. We compare the numerical solution of \Eq{eq:chi_linear_general} (solid blue line) and its poles $\Lambda_n$ (labeled with blue stars) with the ``almost-constant'' approximation~\eqref{eq:chiN:cosh:ext} (orange dashed line) and its approximate poles $\Lambda^{\mathrm{wide}}_n$ in the ``wide-well approximation''~\eqref{eq:lambda_almost_const} (labeled with orange triangles). 
These two approximations are valid when $\phiuv/\Mp\ll1/\alpha$, and $\phiuv/\Mp\gg v_0/\alpha$, respectively. 
The characteristic function is evaluated at $\phi=\phiuv/10$, with $\alpha=0.1$, and $v_0$ and $\phiuv$ have been set such that $\phiuv/\Mp=0.01/\alpha$ and $\phiuv/\Mp = 20 v_0/\alpha$. Increasing $1/\alpha$ and decreasing $v_0/\alpha$ improves the agreement between the three results, but quickly makes the separation between the $\Lambda_n$ impossible to resolve by eye, which is why somewhat intermediate values have been used here, for illustrative purpose.
}
 \label{fig:ichi_linear}
\end{figure}

The poles of the characteristic function at $t=-i\Lambda_n$ are determined by the equation
\bea \label{eq:trascendental_linear}
\tan\left(\sqrt{\Lambda_n-\frac{\alpha^2}{4v_0}}\mu\right)=- 2   \frac{v_0}{\alpha}\frac{\Mp}{\phiuv}     \sqrt{\Lambda_n-\frac{\alpha^2}{4v_0}}\mu\, .
\eea
This equation is of the form $\tan(z)=-2 a z$, with $z=\sqrt{\Lambda_n-\alpha^2/(4v_0)}\mu$ and $a=v_0\Mp/(\alpha\phiuv)$. It has one obvious solution, namely $z=0$, which would lead to $\Lambda=\alpha^2/(4 v_0)$. However, the numerator of \Eq{eq:chiN:cosh:ext} also vanishes at $t=-i \alpha^2/(4 v_0)$, and by carefully expanding the characteristic function around that value, one can see that it is in fact regular, and does not possess a pole. The case $z=0$ can therefore be safely discarded, and for $z>0$, one has to solve a transcendental equation, which cannot be done analytically. However, approximate solutions can be found in the two limits $a\ll 1$ and $a\gg 1$, which we dub the ``wide-well'' and the ``narrow-well'' regimes respectively, since they imply a lower bound and an upper bound on $\phiuv$ respectively.
\subparagraph{Wide-well limit $\phiuv/\Mp \gg v_0/\alpha$}
In this case $a\ll 1$, hence $\vert \tan(z)/z \vert \ll 1$, which implies that $z$ is close to $(n+1)\pi$, with $n$ an integer number. One can write $z=(n+1)\pi+\delta z$, and expand $\tan(z)\simeq \delta z + \delta z^3/3+\cdots$. Plugging this formula into the transcendental equation, and expanding in $a$, one obtains $\delta z\simeq -2 a (n+1)\pi[1-2a (n+1) \pi + \cdots]$, which gives rise to
\bea
\label{eq:lambda_almost_const}
\Lambda_n^\mathrm{wide} = \frac{\alpha^2}{4 v_0}+ \frac{\left(n+1\right)^2 \pi^2}{\mu^2}\left(1-4 \frac{v_0 \Mp}{\alpha\phiuv} + \cdots \right)
\eea
where ``$\cdots$'' denotes higher powers of $v_0\Mp/(\alpha\phiuv)$, so this approximation is indeed valid in the regime
\bea
\label{eq:constant:slope:expansion:transcendental:condition}
\frac{\phiuv}{\Mp} \gg \frac{v_0}{\alpha} \, .
\eea
If one takes $\phiuv$ to its maximal allowed value, $\phiuv\sim \Mp/\alpha$, this condition is satisfied as soon as $v_0\ll 1$, which is always the case. 

\begin{figure}[t!]
\centering 
\includegraphics[width=.59\textwidth]{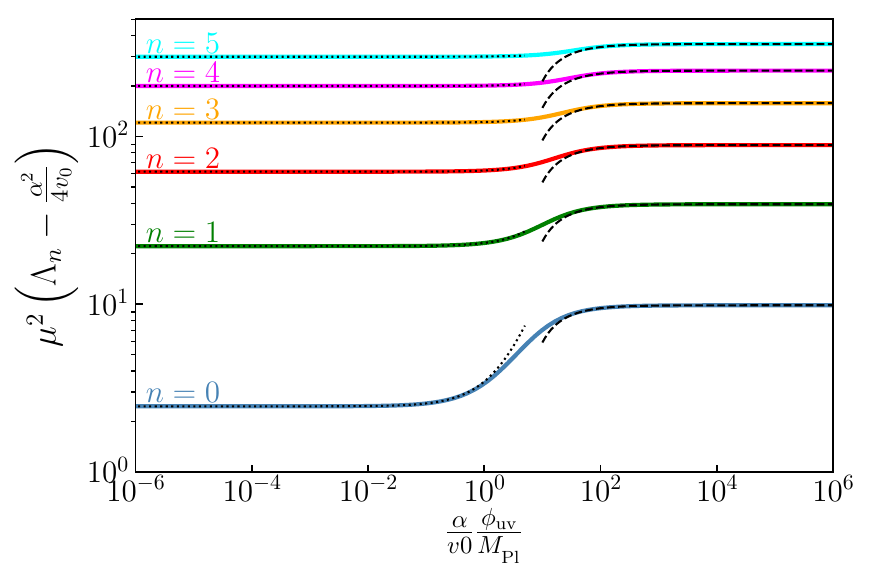}
\caption{Decay rates $\Lambda_n$ for the constant-slope potential~\eqref{eq:pot:constant:slope}, as a function of $\alpha\phiuv/(v_0\Mp)$. The coloured lines correspond to numerical solutions of the transcendental equation~\eqref{eq:trascendental_linear}, the black dashed lines display the wide-well approximation~\eqref{eq:lambda_almost_const}, and the black dotted lines stand for the narrow-well approximation~\eqref{eq:lambda_almost_const:narrow}.}
 \label{fig:constantslope:Lambdan}
\end{figure}

In \Fig{fig:ichi_linear} we show the inverse characteristic function obtained by solving numerically \Eq{eq:chi_linear_general} (solid blue line) and the analytical solution~\eqref{eq:chiN:cosh:ext} in the almost-constant approximation (orange dashed line). We also include the approximate values of $\Lambda_n^\mathrm{wide}$ in the wide-well limit given in \Eq{eq:lambda_almost_const}. In order to remain in the regime of validity of this approximation but to make visible the differences between these different estimates, we set the parameters such that $\phiuv/\Mp=0.01/\alpha$ and $\phiuv/\Mp = 20 v_0/\alpha$. By increasing $1/\alpha$ and decreasing $v_0/\alpha$, the agreement between the three results largely improves, but quickly makes the separation between the $\Lambda_n$ impossible to resolve by eye, which is why intermediate values have been used here, for illustrative purpose. 
To test further the consistency of these results, the formula~\eqref{eq:lambda_almost_const} is compared with a numerical solution of the transcendental equation~\eqref{eq:trascendental_linear} in \Fig{fig:constantslope:Lambdan}, where one can check that, as long as \Eq{eq:constant:slope:expansion:transcendental:condition} is valid, it provides indeed a good approximation.

By comparing \Eqs{eq:flat_poles} and~(\ref{eq:lambda_almost_const}), one can check that, as in the flat case, $\Lambda_n$ receives a $n$-dependent contribution proportional to $\pi^2/\mu^2$, but it is also shifted by a fixed quantity, namely $\alpha^2/(4 v_0)$, which dominates over the $n$-dependent contribution, because of the condition~(\ref{eq:constant:slope:expansion:transcendental:condition}). One concludes that, in the wide-well regime, adding a small slope to the potential is enough to highly suppress the tails.

Finally, the $a_n$ functions can be approximated as follows. One needs to expand \Eq{eq:chiN:cosh:ext} around the poles $t=-i\Lambda_n$ in order to extract the residues. Since the decay rates $\Lambda_n$ are not known exactly, this expansion cannot be done directly. However, writing $\Lambda_n = \Lambda_n^{(0)} + \delta \Lambda_n $, where $\Lambda_n^{(0)}$ corresponds to the approximation~(\ref{eq:lambda_almost_const}), one can parametrise $t$ in the neighbourhood of the poles as $t=- i  \Lambda_n^{(0)}  - i  \delta \Lambda_n + \delta t$, and expand the characteristic function in $\delta t$. Obviously, the function cannot be probed on scales smaller than $\delta \Lambda_n$, so one assumes in fact $\delta \Lambda_n \ll \delta t \ll  \Lambda_n^{(0)}$, and performs a double expansion in $ \delta \Lambda_n$ and $\delta t$ under these conditions. By identification with \Eq{eq:chi:pole:expansion}, the residues can then be extracted, and one obtains
\begin{align}
a_n^{\mathrm{wide}}(\phi) = -(-1)^n \frac{\pi}{\mu^2}2\left(n+1\right) e^{\frac{\alpha\phiuv}{2v_0\Mp} x} \sin\left[{\pi}\left(n+1\right)\left(x-1\right)\right] .
\label{eq:a:constant:slope}
\end{align}
One notices that the structure is similar to, though different from, the one for a flat potential~\eqref{eq:an:flat}. In particular, the exponential term gives a strong enhancement, because of \Eq{eq:constant:slope:expansion:transcendental:condition}.
\subparagraph{Narrow-well limit $\phiuv/\Mp \ll v_0/\alpha$}
In the opposite limit where $a\gg 1$, $\vert \tan(z)/z \vert \gg 1$, which implies that $z$ must be close to $\pi/2+n\pi$, where $n$ is an integer number. One can write $z=\pi/2+n\pi+\delta z$, and expand $\tan(z)\simeq -1/\delta z+\delta z/3+\cdots$. Plugging this formula into the transcendental equation, and expanding in powers of $1/a$, one obtains $\delta z \simeq 1/[2\pi a(n+1/2)]+\cdots$, which gives rise to
\bea
\label{eq:lambda_almost_const:narrow}
\Lambda_{n}^\mathrm{narrow} = \frac{\pi^2}{\mu^2}\left[\left(n+\frac{1}{2}\right)^2+\frac{\alpha\phiuv}{\pi^2 v_0\Mp}+\cdots\right],
\eea
where ``$\cdots$'' denotes higher powers in $\alpha\phiuv/(v_0\Mp)$, so this approximation indeed holds in the regime 
\bea
\label{eq:small:alpha:limit:def}
\frac{\phiuv}{\Mp} \ll \frac{v_0}{\alpha} \, .
\eea
The formula~\eqref{eq:lambda_almost_const:narrow} is compared with a numerical solution of \Eq{eq:trascendental_linear} in \Fig{fig:constantslope:Lambdan}, where one can check that, as long as \Eq{eq:small:alpha:limit:def} is valid, it indeed provides a good approximation.

By comparing \Eqs{eq:flat_poles} and~(\ref{eq:lambda_almost_const:narrow}), one can see that the difference with the flat-potential case is negligible: adding a slope in the narrow-well regime only shifts the spectrum by a correction, $\alpha\phiuv/(v_0\Mp)$, which is, by definition, tiny in that regime. The same procedure as the one outlined above \Eq{eq:alphan:constant:slope} can also be performed in order to extract the $a_n$ functions. At leading order in $\alpha\phiuv/(v_0\Mp)$, one exactly recovers \Eq{eq:alphan:flat}, which finishes to prove that the narrow-well regime in fact corresponds to the flat-potential limit. Since \Eq{eq:small:alpha:limit:def} can also be interpreted as an upper bound on $\alpha$, this result makes sense.
\subparagraph{Comparison with the classical result}
The classical value of the power spectrum, $\calP_{\zeta,\mathrm{cl}}=2v^3/(\Mp^2v'^2)$, in this model, is given by $\calP_{\zeta,\mathrm{cl}}\simeq 2v_0/\alpha^2$. Since $\Lambda_0= \alpha^2/(4 v_0)$ in the wide-well regime, the dominant behaviour on the tail can thus be written as 
\bea
\label{eq:constant:slope:wide:well:tail:comp:class}
P_\phi^{\mathrm{wide}}(\mathcal{N}) \underset{\mathcal{N}\gg 1}{\propto} \exp\left(-\frac{1}{2}\frac{\mathcal{N}}{\calP_{\zeta,\mathrm{cl}}}\right)\, .
\eea
By comparison, in the classical picture, \Eq{eq:tail_expansion:classical} gives rise to
\bea
\label{eq:constant:slope:class:tail}
\left. P_\phi(\mathcal{N}) \right\vert_{\mathrm{cl}} 
\underset{\mathcal{N}\gg 1}{\propto}
 \exp\left(-\frac{1}{2}\frac{\mathcal{N}^2}{\calP_{\zeta,\mathrm{cl}} N_\ucl}\right)\, ,
\eea
where $N_\ucl$ is the classical number of \efolds~that arises from the integration over $k$ in \Eq{eq:tail_expansion:classical}, which can trivially be performed since $\calP_{\zeta,\mathrm{cl}}$ is independent of $\phi$ in the almost-constant approximation. Two remarks are in order. First, as mentioned above, as soon as $\N\gg N_\ucl$, the amount of power on the tail is greatly enhanced in the full stochastic theory compared to the classical, Gaussian approximation. Second, the similarity between \Eqs{eq:constant:slope:wide:well:tail:comp:class} and~(\ref{eq:constant:slope:class:tail}), which coincide for $\N=N_\ucl$, is an illustration of the resemblance between the classical and the full stochastic theory for a linear potential. Indeed, as shown in \Sec{sec:stochastic:delta:N}, the classical limit can be obtained from the full stochastic PDF by a saddle-point expansion, where higher-order corrections involve either $v$, which is always small, or derivatives of the potential of order 2 or higher [see for instance \Eqs{eq:Nmean:vll1limit} and~\eqref{eq:PS:vll1}], which vanish in the present case. For instance, the classical number of \efolds~is given by \Eq{eq:stocha:meanN:classtraj}, which, in the almost-constant approximation, reduces to 
\bea
\label{eq:Ncl:constant:slope}
N_\ucl = \frac{\phi-\phi_\uend}{\alpha \Mp} .
\eea
In the full stochastic theory, plugging \Eq{eq:chiN:cosh:ext} into \Eq{eq:meanN:chi} gives rise to
\begin{align}
\left\langle \N \right\rangle &= 
N_\ucl + \frac{v_0}{\alpha^2} \left[ \ee^{-\frac{\alpha}{v_0\Mp}\left(\phiuv-\phi_\uend\right)}- \ee^{-\frac{\alpha}{v_0\Mp}\left(\phiuv-\phi\right)}\right] .
\label{eq:mean:N:constant:slope}
\end{align}
In the wide-well regime, \ie when the condition~(\ref{eq:constant:slope:expansion:transcendental:condition}) is satisfied, the second term in the above expression is exponentially suppressed (unless one starts at a value of $\phi$ very close to $\phiuv$), and 
\bea
\langle \N \rangle_{\mathrm{wide}} \simeq N_\ucl .
\eea
In the narrow-well regime however, the effect of the boundary located at $\phiuv$ is not negligible anymore, and  expanding \Eq{eq:mean:N:constant:slope} in the limit~(\ref{eq:small:alpha:limit:def}), one finds
\bea
\left\langle \mathcal{N} \right\rangle_\mathrm{narrow} \simeq \frac{\left(\phi-\phi_\uend\right)\left(2\phiuv-\phi-\phi_\uend\right)}{2\Mp^2 v_0},
\eea
which is very different from, and in fact much smaller than, its classical counterpart~\eqref{eq:Ncl:constant:slope}.

\paragraph{Eigenvalue problem} $ $ \\
The eigenvalue problem is analogous to solving the equation for $\chi_\N(t,\phi)$, \Eq{eq:chi_linear_general}; there is no general solution. In the almost-constant approximation,
\bea \label{eq:eigenequation_linear}
\Psi_n''\left(\phi\right) - \frac{\alpha}{\Mp v_0}\Psi_n'\left(\phi\right)+ \frac{\Lambda_n}{v_0\Mp^2} \Psi_n\left(\phi\right) = 0\,,
\eea
the solution reads
\bea
\displaystyle
\Psi_n\left(\phi\right) = \ee^{\frac{\alpha\phi}{2\Mp v_0}}\left(A_n \exp^{i\sqrt{v_0 \Lambda_n - \frac{\alpha^2}{4}}\frac{\phi}{v_0\Mp}}+B_n \ee^{-i\sqrt{v_0\Lambda_n - \frac{\alpha^2}{4}}\frac{\phi}{v_0\Mp}}\right) .
\eea
The first boundary condition imposes $B_n=-A_n$, and the second one implies the transcendental equation \eqref{eq:trascendental_linear}, which has solution~\eqref{eq:lambda_almost_const} in the wide-well limit~\eqref{eq:constant:slope:expansion:transcendental:condition}, and solution~\eqref{eq:lambda_almost_const:narrow} in the narrow-well limit~\eqref{eq:small:alpha:limit:def}. The eigenfunctions are thus given by
\bea
\label{eq:constant:slope:vaccum:dom:Phin}
\Psi_n\left(\phi\right)=\sqrt{\frac{2}{\phiuv}}\exp\left(\frac{\alpha\phi}{2\Mp v_0}\right)\sin\left(\sqrt{v_0 \Lambda_n - \frac{\alpha^2}{4}}\frac{\phi}{v_0\Mp}\right)\, ,
\eea
which boils down to \Eq{eq:phin_flat} if $\alpha=0$, and where the eigenfunctions have been normalised in the (extended) sense that $\langle \Phi_n^{(-\alpha)} , \Phi_m^{(\alpha)}\rangle  = \delta_{n,m}$. 

The coefficients in the expansion~(\ref{eq:tail_expansion}), $\alpha_n$, can be determined by following the procedure outlined at the end of \Sec{sec:eigenvalues}, where in \Eq{eq:alphan:from:an}, our extended scalar product has to be used, \ie $ \alpha_n = \langle \Psi_n^{(-\alpha)}(\phi) , a_n^{(\alpha)}(\phi)\rangle $. In the wide-well regime, this leads to
\bea
\label{eq:alphan:constant:slope}
\alpha_n^\mathrm{wide} = \frac{2\pi}{\mu^2}\left(n+1\right) \sqrt{\frac{\phiuv}{2}}\, .
\eea
Notice that it is similar to \Eq{eq:alphan:flat} for a flat potential, the only difference being that $n+1/2$ is replaced by $n+1$. Combining the above results, the PDF can be approximated as
\bea
P_\phi^{\mathrm{wide}}(\N) = 2\frac{\pi}{\mu^2} e^{\frac{\alpha \phiuv}{2 v0 \Mp}\frac{\phi}{\phiuv}} \sum_{n=0}^\infty n \sin\left(\pi n \frac{\phi}{\phiuv}\right)\ee^{-\left(\frac{\alpha^2}{4 v_0}+n^2 \frac{\pi^2}{\mu^2}\right)\N}\, .
\eea
One notices that the structure of the result is very similar to the one for a flat potential, \Eq{eq:stocha:HeatMethod:PDF:expansion}, if one replaces $n+1/2$ by $n$ in the sum, with the crucial difference that, now, the tails are suppressed by an additional $\ee^{-\alpha^2\N/(4 v_0)}$ factor. As in \Eq{eq:PDF:thetatwo}, the result can be expressed in terms of the derivative of an elliptic theta function 
\bea
\label{eq:constant:slope:PDF:elliptic}
P_\phi^\mathrm{wide}(\N)= -\frac{\pi}{2\mu^2}
\ee^{\frac{\alpha \phiuv}{2 v0 \Mp}\frac{\phi}{\phiuv}}
\ee^{-\frac{\alpha^2}{4 v_0}\N}
\vartheta_3'\left(\frac{\pi\phi}{2\phiuv},\ee^{-\frac{\pi^2}{\mu^2}\N}\right)\, .
\eea
In the narrow-well limit, as explained above, the flat-potential formulas are recovered, hence one obtains \Eq{eq:alphan:flat}, and the flat-potential PDF~\eqref{eq:stocha:HeatMethod:PDF:expansion} is found, up to small corrections suppressed by $\phiuv\alpha/(\Mp v_0)$.

\paragraph{WKB approach} $ $ \\
The above results have been derived in the almost-constant approximation, which holds when $\phiuv\ll \Mp/\alpha$. To go beyond, one can employ the following adiabatic (or WKB) approach, in which a slightly different transcendental equation has to be solved. For a generic single-field slow-roll potential, the eigenvalue problem \eqref{eq:eigen:value:problem} reads
\bea
\label{eq:eigen:equation:app}
\Psi''_n-\frac{v_\phi}{v^2}\Psi'_n+\frac{\Lambda_n}{v}\Psi_n=0\, ,
\eea
with boundary conditions $\Psi_n(\phiend)=\Psi'_n(\phiuv)=0$. A first remark is that the friction term, proportional to $v_\phi$, can be absorbed through the field redefinition
\bea
\Psi_n=\exp\left[{\frac{1}{2}\int\frac{v_\phi}{v^2}\dd\phi}\right]\tilde{\Psi}_n\,.
\eea
In this way, \Eq{eq:eigen:equation:app} becomes
\bea
\tilde{\Psi}''_n+\left[\frac{\Lambda_n}{v}-\frac{1}{4}\lp\frac{v_\phi}{v^2}\rp^2+\frac{1}{2}\lp\frac{v_\phi}{v^2}\rp'\right]\tilde{\Psi}_n=0\,,
\eea
which can be solved in plane waves whenever the frequency is slowly varying. In that regime, the solution reads
\bea
\Psi_n=\exp\left[{\frac{1}{2}\int\frac{v_\phi}{v^2}\dd\phi}\right]
\lp\alpha_n e^{i\int\theta_n\dd\phi}+\beta_ne^{-i\int\theta_n\dd\phi}\rp\,,
\eea  
where the phase $\theta_n$ reads
\bea
\theta_n^2=\frac{\Lambda_n}{v}-\frac{1}{4}\lp\frac{v_\phi}{v^2}\rp^2+\frac{1}{2}\lp\frac{v_\phi}{v^2}\rp'\,.
\eea
The first boundary condition imposes $\alpha_n=-\beta_n$, so that the eigenfunctions are given by
\bea
\Psi_n=2i\alpha_n\, \exp\left[{\frac{1}{2}\int\frac{v_\phi}{v^2}\dd\phi}\right]
\sin\left(\int\theta_n \dd\phi\right)\,.
\eea 
The second boundary condition determines a transcendental equation for the eigenvalues $\Lambda_n$,
\bea
\tan\lb\int_{\phiend}^{\phiuv}\theta_n(\phi) \dd\phi\rb=-2\frac{v^2(\phiuv)}{v_\phi(\phiuv)}\theta_n(\phiuv)\,.
\eea
In practice, one can check numerically that this expression provides a better approximation than the almost constant approximation leading to \Eq{eq:trascendental_linear}. However, one does not avoid having to solve a transcendental equation to obtain the eigenvalues.

In summary, for a potential with a constant slope $\alpha$ over a certain field range $\phiuv$, either the range is narrow in the sense that $\phiuv/\Mp\ll v_0/\alpha$, and the potential can be approximated as constant, such that the results of \Sec{sec:flat_potential} can be used; or the range is wide in the sense that $\phiuv/\Mp\gg v_0/\alpha$, and the PDF receives a strong suppression $\ee^{-\alpha^2\N/(4 v_0)}$ on its tail compared to the flat potential case.

\begin{figure}[t!]
\centering 
\includegraphics[width=.49\textwidth]{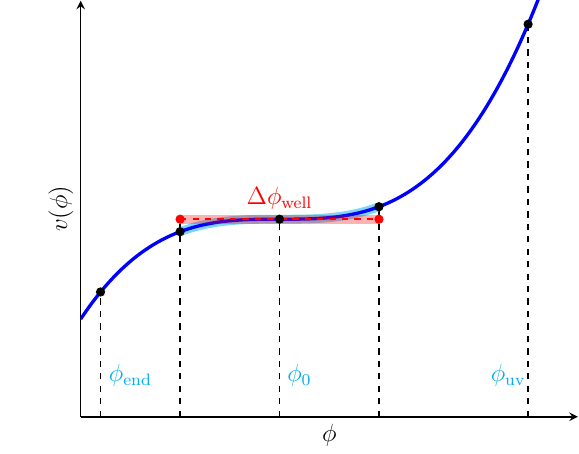}
\includegraphics[width=.49\textwidth]{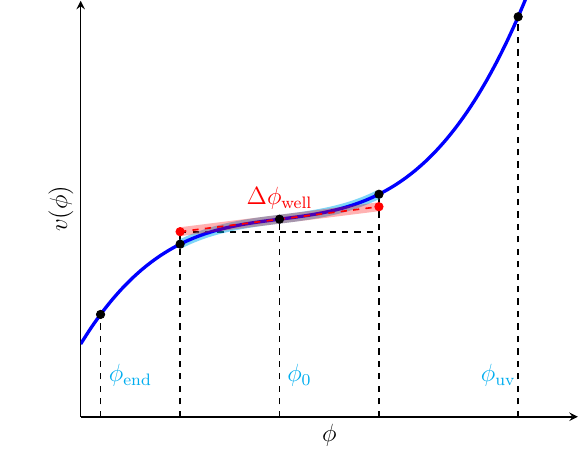}
 \caption{Schematic representation of the inflection (left) and tilted inflection (right) point potentials studied in \Secs{sec:inflection_potential} and \ref{sec:inflection_linear_potential} respectively. We solve the stochastic evolution between $\phiend$ and $\phiuv$. The region where quantum diffusion dominates, of width $\dphiwell$, can be approximated by a flat potential or a constant-slope potential respectively, and is determined by the non-classicality criterion (\ref{eq:classicality_criterion}).}
 \label{fig:inflection_linear_potential}
\end{figure}

\subsubsection{Inflection point potentials}
\label{sec:inflection_potential}

The toy models analysed in the two previous sections, the flat potential in \Sec{sec:flat_potential} and the constant-slope potential in \Sec{sec:linear_potential}, can serve as building blocks to study more realistic scenarios, that we now investigate in the two following sections. The first one is a potential with a flat inflection point located at $\phi_0$, as schematically displayed in the left panel of \Fig{fig:inflection_linear_potential}. In practice, we consider for simplicity a cubic potential
\bea
\label{eq:pot:inflection}
v\left(\phi\right) = v_0\left[1+\beta\lp\frac{\phi-\phi_0}{\Mp}\rp^{3}\right]\,,
\eea
although our conclusions can be easily generalised to other odd powers. Without loss of generality, we assume $\beta>0$, so the potential is positive when $x>-\beta^{-1/3}$, where we have defined 
\bea
x\equiv \frac{\phi-\phi_0}{\Mp}\, .
\eea
The inflaton is assumed to evolve in the field range comprised between $\phi_\uend$, where inflation ends, and an upper bound value $\phiuv$.

The potential being exactly flat around $\phi_0$, stochastic effects dominate in the neighbourhood of the inflection point. More precisely, as explained in \Sec{sec:stochastic:delta:N}, see \Eq{eq:classicalcriterion:def}, and further exemplified in \Sec{sec:example:1_plus_phi_to_the_p}, when the condition
\bea \label{eq:classicality_criterion}
\frac{v''\,v^2}{v'^2}\gg 1
\eea 
holds, the potential can be assumed to be exactly flat (hence dominated by quantum diffusion), while when the opposite condition applies, the dynamics of the inflaton is essentially classical. The condition~\eqref{eq:classicality_criterion} is saturated at a value of $x$ such that $\vert x \vert = [2 v_0/(3\beta)]^{1/3}\equiv \dphiwell/(2\Mp)$, which defines the width
\bea \label{eq:dphiwell_cubic}
\frac{\dphiwell}{\Mp} = 2 \lp\frac{2}{3}\frac{v_0}{\beta}\rp^{1/3}
\eea 
 of the field range dominated by stochastic diffusion. In other words, the inflection point potential can be approximated as being exactly flat over a field interval (that we call the ``quantum well'') centred at the inflection point and of width given by $\dphiwell$, and outside this range, as giving rise to purely classical dynamics. The situation is depicted in \Fig{fig:inflection_linear_potential}.

As a consequence, if $\phiuv$ is set far outside the quantum well, which is what we do in practice, it does not affect the PDF of the number of \efolds~and it becomes an irrelevant parameter. If $\phiend$ lies outside the quantum well, a constant, deterministic number of \efolds~is realised between the exit of the quantum well and $\phiend$, so it only shifts the PDF of the number of \efolds~by a constant value. 

\paragraph{Slow-roll conditions}$ $\\
A few words are in order regarding the slow-roll conditions. Since $\beta x^3 \propto v_0$ at the boundaries of the quantum well, and given that $v_0$, which measures the potential energy in Planckian units, must be much smaller than $1$, the potential is almost constant over the quantum well, and the potential slow roll parameters in that region are controlled by $v_{x}/v \simeq 3\beta x^2$, and $v_{xx}/v\simeq 6\beta x$. Computing these two quantities at the edges of the quantum well, one finds $\beta^{1/3} v_0^{2/3}$ and $\beta^{2/3} v_0^{1/3}$ respectively, so the quantum well is within the slow-roll regime as long as
\bea
\label{eq:flat:inflection:point:SR:condition}
\beta\ll \frac{1}{\sqrt{v_0}}\, .
\eea
This does not guarantee that the full field range comprised between $\phi_\uend$ and $\phiuv$ is within the slow roll regime, but given that, as soon as $\phi_\uend$ and $\phiuv$ are outside the quantum well, they do not (or only trivially) affect the PDF we are aiming to compute, the condition~\eqref{eq:flat:inflection:point:SR:condition} is sufficient in practice. 

Let us also mention that, if $\beta\ll 1$, then the potential slow-roll conditions are always satisfied above the inflection point, \ie for all $x\geq 0$. If this is not the case however, \ie if $1\ll \beta \ll 1/\sqrt{v_0}$, slow roll is strongly violated around $x\sim \beta^{-1/3}$, so starting from an initial large-field value, one could enter the quantum well away from the slow-roll attractor, even though a slow-roll solution exists there~\cite{Pattison:2018bct} (this will be further discussed in \Sec{sec:USR:stability}). This is why, here, we view \Eq{eq:pot:inflection} only as an expansion of the potential around the flat inflection point, and assume that, at large-field values, the potential is modified such that one always approaches the quantum well with initial conditions located on the slow-roll attractor. In setups where slow roll is explicitly violated, the full phase-space formulation of stochastic inflation developed in \Sec{sec:StochasticInflation} must be employed, which will be discussed in \Sec{sec:BeyondSlowRoll}.
\paragraph{Flat quantum-well approximation}$ $\\
The PDF of the number of \efolds~realised in the potential~(\ref{eq:pot:inflection}) can be computed numerically, by solving \Eq{eq:ODE:chi} for the characteristic function and Fourier transforming the result along \Eq{eq:pdf:chi}. Below, we will compare this result with the approximation outlined above, where the dynamics is classical outside the quantum well, and undergoes pure quantum diffusion inside the well. In this approximation, starting from a certain initial field value $\phi$ inside the quantum well, one can write the realised number of \efolds~as
\bea
\N = \N_\mathrm{well} \left(\phi\right) + N_\ucl\left(\phi_0-\dphiwell/2 \to \phiend\right),
\eea 
where the PDF of $\N_\mathrm{well} (\phi)$ has been computed in \Sec{sec:flat_potential}, and $N_\ucl\left(\phi_0-\dphiwell/2 \to \phiend\right)$, which we will simply denote $N_\ucl$ in what follows, stands for the classical, deterministic number of \efolds~realised between the exit of the quantum well, at $\phi=\phi_0-\dphiwell/2$, and the end of inflation. Therefore, the PDF of $\N$ is given by
\bea
P_\phi\left(\N\right) = P_\phi^\mathrm{flat}\left(\N- N_\ucl\right), 
\eea
where $P_\phi^\mathrm{flat}$ is given by \Eq{eq:stocha:HeatMethod:PDF:expansion} with $\dphiwell$ given by \Eq{eq:dphiwell_cubic}. If one starts from an initial value of $\phi$ located beyond the quantum well, \ie $\phi>\phi_0+\dphiwell/2$, then one simply has to add another classical contribution to the total number of \efolds, \ie 
\bea
\N = N_\ucl\left(\phi \to \phi_0+\dphiwell/2\right) + \N_\mathrm{well} \left(\phi_0+\dphiwell/2\right) + N_\ucl\left(\phi_0-\dphiwell/2 \to \phiend\right),
\eea
which we simply write as $\N = N_\ucl + \N_\mathrm{well} (\phi_0+\dphiwell/2)$, and this gives rise to
\bea
P_\phi\left(\N\right) = P_{\phi_0+\dphiwell/2}^\mathrm{flat}\left[\N- N_\ucl\left(\phi\right)\right] .
\eea

Shifting the PDF by a constant number of \efolds~does not change its decay rates, so the eigenvalues $\Lambda_n$ are given by \Eq{eq:flat_poles}, and making use of \Eqs{eq:def:mu} and~\eqref{eq:dphiwell_cubic}, one obtains
\bea \label{eq:inflection_poles}
\Lambda_n=
\left(\frac{9v_0\beta^2}{4}\right)^{1/3} \frac{\pi^2}{4} \left(n+\frac{1}{2}\right)^2.
\eea
Let us stress that, because of the slow-roll condition~\eqref{eq:flat:inflection:point:SR:condition}, $v_0\beta^2\ll 1$, the first eigenvalues are necessarily small, which means that the tails are very much unsuppressed in this model. This will have strong consequences for PBH formation, that we will discuss in \Sec{sec:pbh}.  The way that the coefficients $a_n$ in the expansion~\eqref{eq:tail_expansion} change under a constant shift in the number of \efolds~is also trivial to establish, and this leads to 
\bea
\label{eq:flat:inflection:an:appr}
a_n\left(\phi\right) &=& a_n^\mathrm{flat}\left(\phi\right) \ee^{\Lambda_n N_\ucl}\\
&=& (-1)^n \frac{\pi}{4}\left(\frac{9}{4}v_0\beta^2\right)^{1/3}\left(2n+1\right) \cos\left[\frac{\pi}{2}\left(2n+1\right)\left(\frac{\phi-\phi_0}{\dphiwell}-\frac{1}{2}\right)\right] \ee^{\Lambda_n N_\ucl},
\eea
where we have made use of \Eq{eq:an:flat} to evaluate $a_n^\mathrm{flat}$. This gives rise to
\bea
\label{eq:PDF:flat:inflection:point}
P_\phi\left(\N\right) & = &\frac{\pi}{4}\left(\frac{9}{4}v_0\beta^2\right)^{1/3}  \sum_n (-1)^n \left(2n+1\right) 
\\ &  & \times
\cos\left[\frac{\pi}{2}\left(2n+1\right)\left(\frac{\phi-\phi_0}{\dphiwell}-\frac{1}{2}\right)\right] \ee^{- \left(\frac{9v_0\beta^2}{4}\right)^{1/3} \frac{\pi^2}{4} \left(n+\frac{1}{2}\right)^2 \left(\N-N_\ucl\right)}\, ,
\eea
which can be rewritten in terms of the first elliptic theta function,
\bea
P_\phi\left(\N\right) = &\frac{\pi}{8}\left(\frac{9}{4}v_0\beta^2\right)^{1/3} 
\vartheta_1^\prime\left[\frac{\pi}{2}\left(\frac{\phi-\phi_0}{\dphiwell}-\frac{1}{2}\right),\ee^{- \left(\frac{9v_0\beta^2}{4}\right)^{1/3} \frac{\pi^2}{4}  \left(\N-N_\ucl\right)}\right]\, .
\eea
The above expressions are derived assuming that one starts from inside the well, otherwise, $\phi$ has to be replaced with $\phi_0+\dphiwell/2$ if $\phi>\phi_0+\dphiwell/2$. 

\begin{figure}[t]
\centering 
\includegraphics[width=.49\textwidth]{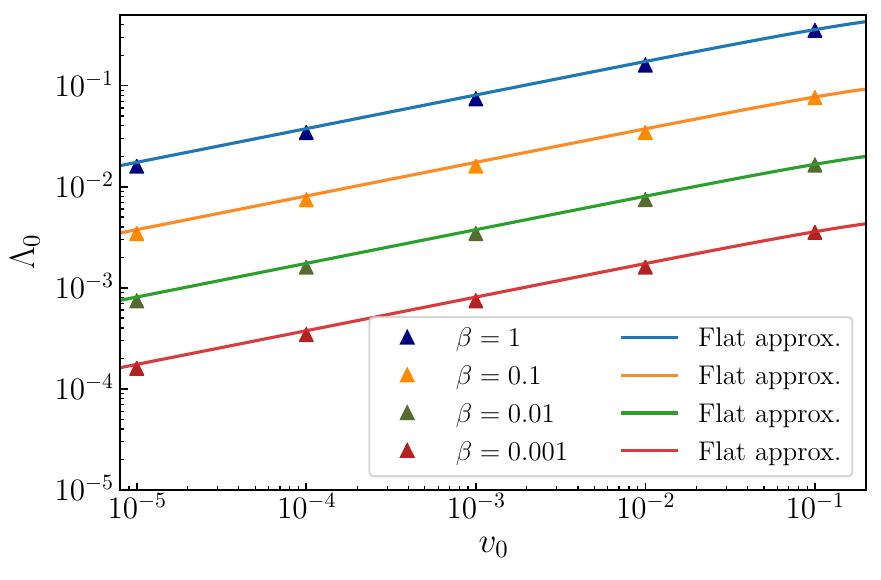}
\includegraphics[width=.49\textwidth]{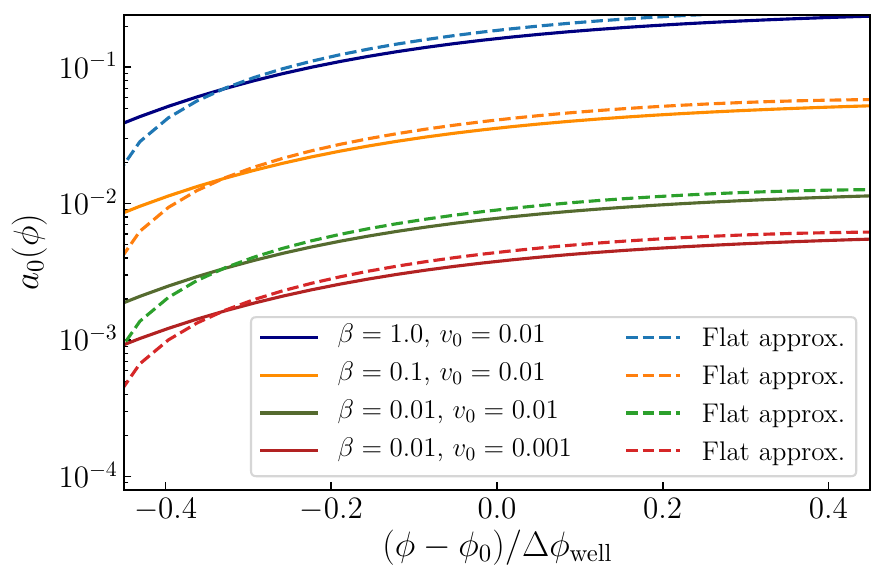}
 \caption{Tail of the PDF of the number of \efolds, $P_\phi(\N)\simeq a_0(\phi) \ee^{-\Lambda_0 \N}$, for the flat inflection-point potential~(\ref{eq:pot:inflection}). The left panel displays the decay rate $\Lambda_0$ as a function of $v_0$ and for a few values of $\beta$. The symbols stand for the full numerical results, while the solid lines stand for the flat quantum-well approximation, \Eq{eq:inflection_poles}.  The right panel shows the amplitude $a_0(\phi)$ for different $v_0$ and  $\beta$. There, the solid lines stand for the full numerical result, and the dashed lines for the flat quantum-well approximation, \Eq{eq:flat:inflection:an:appr}.  In order to satisfy the slow-roll conditions we choose $(\phiuv-\phi_0)/\Mp=1/\beta^{1/3}$, $\phiend=0$ and $\phi_0/\Mp=\dphiwell/2 +0.3/\sqrt{\beta}$.
}
 \label{fig:L0v0_cubic}
\end{figure}
These expressions are compared with a full numerical result in \Fig{fig:L0v0_cubic}. One can see that the leading decay rate, $\Lambda_0$, and the behaviour of $a_0(\phi)$ close to the flat inflection point at $\phi\sim\phi_0$, are accurately reproduced by our approximations~\eqref{eq:inflection_poles} and~\eqref{eq:flat:inflection:an:appr}. At the edges of the quantum well, \ie when $\phi-\phi_0 = \pm \dphiwell/2$, the approximation for $a_0$ starts to deviate from the numerical result, as expected. This otherwise confirms the validity of the approach presented here. Notice that since the expression for $\dphiwell$ comes from saturating the condition~\eqref{eq:classicality_criterion}, $\dphiwell$ in \Eq{eq:dphiwell_cubic} is only defined up to an overall constant of order one, hence so is the case of $\Lambda_n$ in \Eq{eq:inflection_poles} and of $a_n$ in \Eq{eq:flat:inflection:an:appr}. 

\begin{figure}[t]
\centering 
\includegraphics[width=.59\textwidth]{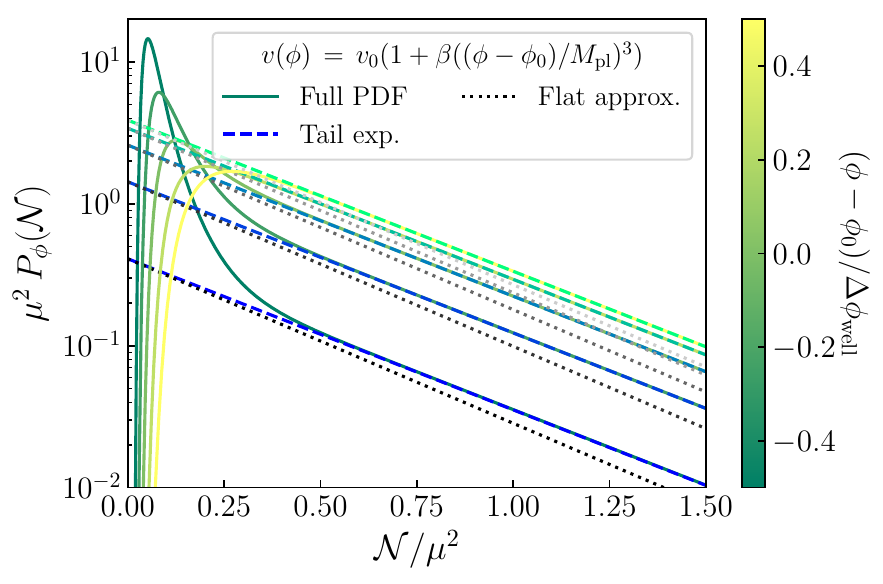}
 \caption{Probability distribution function of the number of \efolds~$\N$ realised in the flat inflection-point potential~(\ref{sec:inflection_potential}), starting from different initial field values $\phi$ labeled in the colour bar. The solid lines stand for the full PDF, while the dashed lines correspond to the leading term in the tail expansion, $P_\phi(\N)\simeq a_0(\phi) \ee^{-\Lambda_0 \N}$. Both are obtained from numerically solving \Eq{eq:ODE:chi}. On the contrary, the dotted lines represent the leading term in the flat approximation using the analytical expression for $a_0(\phi)$ and $\Lambda_0$ given in \Eqs{eq:flat:inflection:an:appr} and \eqref{eq:inflection_poles} respectively. To make the comparison with \Fig{fig:PDF_flat} easy, the number of \efolds~is rescaled by $\mu^2=\dphiwell^2/(v_0 \Mp^2)$, where $\dphiwell$ is the width of the region where quantum diffusion dominates, and is given by \Eq{eq:dphiwell_cubic}.
 In order to satisfy the slow-roll condition (\ref{eq:flat:inflection:point:SR:condition}), we have chosen $v_0=0.01$, $\beta=0.01$, $(\phiuv-\phi_0)/\Mp=1/\beta^{1/3}$, $\phiend=0$ and $\phi_0/\Mp=\dphiwell/2 +0.3/\sqrt{\beta}$. 
 One can see that the analytic, flat approximation accurately predicts the amplitude of the tail with a small deviation in the slope.
}
 \label{fig:PDF_cubic}
\end{figure}

In \Fig{fig:PDF_cubic}, we also display the full PDF, computed numerically from solving \Eq{eq:ODE:chi} for the characteristic function and Fourier transforming the result along \Eq{eq:pdf:chi}. The result is compared with the leading-tail expansion $P_\phi(\N)\simeq a_0(\phi) \ee^{-\Lambda_0 \N}$, where $\Lambda_0$ and $a_0$ are obtained numerically from searching for the first pole of the solution to \Eq{eq:ODE:chi}. Let us note that, when doing so, the fact that \Eq{eq:inflection_poles} provides a good approximation to the pole location $\Lambda_0$ turns out to be very convenient, since it sets an initial value around which to look for the pole, which greatly simplifies the computational problem. One can check that the leading-tail expansion provides an excellent approximation to the full PDF on its tail, as expected. 
On the other hand, the dotted lines, representing the flat approximation given by $a_0(\phi)$ and $\Lambda_0$ in \Eqs{eq:flat:inflection:an:appr} and \eqref{eq:inflection_poles} respectively, show that these simple, analytical formulas provide the right order of magnitude for the amplitude and decay rate of the tail. 

\subsubsection{Tilted inflection-point potentials}
\label{sec:inflection_linear_potential}

We now consider the possibility that the inflection point is not exactly flat, \ie $v''=0$ at $\phi=\phi_0$ but $v'\neq 0$. Tilted-inflection point potentials of this class~\cite{Garcia-Bellido:2017mdw, Ezquiaga:2017fvi} can be constructed by adding a linear slope to our previous cubic potential~\eqref{eq:pot:inflection}, \ie 
\bea
\label{eq:pot:inflection_linear}
v\left(\phi\right) = v_0\left[1+\alpha\lp\frac{\phi-\phi_0}{\Mp}\rp+\beta\lp\frac{\phi-\phi_0}{\Mp}\rp^{3}\right]\, ,
\eea
where we assume $\alpha\geq 0$ and $\beta\geq 0$. The region of the potential where quantum diffusion dominates still has to be determined from the criterion~\eqref{eq:classicality_criterion}. The quantity $v'' v^2/(v')^2$ vanishes at the inflection point $x=0$. Around this point, if $\alpha$ and $\beta$ are small, there is always a slow-roll region where $v\simeq v_0$. In this regime, $v'' v^2/(v')^2$ is maximal at $x=\pm \sqrt{\alpha/\beta}/3$, where its value is $9v_0\sqrt{\beta}/(8 \alpha^{3/2})$. Two cases need therefore to be distinguished, depending on whether this quantity is smaller or larger than one.
\paragraph{A single constant-slope well}$ $ \\
In the case where
\bea
\label{eq:quasi:flat:inflection:point:cond:alpha:nowell}
\alpha\gg (v_0^2 \beta)^{1/3},
\eea 
there is no region where the potential can be approximated as quasi constant, since $v'' v^2/(v')^2$ is never larger than one. So there is no almost-constant quantum well of the kind studied in \Sec{sec:flat_potential}. When $\vert x \vert \ll \sqrt{ \alpha/(3\beta) }$ however, the potential slope is almost constant, so the results derived in \Sec{sec:linear_potential} can be applied, over a field range of width 
\bea
\label{eq:dphiwell:tilted:inflection:point}
\dphiwell \simeq 2 \Mp \sqrt{\frac{ \alpha}{3\beta}}.
\eea 
Let us note that this well falls far within the almost constant regime, where both $\alpha x$ and $\beta x^3$ are much less than one, if $\alpha\ll \beta^{1/3}$, and the potential slow-roll conditions, $\Mp v'/v \ll 1$ and $\Mp^2v''/v\ll 1$, reduce to
\bea
\label{eq:cond:sr:tilted:inflection:case:constant:slope}
\alpha\ll 1\, ,\quad\quad \alpha\beta\ll 1\, ,
\eea
 inside the wells. 

\begin{figure}[t]
\centering 
\includegraphics[width=.49\textwidth]{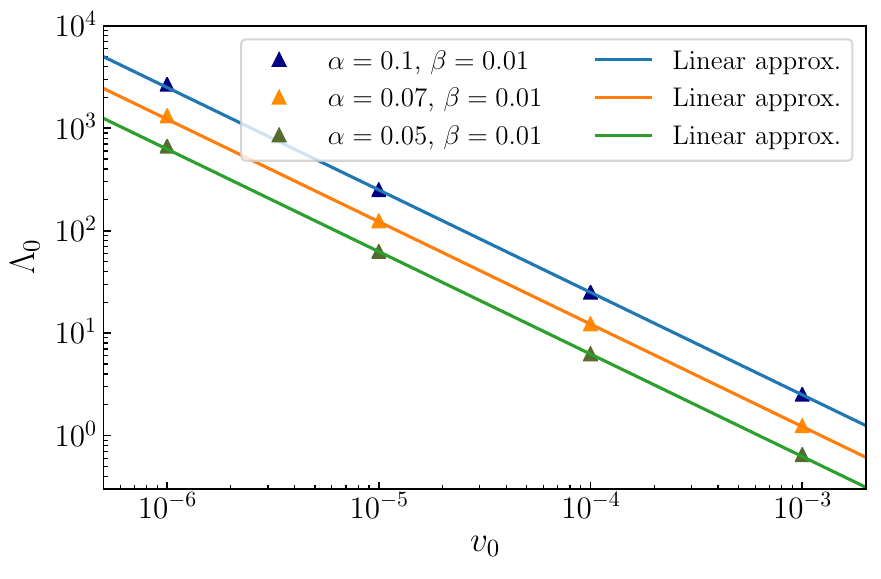}
\includegraphics[width=.48\textwidth]{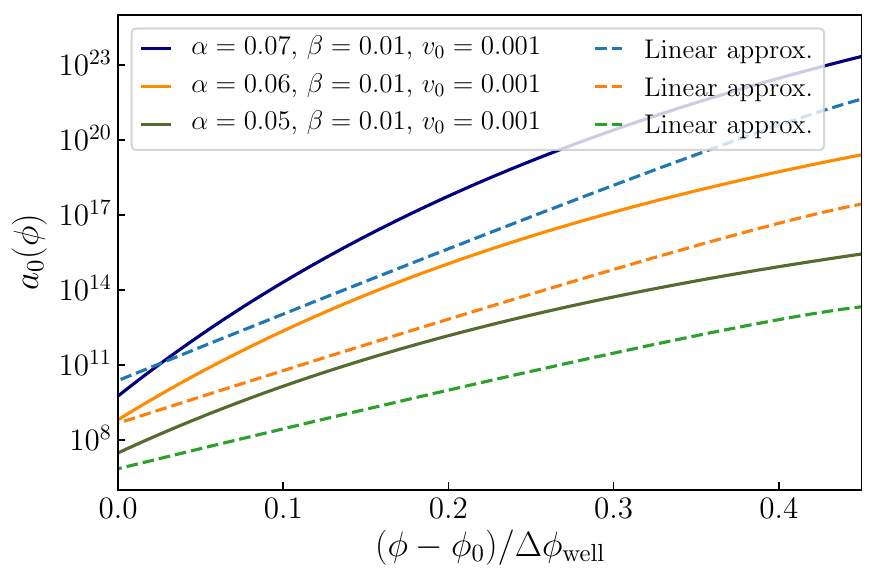}
 \caption{Tail of the PDF of the number of \efolds, $P_\phi(\N)\simeq a_0(\phi) \ee^{-\Lambda_0 \N}$, for the tilted inflection-point potential~(\ref{eq:pot:inflection_linear}). The left panel displays the decay rate $\Lambda_0$ as a function of $v_0$ for a few values of $\alpha$ and $\beta$. The symbols stand for the full numerical results, while the solid lines stand for the linear wide-well approximation, \Eq{eq:quasi_poles}.  The right panel shows the amplitude $a_0(\phi)$ for different $v_0$, $\alpha$ and  $\beta$. There, the solid lines stand for the full numerical result, and the dashed lines for the linear wide-well approximation, \Eq{eq:an:inflection:wide:constant:well}.  In order to satisfy the slow-roll conditions we choose $(\phiuv-\phi_0)/\Mp=\phi_0/\Mp=0.1/\alpha$ and $\phiend=0$.
}
 \label{fig:L0v0_quasi}
\end{figure}

The relation~\eqref{eq:dphiwell:tilted:inflection:point} gives rise to $\dphiwell\alpha/(\Mp v_0) = 2 \alpha^{3/2}/(3 \beta v_0)^{1/2}$, which is much larger than one because of \Eq{eq:quasi:flat:inflection:point:cond:alpha:nowell}. This means that the condition~\eqref{eq:constant:slope:expansion:transcendental:condition} is satisfied, hence we are in the wide-well regime. This implies that \Eq{eq:lambda_almost_const} applies, namely
\bea \label{eq:quasi_poles}
\Lambda_n \simeq \frac{\alpha^2}{4v_0}+\frac{\pi^2 v_0\Mp^2}{\dphiwell^2}\left(n+1\right)^2\, ,
\eea
together with \Eq{eq:a:constant:slope}, namely
\begin{align}
a_n(\phi) \simeq -(-1)^n \frac{\pi v_0 \Mp^2 }{\dphiwell^2 }2\left(n+1\right) \ee^{\frac{\alpha\dphiwell}{2 v_0\Mp} \left(\frac{\phi-\phi_0}{\dphiwell}+\frac{1}{2}\right)} \sin\left[{\pi}\left(n+1\right)\left(\frac{\phi-\phi_0}{\dphiwell}-\frac{1}{2}\right)\right] ,
\label{eq:an:inflection:wide:constant:well}
\end{align}
with a possible additional correction $\ee^{\Lambda_n N_\ucl}$ if a classical number of \efolds~is realised before of after the well, as in \Eq{eq:flat:inflection:an:appr}. Combined together, \Eqs{eq:quasi_poles} and~\eqref{eq:an:inflection:wide:constant:well} lead to the PDF
\bea
P_\phi(\N) = - \frac{\pi v_0 \Mp^2 }{2 \dphiwell^2 } \ee^{\frac{\alpha\dphiwell}{2 v_0\Mp} \left(\frac{\phi-\phi_0}{\dphiwell}+\frac{1}{2}\right)}
\ee^{-\frac{\alpha^2}{4v_0} \left(\N-N_\ucl\right)} 
{\vartheta_4}^\prime \left[\frac{\pi}{2}\left(\frac{\phi-\phi_0}{\dphiwell}-\frac{1}{2}\right),\ee^{-\frac{\pi^2 v_0\Mp^2}{\dphiwell^2}\left(\N-N_\ucl\right)}\right]\, .
\eea

The above approximated formulas for $\Lambda_0$ and $a_0$ are compared with a full numerical solution in \Fig{fig:L0v0_quasi}. On the left panel, one can see that $\Lambda_0$ is accurately reproduced, while on the right panel, only the generic trend and order of magnitude of $a_0$ are accounted for. 
This is because, as already mentioned, the effective values of $\dphiwell$, derived from the saturation of the non-classicality criterion~\eqref{eq:classicality_criterion}, provide estimates up to factors of order one only. The way they enter the PDF for a flat inflection point, \ie the way $\dphiwell$ appears in \Eq{eq:PDF:flat:inflection:point}, is such that this uncertainty produces order-of-one errors in the amplitude of the PDF. However, for a tilted inflection-point in the regime of \Eq{eq:quasi:flat:inflection:point:cond:alpha:nowell}, $\dphiwell$ enters exponentially in the amplitude of the PDF, see \Eq{eq:an:inflection:wide:constant:well}. This implies that these order-one corrections are exponentiated, potentially leading to more substantial corrections in the amplitude of the tail. Let us however stress that our determination of $\Lambda_n$ does not suffer from this issue, and that, as mentioned above, since it provides a first guess for the location of the pole, it plays a crucial role in the numerical determination of the poles and of their residues.

\paragraph{From two quantum wells separated by a constant slope, to a single quantum well}$ $ \\
If the condition
\bea
\label{eq:quasi:flat:inflection:point:cond:alpha}
\alpha\ll (v_0^2 \beta)^{1/3}
\eea
is realised, there exist two regions where $v'' v^2/(v')^2$ is larger than one, namely for $x\in[-x_+,-x_-]$ and $x\in [x_-,x_+]$, where 
\bea
\displaystyle
x_- &\simeq& \frac{\alpha^2}{6\beta v_0} \left(1+\frac{\alpha^3}{6\beta v_0^2}+\cdots\right),\\
x_+ &\simeq& \frac{\dphiwell}{2}\left[1-\left(\frac{2\alpha^3}{3^4\beta v_0^2}\right)^{1/3}+\cdots\right],
\eea
with $\dphiwell$ given in \Eq{eq:dphiwell_cubic} and where ``$\cdots$'' denotes higher powers of $\alpha^3/(\beta v_0^2)$.  In the limit of \Eq{eq:quasi:flat:inflection:point:cond:alpha}, one has $x_- \ll x_+$, so the two wells are almost adjacent. By computing the relative importance of the terms $\beta x^3$ and $\alpha x$ at the point $\pm x_-$, one notices that it is proportional to $\alpha^3/(\beta v_0^2)$, hence it is very small because of \Eq{eq:quasi:flat:inflection:point:cond:alpha}. Therefore, in the interval $[x_-,x_+]$, the potential is of the quasi constant-slope type. One has therefore three wells in series: a first quasi-constant well between $x_+$ and $x_-$, a quasi constant-slope well between $x_-$ and $-x_-$, and a second quasi-constant well between $-x_-$ and $-x_+$. 

Let us note that, if \Eq{eq:quasi:flat:inflection:point:cond:alpha} is satisfied, these wells are far within the almost constant regime where both $\alpha x$ and $\beta x^3$ are much less than one, and the potential slow-roll conditions, $\Mp v'/v \ll 1$ and $\Mp^2v''/v\ll 1$, reduce to 
\bea
\label{eq:cond:alpha:lt:1}
\alpha\ll 1
\eea
and to \Eq{eq:flat:inflection:point:SR:condition} inside the wells. 

The width of the constant-slope well is given by $\dphiwell/\Mp=2 x_- = \alpha^2/(3 \beta v_0)$. As a consequence, one has $\dphiwell \alpha/ ( \Mp v_0 ) = \alpha^4 / ( 3 \beta v_0^2) $, which is much smaller than one because of \Eqs{eq:quasi:flat:inflection:point:cond:alpha} and~\eqref{eq:cond:alpha:lt:1}. Therefore, the constant-slope well is in the narrow-well regime, in the sense of \Eq{eq:small:alpha:limit:def}. According to the considerations of \Sec{sec:linear_potential}, this means that we are in fact in the presence of a quasi-constant potential, so the three wells in series are in effect a single, almost-constant well, with a width given by $\dphiwell = 2\Mp x_+$, \ie by \Eq{eq:dphiwell_cubic}. One concludes that, in that case, the same results as those derived in \Sec{sec:inflection_potential} apply. 

\begin{figure}[t]
\centering 
\includegraphics[width=.59\textwidth]{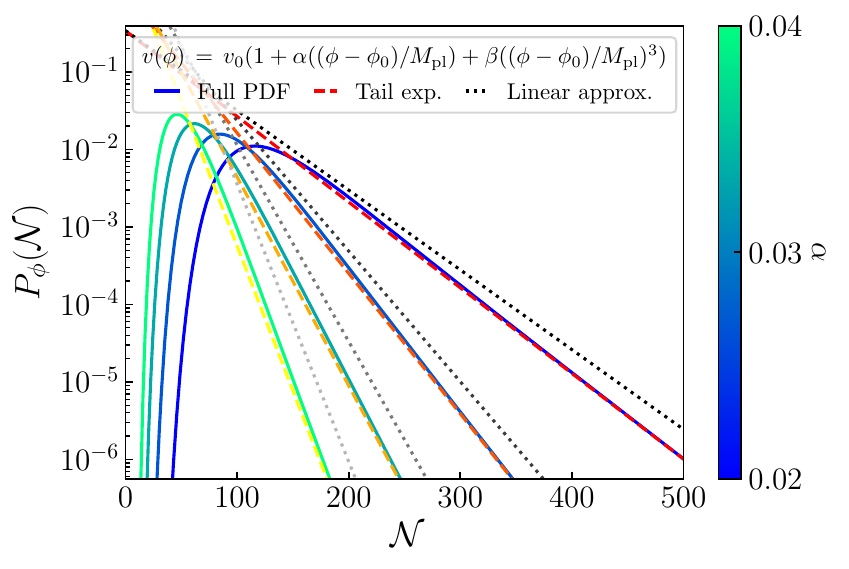}
 \caption{Probability distribution function of the number of \efolds~$\N$ realised in the tilted inflection-point potential~(\ref{eq:pot:inflection_linear}), starting from the inflection point $\phi=\phi_0$, as a function of the linear slope $\alpha$. We compare the full PDF (solid lines) with the leading tail expansion (dashed lines), $P_\phi(\N)\simeq a_0(\phi) \ee^{-\Lambda_0 \N}$, and the linear wide-well approximation given by \Eq{eq:quasi_poles} and \Eq{eq:an:inflection:wide:constant:well}. The $\alpha$-term suppresses the tail of the PDF at large $\N$. 
 We choose $v_0=5\cdot10^{-3}$ and $\beta=10^{-3}$, such that the condition~\eqref{eq:quasi:flat:inflection:point:cond:alpha:nowell} applies, and $(\phiuv-\phi_0)/\Mp=\phi_0/\Mp=0.1/\alpha$ and $\phiend=0$.
 }
 \label{fig:PDF_inflection_linear}
\end{figure}

In summary, when a tilt is introduced into  a flat inflection-point model, as long as the slope $\alpha$ is smaller than the bound~\eqref{eq:quasi:flat:inflection:point:cond:alpha}, it has no effect. When it is larger, it changes the almost constant well into an almost constant-slope well, and adds a contribution $\alpha^2/(4 v_0)$ to the eigenvalues, hence suppresses the tails. This can be clearly seen in \Fig{fig:PDF_inflection_linear}, where the PDF of the number of \efolds~is shown for various values of the slope $\alpha$ [notice that, as in \Fig{fig:L0v0_quasi}, in the linear wide-well approximation, only the order of magnitude of the amplitude of the tail is correctly reproduced, while its decay rate is accurately accounted for, see the discussion below \Eq{eq:an:inflection:wide:constant:well}]. 
%

\subsection{Implications for primordial black hole formation}
\label{sec:pbh}
We have seen that quantum diffusion makes the tail of the PDF of the duration of inflation decay exponentially with the number of $e$-folds. We have exemplified this phenomenon with several toy models including flat, linear, flat inflection-point and tilted inflection-point potentials. The non-Gaussian nature of the tail of the PDF introduces important differences with the standard classical picture of quasi-Gaussian distributions, which translates into important differences for the predicted amount of PBHs, that we now discuss. 

The coarse-grained curvature perturbation is related with the number of \efolds~through \Eq{eq:deltaNcg:zeta}, where the mean number of \efolds~can be computed directly from the characteristic function by making use of \Eq{eq:meanN:chi}.\footnote{Alternatively, as shown in \Sec{sec:meanN}, the mean number of \efolds~can also be obtained by solving the differential equation
\bea
\langle\N\rangle''-\frac{v_\phi}{v^2}\langle\N\rangle'+\frac{1}{v\Mp^2}=0\, ,
\eea
with boundary conditions $\langle\N\rangle(\phiend)=\langle\N\rangle'(\phiuv)=0$, the solution of which is given by \Eq{eq:f:sol}.  Combining \Eqs{eq:chi:pole:expansion} and~\eqref{eq:meanN:chi}, one also has
\bea
\langle \N(\phi) \rangle = \sum_n\frac{a_n(\phi)}{\Lambda_n^2}\, .
\eea}  
By integrating the PDF~\eqref{eq:tail_expansion} above the threshold $\N_\uc = \langle \N \rangle + \zeta_\uc$, one obtains from \Eq{eq:beta:def} the mass fraction of PBHs,
\bea
\label{eq:beta:sum}
\beta_\mathrm{f}(\phi)=\sum_{n}\frac{1}{\Lambda_n}\,a_n(\phi)\,e^{-\Lambda_n\left[\zeta_\uc+\langle\N\rangle(\phi)\right]}\, .
\eea
This should be contrasted with the standard classical result, \Eq{eq:beta:erfc}, which on the tail can be expanded as
\bea
\label{eq:beta:class}
\beta_{\mathrm{f}}^\ucl(\phi)=\frac{\sqrt{\int_{\bar{k}(\phi)}^{k_\uend}\calP_{\zeta,\mathrm{cl}}\dd\ln k}}{\sqrt{2\pi}\zeta_\uc}\,\exp\left[-\dfrac{\zeta_\uc^2}{2\int_{\bar{k}(\phi)}^{k_\uend}\calP_{\zeta,\mathrm{cl}}\dd\ln k}\right]\, .
\eea
This depends exponentially on the square of $\zeta_\uc$, rather than on $\zeta_\uc$ directly as in \Eq{eq:beta:sum}, and it leads to estimates of the mass fraction that can be orders of magnitude away from the actual result~\eqref{eq:beta:sum}.
Let us now review the potentials discussed in \Sec{sec:applications}. 
\paragraph{Flat potential} $ $\\
In a flat potential, in the notations of \Sec{sec:flat_potential}, the mean number of \efolds~is given by $\langle \N \rangle = \mu^2 x(1-x/2)$, which also corresponds to the $\alpha\to 0$ limit of \Eq{eq:mean:N:constant:slope}. Using the formulas derived in \Sec{sec:flat_potential}, \Eq{eq:beta:sum} gives rise to
\bea
\label{eq:beta:zetac:flat:potential}
\beta_\mathrm{f}(\phi)=\sum_n \frac{4}{(2n+1)\pi}\sin\left[\frac{\pi}{2}\left(2n+1\right)x\right]\,\ee^{-\pi^2\left(n+\frac{1}{2}\right)^2\left[\frac{\zeta_\uc}{\mu^2}+  x(1-\frac{x}{2})\right]}\, .
\eea
This expression is always well approximated by its first term, and one can see that for PBHs not to be over produced, one needs to impose
\bea
\label{eq:def:mu:cond:PBH}
\mu\ll \sqrt{\zeta_\uc}\, ,
\eea
in agreement with the conclusions of \Sec{sec:pbh:stochastic:limit:pbh}. This places an upper bound on the width, or a lower bound on the height, of flat sections in the potential. 
\paragraph{Constant-slope potential} $ $\\
In a constant-slope potential, two regimes have to be distinguished. In the narrow-well regime, defined by \Eq{eq:small:alpha:limit:def}, the same results as for the flat potential apply, and one recovers \Eq{eq:def:mu:cond:PBH}. In the wide-well regime, defined by \Eq{eq:constant:slope:expansion:transcendental:condition}, the mean number of \efolds~is given by \Eq{eq:Ncl:constant:slope}, and using the results of \Sec{sec:linear_potential}, \Eq{eq:beta:sum} gives rise to
\bea
\displaystyle
\beta_\mathrm{f}^\mathrm{wide}(\phi) =
\frac{8 v_0 \pi}{\alpha^2 \mu^2}
\frac{ \ee^{\frac{\alpha}{4 v_0}\left(\frac{\phiuv}{\Mp}x-\alpha\zeta_\uc\right)}}{1+4 \pi^2\frac{ v_0^2  \Mp^2}{\alpha^2\phiuv^2}(n+1)^2}
\sum_n (-1)^{n+1} (n+1)\sin\left[\pi(n+1)(x-1)\right] \ee^{-\pi^2 (n+1)^2\left(\frac{\zeta_\uc}{\mu^2}+x \frac{v_0\Mp}{\alpha\phiuv}\right)}
\, .
\eea
In the limit of \Eq{eq:constant:slope:expansion:transcendental:condition}, the argument of the overall exponential always dominates over the one of the exponential in the sum (at least for the first few terms), so in order to avoid overproduction of PBHs in this model, one must have
\bea
\label{eq:alpha:min:overproduction} 
\alpha \gg  \max\left(\sqrt{v_0},\frac{\phiuv}{\Mp}\right)\, .
\eea

\begin{figure}[t]
\centering 
\includegraphics[width=.49\textwidth]{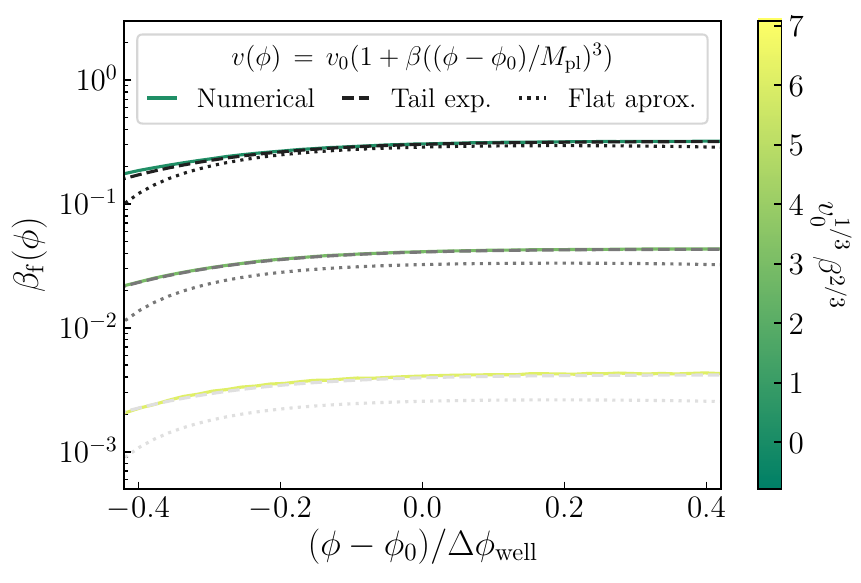}
\includegraphics[width=.49\textwidth]{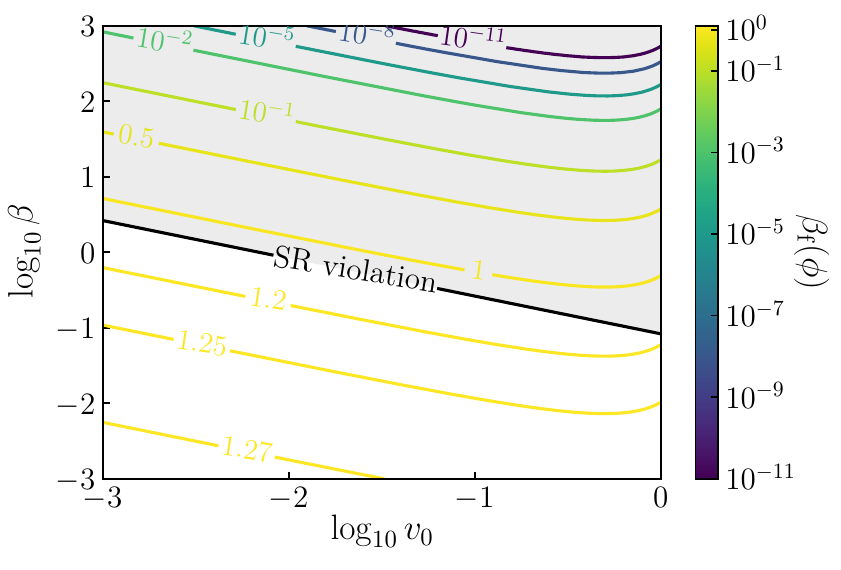}
\caption{PBH abundance $\beta_\mathrm{f}(\phi)$ in a flat inflection-point potential. Left panel: $\beta_\mathrm{f}(\phi)$ is displayed as a function of the initial field value $\phi$ for $\zeta_\uc=1$ and different choices of the combination of parameters $v_0\beta^2$, which controls the tail of the PDF, see \Eq{eq:def:mu:cubic:inflection:point}. The solid lines stand for a full numerical result, the dashed lines for the numerical result if only the dominant term is kept in the tail expansion (\ref{eq:beta:sum}), and the dotted lines to the constant-well approximation, \ie to \Eq{eq:beta:zetac:flat:potential} with $\mu$ given by \Eq{eq:def:mu:cubic:inflection:point}. 
For the plot we fix $v_0=10^{-2}$ and vary $\beta$.  
Right panel: Contour plot of $\beta_\text{f}(\phi)$ as a function of the parameters $v_0$ and $\beta$ for $\zeta_c=1$. The grey shaded region corresponds to where the slow-roll approximation does not hold across the entire range $\phi_0-\dphiwell/2$ to $\phi_0+\dphiwell/2$, \ie to where \Eq{eq:flat:inflection:point:SR:condition} is not satisfied. One can see that, when slow roll is satisfied, PBHs are overproduced.
}
 \label{fig:beta_inflection}
\end{figure}
\paragraph{Flat inflection-point potential}$ $\\
As explained in \Sec{sec:inflection_potential}, a flat inflection-point potential is equivalent to a flat potential with $\dphiwell$ given by \Eq{eq:dphiwell_cubic}, \ie $\mu$ given by 
\bea
\label{eq:def:mu:cubic:inflection:point}
\mu^2=4  \left(\frac{2}{3\beta\sqrt{v_0}}\right)^{2/3}\, .
\eea
This parameter is necessarily large because of the slow-roll condition~\eqref{eq:flat:inflection:point:SR:condition}, so according to the above considerations, see \Eq{eq:def:mu:cond:PBH}, $\beta_\mathrm{f}$ is large in this model, which is confirmed by the numerical results displayed in \Fig{fig:beta_inflection}. We therefore reach the interesting conclusion that PBHs are always overproduced in a flat inflection point potential, if one approaches the inflection point along the slow-roll attractor. 

This is also consistent with the results of \Sec{sec:example:1_plus_phi_to_the_p}, where the same conclusion was reached for potentials of the type $v=v_0(1+\alpha\phi^p)$, although these potentials were restricted to positive field values. This suggests that our findings are independent of the order of the polynomial (here cubic) that realises the flat inflection point.

\begin{figure}[t]
\centering 
\includegraphics[width=.49\textwidth]{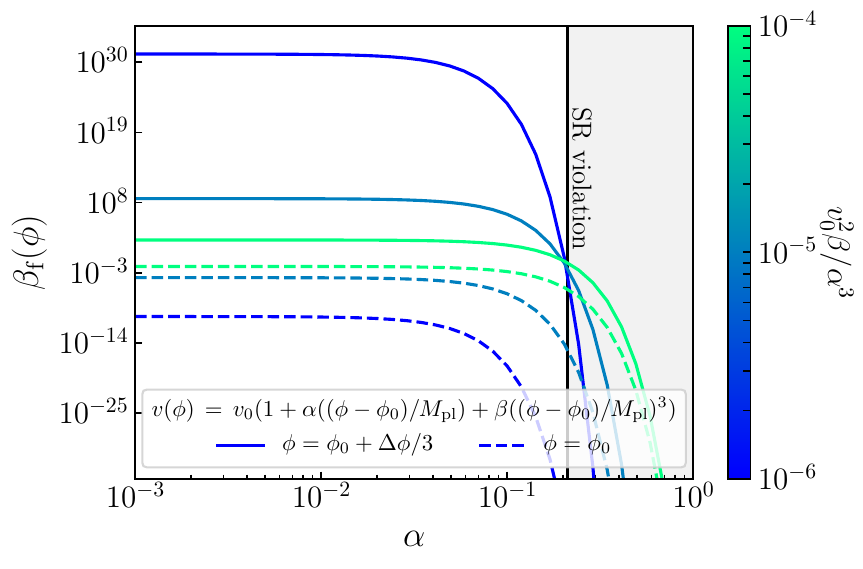}
 \caption{PBH abundance $\beta_\text{f}(\phi)$ for a tilted-inflection point potential as a function of the slope $\alpha$ for $\zeta_c=1$. The combination $v_0\beta/\alpha^3$ is shown in the colour bar, and is such that the condition~\eqref{eq:quasi:flat:inflection:point:cond:alpha:nowell} applies (otherwise, the model would be equivalent to a flat inflection-point potential). We set $\beta=50\alpha^3$, such that the slow-roll validity condition, \Eq{eq:cond:sr:tilted:inflection:case:constant:slope}, reads $\alpha\ll 0.3$, and the region violating slow roll is shaded in grey. The bounds $\phiend$ and $\phiuv$ are set according to $\phiuv-\phi_0=\phi_0=0.1\Mp/\alpha$ and $\phiend=0$. We also evaluate $\beta_\text{f}(\phi)$ at different initial field values $\phi$ indicated by the solid/dashed lines. When $\phi>\phi_0$ one either overproduces PBHs or violates the slow-roll condition.
}
 \label{fig:beta_inflection_linear}
\end{figure}

\paragraph{Tilted inflection-point potential} $ $\\
If the inflection-point potential is tilted with a slope $\alpha$ smaller than the upper bound~\eqref{eq:quasi:flat:inflection:point:cond:alpha}, the effect of the slope is negligible and one recovers a quasi flat inflection-point potential, which we just saw overproduces PBHs. If $\alpha$ is larger however, such that the condition~\eqref{eq:quasi:flat:inflection:point:cond:alpha:nowell} is satisfied, one recovers an almost constant-slope potential in the wide-well regime, with $\dphiwell$ given by \Eq{eq:dphiwell:tilted:inflection:point}. This implies that $\alpha\Mp/\dphiwell \propto \sqrt{\alpha\beta} $, which is much smaller than one because of \Eq{eq:cond:sr:tilted:inflection:case:constant:slope}. Therefore, the second of the conditions~\eqref{eq:alpha:min:overproduction} is not satisfied, and PBHs are overproduced too. This is confirmed by the numerical results of \Fig{fig:beta_inflection_linear} where one can check that, when $\phi\neq\phi_0$, $\beta_\mathrm{f}>1$ as soon as the slow-roll conditions are satisfied (this explains why all the solid curves intersect at around the same point, where $\beta_\mathrm{f}\sim 1$, and at the boundary of the slow-roll condition). As a consequence, the previous conclusion extends to tilted inflection-point models: PBHs are always overproduced if one approaches the inflection point without violating slow roll. 

\paragraph{Discussion} $ $\\
Some inflationary models are already known to display a $\chi$-square statistics for the curvature perturbation, such as models of axion inflation in which the gauge field sources the curvature perturbations~\cite{Garcia-Bellido:2016dkw}, and that the implications of a $\chi$-square distribution for the gravitational wave signature have been studied in detail in \Refa{Garcia-Bellido:2017aan}. However, here, the presence of exponential tails has been found even in single-field, slow-roll inflationary scenarios. For such models, we have found simple analytical approximations that capture the behaviour of the tails of the PDF, as well as developed efficient numerical techniques to compute them precisely. This allowed us to properly estimate the abundance of PBHs associated with each model. 

We have found that potentials featuring regions where quantum diffusion dominates over the classical roll of the field can be either approximated by locally constant potentials, or by locally constant-slope potentials. In the first case, the requirement that PBHs are not overproduced places an upper bound on the squared width of the flat region, divided by its height, see \Eq{eq:def:mu:cond:PBH}. In the second case, both the width and the height are bounded from above, see \Eq{eq:alpha:min:overproduction}. When applied to inflection-point potentials, regardless of whether the inflection point is flat or tilted, these conditions cannot be satisfied unless slow roll is violated when approaching the inflection point. This is therefore one of the main results of this section: inflection-point models that do not feature slow-roll violations overproduce PBHs.

It is also important to emphasise the universality of these results. The tail of the distribution function of the coarse-grained curvature perturbation $\zeta_{\text{cg}}$ induces non-Gaussian deviations, in the form of exponential tails, on {\em all scales}, with amplitude $a_n(\phi)$ and exponential decay $\Lambda_n$. This is because, in a given inflationary model, $\Lambda_n$ is fixed (it does not depend on $\phi$) while $a_n(\phi)$ depends on the scale at which $\zeta_{\text{cg}}$ exits the Hubble radius. Therefore, there is an exponential tail across the whole spectrum of modes, with the same decay rate, although its amplitude depends on the specific inflationary dynamics associated with each scale. 
For plateau-like potentials for instance, these non-Gaussian effects may be significantly relevant, as in the case of quasi-inflection point models for PBH production.

Even if on large scales (probed in the CMB and in the large-scale structures), the effect of exponential tails may be negligible in most models (although this remains to be checked explicitly), on intermediate scales, corresponding to small halos, \eg Lyman-alpha scales, or even smaller, like ultra-compact mini halos, the exponential tail effects may become very relevant. In particular, they could induce an enhancement of the non-linear collapse of structures on small scales that could have important consequences for large-scale structure formation, and thus for interpreting data from future surveys like \MYhref[violet]{https://www.desi.lbl.gov/}{DESI}, \MYhref[violet]{https://www.euclid-ec.org/}{Euclid} and \MYhref[violet]{https://www.lsst.org/}{LSST}.

\section{Beyond the slow-roll attractor}
\label{sec:BeyondSlowRoll}

Primordial black holes require large quantum fluctuations to be produced, which in turn require a very flat potential during inflation. We have argued that with such a flat potential, stochastic fluctuations of the inflaton field may overtake the classical, slow-roll drift, which implies that stochastic effects have to be taken into account in order to properly assess the abundance of PBHs. In fact, if the potential is very flat, deviations from slow roll can also be encountered. Indeed, in the limit where the potential is exactly flat, the potential gradient can be neglected, and the field acceleration parameter $f$ in \Eq{eq:def:f} becomes close to one, \ie different from the slow-roll value that satisfies $\vert f_\SR \vert \ll 1$, see \Eq{eq:f:SR}. 
Moreover, in \Secs{sec:PBHs} and~\ref{sec:tail:expansion}, it was shown that for models featuring an inflection point, or if inflation proceeds towards an uplifted local minimum of its potential, PBHs are overproduced unless slow roll is violated. This is why, in this section, we extend our previous results to situations where inflation proceeds away from the slow-roll attractor.
 
As mentioned below \Eq{eq:phidotdotSR}, the regime where $f\simeq 1$ is called ``ultra slow roll''. In order to avoid any confusion, let us recall that, if the slow-roll conditions~\eqref{eq:sr:consistency} are satisfied, there always exists a slow-roll solution to the equations of motion, and this solution is always an attractor. However, the inflaton might enter the very flat region of the potential with initial conditions displaced from that attractor. In that case, either initial conditions fall in the basin of attraction of slow roll, and there is a transient, non slow-roll phase until the system relaxes to slow roll, or the system falls outside the basin of attraction of slow roll and asymptotes to another, non slow-roll stable regime (typically ultra slow roll).

We therefore start this section by investigating the regime of ultra slow roll inflation, identifying the conditions under which it is stable or unstable. This follows \Refa{Pattison:2018bct}. We then move on to compute the gauge corrections of \Sec{sec:uniformexpansion} in this regime, following \Refa{Pattison:2019hef}, and show that they can be neglected. This leads us to deriving a stochastic formalism for ultra slow roll inflation, of which we give a preliminary analysis in \Sec{sec:USR:stochastic}. 
\subsection{(In)stability of ultra-slow roll inflation}
\label{sec:USR:stability}

In the ultra-slow roll regime, the driving term, corresponding to the gradient of the potential, is neglected in the Klein-Gordon equation \eqref{eq:kleingordon}, rather than the field acceleration. This corresponds to the relative field acceleration $f\approx 1$ in \Eq{eq:def:f}, leading to 
\bea
\label{eq:eom:KG:usr}
\ddot{\phi}_\USR\simeq-3H\dot{\phi}_\USR\, .
\eea
Note that if $\phi$ follows the gradient of its potential, then $\dot{\phi} V' = \dot{V}<0$ and conversely $\dot{\phi} V'>0$ if the field evolves in the opposite direction. From \Eq{eq:def:f}, one can thus see that $f<1$ corresponds to situations where the inflaton rolls down its potential and $f>1$ to cases where the field climbs up its potential.

Integrating \Eq{eq:eom:KG:usr} leads to the USR solution
\bea
\label{eq:USR:phidot:N}
\dot{\phi}_{\USR} \propto \ee^{-3N} \, .
\eea
Instead of being driven by $V'$ as in the SR case \eqref{eq:slowroll}, here the time derivative $\dot{\phi}_{\USR}$ is exponentially decreasing with the number of \efolds. If we also assume quasi-de Sitter ($\epsilon_1\ll1$), the above can be integrated as
\bea
\label{eq:USR:traj}
\phi_\USR - \phi_{\USR,*} \simeq \frac13 \frac{\dot{\phi}_{\USR,*}}{H_*}  \left[ 1-\ee^{-3\left(N-N_*\right)} \right]  \, ,
\eea
where the star denotes some reference time. Thus the USR solution may be thought of as the free or transient response of the scalar field in an expanding FRLW cosmology. It is independent of the shape of the potential, but depends instead on the initial value of the field and its time derivative, $\phi_*$ and $\dot\phi_*$.

Despite the different background evolution, linear fluctuations of a massless field about ultra-slow-roll inflation have the same scale-invariant form as during slow-roll inflation \cite{Seto:1999jc,Leach:2001zf,Kinney:2005vj}. This is a striking example of the invariance of field perturbations under ``duality'' transformations \cite{Wands:1998yp,Biagetti:2018pjj}. 

The condition under which USR takes place reads $\vert f-1 \vert \ll 1$, which implies that $3H \vert\dot{\phi}\vert \gg \vert V' \vert$. Clearly this is possible for any finite potential gradient so long as we have a sufficiently large field kinetic energy.
However, in order to have inflation we also need to have $\epsilon_{1} < 1$, which from \Eq{eq:inflation:condition} corresponds to
\bea
\label{eq:consistency:usr}
\epsilon_{1V} < \frac94 \left(1-f \right)^2
 \, ,
\eea
where $\epsilon_{1V}$ is the first potential slow-roll parameter given in \Eq{eq:epsV}. The quasi-de Sitter approximation $\epsilon_{1} \ll 1$ simply corresponds to $\epsilon_{V} \ll (1-f)^2$. Comparing this relation with \Eq{eq:sr:consistency}, one can see that USR inflation requires a potential that is even flatter than what SR imposes at the level of $\epsilon_V$ (hence the name ``ultra''-slow roll, which is otherwise not so apt since SR and USR are disjoint regimes); but that no constraint is required on $\eta_V$, \ie on the second derivative of the potential. 

In the following, we thus distinguish two regimes: USR, that corresponds to $\vert f-1\vert \ll 1$, and USR \emph{inflation}, that corresponds to $\sqrt{\epsilon_V}\ll\vert f-1\vert \ll 1$.

USR is often referred to as a transient or non-attractor solution during inflation~\cite{Cai:2017bxr, Dimopoulos:2017ged, Anguelova:2017djf, Biagetti:2018pjj, Morse:2018kda}. This is because of results in constant-roll models~\cite{Martin:2012pe, Motohashi:2014ppa, Yi:2017mxs, Morse:2018kda}, where the field acceleration parameter $f$ defined in \Eq{eq:def:f} is taken to be a constant. In the Hamilton-Jacobi formalism, this corresponds to taking $H(\phi)\propto \exp (\pm\sqrt{3f/2}\phi/\Mp)$, and the potentials that support such a phase of constant roll can be obtained from $V=3\Mp^2H^2-2\Mp^4 H'^2$. In these potentials, the constant-roll solution is only one possible trajectory in phase space and one can study its stability. One finds that the constant-roll solution is an attractor if $f<1/2$~\cite{Motohashi:2014ppa}. This excludes the USR limit $f\simeq 1$, which could lead to the incorrect conclusion that USR is always unstable. However this result only applies to the family of potentials mentioned above. Moreover, it is singular in the limit $f \rightarrow 1$ since combining the equations above, one finds $V\equiv \mathrm{constant}$ in that case, for which $f=1$ is the only solution so nothing can be concluded about its attractive or non-attractive behaviour.

This motivates us to go beyond these considerations and to study the phase-space stability of USR in a generic potential.

\subsubsection{Stability analysis}
\label{sec:stability:USR}

In the ultra-slow-roll limit we have $f=1$, which we can readily see is a fixed point of \Eq{eq:f:dynamical} for any potential. We can therefore carry out a generic stability analysis of this fixed point that is valid for any potential. The results will be then illustrated with two specific models.

The strategy is to linearise \Eq{eq:f:dynamical} around $f=1$ by parameterising 
\bea
f=1-\delta\, ,
\eea
where we assume $\vert \delta \vert \ll 1$ in order to study small deviations from USR. The only ambiguity is in the argument of the square root in \Eq{eq:f:dynamical}, that reads $1+\epsilon_V/(6\delta^2)$, since both $\epsilon_V$ and $\delta$ are small numbers. However, from \Eq{eq:consistency:usr} and the discussion below it, one recalls that inflation requires $\epsilon_V<9\delta^2/4$, and $\epsilon_V\ll \delta^2$ ensures quasi de-Sitter inflation $\epsilon_1\ll 1$. This is why \Eq{eq:f:dynamical} should be expanded in the USR \emph{inflation} limit $\sqrt{\epsilon_V}\ll\vert\delta\vert\ll 1$,\footnote{\label{footnote:USR:Non:Inflation}An expansion in the USR \emph{non-inflating} limit, $\vert \delta \vert \ll \sqrt{\epsilon_V}$ and $\vert \delta \vert \ll 1$, can also be performed along similar lines.  At linear order in $\delta$, \Eq{eq:f:dynamical} gives rise to
\bea
\frac{\dd \delta}{\dd \phi} \simeq \left[-\frac{\sqrt{6}}{\Mp}\mathrm{sign}\left(V'\delta \right)+\frac{V''}{V'}\right]\delta\, .
\label{eq:f:dynamical:USR:noninflating}
\eea
If the field follows the gradient of its potential, one obtains the stability condition
\bea
\frac{V''}{\left\vert V' \right\vert }>\frac{\sqrt{6}}{\Mp}\, ,
\label{eq:USR:noninflating:stability:condition}
\eea
and conversely, if the field climbs up the potential, one gets $V''/|V'|<-\sqrt{6}/\Mp$. The solution to \Eq{eq:f:dynamical:USR:noninflating} reads
\bea
\delta \simeq \delta_\uin \frac{V'(\phi)}{V'\left(\phi_\uin\right)}\exp\left(\sqrt{6}\frac{|\phi-\phi_\uin|}{\Mp}\right) \, .
\label{eq:delta:sol:USRnoninflating}
\eea
To determine how $\epsilon_1$ varies, one can plug \Eq{eq:delta:sol:USRnoninflating} into \Eq{eq:eps1:f:phi}. One finds that if the field follows the gradient of its potential, then $\epsilon_1$ decreases if $\epsilon_V<3$ and increases otherwise, and it always decreases if the field climbs up its potential. When $\epsilon_1$ decreases, it may become smaller than one at some point, and a phase of USR \emph{inflation} starts, whose stability properties are discussed in the main text.} which gives rise to
\bea
\label{eq:f:dynamical:USR:new}
\frac{\dd\delta}{\dd\phi}\simeq -\frac{3}{\Mp^2}\frac{V}{V'}\delta^2+\frac{V''}{V'}\delta\, .
\eea
The right-hand side of this expression is proportional to $\delta-\eta_V/3$, so which term dominates depends on the magnitude of $\vert\delta\vert$ with respect to $\vert\eta_V\vert$. Since $\delta$ must be larger than $\sqrt{\epsilon_V}$, two possibilities have to be distinguished.

\paragraph{Case $\eta_V^2<\epsilon_V$}
\label{sec:stability:USR:inflation:case1}

In this case the condition for USR inflation, $\sqrt{\epsilon_V}\ll\vert\delta\vert$, guarantees that $\vert\delta\vert \gg \eta_V$ and the first term on the right-hand side of \Eq{eq:f:dynamical:USR:new} dominates,
\bea
\label{eq:f:dynamical:USR:case1}
\frac{\dd\delta}{\dd\phi}\simeq -\frac{3}{\Mp^2}\frac{V}{V'}\delta^2\, .
\eea
As explained above, if the field follows the gradient of its potential, then $f<1$ and $\delta>0$. If $V'>0$ and $\phi$ decreases with time, then from \Eq{eq:f:dynamical:USR:case1} $\delta$ increases with time and USR is unstable. If $V'<0$ and $\phi$ increases with time, then again $\delta$ increases with time and USR is still unstable. Conversely, if we have $f>1$ and $\delta<0$, so that the field climbs up the potential, if $V'>0$ then $\phi$ increases with time and so does $\delta$, so USR is unstable, and if $V'<0$ then $\phi$ decreases with time and USR is still unstable.

We conclude that USR inflation is always unstable in that case.  The example discussed in \Sec{sec:Staro} corresponds to this situation.

\paragraph{Case $\epsilon_V<\eta_V^2$}
In this case, which term dominates in \Eq{eq:f:dynamical:USR:new} depends on the magnitude of $\delta$. However, strictly speaking, a stability analysis of the fixed point $\delta\sim 0$ should only deal with its immediate neighbourhood, \ie with the smallest possible values of $\vert \delta \vert$ which in this case are smaller than $\vert \eta_V \vert$ (notice that if $\vert \eta_V \vert \gtrsim 1$  this becomes true for all  $\sqrt{\epsilon_V}\ll \vert \delta \vert \ll 1$). The second term in \Eq{eq:f:dynamical:USR:new} then dominates and one has
\bea
\label{eq:f:dynamical:USR:case2}
\frac{\dd\delta}{\dd\phi}\simeq \frac{V''}{V'} \delta\, .
\eea
A similar discussion as in the previous case can be carried out, by first considering the situation where the field follows the gradient of its potential, so $f<1$ and $\delta>0$. If $V'>0$ and $\phi$ decreases with time then $\vert \delta\vert$ decreases if $V''>0$. If $V'<0$ and $\phi$ increases with time then $\vert \delta\vert$ decreases under the same condition 
\bea
V''>0\, .
\eea
Thus we conclude that USR is stable for a scalar field rolling down a convex potential.
Conversely, we find that USR is stable for a scalar field rolling up a concave potential,
$V''<0$. 

The fact that $\vert \delta\vert$ decreases with time is a necessary condition for USR inflation stability but not a sufficient one, since one also has to check that $\vert\delta\vert$ remains much larger than $\sqrt{\epsilon_V}$, \ie that the system remains inflating. To this end, let us notice that \Eq{eq:f:dynamical:USR:case2} can be integrated and gives
\bea
\delta \simeq \delta_\uin \frac{V'(\phi)}{V'\left(\phi_\uin\right)} \, .
\label{eq:delta:sol}
\eea
This confirms that $\vert \delta \vert$ decreases with time when $\vert V' \vert$ decreases. Substituting \Eq{eq:delta:sol} into \Eq{eq:eps1:f:phi} (expanded in the $\sqrt{\epsilon_V}\ll\vert\delta\vert\ll 1$ limit), one obtains
\bea
\label{eq:USR:stable:eps1:appr}
\epsilon_1\simeq\epsilon_{1,\uin}\left( \frac{V_\uin}{V}\right)^2\, .
\eea
Therefore, $\epsilon_1$ increases if the field follows the gradient of its potential and decreases otherwise. Whether or not this increase can stop inflation in the former case depends on the potential. If the relative variations of the potential are bounded this may never happen if $\epsilon_1$ has a sufficiently small value initially. 

One can also use \Eq{eq:delta:sol} to compute the number of \efolds~spent in the USR regime. Since $\dd N/\dd\phi = H/\dot{\phi} = -3H^2\delta/V'$ where we have used the definition~(\ref{eq:def:f}), in the quasi de-Sitter limit where $H^2\simeq V/(3\Mp^2)$, one obtains $\dd N/\dd \phi = -V\delta/(V'\Mp^2)$. Making use of \Eq{eq:delta:sol}, this gives rise to $\dd N/\dd \phi = - V \delta_\uin/(V'_\uin \Mp^2)$, and hence
\bea
\label{eq:DeltaN:USR}
\Delta N_\USR = -\frac{\delta_\uin}{\Mp^2V'_\uin}\int_{\phi_\uin}^\phi V(\tilde{\phi})\dd\tilde{\phi}\, .
\eea
This should be compared with the slow-roll formula $\Delta N_\SR = - 1/\Mp^2\int_{\phi_\uin}^\phi V(\tilde{\phi})/V'(\tilde{\phi})\dd\tilde{\phi} $, which shows that in general fewer \efolds~are realised between two given field values in the USR regime than in standard slow roll. 

Note that when we find USR to be a local attractor ($|\delta|$ decreases), it is so for sufficiently small values of $|\delta|<\vert \eta_V \vert$ only. If $|\delta| > \vert \eta_V \vert$ initially, then the first term on the right-hand side of \Eq{eq:f:dynamical:USR:new} dominates even if $\epsilon_V<\eta_V^2$ and the analysis of section~\Sec{sec:stability:USR:inflation:case1} shows that USR becomes unstable. In this case trajectories diverge from USR ($f\sim 1$) to approach the standard slow roll ($f\ll1$) for $\epsilon_V\ll1$ and $|\eta_V|\ll1$. This shows that, if $\epsilon_V<\eta_V^2 \ll 1$, the boundary between the SR and the USR basins of attraction is located around the line $\vert\delta\vert\sim\vert\eta_V\vert$. This will be checked explicitly in the example presented in \Sec{sec:example:inflectionPoint}.

In summary, we find that if the inflaton rolls down its potential, USR inflation is stable if $V''>0$ and $\eta_V^2>\epsilon_V$, which can be combined into the condition
\bea
\label{eq:USR:stable:criterion}
\eta_V>\sqrt{\epsilon_V}\, ,
\eea
and continues to inflate provided $V/V_\uin$ remains larger than $\sqrt{\epsilon_{1,\uin}}$. 

Let us now illustrate the stability analysis performed in the previous section with two examples. In the first one, the potential has a discontinuity in its slope which produces a transient regime of USR inflation. In the second one, the potential has a flat infection point around which the inflaton field evolves in the USR regime.
\subsubsection{Example 1: Starobinsky inflation}
\label{sec:Staro}
\begin{figure}[t]
\begin{center}
\includegraphics[width=0.65\textwidth]{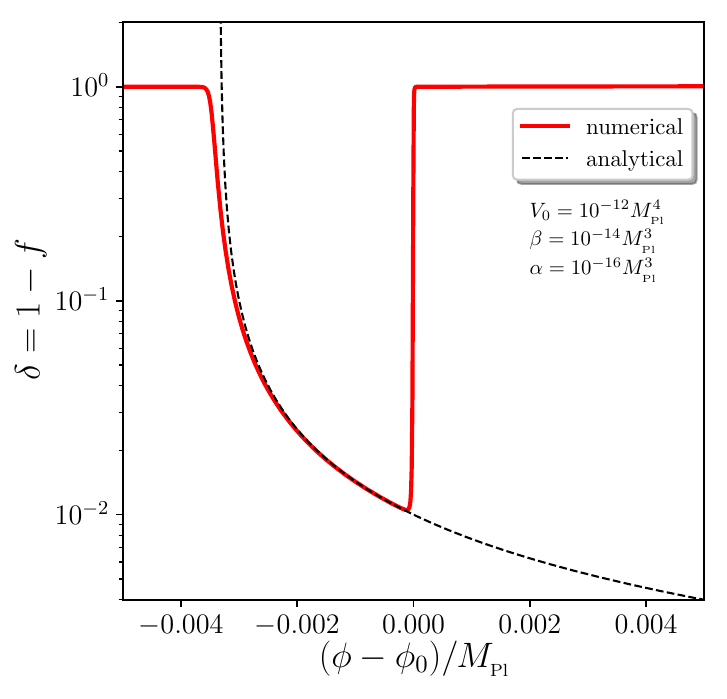}
\caption{Field acceleration parameter $\delta=1-f$ in the Starobinsky model~(\ref{eq:def:pot:strobinsky}) as a function of the field value. Before crossing the discontinuity point, a regime of SR inflation takes place where $\delta\simeq 1$. Right after crossing $\phi=\phi_0$, $\delta$ drops to small values which signals the onset of a USR phase of inflation, that quickly transitions towards a new SR phase. The solid red curve is obtained from numerically solving \Eqs{eq:kleingordon} and~(\ref{eq:Friedmann:cosmic:time}) and making use of \Eq{eq:def:f}, while the black dashed curve corresponds to the analytical approximation~(\ref{eq:Staro:delta:appr}). One can check that it provides a good fit to the numerical result when $\vert\delta\vert\ll 1$.}
\label{fig:Staro}
\end{center}
\end{figure}
Let us first analyse the Starobinsky model~\cite{Starobinsky:1992ts}, where the potential is made up of two linear segments with different gradients,
\bea
\label{eq:def:pot:strobinsky}
V(\phi) = \begin{cases}
V_0 + \alpha \left( \phi-\phi_0 \right) &\text{for $\phi < \phi_0$}\\
V_0 + \beta \left( \phi-\phi_0 \right) &\text{for $\phi > \phi_0$} 
\end{cases}
\, ,
\eea
where $\beta>\alpha>0$.

Starting with $\phi>\phi_0$, the inflaton quickly relaxes to the slow-roll attractor for $V_0\gg\beta \Mp$ (corresponding to $\epsilon_V\ll1$), where, according to \Eq{eq:slowroll}, $3H\dot{\phi}\simeq-\beta$. Right after crossing $\phi=\phi_0$ where the gradient of the potential is discontinuous, $\dot{\phi}$ is still given by the same value (since the equation of motion~(\ref{eq:kleingordon}) for $\phi$ is second order, $\dot{\phi}$ is continuous through the discontinuity point) but the value of $V'$ is now different, such that $f$ given by \Eq{eq:def:f} reads
\bea
\label{eq:Staro:fminus}
f_{-}= 1+\frac{V'_-}{3(H\dot{\phi})_-} = 1+\frac{V'_-}{3(H\dot{\phi})_+}\simeq 1-\frac{V'_-}{V'_+}
= 1-\frac{\alpha}{\beta}
\, .
\eea
In this expression, a subscript ``$-$'' (or ``$+$'') means that the quantity is evaluated at $\phi\rightarrow \phi_0$ with $\phi<\phi_0$ (or $\phi>\phi_0$, respectively). If $\alpha\ll \beta$, $f_-\simeq 1$ and a phase of USR is triggered.

The analysis of \Sec{sec:stability:USR} revealed that the stability of USR inflation depends on whether $\epsilon_V$ is smaller or larger than $\eta_V^2$. In the present model, since $\eta_V$ exactly vanishes, one necessarily falls in the later case, \ie the case discussed in \Sec{sec:stability:USR:inflation:case1} where it was shown that USR inflation is always unstable. Let us also notice that in the Starobinsky model, \Eq{eq:f:dynamical:USR:case1} can be integrated analytically, and making use of \Eq{eq:Staro:fminus} for the initial condition, one finds
\bea
\label{eq:Staro:delta:appr}
\delta\simeq \dfrac{\alpha}{\beta+\frac{3V_0}{\Mp^2}\left(\phi-\phi_0\right)}
\, .
\eea
Since $\phi$ decreases as a function of time, $\delta$ increases, and this confirms that USR is unstable in the Starobinsky model.

These considerations are numerically checked in \Fig{fig:Staro}. One can see that when the inflaton field crosses the discontinuity point at $\phi=\phi_0$, a phase of USR inflation with small values of $\delta$ starts, which \Eq{eq:Staro:delta:appr} accurately describes. This regime is however unstable and when the inflaton field crosses the value
\bea
\label{eq:phiUSR:SR:Staro}
\phi_{\USR\rightarrow\SR}=\phi_0-\frac{\Mp^2\left(\beta-\alpha\right)}{3V_0}
\, ,
\eea 
$\delta\simeq 1$ and the system relaxes back to SR. Making use of \Eq{eq:USR:traj}, one can also estimate the number of \efolds~spent in the USR regime between the field values $\phi_0$ and $\phi_{\USR\rightarrow\SR}$, and one finds
\bea
\label{eq:NUSR:Staro}
N_\USR\simeq\frac{1}{3}\ln\left(\frac{\beta}{\alpha}\right)\, .
\eea
The number of USR \efolds~is therefore of order a few or less in this model.
\subsubsection{Example 2: Cubic inflection point potential}
\label{sec:example:inflectionPoint}
Let us now consider the case where the potential contains a flat inflection point at $\phi=0$, around which it can be expanded as
\bea 
\label{eq:pot:inflectionPoint:cubic}
V(\phi) = V_0 \left[1+\left(\frac{\phi}{\phi_0}\right)^3\right]\, .
\eea
One could parametrise the potential with a higher odd power of the field, say $V\propto 1+(\phi/\phi_0)^5$, but this would not change the qualitative conclusions that we draw below. The potential~(\ref{eq:pot:inflectionPoint:cubic}) has a flat inflection point at $\phi=0$, where $V'=V''=0$. This is a seemingly simple model but we will show that it displays a lot of interesting phenomenology. 

As explained in \Eq{eq:sr:consistency}, SR inflation requires $\epsilon_V\ll 1$ and $\vert\eta_V\vert\ll 1$, where the potential slow-roll parameters \eqref{eq:epsV} are here given by
\bea
\label{eq:InflectionPoint:SRpotParam}
\epsilon_V &=& \dfrac92 \frac{\Mp^2}{\phi_0^2} \frac{(\phi/\phi_0)^4}{\left[ 1+(\phi/\phi_0)^3 \right]^2} \,, \\
\eta_V &=& 6 \dfrac{\Mp^2}{\phi_0^2} \frac{\phi/\phi_0}{1+(\phi/\phi_0)^3} \,. 
\eea
Let us first focus on the part of the potential located before the inflection point, \ie at $\phi>0$. The parameter $\epsilon_V$ vanishes at $\phi=0$ and at $\phi\rightarrow\infty$, and in between it reaches a maximum at $\phi=2^{1/3}\phi_0$ where its value is $\epsilon_{V,\umax}=2^{1/3}\Mp^2/\phi_0^2$. The parameter $\eta_V$ has a similar behaviour, with a maximum at $\phi=2^{-1/3}\phi_0$ where its value is $\eta_{V,\umax}=2^{5/3}\Mp^2/\phi_0^2$. 

Two regimes need therefore to be distinguished: (i) if $\phi_0\gg \Mp$, SR inflation can be realised for all $\phi>0$, while (ii) if $\phi_0\ll \Mp$, SR inflation only takes place at sufficiently large ($\phi\gg\Mp$) or sufficiently small ($\phi\ll \phi_0^3/\Mp^2$) field values. After crossing the inflection point at  $\phi=0$, the potential decreases towards zero and the potential slow-roll parameter, $\epsilon_V$, diverges, signalling the end of inflation, so we restrict our analysis to the field values $\phi>-\phi_0$.

USR inflation can be studied making use of the results of \Sec{sec:stability:USR}, where it was shown that USR inflation is stable if $\eta_V>\sqrt{\epsilon_V}$, see \Eq{eq:USR:stable:criterion}. Together with \Eq{eq:InflectionPoint:SRpotParam}, this gives rise to the USR stability condition
\bea
\label{eq:InflectionPoint:USR:stability:Condition}
0<\phi<2\sqrt{2}\Mp\, .
\eea

We shall now study the two regimes $\phi_0\gg\Mp$ and $\phi_0\ll\Mp$ separately.

\paragraph{Case $\phi_0\gg \Mp$}
\begin{figure}
\begin{center}
\includegraphics[width=0.45\textwidth, height=6cm]{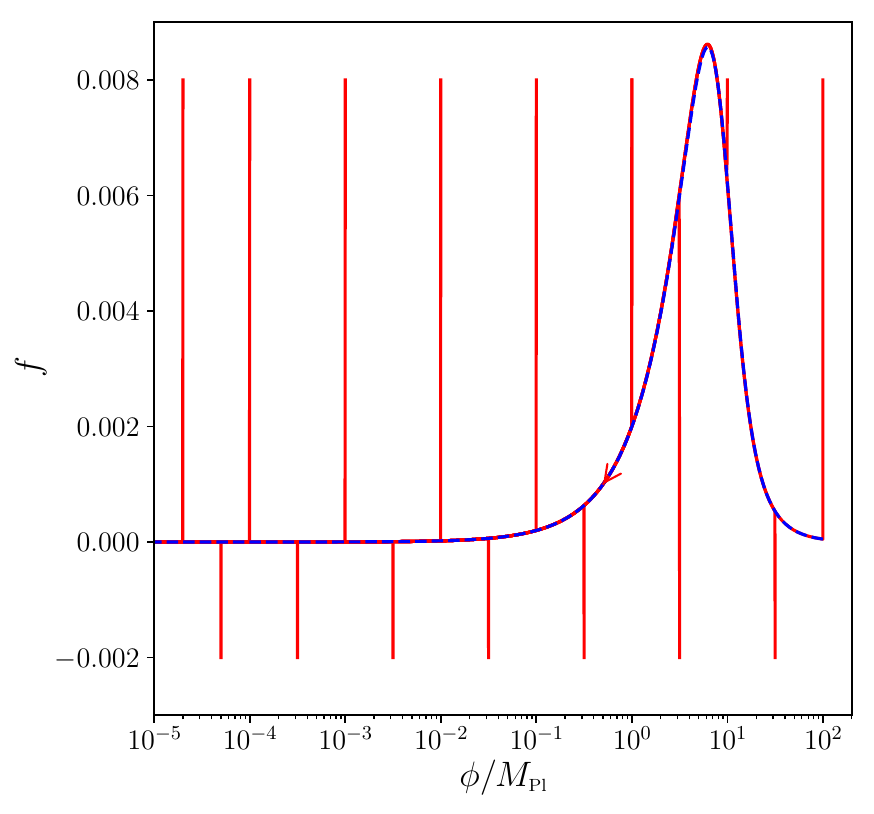} 
\includegraphics[width=0.45\textwidth, height=6.1cm]{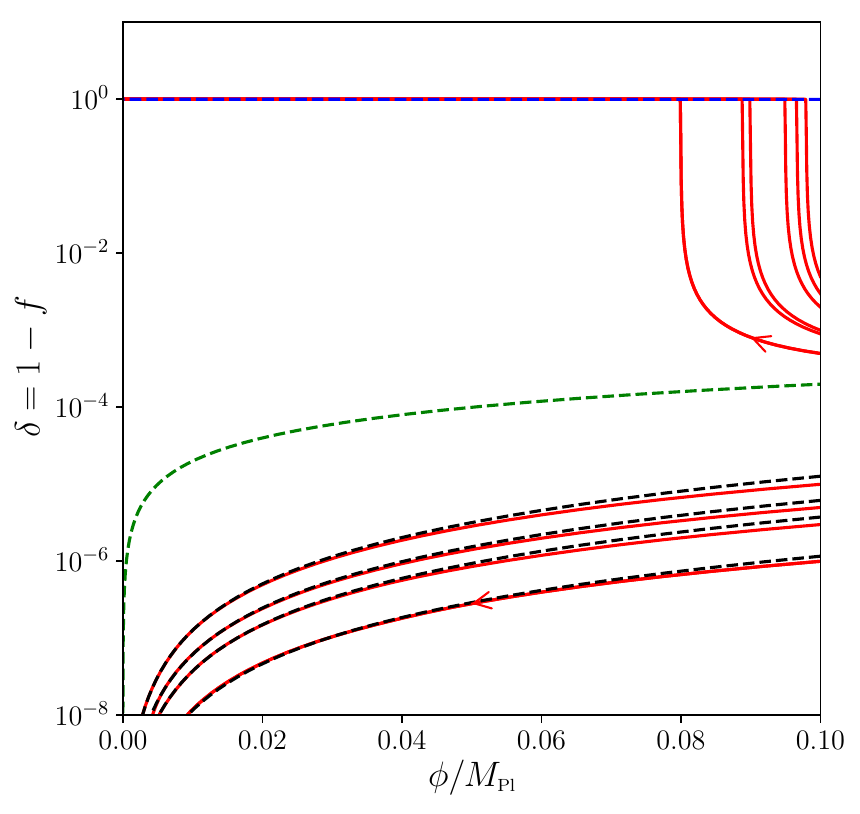} 
\caption{Field acceleration parameter in the cubic inflection point model~(\ref{eq:pot:inflectionPoint:cubic}) as a function of the field value, for $\phi_0=10\Mp$ and $V_0=4.2\times 10^{-11}$. The red lines correspond to numerical solutions of \Eq{eq:f:dynamical} and the dashed blue line stands for the slow-roll limit~(\ref{eq:f:SR}). The left panel zooms in on the region $f\simeq 0$ where one can see that SR is an attractor. The right panel uses a logarithmic scale on $1-f=\delta$, such that it zooms in on the USR regime $f\simeq 1$. If $\delta$ is initially smaller than $\vert \eta_V \vert/3$, represented with the dashed green line, the trajectories evolve towards $\delta=0$, otherwise they evolve to reach the SR attractor. The black dotted lines correspond to the analytical USR approximation~(\ref{eq:delta:sol}).
}
\label{fig:InflectionPoint:fsol:phi0GTMp}
\end{center}
\end{figure}
In this case, as already mentioned, SR inflation is an attractor over the entire range $\phi>0$ (until inflation stops when $\phi$ approaches $-\phi_0$). This implies that if one starts from an initial field value that is larger than the USR stability upper bound given in \Eq{eq:InflectionPoint:USR:stability:Condition}, $\phi=2\sqrt{2}\Mp$, the system relaxes towards the SR attractor (SR is the only stable attractor at $\phi>2\sqrt{2}\Mp$) and stays in SR until the end of inflation. In this scenario, even though USR inflation is also a local attractor at $\phi<2\sqrt{2}\Mp$, the inflaton field never drives a phase of USR inflation. 

The only way to get a period of USR inflation is therefore to start with $\phi<2\sqrt{2}\Mp$. There, as explained in \Sec{sec:stability:USR}, USR inflation is stable and its basin of attraction is bounded by the condition $\sqrt{\epsilon_V}<|\delta|<\eta_V$.

These considerations are numerically verified in \Fig{fig:InflectionPoint:fsol:phi0GTMp}. In the left panel, the SR region $\vert f\vert\ll 1 $ is displayed, where one can check that the numerical solutions of \Eq{eq:f:dynamical} (red curves) all converge towards the SR attractor~(\ref{eq:f:SR}) (dashed blue curve). In the right panel, a logarithmic scale is used on $1-f$, which allows one to zoom in on the USR region $f\simeq 1$. Since the initial values for $\phi$ satisfy \Eq{eq:InflectionPoint:USR:stability:Condition}, one can check that the trajectories with $\delta<\eta_V$ converge towards USR, while the ones for which $\delta>\eta_V$ approach the SR attractor. This confirms that the boundary between the two basins of attraction is located around the line $|\delta|=|\eta_V|$. The analytical approximation~(\ref{eq:delta:sol}) is displayed with the black dotted lines and one can check that it provides a good fit to the numerical result in the USR regime.

Let us finally estimate the number of \efolds~that is typically realised in the USR inflating regime. In the stability range~(\ref{eq:InflectionPoint:USR:stability:Condition}) of USR inflation, the potential is dominated by its constant piece since $\phi_0\gg\Mp$. The first Hubble-flow parameter is therefore roughly constant during the USR epoch, see \Eq{eq:USR:stable:eps1:appr}. Starting USR inflation at $\phi_\uin\sim \Mp$ with $\delta=\delta_\uin$, \Eq{eq:delta:sol} implies that $\delta$ goes back to its initial value $\delta_\uin$ at around $\phi\sim -\Mp$.  Plugging these values into \Eq{eq:DeltaN:USR}, one obtains
\bea
\label{eq:FlatInflectionPoint:NUSR:phi0GTMp}
\Delta N_\USR\simeq \frac{2\delta_\uin}{3}\left(\frac{\phi_0}{\Mp}\right)^3.
\eea
This shows that, in the regime $\phi_0\gg \Mp$, a large number of USR~\efolds~can be realised.

This large number of USR inflationary \efolds~is however derived under the assumption that the field behaves classically all the way down to the inflection point, while stochastic diffusion is expected to play a role when the potential becomes very flat. Although this requires to employ a full USR stochastic formalism, such as developed below in \Sec{sec:USR:stochastic}, let us estimate how this changes the above result.

Starting from $\phi_\uin=\Mp$ and $\delta=\delta_\uin$ as explained above, \Eq{eq:delta:sol} leads to $\delta(\phi)\simeq \delta_\uin(\phi/\Mp)^2$ (where we assume $\phi<\Mp$). Then, making use of \Eq{eq:DeltaN:USR}, if the field behaved in a purely classical manner, the number of \efolds~realised between $\phi$ and $-\phi$ would be given by $\Delta N_\USR(\phi)\simeq 2\delta_\uin/3 (\phi_0/\Mp)^3(\phi/\Mp)$.

On the other hand, if the field was only driven by stochastic noise, as shown in \Sec{sec:USR:stochastic}, its equation of motion would be given by $\dd\phi/\dd N=H/(2\pi)\xi$, where $\xi$ is a white Gaussian noise with vanishing mean and unit variance, such that $\langle \xi(N) \xi(N') \rangle = \delta(N-N')$.  Assuming that $H$ is roughly constant, this leads to $\langle \phi^2 \rangle = H^2/(2\pi)^2 N$, hence the typical number of \efolds~required for the inflaton field value to go from $\phi$ to $-\phi$ is given by $\Delta N_{\mathrm{sto}}= 48\pi^2\phi^2\Mp^2/V_0$. Notice that this can also be more precisely obtained using the first-passage-time techniques introduced in \Sec{sec:FirstPassageTime}. Setting a reflective boundary condition at $\phi$ and an absorbing one at $-\phi$, one then finds that the mean number of \efolds~required to reach $-\phi$ starting from $\phi$ exactly coincides with the expression we just wrote for  $\Delta N_{\mathrm{sto}}$.

Since $\Delta N_\USR$ scales as $\phi$ and $\Delta N_{\mathrm{sto}}$ as $\phi^2$, two regimes need to be distinguished. When $\phi>\phi_{\mathrm{sto}}$, where 
\bea
\frac{\phi_{\mathrm{sto}}}{\phi_0} = \frac{\delta_\uin}{72\pi^2}\left(\frac{\phi_0}{\Mp}\right)^2\frac{V_0}{\Mp^4}
\eea
is the solution of $\Delta N_\USR(\phi_{\mathrm{sto}})=\Delta N_{\mathrm{sto}}(\phi_{\mathrm{sto}})$, one has $\Delta N_\USR<\Delta N_{\mathrm{sto}}$, which means that classical USR is more efficient at driving the field than stochastic diffusion, hence the dynamics of the field are essentially classical. When $\phi<\phi_{\mathrm{sto}}$ on the other hand, stochastic diffusion takes over, which means that the part of the potential where $-\phi_{\mathrm{sto}}<\phi<\phi_{\mathrm{sto}}$ is dominated by quantum diffusion.  

This is why, for classical USR to take place, one needs to impose $\phi_{\mathrm{sto}}<\Mp$, which means that the potential energy cannot be too large,
\bea
\frac{V_0}{\Mp^4}\ll 72\pi^2\left(\frac{\Mp}{\phi_0}\right)^3
\eea
(recall that $\phi_0\gg \Mp$ so this is not necessarily guaranteed). When this is the case, the number of \emph{classical} USR inflationary \efolds~is given by $\Delta N_{\USR {}_{,\mathrm{class}}} = \Delta N_{\USR}(\Mp)-\Delta N_{\USR}(\phi_{\mathrm{sto}} )$, where $\Delta N_{\USR}(\Mp)$ was given in \Eq{eq:FlatInflectionPoint:NUSR:phi0GTMp}, and one obtains
\bea
\Delta N_{\USR{}_{,\mathrm{class}}} = \frac{2\delta_\uin}{3}\left(\frac{\phi_0}{\Mp}\right)^3 \left[1 - \frac{\delta_\uin}{72\pi^2}\left(\frac{\phi_0}{\Mp}\right)^3 \frac{V_0}{\Mp^4}\right]\, .
\eea
If the parameter $\phi_0$ is chosen such that $1\ll \phi_0/\Mp \ll (72\pi^2V_0/\Mp^4)^{-1/3}$, this number can still be very large and a sustained phase of classical USR inflation takes place.
\paragraph{Case $\phi_0\ll \Mp$}
\begin{figure}
\begin{center}
\includegraphics[width=0.45\textwidth, height=6cm]{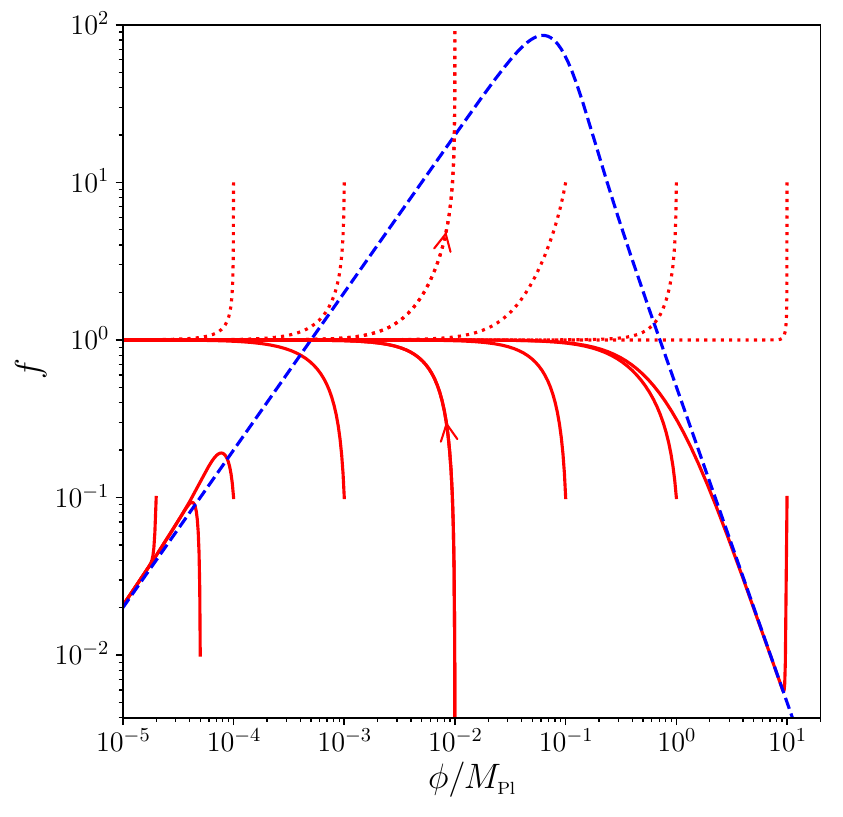} 
\includegraphics[width=0.45\textwidth, height=6.1cm]{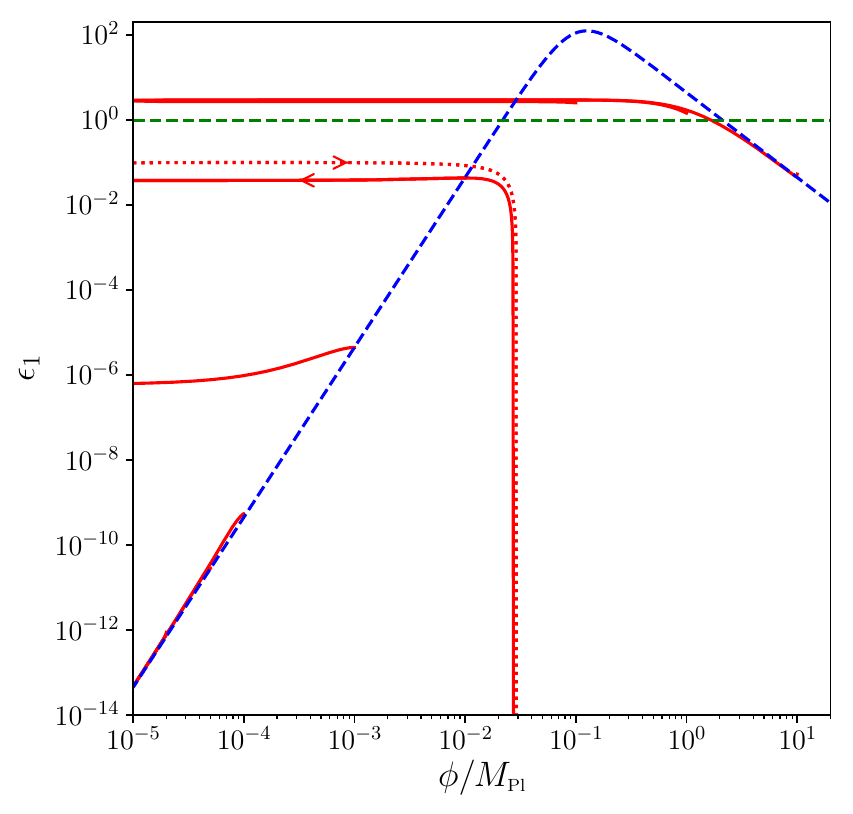} 
\caption{Left panel: Field acceleration parameter in the cubic inflection point model~(\ref{eq:pot:inflectionPoint:cubic}) as a function of the field value, for $\phi_0=0.1\Mp$ and $V_0=4.2\times 10^{-11}$, with the same conventions as in \Fig{fig:InflectionPoint:fsol:phi0GTMp}. The solid part of the red curves correspond to when $f<1$ and $\phi$ decreases with time, while the dotted parts are for $f>1$ and $\phi$ increases (as indicated by the arrows). Right panel: first Hubble-flow parameter $\epsilon_1$ as a function of the field value for the same solid trajectories and the dotted trajectory with an arrow in the left panel. The dashed green line stands for $\epsilon_1=1$ below which inflation proceeds. The trajectories that have both $\epsilon_1 \ll 1$ and $f \to 1$ drive a phase of USR inflation. 
}
\label{fig:InflectionPoint:fsol:phi0LTMp}
\end{center}
\end{figure}
In this case, SR inflation can only occur at $\phi\gg\Mp$ or $\phi\ll\phi_0^3/\Mp^2$. One therefore has three regions: if $\phi\gg\Mp$, SR is the only attractor, if $\phi_0^3/\Mp^2\ll\phi\ll\Mp$, USR is the only attractor, and if $\phi\ll \phi_0^3/\Mp^2$, both SR and USR are local attractors. These three regimes can be clearly seen in the left panel of \Fig{fig:InflectionPoint:fsol:phi0LTMp}, where the same colour code as in \Fig{fig:InflectionPoint:fsol:phi0GTMp} is employed. In the right panel of \Fig{fig:InflectionPoint:fsol:phi0LTMp}, the first Hubble-flow parameter is displayed for the same trajectories. The solid curves have $f<1$ for which $\phi$ decreases with time and the dotted curves have $f>1$ for which $\phi$ increases with time. We shall now discuss each of these three regimes in more detail.

Firstly, if one starts with an initial value of $\phi$ that is super Planckian, one quickly reaches the SR attractor. Then when $\phi$ becomes of order $\Mp$, $f_\SR$ becomes of order one which signals the breakdown of SR and one leaves the SR line to settle down to $f\simeq 1$, \ie in the USR regime. However, as can be seen on the right panel of \Fig{fig:InflectionPoint:fsol:phi0LTMp}, the first Hubble-flow parameter converges towards $\epsilon_1\simeq 3$, so inflation stops around $\phi\simeq\Mp$ and does not resume afterwards. In this case, for $\phi<\Mp$, we have USR but not USR inflation, and this non-inflating USR regime is stable due to the considerations of footnote~\ref{footnote:USR:Non:Inflation}.

Secondly, if one starts with an initial field value between $\phi_0^3/\Mp^2$ and $\Mp$ and with $\dot\phi<0$ (rolling down the potential), the field converges towards USR, since it is the only stable solution. This is the case for the trajectory with $f<1$ on which an arrow has been added in \Fig{fig:InflectionPoint:fsol:phi0LTMp}. Let us recall that the dotted part of the trajectory corresponds to $f>1$ and the inflaton climbs up its potential ($\dot\phi>0$), until its velocity changes vanishes at which point $f$ diverges and $\dot\phi$ changes sign. The inflaton then rolls down its potential starting from very negative values for $f$ (solid part of the curve) and quickly reaches USR. On the right panel of \Fig{fig:InflectionPoint:fsol:phi0LTMp} one can see that $\epsilon_1$ is roughly constant in the rolling down phase, which is consistent with \Eq{eq:USR:stable:eps1:appr} since the potential is dominated by its constant piece $V\simeq V_0$ when $\phi\ll\phi_0$.
\begin{figure}
\begin{center}
\includegraphics[width=0.65\textwidth]{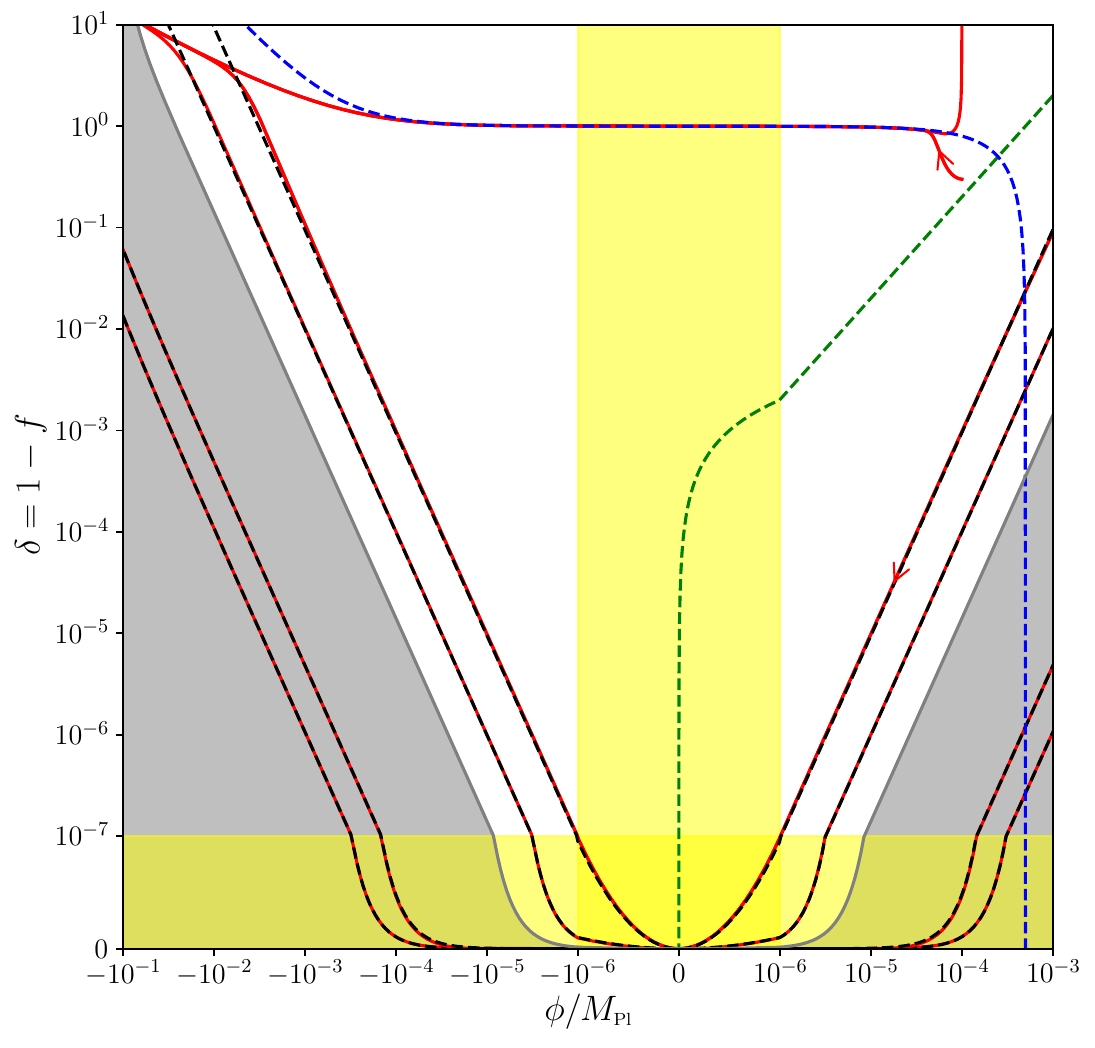} 
\caption{Same as in the left panel of \Fig{fig:InflectionPoint:fsol:phi0LTMp} for the region $-\phi_0<\phi<\phi_0^3/\Mp^2$. Yellow shading denotes regions when the axis scale is linear rather than logarithmic, and the grey shaded region is where inflation in not happening $\epsilon_1>1$, as given by \eqref{eq:inflation:condition}. The black dashed lines stand for the analytical USR inflation approximation~(\ref{eq:delta:sol}) in the inflating part, and to the USR \emph{non}-inflation approximation~(\ref{eq:delta:sol:USRnoninflating}) in the non-inflating part (the field excursion being sub Planckian, the two behaviours are very much similar). The dashed green line stands for $\delta=\vert\eta_V\vert/3$, which in the $\phi>0$ region corresponds to the boundary between the SR and the USR basins of attraction. For $\phi<0$, only SR is an attractor which explains why those trajectories that reach the USR attractor in the $\phi>0$ region have $\delta$ increasing with time in the $\phi<0$ region where USR is unstable. 
}
\label{fig:InflectionPoint:fsol:phi0LTMp:phiLTphi0power3}
\end{center}
\end{figure}

Lastly, if one starts with $\phi\ll \phi_0^3/\Mp^2$, one either reaches the SR attractor if $|\delta|>|\eta_V|$ or the USR attractor if $|\delta|<|\eta_V|$. This can be more clearly seen in \Fig{fig:InflectionPoint:fsol:phi0LTMp:phiLTphi0power3}, where the whole region $-\phi_0<\phi<\phi_0^3/\Mp^2$ is displayed. One can check that the inflating USR approximation~(\ref{eq:delta:sol}) provides a good approximation to the numerical solutions of \Eq{eq:f:dynamical} in the inflating part of phase space for those trajectories that reach the USR attractor, while the \emph{non}-inflating USR approximation~(\ref{eq:delta:sol:USRnoninflating}) correctly describes the non-inflating trajectories (in the grey shaded region of the plot). Here, because the field excursion is sub-Planckian (since $\phi_0\ll\Mp$), these two behaviours are almost identical. As in the right panel of \Fig{fig:InflectionPoint:fsol:phi0GTMp}, one can also check that the line $\vert\delta\vert\sim\vert\eta_V\vert$ correctly delimitates the boundary between the two basins of attraction when $\phi>0$. If $\phi<0$, USR becomes unstable and only SR remains as an attractor, which explains why $\delta$ increases with time for those trajectories that reached the USR attractor before crossing the flat inflection point. However, one should note that those trajectories do not have time to reach the SR attractor before the potential becomes too steep and SR is violated.
Similarly, for $\phi > \phi_0^3/\Mp^2$ the potential is too steep and SR is not a valid approximation.

Let us also estimate the number of \efolds~that is typically realised in the USR inflating regime. USR is stable in the range~(\ref{eq:InflectionPoint:USR:stability:Condition}). However, for $\phi>\phi_0$, the potential is not dominated by its constant piece so $\epsilon_1$ can substantially increase because of \Eq{eq:USR:stable:eps1:appr}. Whether or not USR \emph{inflation} is maintained depends on the initial value of $\epsilon_1$ (see the right panel of \Fig{fig:InflectionPoint:fsol:phi0LTMp}) and to avoid this initial condition dependence, let us consider the case where we start USR inflation around $\phi\sim\phi_0$. Starting with $\delta=\delta_\uin$, \Eq{eq:delta:sol} implies that $\delta$ goes back to its initial value $\delta_\uin$ at around $\phi\sim -\phi_0$.  Making use of \Eq{eq:DeltaN:USR}, this gives rise to
\bea
\Delta N_\USR \simeq \frac{2\delta_\uin}{3}\left(\frac{\phi_0}{\Mp}\right)^2\, .
\eea
This shows that, in the regime $\phi_0\ll\Mp$, the number of \efolds~realised in the USR regime is necessarily small, contrary to the case $\phi_0\gg\Mp$, see \Eq{eq:FlatInflectionPoint:NUSR:phi0GTMp}.
\begin{figure}
\begin{center}
\includegraphics[width=0.49\textwidth]{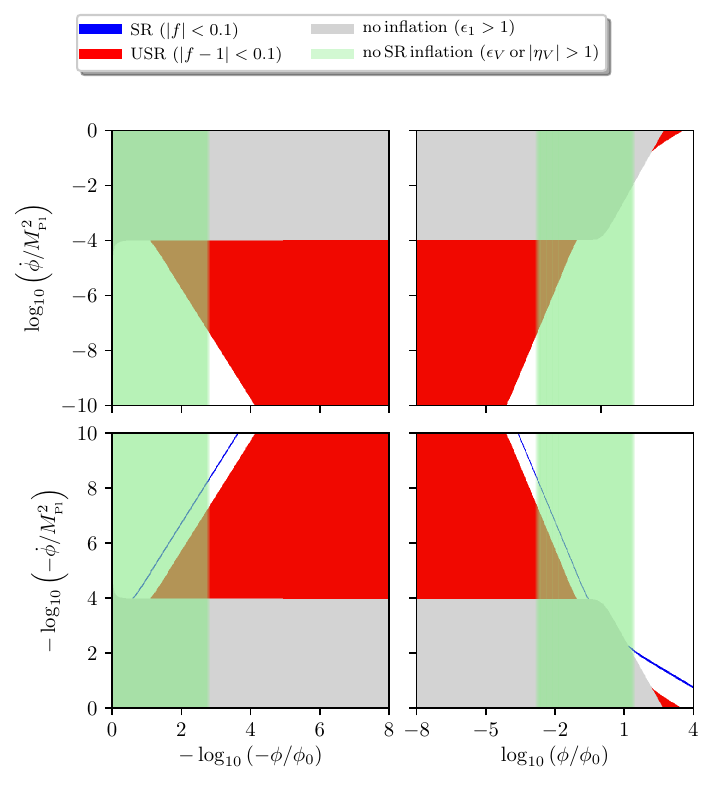} 
\includegraphics[width=0.49\textwidth]{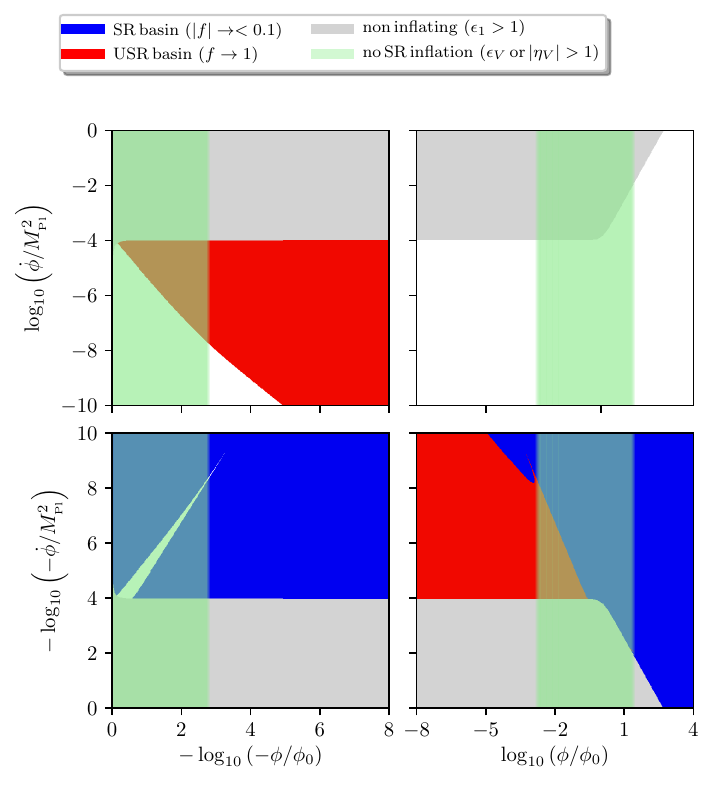} 
\caption{Regions in phase space for the cubic inflection point model~(\ref{eq:pot:inflectionPoint:cubic}) with $V_0=4.2\times 10^{-11}$ and $\phi_0=0.1\Mp$ where SR and USR solutions exist (left panel), and (right panel) basins of attraction for SR ($f<1$ and $|f|$ decreasing) and USR ($|1-f|$ decreasing).}
\label{fig:phasespace_phi0eq0p1}
\end{center}
\end{figure}

\begin{figure}
\begin{center}
\includegraphics[width=0.49\textwidth]{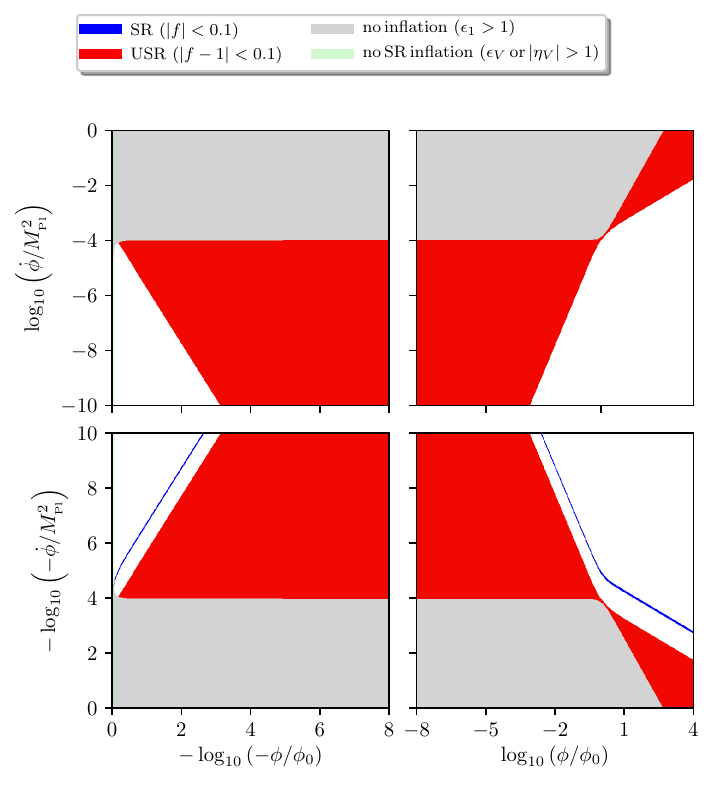} 
\includegraphics[width=0.49\textwidth]{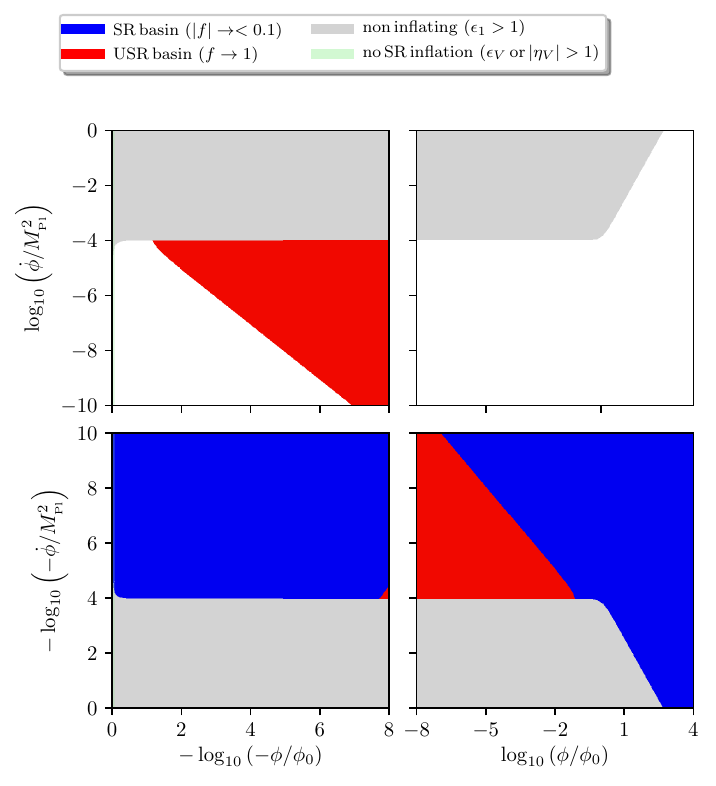} 
\caption{Regions in phase space for the cubic inflection point model~(\ref{eq:pot:inflectionPoint:cubic}) with $V_0=4.2\times 10^{-11}$ and $\phi_0=10\Mp$ where SR and USR solutions exist (left panel), and regions of stability for SR and USR (right panel).}
\label{fig:phasespace_phi0e10}
\end{center}
\end{figure}
Finally we plot in \Figs{fig:phasespace_phi0eq0p1} and~\ref{fig:phasespace_phi0e10} the phase space $(\phi,\dot\phi)$ for the cubic inflection point model~(\ref{eq:pot:inflectionPoint:cubic}), for $\phi_0=0.1\Mp$ and $\phi_0=10\Mp$ respectively. In the left panels, the blue region corresponds to SR solutions (defined as $|f|<0.1$) and the red region to USR (defined as $|1-f|<0.1$). In the right panels we show the basins of attraction of SR and USR, defined by the behaviour of $f$.

One can see that SR corresponds to a thin line in phase space while USR spans a larger region. This is due to the fact that in USR inflation, there is no unique USR trajectory in phase space and solutions retain a dependence on initial conditions as can be seen \eg in \Eq{eq:delta:sol}. This is not the case for SR that singles out a unique phase-space trajectory, see \Eq{eq:f:SR}. Note also that SR solutions only exist in the quadrants where the field velocity is aligned with the potential gradient while USR exists in every quadrant.

The right-hand plots show the basins of attraction of SR (\ie where $f<1$ and $|f|$ decreases) and USR (where $|1-f|$ decreases). When $\phi>0$, if the field goes up the potential ($\dot{\phi} >0$) then  USR is unstable, and there is no SR regime and hence no SR basin of attraction either. If the field rolls down the potential ($\dot{\phi}<0$), when $\phi\gg \Mp$ or $\phi<0$ we see that only SR is an attractor as discussed above, and when $0<\phi\ll\Mp$ both SR and USR can be attractors. When $\phi<0$ and the field goes up the potential, USR is an attractor in some region of the phase space.
This corresponds to initial conditions where the field arrives at the inflection point with an almost vanishing velocity and inflates in the USR regime.
\subsection{Gauge correction in ultra-slow roll}
\label{sec:USR:GaugeCorrections}
Having determined under which conditions a phase of ultra-slow roll inflation can take place, and in which cases it is stable or unstable, we now aim at developing a stochastic formalism for ultra-slow roll. As explained in \Sec{sec:StochasticInflation}, the first step is to compute the gauge corrections, along the procedure outlined in \Sec{sec:uniformexpansion}. This is the goal of this section, which follows \Refa{Pattison:2019hef}. 
\subsubsection{Gauge corrections}
\label{sec:USR:gauge:correction}

As explained in \Sec{sec:USR:stability}, the phase-space trajectory of ultra-slow roll, \Eq{eq:USR:traj}, carries a dependence on initial conditions that is not present in slow roll, which explains why ultra-slow roll is not a dynamical attractor (though it can be stable) while slow roll is. We therefore expect the non-adiabatic pressure perturbation not to vanish in ultra-slow roll, which may lead to some non-trivial gauge corrections. In ultra-slow roll,  as explained in \Sec{sec:USR:stability}, the field acceleration parameter $f$ introduced in \Eq{eq:def:f} is close to one (while it is close to zero in slow roll), so $\delta\equiv 1-f$ quantifies how deep in the ultra-slow-roll regime one is. In the limit where $\delta=0$,  \Eq{eq:USR:phidot:N} gives rise to $\epsilon_1^{\mathrm{USR}}\propto \ee^{-6N}/H^2$, hence
\bea \label{eq:eps:USR}
\epsilon_n^{\mathrm{USR}} = \begin{cases}
-6 + 2\epsilon_{1} &\text{if $n$ is even}\\
2\epsilon_{1} &\text{if $n>1$ is odd} \, .
\end{cases}
\eea 
The even slow-roll parameters are therefore large in ultra-slow roll. When $\delta$ does not strictly vanish, these expressions can be corrected, and for the second and the third slow-roll parameters, one finds
\begin{align}
\label{eq:eps2:USR}
\epsilon_{2}&= -6  + 2\epsilon_{1}+6\delta \,, \\
\epsilon_{3}&= 2\epsilon_{1} - \frac{\dd \delta}{\dd N}\frac{6}{6 - 2\epsilon_{1} - 6\delta} \, ,
\end{align}
which are exact formulas. One can then calculate
\bea \label{eq:ddeltadt:USR}
\frac{\dd \delta}{\dd N} &=  -\eta + 3\delta - 3\delta^2 + \delta\epsilon_{1} \, ,
\eea 
where $\eta$ is the dimensionless mass parameter defined in \Eq{eta:def}. For small $\delta$ and $\epsilon_{1}$, one then has
\bea 
\epsilon_{3}^{\mathrm{USR}} \simeq 2\epsilon_{1} + \eta - 3\delta + \eta\left( \frac{2\epsilon_{1}+6\delta}{6} \right) \, .
\eea 
There is no reason, \textit{a priori}, that $\eta$ needs to be small, and hence these corrections can be large for models with $V_{,\phi\phi}\neq 0$.
Note also that \Eq{eq:ddeltadt:USR} provides a criterion for the stability of ultra-slow roll, which is stable when the right-hand side of this equation is negative, in agreement with the results of \Sec{sec:stability:USR}. 

Let us now derive the gauge corrections in ultra-slow roll. We perform a calculation at leading order in $\epsilon_1$, $\delta$ and $\eta$ (see \Refa{Pattison:2019hef} for an extension to next-to-leading order in $\epsilon_1$). At leading order, one simply has $z''/z\simeq 2 \calH^2$, hence \Eq{eq:MSequn} is solved according to
\bea \label{eq:v:nu=3/2}
v_k = \frac{1}{\sqrt{2k}}\ee^{-ik\eta}\left( 1 - \frac{i}{k\eta}\right) \, .
\eea 
Since $a = -1/(H_{*}\eta)$ at leading order, this gives rise to
\bea \label{eq:Q'overQ:USR:nu=3over2}
\frac{Q'_k}{Q_k} = \frac{-ik^2\eta}{k\eta-i} \, ,
\eea 
and the source function~\eqref{eq:sourcefunction:general} reads
\bea \label{eq:source:USR:nu=3/2}
S_k &= \frac{H_{*}}{2\Mp}\sqrt{\frac{\epsilon_{1}}{k}}\ee^{-ik\eta}\left( 3 - \frac{3i}{k\eta} + ik\eta \right) \mathrm{sign}\left(\dot{\phi}\right) \, .
\eea
Since $\epsilon_{1}\simeq \epsilon_{1*}(a/a_*)^{-6}$, the gauge transformation parameter $\alpha$ can be obtained from \Eq{eq:alphaintegral1:general} and is given by
\bea \label{eq:alpha:USR:nu=3/2}
\alpha_k = \frac{iH_{*}\sqrt{\epsilon_{1*}}}{6\Mp}{k^{-\frac{5}{2}}}(k\eta)^4\mathrm{sign}\left(\dot{\phi}\right) \left[  1+ \mathcal{O}(k\eta)^2\right] \, .
\eea
Comparing this expression with \Eq{eq:alphaLO:SR}, one can see that the gauge correction decays even faster than in the slow-roll regime, hence is even more suppressed. This is because, although slow roll is a dynamical attractor while ultra-slow roll is not, the field velocity (hence the conjugate momentum) decays very quickly in ultra-slow roll, and this also damps away one of the two dynamical degrees of freedom. Finally, the gauge transformation \eqref{eq:transform:phi} gives rise to
\bea \label{eq:deltaphi:USRtransform}
\widetilde{\delta\phi_k} = Q_k\left[ 1 + \frac{\epsilon_{1*}}{3}\left(-k\eta\right)^6\right]\, .
\eea
The relative corrections to the noises correlators scale as $\epsilon_{1*}\sigma^6$ and can therefore be neglected, even more accurately than in slow roll. 
\subsubsection{Example: Starobinsky model}
\label{sec:starobinsky}
In \Secs{sec:slowroll:gauge:corrections} and~\ref{sec:USR:gauge:correction}, we have shown that the gauge corrections to the noise correlators are negligible both in slow-roll and in ultra-slow-roll inflation. In this section, we consider again the model of \Sec{sec:Staro}, that interpolates between these two limits. This allows us to study a regime that is neither slow roll nor ultra-slow roll, but for which the early-time (ultra-slow roll) and the late-time (slow roll) limits are under control.

The potential is given by \Eq{eq:def:pot:strobinsky}, where, without loss of generality, we set $\phi_0=0$. We also introduce the two dimensionless parameters $a_+$ and $a_-$, related to $\alpha$ and $\beta$ through
\bea
a_- = \frac{\Mp}{V_0} \alpha\, , \quad a_+ = \frac{\Mp}{V_0} \beta,
\eea
such that the potential~\eqref{eq:def:pot:strobinsky} can be rewritten as
\bea 
V(\phi) = \begin{cases} V_{0}\left(1+a_+\frac{\phi}{\Mp}\right) & \mathrm{for}\, \phi> 0 \\ 
V_{0}\left(1+a_-\frac{\phi}{\Mp}\right) & \mathrm{for}\, \phi< 0 
\end{cases}
\, .
\eea
For a transient phase of ultra slow roll to take place, we assume that $a_+\gg a_- > 0$, and in order to ensure both parts of the potential are able to support slow-roll inflation, we require $a_{\pm}\ll 1$. 

The dynamics of the inflaton, as it evolves across the discontinuity in the potential gradient at $\phi=0$, can be split into three phases. The first phase, which we label $\mathrm{SR}_{+}$, is a slow-roll phase for $\phi>0$ and $\dot\phi<0$. When the inflaton crosses $\phi=0$, it then starts down the $\phi<0$ part of the potential with an initial velocity inherited from the first slow-roll phase $\mathrm{SR}_{+}$ that is much larger than the slow-roll velocity for $\phi<0$. The second phase thus starts in an ultra-slow-roll regime and is denoted USR. It corresponds to the field range $\phi_{\mathrm{USR}\to\mathrm{SR}}<\phi<0$. Finally, the inflaton relaxes back to slow roll for $\phi<\phi_{\mathrm{USR}\to\mathrm{SR}}$, and we call this third phase $\mathrm{SR}_{-}$. 

\begin{figure}
    \centering
    \includegraphics[width=0.50\textwidth]{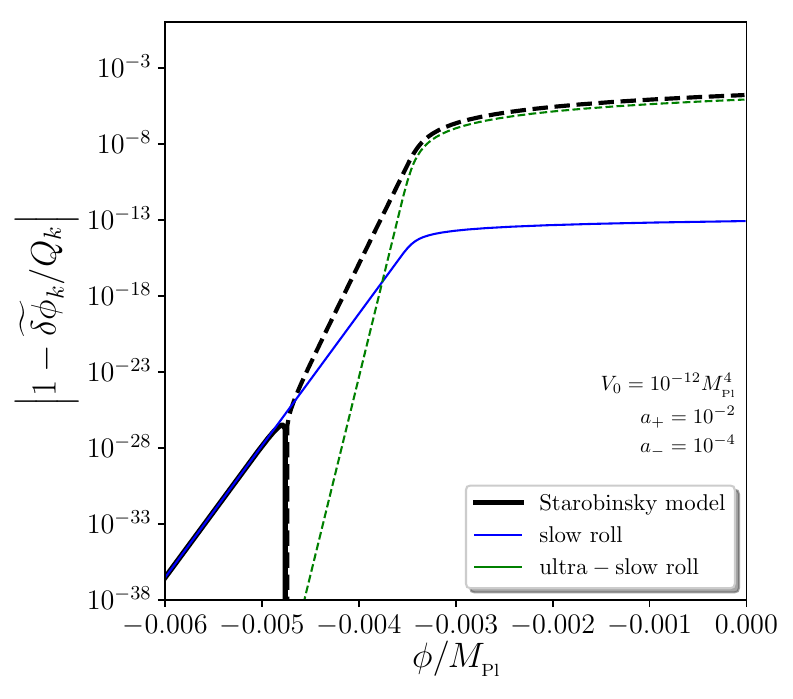}
    \caption{Fractional gauge correction to the field perturbation in the uniform-$N$ gauge in the Starobinsky model, for a mode such that $k/aH=10^{-2}$ at the transition time $t=0$. The black line corresponds to the full result~(\ref{eq:Staro:gaugeCorr}), the blue line stands for the slow-roll result~\eqref{eq:GaugeCorr:SR}, and the green line is the ultra-slow-roll result~\eqref{eq:deltaphi:USRtransform}. Solid lines are such that $1-\widetilde{\delta\phi}_k/Q_k>0$ and dashed lines are such that $1-\widetilde{\delta\phi}_k/Q_k<0$. }
    \label{fig:Staro:GaugeCorrection}
\end{figure}
During the USR phase the Hubble parameter can be taken as approximately constant, $H\simeq H_0= \sqrt{V_0/(3\Mp^2)}$; the consistency of that assumption will be checked below. The Klein--Gordon equation~(\ref{eq:kleingordon}) then becomes $\ddot{\phi} + 3H_{0}\dot{\phi} + V_0 a_-/\Mp = 0$, and can be solved to give
\bea
\label{eq:phi:t:starobinsky}
\frac{\phi(t)}{\Mp}=\frac{a_+-a_-}{3}\left(\ee^{-3H_0t}-1\right)-a_- H_0 t\, ,
\eea
where we choose $t=0$ to denote the time when $\phi=0$, and the initial velocity is set such that its value at the transition point is given by its slow-roll counterpart in the $\phi>0$ branch of the potential, \ie $\dot{\phi}(\phi=0^-)=\dot{\phi}(\phi=0^+)=-H_0 a_+$. The acceleration parameter defined in \Eq{eq:def:f} is then given by
\bea 
\label{eq:Staro:f(t)}
f(t) =  1 - \frac{a_-}{a_-+(a_+-a_-)\ee^{-3H_0t}} \, .
\eea 
At the transition time, it reads $f(t=0)= 1-\frac{a_-}{a_+}$, so if $a_-/a_+\ll 1$, $f\simeq 1$ and ultra-slow roll takes place. At late time, however, $f$ is damped so that the system relaxes back to a phase of slow-roll inflation. Note that the solution~\eqref{eq:phi:t:starobinsky} can be inverted, 
\bea 
\label{eq:Staro:traj:inverted}
H_0 t(\phi) = \frac{1}{3}\left(1-\frac{a_+}{a_-}\right) - \frac{\phi}{\Mp a_-} + \frac{1}{3}W_{0}\left[\frac{a_+-a_-}{a_-}\exp\left(
\frac{a_+}{a_-}-1+3 \frac{\phi}{\Mp a_-}\right)\right] \, ,
\eea 
where $W_{0}(x)$ is the $0$-branch of the Lambert function, which leads to the phase-space trajectory
\bea 
\label{eq:Staro:phidot(phi)}
\dot{\phi}(\phi) = -\frac{\Mp }{H} H_0^2 a_- \left\lbrace 1+W_{0}\left[\frac{a_+-a_-}{a_-}\exp\left(\frac{a_+}{a_-}-1+3 \frac{\phi}{\Mp a_-}\right)\right] \right\rbrace \, .
\eea 
In the denominator of the first term in the right-hand side, $H$ is left to vary~\cite{Martin:2011sn}, in such a way that at late time, \ie when $\phi$ goes to $-\infty$, one recovers the slow-roll result $\dot{\phi}=-\Mp H_0^2 a_-/H$. Plugging \Eq{eq:Staro:traj:inverted} into \Eq{eq:Staro:f(t)} also leads to 
\bea \label{eq:f:phi:starobinsky}
f(\phi) = 1 - \frac{1}{1+W_{0}\left[\frac{a_+-a_-}{a_-}\exp\left( 
\frac{a_+}{a_-}-1+3 \frac{\phi}{\Mp a_-}\right)\right]} \, ,
\eea 
which perfectly reproduces the numerical solution of the Klein--Gordon equation in \Fig{fig:Staro}. One can check that $f$ starts from a value close to one at early time and approaches zero at late time. If one expands \Eq{eq:f:phi:starobinsky} around $\phi=0$, one obtains
\bea \label{eq:f:phi:starobinsky:appr}
f\simeq 1-\frac{a_-}{a_++3\frac{\phi}{\Mp}}\, ,
\eea
which matches \Eq{eq:Staro:delta:appr}. This approximation is also shown in \Fig{fig:Staro}, with the black dashed line. 

From \Eq{eq:f:phi:starobinsky}, the transition time between USR and $\mathrm{SR}_{-}$, defined as the time when $f=1/2$, is found to be
\bea 
t_{\mathrm{USR}\to\mathrm{SR}} = \frac{1}{3 H_0}\ln\left(\frac{a_+-a_-}{a_-}\right) \, ,
\eea 
which is consistent with \Eq{eq:NUSR:Staro}. Making use of \Eq{eq:phi:t:starobinsky}, the field value at which this happens is given by
\bea 
\phi_{\mathrm{USR}\to\mathrm{SR}} = -\frac{\Mp }{3}\left[ a_+ - 2 a_-+a_-\ln\left(\frac{a_+-a_-}{a_-}\right)\right]
 \simeq -\frac{a_+}{3}\Mp \, ,
\eea 
where the last expression is derived in the limit $a_-/a_+\ll 1$ and agrees with \Eq{eq:phiUSR:SR:Staro}. This allows us to test the assumption made above that the potential, hence the Hubble parameter, does not vary much during the USR phase. The relative shift in the potential value between $\phi=0$ and $\phi_{\mathrm{USR}\to\mathrm{SR}}$ is indeed given by
\bea 
\frac{\Delta V}{V} = \frac{a_-(a_+-a_-)}{3} 
\ll 1 \, ,
\eea
which justifies the above assumption.

Let us now calculate the gauge transformation from the spatially-flat to uniform-$N$ gauge in this model. As explained above, combining \Eq{eq:Staro:phidot(phi)} and~(\ref{eq:phi:t:starobinsky}) leads to
\bea
\dot{\phi}(t) = \frac{H_0^2\Mp}{H}\left[ (a_--a_+)\ee^{-3H_{0}t} -  a_- \right] ,
\eea 
that allows us to both describe the USR and the $\mathrm{SR}_-$ phases, as well as the transition between the two. Making use of the relation $\epsilon_{1}=\dot{\phi}^2/(2\Mp^2H^2)$, one obtains
\bea 
\epsilon_{1}(t) &=& \dfrac{1}{2}\left(\frac{H_0}{H}\right)^4\left[  a_- - (a_--a_+)\ee^{-3H_{0}t}\right]^2 \\
\epsilon_{2}(t) &= &-\dfrac{6(a_--a_+)\ee^{-3H_0t}}{(a_--a_+)\ee^{-3H_0t}-a_-} + 4\epsilon_{1}(t) \, .
\eea 
One can check that, at late times, one recovers $\epsilon_2=4\epsilon_1$, which is indeed satisfied in slow roll for linear potentials, see \Eqs{eq:eps1:SR} and~(\ref{eq:f:SR}).

Since $\eta=0$ in this model, the fact that $\epsilon_1$ remains small implies that \Eq{eq:z''overz:general} is close to its de-Sitter limit. Moreover, one can check that, at early times, the term $Q'_k/Q_k$ in \Eq{eq:sourcefunction:general} provides a subdominant contribution, hence it is sufficient to evaluate $Q'_k/Q_k$ at late time and use the result of \Eq{eq:Q'overQ:SR}, $Q'_k/Q_k\simeq - \epsilon_{2*}/(2\eta)+k^2\eta = a_-^2/\eta+k^2\eta$. One then obtains
\bea 
S &=& \frac{i  H_0}{\Mp}\frac{\mathrm{sign}\left(\dot{\phi}\right)}{(2k)^{\frac{3}{2}}\eta}\Bigg\{ 3(a_--a_+)\ee^{-3H_{0}t}\left(1+\frac{a_-^2}{3}\right)
 -  a_-^3 \nonumber \\
& &\hspace{1cm} 
 + \left[{a_-+(a_+-a_-)\ee^{-3H_{0}t}}\right]^3 - k^2\eta^2\left[ \left(a_--a_+\right)\ee^{-3H_{0}t} - a_-\right]\Bigg\} \, .
\eea 
From \Eq{eq:alphaintegral1:general}, we then find the gauge transformation parameter to be
\bea \label{eq:gaugetransformation:starobinsky}
\alpha &\simeq & \frac{-i\eta H_0}{3(2k)^{\frac{3}{2}}\Mp}\mathrm{sign}\left(\dot{\phi}\right)
{\Bigg[}  
\frac{(k\eta)^2}{2}a_- 
+ \left(a_--a_+\right)\ee^{-3H_0t}\left(1+\frac{a_-^3}{2}\right) \nonumber  \\
& &\hspace{5mm} 
+ a_-^2\left(a_+-a_-\right)\ee^{-3H_0t} + \frac{a_-\left(a_+-a_-\right)^2}{2}\ee^{-6H_0t} + \frac{\left(a_+-a_-\right)^3}{9}\ee^{-9H_0t} {\Bigg]}  \, ,
\eea 
where only the $(k\eta)^2$-suppressed term that becomes dominant at late times has been kept, \ie there are other $(k\eta)^2$ terms that have been dropped for consistency since they always provide sub-dominant contributions. One can check that at early time, \ie when $t\to 0$, the ultra-slow-roll expression~(\ref{eq:alpha:USR:nu=3/2}) is recovered if $a_-/a_+\ll 1$, while at late time, \ie when $t \to \infty$, the slow-roll expression~(\ref{eq:alphaLO:SR}) is recovered. This gives rise to the gauge correction
\bea
\label{eq:Staro:gaugeCorr}
\frac{\widetilde{\delta\phi}_k}{Q_k}  \kern-0.1em &=& \kern-0.1em 1 \kern-0.2em + \kern-0.2em \frac{1}{6}\left(\frac{H_0}{H}\right)^3\left[\left(a_--a_+\right)\ee^{-3H_0 t}-a_-\right]\kern-0.2em
{\Bigg[}  
\frac{(k\eta)^2}{2}a_- 
+ \left(a_--a_+\right)\ee^{-3H_0t}\left(1+\frac{a_-^2}{3}\right) 
\nonumber \\ & &
+ a_-^2\left(a_+-a_-\right)\ee^{-3H_0t} + \frac{a_-\left(a_+-a_-\right)^2}{2}\ee^{-6H_0t} + \frac{\left(a_+-a_-\right)^3}{9}\ee^{-9H_0t} {\Bigg]}\, ,
\eea
which is displayed in the right panel of \Fig{fig:Staro:GaugeCorrection} for a mode such that $k/aH=10^{-2}$ at the transition time $t=0$. Right after the transition point, one can check that the ultra-slow-roll result~\eqref{eq:deltaphi:USRtransform} is recovered (the slight discrepancy is due to the finite value of $a_-/a_+$, \ie the finite initial value of $\delta$, we work with in \Fig{fig:Staro:GaugeCorrection}), and at late time, the slow-roll result~\eqref{eq:GaugeCorr:SR} is obtained. In between, the gauge correction to the noise correlators remains tiny and can therefore be safely neglected.

\subsection{Stochastic ultra-slow-roll inflation}
\label{sec:USR:stochastic}
As we have explained above, PBHs require large quantum fluctuations to be produced, which in turn require a very flat potential during inflation, where deviations from slow roll are likely to be encountered. This is for instance the case in inflection point models of inflation~\cite{Garcia-Bellido:2017mdw,Germani:2017bcs,Ezquiaga:2018gbw,Biagetti:2018pjj}. Moreover, in \Secs{sec:PBHs} and~\ref{sec:tail:expansion}, we have shown that for models featuring an inflection point, or if inflation proceeds towards an uplifted local minimum of the potential, PBHs are overproduced unless slow roll is violated. In inflection-point models, if the slow-roll conditions are violated as one approaches the inflection point, inflation usually proceeds along the ultra-slow-roll regime, the classical stability of which has been studied in \Sec{sec:USR:stability}, and in \Sec{sec:USR:GaugeCorrections} we have shown that the gauge corrections of \Sec{sec:uniformexpansion} are negligible in ultra slow roll (and in fact, even more so than in slow roll).

We are thus now in a position where we can apply the stochastic-$\delta N$ programme to the ultra-slow-roll setup. In practice, we consider the situation where ultra slow roll is exact at the classical level, \ie the potential is exactly flat between $\phi=0$ (which we set without loss of generality) and $\phi=\dphiwell$. The situation is therefore exactly the same as the one investigated in \Sec{sec:StochasticLimit}, see \Fig{fig:sketch2}, except that we now account for the possible classical velocity of the field, inherited from previous dynamics, when it enters the flat region of the potential, at $\phi=\dphiwell$. Our goal is to determine how this inherited velocity changes the results of \Secs{sec:StochasticLimit} and~\ref{sec:flat_potential}. As before, we assume that the potential becomes steeper at $\phi\geq\dphiwell$ such that the classical drift prevents the inflaton from exploring that region of the potential once fallen in the well, which is modelled by implementing a reflective wall at $\phi=\dphiwell$. Similarly, we consider the case where the classical drift dominates again the inflaton dynamics at $\phi\leq 0$, so an absorbing wall can be placed at $\phi=0$. Finally, we assume that the initial velocity of the inflaton is smaller than the Hubble scale in Planckian units, $\dot\phi\ll \Mp$, such that $\epsilon_1\ll 1$ and the dynamics of expansion is close to de-Sitter (in practice, since $\dot{\phi}\propto\ee^{-3 N}$, see \Eq{eq:USR:phidot:N}, this regime is always quickly reached and $3H^2\Mp^2$ is soon dominated by the potential energy).

The correlation matrix of the noise can be calculated as explained in \Sec{ssec:noise}. Working with the number of \efolds~as the time variable, combining \Eqs{eq:deltaphi:USRtransform},~\eqref{eq:MS:SRsolution} and~\eqref{eq:v:nu=3/2} leads to $\widetilde{\delta\phi_k} \simeq i H/( k \sqrt{2k})$, and \Eq{eq:GaugeTransf:pi} to $\dd \widetilde{\delta\phi_k}/(\dd N) \simeq -1/(a \sqrt{2 k})$, on super-Hubble scales. From \Eq{eq:calP:def}, the power spectra thus read $\mathcal{P}_{\delta\phi,\delta\phi} = H^2/(2\pi)^2$, $\mathcal{P}_{\delta\phi,\dd\delta\phi/\dd N} = \mathcal{P}_{\dd\delta\phi/\dd N,\delta\phi}^*=-i H^2/(2\pi)^2 k/(aH)$ and $\mathcal{P}_{\dd\delta\phi/\dd N,\dd\delta\phi/\dd N} = H^2/(2\pi)^2 k^2/(aH)^2 $. Making use of the coarse-graining scale~\eqref{eq:ksigma}, the diffusion matrix $\boldsymbol{D}$, which, we recall, is the symmetric part of the correlation matrix $\boldsymbol{\Xi}$ defined in \Eq{eq:noisecorrel_Pk}, see the discussion above \Eq{eq:fokker}, is given by $D_{\phi,\phi}=H^2/(2\pi)^2$, $D_{\phi,\dd\phi/\dd N}=\order{\sigma^2}$ and $D_{\dd\phi/\dd N,\dd\phi/\dd N}=\sigma^2 H^2/(2\pi)^2 $. The situation is therefore similar as for slow roll: at leading order in $\sigma$, the noise in the velocity direction $\gamma=\dd\phi/\dd N$ can be neglected (we use the same notation for $\gamma$ as in \Sec{sec:Stochastic:Non:test:fields}), and the Langevin equations~(\ref{eq:conjmomentum:langevin}) and~\eqref{eq:KG:efolds:langevin} read
\begin{align}
 \label{eq:conjmomentum:langevin:USR}
\frac{\dd {{\phi}}}{\dd N} &= {{\gamma}} + \frac{H}{2\pi} {\xi}(N) \, ,\\
\frac{\dd {{\gamma}}}{\dd N} &= -\left(3-\frac{{\gamma}^2}{2\Mp^2}\right){{\gamma}}  \, ,
\label{eq:KG:efolds:langevin:USR}
\end{align}
where $\xi$ is a white Gaussian noise with vanishing mean and unit variance, $H$ is a function of $\phi$ and $\gamma$ through \Eq{eq:friedmann}, and the bars used in \Sec{sec:Stochastic:Non:test:fields} have been dropped to lighten the notation of the coarse-grained fields. The main simplification compared to the most generic case is that \Eq{eq:KG:efolds:langevin:USR} is deterministic, since it does not contain a noise term. In fact, it can even be integrated analytically, and one obtains
\bea
\label{eq:USR:gamma(N):sol}
\gamma(N)= \sqrt{6}\Mp\left[ 1+\left(\frac{6\Mp^2}{\gamma_\uin^2}-1\right) \ee^{-6\left(N-N_\uin\right)} \right]^{-1/2}\, .
\eea 
In practice, as mentioned above, we restrict the analysis to cases where the dynamics of space-time expansion is close to de-Sitter, so $\gamma^2/\Mp^2 = 2\epsilon_1\ll 1 $ (this is also the regime in which the correlation matrix of the noise was evaluated anyway). In this limit, \Eq{eq:USR:gamma(N):sol} boils down to $\gamma(N)=\gamma_\uin \ee^{-3(N-N_\uin)}$. If one further introduces $x=\phi/\dphiwell$ and $\mu=\dphiwell^2/(v_0\Mp^2)$ as in \Sec{sec:StochasticLimit}, see \Eq{eq:def:mu}, the Langevin equations~\eqref{eq:conjmomentum:langevin:USR}-\eqref{eq:KG:efolds:langevin:USR} become
\begin{align}
 \label{eq:conjmomentum:langevin:USR:rescaled}
\frac{\dd x}{\dd N} &= -3y + \frac{\sqrt{2}}{\mu} {\xi}(N) \, ,\\
\frac{\dd y}{\dd N} &= -3y  \, ,
\label{eq:KG:efolds:langevin:USR:rescaled}
\end{align}
where we have also defined $y\equiv -\gamma/(3\dphiwell)$ (the minus sign is such that $y>0$ in the present setup). 

Let us first note that, in the limit where $y_\uin\to 0$, \ie when the system enters the well at $\phi=\dphiwell$ with no classical velocity, one recovers the Langevin equation that was solved in \Sec{sec:StochasticLimit}, \ie the slow-roll Langevin equation. This confirms that, as announced before and contrary to what one may have been concerned with, using the slow-roll Langevin equation on an exactly flat potential is not inconsistent, and should rather be viewed as working in the small velocity limit of the full ultra-slow-roll dynamics. At the classical level, this limit is ill-defined since the field comes to a complete rest in the absence of classical velocity, but it becomes perfectly regular once stochastic effects are incorporated. This will be confirmed below, where we will show that for all quantities of interest, the results of \Secs{sec:StochasticLimit} and~\ref{sec:flat_potential} are recovered in the limit $y_\uin\to 0$, which further establishes the validity of the treatment of these sections.

One can also see that the amplitude of the stochastic effects is still controlled by the parameter $\mu$. The field coordinate $x$ starts at $x_\uin=1$ and varies between $0$ and $1$ afterwards, until it reaches the absorbing boundary at $x=0$, and the initial rescaled velocity $y_\uin$ is the second parameter of the model. The value $y_\uin =x_\uin$ plays a critical role (hereafter, $x_\uin$ is left unspecified to keep our calculation generic, and we will set $x_\uin =1$ only in practical applications), since it corresponds to the minimum initial velocity required to cross over the well by means of the classical drift only. Indeed, in the absence of boundary conditions, the solution to \Eqs{eq:conjmomentum:langevin:USR:rescaled}-\eqref{eq:KG:efolds:langevin:USR:rescaled} can be written as
\bea
x_{\mathrm{no\ boundaries}}(N)=x_\uin - y_\uin\left[1-\ee^{-3\left(N-N_\uin\right)}\right]+\frac{\sqrt{2}}{\mu}\int_{N_\uin}^N \xi ({\tilde{N}})\dd \tilde{N}\, .
\eea
This implies that the mean field value coincides with the classical trajectory, $\left\langle x(N) \right\rangle = x_\ucl(N)$, and that its variance is given by $(\Delta x )^2 \equiv \left\langle x^2(N) \right\rangle -\left\langle x(N) \right\rangle^2=2(N-N_\uin)/\mu^2 $. If the stochastic noise is discarded ($\xi=0$), when $N\to\infty$, $x$ asymptotes a positive value if $y_\uin\leq x_\uin$, and the system never escapes the flat well, while for $y_\uin>x_\uin$, $x$ exits the well after $N_\ucl=-\ln(1-x_\uin/y_\uin)/3$ \efolds. We therefore expect that, when $y_\uin< x_\uin$, stochastic effects will be important (since they are necessary to exit the flat well). When $y_\uin \gg x_\uin$, the number of \efolds~required to exit the well classically becomes $N_\ucl \simeq x_\uin/(3 y_\uin)$. At that point, the typical spread of the field coordinate $x$ is, according to the above, $\Delta x \simeq  \sqrt{2 N_\ucl}/\mu = \sqrt{2 x_\uin}/(\sqrt{3y_\uin} \mu)$.


Let us now apply the stochastic-$\delta N$ program to the problem at hand. As explained in \Sec{sec:FP}, the Langevin equations~\eqref{eq:conjmomentum:langevin:USR:rescaled}-\eqref{eq:KG:efolds:langevin:USR:rescaled} give rise to the Fokker-Planck equation $\partial P(x,y,N)/ \partial N = \mathcal{L}_\mathrm{FP} P(x,y,N)$, where the Fokker-Planck operator is given by
\bea
\label{eq:USR:FokkerPlanckOperator}
\mathcal{L}_\mathrm{FP} = \frac{1}{\mu^2}\frac{\partial^2}{\partial x^2} + 3y \left(\frac{\partial}{\partial x} + \frac{\partial}{\partial y}\right) +3\, ,
\eea
see \Eq{eq:Fokker:Planck:operator}. Here, since the amplitude of the noise term, $\sqrt{2}/\mu$, does not depend on the field coordinates $x$ and $y$, the resulting Fokker-Planck equation is independent of the prescription parameter $\alpha$. The characteristic function introduced in \Sec{sec:FPT:full:PDF} obeys the partial differential equation~\eqref{eq:ODE:chi}, which, upon using \Eq{eq:USR:FokkerPlanckOperator}, reads
\bea
\label{eq:USR:chi:PDE}
\left[\frac{1}{\mu^2}\frac{\partial^2}{\partial x^2}-3y\left(\frac{\partial}{\partial x} + \frac{\partial}{\partial y}\right) + i t\right]\chi_\N\left(x,y,t\right)=0\, ,
\eea
with the boundary conditions $\partial \chi_\N/\partial x = 0$ when $x=1$, and $\chi_\N =1$ when $x=0$. Let us now study the two regimes arising from the above condition, namely the small velocity limit $y_\uin \ll x_\uin$ where stochastic effects are expected to dominate, and the large velocity limit $y_\uin\gg x_\uin$ where one expects to recover the classical behaviour.
\subsubsection{Classical limit}
\label{sec:USR:Sto:Class:Lim}
At leading order in the classical limit, the first term in \Eq{eq:USR:chi:PDE} can be neglected, and one obtains a first-order partial differential equation. Such equations can be solved with the method of characteristics as follows. Let us parametrise a line in phase space by the functions $x(u)$ and $y(u)$, where $u$ is the affine parameter along that line. Setting $t$ to a fixed value, we study the value of the characteristic function along the line, which evolves according to
\bea
\frac{\dd}{\dd u} \chi_\N\left[x(u),y(u),t\right] = \frac{\partial \chi_\N}{\partial x}\left[x(u),y(u),t\right] x'(u)+\frac{\partial }{\partial y}\chi_\N\left[x(u),y(u),t\right] y'(u)\, .
\eea
From \Eq{eq:USR:chi:PDE}, one has $(\partial/\partial x + \partial/\partial y)\chi_\N=it/(3 y)\chi_\N$, so it is convenient to choose $x'(u)=y'(u)=1$, \ie $x(u)=u$ and $y(u)=y_0+u$ (where we absorb in the definition of $u$ the integration constant coming from the equation $x'(u)=1$). This leads to
\bea
\frac{\dd}{\dd u} \chi_\N\left[x(u),y(u),t\right] = \frac{it}{3y(u)}\chi_\N\left[x(u),y(u),t\right]
= \frac{it}{3\left(y_0+u\right)}\chi_\N\left[x(u),y(u),t\right]\, .
\eea
This is an ordinary first-order differential equation in the $u$ variable, that can be easily solved according to
\bea
 \chi_\N\left(u,y_0+u,t\right)  = \chi_0\left(y_0\right) \left(1+\frac{u}{y_0}\right)^{it/3}\, .
\eea
When $u$ and $y_0$ vary, the whole plane $(x=u,y=y_0+u)$ is being described, so one can write
\bea
\chi_\N(x,y,t)= \chi_0\left(y-x\right) \left(\frac{y}{y-x}\right)^{it/3}\, ,
\eea
where only the function $\chi_0(z)$ remains to be determined. This can be done by using the boundary condition $\chi_\N(0,y,t)=1$, which gives rise to $\chi_0(z) = 1$, and one obtains
\bea
\label{USR:chi:class:lo}
\chi_\N(x,y,t)= \left(\frac{y}{y-x}\right)^{it/3}\, .
\eea
The mean number of \efolds~can then be evaluated by means of \Eq{eq:meanN:chi}, and one finds
\bea
\label{eq:USR:sto:Nclass:FlatWell}
\left\langle \N \right\rangle_\ucl = -\frac{1}{3}\ln\left(1-\frac{x_\uin}{y_\uin}\right)\, ,
\eea
which matches the expression given above for $N_\ucl$. The full PDF of the number of \efolds~can be obtained by Fourier transforming \Eq{USR:chi:class:lo} according to \Eq{eq:PDF:chi}, which gives rise to
\bea
P\left(\N; x,y\right)= \delta\left(\N-\left\langle \N \right\rangle_\ucl\right)\, .
\eea
This is expected since stochastic diffusion has been entirely neglected at that order, so all realisations of the stochastic process realise the same, deterministic number of \efolds.

At next-to-leading order, stochastic corrections can be incorporated by evaluating the first term of \Eq{eq:USR:chi:PDE} with the leading order solution,
\bea
\left[-3y\left(\frac{\partial}{\partial x} + \frac{\partial}{\partial y}\right) + i t\right]\chi_\N\left(x,y,t\right)=-\frac{1}{\mu^2}\frac{\partial^2}{\partial x^2} \left(\frac{y}{y-x}\right)^{it/3}\, .
\eea
This first-order partial differential equation can be solved along the same characteristics $x(u)=u$ and $y(u)=y_0+u$ as before, and one finds
\bea
\frac{\dd}{\dd u} \chi_\N\left[x(u),y(u),t\right] 
= \frac{it}{3\left(y_0+u\right)}\chi_\N\left[x(u),y(u),t\right]
+\frac{1}{3\mu^2y_0^{2+it/3}}  \left(y_0+u\right)^{it/3-1} \frac{i t}{3}\left(\frac{i t}{3}+1\right)\, .
\eea
One again obtains an ordinary, first-order differential equation, the solution of which is given by 
\bea
\chi_\N\left[x(u),y(u),t\right]= \left(1+\frac{u}{y_0}\right)^{it/3}\left[\chi_0(y_0)+\frac{1}{3 \mu^2 y_0^2}\frac{i t}{3}\left(\frac{i t}{3}+1\right) \ln\left(1+\frac{u}{y_0}\right)\right]\, ,
\eea
where $\chi_0(y_0)$ is an integration constant. When $u$ and $y_0$ vary, the whole plane $(x,y)$ is being described, and one can write
\bea
\chi_\N\left(x,y,t\right)= \left(\frac{y}{y-x}\right)^{it/3}\left[\chi_0(y-x)+\frac{1}{3 \mu^2 \left(y-x\right)^2}\frac{i t}{3}\left(\frac{i t}{3}+1\right) \ln\left(\frac{y}{y-x}\right)\right]\, .
\eea
As before, the function $\chi_0(z)$ can be determined using the boundary condition $\chi_\N(0,y,t)=1$, which gives rise to $\chi_0(z) = 1$, and one obtains
\bea
\label{USR:chi:class:lo}
\chi_\N\left(x,y,t\right)= \left(\frac{y}{y-x}\right)^{it/3}\left[1+\frac{1}{3 \mu^2 \left(y-x\right)^2}\frac{i t}{3}\left(\frac{i t}{3}+1\right) \ln\left(\frac{y}{y-x}\right)\right]\, .
\eea
Let us note that the same procedure can be iterated again at higher orders, using the expression of the characteristic function at order $n$ to evaluate the first term of \Eq{eq:USR:chi:PDE} and solve for the characteristic function at order $n+1$. 

The mean number of \efolds~can again be evaluated using \Eq{eq:meanN:chi}, and one finds
\bea
\label{eq:USR:meanN:classical:nlo}
\left\langle \N \right\rangle_\ucl = -\frac{1}{3}\ln\left(1-\frac{x_\uin}{y_\uin}\right)\left[1+\frac{1}{3\mu^2\left(x_\uin-y_\uin\right)^2}\right]\, ,
\eea
where one can see that stochastic effects tend to increase the number of \efolds, at least in the classical regime. This result is consistent with \Refa{Firouzjahi:2018vet}, see Eq.~(4.15) of that reference, although it is derived here using a different method. The relative correction to the fully classical result scales as $\mu^{-2}(x_\uin-y_\uin)^{-2}$, so one can see that when $\mu\gg 1$, the stochastic correction to the mean number of \efolds~spent in the well (\ie starting from $x_\uin=1$) is always small in the large velocity limit $y_\uin\gg x_\uin$. When $\mu\ll 1$, it takes a very large initial velocity, $y_\uin\gg 1/\mu$, to suppress stochastic corrections. In summary, stochastic corrections are small as long as
\bea
\label{eq:USR:sto:Class:Cond}
y_\uin\gg\max\left(1,\frac{1}{\mu}\right).
\eea

The second moment of the number of \efolds~can also be evaluated from the characteristic function by noting that 
\bea
\left\langle \N^2 \right\rangle_\ucl &=& - \left.\frac{\partial^2}{\partial t^2}\chi_\N(x,y,t)\right\vert_{t=0}\\
&=&\frac{1}{27 \mu^2 \left(x_\uin-y_\uin\right)^2}\ln\left(1-\frac{x_\uin}{y_\uin}\right)\left\lbrace \left[2+3\mu^2 \left(x_\uin-y_\uin\right)^2\right]\ln\left(1-\frac{x_\uin}{y_\uin}\right)-2   \right\rbrace
\, ,
\eea
see \Eq{eq:characteristicFunction:def}. From here, the variance in the number of \efolds, $\langle \delta \N^2\rangle = \langle \N^2\rangle - \langle \N \rangle^2$, can be computed, 
\bea
\label{eq:USR:deltaN2:classical:nlo}
\langle \delta \N^2 \rangle = -\frac{\ln\left(1-\frac{x_\uin}{y_\uin}\right)}{81\mu^4\left(x_\uin-y_\uin\right)^4}\left[6\mu^2\left(x_\uin-y_\uin\right)^2+\ln\left(1-\frac{x_\uin}{y_\uin}\right)\right], 
\eea
and the power spectrum can be obtained from \Eq{eq:Pzeta:stochaDeltaN}. Let us recall that the derivatives in \Eq{eq:Pzeta:stochaDeltaN} need to be evaluated along a reference phase-space trajectory. In the classical limit, at the order at which the calculation is performed here, this reference trajectory is nothing but the classical trajectory, $x=x_\uin - y_\uin\left[1-\ee^{-3(N-N_\uin)}\right]$ and $y=y_\uin\ee^{-3(N-N_\uin)}$, and this leads to
\bea
\calP_{\zeta,\, \ucl} = 
\frac{6\mu^2\left(x_\uin-y_\uin\right)^2+2\ln\left(1-\frac{x_\uin}{y_\uin}\right)}{9\mu^2\left(x_\uin-y_\uin\right)^2\left[1+3\mu^2\left(x_\uin-y_\uin\right)^2\right]}
\simeq \frac{2}{9\mu^2 y_\uin^2}\, ,
\eea
where in the second expression, we have taken the large velocity limit $y_\uin\gg x_\uin$.
Let us note that the amplitude of the power spectrum is of the same order as the relative correction to the mean number of \efolds, see \Eq{eq:USR:meanN:classical:nlo}. In this context, the ``classical'' regime is therefore one where the classical power spectrum remains small. One can also check that the same result would be obtained at leading order with a direct classical $\delta N$ calculation, $\calP_\zeta = (\sqrt{2}/\mu)^2 (\partial N_\ucl/\partial x)^2$, which is expected. 
\subsubsection{Stochastic limit}
\label{sec:USR:Stochastic:Limit}
\begin{figure}
    \centering
    \includegraphics[width=0.50\textwidth]{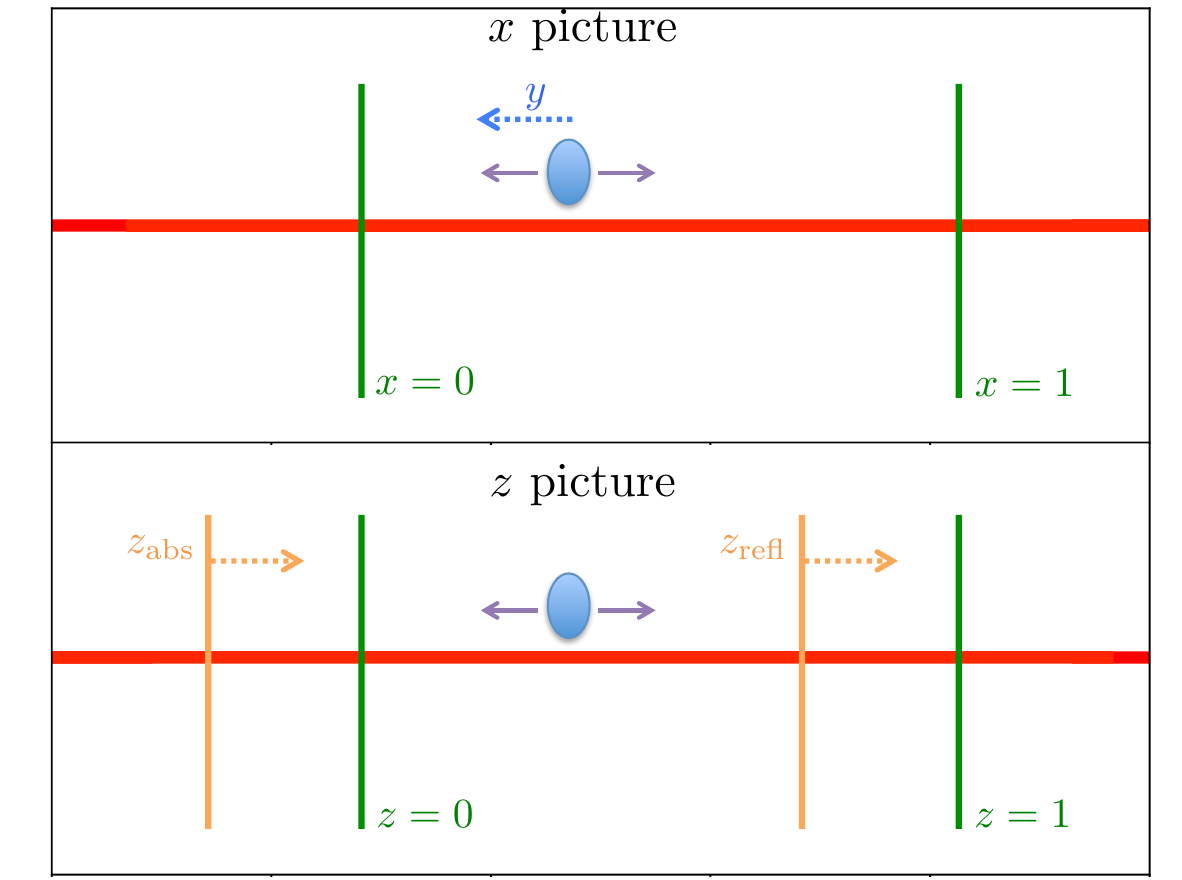}
    \caption{Sketch of the stochastic system solved in \Sec{sec:USR:stochastic}. The $x$ coordinate undergoes stochastic fluctuations added to a decaying classical velocity, in a fixed potential well. Equivalently, the $z$ coordinate is subject to stochastic fluctuations only (no classical velocity), but in a well with moving barriers.}
    \label{fig:USR:MovingBarrier:Sketch}
\end{figure}

Let us now investigate the opposite limit, namely the regime where the initial velocity is small, and stochastic effects are expected to play an important role. As already noted, at leading order in that limit, the initial velocity can be entirely neglected and the system becomes identical to the one solved in \Sec{sec:StochasticLimit}. In that case, the first passage time problem has already been solved, and the distribution function of the number of \efolds~is given by \Eq{eq:stocha:HeatMethod:PDF:expansion}. In this section, we aim at computing the first corrections in the initial velocity to this result, using perturbative techniques.

Before doing so, let us explain why the decay rate on the tail of the PDF of the number of \efolds, \ie the quantity noted $\Lambda_0$ in \Sec{sec:tail:expansion}, should be independent of the initial velocity. To this end, let us note that the stochastic system~\eqref{eq:conjmomentum:langevin:USR:rescaled}-\eqref{eq:KG:efolds:langevin:USR:rescaled} can be reformulated in terms of a pure diffusion problem (\ie without classical velocity), but with time-dependent boundary conditions. This can be done by introducing the variable $z=x-y$, which evolves according to
\bea
\frac{\dd z}{\dd N} = \frac{\sqrt{2}}{\mu} \xi(N)
\eea
in a well defined by an absorbing boundary condition at $z_{\mathrm{abs}}(N)=-y_\uin\ee^{-3(N-N_\uin)}$ and a reflective boundary condition at $z_{\mathrm{refl}}(N)=1-y_\uin\ee^{-3(N-N_\uin)}$. The width of the well, $z_{\mathrm{refl}} - z_{\mathrm{abs}}$, remains constant, but its overall location evolves with time. The situation is depicted in \Fig{fig:USR:MovingBarrier:Sketch}. At late time, the two boundaries approach the asymptotic values $z_{\mathrm{abs}}=0$ and $z_{\mathrm{refl}}=1$ and come to a rest. The realisations of the stochastic process that give rise to the tail of the PDF exit the well at late time, precisely when the boundaries have stopped moving, and where the problem for the $z$ variable becomes the same as in the absence of classical velocity. The asymptotic shape of the tail of the PDF should therefore not depend on the presence of initial velocity. In fact, this is merely an illustration of the more generic argument developed in \Sec{sec:pole:chi}, where it was shown that the decay rates $\Lambda_n$ are universal in a given inflationary potential, and cannot depend on the initial phase-space coordinates. Obviously, the overall amplitude does, and we now try to determine in which way. 

\begin{figure}
    \centering
    \includegraphics[width=0.49\textwidth]{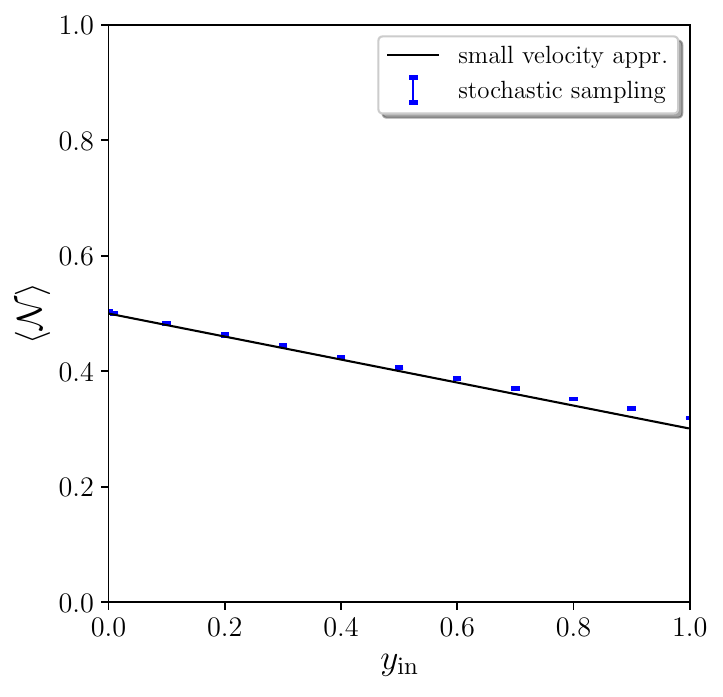}
        \includegraphics[width=0.49\textwidth]{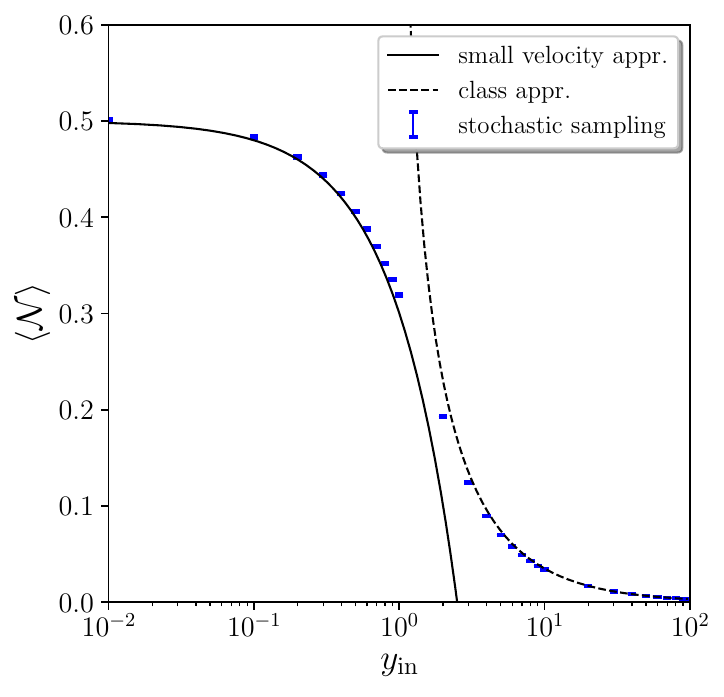}
    \caption{Mean number of \efolds~realised in a flat potential well with $\mu=1$, as a function of the initial field velocity. The blue bars correspond to averages over a large number of numerical realisations of the Langevin equations~\eqref{eq:conjmomentum:langevin:USR:rescaled}-\eqref{eq:KG:efolds:langevin:USR:rescaled}, where the size of the bars is a jackknife resampling estimate for the $2\sigma$-statistical errors (see main text). The solid lines correspond to the small velocity analytical expansion, \Eq{eq:USR:sto:StoLim:meanN}, while the dashed line stands for the classical result~(\ref{eq:USR:sto:Nclass:FlatWell}).}
    \label{fig:USR:meanN}
\end{figure}

Since we want to study the situation where the classical velocity is small, our starting point is to Taylor expand the characteristic function in $y$ at leading order,
\bea
\label{eq:USR:sto:chi:exp:y}
\chi_\N(x,y,t) = \chi_\N(x,0,t) + y f(x,t)\, ,
\eea
where $\chi_\N(x,0,t)$ was already computed in \Eq{eq:chiN:cosh}. Plugging this expansion into \Eq{eq:USR:chi:PDE}, at leading order in $y$, one obtains
\bea
\frac{1}{\mu^2}\frac{\partial f(x,t)}{\partial x^2} + \left(i t-3\right)f(x,t)=3\frac{\partial \chi_\N(x,0,t)}{\partial x}\, .
\eea
This provides an ordinary, linear differential equation for $f(x)$ (where $t$ and $y$ are dummy parameters) that can be solved exactly, and one obtains
\bea
f(x,t) = C_1(t) \ee^{\sqrt{3-it}\mu x}+ C_2(t) \ee^{-\sqrt{3-it}\mu x}-\alpha\sqrt{t}\mu\frac{\sinh\left[\alpha \sqrt{t}\mu (x-1)\right]}{\cosh\left(\alpha \sqrt{t}\mu\right)}\, ,
\eea
where we recall that $\alpha = (i-1)/\sqrt{2}$. The integration constants $C_1$ and $C_2$ can be determined by imposing the boundary conditions given below \Eq{eq:USR:chi:PDE}, which imply that $f(x=0,t)=0$ and $\partial f(x=1,t)/\partial x=0$. This provides a linear system for $C_1$ and $C_2$ that can be readily solved, and one obtains
\bea
f(x,t) & = & \frac{-i \sqrt{t} \mu}{\sqrt{3-i t}\cosh\left(\alpha \sqrt{t}\mu\right)\cosh\left(\sqrt{3-i t}\mu\right)}
\left\lbrace
\sqrt{t}\sinh\left(\sqrt{3-i t}\mu x\right)
\right. \nonumber \\ & & \left.
-\alpha^* \sqrt{3-i t}\sinh\left(\alpha\sqrt{t}\mu\right)\cosh\left[\sqrt{3-i t}\mu \left(x-1\right)\right]
\right. \nonumber \\ & & \left.
-\alpha^* \sqrt{3-i t} \cosh\left(\sqrt{3-i t}\mu\right)\sinh\left[\alpha\sqrt{t}\mu\left(x-1\right)\right]
\right\rbrace  .
\label{eq:USR:stocha:stoLim:f}
\eea
The characteristic function at first order in $y$ is thus given by \Eq{eq:USR:sto:chi:exp:y}, where the two terms on the right-hand side are  given by \Eqs{eq:chiN:cosh} and~\eqref{eq:USR:stocha:stoLim:f} respectively. This allows one to evaluate the mean number of \efolds~by making use of \Eq{eq:meanN:chi}, which gives rise to
\bea
\label{eq:USR:sto:StoLim:meanN}
\left\langle \N \right\rangle \simeq
\frac{\mu^2}{2}\left(2x_\uin-x_\uin^2-2y_\uin+2x_\uin y_\uin \right)
+\frac{\mu y_\uin}{ \cosh\left(\sqrt{3}\mu\right)}
\left\lbrace
\mu\cosh\left[\sqrt{3}\mu\left(x_\uin-1\right)\right] - \frac{1}{\sqrt{3}} \sinh\left(\sqrt{3} x_\uin \mu\right)
\right\rbrace .
\eea
One can check that the relative correction to the result without classical velocity is of order $y_\uin$ if $\mu\gg 1$, and of order $\mu^2 y_\uin $ is $\mu\ll 1$. The present expansion is therefore under control if the condition opposite to \Eq{eq:USR:sto:Class:Cond} is satisfied. This is why we dub the large-velocity and the small-velocity expansions the ``classical'' and ``stochastic'' expansions respectively.

This formula~\eqref{eq:USR:sto:StoLim:meanN} is compared with a numerical solution of the Langevin equations~\eqref{eq:conjmomentum:langevin:USR:rescaled}-\eqref{eq:KG:efolds:langevin:USR:rescaled} in \Fig{fig:USR:meanN}, for $\mu=1$, $x_\uin=1$ and as a function of the initial velocity $y_\uin$. A large number (between $10^6$ and $10^8$, depending on the value of $y_\uin$) of realisations have been simulated numerically, and the ensemble average of the number of \efolds~elapsed in these realisations is displayed with the blue bars. The sizes of the bars correspond to an estimate of the $2\sigma$-statistical error (due to having simulated a finite number of realisations only), which is obtained using the jackknife resampling method. For a sample of $n$ trajectories, according to the central limit theorem, the statistical error scales as $1/\sqrt{n}$, so one can write $\sigma_n = \lambda/\sqrt{n}$. In order to determine $\lambda$, we divide our set of realisations into $n_\mathrm{sub}$ subsamples of size $n/n_\mathrm{sub}$ each. In each subsample, one can compute the mean number of \efolds, and compute the standard deviation $\sigma_{n/n_\mathrm{sub}}$ across the set of values that are thus obtained. This gives rise to $\lambda = \sqrt{n/n_\mathrm{sub}}\sigma_{n/n_\mathrm{sub}}$, and one has $\sigma_n =\sigma_{n_\mathrm{sub}} /\sqrt{n_\mathrm{sub}} $. In practice, we use $n_\mathrm{sub}=100$ and assess statistical error with this formula in \Fig{fig:USR:meanN} and following figures. One can check that, when $y_\uin\ll 1$, \Eq{eq:USR:sto:StoLim:meanN} provides a good approximation to the numerical results. In the opposite limit, $y_\uin\gg 1$, one can also see that the classical formula~\eqref{eq:USR:sto:Nclass:FlatWell}, which is displayed with the black dashed line, is recovered.  
\subsubsection{Primordial black holes}
\begin{figure}
    \centering
    \includegraphics[width=0.49\textwidth]{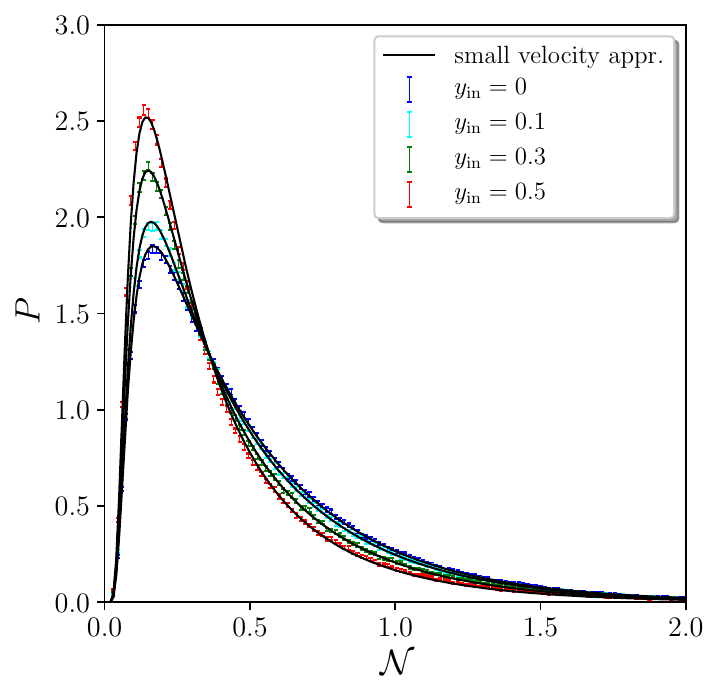}
        \includegraphics[width=0.49\textwidth]{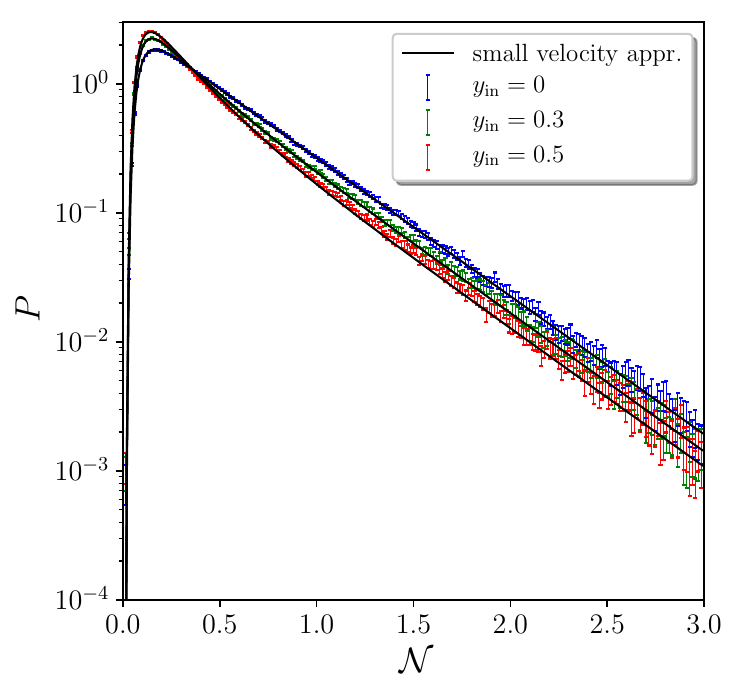}
    \caption{Distribution function of the number of \efolds~realised in a flat potential well with $\mu=1$, for a few values of the initial field velocity $y_{\uin}$, with a linear scale on the vertical axis in the left panel and a logarithmic scale in the right panel (the value $y_\uin=0.1$ is not shown in the right panel for display convenience). The blue bars are reconstructed from a large number of numerical realisations of the Langevin equations~\eqref{eq:conjmomentum:langevin:USR:rescaled}-\eqref{eq:KG:efolds:langevin:USR:rescaled}, with Gaussian kernel density estimation of width $\sigma_\N=0.005$. The size of the bars stand for $2\sigma$ estimates of the statistical error, which are obtained following the jackknife procedure explained around \Fig{fig:USR:meanN}. The black lines correspond to the small velocity analytical formula~(\ref{eq:USR:sto:StoLim:PDF}), which provide good fits to the simulations even for sizeable values of $y_\uin$.}
    \label{fig:USR:Sto:PDF}
\end{figure}
In the small velocity limit, the expression we have obtained for the characteristic function allows one to discuss how the initial velocity affects the tail expansion performed in \Sec{sec:flat_potential}. A first remark is that the poles of the first term in \Eq{eq:USR:sto:chi:exp:y}, $\chi_\N(x,0,t)$, are also poles of the second term, $yf(x,t)$, see \Eq{eq:USR:stocha:stoLim:f}, since both feature $\cosh(\alpha \sqrt{t} \mu)$ in their denominator. There is therefore a first set of eigenvalues $\Lambda_n^{(1)}$ given by \Eq{eq:flat_poles}, namely
\bea 
\label{eq:USR:sto:StoLim:spectrum:1}
\Lambda_n^{(1)}=\frac{\pi^2}{\mu^2}\lp n+\frac{1}{2}\rp^2\,.
\eea
By expanding the characteristic function around the poles $t=-i\Lambda_n^{(1)}$, one obtains the residues $a_n^{(1)}$ defined in \Eq{eq:chi:pole:expansion}, which read
\bea
a_n^{(1)}(x,y) &=& 
 \frac{\pi}{\mu^2}\left(n+\frac{1}{2}\right)
 \left\lbrace 2 \sin \left[ \pi  \left( n+\frac{1}{2}\right) x\right]-\pi  \left(2 n+1\right) y \cos \left[ \pi \left(n+\frac{1}{2}\right) x\right]\right\rbrace
 \nonumber \\ & &
+\frac{ (-1)^n 2 \pi^2  \left( n+\frac{1}{2}\right)^2 y}{\mu^2\sqrt{3 \mu ^2-  \pi^2\left(n+\frac{1}{2} \right)^2}  \cosh\left[\mu  \sqrt{3-\frac{\pi^2\left(n+\frac{1}{2}\right )^2}{\mu ^2}}\right] }
 \Bigg\lbrace(-1)^n  \sqrt{3 \mu ^2- \pi^2\left(  n+\frac{1}{2} \right)^2} 
 \nonumber \\ & &
  \cosh \left[\mu  (x-1) \sqrt{3-\frac{ \pi^2\left(  n+\frac{1}{2} \right)^2}{ \mu ^2}}\right]- \pi  \left( n+\frac{1}{2}\right) \sinh \left[\mu  x \sqrt{3-\frac{\pi^2\left(   n+\frac{1}{2} \right)^2}{ \mu ^2}}\right]\Bigg\rbrace\, .
\eea
One can check that, when $y=0$, \Eq{eq:an:flat} is recovered. In \Eq{eq:USR:stocha:stoLim:f}, one can see that there is an additional term in the denominator, namely $\cosh\left(\sqrt{3-i t}\mu\right)$. This gives rise to a second set of poles, namely
\bea
\label{eq:USR:sto:StoLim:spectrum:2}
{\Lambda}^{(2)}_n = 3+\frac{\pi^2}{\mu^2} \left( n+\frac{1}{2}\right)^2 = \Lambda_n^{(1)}+3\, ,
\eea
which is a simple translation by $3$ of the spectrum~\eqref{eq:USR:sto:StoLim:spectrum:1}. By expanding again the characteristic function around the poles $t=-i\Lambda_n^{(2)} $, one obtains the associated residues,
\bea
a_n^{(2)}(x,y) &=& 
\frac{2 (-1)^n y}{\mu^2}\frac{\sin\left[\left(n+\frac{1}{2}\right)\pi x\right]}{\cos\left[\sqrt{3\mu^2+\pi^2 \left(n+\frac{1}{2}\right)^2}\right]}
\left\lbrace
-3 \mu ^2-\pi^2\left(n+\frac{1}{2}\right)^2
\nonumber\right. \\ & &\left.
+ \pi   (-1)^n \left( n+\frac{1}{2}\right) \sqrt{3 \mu ^2+\pi^2\left(n+\frac{1}{2}\right)^2} \sin \left[\sqrt{3\mu^2+\pi^2 \left(n+\frac{1}{2}\right)^2}\right]
 \right\rbrace\, .
\eea
The PDF of the first passage times is then given by \Eq{eq:tail_expansion} with the two sets of poles, namely
\bea
\label{eq:USR:sto:StoLim:PDF}
P\left(\N;x,y\right)=\sum_{i=1,2}\, \sum_{n=0}^\infty a_n^{(i)}(x,y)\ee^{-\Lambda_n^{(i)} \N}=
 \sum_{n=0}^\infty \left[ a_n^{(1)}+a_n^{(2)}\ee^{-3\N}\right]\ee^{-\Lambda_n^{(1)}\N}\, .
\eea
Let us note that, as announced above, the presence of classical velocity does not change the location of the poles (hence the decay rates of the PDF). It only adds a second set of eigenvalues, with associated residues that vanish when $y=0$. This second set of eigenvalues provide a small contribution to the tail, because of the suppression factor $\ee^{-3\N}$ in \Eq{eq:USR:sto:StoLim:PDF}. The formula~\eqref{eq:USR:sto:StoLim:PDF} is compared with numerical reconstructions of the PDF in \Fig{fig:USR:Sto:PDF}. One can see that it provides a very good approximation of the distribution functions, even for sizeable values of $y_\uin$. On the tail, the statistical error bars become larger since realisations upon which the statistics can be computed become more scarce, but one can see that the decay rate is indeed independent of $y_\uin$. 
 
\begin{figure}
    \centering
    \includegraphics[width=0.49\textwidth]{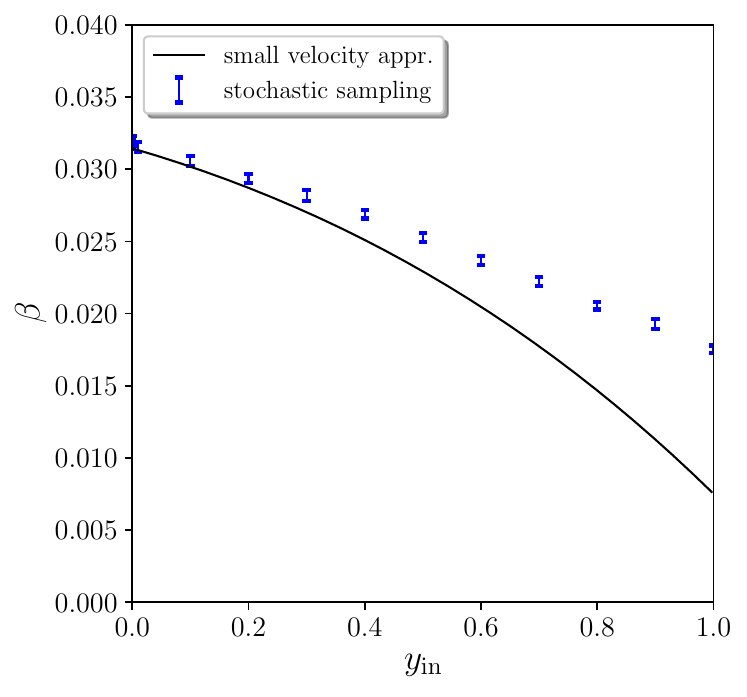}
        \includegraphics[width=0.49\textwidth]{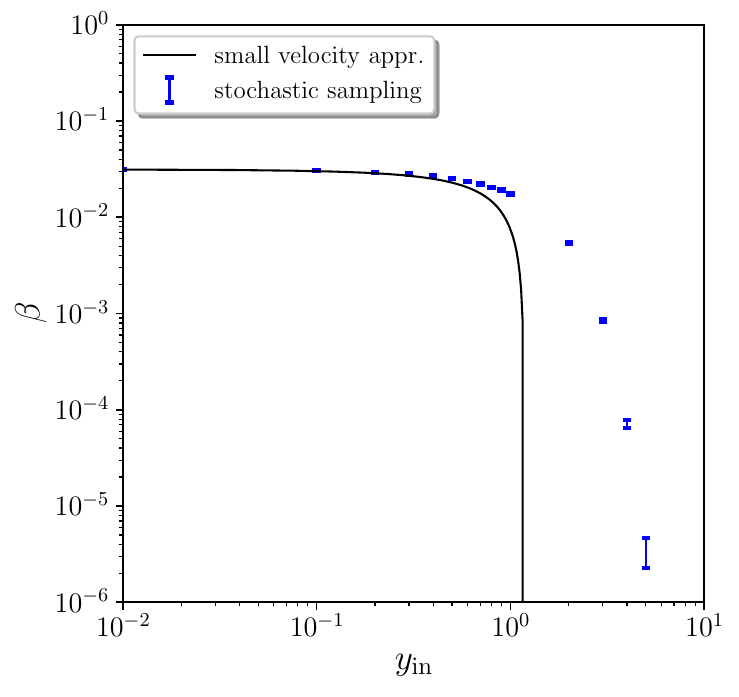}
    \caption{PBH mass fraction in a flat potential well with $\mu=1$, for a few values of the initial field velocity $y_{\uin}$, with a linear scale on the vertical axis in the left panel and a logarithmic scale in the right panel. The blue bars are reconstructed from a large number of numerical realisations of the Langevin equations~\eqref{eq:conjmomentum:langevin:USR:rescaled}-\eqref{eq:KG:efolds:langevin:USR:rescaled}, where the size of the bars stand for $2\sigma$-estimates of the statistical error, which are obtained following the jackknife procedure explained around \Fig{fig:USR:meanN}. The black lines correspond to the small velocity analytical formula~(\ref{eq:USR:sto:StoLim:beta}).}
    \label{fig:USR:Sto:beta}
\end{figure}
The mass fraction can then be obtained by plugging \Eq{eq:USR:sto:StoLim:PDF} into \Eq{eq:beta:def}, and one obtains
\bea
\label{eq:USR:sto:StoLim:beta}
\beta=\sum_{n=0}^\infty \left[\frac{a_n^{(1)}}{\Lambda_n^{(1)}}+\frac{a_n^{(2)}}{\Lambda_n^{(1)}+3}\ee^{-3\left(\left\langle \N \right\rangle + \zeta_\uc\right)}\right]\ee^{-\Lambda_n^{(1)}\left(\left\langle \N \right\rangle + \zeta_\uc\right)}\, ,
\eea
where we recall that $\left\langle \N \right\rangle$ is given in \Eq{eq:USR:sto:StoLim:meanN}.  This formula is again compared with the result from numerical realisations of the Langevin equations in \Fig{fig:USR:Sto:beta}. In practice, the number of realisations that produce more \efolds~than the mean value plus $\zeta_\uc=1$ is divided by the total number of realisations, which gives $\beta$. When $\beta$ becomes very small, this implies that a very large number of realisations must be produced (in practice, many more realisations than $1/\beta$) in order not to be dominated by statistical noise, which is numerically expensive and explains why we do not compute values of $\beta$ smaller than $\sim 10^{-6}$. In practice, when $y_\uin>6$, $10^7$ realisations were produced and not a single one produced more than $\langle \N \rangle +1$ \efolds, which only allow us to place an upper bound around $10^{-7}$ for these values of the initial velocity. 

In the small velocity limit, in order to obtain a more insightful formula, let us evaluate the mass fraction~\eqref{eq:USR:sto:StoLim:beta} with the leading pole $\Lambda_0^{(1)}$ only, when $x=1$. One obtains
\bea
\beta\simeq 
\left[ \frac{4}{\pi}+\frac{2y}{\cosh\left(\sqrt{3\mu^2-\frac{\pi^2}{4}}\right)} - 2y\frac{\tanh\left(\sqrt{3\mu^2-\frac{\pi^2}{4}}\right)}{\sqrt{12\frac{\mu^2}{\pi^2}-1}}\right]
\ee^{-\frac{\pi^2}{4\mu^2}\left[\frac{\mu^2}{2}+1+\frac{y\mu^2}{\cosh\left(\sqrt{3}\mu\right)}- \frac{y\mu \tanh\left(\sqrt{3}\mu\right)}{\sqrt{3}}\right]}\, .
\eea
Interestingly, both in the limits $\mu\ll 1$ and $\mu\gg 1$, the effect of velocity disappears and one recovers the slow-roll result $\beta\simeq 4/\pi \ee^{-\pi^2/(4\mu^2)-\pi^2/8}$. When $\mu$ is of order one, the situation is displayed in \Fig{fig:USR:Sto:beta} where one can see that values of $y_\uin<1$ only mildly affect the mass fraction. The conclusion of \Sec{sec:PBHs} that PBHs are overproduced when $\mu\gg 1$ and highly suppressed if $\mu\ll 1$ is therefore robust to the introduction of classical velocity, as long as it remains below $y_\uin<1$. 

For values of $y_\uin>1$, as mentioned above, a direct simulation of realisations of the Langevin equations is numerically very expensive and cannot be performed to a satisfactory level of statistical error. In the classical approximation of \Sec{sec:USR:Sto:Class:Lim}, the PDF of the number of \efolds~is Gaussian, and the mass fraction reads $\beta\simeq \erfc[1/(\sqrt{2 \langle \delta \N^2 \rangle})]/2$, where $\langle \delta \N^2 \rangle$ is given in \Eq{eq:USR:deltaN2:classical:nlo}. This formula however fails to reproduce the results obtained in \Fig{fig:USR:Sto:beta} for the values of $y_\uin$ slightly above one, where it vastly underestimates the mass fraction, and may only be reliable for much larger values of $y_\uin$. Let us finally mention that although the regime $y_\uin>1$ cannot be properly discussed with the above results, since the location of the pole is independent of the initial phase-space conditions, they are still given by \Eqs{eq:USR:sto:StoLim:spectrum:1} and~\eqref{eq:USR:sto:StoLim:spectrum:2} for sizeable or large values of $y_\uin$. This means that the numerical techniques used in \Sec{sec:tail:expansion} could be applied to the present setup: one could solve the partial differential equation that the characteristic function satisfies, \Eq{eq:USR:chi:PDE}, numerically, and compute the residues around the first poles, in order to extract the PDF and the mass fraction. Another approach is to use the formulation depicted in \Fig{fig:USR:MovingBarrier:Sketch} of a one-dimensional diffusion problem with time-dependent boundaries, and employ the Volterra integral equation for first-passage-time distribution functions~\cite{10.2307/3213641}. We plan to follow these directions
in a work in preparation.


\section{Conclusion}

Primordial cosmological perturbations, born out of vacuum quantum fluctuations in the early universe, set the initial conditions for the standard model of cosmology. At large scales, they determine the seeds of the cosmic structures we observe in the anisotropies of the CMB and surveys of the large-scale structure. This gives us precise measurements of the amplitude and tilt of the primordial power spectrum, as well as constraints on the leading-order non-Gaussian corrections, across a $\sim 7$ \efolds~range of scales. At small scales, however, non-linear structures in the universe prevent us from probing the nature of the primordial fluctuations. This leaves most of the inflationary potential, after the CMB scales are generated, unconstrained.  

One possible tracer of the late-time inflationary evolution is primordial black holes. If the cosmological perturbations are made sufficiently large, they collapse into black holes upon horizon reentry during the radiation-dominated era. Their mass is associated with the horizon size at the time of reentry, which can be linked to the time when the perturbations are generated during inflation. Therefore, by measuring or constraining the abundance of PBHs and their mass spectrum, one could reconstruct the behaviour of primordial perturbations at scales much smaller than those probed by the CMB and by large-scale structure surveys, and constrain the dynamics of inflation when those scales were generated.

Since PBHs form from rare, large perturbations, they require the existence during inflation of phases featuring large quantum fluctuations. These quantum fluctuations backreact on the dynamics of space-time expansion, and this has a number of consequences, for inflation in general and for PBHs in particular, that this manuscript has explored. The stochastic-$\delta N$ formalism was derived and employed to describe the modifications to the large-scale dynamics that quantum fluctuations induce as they cross out the Hubble radius during inflation, and to compute the statistics of perturbations in this modified background. 

Contrary to the standard, classical calculation, in which these statistics are Gaussian (with possible polynomial modulations, if $\fnl$ or $\gnl$-type corrections are included), we have found that the distribution functions of inflationary fluctuations have exponential tails. This feature is fully generic, and operates at all scales. At large scales (probed in the CMB and in the large-scale structures), the effect of exponential tails may be negligible in most models, although this remains to be checked explicitly. At intermediate scales, corresponding to small halos, \eg Lyman-alpha scales, or ultra-compact mini halos scales, the exponential tail effects may become very relevant. In particular, they could induce an enhancement of the non-linear collapse of structures on small scales that could have important consequences for large-scale structure formation, and thus for interpreting data from future surveys like \MYhref[violet]{https://www.desi.lbl.gov/}{DESI}, \MYhref[violet]{https://www.euclid-ec.org/}{Euclid} and \MYhref[violet]{https://www.lsst.org/}{LSST}. At small scales, they have important consequences for PBHs that we have studied. Our main conclusion is that, in the models of single-field inflation where PBHs are likely to be produced, such as in inflection-point potentials, PBHs are overproduced unless slow roll is violated. If slow roll is violated, one has to extend the stochastic-$\delta N$ formalism to the full phase space, and we have found that the abundance of PBHs is mostly determined with the width of the flat region in the potential, divided by the square root of the potential energy, in Planckian units. If this ratio is large, PBHs are overproduced, if it is small, they are highly suppressed, so getting an amount of PBHs that is cosmologically relevant requires some level of fine tuning on this parameter.

These results open up a number of new questions that I plan to investigate in the future, among which I now give a few examples. The formation of PBHs from local large inhomogeneities, the conditions that these inhomogeneities must satisfy in order to give rise to a black hole, and the mass of the resulting black hole are difficult topics of ongoing research. Various analytical and numerical tools are employed: peak theory, the excursion set approach, the Press-Schechter formalism, numerical relativity codes, \etc. Most of these  approaches are applied to Gaussian fluctuations, and give rise to criteria on the power spectra of these fluctuations. Our results suggests that they should instead be used with different statistics, namely those featuring exponential tails, for which the conditions for PBH formation have to be re-examined. 

We have also seen that stochastic effects are such that the classical one-to-one correspondence between a given physical scale measured today and a given location in the inflationary potential, is blurred by stochastic effects. In the classical picture, indeed, by measuring the statistics of perturbations at a given scale, one only has access to the local shape of the potential (namely its value and the value of its first derivative) at the point where the inflaton is when this scale crosses out the Hubble radius. In the stochastic picture, however, each scale carries information about the entire potential, see for instance \Eq{eq:PS:fullstocha}. If stochastic corrections are negligible across the full potential, this dependence is however suppressed, and the classical correspondence is recovered. But if the potential features a flat portion, say towards the end of inflation where it could lead to PBHs, it could thus affect the prediction for perturbations at all scales. In single-field models, since the curvature perturbation is conserved on large scales, its dynamics is immune to stochastic corrections if they arise later on. However, those stochastic corrections still modify the classical correspondence between scales and locations in the potential: for a local observer at the end of inflation, a physical scale can be related to a given number of \efolds~counted backwards from the end of inflation~\cite{Tada:2016pmk}. If the inflaton crosses a flat potential region between the CMB scales are being produced and the end of inflation, different patches in the universe can realise very different numbers of \efolds~across that region, so in each patch, a different region of the potential gives rise to the statistics of a given measured scale. This implies that even the most standard predictions for the CMB should be reconsidered in such cases, and that models producing PBHs may be associated to specific, testable features in the CMB. 

The blurring of the classical correspondence should be even more important in models where, towards the end of inflation, multiple-field effects also become relevant, such as in models where the end of inflation is triggered by additional scalar fields that become unstable, for instance in the context of hybrid inflation~\cite{GarciaBellido:1996qt} or geometrical destabilisation~\cite{Renaux-Petel:2015mga}. This is because, in such setups, stochastic diffusion does not only induce fluctuations along a reference phase-space trajectory, but also makes the system spread over different trajectories, hence combining different regions of both the potential and phase space inside a same measured scale.

Inflation is sometimes assumed to be preceded by a phase of slow contraction, followed by a bounce. This is indeed naturally expected in most theories of quantum gravity, and avoids the initial singularity that is otherwise still present in inflation. In contracting cosmologies, quantum fluctuations are still stretched out of the Hubble radius, hence backreact of the background dynamics. A stochastic formalism for contracting cosmologies could therefore be also derived. The absence of the slow-roll attractor in contracting cosmologies makes it necessary to track the stochastic dynamics in full phase space, so the extensions of the stochastic-$\delta N$ formalism presented here would be very useful in this approach. Close to the bounce, at very-high energy densities, cosmological inhomogeneities become large and it is natural to expect PBHs to be seeded~\cite{Chen:2016kjx,Quintin:2016qro}. Developing a ``stochastic contraction'' formalism would also allow one to study how quantum diffusion changes the dynamics of the universe as one approaches a bounce. In particular, the growth of shear-driven anisotropies from stochastic fluctuations could be studied, given that those are known to be problematic at the classical level for some contracting models. This may require to generalise the stochastic formalism to Bianchi (\textit{i.e.}~anisotropic) universes. Such a generalisation could also be useful to study the effect of quantum fluctuations on the onset of inflation with anisotropic initial conditions.  

These prospects illustrate the variety of topics early universe cosmology has to address. Primordial cosmology is one of the only places in physics where an effect based on general relativity (accelerated expansion) and quantum mechanics (parametric amplification of vacuum quantum fluctuations) leads to predictions that can be tested experimentally. In this context, various fundamental questions related to both these theories and how they behave when combined together can be studied. This drives my enthusiasm for this very lively and enjoyable field of research, and, as Isaac Asimov put it, ``The most exciting phrase to hear in Science, the one that heralds the most discoveries, is not `Eureka', but `That's funny...' ''.

\clearpage
\section{List of scientific publications submitted as part of the habilitation}
\label{sec:hdr:papers}

A full list of my research articles can be found \MYhref[violet]{http://inspirehep.net/author/profile/V.Vennin.1}{here}. Below are the publications submitted as part of the HDR.\\

\CVpaper{1}{https://arxiv.org/abs/1912.05399}{$\!\!\!\!\!\!\!\!\!\!\,$The exponential tail of inflationary fluctuations: consequences for primordial black holes}\\
Jose Mar\'ia Ezquiaga, Juan Garc\'ia-Bellido, Vincent Vennin\\
 \href{https://arxiv.org/abs/1912.05399}{arXiv:1912.05399} \\ 
\SmallSep

\CVpaper{1}{https://doi.org/10.1088/1475-7516/2019/07/031}{$\!\!\!\!\!\!\!\!\!\!\,$Stochastic inflation beyond slow roll}\\
Chris Pattison, Vincent Vennin, Hooshyar Assadullahi, David Wands\\
\textit{JCAP} 1907 (2019) 07, 031\\
 \href{https://arxiv.org/abs/1905.06300}{arXiv:1905.06300} \\ 
\SmallSep

\CVpaper{2}{https://doi.org/10.1088/1475-7516/2017/10/046}{$\!\!\!\!\!\!\!\!\!\!\,$Quantum diffusion during inflation and primordial black holes}\\
Chris Pattison, Vincent Vennin, Hooshyar Assadullahi, David Wands\\
\textit{JCAP} 1710 (2017) 046\\
 \href{https://arxiv.org/abs/1707.00537}{arXiv:1707.00537} \\
\SmallSep

\CVpaper{3}{https://doi.org/10.1088/1475-7516/2017/05/045}{$\!\!\!\!\!\!\!\!\!\!\,$Stochastic inflation in phase space: Is slow roll a stochastic attractor?}\\
Julien Grain, Vincent Vennin\\
\textit{JCAP} 1705 (2017) 045\\
 \href{https://arxiv.org/abs/1703.00447}{arXiv:1703.00447}\\
\SmallSep

\CVpaper{4}{https://doi.org/10.1103/PhysRevLett.118.031301}{$\!\!\!\!\!\!\!\!\!\!\,$Critical Number of Fields in Stochastic Inflation}\\
Vincent Vennin, Hooshyar Assadullahi, Hassan Firouzjahi, Mahdiyar Noorbala, David Wands\\
 \textit{Phys. Rev.} L 118 (2017) 031301\\
 \href{https://arxiv.org/abs/1604.06017}{arXiv:1604.06017}\\
\SmallSep

\CVpaper{5}{https://doi.org/10.1088/1475-7516/2016/06/043}{$\!\!\!\!\!\!\!\!\!\!\,$Multiple Fields in Stochastic Inflation}\\
Hooshyar Assadullahi, Hassan Firouzjahi, Mahdiyar Noorbala, Vincent Vennin, David Wands\\
\textit{JCAP} 1606 (2016) 043\\
 \href{https://arxiv.org/abs/1604.04502}{arXiv:1604.04502}\\
\SmallSep

\CVpaper{6}{https://doi.org/10.1140/epjc/s10052-015-3643-y}{$\!\!\!\!\!\!\!\!\!\!\,$Correlation Functions in Stochastic Inflation}\\
Vincent Vennin, Alexei A. Starobinsky\\
 \textit{Eur. Phys. J.} C (2015) 75:413\\
\href{http://arxiv.org/abs/1506.04732}{arXiv:1506.04732} \\
\SmallSep

\clearpage
\addcontentsline{toc}{section}{\textsc{References}}

\bibliographystyle{JHEP}
\bibliography{HdR}

\providecommand{\href}[2]{#2}\begingroup\raggedright\begin{thebibliography}{100}

\bibitem{Starobinsky:1980te}
A.A.~Starobinsky, \emph{{A New Type of Isotropic Cosmological Models Without
  Singularity}},
  \href{https://doi.org/10.1016/0370-2693(80)90670-X}{\emph{Phys. Lett.}
  {\bfseries B91} (1980) 99}.

\bibitem{Guth:1980zm}
A.H.~Guth, \emph{{The Inflationary Universe: A Possible Solution to the Horizon
  and Flatness Problems}},
  \href{https://doi.org/10.1103/PhysRevD.23.347}{\emph{Phys. Rev.} {\bfseries
  D23} (1981) 347}.

\bibitem{Baumann:2014nda}
D.~Baumann and L.~McAllister, \emph{{Inflation and String Theory}}, Cambridge
  Monographs on Mathematical Physics, Cambridge University Press (2015),
  \href{https://doi.org/10.1017/CBO9781316105733}{10.1017/CBO9781316105733},
  [\href{https://arxiv.org/abs/1404.2601}{{\ttfamily 1404.2601}}].

\bibitem{Kawasaki:2000yn}
M.~Kawasaki, M.~Yamaguchi and T.~Yanagida, \emph{{Natural chaotic inflation in
  supergravity}},
  \href{https://doi.org/10.1103/PhysRevLett.85.3572}{\emph{Phys. Rev. Lett.}
  {\bfseries 85} (2000) 3572}
  [\href{https://arxiv.org/abs/hep-ph/0004243}{{\ttfamily hep-ph/0004243}}].

\bibitem{Bezrukov:2007ep}
F.L.~Bezrukov and M.~Shaposhnikov, \emph{{The Standard Model Higgs boson as the
  inflaton}}, \href{https://doi.org/10.1016/j.physletb.2007.11.072}{\emph{Phys.
  Lett.} {\bfseries B659} (2008) 703}
  [\href{https://arxiv.org/abs/0710.3755}{{\ttfamily 0710.3755}}].

\bibitem{Akrami:2018vks}
{\scshape Planck} collaboration, \emph{{Planck 2018 results. I. Overview and
  the cosmological legacy of Planck}},
  \href{https://arxiv.org/abs/1807.06205}{{\ttfamily 1807.06205}}.

\bibitem{Martin:2013tda}
J.~Martin, C.~Ringeval and V.~Vennin, \emph{{Encyclopaedia Inflationaris}},
  \href{https://doi.org/10.1016/j.dark.2014.01.003}{\emph{Phys. Dark Univ.}
  {\bfseries 5-6} (2014) 75} [\href{https://arxiv.org/abs/1303.3787}{{\ttfamily
  1303.3787}}].

\bibitem{Martin:2013nzq}
J.~Martin, C.~Ringeval, R.~Trotta and V.~Vennin, \emph{{The Best Inflationary
  Models After Planck}},
  \href{https://doi.org/10.1088/1475-7516/2014/03/039}{\emph{JCAP} {\bfseries
  1403} (2014) 039} [\href{https://arxiv.org/abs/1312.3529}{{\ttfamily
  1312.3529}}].

\bibitem{Martin:2014nya}
J.~Martin, C.~Ringeval and V.~Vennin, \emph{{Observing Inflationary
  Reheating}},
  \href{https://doi.org/10.1103/PhysRevLett.114.081303}{\emph{Phys. Rev. Lett.}
  {\bfseries 114} (2015) 081303}
  [\href{https://arxiv.org/abs/1410.7958}{{\ttfamily 1410.7958}}].

\bibitem{Akrami:2018odb}
{\scshape Planck} collaboration, \emph{{Planck 2018 results. X. Constraints on
  inflation}},  \href{https://arxiv.org/abs/1807.06211}{{\ttfamily
  1807.06211}}.

\bibitem{Martin:2016oyk}
J.~Martin, C.~Ringeval and V.~Vennin, \emph{{Information Gain on Reheating: the
  One Bit Milestone}},
  \href{https://doi.org/10.1103/PhysRevD.93.103532}{\emph{Phys. Rev.}
  {\bfseries D93} (2016) 103532}
  [\href{https://arxiv.org/abs/1603.02606}{{\ttfamily 1603.02606}}].

\bibitem{Martin:2019nuw}
J.~Martin, T.~Papanikolaou and V.~Vennin, \emph{{Primordial black holes from
  the preheating instability}},
  \href{https://arxiv.org/abs/1907.04236}{{\ttfamily 1907.04236}}.

\bibitem{Martin:2020fgl}
J.~Martin, T.~Papanikolaou, L.~Pinol and V.~Vennin, \emph{{Metric preheating
  and radiative decay in single-field inflation}},
  \href{https://arxiv.org/abs/2002.01820}{{\ttfamily 2002.01820}}.

\bibitem{Sudarsky:2009za}
D.~Sudarsky, \emph{{Shortcomings in the Understanding of Why Cosmological
  Perturbations Look Classical}},
  \href{https://doi.org/10.1142/S0218271811018937}{\emph{Int. J. Mod. Phys.}
  {\bfseries D20} (2011) 509}
  [\href{https://arxiv.org/abs/0906.0315}{{\ttfamily 0906.0315}}].

\bibitem{Goldstein:2015mha}
S.~Goldstein, W.~Struyve and R.~Tumulka, \emph{{The Bohmian Approach to the
  Problems of Cosmological Quantum Fluctuations}},
  \href{https://arxiv.org/abs/1508.01017}{{\ttfamily 1508.01017}}.

\bibitem{Hartle:2019hae}
J.B.~Hartle, \emph{{The Impact of Cosmology on Quantum Mechanics}},
  \href{https://arxiv.org/abs/1901.03933}{{\ttfamily 1901.03933}}.

\bibitem{Kiefer:2006je}
C.~Kiefer, I.~Lohmar, D.~Polarski and A.A.~Starobinsky, \emph{{Pointer states
  for primordial fluctuations in inflationary cosmology}},
  \href{https://doi.org/10.1088/0264-9381/24/7/002}{\emph{Class. Quant. Grav.}
  {\bfseries 24} (2007) 1699}
  [\href{https://arxiv.org/abs/astro-ph/0610700}{{\ttfamily
  astro-ph/0610700}}].

\bibitem{Perez:2005gh}
A.~Perez, H.~Sahlmann and D.~Sudarsky, \emph{{On the quantum origin of the
  seeds of cosmic structure}},
  \href{https://doi.org/10.1088/0264-9381/23/7/008}{\emph{Class. Quant. Grav.}
  {\bfseries 23} (2006) 2317}
  [\href{https://arxiv.org/abs/gr-qc/0508100}{{\ttfamily gr-qc/0508100}}].

\bibitem{DeUnanue:2008fw}
A.~De~Unanue and D.~Sudarsky, \emph{{Phenomenological analysis of quantum
  collapse as source of the seeds of cosmic structure}},
  \href{https://doi.org/10.1103/PhysRevD.78.043510}{\emph{Phys. Rev.}
  {\bfseries D78} (2008) 043510}
  [\href{https://arxiv.org/abs/0801.4702}{{\ttfamily 0801.4702}}].

\bibitem{Leon:2010fi}
G.~Leon and D.~Sudarsky, \emph{{The Slow roll condition and the amplitude of
  the primordial spectrum of cosmic fluctuations: Contrasts and similarities of
  standard account and the 'collapse scheme'}},
  \href{https://doi.org/10.1088/0264-9381/27/22/225017}{\emph{Class. Quant.
  Grav.} {\bfseries 27} (2010) 225017}
  [\href{https://arxiv.org/abs/1003.5950}{{\ttfamily 1003.5950}}].

\bibitem{Canate:2012ua}
P.~Canate, P.~Pearle and D.~Sudarsky, \emph{{Continuous spontaneous
  localization wave function collapse model as a mechanism for the emergence of
  cosmological asymmetries in inflation}},
  \href{https://doi.org/10.1103/PhysRevD.87.104024}{\emph{Phys. Rev.}
  {\bfseries D87} (2013) 104024}
  [\href{https://arxiv.org/abs/1211.3463}{{\ttfamily 1211.3463}}].

\bibitem{Das:2013qwa}
S.~Das, K.~Lochan, S.~Sahu and T.P.~Singh, \emph{{Quantum to classical
  transition of inflationary perturbations: Continuous spontaneous localization
  as a possible mechanism}}, \href{https://doi.org/10.1103/PhysRevD.89.109902,
  10.1103/PhysRevD.88.085020}{\emph{Phys. Rev.} {\bfseries D88} (2013) 085020}
  [\href{https://arxiv.org/abs/1304.5094}{{\ttfamily 1304.5094}}].

\bibitem{Martin:2012pea}
J.~Martin, V.~Vennin and P.~Peter, \emph{{Cosmological Inflation and the
  Quantum Measurement Problem}},
  \href{https://doi.org/10.1103/PhysRevD.86.103524}{\emph{Phys. Rev.}
  {\bfseries D86} (2012) 103524}
  [\href{https://arxiv.org/abs/1207.2086}{{\ttfamily 1207.2086}}].

\bibitem{Martin:2019jye}
J.~Martin and V.~Vennin, \emph{{A cosmic shadow on CSL}},
  \href{https://arxiv.org/abs/1906.04405}{{\ttfamily 1906.04405}}.

\bibitem{Peter:2006hx}
P.~Peter, E.J.C.~Pinho and N.~Pinto-Neto, \emph{{A Non inflationary model with
  scale invariant cosmological perturbations}},
  \href{https://doi.org/10.1103/PhysRevD.75.023516}{\emph{Phys. Rev.}
  {\bfseries D75} (2007) 023516}
  [\href{https://arxiv.org/abs/hep-th/0610205}{{\ttfamily hep-th/0610205}}].

\bibitem{PintoNeto:2011ui}
N.~Pinto-Neto, G.~Santos and W.~Struyve, \emph{{Quantum-to-classical transition
  of primordial cosmological perturbations in de Broglie--Bohm quantum
  theory}}, \href{https://doi.org/10.1103/PhysRevD.85.083506}{\emph{Phys. Rev.}
  {\bfseries D85} (2012) 083506}
  [\href{https://arxiv.org/abs/1110.1339}{{\ttfamily 1110.1339}}].

\bibitem{Peter:2016kan}
P.~Peter and S.D.P.~Vitenti, \emph{{The simplest possible bouncing quantum
  cosmological model}},
  \href{https://doi.org/10.1142/S021773231640006X}{\emph{Mod. Phys. Lett.}
  {\bfseries A31} (2016) 1640006}
  [\href{https://arxiv.org/abs/1603.02342}{{\ttfamily 1603.02342}}].

\bibitem{Polarski:1995jg}
D.~Polarski and A.A.~Starobinsky, \emph{{Semiclassicality and decoherence of
  cosmological perturbations}},
  \href{https://doi.org/10.1088/0264-9381/13/3/006}{\emph{Class. Quant. Grav.}
  {\bfseries 13} (1996) 377}
  [\href{https://arxiv.org/abs/gr-qc/9504030}{{\ttfamily gr-qc/9504030}}].

\bibitem{Kiefer:1998qe}
C.~Kiefer, D.~Polarski and A.A.~Starobinsky, \emph{{Quantum to classical
  transition for fluctuations in the early universe}},
  \href{https://doi.org/10.1142/S0218271898000292}{\emph{Int. J. Mod. Phys.}
  {\bfseries D7} (1998) 455}
  [\href{https://arxiv.org/abs/gr-qc/9802003}{{\ttfamily gr-qc/9802003}}].

\bibitem{Martin:2015qta}
J.~Martin and V.~Vennin, \emph{{Quantum Discord of Cosmic Inflation: Can we
  Show that CMB Anisotropies are of Quantum-Mechanical Origin?}},
  \href{https://doi.org/10.1103/PhysRevD.93.023505}{\emph{Phys. Rev.}
  {\bfseries D93} (2016) 023505}
  [\href{https://arxiv.org/abs/1510.04038}{{\ttfamily 1510.04038}}].

\bibitem{Martin:2016tbd}
J.~Martin and V.~Vennin, \emph{{Bell inequalities for continuous-variable
  systems in generic squeezed states}},
  \href{https://doi.org/10.1103/PhysRevA.93.062117}{\emph{Phys. Rev.}
  {\bfseries A93} (2016) 062117}
  [\href{https://arxiv.org/abs/1605.02944}{{\ttfamily 1605.02944}}].

\bibitem{Martin:2017zxs}
J.~Martin and V.~Vennin, \emph{{Obstructions to Bell CMB Experiments}},
  \href{https://doi.org/10.1103/PhysRevD.96.063501}{\emph{Phys. Rev.}
  {\bfseries D96} (2017) 063501}
  [\href{https://arxiv.org/abs/1706.05001}{{\ttfamily 1706.05001}}].

\bibitem{Martin:2016nrr}
J.~Martin and V.~Vennin, \emph{{Leggett-Garg Inequalities for Squeezed
  States}}, \href{https://doi.org/10.1103/PhysRevA.94.052135}{\emph{Phys. Rev.}
  {\bfseries A94} (2016) 052135}
  [\href{https://arxiv.org/abs/1611.01785}{{\ttfamily 1611.01785}}].

\bibitem{Martin:2018zbe}
J.~Martin and V.~Vennin, \emph{{Observational constraints on quantum
  decoherence during inflation}},
  \href{https://doi.org/10.1088/1475-7516/2018/05/063}{\emph{JCAP} {\bfseries
  1805} (2018) 063} [\href{https://arxiv.org/abs/1801.09949}{{\ttfamily
  1801.09949}}].

\bibitem{Espinosa-Portales:2019peb}
L.~Espinosa-Portales and J.~Garcia-Bellido, \emph{{Entanglement entropy of
  Primordial Black Holes after inflation}},
  \href{https://arxiv.org/abs/1907.07601}{{\ttfamily 1907.07601}}.

\bibitem{Goldwirth:1989pr}
D.S.~Goldwirth and T.~Piran, \emph{{Inhomogeneity and the Onset of Inflation}},
  \href{https://doi.org/10.1103/PhysRevLett.64.2852}{\emph{Phys. Rev. Lett.}
  {\bfseries 64} (1990) 2852}.

\bibitem{Goldwirth:1991rj}
D.S.~Goldwirth and T.~Piran, \emph{{Initial conditions for inflation}},
  \href{https://doi.org/10.1016/0370-1573(92)90073-9}{\emph{Phys. Rept.}
  {\bfseries 214} (1992) 223}.

\bibitem{Chowdhury:2019otk}
D.~Chowdhury, J.~Martin, C.~Ringeval and V.~Vennin, \emph{{Inflation after
  Planck: Judgment Day}},  \href{https://arxiv.org/abs/1902.03951}{{\ttfamily
  1902.03951}}.

\bibitem{Goldwirth:1989vz}
D.S.~Goldwirth and T.~Piran, \emph{{Spherical Inhomogeneous Cosmologies and
  Inflation. 1. Numerical Methods}},
  \href{https://doi.org/10.1103/PhysRevD.40.3263}{\emph{Phys. Rev.} {\bfseries
  D40} (1989) 3263}.

\bibitem{Goldwirth:1990pm}
D.S.~Goldwirth, \emph{{On inhomogeneous initial conditions for inflation}},
  \href{https://doi.org/10.1103/PhysRevD.43.3204}{\emph{Phys. Rev.} {\bfseries
  D43} (1991) 3204}.

\bibitem{KurkiSuonio:1987pq}
H.~Kurki-Suonio, R.A.~Matzner, J.~Centrella and J.R.~Wilson, \emph{{Inflation
  From Inhomogeneous Initial Data in a One-dimensional Back Reacting
  Cosmology}}, \href{https://doi.org/10.1103/PhysRevD.35.435}{\emph{Phys. Rev.}
  {\bfseries D35} (1987) 435}.

\bibitem{Laguna:1991zs}
P.~Laguna, H.~Kurki-Suonio and R.A.~Matzner, \emph{{Inhomogeneous inflation:
  The Initial value problem}},
  \href{https://doi.org/10.1103/PhysRevD.44.3077}{\emph{Phys. Rev.} {\bfseries
  D44} (1991) 3077}.

\bibitem{KurkiSuonio:1993fg}
H.~Kurki-Suonio, P.~Laguna and R.A.~Matzner, \emph{{Inhomogeneous inflation:
  Numerical evolution}},
  \href{https://doi.org/10.1103/PhysRevD.48.3611}{\emph{Phys. Rev.} {\bfseries
  D48} (1993) 3611} [\href{https://arxiv.org/abs/astro-ph/9306009}{{\ttfamily
  astro-ph/9306009}}].

\bibitem{East:2015ggf}
W.E.~East, M.~Kleban, A.~Linde and L.~Senatore, \emph{{Beginning inflation in
  an inhomogeneous universe}},
  \href{https://doi.org/10.1088/1475-7516/2016/09/010}{\emph{JCAP} {\bfseries
  1609} (2016) 010} [\href{https://arxiv.org/abs/1511.05143}{{\ttfamily
  1511.05143}}].

\bibitem{Clough:2016ymm}
K.~Clough, E.A.~Lim, B.S.~DiNunno, W.~Fischler, R.~Flauger and S.~Paban,
  \emph{{Robustness of Inflation to Inhomogeneous Initial Conditions}},
  \href{https://doi.org/10.1088/1475-7516/2017/09/025}{\emph{JCAP} {\bfseries
  1709} (2017) 025} [\href{https://arxiv.org/abs/1608.04408}{{\ttfamily
  1608.04408}}].

\bibitem{Bloomfield:2019rbs}
J.K.~Bloomfield, P.~Fitzpatrick, K.~Hilbert and D.I.~Kaiser, \emph{{Onset of
  inflation amid backreaction from inhomogeneities}},
  \href{https://doi.org/10.1103/PhysRevD.100.063512}{\emph{Phys. Rev.}
  {\bfseries D100} (2019) 063512}
  [\href{https://arxiv.org/abs/1906.08651}{{\ttfamily 1906.08651}}].

\bibitem{Calzetta:1992gv}
E.~Calzetta and M.~Sakellariadou, \emph{{Inflation in inhomogeneous
  cosmology}}, \href{https://doi.org/10.1103/PhysRevD.45.2802}{\emph{Phys.
  Rev.} {\bfseries D45} (1992) 2802}.

\bibitem{Perez:2012pn}
R.S.~Perez and N.~Pinto-Neto, \emph{{Spherically Symmetric Inflation}},
  \href{https://doi.org/10.1134/S0202289311020174}{\emph{Grav. Cosmol.}
  {\bfseries 17} (2011) 136} [\href{https://arxiv.org/abs/1205.3790}{{\ttfamily
  1205.3790}}].

\bibitem{Liddle:1994dx}
A.R.~Liddle, P.~Parsons and J.D.~Barrow, \emph{{Formalizing the slow roll
  approximation in inflation}},
  \href{https://doi.org/10.1103/PhysRevD.50.7222}{\emph{Phys. Rev.} {\bfseries
  D50} (1994) 7222} [\href{https://arxiv.org/abs/astro-ph/9408015}{{\ttfamily
  astro-ph/9408015}}].

\bibitem{Bunch:1978yq}
T.S.~Bunch and P.C.W.~Davies, \emph{{Quantum Field Theory in de Sitter Space:
  Renormalization by Point Splitting}},
  \href{https://doi.org/10.1098/rspa.1978.0060}{\emph{Proc. Roy. Soc. Lond.}
  {\bfseries A360} (1978) 117}.

\bibitem{Grain:2019vnq}
J.~Grain and V.~Vennin, \emph{{Squeezing formalism and canonical
  transformations in cosmology}},
  \href{https://arxiv.org/abs/1910.01916}{{\ttfamily 1910.01916}}.

\bibitem{PhysRevD.32.3136}
B.~Allen, \emph{Vacuum states in de sitter space},
  \href{https://doi.org/10.1103/PhysRevD.32.3136}{\emph{Phys. Rev. D}
  {\bfseries 32} (1985) 3136}.

\bibitem{Brandenberger:1985fc}
R.H.~Brandenberger and C.T.~Hill, \emph{{Energy Density Fluctuations in de
  Sitter Space}},
  \href{https://doi.org/10.1016/0370-2693(86)90430-2}{\emph{Phys. Lett.}
  {\bfseries B179} (1986) 30}.

\bibitem{Kaloper:2018zgi}
N.~Kaloper and J.~Scargill, \emph{{Quantum Cosmic No-Hair Theorem and
  Inflation}}, \href{https://doi.org/10.1103/PhysRevD.99.103514}{\emph{Phys.
  Rev.} {\bfseries D99} (2019) 103514}
  [\href{https://arxiv.org/abs/1802.09554}{{\ttfamily 1802.09554}}].

\bibitem{Martin:2000xs}
J.~Martin and R.H.~Brandenberger, \emph{{The TransPlanckian problem of
  inflationary cosmology}},
  \href{https://doi.org/10.1103/PhysRevD.63.123501}{\emph{Phys. Rev.}
  {\bfseries D63} (2001) 123501}
  [\href{https://arxiv.org/abs/hep-th/0005209}{{\ttfamily hep-th/0005209}}].

\bibitem{Brandenberger:2000wr}
R.H.~Brandenberger and J.~Martin, \emph{{The Robustness of inflation to changes
  in superPlanck scale physics}},
  \href{https://doi.org/10.1142/S0217732301004170}{\emph{Mod. Phys. Lett.}
  {\bfseries A16} (2001) 999}
  [\href{https://arxiv.org/abs/astro-ph/0005432}{{\ttfamily
  astro-ph/0005432}}].

\bibitem{Brandenberger:2004kx}
R.H.~Brandenberger and J.~Martin, \emph{{Back-reaction and the trans-Planckian
  problem of inflation revisited}},
  \href{https://doi.org/10.1103/PhysRevD.71.023504}{\emph{Phys. Rev.}
  {\bfseries D71} (2005) 023504}
  [\href{https://arxiv.org/abs/hep-th/0410223}{{\ttfamily hep-th/0410223}}].

\bibitem{Martin:2002kt}
J.~Martin and R.H.~Brandenberger, \emph{{The Corley-Jacobson dispersion
  relation and transPlanckian inflation}},
  \href{https://doi.org/10.1103/PhysRevD.65.103514}{\emph{Phys. Rev.}
  {\bfseries D65} (2002) 103514}
  [\href{https://arxiv.org/abs/hep-th/0201189}{{\ttfamily hep-th/0201189}}].

\bibitem{Carr:1974nx}
B.J.~Carr and S.W.~Hawking, \emph{{Black holes in the early Universe}},
  {\emph{Mon. Not. Roy. Astron. Soc.} {\bfseries 168} (1974) 399}.

\bibitem{Tada:2019amh}
Y.~Tada and S.~Yokoyama, \emph{{Primordial black hole tower: Dark matter,
  earth-mass, and LIGO black holes}},
  \href{https://doi.org/10.1103/PhysRevD.100.023537}{\emph{Phys. Rev.}
  {\bfseries D100} (2019) 023537}
  [\href{https://arxiv.org/abs/1904.10298}{{\ttfamily 1904.10298}}].

\bibitem{Carr:2009jm}
B.J.~Carr, K.~Kohri, Y.~Sendouda and J.~Yokoyama, \emph{{New cosmological
  constraints on primordial black holes}},
  \href{https://doi.org/10.1103/PhysRevD.81.104019}{\emph{Phys. Rev.}
  {\bfseries D81} (2010) 104019}
  [\href{https://arxiv.org/abs/0912.5297}{{\ttfamily 0912.5297}}].

\bibitem{Graham:2015apa}
P.W.~Graham, S.~Rajendran and J.~Varela, \emph{{Dark Matter Triggers of
  Supernovae}}, \href{https://doi.org/10.1103/PhysRevD.92.063007}{\emph{Phys.
  Rev.} {\bfseries D92} (2015) 063007}
  [\href{https://arxiv.org/abs/1505.04444}{{\ttfamily 1505.04444}}].

\bibitem{Niikura:2017zjd}
H.~Niikura et~al., \emph{{Microlensing constraints on primordial black holes
  with Subaru/HSC Andromeda observations}},
  \href{https://doi.org/10.1038/s41550-019-0723-1}{\emph{Nat. Astron.}
  {\bfseries 3} (2019) 524} [\href{https://arxiv.org/abs/1701.02151}{{\ttfamily
  1701.02151}}].

\bibitem{Griest:2013esa}
K.~Griest, A.M.~Cieplak and M.J.~Lehner, \emph{{New Limits on Primordial Black
  Hole Dark Matter from an Analysis of Kepler Source Microlensing Data}},
  \href{https://doi.org/10.1103/PhysRevLett.111.181302}{\emph{Phys. Rev. Lett.}
  {\bfseries 111} (2013) 181302}.

\bibitem{Tisserand:2006zx}
{\scshape EROS-2} collaboration, \emph{{Limits on the Macho Content of the
  Galactic Halo from the EROS-2 Survey of the Magellanic Clouds}},
  \href{https://doi.org/10.1051/0004-6361:20066017}{\emph{Astron. Astrophys.}
  {\bfseries 469} (2007) 387}
  [\href{https://arxiv.org/abs/astro-ph/0607207}{{\ttfamily
  astro-ph/0607207}}].

\bibitem{Brandt:2016aco}
T.D.~Brandt, \emph{{Constraints on MACHO Dark Matter from Compact Stellar
  Systems in Ultra-Faint Dwarf Galaxies}},
  \href{https://doi.org/10.3847/2041-8205/824/2/L31}{\emph{Astrophys. J.}
  {\bfseries 824} (2016) L31}
  [\href{https://arxiv.org/abs/1605.03665}{{\ttfamily 1605.03665}}].

\bibitem{Ali-Haimoud:2016mbv}
Y.~Ali-Haimoud and M.~Kamionkowski, \emph{{Cosmic microwave background limits
  on accreting primordial black holes}},
  \href{https://doi.org/10.1103/PhysRevD.95.043534}{\emph{Phys. Rev.}
  {\bfseries D95} (2017) 043534}
  [\href{https://arxiv.org/abs/1612.05644}{{\ttfamily 1612.05644}}].

\bibitem{Blum:2016cjs}
D.~Aloni, K.~Blum and R.~Flauger, \emph{{Cosmic microwave background
  constraints on primordial black hole dark matter}},
  \href{https://doi.org/10.1088/1475-7516/2017/05/017}{\emph{JCAP} {\bfseries
  1705} (2017) 017} [\href{https://arxiv.org/abs/1612.06811}{{\ttfamily
  1612.06811}}].

\bibitem{Horowitz:2016lib}
B.~Horowitz, \emph{{Revisiting Primordial Black Holes Constraints from
  Ionization History}},  \href{https://arxiv.org/abs/1612.07264}{{\ttfamily
  1612.07264}}.

\bibitem{Abbott:2016blz}
{\scshape LIGO Scientific, Virgo} collaboration, \emph{{Observation of
  Gravitational Waves from a Binary Black Hole Merger}},
  \href{https://doi.org/10.1103/PhysRevLett.116.061102}{\emph{Phys. Rev. Lett.}
  {\bfseries 116} (2016) 061102}
  [\href{https://arxiv.org/abs/1602.03837}{{\ttfamily 1602.03837}}].

\bibitem{Bean:2002kx}
R.~Bean and J.~Magueijo, \emph{{Could supermassive black holes be
  quintessential primordial black holes?}},
  \href{https://doi.org/10.1103/PhysRevD.66.063505}{\emph{Phys. Rev.}
  {\bfseries D66} (2002) 063505}
  [\href{https://arxiv.org/abs/astro-ph/0204486}{{\ttfamily
  astro-ph/0204486}}].

\bibitem{Meszaros:1975ef}
P.~Meszaros, \emph{{Primeval black holes and galaxy formation}}, {\emph{Astron.
  Astrophys.} {\bfseries 38} (1975) 5}.

\bibitem{Carr:2018rid}
B.~Carr and J.~Silk, \emph{{Primordial Black Holes as Generators of Cosmic
  Structures}}, \href{https://doi.org/10.1093/mnras/sty1204}{\emph{Mon. Not.
  Roy. Astron. Soc.} {\bfseries 478} (2018) 3756}
  [\href{https://arxiv.org/abs/1801.00672}{{\ttfamily 1801.00672}}].

\bibitem{Clesse:2017bsw}
S.~Clesse and J.~Garcia-Bellido, \emph{{Seven Hints for Primordial Black Hole
  Dark Matter}}, \href{https://doi.org/10.1016/j.dark.2018.08.004}{\emph{Phys.
  Dark Univ.} {\bfseries 22} (2018) 137}
  [\href{https://arxiv.org/abs/1711.10458}{{\ttfamily 1711.10458}}].

\bibitem{Kashlinsky:2016sdv}
A.~Kashlinsky, \emph{{LIGO gravitational wave detection, primordial black holes
  and the near-IR cosmic infrared background anisotropies}},
  \href{https://doi.org/10.3847/2041-8205/823/2/L25}{\emph{Astrophys. J.}
  {\bfseries 823} (2016) L25}
  [\href{https://arxiv.org/abs/1605.04023}{{\ttfamily 1605.04023}}].

\bibitem{Markov:1984xd}
M.A.~Markov and P.C.~West, eds., \emph{{QUANTUM GRAVITY. PROCEEDINGS, 2ND
  SEMINAR, MOSCOW, USSR, OCTOBER 13-15, 1981}}, 1984.

\bibitem{Coleman:1991ku}
S.R.~Coleman, J.~Preskill and F.~Wilczek, \emph{{Quantum hair on black holes}},
  \href{https://doi.org/10.1016/0550-3213(92)90008-Y}{\emph{Nucl. Phys.}
  {\bfseries B378} (1992) 175}
  [\href{https://arxiv.org/abs/hep-th/9201059}{{\ttfamily hep-th/9201059}}].

\bibitem{Barrau:2003xp}
A.~Barrau, D.~Blais, G.~Boudoul and D.~Polarski, \emph{{Peculiar relics from
  primordial black holes in the inflationary paradigm}},
  \href{https://doi.org/10.1002/andp.200310067}{\emph{Annalen Phys.} {\bfseries
  13} (2004) 115} [\href{https://arxiv.org/abs/astro-ph/0303330}{{\ttfamily
  astro-ph/0303330}}].

\bibitem{Khlopov:2004tn}
M.Y.~Khlopov, A.~Barrau and J.~Grain, \emph{{Gravitino production by primordial
  black hole evaporation and constraints on the inhomogeneity of the early
  universe}}, \href{https://doi.org/10.1088/0264-9381/23/6/004}{\emph{Class.
  Quant. Grav.} {\bfseries 23} (2006) 1875}
  [\href{https://arxiv.org/abs/astro-ph/0406621}{{\ttfamily
  astro-ph/0406621}}].

\bibitem{Carr:2017jsz}
B.~Carr, M.~Raidal, T.~Tenkanen, V.~Vaskonen and H.~Veermae, \emph{{Primordial
  black hole constraints for extended mass functions}},
  \href{https://doi.org/10.1103/PhysRevD.96.023514}{\emph{Phys. Rev.}
  {\bfseries D96} (2017) 023514}
  [\href{https://arxiv.org/abs/1705.05567}{{\ttfamily 1705.05567}}].

\bibitem{Blais:2002nd}
D.~Blais, C.~Kiefer and D.~Polarski, \emph{{Can primordial black holes be a
  significant part of dark matter?}},
  \href{https://doi.org/10.1016/S0370-2693(02)01803-8}{\emph{Phys. Lett.}
  {\bfseries B535} (2002) 11}
  [\href{https://arxiv.org/abs/astro-ph/0203520}{{\ttfamily
  astro-ph/0203520}}].

\bibitem{Chluba:2012gq}
J.~Chluba, R.~Khatri and R.A.~Sunyaev, \emph{{CMB at 2x2 order: The dissipation
  of primordial acoustic waves and the observable part of the associated energy
  release}}, \href{https://doi.org/10.1111/j.1365-2966.2012.21474.x}{\emph{Mon.
  Not. Roy. Astron. Soc.} {\bfseries 425} (2012) 1129}
  [\href{https://arxiv.org/abs/1202.0057}{{\ttfamily 1202.0057}}].

\bibitem{Pajer:2012vz}
E.~Pajer and M.~Zaldarriaga, \emph{{A New Window on Primordial
  non-Gaussianity}},
  \href{https://doi.org/10.1103/PhysRevLett.109.021302}{\emph{Phys. Rev. Lett.}
  {\bfseries 109} (2012) 021302}
  [\href{https://arxiv.org/abs/1201.5375}{{\ttfamily 1201.5375}}].

\bibitem{Dalal:2007cu}
N.~Dalal, O.~Dore, D.~Huterer and A.~Shirokov, \emph{{The imprints of
  primordial non-gaussianities on large-scale structure: scale dependent bias
  and abundance of virialized objects}},
  \href{https://doi.org/10.1103/PhysRevD.77.123514}{\emph{Phys. Rev.}
  {\bfseries D77} (2008) 123514}
  [\href{https://arxiv.org/abs/0710.4560}{{\ttfamily 0710.4560}}].

\bibitem{Chiba:2017rvs}
T.~Chiba and S.~Yokoyama, \emph{{Spin Distribution of Primordial Black Holes}},
  \href{https://doi.org/10.1093/ptep/ptx087}{\emph{PTEP} {\bfseries 2017}
  (2017) 083E01} [\href{https://arxiv.org/abs/1704.06573}{{\ttfamily
  1704.06573}}].

\bibitem{Kinugawa:2016ect}
T.~Kinugawa, H.~Nakano and T.~Nakamura, \emph{{Gravitational wave quasinormal
  mode from Population III massive black hole binaries in various models of
  population synthesis}},
  \href{https://doi.org/10.1093/ptep/ptw143}{\emph{PTEP} {\bfseries 2016}
  (2016) 103E01} [\href{https://arxiv.org/abs/1606.00362}{{\ttfamily
  1606.00362}}].

\bibitem{Kocsis:2017yty}
B.~Kocsis, T.~Suyama, T.~Tanaka and S.~Yokoyama, \emph{{Hidden universality in
  the merger rate distribution in the primordial black hole scenario}},
  \href{https://doi.org/10.3847/1538-4357/aaa7f4}{\emph{Astrophys. J.}
  {\bfseries 854} (2018) 41}
  [\href{https://arxiv.org/abs/1709.09007}{{\ttfamily 1709.09007}}].

\bibitem{Bartolo:2016ami}
N.~Bartolo et~al., \emph{{Science with the space-based interferometer LISA. IV:
  Probing inflation with gravitational waves}},
  \href{https://doi.org/10.1088/1475-7516/2016/12/026}{\emph{JCAP} {\bfseries
  1612} (2016) 026} [\href{https://arxiv.org/abs/1610.06481}{{\ttfamily
  1610.06481}}].

\bibitem{Zhao:2013bba}
W.~Zhao, Y.~Zhang, X.-P.~You and Z.-H.~Zhu, \emph{{Constraints of relic
  gravitational waves by pulsar timing arrays: Forecasts for the FAST and SKA
  projects}}, \href{https://doi.org/10.1103/PhysRevD.87.124012}{\emph{Phys.
  Rev.} {\bfseries D87} (2013) 124012}
  [\href{https://arxiv.org/abs/1303.6718}{{\ttfamily 1303.6718}}].

\bibitem{Garcia-Bellido:2017qal}
J.~Garcia-Bellido and S.~Nesseris, \emph{{Gravitational wave bursts from
  Primordial Black Hole hyperbolic encounters}},
  \href{https://doi.org/10.1016/j.dark.2017.10.002}{\emph{Phys. Dark Univ.}
  {\bfseries 18} (2017) 123}
  [\href{https://arxiv.org/abs/1706.02111}{{\ttfamily 1706.02111}}].

\bibitem{Starobinsky:1982ee}
A.A.~Starobinsky, \emph{{Dynamics of Phase Transition in the New Inflationary
  Universe Scenario and Generation of Perturbations}},
  \href{https://doi.org/10.1016/0370-2693(82)90541-X}{\emph{Phys.Lett.}
  {\bfseries B117} (1982) 175}.

\bibitem{Starobinsky:1986fxa}
A.A.~Starobinsky, \emph{{Multicomponent de Sitter (Inflationary) Stages and the
  Generation of Perturbations}}, {\emph{JETP Lett.} {\bfseries 42} (1985) 152}.

\bibitem{Bachelier1900}
L.~Bachelier, \emph{Th\'eorie de la sp\'eculation}, {\emph{Annales
  Scientifiques de L'Ecole Normale Sup\'erieure} {\bfseries 17} (1900) 21}.

\bibitem{Vennin:2015hra}
V.~Vennin and A.A.~Starobinsky, \emph{{Correlation Functions in Stochastic
  Inflation}}, \href{https://doi.org/10.1140/epjc/s10052-015-3643-y}{\emph{Eur.
  Phys. J.} {\bfseries C75} (2015) 413}
  [\href{https://arxiv.org/abs/1506.04732}{{\ttfamily 1506.04732}}].

\bibitem{Grishchuk:1990bj}
L.~Grishchuk and Y.~Sidorov, \emph{{Squeezed quantum states of relic gravitons
  and primordial density fluctuations}},
  \href{https://doi.org/10.1103/PhysRevD.42.3413}{\emph{Phys. Rev.} {\bfseries
  D42} (1990) 3413}.

\bibitem{Grishchuk:1992tw}
L.~Grishchuk, H.~Haus and K.~Bergman, \emph{{Generation of squeezed radiation
  from vacuum in the cosmos and the laboratory}},
  \href{https://doi.org/10.1103/PhysRevD.46.1440}{\emph{Phys. Rev.} {\bfseries
  D46} (1992) 1440}.

\bibitem{Lesgourgues:1996jc}
J.~Lesgourgues, D.~Polarski and A.A.~Starobinsky, \emph{{Quantum to classical
  transition of cosmological perturbations for nonvacuum initial states}},
  \href{https://doi.org/10.1016/S0550-3213(97)00224-1}{\emph{Nucl. Phys.}
  {\bfseries B497} (1997) 479}
  [\href{https://arxiv.org/abs/gr-qc/9611019}{{\ttfamily gr-qc/9611019}}].

\bibitem{Polarski:2001yn}
D.~Polarski, \emph{{Classicality of primordial fluctuations and primordial
  black holes}}, \href{https://doi.org/10.1142/S021827180100161X}{\emph{Int. J.
  Mod. Phys.} {\bfseries D10} (2001) 927}
  [\href{https://arxiv.org/abs/astro-ph/0109388}{{\ttfamily
  astro-ph/0109388}}].

\bibitem{Kiefer:2008ku}
C.~Kiefer and D.~Polarski, \emph{{Why do cosmological perturbations look
  classical to us?}}, \href{https://doi.org/10.1166/asl.2009.1023}{\emph{Adv.
  Sci. Lett.} {\bfseries 2} (2009) 164}
  [\href{https://arxiv.org/abs/0810.0087}{{\ttfamily 0810.0087}}].

\bibitem{Burgess:2014eoa}
C.P.~Burgess, R.~Holman, G.~Tasinato and M.~Williams, \emph{{EFT Beyond the
  Horizon: Stochastic Inflation and How Primordial Quantum Fluctuations Go
  Classical}}, \href{https://doi.org/10.1007/JHEP03(2015)090}{\emph{JHEP}
  {\bfseries 03} (2015) 090} [\href{https://arxiv.org/abs/1408.5002}{{\ttfamily
  1408.5002}}].

\bibitem{Starobinsky:1986fx}
A.A.~Starobinsky, \emph{{Stochastic de Sitter (inflationary) stage in the early
  universe}}, \href{https://doi.org/10.1007/3-540-16452-9_6}{\emph{Lect.Notes
  Phys.} {\bfseries 246} (1986) 107}.

\bibitem{Nambu:1987ef}
Y.~Nambu and M.~Sasaki, \emph{{Stochastic Stage of an Inflationary Universe
  Model}},
  \href{https://doi.org/10.1016/0370-2693(88)90974-4}{\emph{Phys.Lett.}
  {\bfseries B205} (1988) 441}.

\bibitem{Nambu:1988je}
Y.~Nambu and M.~Sasaki, \emph{{Stochastic Approach to Chaotic Inflation and the
  Distribution of Universes}},
  \href{https://doi.org/10.1016/0370-2693(89)90385-7}{\emph{Phys.Lett.}
  {\bfseries B219} (1989) 240}.

\bibitem{Kandrup:1988sc}
H.E.~Kandrup, \emph{{Stochastic inflation as a time dependent random walk}},
  \href{https://doi.org/10.1103/PhysRevD.39.2245}{\emph{Phys.Rev.} {\bfseries
  D39} (1989) 2245}.

\bibitem{Nakao:1988yi}
K.-i.~Nakao, Y.~Nambu and M.~Sasaki, \emph{{Stochastic Dynamics of New
  Inflation}},
  \href{https://doi.org/10.1143/PTP.80.1041}{\emph{Prog.Theor.Phys.} {\bfseries
  80} (1988) 1041}.

\bibitem{Nambu:1989uf}
Y.~Nambu, \emph{{Stochastic Dynamics of an Inflationary Model and Initial
  Distribution of Universes}},
  \href{https://doi.org/10.1143/PTP.81.1037}{\emph{Prog.Theor.Phys.} {\bfseries
  81} (1989) 1037}.

\bibitem{Mollerach:1990zf}
S.~Mollerach, S.~Matarrese, A.~Ortolan and F.~Lucchin, \emph{{Stochastic
  inflation in a simple two field model}},
  \href{https://doi.org/10.1103/PhysRevD.44.1670}{\emph{Phys.Rev.} {\bfseries
  D44} (1991) 1670}.

\bibitem{Linde:1993xx}
A.D.~Linde, D.A.~Linde and A.~Mezhlumian, \emph{{From the Big Bang theory to
  the theory of a stationary universe}},
  \href{https://doi.org/10.1103/PhysRevD.49.1783}{\emph{Phys.Rev.} {\bfseries
  D49} (1994) 1783} [\href{https://arxiv.org/abs/gr-qc/9306035}{{\ttfamily
  gr-qc/9306035}}].

\bibitem{Starobinsky:1994bd}
A.A.~Starobinsky and J.~Yokoyama, \emph{{Equilibrium state of a selfinteracting
  scalar field in the De Sitter background}},
  \href{https://doi.org/10.1103/PhysRevD.50.6357}{\emph{Phys. Rev.} {\bfseries
  D50} (1994) 6357} [\href{https://arxiv.org/abs/astro-ph/9407016}{{\ttfamily
  astro-ph/9407016}}].

\bibitem{Finelli:2008zg}
F.~Finelli, G.~Marozzi, A.~Starobinsky, G.~Vacca and G.~Venturi,
  \emph{{Generation of fluctuations during inflation: Comparison of stochastic
  and field-theoretic approaches}},
  \href{https://doi.org/10.1103/PhysRevD.79.044007}{\emph{Phys.Rev.} {\bfseries
  D79} (2009) 044007} [\href{https://arxiv.org/abs/0808.1786}{{\ttfamily
  0808.1786}}].

\bibitem{Finelli:2010sh}
F.~Finelli, G.~Marozzi, A.~Starobinsky, G.~Vacca and G.~Venturi,
  \emph{{Stochastic growth of quantum fluctuations during slow-roll
  inflation}},
  \href{https://doi.org/10.1103/PhysRevD.82.064020}{\emph{Phys.Rev.} {\bfseries
  D82} (2010) 064020} [\href{https://arxiv.org/abs/1003.1327}{{\ttfamily
  1003.1327}}].

\bibitem{Burgess:2015ajz}
C.P.~Burgess, R.~Holman and G.~Tasinato, \emph{{Open EFTs, IR effects \&
  late-time resummations: systematic corrections in stochastic inflation}},
  \href{https://doi.org/10.1007/JHEP01(2016)153}{\emph{JHEP} {\bfseries 01}
  (2016) 153} [\href{https://arxiv.org/abs/1512.00169}{{\ttfamily
  1512.00169}}].

\bibitem{Morikawa:1989xz}
M.~Morikawa, \emph{{Dissipation and Fluctuation of Quantum Fields in Expanding
  Universes}}, \href{https://doi.org/10.1103/PhysRevD.42.1027}{\emph{Phys.
  Rev.} {\bfseries D42} (1990) 1027}.

\bibitem{Hu:1994dka}
B.L.~Hu and S.~Sinha, \emph{{A Fluctuation - dissipation relation for
  semiclassical cosmology}},
  \href{https://doi.org/10.1103/PhysRevD.51.1587}{\emph{Phys. Rev.} {\bfseries
  D51} (1995) 1587} [\href{https://arxiv.org/abs/gr-qc/9403054}{{\ttfamily
  gr-qc/9403054}}].

\bibitem{Matarrese:2003ye}
S.~Matarrese, M.A.~Musso and A.~Riotto, \emph{{Influence of superhorizon scales
  on cosmological observables generated during inflation}},
  \href{https://doi.org/10.1088/1475-7516/2004/05/008}{\emph{JCAP} {\bfseries
  0405} (2004) 008} [\href{https://arxiv.org/abs/hep-th/0311059}{{\ttfamily
  hep-th/0311059}}].

\bibitem{Grain:2017dqa}
J.~Grain and V.~Vennin, \emph{{Stochastic inflation in phase space: Is slow
  roll a stochastic attractor?}},
  \href{https://doi.org/10.1088/1475-7516/2017/05/045}{\emph{JCAP} {\bfseries
  1705} (2017) 045} [\href{https://arxiv.org/abs/1703.00447}{{\ttfamily
  1703.00447}}].

\bibitem{Pattison:2019hef}
C.~Pattison, V.~Vennin, H.~Assadullahi and D.~Wands, \emph{{Stochastic
  inflation beyond slow roll}},
  \href{https://doi.org/10.1088/1475-7516/2019/07/031}{\emph{JCAP} {\bfseries
  1907} (2019) 031} [\href{https://arxiv.org/abs/1905.06300}{{\ttfamily
  1905.06300}}].

\bibitem{thieman_book}
T.~Thiemann, \emph{{Modern Canonical Quantum General Relativity}}, Cambridge
  University Press (2008).

\bibitem{Langlois:1994ec}
D.~Langlois, \emph{{Hamiltonian formalism and gauge invariance for linear
  perturbations in inflation}},
  \href{https://doi.org/10.1088/0264-9381/11/2/011}{\emph{Class. Quant. Grav.}
  {\bfseries 11} (1994) 389}.

\bibitem{Lyth:2001nq}
D.H.~Lyth and D.~Wands, \emph{{Generating the curvature perturbation without an
  inflaton}}, \href{https://doi.org/10.1016/S0370-2693(01)01366-1}{\emph{Phys.
  Lett.} {\bfseries B524} (2002) 5}
  [\href{https://arxiv.org/abs/hep-ph/0110002}{{\ttfamily hep-ph/0110002}}].

\bibitem{birrell1982}
N.D.~Birrell and P.C.W.~Davies, \emph{Quantum Fields in Curved Space:},
  Cambridge University Press, Cambridge (002, 1982),
  \href{https://doi.org/10.1017/CBO9780511622632}{10.1017/CBO9780511622632}.

\bibitem{PhysRevD.46.2408}
S.~Habib, \emph{Stochastic inflation: Quantum phase-space approach},
  \href{https://doi.org/10.1103/PhysRevD.46.2408}{\emph{Phys. Rev. D}
  {\bfseries 46} (1992) 2408}.

\bibitem{Rigopoulos:2005xx}
G.I.~Rigopoulos, E.P.S.~Shellard and B.J.W.~van Tent, \emph{{Non-linear
  perturbations in multiple-field inflation}},
  \href{https://doi.org/10.1103/PhysRevD.73.083521}{\emph{Phys. Rev.}
  {\bfseries D73} (2006) 083521}
  [\href{https://arxiv.org/abs/astro-ph/0504508}{{\ttfamily
  astro-ph/0504508}}].

\bibitem{Tolley:2008na}
A.J.~Tolley and M.~Wyman, \emph{{Stochastic Inflation Revisited: Non-Slow Roll
  Statistics and DBI Inflation}},
  \href{https://doi.org/10.1088/1475-7516/2008/04/028}{\emph{JCAP} {\bfseries
  0804} (2008) 028} [\href{https://arxiv.org/abs/0801.1854}{{\ttfamily
  0801.1854}}].

\bibitem{Weenink:2011dd}
J.~Weenink and T.~Prokopec, \emph{{On decoherence of cosmological perturbations
  and stochastic inflation}},
  \href{https://arxiv.org/abs/1108.3994}{{\ttfamily 1108.3994}}.

\bibitem{risken1996fokker}
H.~Risken and T.~Frank, \emph{The Fokker-Planck Equation: Methods of Solution
  and Applications}, Springer Series in Synergetics, Springer Berlin Heidelberg
  (1996).

\bibitem{2005PhRvA..71b2103R}
M.~{Revzen}, P.A.~{Mello}, A.~{Mann} and L.M.~{Johansen}, \emph{{Bell's
  inequality violation with non-negative Wigner functions}},
  \href{https://doi.org/10.1103/PhysRevA.71.022103}{\emph{Phys.Rev.} {\bfseries
  A71} (2005) 022103} [\href{https://arxiv.org/abs/quant-ph/0405100}{{\ttfamily
  quant-ph/0405100}}].

\bibitem{Salopek:1990jq}
D.S.~Salopek and J.R.~Bond, \emph{{Nonlinear evolution of long wavelength
  metric fluctuations in inflationary models}},
  \href{https://doi.org/10.1103/PhysRevD.42.3936}{\emph{Phys. Rev.} {\bfseries
  D42} (1990) 3936}.

\bibitem{Sasaki:1995aw}
M.~Sasaki and E.D.~Stewart, \emph{{A General analytic formula for the spectral
  index of the density perturbations produced during inflation}},
  \href{https://doi.org/10.1143/PTP.95.71}{\emph{Prog.Theor.Phys.} {\bfseries
  95} (1996) 71} [\href{https://arxiv.org/abs/astro-ph/9507001}{{\ttfamily
  astro-ph/9507001}}].

\bibitem{Wands:2000dp}
D.~Wands, K.A.~Malik, D.H.~Lyth and A.R.~Liddle, \emph{{A New approach to the
  evolution of cosmological perturbations on large scales}},
  \href{https://doi.org/10.1103/PhysRevD.62.043527}{\emph{Phys. Rev.}
  {\bfseries D62} (2000) 043527}
  [\href{https://arxiv.org/abs/astro-ph/0003278}{{\ttfamily
  astro-ph/0003278}}].

\bibitem{Lyth:2003im}
D.H.~Lyth and D.~Wands, \emph{{Conserved cosmological perturbations}},
  \href{https://doi.org/10.1103/PhysRevD.68.103515}{\emph{Phys.Rev.} {\bfseries
  D68} (2003) 103515} [\href{https://arxiv.org/abs/astro-ph/0306498}{{\ttfamily
  astro-ph/0306498}}].

\bibitem{Rigopoulos:2003ak}
G.I.~Rigopoulos and E.P.S.~Shellard, \emph{{The separate universe approach and
  the evolution of nonlinear superhorizon cosmological perturbations}},
  \href{https://doi.org/10.1103/PhysRevD.68.123518}{\emph{Phys. Rev.}
  {\bfseries D68} (2003) 123518}
  [\href{https://arxiv.org/abs/astro-ph/0306620}{{\ttfamily
  astro-ph/0306620}}].

\bibitem{Lyth:2005fi}
D.H.~Lyth and Y.~Rodriguez, \emph{{The Inflationary prediction for primordial
  non-Gaussianity}},
  \href{https://doi.org/10.1103/PhysRevLett.95.121302}{\emph{Phys.Rev.Lett.}
  {\bfseries 95} (2005) 121302}
  [\href{https://arxiv.org/abs/astro-ph/0504045}{{\ttfamily
  astro-ph/0504045}}].

\bibitem{Lifshitz:1960}
E.M.~Lifshitz and I.M.~Khalatnikov, \emph{{About singularities of cosmological
  solutions of the gravitational equations. I}}, {\emph{ZhETF} {\bfseries 39}
  (1960) 149}.

\bibitem{Starobinsky:1982mr}
A.A.~Starobinsky, \emph{{Isotropization of arbitrary cosmological expansion
  given an effective cosmological constant}}, {\emph{JETP Lett.} {\bfseries 37}
  (1983) 66}.

\bibitem{Comer:1994np}
G.~Comer, N.~Deruelle, D.~Langlois and J.~Parry, \emph{{Growth or decay of
  cosmological inhomogeneities as a function of their equation of state}},
  \href{https://doi.org/10.1103/PhysRevD.49.2759}{\emph{Phys.Rev.} {\bfseries
  D49} (1994) 2759}.

\bibitem{Khalatnikov:2002kn}
I.~Khalatnikov, A.Y.~Kamenshchik and A.A.~Starobinsky, \emph{{Comment about
  quasiisotropic solution of Einstein equations near cosmological
  singularity}},
  \href{https://doi.org/10.1088/0264-9381/19/14/322}{\emph{Class.Quant.Grav.}
  {\bfseries 19} (2002) 3845}
  [\href{https://arxiv.org/abs/gr-qc/0204045}{{\ttfamily gr-qc/0204045}}].

\bibitem{Cruces:2018cvq}
D.~Cruces, C.~Germani and T.~Prokopec, \emph{{Failure of the stochastic
  approach to inflation in constant-roll and ultra-slow-roll}},
  \href{https://arxiv.org/abs/1807.09057}{{\ttfamily 1807.09057}}.

\bibitem{Gordon:2000hv}
C.~Gordon, D.~Wands, B.A.~Bassett and R.~Maartens, \emph{{Adiabatic and entropy
  perturbations from inflation}},
  \href{https://doi.org/10.1103/PhysRevD.63.023506}{\emph{Phys. Rev.}
  {\bfseries D63} (2001) 023506}
  [\href{https://arxiv.org/abs/astro-ph/0009131}{{\ttfamily
  astro-ph/0009131}}].

\bibitem{Malik:2008im}
K.A.~Malik and D.~Wands, \emph{{Cosmological perturbations}},
  \href{https://doi.org/10.1016/j.physrep.2009.03.001}{\emph{Phys. Rept.}
  {\bfseries 475} (2009) 1} [\href{https://arxiv.org/abs/0809.4944}{{\ttfamily
  0809.4944}}].

\bibitem{Sasaki:1986hm}
M.~Sasaki, \emph{{Large Scale Quantum Fluctuations in the Inflationary
  Universe}}, \href{https://doi.org/10.1143/PTP.76.1036}{\emph{Prog. Theor.
  Phys.} {\bfseries 76} (1986) 1036}.

\bibitem{Mukhanov:1988jd}
V.F.~Mukhanov, \emph{{Quantum Theory of Gauge Invariant Cosmological
  Perturbations}}, {\emph{Sov. Phys. JETP} {\bfseries 67} (1988) 1297}.

\bibitem{Fujita:2013cna}
T.~Fujita, M.~Kawasaki, Y.~Tada and T.~Takesako, \emph{{A new algorithm for
  calculating the curvature perturbations in stochastic inflation}},
  \href{https://doi.org/10.1088/1475-7516/2013/12/036}{\emph{JCAP} {\bfseries
  1312} (2013) 036} [\href{https://arxiv.org/abs/1308.4754}{{\ttfamily
  1308.4754}}].

\bibitem{Levasseur:2013tja}
L.~Perreault~Levasseur, V.~Vennin and R.~Brandenberger, \emph{{Recursive
  Stochastic Effects in Valley Hybrid Inflation}},
  \href{https://doi.org/10.1103/PhysRevD.88.083538}{\emph{Phys. Rev.}
  {\bfseries D88} (2013) 083538}
  [\href{https://arxiv.org/abs/1307.2575}{{\ttfamily 1307.2575}}].

\bibitem{Stewart:1993bc}
E.D.~Stewart and D.H.~Lyth, \emph{A more accurate analytic calculation of the
  spectrum of cosmological perturbations produced during inflation},
  {\emph{Phys. Lett.} {\bfseries B302} (1993) 171}
  [\href{https://arxiv.org/abs/gr-qc/9302019}{{\ttfamily gr-qc/9302019}}].

\bibitem{Linde:1996gt}
A.D.~Linde and V.F.~Mukhanov, \emph{{Nongaussian isocurvature perturbations
  from inflation}}, \href{https://doi.org/10.1103/PhysRevD.56.R535}{\emph{Phys.
  Rev.} {\bfseries D56} (1997) R535}
  [\href{https://arxiv.org/abs/astro-ph/9610219}{{\ttfamily
  astro-ph/9610219}}].

\bibitem{Enqvist:2001zp}
K.~Enqvist and M.S.~Sloth, \emph{{Adiabatic CMB perturbations in pre - big bang
  string cosmology}},
  \href{https://doi.org/10.1016/S0550-3213(02)00043-3}{\emph{Nucl. Phys.}
  {\bfseries B626} (2002) 395}
  [\href{https://arxiv.org/abs/hep-ph/0109214}{{\ttfamily hep-ph/0109214}}].

\bibitem{Moroi:2001ct}
T.~Moroi and T.~Takahashi, \emph{{Effects of cosmological moduli fields on
  cosmic microwave background}},
  \href{https://doi.org/10.1016/S0370-2693(02)02070-1,
  10.1016/S0370-2693(01)01295-3}{\emph{Phys. Lett.} {\bfseries B522} (2001)
  215} [\href{https://arxiv.org/abs/hep-ph/0110096}{{\ttfamily
  hep-ph/0110096}}].

\bibitem{Bartolo:2002vf}
N.~Bartolo and A.R.~Liddle, \emph{{The Simplest curvaton model}},
  \href{https://doi.org/10.1103/PhysRevD.65.121301}{\emph{Phys. Rev.}
  {\bfseries D65} (2002) 121301}
  [\href{https://arxiv.org/abs/astro-ph/0203076}{{\ttfamily
  astro-ph/0203076}}].

\bibitem{Vennin:2015vfa}
V.~Vennin, K.~Koyama and D.~Wands, \emph{{Encycloaedia curvatonis}},
  \href{https://doi.org/10.1088/1475-7516/2015/11/008}{\emph{JCAP} {\bfseries
  1511} (2015) 008} [\href{https://arxiv.org/abs/1507.07575}{{\ttfamily
  1507.07575}}].

\bibitem{Vennin:2015egh}
V.~Vennin, K.~Koyama and D.~Wands, \emph{{Inflation with an extra light scalar
  field after Planck}},
  \href{https://doi.org/10.1088/1475-7516/2016/03/024}{\emph{JCAP} {\bfseries
  1603} (2016) 024} [\href{https://arxiv.org/abs/1512.03403}{{\ttfamily
  1512.03403}}].

\bibitem{Assadullahi:2016gkk}
H.~Assadullahi, H.~Firouzjahi, M.~Noorbala, V.~Vennin and D.~Wands,
  \emph{{Multiple Fields in Stochastic Inflation}},
  \href{https://doi.org/10.1088/1475-7516/2016/06/043}{\emph{JCAP} {\bfseries
  1606} (2016) 043} [\href{https://arxiv.org/abs/1604.04502}{{\ttfamily
  1604.04502}}].

\bibitem{Vennin:2016wnk}
V.~Vennin, H.~Assadullahi, H.~Firouzjahi, M.~Noorbala and D.~Wands,
  \emph{{Critical Number of Fields in Stochastic Inflation}},
  \href{https://doi.org/10.1103/PhysRevLett.118.031301}{\emph{Phys. Rev. Lett.}
  {\bfseries 118} (2017) 031301}
  [\href{https://arxiv.org/abs/1604.06017}{{\ttfamily 1604.06017}}].

\bibitem{Sasaki:1998ug}
M.~Sasaki and T.~Tanaka, \emph{{Superhorizon scale dynamics of multiscalar
  inflation}},
  \href{https://doi.org/10.1143/PTP.99.763}{\emph{Prog.Theor.Phys.} {\bfseries
  99} (1998) 763} [\href{https://arxiv.org/abs/gr-qc/9801017}{{\ttfamily
  gr-qc/9801017}}].

\bibitem{Lyth:2004gb}
D.H.~Lyth, K.A.~Malik and M.~Sasaki, \emph{{A General proof of the conservation
  of the curvature perturbation}},
  \href{https://doi.org/10.1088/1475-7516/2005/05/004}{\emph{JCAP} {\bfseries
  0505} (2005) 004} [\href{https://arxiv.org/abs/astro-ph/0411220}{{\ttfamily
  astro-ph/0411220}}].

\bibitem{Creminelli:2004yq}
P.~Creminelli and M.~Zaldarriaga, \emph{{Single field consistency relation for
  the 3-point function}},
  \href{https://doi.org/10.1088/1475-7516/2004/10/006}{\emph{JCAP} {\bfseries
  0410} (2004) 006} [\href{https://arxiv.org/abs/astro-ph/0407059}{{\ttfamily
  astro-ph/0407059}}].

\bibitem{Enqvist:2008kt}
K.~Enqvist, S.~Nurmi, D.~Podolsky and G.I.~Rigopoulos, \emph{{On the
  divergences of inflationary superhorizon perturbations}},
  \href{https://doi.org/10.1088/1475-7516/2008/04/025}{\emph{JCAP} {\bfseries
  0804} (2008) 025} [\href{https://arxiv.org/abs/0802.0395}{{\ttfamily
  0802.0395}}].

\bibitem{Fujita:2014tja}
T.~Fujita, M.~Kawasaki and Y.~Tada, \emph{{Non-perturbative approach for
  curvature perturbations in stochastic $\delta N$ formalism}},
  \href{https://doi.org/10.1088/1475-7516/2014/10/030}{\emph{JCAP} {\bfseries
  1410} (2014) 030} [\href{https://arxiv.org/abs/1405.2187}{{\ttfamily
  1405.2187}}].

\bibitem{Kawasaki:2015ppx}
M.~Kawasaki and Y.~Tada, \emph{{Can massive primordial black holes be produced
  in mild waterfall hybrid inflation?}},
  \href{https://doi.org/10.1088/1475-7516/2016/08/041}{\emph{JCAP} {\bfseries
  1608} (2016) 041} [\href{https://arxiv.org/abs/1512.03515}{{\ttfamily
  1512.03515}}].

\bibitem{Risken:1984book}
H.~Risken, \emph{{The Fokker-Planck Equation}}, vol.~18, Spinger Series in
  Synergetics (1984).

\bibitem{Pinol:2018euk}
L.~Pinol, S.~Renaux-Petel and Y.~Tada, \emph{{Inflationary stochastic
  anomalies}}, \href{https://doi.org/10.1088/1361-6382/ab097f}{\emph{Class.
  Quant. Grav.} {\bfseries 36} (2019) 07LT01}
  [\href{https://arxiv.org/abs/1806.10126}{{\ttfamily 1806.10126}}].

\bibitem{ito1944}
K.~It\^o, \emph{Stochastic integral},
  \href{https://doi.org/10.3792/pia/1195572786}{\emph{Proceedings of the
  Imperial Academy} {\bfseries 20} (1944) 519}.

\bibitem{Noorbala:2018zlv}
M.~Noorbala, V.~Vennin, H.~Assadullahi, H.~Firouzjahi and D.~Wands,
  \emph{{Tunneling in Stochastic Inflation}},
  \href{https://arxiv.org/abs/1806.09634}{{\ttfamily 1806.09634}}.

\bibitem{Pattison:2017mbe}
C.~Pattison, V.~Vennin, H.~Assadullahi and D.~Wands, \emph{{Quantum diffusion
  during inflation and primordial black holes}},
  \href{https://doi.org/10.1088/1475-7516/2017/10/046}{\emph{JCAP} {\bfseries
  1710} (2017) 046} [\href{https://arxiv.org/abs/1707.00537}{{\ttfamily
  1707.00537}}].

\bibitem{Maldacena:2002vr}
J.M.~Maldacena, \emph{{Non-Gaussian features of primordial fluctuations in
  single field inflationary models}},
  \href{https://doi.org/10.1088/1126-6708/2003/05/013}{\emph{JHEP} {\bfseries
  05} (2003) 013} [\href{https://arxiv.org/abs/astro-ph/0210603}{{\ttfamily
  astro-ph/0210603}}].

\bibitem{Allen:2005ye}
L.E.~Allen, S.~Gupta and D.~Wands, \emph{{Non-gaussian perturbations from
  multi-field inflation}},
  \href{https://doi.org/10.1088/1475-7516/2006/01/006}{\emph{JCAP} {\bfseries
  0601} (2006) 006} [\href{https://arxiv.org/abs/astro-ph/0509719}{{\ttfamily
  astro-ph/0509719}}].

\bibitem{Martin:2011ib}
J.~Martin and V.~Vennin, \emph{{Stochastic Effects in Hybrid Inflation}},
  \href{https://doi.org/10.1103/PhysRevD.85.043525}{\emph{Phys. Rev.}
  {\bfseries D85} (2012) 043525}
  [\href{https://arxiv.org/abs/1110.2070}{{\ttfamily 1110.2070}}].

\bibitem{Linde:2005ht}
A.D.~Linde, \emph{{Particle physics and inflationary cosmology}},
  {\emph{Contemp.Concepts Phys.} {\bfseries 5} (1990) 1}
  [\href{https://arxiv.org/abs/hep-th/0503203}{{\ttfamily hep-th/0503203}}].

\bibitem{Smolin:1979ca}
L.~Smolin, \emph{{Gravitational Radiative Corrections as the Origin of
  Spontaneous Symmetry Breaking!}},
  \href{https://doi.org/10.1016/0370-2693(80)90103-3}{\emph{Phys.Lett.}
  {\bfseries B93} (1980) 95}.

\bibitem{Bardeen:1983st}
W.A.~Bardeen and M.~Moshe, \emph{{Phase Structure of the O(N) Vector Model}},
  \href{https://doi.org/10.1103/PhysRevD.28.1372}{\emph{Phys.Rev.} {\bfseries
  D28} (1983) 1372}.

\bibitem{Seery:2007we}
D.~Seery, \emph{{One-loop corrections to a scalar field during inflation}},
  \href{https://doi.org/10.1088/1475-7516/2007/11/025}{\emph{JCAP} {\bfseries
  0711} (2007) 025} [\href{https://arxiv.org/abs/0707.3377}{{\ttfamily
  0707.3377}}].

\bibitem{Dimastrogiovanni:2008af}
E.~Dimastrogiovanni and N.~Bartolo, \emph{{One-loop graviton corrections to the
  curvature perturbation from inflation}},
  \href{https://doi.org/10.1088/1475-7516/2008/11/016}{\emph{JCAP} {\bfseries
  0811} (2008) 016} [\href{https://arxiv.org/abs/0807.2790}{{\ttfamily
  0807.2790}}].

\bibitem{Seery:2010kh}
D.~Seery, \emph{{Infrared effects in inflationary correlation functions}},
  \href{https://doi.org/10.1088/0264-9381/27/12/124005}{\emph{Class.Quant.Grav.}
  {\bfseries 27} (2010) 124005}
  [\href{https://arxiv.org/abs/1005.1649}{{\ttfamily 1005.1649}}].

\bibitem{Byrnes:2007tm}
C.T.~Byrnes, K.~Koyama, M.~Sasaki and D.~Wands, \emph{{Diagrammatic approach to
  non-Gaussianity from inflation}},
  \href{https://doi.org/10.1088/1475-7516/2007/11/027}{\emph{JCAP} {\bfseries
  0711} (2007) 027} [\href{https://arxiv.org/abs/0705.4096}{{\ttfamily
  0705.4096}}].

\bibitem{Hardwick:2018sck}
R.J.~Hardwick, \emph{{Multiple spectator condensates from inflation}},
  \href{https://doi.org/10.1088/1475-7516/2018/05/054}{\emph{JCAP} {\bfseries
  1805} (2018) 054} [\href{https://arxiv.org/abs/1803.03521}{{\ttfamily
  1803.03521}}].

\bibitem{Malik:1998gy}
K.A.~Malik and D.~Wands, \emph{{Dynamics of assisted inflation}},
  \href{https://doi.org/10.1103/PhysRevD.59.123501}{\emph{Phys. Rev.}
  {\bfseries D59} (1999) 123501}
  [\href{https://arxiv.org/abs/astro-ph/9812204}{{\ttfamily
  astro-ph/9812204}}].

\bibitem{Saffin:2012et}
P.M.~Saffin, \emph{{The covariance of multi-field perturbations, pseudo-susy
  and fNL}}, \href{https://doi.org/10.1088/1475-7516/2012/09/002}{\emph{JCAP}
  {\bfseries 1209} (2012) 002}
  [\href{https://arxiv.org/abs/1203.0397}{{\ttfamily 1203.0397}}].

\bibitem{Steinhardt:1982kg}
P.J.~Steinhardt, \emph{{NATURAL INFLATION}},  in \emph{{Nuffield Workshop on
  the Very Early Universe Cambridge, England, June 21-July 9, 1982}},
  pp.~251--266, 1982.

\bibitem{Vilenkin:1983xq}
A.~Vilenkin, \emph{{The Birth of Inflationary Universes}},
  \href{https://doi.org/10.1103/PhysRevD.27.2848}{\emph{Phys. Rev.} {\bfseries
  D27} (1983) 2848}.

\bibitem{Guth:1985ya}
A.H.~Guth and S.-Y.~Pi, \emph{{The Quantum Mechanics of the Scalar Field in the
  New Inflationary Universe}},
  \href{https://doi.org/10.1103/PhysRevD.32.1899}{\emph{Phys. Rev.} {\bfseries
  D32} (1985) 1899}.

\bibitem{Linde:1986fc}
A.D.~Linde, \emph{{ETERNAL CHAOTIC INFLATION}},
  \href{https://doi.org/10.1142/S0217732386000129}{\emph{Mod. Phys. Lett.}
  {\bfseries A1} (1986) 81}.

\bibitem{Barenboim:2016mmw}
G.~Barenboim, W.-I.~Park and W.H.~Kinney, \emph{{Eternal Hilltop Inflation}},
  \href{https://doi.org/10.1088/1475-7516/2016/05/030}{\emph{JCAP} {\bfseries
  1605} (2016) 030} [\href{https://arxiv.org/abs/1601.08140}{{\ttfamily
  1601.08140}}].

\bibitem{Polya:1921}
G.~P\'olya, \emph{{\"Uber eine aufgabe betreffend die irrfahrt im
  strassennetz}}, {\emph{Math. Ann.} {\bfseries 84} (1921) 149–160}.

\bibitem{Polarski:1992dq}
D.~Polarski and A.A.~Starobinsky, \emph{{Spectra of perturbations produced by
  double inflation with an intermediate matter dominated stage}},
  \href{https://doi.org/10.1016/0550-3213(92)90062-G}{\emph{Nucl. Phys.}
  {\bfseries B385} (1992) 623}.

\bibitem{Ringeval:2010hf}
C.~Ringeval, T.~Suyama, T.~Takahashi, M.~Yamaguchi and S.~Yokoyama, \emph{{Dark
  energy from primordial inflationary quantum fluctuations}},
  \href{https://doi.org/10.1103/PhysRevLett.105.121301}{\emph{Phys. Rev. Lett.}
  {\bfseries 105} (2010) 121301}
  [\href{https://arxiv.org/abs/1006.0368}{{\ttfamily 1006.0368}}].

\bibitem{Enqvist:2012xn}
K.~Enqvist, R.N.~Lerner, O.~Taanila and A.~Tranberg, \emph{{Spectator field
  dynamics in de Sitter and curvaton initial conditions}},
  \href{https://doi.org/10.1088/1475-7516/2012/10/052}{\emph{JCAP} {\bfseries
  1210} (2012) 052} [\href{https://arxiv.org/abs/1205.5446}{{\ttfamily
  1205.5446}}].

\bibitem{Kitajima:2019ibn}
N.~Kitajima, Y.~Tada and F.~Takahashi, \emph{{Stochastic inflation with an
  extremely large number of $e$-folds}},
  \href{https://arxiv.org/abs/1908.08694}{{\ttfamily 1908.08694}}.

\bibitem{Ade:2015xua}
{\scshape Planck} collaboration, \emph{{Planck 2015 results. XIII. Cosmological
  parameters}},
  \href{https://doi.org/10.1051/0004-6361/201525830}{\emph{Astron. Astrophys.}
  {\bfseries 594} (2016) A13}
  [\href{https://arxiv.org/abs/1502.01589}{{\ttfamily 1502.01589}}].

\bibitem{Broy:2014sia}
B.J.~Broy, D.~Roest and A.~Westphal, \emph{{Power Spectrum of Inflationary
  Attractors}}, \href{https://doi.org/10.1103/PhysRevD.91.023514}{\emph{Phys.
  Rev.} {\bfseries D91} (2015) 023514}
  [\href{https://arxiv.org/abs/1408.5904}{{\ttfamily 1408.5904}}].

\bibitem{Coone:2015fha}
D.~Coone, D.~Roest and V.~Vennin, \emph{{The Hubble Flow of Plateau
  Inflation}}, \href{https://doi.org/10.1088/1475-7516/2015/11/010}{\emph{JCAP}
  {\bfseries 1511} (2015) 010}
  [\href{https://arxiv.org/abs/1507.00096}{{\ttfamily 1507.00096}}].

\bibitem{Hardwick:2017qcw}
R.J.~Hardwick, V.~Vennin and D.~Wands, \emph{{A Quantum Window Onto Early
  Inflation}}, \href{https://doi.org/10.1142/S0218271817430258}{\emph{Int. J.
  Mod. Phys.} {\bfseries D26} (2017) 1743025}
  [\href{https://arxiv.org/abs/1705.05746}{{\ttfamily 1705.05746}}].

\bibitem{Dvali:2003ar}
G.~Dvali, A.~Gruzinov and M.~Zaldarriaga, \emph{{Cosmological perturbations
  from inhomogeneous reheating, freezeout, and mass domination}},
  \href{https://doi.org/10.1103/PhysRevD.69.083505}{\emph{Phys. Rev.}
  {\bfseries D69} (2004) 083505}
  [\href{https://arxiv.org/abs/astro-ph/0305548}{{\ttfamily
  astro-ph/0305548}}].

\bibitem{Abbott:1984fp}
L.F.~Abbott and M.B.~Wise, \emph{{Constraints on Generalized Inflationary
  Cosmologies}},
  \href{https://doi.org/10.1016/0550-3213(84)90329-8}{\emph{Nucl. Phys.}
  {\bfseries B244} (1984) 541}.

\bibitem{Ade:2015lrj}
{\scshape Planck} collaboration, \emph{{Planck 2015 results. XX. Constraints on
  inflation}},  \href{https://arxiv.org/abs/1502.02114}{{\ttfamily
  1502.02114}}.

\bibitem{Hawking:1971ei}
S.~Hawking, \emph{{Gravitationally collapsed objects of very low mass}},
  {\emph{Mon. Not. Roy. Astron. Soc.} {\bfseries 152} (1971) 75}.

\bibitem{Carr:1975qj}
B.J.~Carr, \emph{{The Primordial black hole mass spectrum}},
  \href{https://doi.org/10.1086/153853}{\emph{Astrophys. J.} {\bfseries 201}
  (1975) 1}.

\bibitem{Zaballa:2006kh}
I.~Zaballa, A.M.~Green, K.A.~Malik and M.~Sasaki, \emph{{Constraints on the
  primordial curvature perturbation from primordial black holes}},
  \href{https://doi.org/10.1088/1475-7516/2007/03/010}{\emph{JCAP} {\bfseries
  0703} (2007) 010} [\href{https://arxiv.org/abs/astro-ph/0612379}{{\ttfamily
  astro-ph/0612379}}].

\bibitem{Harada:2013epa}
T.~Harada, C.-M.~Yoo and K.~Kohri, \emph{{Threshold of primordial black hole
  formation}}, \href{https://doi.org/10.1103/PhysRevD.88.084051,
  10.1103/PhysRevD.89.029903}{\emph{Phys. Rev.} {\bfseries D88} (2013) 084051}
  [\href{https://arxiv.org/abs/1309.4201}{{\ttfamily 1309.4201}}].

\bibitem{Young:2014ana}
S.~Young, C.T.~Byrnes and M.~Sasaki, \emph{{Calculating the mass fraction of
  primordial black holes}},
  \href{https://doi.org/10.1088/1475-7516/2014/07/045}{\emph{JCAP} {\bfseries
  1407} (2014) 045} [\href{https://arxiv.org/abs/1405.7023}{{\ttfamily
  1405.7023}}].

\bibitem{Blais:2002gw}
D.~Blais, T.~Bringmann, C.~Kiefer and D.~Polarski, \emph{{Accurate results for
  primordial black holes from spectra with a distinguished scale}},
  \href{https://doi.org/10.1103/PhysRevD.67.024024}{\emph{Phys. Rev.}
  {\bfseries D67} (2003) 024024}
  [\href{https://arxiv.org/abs/astro-ph/0206262}{{\ttfamily
  astro-ph/0206262}}].

\bibitem{Ando:2018qdb}
K.~Ando, K.~Inomata and M.~Kawasaki, \emph{{Primordial black holes and
  uncertainties in the choice of the window function}},
  \href{https://doi.org/10.1103/PhysRevD.97.103528}{\emph{Phys. Rev.}
  {\bfseries D97} (2018) 103528}
  [\href{https://arxiv.org/abs/1802.06393}{{\ttfamily 1802.06393}}].

\bibitem{Musco:2004ak}
I.~Musco, J.C.~Miller and L.~Rezzolla, \emph{{Computations of primordial black
  hole formation}},
  \href{https://doi.org/10.1088/0264-9381/22/7/013}{\emph{Class. Quant. Grav.}
  {\bfseries 22} (2005) 1405}
  [\href{https://arxiv.org/abs/gr-qc/0412063}{{\ttfamily gr-qc/0412063}}].

\bibitem{Polnarev:2006aa}
A.G.~Polnarev and I.~Musco, \emph{{Curvature profiles as initial conditions for
  primordial black hole formation}},
  \href{https://doi.org/10.1088/0264-9381/24/6/003}{\emph{Class. Quant. Grav.}
  {\bfseries 24} (2007) 1405}
  [\href{https://arxiv.org/abs/gr-qc/0605122}{{\ttfamily gr-qc/0605122}}].

\bibitem{Musco:2018rwt}
I.~Musco, \emph{{Threshold for primordial black holes: Dependence on the shape
  of the cosmological perturbations}},
  \href{https://doi.org/10.1103/PhysRevD.100.123524}{\emph{Phys. Rev.}
  {\bfseries D100} (2019) 123524}
  [\href{https://arxiv.org/abs/1809.02127}{{\ttfamily 1809.02127}}].

\bibitem{Kalaja:2019uju}
A.~Kalaja, N.~Bellomo, N.~Bartolo, D.~Bertacca, S.~Matarrese, I.~Musco et~al.,
  \emph{{From Primordial Black Holes Abundance to Primordial Curvature Power
  Spectrum (and back)}},
  \href{https://doi.org/10.1088/1475-7516/2019/10/031}{\emph{JCAP} {\bfseries
  1910} (2019) 031} [\href{https://arxiv.org/abs/1908.03596}{{\ttfamily
  1908.03596}}].

\bibitem{Casini:1998wr}
H.~Casini, R.~Montemayor and P.~Sisterna, \emph{{Stochastic approach to
  inflation. 2. Classicality, coarse graining and noises}},
  \href{https://doi.org/10.1103/PhysRevD.59.063512}{\emph{Phys. Rev.}
  {\bfseries D59} (1999) 063512}
  [\href{https://arxiv.org/abs/gr-qc/9811083}{{\ttfamily gr-qc/9811083}}].

\bibitem{Winitzki:1999ve}
S.~Winitzki and A.~Vilenkin, \emph{{Effective noise in stochastic description
  of inflation}}, \href{https://doi.org/10.1103/PhysRevD.61.084008}{\emph{Phys.
  Rev.} {\bfseries D61} (2000) 084008}
  [\href{https://arxiv.org/abs/gr-qc/9911029}{{\ttfamily gr-qc/9911029}}].

\bibitem{Liguori:2004fa}
M.~Liguori, S.~Matarrese, M.~Musso and A.~Riotto, \emph{{Stochastic inflation
  and the lower multipoles in the CMB anisotropies}},
  \href{https://doi.org/10.1088/1475-7516/2004/08/011}{\emph{JCAP} {\bfseries
  0408} (2004) 011} [\href{https://arxiv.org/abs/astro-ph/0405544}{{\ttfamily
  astro-ph/0405544}}].

\bibitem{1975ApJ...201....1C}
B.J.~{Carr}, \emph{{The primordial black hole mass spectrum}},
  \href{https://doi.org/10.1086/153853}{\emph{ApJ} {\bfseries 201} (1975) 1}.

\bibitem{Choptuik:1992jv}
M.W.~Choptuik, \emph{{Universality and scaling in gravitational collapse of a
  massless scalar field}},
  \href{https://doi.org/10.1103/PhysRevLett.70.9}{\emph{Phys. Rev. Lett.}
  {\bfseries 70} (1993) 9}.

\bibitem{Niemeyer:1997mt}
J.C.~Niemeyer and K.~Jedamzik, \emph{{Near-critical gravitational collapse and
  the initial mass function of primordial black holes}},
  \href{https://doi.org/10.1103/PhysRevLett.80.5481}{\emph{Phys. Rev. Lett.}
  {\bfseries 80} (1998) 5481}
  [\href{https://arxiv.org/abs/astro-ph/9709072}{{\ttfamily
  astro-ph/9709072}}].

\bibitem{Kuhnel:2015vtw}
F.~Kuhnel, C.~Rampf and M.~Sandstad, \emph{{Effects of Critical Collapse on
  Primordial Black-Hole Mass Spectra}},
  \href{https://doi.org/10.1140/epjc/s10052-016-3945-8}{\emph{Eur. Phys. J.}
  {\bfseries C76} (2016) 93}
  [\href{https://arxiv.org/abs/1512.00488}{{\ttfamily 1512.00488}}].

\bibitem{Mukhanov:1985rz}
V.F.~Mukhanov, \emph{{Gravitational Instability of the Universe Filled with a
  Scalar Field}}, {\emph{JETP Lett.} {\bfseries 41} (1985) 493}.

\bibitem{Young:2015cyn}
S.~Young, D.~Regan and C.T.~Byrnes, \emph{{Influence of large local and
  non-local bispectra on primordial black hole abundance}},
  \href{https://doi.org/10.1088/1475-7516/2016/02/029}{\emph{JCAP} {\bfseries
  1602} (2016) 029} [\href{https://arxiv.org/abs/1512.07224}{{\ttfamily
  1512.07224}}].

\bibitem{Olver:2010:NHM:1830479:theta}
F.W.~Olver, D.W.~Lozier, R.F.~Boisvert and C.W.~Clark, \emph{Ch. 20 in NIST
  Handbook of Mathematical Functions}, Cambridge University Press, New York,
  NY, USA, 1st~ed. (2010).

\bibitem{Abramovitz:1970aa:theta}
M.~Abramowitz and I.A.~Stegun, \emph{Ch. 16 in Handbook of mathematical
  functions with formulas, graphs, and mathematical tables}, National Bureau of
  Standards, Washington, US, ninth~ed. (1970).

\bibitem{NIST:DLMF}
``{NIST Digital Library of Mathematical Functions}.''
  \url{http://dlmf.nist.gov/20.7E30}, Release 1.0.14 of 2016-12-21.

\bibitem{Young:2013oia}
S.~Young and C.T.~Byrnes, \emph{{Primordial black holes in non-Gaussian
  regimes}}, \href{https://doi.org/10.1088/1475-7516/2013/08/052}{\emph{JCAP}
  {\bfseries 1308} (2013) 052}
  [\href{https://arxiv.org/abs/1307.4995}{{\ttfamily 1307.4995}}].

\bibitem{GarciaBellido:1996qt}
J.~Garcia-Bellido, A.D.~Linde and D.~Wands, \emph{{Density perturbations and
  black hole formation in hybrid inflation}},
  \href{https://doi.org/10.1103/PhysRevD.54.6040}{\emph{Phys. Rev.} {\bfseries
  D54} (1996) 6040} [\href{https://arxiv.org/abs/astro-ph/9605094}{{\ttfamily
  astro-ph/9605094}}].

\bibitem{Linde:1991km}
A.D.~Linde, \emph{{Axions in inflationary cosmology}},
  \href{https://doi.org/10.1016/0370-2693(91)90130-I}{\emph{Phys. Lett.}
  {\bfseries B259} (1991) 38}.

\bibitem{Linde:1993cn}
A.D.~Linde, \emph{{Hybrid inflation}},
  \href{https://doi.org/10.1103/PhysRevD.49.748}{\emph{Phys.Rev.} {\bfseries
  D49} (1994) 748} [\href{https://arxiv.org/abs/astro-ph/9307002}{{\ttfamily
  astro-ph/9307002}}].

\bibitem{Copeland:1994vg}
E.J.~Copeland, A.R.~Liddle, D.H.~Lyth, E.D.~Stewart and D.~Wands, \emph{{False
  vacuum inflation with Einstein gravity}},
  \href{https://doi.org/10.1103/PhysRevD.49.6410}{\emph{Phys. Rev.} {\bfseries
  D49} (1994) 6410} [\href{https://arxiv.org/abs/astro-ph/9401011}{{\ttfamily
  astro-ph/9401011}}].

\bibitem{Renaux-Petel:2015mga}
S.~Renaux-Petel and K.~Turzy?ski, \emph{{Geometrical Destabilization of
  Inflation}},
  \href{https://doi.org/10.1103/PhysRevLett.117.141301}{\emph{Phys. Rev. Lett.}
  {\bfseries 117} (2016) 141301}
  [\href{https://arxiv.org/abs/1510.01281}{{\ttfamily 1510.01281}}].

\bibitem{Renaux-Petel:2017dia}
S.~Renaux-Petel, K.~Turzyński and V.~Vennin, \emph{{Geometrical
  destabilization, premature end of inflation and Bayesian model selection}},
  \href{https://arxiv.org/abs/1706.01835}{{\ttfamily 1706.01835}}.

\bibitem{Akrami:2019izv}
{\scshape Planck} collaboration, \emph{{Planck 2018 results. IX. Constraints on
  primordial non-Gaussianity}},
  \href{https://arxiv.org/abs/1905.05697}{{\ttfamily 1905.05697}}.

\bibitem{Byrnes:2012yx}
C.T.~Byrnes, E.J.~Copeland and A.M.~Green, \emph{{Primordial black holes as a
  tool for constraining non-Gaussianity}},
  \href{https://doi.org/10.1103/PhysRevD.86.043512}{\emph{Phys. Rev.}
  {\bfseries D86} (2012) 043512}
  [\href{https://arxiv.org/abs/1206.4188}{{\ttfamily 1206.4188}}].

\bibitem{Garcia-Bellido:2016dkw}
J.~Garcia-Bellido, M.~Peloso and C.~Unal, \emph{{Gravitational waves at
  interferometer scales and primordial black holes in axion inflation}},
  \href{https://doi.org/10.1088/1475-7516/2016/12/031}{\emph{JCAP} {\bfseries
  1612} (2016) 031} [\href{https://arxiv.org/abs/1610.03763}{{\ttfamily
  1610.03763}}].

\bibitem{Garcia-Bellido:2017aan}
J.~Garcia-Bellido, M.~Peloso and C.~Unal, \emph{{Gravitational Wave signatures
  of inflationary models from Primordial Black Hole Dark Matter}},
  \href{https://doi.org/10.1088/1475-7516/2017/09/013}{\emph{JCAP} {\bfseries
  1709} (2017) 013} [\href{https://arxiv.org/abs/1707.02441}{{\ttfamily
  1707.02441}}].

\bibitem{Ezquiaga:2018gbw}
J.M.~Ezquiaga and J.~Garcia-Bellido, \emph{{Quantum diffusion beyond slow-roll:
  implications for primordial black-hole production}},
  \href{https://doi.org/10.1088/1475-7516/2018/08/018}{\emph{JCAP} {\bfseries
  1808} (2018) 018} [\href{https://arxiv.org/abs/1805.06731}{{\ttfamily
  1805.06731}}].

\bibitem{Franciolini:2018vbk}
G.~Franciolini, A.~Kehagias, S.~Matarrese and A.~Riotto, \emph{{Primordial
  Black Holes from Inflation and non-Gaussianity}},
  \href{https://doi.org/10.1088/1475-7516/2018/03/016}{\emph{JCAP} {\bfseries
  1803} (2018) 016} [\href{https://arxiv.org/abs/1801.09415}{{\ttfamily
  1801.09415}}].

\bibitem{Cai:2018dig}
R.-g.~Cai, S.~Pi and M.~Sasaki, \emph{{Gravitational Waves Induced by
  non-Gaussian Scalar Perturbations}},
  \href{https://doi.org/10.1103/PhysRevLett.122.201101}{\emph{Phys. Rev. Lett.}
  {\bfseries 122} (2019) 201101}
  [\href{https://arxiv.org/abs/1810.11000}{{\ttfamily 1810.11000}}].

\bibitem{Passaglia:2018ixg}
S.~Passaglia, W.~Hu and H.~Motohashi, \emph{{Primordial black holes and local
  non-Gaussianity in canonical inflation}},
  \href{https://doi.org/10.1103/PhysRevD.99.043536}{\emph{Phys. Rev.}
  {\bfseries D99} (2019) 043536}
  [\href{https://arxiv.org/abs/1812.08243}{{\ttfamily 1812.08243}}].

\bibitem{Young:2019yug}
S.~Young, I.~Musco and C.T.~Byrnes, \emph{{Primordial black hole formation and
  abundance: contribution from the non-linear relation between the density and
  curvature perturbation}},
  \href{https://doi.org/10.1088/1475-7516/2019/11/012}{\emph{JCAP} {\bfseries
  1911} (2019) 012} [\href{https://arxiv.org/abs/1904.00984}{{\ttfamily
  1904.00984}}].

\bibitem{DeLuca:2019qsy}
V.~De~Luca, G.~Franciolini, A.~Kehagias, M.~Peloso, A.~Riotto and C.~Unal,
  \emph{{The Ineludible non-Gaussianity of the Primordial Black Hole
  Abundance}}, \href{https://doi.org/10.1088/1475-7516/2019/07/048}{\emph{JCAP}
  {\bfseries 1907} (2019) 048}
  [\href{https://arxiv.org/abs/1904.00970}{{\ttfamily 1904.00970}}].

\bibitem{Panagopoulos:2019ail}
G.~Panagopoulos and E.~Silverstein, \emph{{Primordial Black Holes from
  non-Gaussian tails}},  \href{https://arxiv.org/abs/1906.02827}{{\ttfamily
  1906.02827}}.

\bibitem{Yoo:2019pma}
C.-M.~Yoo, J.-O.~Gong and S.~Yokoyama, \emph{{Abundance of primordial black
  holes with local non-Gaussianity in peak theory}},
  \href{https://doi.org/10.1088/1475-7516/2019/09/033}{\emph{JCAP} {\bfseries
  1909} (2019) 033} [\href{https://arxiv.org/abs/1906.06790}{{\ttfamily
  1906.06790}}].

\bibitem{Carr:2019hud}
B.~Carr, S.~Clesse and J.~Garcia-Bellido, \emph{{Primordial black holes, dark
  matter and hot-spot electroweak baryogenesis at the quark-hadron epoch}},
  \href{https://arxiv.org/abs/1904.02129}{{\ttfamily 1904.02129}}.

\bibitem{Gangui:1993tt}
A.~Gangui, F.~Lucchin, S.~Matarrese and S.~Mollerach, \emph{{The Three point
  correlation function of the cosmic microwave background in inflationary
  models}}, \href{https://doi.org/10.1086/174421}{\emph{Astrophys. J.}
  {\bfseries 430} (1994) 447}
  [\href{https://arxiv.org/abs/astro-ph/9312033}{{\ttfamily
  astro-ph/9312033}}].

\bibitem{Ezquiaga:2019ftu}
J.M.~Ezquiaga, J.~Garcia-Bellido and V.~Vennin, \emph{{The exponential tail of
  inflationary fluctuations: consequences for primordial black holes}},
  \href{https://arxiv.org/abs/1912.05399}{{\ttfamily 1912.05399}}.

\bibitem{Pattison:2018bct}
C.~Pattison, V.~Vennin, H.~Assadullahi and D.~Wands, \emph{{The attractive
  behaviour of ultra-slow-roll inflation}},
  \href{https://doi.org/10.1088/1475-7516/2018/08/048}{\emph{JCAP} {\bfseries
  1808} (2018) 048} [\href{https://arxiv.org/abs/1806.09553}{{\ttfamily
  1806.09553}}].

\bibitem{Garcia-Bellido:2017mdw}
J.~Garcia-Bellido and E.~Ruiz~Morales, \emph{{Primordial black holes from
  single field models of inflation}},
  \href{https://doi.org/10.1016/j.dark.2017.09.007}{\emph{Phys. Dark Univ.}
  {\bfseries 18} (2017) 47} [\href{https://arxiv.org/abs/1702.03901}{{\ttfamily
  1702.03901}}].

\bibitem{Ezquiaga:2017fvi}
J.M.~Ezquiaga, J.~Garcia-Bellido and E.~Ruiz~Morales, \emph{{Primordial Black
  Hole production in Critical Higgs Inflation}},
  \href{https://doi.org/10.1016/j.physletb.2017.11.039}{\emph{Phys. Lett.}
  {\bfseries B776} (2018) 345}
  [\href{https://arxiv.org/abs/1705.04861}{{\ttfamily 1705.04861}}].

\bibitem{Seto:1999jc}
O.~Seto, J.~Yokoyama and H.~Kodama, \emph{{What happens when the inflaton stops
  during inflation}},
  \href{https://doi.org/10.1103/PhysRevD.61.103504}{\emph{Phys. Rev.}
  {\bfseries D61} (2000) 103504}
  [\href{https://arxiv.org/abs/astro-ph/9911119}{{\ttfamily
  astro-ph/9911119}}].

\bibitem{Leach:2001zf}
S.M.~Leach, M.~Sasaki, D.~Wands and A.R.~Liddle, \emph{{Enhancement of
  superhorizon scale inflationary curvature perturbations}},
  \href{https://doi.org/10.1103/PhysRevD.64.023512}{\emph{Phys. Rev.}
  {\bfseries D64} (2001) 023512}
  [\href{https://arxiv.org/abs/astro-ph/0101406}{{\ttfamily
  astro-ph/0101406}}].

\bibitem{Kinney:2005vj}
W.H.~Kinney, \emph{{Horizon crossing and inflation with large eta}},
  \href{https://doi.org/10.1103/PhysRevD.72.023515}{\emph{Phys. Rev.}
  {\bfseries D72} (2005) 023515}
  [\href{https://arxiv.org/abs/gr-qc/0503017}{{\ttfamily gr-qc/0503017}}].

\bibitem{Wands:1998yp}
D.~Wands, \emph{{Duality invariance of cosmological perturbation spectra}},
  \href{https://doi.org/10.1103/PhysRevD.60.023507}{\emph{Phys. Rev.}
  {\bfseries D60} (1999) 023507}
  [\href{https://arxiv.org/abs/gr-qc/9809062}{{\ttfamily gr-qc/9809062}}].

\bibitem{Biagetti:2018pjj}
M.~Biagetti, G.~Franciolini, A.~Kehagias and A.~Riotto, \emph{{Primordial Black
  Holes from Inflation and Quantum Diffusion}},
  \href{https://doi.org/10.1088/1475-7516/2018/07/032}{\emph{JCAP} {\bfseries
  1807} (2018) 032} [\href{https://arxiv.org/abs/1804.07124}{{\ttfamily
  1804.07124}}].

\bibitem{Cai:2017bxr}
Y.-F.~Cai, X.~Chen, M.H.~Namjoo, M.~Sasaki, D.-G.~Wang and Z.~Wang,
  \emph{{Revisiting non-Gaussianity from non-attractor inflation models}},
  \href{https://doi.org/10.1088/1475-7516/2018/05/012}{\emph{JCAP} {\bfseries
  1805} (2018) 012} [\href{https://arxiv.org/abs/1712.09998}{{\ttfamily
  1712.09998}}].

\bibitem{Dimopoulos:2017ged}
K.~Dimopoulos, \emph{{Ultra slow-roll inflation demystified}},
  \href{https://doi.org/10.1016/j.physletb.2017.10.066}{\emph{Phys. Lett.}
  {\bfseries B775} (2017) 262}
  [\href{https://arxiv.org/abs/1707.05644}{{\ttfamily 1707.05644}}].

\bibitem{Anguelova:2017djf}
L.~Anguelova, P.~Suranyi and L.C.R.~Wijewardhana, \emph{{Systematics of
  Constant Roll Inflation}},
  \href{https://doi.org/10.1088/1475-7516/2018/02/004}{\emph{JCAP} {\bfseries
  1802} (2018) 004} [\href{https://arxiv.org/abs/1710.06989}{{\ttfamily
  1710.06989}}].

\bibitem{Morse:2018kda}
M.J.P.~Morse and W.H.~Kinney, \emph{{Large-$\eta$ Constant-Roll Inflation Is
  Never An Attractor}},  \href{https://arxiv.org/abs/1804.01927}{{\ttfamily
  1804.01927}}.

\bibitem{Martin:2012pe}
J.~Martin, H.~Motohashi and T.~Suyama, \emph{{Ultra Slow-Roll Inflation and the
  non-Gaussianity Consistency Relation}},
  \href{https://doi.org/10.1103/PhysRevD.87.023514}{\emph{Phys.Rev.} {\bfseries
  D87} (2013) 023514} [\href{https://arxiv.org/abs/1211.0083}{{\ttfamily
  1211.0083}}].

\bibitem{Motohashi:2014ppa}
H.~Motohashi, A.A.~Starobinsky and J.~Yokoyama, \emph{{Inflation with a
  constant rate of roll}},
  \href{https://doi.org/10.1088/1475-7516/2015/09/018}{\emph{JCAP} {\bfseries
  1509} (2015) 018} [\href{https://arxiv.org/abs/1411.5021}{{\ttfamily
  1411.5021}}].

\bibitem{Yi:2017mxs}
Z.~Yi and Y.~Gong, \emph{{On the constant-roll inflation}},
  \href{https://doi.org/10.1088/1475-7516/2018/03/052}{\emph{JCAP} {\bfseries
  1803} (2018) 052} [\href{https://arxiv.org/abs/1712.07478}{{\ttfamily
  1712.07478}}].

\bibitem{Starobinsky:1992ts}
A.A.~Starobinsky, \emph{{Spectrum of adiabatic perturbations in the universe
  when there are singularities in the inflation potential}}, {\emph{JETP Lett.}
  {\bfseries 55} (1992) 489}.

\bibitem{Martin:2011sn}
J.~Martin and L.~Sriramkumar, \emph{{The scalar bi-spectrum in the Starobinsky
  model: The equilateral case}},
  \href{https://doi.org/10.1088/1475-7516/2012/01/008}{\emph{JCAP} {\bfseries
  1201} (2012) 008} [\href{https://arxiv.org/abs/1109.5838}{{\ttfamily
  1109.5838}}].

\bibitem{Germani:2017bcs}
C.~Germani and T.~Prokopec, \emph{{On primordial black holes from an inflection
  point}}, \href{https://doi.org/10.1016/j.dark.2017.09.001}{\emph{Phys. Dark
  Univ.} {\bfseries 18} (2017) 6}
  [\href{https://arxiv.org/abs/1706.04226}{{\ttfamily 1706.04226}}].

\bibitem{Firouzjahi:2018vet}
H.~Firouzjahi, A.~Nassiri-Rad and M.~Noorbala, \emph{{Stochastic Ultra Slow
  Roll Inflation}},
  \href{https://doi.org/10.1088/1475-7516/2019/01/040}{\emph{JCAP} {\bfseries
  1901} (2019) 040} [\href{https://arxiv.org/abs/1811.02175}{{\ttfamily
  1811.02175}}].

\bibitem{10.2307/3213641}
L.M.~Ricciardi, L.~Sacerdote and S.~Sato, \emph{On an integral equation for
  first-passage-time probability densities}, {\emph{Journal of Applied
  Probability} {\bfseries 21} (1984) 302}.

\bibitem{Tada:2016pmk}
Y.~Tada and V.~Vennin, \emph{{Squeezed bispectrum in the $\delta N$ formalism:
  local observer effect in field space}},
  \href{https://doi.org/10.1088/1475-7516/2017/02/021}{\emph{JCAP} {\bfseries
  1702} (2017) 021} [\href{https://arxiv.org/abs/1609.08876}{{\ttfamily
  1609.08876}}].

\bibitem{Chen:2016kjx}
J.-W.~Chen, J.~Liu, H.-L.~Xu and Y.-F.~Cai, \emph{{Tracing Primordial Black
  Holes in Nonsingular Bouncing Cosmology}},
  \href{https://doi.org/10.1016/j.physletb.2017.03.036}{\emph{Phys. Lett.}
  {\bfseries B769} (2017) 561}
  [\href{https://arxiv.org/abs/1609.02571}{{\ttfamily 1609.02571}}].

\bibitem{Quintin:2016qro}
J.~Quintin and R.H.~Brandenberger, \emph{{Black hole formation in a contracting
  universe}}, \href{https://doi.org/10.1088/1475-7516/2016/11/029}{\emph{JCAP}
  {\bfseries 1611} (2016) 029}
  [\href{https://arxiv.org/abs/1609.02556}{{\ttfamily 1609.02556}}].

\end{thebibliography}\endgroup

\label{this page}
\clearpage

\end{document}